\documentclass[a4paper,10pt]{article}
\usepackage[utf8]{inputenc}

\usepackage[T1]{fontenc}
\usepackage{textcomp}
\usepackage{fancyhdr}
\usepackage{fullpage}
\usepackage{lmodern}
\usepackage{amsmath,amssymb,bm,bbm}
\usepackage{algorithm}
\usepackage{algorithmic}
\usepackage{float}
\usepackage{graphicx}
\usepackage{extarrows}
\usepackage{caption}
\usepackage{graphicx, subfig}
\usepackage{titling}
\usepackage{yhmath}
\usepackage{mathrsfs}
\usepackage{footnote}

\usepackage{amsthm}
\usepackage{color}
\usepackage{slashed}
\usepackage{verbatim}
\usepackage[colorlinks=false,linkcolor=blue,citecolor=blue]{hyperref}
\usepackage{amsfonts,euscript}
\usepackage{amssymb, multicol}
\usepackage{epstopdf}
\usepackage{psfrag}
\usepackage{xcolor}
\usepackage{comment}
\usepackage{authblk}
\usepackage{indentfirst}

\usepackage{lipsum}

\usepackage[
backend=biber,
style=numeric,
maxcitenames=5,
maxbibnames=9]{biblatex}

\renewbibmacro{in:}{}

\DeclareFontFamily{U}{mathx}{\hyphenchar\font45}
\DeclareFontShape{U}{mathx}{m}{n}{
	<5> <6> <7> <8> <9> <10>
	<10.95> <12> <14.4> <17.28> <20.74> <24.88>
	mathx10
}{}
\DeclareSymbolFont{mathx}{U}{mathx}{m}{n}
\DeclareFontSubstitution{U}{mathx}{m}{n}
\DeclareMathAccent{\widecheck}{0}{mathx}{"71}

\numberwithin{equation}{subsection}
\newtheorem{thm}{Theorem}[subsection]
\newtheorem{prop}[thm]{Proposition}
\newtheorem{lem}[thm]{Lemma}
\newtheorem{cor}[thm]{Corollary}
\newtheorem{defi}[thm]{Definition}

\newtheorem*{conjecture}{Conjecture}

\newtheorem{rem}[thm]{Remark}

\newcommand{\mcw}{\mathcal{W}}
\newcommand{\Jplus}{J^{(+)}}
\newcommand{\Jmoins}{J^{(-)}}
\newcommand{\Jzero}{J^{(0)}}
\newcommand{\Bbar}{\underline{B}}

\newcommand{\curl}{\operatorname{curl}}
\newcommand{\lambdabar}{\underline{\lambda}}

\newcommand{\Mcheck}{\widecheck{M}}
\newcommand{\Z}{\mathbf{Z}}
\newcommand{\mca}{\mathcal{A}}
\newcommand{\nablaslash}{\nabla\hspace{-7.8pt}/\hspace{3pt}}

\newcommand{\shl}[2]{#1\hspace{-#2pt}/\:}
\newcommand{\mcv}{\mathcal{V}}
\newcommand{\Hbar}{\underline{H}}
\newcommand{\Xbar}{\underline{X}}
\newcommand{\Xibar}{\underline{\Xi}}
\newcommand{\xibar}{\underline{\xi}}
\newcommand{\diver}{\operatorname{div}}
\newcommand{\barre}[1]{#1\hspace{-5.3pt}/}
\newcommand{\barred}[1]{#1\hspace{-5.8pt}/}

\newcommand{\uring}{\mathring{u}}
\newcommand{\ubarring}{\mathring{\ubar}}
\newcommand{\mcb}{\mathcal{B}}

\newcommand{\T}{\mathbf{T}}
\newcommand{\Ener}{\mathbf{E}_{deg}}
\newcommand{\hatpsim}{{\Psi}_m}

\newcommand{\lieT}{\barre{\mcl}_\mathbf{T}}
\newcommand{\df}{\mathfrak{d}}
\newcommand{\Ybar}{\underline{Y}}
\newcommand{\gbarc}{\widecheck{\gbar}}
\newcommand{\LN}[1]{\lieT^{j_{#1}}\nabring^{i_{#1}}}

\newcommand{\lambdacheck}{\widecheck{\lambda}}
\newcommand{\gslash}{g\hspace{-5pt}/}
\newcommand{\lie}{\barre{\mcl}}
\newcommand{\atrchi}{{}^{(a)}tr\chi}
\newcommand{\atrchibar}{{}^{(a)}tr\chibar}

\newcommand{\mcf}{\mathcal{F}}
\newcommand{\g}{{\mathbf{g}}}
\newcommand{\qbar}{\overline{q}}

\newcommand{\hodge}[1]{{}^*\!#1}
\newcommand{\wh}{\widehat}
\newcommand{\frakJ}{\mathfrak{J}}
\newcommand{\vol}{\operatorname{Vol}_\mathbf{g}}
\newcommand{\wt}{\widetilde}
\newcommand{\mcn}{\mathcal{N}}
\newcommand{\fbar}{\underline{f}}

\newcommand{\Abar}{\underline{A}}

\newcommand{\mct}{\mathcal{T}}

\newcommand{\nn}{\nonumber}
\newcommand{\fc}{\widecheck{f}}

\newcommand{\D}{\mathbf{D}}
\newcommand{\divc}{\overline{\mcd}\cdot}

\newcommand{\parentheses}[1]{\left( #1 \right)}

\newcommand{\hhbar}{\underline{h}}

\newcommand{\Ldeux}[1]{\|#1\|_{L^2(S(u,\ubar))}}

\newcommand{\intS}{\int_{S(u,\ubar)}}

\newcommand{\hot}{\widehat{\otimes}}

\newcommand{\err}{\mathrm{Err}}

\newcommand{\Real}{\mathfrak{R}}
\newcommand{\Imag}{\mathfrak{I}}
\newcommand{\atr}{{}^{(a)}tr}
\newcommand{\gbar}{\underline{g}}

\newcommand{\mch}{\mathcal{H}}
\newcommand{\mcm}{\mathcal{M}}
\newcommand{\mck}{\mathcal{K}}
\newcommand{\mcq}{\mathcal{Q}}
\newcommand{\mcu}{\mathcal{U}}

\newcommand{\chihat}{\wh{\chi}}
\newcommand{\Deltahat}{\hat{\Delta}}

\newcommand{\mcd}{\mathcal{D}}
\newcommand{\mcr}{\mathcal{R}}
\newcommand{\dgras}{\mathbf{D}}

\newcommand{\chibar}{\underline{\chi}}
\newcommand{\etabar}{\underline{\eta}}

\newcommand{\wbar}{\underline{w}}

\newcommand{\ubar}{{\underline{u}}}

\newcommand{\lambdabarcheck}{\widecheck{\lambdabar}}
\newcommand{\mci}{\mathcal{I}}

\newcommand{\C}{\mathbb C}

\newcommand{\ering}{\mathring{e}}
\newcommand{\nabring}{\mathring{\nabla}}

\newcommand{\betabar}{\underline{\beta}}

\newcommand{\omegabar}{\underline{\omega}}
\newcommand{\alphabar}{\underline{\alpha}}

\newcommand{\gcheck}{\widecheck{g}}
\newcommand{\gammacheck}{\widecheck{\gamma}}
\newcommand{\logomegacheck}{\widecheck{\log\Omega}}
\newcommand{\bcheck}{\widecheck{b}}
\newcommand{\psicheck}{\widecheck{\psi}}
\newcommand{\etabarcheck}{\widecheck{\etabar}}
\newcommand{\etacheck}{\widecheck{\eta}}
\newcommand{\zetacheck}{\widecheck{\zeta}}

\newcommand{\betacheck}{\widecheck{\beta}}
\newcommand{\betabarcheck}{\widecheck{\betabar}}
\newcommand{\Kcheck}{\widecheck{K}}

\newcommand{\mubar}{\underline{\mu}}
\newcommand{\mucheck}{\widecheck{\mu}}
\newcommand{\mubarcheck}{\widecheck{\mubar}}

\newcommand{\dee}{\mathrm{d}}

\newcommand{\ch}{\mathcal{CH_+}}
\newcommand{\eh}{\mathcal{H_+}}

\newcommand{\mco}{\mathcal{O}}
\newcommand{\mcl}{\mathcal{L}}

\newcommand{\drond}{\mathring{\eth}}

\newcommand{\Psihat}{\widehat{\psi}}
\newcommand{\psihat}{\widehat{\psi}}

\newcommand{\R}{\mathbb{R}}
\newcommand{\Gammacheck}{\widecheck{\Gamma}}
\newcommand{\Rcheck}{\widecheck{R}}
\newcommand{\Eun}{\mathbf{E}_{\un}}

\newcommand{\un}{\mathbf{I}}
\newcommand{\deux}{\mathbf{II}}
\newcommand{\trois}{\mathbf{III}}

\newcommand{\psip}{\psi_{+2}}
\newcommand{\poids}{\Deltahat\varpi^N}

\newcommand{\Psihatp}{\Psihat_{+2}}

\newcommand{\Psihatm}{\Psihat_{-2}}

\newcommand{\Fhat}{\hat{F}}

\newcommand{\fraks}{\mathfrak{s}}

\title{Non-linear instability of the Kerr Cauchy horizon near $i_+$}

\author{Sebastian Gurriaran\footnote{Email adress : sebastian.gurriaran@sorbonne-universite.fr}\\
	\textit{Laboratoire Jacques-Louis Lions, Sorbonne Universit\'e, 75005 Paris, France}}

\begin{document}
	
	\maketitle
	\begin{abstract}
We consider solutions of the Einstein vacuum equations which arise from smooth initial data on a hypersurface slightly inside a dynamical black hole settling down to a subextremal Kerr black hole, and satisfying a precise non-linear Price's law-type estimate (which we expect to hold generically). We prove that the corresponding maximal globally hyperbolic development admits a non-trivial piece of future null boundary - the Cauchy horizon - emanating from timelike infinity $i_+$, which exhibits a kind of curvature blow-up, and across which the spacetime metric is Lipschitz-inextendible. Our results thus imply a Lipschitz version of Strong Cosmic Censorship for Kerr spacetimes near timelike infinity under this Price's law-type assumption.

The analysis relies on the proof of the $C^0$ stability of the Kerr Cauchy horizon by Dafermos and Luk \cite{stabC0}, on the non-integrable formalism of Giorgi-Klainerman-Szeftel \cite{KSwaveeq} and principal temporal gauge of Klainerman and Szeftel \cite{KS21} used in the proof of the exterior stability of slowly rotating Kerr black holes, on the linearized analysis for the Teukolsky equation inside subextremal Kerr black holes by the author \cite{spin+2} (following an earlier paper on the scalar wave equation by Ma and Zhang \cite{scalarMZ}), and on Sbierski's criterion for Lipschitz inextendibility \cite{sbierskiinextdernier}. More precisely, we proceed by decomposing the black hole interior into different regions equipped with appropriate gauges, allowing for a proof of stability estimates and a thorough analysis of the non-linear analog of the Teukolsky equation, from which we infer our instability results.
	\end{abstract}

	\tableofcontents
	
\section{Introduction}

The stability of black holes as solutions to the Einstein vacuum equations, 
\begin{align}\tag{1}\label{eq:EVE}
	\mathbf{Ric}[\g]=0,
\end{align}
where $\mathbf{Ric}[\g]$ is the Ricci tensor for a Lorentzian metric $\g$, is a central problem in general relativity. In particular, the explicit solution discovered by Roy Kerr \cite{kerrmetric}, called the Kerr metric, is the mathematical model for rotating black holes in the absence of matter, and is expected to describe the final state of generic spacetimes governed by Einstein's equations. The stability of the exterior region of slowly rotating Kerr black holes having recently been proven in the series of works \cite{GCMKS, GCMKS2, dawei, KSwaveeq, KS21}, a natural question concerns the evolution of perturbations which cross the event horizon and enter the black hole region. 

In the present paper, we pursue this direction by investigating the properties of the interior of Kerr black holes perturbations (without restricting to the slowly rotating case), and we prove the instability of the Kerr Cauchy horizon in the Lipschitz regularity class under the assumption of a precise non-linear Price's law-type estimate.

\subsection{Einstein equations and Strong Cosmic Censorship conjecture}
\subsubsection{The Cauchy problem for the Einstein equations}
The Einstein vacuum equations \eqref{eq:EVE} form an evolution problem and hence admit an initial value formulation. The corresponding initial data are given by a triplet $(\Sigma_0,g,k)$, where $(\Sigma_0,g)$ is a Riemannian 3-manifold and where $k$ is a symmetric 2-tensor such that the following {constraint equations} are satisfied:
$$R_g+(tr k)^2-|k|^2=0,\quad \diver k-\overline{\nabla} tr k=0,$$
where $R_g$ is the scalar curvature of $g$, $\overline{\nabla}$ is the covariant derivative of $(\Sigma_0,g)$, and the divergence, trace, and tensor norm above are taken with respect to $g$. 

The landmark papers \cite{YCB52,BG69} by Choquet-Bruhat and Choquet-Bruhat--Geroch ensure that for any smooth initial data $(\Sigma_0,g,k)$ satisfying the constraint equations, there exists a unique smooth maximal globally hyperbolic development $(\mcm,\g)$ solution of the Einstein equations \eqref{eq:EVE}, such that the embedding $\Sigma\subset\mcm$ has induced metric $g$ and second fundamental form $k$.
\subsubsection{The Strong Cosmic Censorship conjecture}\label{section:sccintro}

Some classical solutions of Einstein's equations present a pathological breakdown of determinism, for instance in the interior of Kerr and Reissner-Nordström black holes. More precisely, for these solutions the maximal globally hyperbolic development is bounded in the future by a null hypersurface called \emph{Cauchy horizon}, across which the metric admits infinitely many smooth extensions satisfying Einstein's equations. As a consequence, these spacetimes are not predictable, since the fate of an observer crossing the Cauchy horizon is not determined by its past. Penrose's Strong Cosmic Censorship (SCC) conjecture \cite{sccpenrose}, which may be stated as follows in its modern version (see \cite{christo, chruscc}, and \cite{sccreview} for a recent review on SCC), was formulated as an attempt to save determinism in general relativity.

\begin{conjecture}[SCC conjecture]\label{conj:scc}
	The maximal globally hyperbolic development of generic initial data for the Einstein equations is inextendible as a suitably regular Lorentzian manifold.
\end{conjecture}

The formulation of SCC above states that the failure of determinism found for instance inside Kerr black holes is \emph{non-generic and thus should disappear upon small perturbations}. In particular physically realistic black holes, which correspond to perturbations of Kerr in the vacuum case, should respect determinism. This suggests that the Cauchy horizon inside Kerr black holes is unstable in some sense: generic perturbations coming from the exterior should build up in the interior of the black hole and blow-up at the Cauchy horizon, thereby preventing the unphysical non-unique extensions.

\subsection{Kerr black holes and their internal structure}

\subsubsection{The Kerr metric and the Cauchy horizon}
The subextremal Kerr metrics form a 2-parameter family $(\g_{a,M})$ of stationary and axisymmetric solutions to the Einstein vacuum equations \eqref{eq:EVE}, where $M$ is the mass of the black hole and $a$ is the angular momentum per unit mass, such that $|a|<M$. We further restrict to the range $0<|a|<M$ in this paper. In standard Boyer-Lindquist coordinates $(t,r,\theta,\phi)\in\R^2\times\mathbb{S}^2$,
$$\g_{a,M}=\frac{a^2\sin^2\theta-\Delta}{|q|^2}\dee t^2-\frac{4aMr}{|q|^2}\sin^2\theta\dee t\dee\phi+\frac{|q|^2}{\Delta}\dee r^2+|q|^2\dee\theta^2+\frac{(r^2+a^2)^2-a^2\sin^2\theta\Delta}{|q|^2}\sin^2\theta\dee\phi^2,$$
where, denoting $i$ the imaginary unit and $r_\pm:=M\pm\sqrt{M^2-a^2}$,
\begin{align*}
	\Delta:=r^2+a^2-2Mr=(r-r_+)(r-r_-),\quad\quad q:=r+ia\cos\theta.
\end{align*}
\begin{figure}[h!]
	\centering
	\includegraphics[scale=0.43]{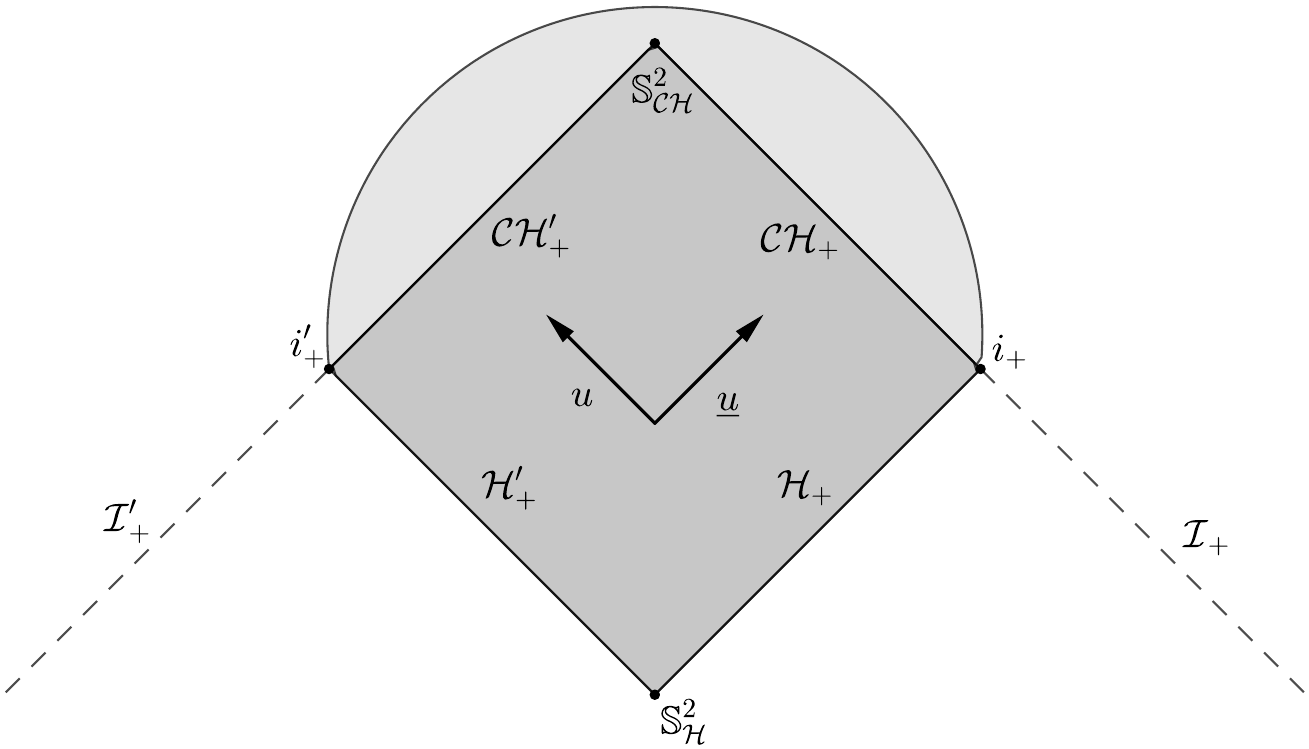}
	\caption{The Penrose diagram of Kerr spacetimes with $0<|a|<M$, with the exterior region in white, the black hole interior in the darker shaded grey region, and the region with infinitely many extensions in the lighter shaded grey region.}
	\label{fig:penrosediagram}
\end{figure}

Denoting $u,\ubar$ the standard Eddington-Finkelstein retarded and advanced time adapted to the black hole interior (see \eqref{eq:defuubarEFkerr}), the event horizon is the bifurcate null hypersurface $\mch=\mch_+\cup\mch_+'$ where
$$\mch_+=\{u=-\infty\},\quad\quad\mch'_+=\{\ubar=-\infty\},$$
and corresponds to the level set $\{r=r_+\}$, while the Cauchy horizon is the bifurcate null hypersurface $\ch\cup\mathcal{CH}'_+$ where 
$$\ch=\{\ubar=+\infty\},\quad\quad\mathcal{CH}'_+=\{u=+\infty\},$$
and corresponds to the level set $\{r=r_-\}$, see Figure \ref{fig:penrosediagram} for the Penrose diagram of Kerr spacetime. {In particular the scalar $\Delta$ vanishes at both the event and Cauchy horizon.} We denote $\text{Kerr}(a,M)$ the Kerr spacetime with black hole parameters $a,M$.
\subsubsection{Conjectured instability of the Kerr Cauchy horizon and blueshift effect}\label{section:blueshift}
\begin{figure}[h!]
	\centering
	\includegraphics[scale=0.5]{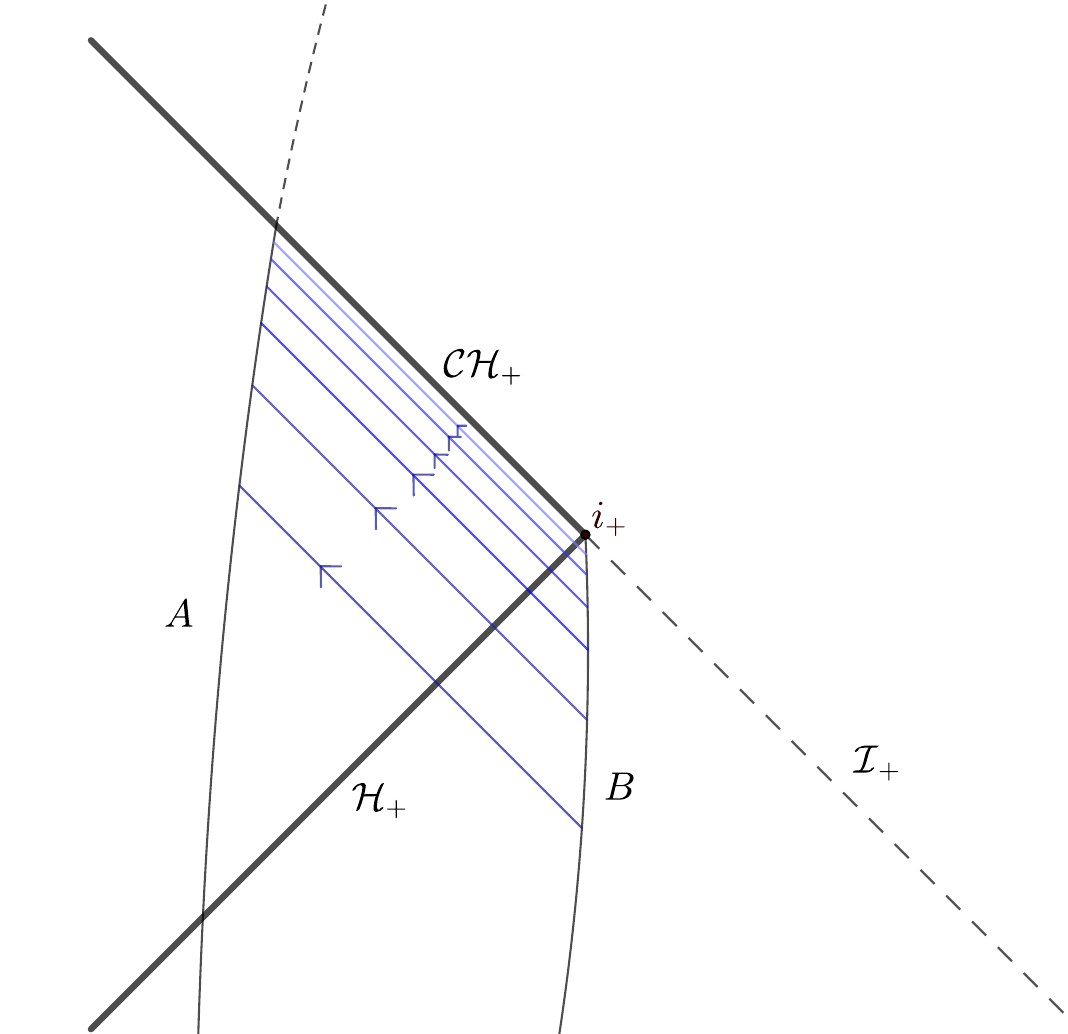}
	\caption{The blueshift effect. Observer $B$ lives outside of the black hole and reaches timelike infinity in infinite proper time, while sending signals to observer $A$ that crosses the event horizon and then the Cauchy horizon in finite proper time. The waves are sent periodically by $B$, but as $A$ approaches $\ch$, the perceived frequency of the waves becomes infinite.}
	\label{fig:blueshift}
\end{figure}

As mentioned in Section \ref{section:sccintro}, the Kerr spacetime is globally hyperbolic only up to the Cauchy horizon since beyond it (in the lighter shaded grey region in Figure \ref{fig:penrosediagram}) there are infinitely many smooth extensions of the Kerr metric which satisfy Einstein's equations. These extensions are however non-physical, and SCC conjecture suggests that for generic perturbations of Kerr black holes the Cauchy horizon becomes singular in a way which prevents such extensions. The heuristic mechanism responsible for this expected instability goes back to Penrose \cite{blueshiftpenrose}, and is called the \emph{blueshift effect}. Roughly stated, the blueshift effect is a consequence of the Kerr spacetime geometry: all radiations entering the black hole must accumulate in finite time at the Cauchy horizon, generically giving rise to the formation of a singularity, see Figure \ref{fig:blueshift}.

This blueshift instability was initially studied in the linearized setting, for solutions of the scalar wave equation $\Box_\g\psi=0$ on fixed stationary black hole backgrounds. In this model, the scalar $\psi$ represents the metric and the wave equation is seen as a linearized version of the Einstein equations in coordinates. In many works in Kerr and Reissner-Nordström (see Section \ref{section:litteraturewaveeq}), it is typically observed that $\psi$ itself remains bounded up to the Cauchy horizon while its derivative \emph{transverse} to the Cauchy horizon blows up. By transferring this heuristic to the non-linear setting, this suggests that the Cauchy horizon inside black holes is $C^0$ stable and that the instability happens at a {higher regularity level} for the metric, giving rise to a so-called \emph{weak null singularity} at the Cauchy horizon, see \cite{weakluk}.   

\begin{conjecture}[Instability of the Kerr Cauchy horizon]
	For generic initial data, perturbations of Kerr black holes governed by the Einstein vacuum equations are inextendible across the (perturbed) Cauchy horizon in a suitable regularity class.
\end{conjecture}

The $C^0$ \emph{stability} of the Kerr Cauchy horizon was proved by Dafermos and Luk \cite{stabC0}, thereby definitely ensuring that the $C^0$ formulation of SCC conjecture is false for Kerr spacetimes, in accordance with the results for the scalar wave equation. The present paper proves a $C^{0,1}_{loc}$ (Lipschitz) version of SCC conjecture for Kerr spacetimes in a neighborhood of timelike infinity\footnote{As in \cite{stabC0}, our results are restricted to a piece of Cauchy horizon emanating from timelike infinity. A complete SCC statement for Kerr spacetimes requires to understand the instability properties of the full Cauchy horizon.}, under a Price's law-type assumption.

\subsection{The Teukolsky equations and Price's law-type results}
\subsubsection{The Teukolsky equations}
The \emph{Teukolsky equations}, discovered by Teukolsky \cite{teukolsky}, play a central role in the study of perturbations of Kerr black holes. These are wave equations satisfied by extremal curvature components with respect to a null frame which is aligned with some special principal null directions of the Kerr spacetime. The Teukolsky equations can be derived directly from the Einstein equations, and their main property is the complete decoupling at the linearized level of the extremal curvature components from the other Ricci and curvature coefficients, relative to the principal null frame. 

In order to write down the Teukolsky equations in Kerr, we introduce the following null pair
$$n:=\frac{r^2+a^2}{2|q|^2}\partial_t-\frac{\Delta}{2|q|^2}\partial_r+\frac{a}{2|q|^2}\partial_\phi,\quad l:=\frac{r^2+a^2}{\Delta}\partial_t+\partial_r+\frac{a}{\Delta}\partial_\phi,$$
which is aligned with the principal null directions of Kerr, and the complex valued vectorfield
$$m:=\frac{1}{\sqrt{2}(r+ia\cos\theta)}\left(\partial_\theta+i\left(\frac{1}{\sin\theta}\partial_\phi+a\sin\theta\partial_t\right)\right).$$
Then, in a linearized perturbation of Kerr, if $\dot{\mathbf{R}}$ is the linearized curvature tensor, the scalars 
\begin{align}\label{eq:teukolskyscalars}
	\Psihatp:=\dot{\mathbf{R}}_{lmlm},\quad\widehat{\psi}_{-2}:=(r-ia\cos\theta)^4\dot{\mathbf{R}}_{n\overline{m}n\overline{m}},
\end{align}
satisfy the Teukolsky equations which write in Boyer-Lindquist coordinates, for $s=\pm2$,
\begin{align*}
	& -\left[\frac{\left(r^2+a^2\right)^2}{\Delta}-a^2 \sin ^2 \theta\right] \partial_t^2\psihat_s -\frac{4 M a r}{\Delta}\partial_t\partial_\phi\psihat_s  -\left[\frac{a^2}{\Delta}-\frac{1}{\sin ^2 \theta}\right] \partial_\phi^2\psihat_s \nonumber \\
	& +\Delta^{-s} \partial_r\left(\Delta^{s+1} \partial_r \psihat_s\right)+\frac{1}{\sin \theta} \partial_\theta\left(\sin \theta \partial_\theta \psihat_s\right)+2 s\left[\frac{a(r-M)}{\Delta}+\frac{i \cos \theta}{\sin ^2 \theta}\right] \partial_\phi\psihat_s  \nonumber\\
	& +2 s\left[\frac{M\left(r^2-a^2\right)}{\Delta}-r-i a \cos \theta\right]\partial_t\psihat_s -\left[\frac{s^2 \cos ^2 \theta}{\sin ^2 \theta}-s\right]\psihat_s=0.
\end{align*}
The Teukolsky equations can be generalized to study non-linear perturbations of Kerr. They were originally introduced to study the stability of the exterior region of black hole spacetimes, see for instance \cite{schlinDHR,mablueandersson,Ma:energymorawetz} in the linearized setting and \cite{schpolarized,schNLDHRT,KSwaveeq} in the non-linear setting. In the present paper, the non-linear analog of the $s=+2$ Teukolsky equation is used to study the instability properties of the interior of Kerr black holes.

 \subsubsection{Price's law for Teukolsky in Kerr}\label{section:pricelaw1}
 
The behavior of non-linear perturbations inside Kerr black holes are intimately connected to the precise polynomial decay rates of perturbations near their event horizons. Such sharp decay rates go under the name of Price's law and are discussed below, emphasizing on the case of Teukolsky in Kerr which is most relevant to the present paper.

Price's law was first stated in the 1972 paper \cite{price2} by Richard H. Price, see also \cite{price3,price4,price1,barackori} for additional results in the physics literature. It is a general statement for solutions to wave equations on black hole spacetimes, asserting that suitably regular solutions arising from localized initial data decay at late times at an inverse polynomial rate along the event horizon. 

 In the case of a fixed background Kerr black hole, a precise version of Price's law for solutions to the Teukolsky equation was proven by Ma and Zhang \cite{pricelaw}\footnote{The version of Price's law stated in that paper holds in the slowly rotating case $|a|\ll M$ and for $|a|<M$ conditionally on an energy-Morawetz type bound, which was later proven by Teixeira da Costa and Shlapentokh-Rothman in \cite{TDCSR2,TDCSR1}.}, see also the work of Millet \cite{millet} in the whole subextremal range. For initially smooth and compactly supported initial data, recalling \eqref{eq:teukolskyscalars} and denoting
$$\psi_{+2}:=\Delta^2\Psihatp,\quad \psi_{-2}:=\Delta^{-2}\widehat{\psi}_{-2},$$
the Price's law stated in \cite{pricelaw} writes
\begin{equation}\label{eq:ansatzlineaires}
	\begin{aligned}
		\psi_{+2}&=\frac{1}{\ubar^7}\sum_{|m|\leq 2}A_m(r)Q_mY_{m,2}^{+2}(\cos\theta)e^{im\phi_+}+O(\ubar^{-7-\delta}),\\
		\psi_{-2}&=\frac{1}{\ubar^7}\sum_{|m|\leq 2}Q_mY_{m,2}^{-2}(\cos\theta)e^{im\phi_+}+O(\ubar^{-7-\delta}),
	\end{aligned}
\end{equation}
on the Kerr event horizon $\mch_+$, where $\delta>0$, $A_m(r)$ is an explicit function (see \eqref{eq:dedfAm(r)}), $Q_m$ are some complex constants which depend on the initial data and are generically non-zero, and $Y_{m,2}^{\pm2}(\cos\theta)e^{im\phi_+}$ are the $(\ell=2,m)$ spin $\pm 2$ spherical harmonics ($\phi_+$ is a renormalization of the coordinate $\phi$ which is regular on $\mch_+$). For a complete review of results related to Price's law, we refer to the introduction in \cite{pricelaw}.

\subsubsection{Expected Price's law-type results for non-linear perturbations}\label{section:assumptionpricelaw}

Concerning the non-linear setting, a general correction to Price's law for wave equations due to non-linear effects was noticed heuristically and numerically in \cite{Bizon} by Bizon and Rostworowski. There has been recent progress on rigorous non-linear Price's law-type results, see \cite{lukohpricenl} by Luk and Oh and \cite{gajickehrb} by Gajic and Kehrberger. Both papers observe that on dynamical black hole backgrounds, the late time tails of solutions to wave equations are in general different from the ones in the stationary case. In particular, their results suggest that for solutions to the Teukolsky equations, the Price's law decay $\ubar^{-7}$ of the linearized setting should be corrected to yield a generic $\ubar^{-6}$ decay near the event horizon of dynamical black holes settling down to a member of the Kerr family, see for example the discussion on page 7 of \cite{lukohpricenl}.

We thus expect that for sufficiently fast decaying initial data, near the event horizon of a dynamical vacuum black hole settling down to a Kerr solution, the ansatz for $\psip$ in \eqref{eq:ansatzlineaires} should be modified as follows
\begin{align}\label{eq:expectedpricelaw}
	\psi_{+2}\sim\frac{1}{\ubar^6}\sum_{|m|\leq 2}A_m(r)Q_{m}Y_{m,2}^{+2}(\cos\theta)e^{im\phi_+},
\end{align}
for some generically non-zero constants $(Q_m)_{|m|\leq 2}$ which depend on the initial data. Moreover, the works \cite{KS21, KSwaveeq} on the exterior stability of slowly rotating Kerr spacetimes naturally provide estimates on a region which slightly penetrates the interior of the black hole, up to a spacelike hypersurface $\mca$ slightly beyond the event horizon. This hypersurface corresponds to a level set $\mca=\{r=r_+(1-\delta_+)\}$ for some small $\delta_+>0$.

\textbf{This discussion motivates us to assume in the present work a precise version of Price's law of the type \eqref{eq:expectedpricelaw} on a hypersurface $\mca=\{r=r_+(1-\delta_+)\}$ slightly inside a dynamical vacuum black hole settling down to a fixed Kerr black hole.}

We expect that such an assumption should hold with $Q_m\neq 0$ for generic initial perturbations, see also Remark \ref{rem:furtherdisc} for further comments.

\subsection{Previous results on perturbations of stationary black hole interiors}

\subsubsection{The wave equation inside Reissner-Nordström and Kerr black holes}\label{section:litteraturewaveeq}

The simplest perturbation model is the scalar wave equation $\Box_\g\psi=0$. The blow-up of the derivative of $\psi$ transverse to the Kerr and Reissner-Nördstrom Cauchy horizons for some well-behaved initial data was studied in \cite{mac} by McNamara. A heuristic power law for scalar waves inside Kerr black holes was then found by Ori in \cite{oriscalar}. The boundedness of scalar waves inside Reissner-Nordström black holes was later rigorously proven by Franzen in \cite{franzen}. This result was then extended to the case of slowly rotating Kerr black hole interiors in \cite{hintzkerrwave} by Hintz, which was then further extended to the full subextremal range in \cite{franzen2} by Franzen. The blow-up of the energy of generic scalar waves at the Reissner-Nordström Cauchy horizon was proven in \cite{RNscalar} by Luk and Oh, and then extended to Kerr black holes in \cite{kerrwave} by Luk and Sbierski. The precise asymptotics (and precise generic energy blow-up) of scalar fields inside Kerr black holes was later proven by Ma and Zhang in \cite{scalarMZ} using a robust physical space method which we adapted to the Teukolsky equations (see Section \ref{section:teukdanskerr}), and that we adapt again in the present paper to the non-linear setting. For a scattering approach to Cauchy horizon instability in Reissner-Nordström and an application to mass inflation, see \cite{lukk} by Luk, Oh and Shlapentokh-Rothman.

\subsubsection{The Teukolsky equations in the interior of Kerr black holes}\label{section:teukdanskerr}
The first result for the Teukolsky equation in the interior of Kerr black holes was the oscillatory blow-up asymptotic behavior for $\Psihatp$ heuristically predicted by Ori \cite{ori}. The first rigorous proof of the linear instability of the Kerr Cauchy horizon was due to Sbierski \cite{sbierski}, where it is proven that a weighted $L^2$ norm of $\Psihatp$ along spacelike hypersurfaces transverse to $\ch$ blows up, ensuring the singularity of $\Psihatp$ at $\ch$. The precise oscillatory blow-up behavior predicted by Ori was then rigorously confirmed by the author in \cite{spin+2}, extending the method of \cite{scalarMZ}.

In our subsequent paper \cite{spin-2}, we proved the precise regular asymptotics of the other Teukolsky scalar, $\Psihatm$, near the Kerr Cauchy horizon. Combining the precise polynomial tail for $\Psihatm$ with the oscillatory blow-up asymptotics for $\Psihatp$ suggests that in the non-linear setting, in generic perturbations of Kerr black holes, the coordinate invariant Kretschmann curvature scalar blows up at the perturbed Cauchy horizon with a precise rate, while oscillating violently\footnote{The precise control of $\Psihatm$ is necessary to control the Kretschmann scalar which is expected to behave like $\Real(\Psihatp\Psihatm)$ at main order, see Section 1.3 in \cite{spin-2} for further discussions.}, supporting a conjecture of Ori \cite{orikretschmann}. While we do not prove this oscillatory Kretschmann blow-up in the present article, we expect that this result could be proven by combining the results of the present paper, where we precisely analyze the analog of the curvature component $\Psihatp$, with a precise analysis of the Teukolsky equation for the analog of $\Psihatm$, which would require to adapt the strategy in \cite{spin-2} to the non-linear setting.

\subsubsection{Other results in the spherically symmetric, cosmological, and extremal cases}
We now present results concerning spherically symmetric perturbations of black hole interiors. The works of Hiscock \cite{HISCOCK1981110}, Poisson-Israel \cite{poissonisrael1, poissonisrael2} and Ori \cite{orimass} show the instability of the Reissner-Nordström Cauchy horizon for the spherically symmetric Einstein-Maxwell-null dust system, and later, Dafermos \cite{dafscc,Dafermoss} and Luk-Oh \cite{RNNLI,RNNLII} proved the instability for the spherically symmetric Einstein-Maxwell-scalar field system. These latter papers by Luk and Oh prove the SCC conjecture for this model in the $C^2$ regularity class for two-ended initial data, and even in the $C^{0,1}_{loc}$ (Lipschitz) regularity class by a work of Sbierski \cite{sbierskiholo}. Mass inflation for this model was then proven combining the work \cite{gautam} of Gautam with \cite{lukk}. See also \cite{vdminsta,vdmmmm} by Van de Moortel for extensions of these results to the spherically symmetric Einstein-Maxwell-Klein-Gordon equations, \cite{KehleVdM} by Kehle and Van de Moortel for results on the validity of the $C^0$ version of SCC conjecture in the presence of matter, and \cite{weakvdm,vdm11,vdm22} by Van de Moortel for results on the breakdown behavior of weak null singularities into a spacelike singularity inside spherically symmetric charged black holes.

We also refer to \cite{GajicextremalI,GajicextremalII} for linear waves, and to \cite{GajicLuk} for spherically symmetric non-linear results, in the interior of extremal black holes, and to \cite{Hintz:2015jkj,DafermosBH,DMSRY,Rossetti1,RossettiII} for results in the cosmological setting with $\Lambda>0$, where in both cases the blueshift instability at the Cauchy horizon is much weaker than in the $\Lambda=0$ subextremal case. For results concerning $\Lambda<0$, we refer to \cite{Kehle22,Kads}.
 
\subsubsection{Known non-linear results on the interior of Kerr black holes perturbations}\label{section:introNLKerr}

For non-linear perturbations of Kerr black hole interiors, the only known results are the $C^0$ stability of the Kerr Cauchy horizon by Dafermos and Luk \cite{stabC0}, and the recent work \cite{sbierskiinextdernier} by Sbierski (see also \cite{weakluk,WNSEinsteinEuler} on weak null singularities outside symmetry, and \cite{CameronSbierski} on the uniqueness of $C^0$ spacetime extensions). This latter work states in particular that if, in the Dafermos-Luk spacetime constructed in \cite{stabC0}, there exist vector fields $\overline{X}_i$, $i=1,2,3,4$ continuous up to $\ch$ (in the $C^0$ coordinates on $\ch$) such that the blow-up
$$\left|\int_{V_k}\mathbf{R}(\hat{X}_1,\hat{X}_2,\hat{X}_3,\hat{X}_4)\right|\underset{k\to+\infty}{\longrightarrow} +\infty,$$
holds where the vectors $\hat{X}_i$ are small perturbations of $\overline{X}_i$, and where $V_k$ is some suitable sequence of spacetime regions approaching $\ch$, then the spacetime is Lipschitz-inextendible across $\ch$.

In the present paper, under our assumptions, we prove that the spacetime can be extended up to the Dafermos-Luk Cauchy horizon where a curvature blow-up occurs, implying that the above inextendibility condition is satisfied, thereby ensuring the Lipschitz inextendibility across $\ch$\footnote{Actually, we prove and use a slightly different version of Sbierski's result in \cite{sbierskiinextdernier}, where the blow-up of the integral of $\mathbf{R}(\hat{X}_1,\hat{X}_2,\hat{X}_3,\hat{X}_4)$ occurs on a different sequence of spacetime regions approaching $\ch$.}.

\subsection{Rough statement of the main theorem}\label{section:roughversion}
\begin{figure}[h!]
	\centering
	\includegraphics[scale=0.45]{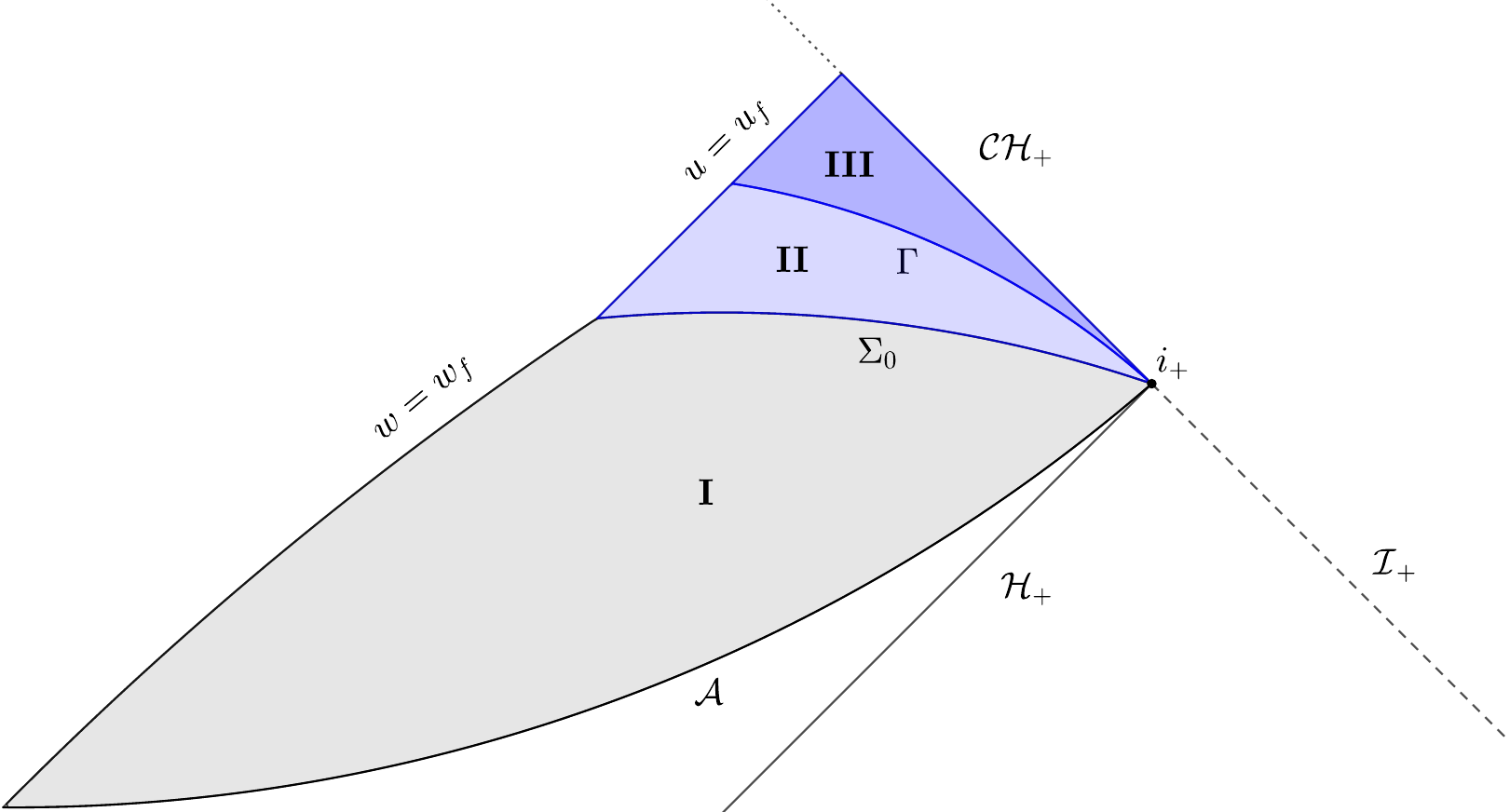}
	\caption{The Penrose diagram of the spacetime $(\mcm,\g)$ constructed in Theorem \ref{thm:roughversion}.}
	\label{fig:regionsNL}
\end{figure}

We now state the rough version of our main results, see Theorem \ref{thm:mainthm} for the precise version.
\begin{thm}[Main theorem, rough version]\label{thm:roughversion}
	We consider initial data for the Einstein vacuum equations on a spacelike hypersurface $\mca$ extending to timelike infinity slightly inside a dynamical vacuum black hole, which settles down to a fixed member of the Kerr family with subextremal parameters $(a,M)$, and such that a precise Price's law-type behavior \eqref{eq:expectedpricelaw} holds on $\mca$. Then, the maximal globally hyperbolic development $(\mcm,\g)$ extends up to a future null boundary $\ch$ (the Cauchy horizon) such that:
	\begin{enumerate}
		\item The spacetime $(\mcm,\g)$ remains close to Kerr$(a,M)$ in a suitable topology and extends continuously beyond $\ch$ in a coordinate system $(x^\mu)$.\label{item:1roughthm}
		\item The Riemann curvature tensor blows up at $\ch$ in the coordinates $(x^\mu)$, while violently oscillating.
		\item The spacetime cannot be extended beyond $\ch$ with a $C^{0,1}_{loc}$ (Lipschitz) metric.
	\end{enumerate}
\end{thm}
The Penrose diagram of the spacetime $(\mcm,\g)$ constructed in Theorem \ref{thm:roughversion} is given in Figure \ref{fig:regionsNL}, where: 
\begin{itemize}
	\item The past boundary of $(\mcm,\g)$ is the spacelike initial hypersurface $\partial_-\mcm=\mca$ which extends to future timelike infinity $i_+$.
	\item The future boundary of $(\mcm,\g)$ is $\partial_+\mcm=\mathcal{H}_f\cup(\ch\cap\{u<u_f\})$, where:
	\begin{itemize}
		\item $\mch_f=\{w=w_f\}\cup\{u=u_f\}$ is a causal hypersurface, where $\{w=w_f\}$ is spacelike and $\{u=u_f\}$ is null with $u_f\ll -1$, $u$ being an optical function,
		\item $\ch\cap\{u<u_f\}$ is the Cauchy horizon restricted to $\{u<u_f\}$, which emanates from future timelike infinity $i_+$.
	\end{itemize}
\end{itemize}

\begin{rem}\label{rem:furtherdisc}
	We make the following remarks on the assumptions of Theorem \ref{thm:roughversion}:
	\begin{itemize}
		\item We assume that the black hole settles down to Kerr$(a,M)$ on $\mca$ at a rate $\ubar^{-3-\delta}$. This decay is faster than the one obtained in \cite{KS21,KSwaveeq}, but is expected to hold for sufficiently fast decaying initial data. We also expect that the Price's law-type assumption \eqref{eq:expectedpricelaw} holds for sufficiently fast decaying initial data on the exterior of the black hole. Moreover, our results generalize to the case where the initial decay exponent $3+\delta$ is replaced by $q$ with $1< q\leq 3+\delta$ and where the exponent 6 in Price's law \eqref{eq:expectedpricelaw} is replaced by $p$ with $1<p<2q$. This would apply to perturbations of Kerr with slower fall-off of the initial data.
		\item Although we expect that the Price's law-type estimate \eqref{eq:expectedpricelaw} generically holds on $\mca$, the exact form of the ansatz in \eqref{eq:expectedpricelaw} is not crucial to the argument. The only relevant property of \eqref{eq:expectedpricelaw} for the proof is that it is a good approximate solution to the Teukolsky equation.
	\end{itemize}
\end{rem}
\begin{rem}
	We make the following remarks on the conclusions of Theorem \ref{thm:roughversion}:
	\begin{itemize}
		\item This is the first proof of instability of a Cauchy horizon with no symmetry assumption, for the full, non-linear, Einstein equations. The results also provide a precise quantitative control of the black hole interior geometry, both concerning stability and instability features.
		\item The Cauchy horizon and the coordinates $(x^\mu)$ in Item \ref{item:1roughthm} of Theorem \ref{thm:roughversion} are the ones constructed in \cite{stabC0}.
		\item The blow-up and inextendibilty results actually hold under the assumption that the constants $Q_{-2}$, $Q_{-1}$, $Q_{+1}$, $Q_{+2}$ in \eqref{eq:expectedpricelaw} do not all vanish, which is expected to hold for generic initial data in the exterior. The $m\neq 0$ non-axisymmetric part of the ansatz \eqref{eq:expectedpricelaw} combined with the rotation of the black hole gives rise to a curvature blow-up which is \underline{highly oscillatory}, as both the amplitude and the oscillation frequency blow-up at the Cauchy horizon. This effect, which was first predicted by Ori \cite{orikretschmann}, was also observed in the linearized setting in our previous works \cite{spin-2,spin+2}, and highlights the fact that it is the \underline{asymmetry} of generic perturbations which ultimately causes the singularity and allows to recover determinism.
		\item The stated inextendibility across $\ch$ is proven near timelike infinity $i_+$, for $u_f\ll -1$. This leaves open the structure of the spacetime near $\ch$ in the region $\{u\geq u_f\}$.
	\end{itemize}
\end{rem}

	The proof of Theorem \ref{thm:roughversion} is an adaptation to the non-linear setting of our previous work \cite{spin+2}. We proceed by choosing appropriate gauges and constructing specific spacetime regions which simplify the analysis.  The non-linear terms that appear in the Teukolsky equation are then proven to be negligible, leading to a blow-up at the Cauchy horizon as in the linearized setting. We detail the precise structure of the proof in Section \ref{section:structureofproof}.
\subsection{Structure of the proof}\label{section:structureofproof}

The proof proceeds by dividing the spacetime $\mcm$ in three regions 
$$\mcm=\un\cup\deux\cup\trois,$$
see the Penrose diagram in Figure \ref{fig:regionsNL} for an illustration of this decomposition. Roughly speaking, $\un$ is the no-shift region bounded away from both $\mch_+$ and $\ch$, and $\deux\cup\trois$ is the blueshift region close to $\ch$ where the blueshift effect described in Section \ref{section:blueshift} occurs. More precisely, region $\deux$ is still relatively far from $\ch$ which allows one to neglect the effect of the weak null singularity by sacrificing some precision in the decay estimates, while region $\trois$ is very close to $\ch$, so that the fate of the already blueshifted perturbations is sealed when entering region $\trois$. Let us mention that such a decomposition in subregions is often used in the study of black hole interior perturbations. Here, our specific choice of subregions follows the one in \cite{scalarMZ}.

 Additionally, the spacetime $\mcm$ is not given a priori and hence the regions have to be constructed dynamically in the course of the proof. We outline below the analysis in each of these regions, distinguishing for clarity between stability and instability estimates.

\subsubsection{Stability estimates}\label{section:stabesti}
\noindent\textbf{Region $\un$.} In region $\un$ we proceed by a standard bootstrap argument using a Principal Temporal (PT) gauge in the ingoing normalization, as introduced in \cite{KS21} (see Section 2.8 in \cite{KS21}) in the proof of stability of the exterior of slowly rotating Kerr black holes. This is because we consider frames in $\un$ which are perturbations of the principal null frames in Kerr, for which the associated horizontal distribution is non-integrable. For this reason we work in the non-integrable formalism of \cite{KSwaveeq}, and in this this context the only known well-posed gauge is the PT gauge introduced in \cite{KS21}.

In this gauge, the stability analysis in region $\un$ becomes particularly simple since $\un$ is bounded away from $\mch_+$ and $\ch$ in the sense that the PT coordinate $r$ is bounded away from $r_+$ and $r_-$. As a consequence, the function $\Delta(r)$ is nowhere vanishing in $\un$, which allows us to obtain non-singular bounds for all the linearized quantities. This considerably simplifies the analysis compared to regions $\deux$ and $\trois$, together with the fact that all linearized quantities decay with respect to $\ubar$ in $\un$\footnote{In $\deux\cup\trois$, some linearized quantities decay with respect to $\ubar$ while some others decay with respect to $|u|$ which complicates the analysis.}. These observations easily deal with the non-linear terms in the null structure and Bianchi identities, and the linear terms are dealt with by means of standard $r^p$ multipliers, where $p$ can chosen arbitrarily large since $r$ is bounded inside the black hole.
\\\\
\noindent\textbf{Region $\deux\cup\trois$.} We define the hypersurface $\Sigma_0\subset\un$ as
$$\Sigma_0=\{\uring+\ubarring=C_R\},$$
where $\uring,\ubarring$ are the Kerr values of the Israel-Pretorius advanced and retarded time \cite{isrealpretorius} with respect to the PT coordinates in $\un$, and where $C_R\gg 1$ is a large constant. From the estimates in the PT gauge we can initialize a double null foliation on $\Sigma_0$, and we show that the induced metric and second fundamental form on $\Sigma_0$ are close to their Kerr values, ensuring that the initial data assumptions of \cite{stabC0} are satisfied on $\Sigma_0$, so that we can apply \cite{stabC0}. Note the following:
\begin{itemize}
	\item The polynomial decay estimates in \cite{stabC0} are stated in weighted Sobolev spaces, which is not sufficient for our purpose. Thus, we re-integrate the system of equations in the double null gauge and use this Sobolev control to deduce pointwise polynomial decay estimates in $\deux\cup\trois$.
	\item  Moreover, we need to control more derivatives, with more decay, than what is achieved in \cite{stabC0}. We note that the estimates in \cite{stabC0} can be generalized to the case of more decay with the same proof. To control additional derivatives, we note that a straightforward adaptation of \cite{stabC0} allows one to control an arbitrary number of angular derivatives. We then commute the equations with the approximate symmetry $\T$ (that coincides with the $\partial_t$ Killing vector field in exact Kerr), which generates commutator terms that are quadratic error terms with good structure, so that the commuted equations can be analyzed with the same estimates as in \cite{stabC0}. Finally, we control the remaining derivatives by induction using the system of equations\footnote{This last step is only done in region $\deux$. Indeed, the induction mixes quantities with $|u|$ and $\ubar$ decay. In region $\deux$ we have $\ubar\sim|u|$ so this is not a problem, while in region $\trois$ we only need the control of some angular derivatives, which have already been appropriately controlled.}.

	\item We choose a large constant $C_R\gg 1$, which implies that $\Sigma_0$ is close to $\ch$ and in turn yields the smallness of $|\Delta|\lesssim e^{-|\kappa_-|C_R}$ in $\deux\cup\trois$\footnote{Here, $\kappa_-$ denotes the surface gravity of $\ch$ in Kerr, see Section \ref{section:BLEFKerr}.}. This smallness simplifies the instability estimates in $\deux\cup\trois$.
\end{itemize} 

\subsubsection{Instability estimates}\label{section:instabesti}

\noindent\textbf{Region $\un$.} The instability estimates in region $\un$ consist in propagating the initial behavior \eqref{eq:expectedpricelaw} on $\mca$ to the whole region $\un$ by analyzing the Teukolsky equation. We use the version of the Teukolsky equation in the non-integrable formalism that is derived in \cite{KSwaveeq}, that concerns the curvature component $A$ which is the analog in the non-integrable formalism of the Newman-Penrose quantity $\psip$. Then, the analysis is similar to the one carried out in the linearized setting in \cite{spin+2} except that:
\begin{enumerate}
		\item The non-linear terms in the Teukolsky equation for $A$ are easily shown to be negligible error terms, by the non-degenerate stability estimates proven in region $\un$.
	\item In \cite{spin+2}, we start from initial data on the event horizon $\mch_+$, and we first analyse the Teukolsky equation for $\psi_{-2}$ (which is the Newman-Penrose equivalent of the curvature component $\Abar$ in the non-integrable formalism) and we then recover $\psi_{+2}$ (which corresponds to the curvature component $A$) via the Teukolsky-Starobinsky identities. Here, we start from initial data on the spacelike hypersurface $\mca$, which implies that we can directly control $A$ from its Teukolsky equation (the analysis for $A$ is easier in the no-shift region than in the redshift region).
		\item In Kerr, we have the stationary and azimuthal Killing symetries $\partial_t,\partial_\phi$. In \cite{spin+2}, we also used the \emph{Carter operator} which is a second-order operator that commutes with the Teukolsky equation. This allowed us to control angular derivatives of $\psi_{+2}$ in $L^2(S(r,\ubar))$, and then $\psi_{+2}$ in $L^\infty$ via Sobolev on the spheres. Here, in the non-linear setting, we only use approximate symmetries $\T,\Z$ which are perturbations of $\partial_t,\partial_\phi$, and instead of using a perturbation of the Carter operator (as is done in \cite{KSwaveeq}), we prefer to commute successively the Teukolsky equation with horizontal Hodge operators. These do not commute with the Teukolsky equation (even in Kerr) but we do so in a specific order such that at every step, the right-hand side of the commuted equation is bounded by the energy of lower-order derivatives, allowing for a control of all such derivatives by induction.\label{item:deuxxxx}
	\item\label{item:troissss} Similarly as in \cite{spin+2}, we use the fact that the Teukolsky operator applied to the ansatz yields an error term, which allows us to do an energy estimate on the inhomogeneous Teukolsky equation satisfied by the difference between $A$ and its ansatz. In \cite{spin+2} we use the Newman-Penrose formalism, and the computation of the Teukolskly operator applied to the ansatz is explicit and quite straightforward. Here, in the non-linear setting, we use the tensorial non-integrable formalism, for which the computation of the Teukolsky operator applied to the ansatz is more involved.
\end{enumerate}
\noindent\textbf{Region $\deux$.} Here, we want to propagate the precise asymptotics for $A$ from region $\un$ to region $\deux$. However the PT frame is  not defined in $\deux$, as for now we only have the double null frame, along with relatively precise estimates on the double null foliation. We define a non-integrable frame in $\deux$ via a frame transformation as defined in \cite{GCMKS} applied to the double null frame. We choose the coefficients of the frame transformation such that:
\begin{itemize}
	\item They are perturbations (with precise decay rates) of the analog coefficients in Kerr spacetime, and they are initialized on $\Sigma_0$ such that the constructed non-integrable frame of region $\deux$ coincides with the PT frame of region $\un$ on $\Sigma_0$. As a consequence, the constructed frame is a perturbation of the principal null frame of Kerr spacetime in the \emph{ingoing normalization} (which is non-regular on $\ch$).
	\item In fact, we construct two frames: the ingoing non-integrable frame $(e_3,e_4,e_1,e_2)$ and the outgoing non-integrable frame $(e_3',e_4',e_1',e_2')$ which are conformally equivalent. We choose the gauge conditions $\D_{e_3}e_3=\D_{e_4'}e_4'=0$, that rewrite as non-linear transport equations for the frame transformation coefficients, which allows us to prove their existence and control.
\end{itemize}
Region $\deux$ is more precisely defined as
$$\deux=\{C_R\leq \uring+\ubarring\leq\ubarring^\gamma\}$$
for some small $\gamma>0$, where $\uring,\ubarring$ correspond here to the double null advanced and retarded time. This implies the estimate $\ubarring\sim|\uring|$ in $\deux$, which simplifies the analysis in $\deux$ since linearized quantities thus all decay with respect to $\ubarring$ there. The definition of $\deux$ also implies that the $\ubarring$-length of region $\deux$ along level sets of $\uring$ is bounded by $\ubarring^\gamma$, and similarly the $\uring$-length of region $\deux$ along level sets of $\ubarring$ is bounded by $\ubarring^\gamma$. This observation simplifies the integration of error terms, sacrificing some $\ubarring^\gamma$ decay, which does not lose much precision as we ultimately choose $\gamma$ sufficiently small with respect to the initial decay parameter $\delta$.

Once the non-integrable frame is constructed, we propagate the precise asymptotics for Teukolsky in $\un$ to region $\deux$ by analyzing the Teukolsky equation with respect to this frame. As region $\deux$ is included in the blueshift region, the analysis is similar to the one done for $\psi_{+2}$ in the Kerr blueshift region in \cite{spin+2}, except for one crucial fact: in region $\deux$, in the energy estimate for the Teukolsky equation we use a multiplier which is $\Delta$-degenerate compared to the one used in the linearized setting in \cite{spin+2}. This is because the method in \cite{spin+2} yields an energy control which is too strong to be compatible with some non-linear terms in the Teukolsky equation that blow-up at the weak null singularity. As a consequence, for the blueshift region in the non-linear setting, we control an energy which is $\Delta$-degenerate compared to the energy controlled in \cite{spin+2} in the linearized setting. Using the above observation on the length of region $\deux$ along level sets of $\uring$ and $\ubarring$, from this degenerate energy control we only lose some $\ubarring^\gamma$ precision in the resulting polynomial decay estimates, which suffices for our purpose choosing $\gamma$ small.

We make some additional comments:
\begin{itemize}
	\item Comments \ref{item:deuxxxx} and \ref{item:troissss} above regarding Teukolsky in region $\un$ also apply in region $\deux$.
	\item In order to capture the precise asymptotics of $A$ in $\deux$, we must improve the bounds for the Teukolsky error term (our bounds in the double null frame only imply that it decays like $\ubarring^{-4-\delta}$) as follows: we first prove non-sharp $\ubarring^{-4-\delta}$ decay for $A$ in $\deux$ by the degenerate energy method. Then, we integrate some Bianchi and null structure equations (which only have $A$ on the right-hand side at the linear level in this gauge) to deduce improved $\ubarring^{-6-\delta}$ decay for the Teukolsky error term in $\deux$. This yields the sharp asymptotics of $A$ in $\deux$ by the degenerate energy method, choosing $\gamma$ small enough ($\gamma<\delta/12$ will be satisfactory).

\end{itemize}

\noindent\textbf{Region $\trois$.} The main property of region $\trois$ is that the double null metric quantity $\Omega^2\sim|\Delta|$ decays (sub-)exponentially towards $\ch$ there:
\begin{align}\label{eq:neglibible}
	\Omega^2\lesssim e^{-|\kappa_-|\ubarring^\gamma}\quad\text{in region }\trois.
\end{align}
This considerably simplifies the analysis. This was already the case in the linearized setting in \cite{spin+2}, but in some sense it is even simpler here in the non-linear setting thanks to the Bianchi equation for $\nabla_3\alpha$ (while in the linearized setting we had to rely only on the Teukolsky equation). First, using the change of frame formulas, we reformulate the estimates proven for $A$ in $\deux$ into precise asymptotics for the outgoing extremal curvature component $\mathring{\alpha}$ in the double null frame on the future boundary $\Gamma$ of region $\deux$ (see Figure \ref{fig:regionsNL}). This is possible because one of the coefficients of the frame transformation is bounded by $\Omega^2$, which implies that the difference between $\Omega^4\mathring{\alpha}$ and (some bounded renormalization of) $A$ is bounded by $\Omega^2$, that decays exponentially on $\Gamma$, yielding a precise $\ubarring^{-6}$ asymptotic behavior for $\Omega^4\mathring{\alpha}$ on $\Gamma$. We then easily propagate this asymptotic behavior to the whole region $\trois$ by integrating the double null Bianchi identity 
	$$\nabring_3(\Omega^4\mathring{\alpha})+\frac{1}{2}\mathring{tr\chibar}\Omega^4\mathring{\alpha}=\Omega^4\left(\mathring{\nabla}\hot\mathring{\beta}+(\mathring{\zeta}+4\mathring{\eta})\hot\mathring{\beta}-3(\mathring{\rho}\mathring{\chihat}+\hodge{\mathring{\rho}}\hodge{\mathring{\chihat}})\right),$$
where the right-hand side is bounded by $\Omega^2$ plus some cubic error term, as can be seen by the bounds in the double null gauge, and hence is negligible by \eqref{eq:neglibible}. Roughly stated, using $\Omega^2\sim e^{-|\kappa_-|(\uring+\ubarring)}$, this implies $|\mathring{\alpha}|\sim e^{2|\kappa_-|(\uring+\ubarring)}\ubarring^{-6}\times\mathrm{osc}$ in $\trois$, where osc is an oscillating factor, hence the stated curvature blow-up on $\ch=\{\ubarring=+\infty\}$.
\subsubsection{Lipschitz inextendibility beyond the Cauchy horizon}
Once we have derived the precise oscillatory blow-up asymptotics for $\mathring{\alpha}$ in $\trois$, we wish to apply Sbierski's result \cite{sbierskiinextdernier} to deduce Lipschitz inextendibility across $\ch$. Recalling the discussion in Section \ref{section:introNLKerr}, this requires to choose vector fields $\overline{X}_i$ such that the integral on some spacetime regions approaching $\ch$ of the curvature contracted with small perturbations of the $\overline{X}_i$ blows up. In the double null gauge, $\mathring{\alpha}_{AB}=\mathbf{R}(\ering_4,\partial_{\theta^A},\ering_4,\partial_{\theta^B})$, where $\ering_4$ is the geodesic outgoing null vector of the double null frame and $\partial_{\theta^A}, \partial_{\theta^B}$ are the coordinate vector fields with respect to spherical coordinates $\theta^A$ on the double null spheres $S(\uring,\ubarring)$. Thus, a natural choice is given by
$$\overline{X}_1=\overline{X}_3=\ering_4,\quad \overline{X}_2=h_2\partial_{\phi_*},\quad\overline{X}_4=h_4\partial_{\phi_*},$$
where $\partial_{\phi_*}\in TS(\uring,\ubarring)$ is the azimuthal coordinate vector field (which coincides with the Killing symmetry $\partial_\phi$ in exact Kerr), and where $h_2$ and $h_4$ are functions to be chosen to both ensure the blow-up and simplify the computation of the integral of $\mathbf{R}(\overline{X}_1,\overline{X}_2,\overline{X}_3,\overline{X}_4)$. Both $\ering_4$ and $\partial_{\phi^*}$ extend continuously to $\ch$ in the $C^0$ coordinates, and we chose $h_2,h_4$ continuous at $\ch$ such that all the $\overline{X}_i$ also do. Once this choice is carefully made, we can compute the integral and conclude to the Lipschitz inextendibility by applying \cite{sbierskiinextdernier} as follows:	
 \begin{itemize}
 	\item We show that some technical assumptions in \cite{sbierskiinextdernier} are satisfied, as a consequence of our precise estimates in the double null gauge.
 	\item We deal with the perturbations of the vectors $\overline{X}_i$ using the previously proven bounds for the curvature components in the double null frame.
 	\item We actually prove blow-up of the integral of the curvature on a slightly modified set of spacetime regions approaching $\ch$, compared to what is assumed in \cite{sbierskiinextdernier}, which we thus cannot apply as a complete black box. More precisely, the spacetime regions in \cite{sbierskiinextdernier} are of the form $\overline{V}\cap\{\ubarring<\ubarring_k\}$ where $\overline{V}$ is some neighborhood of a point on $\ch$, $\ubarring_k\to+\infty$ as $k\to+\infty$, and $\ubarring$ is the double null advanced time; while our spacetime regions take the form $\overline{V}\cap\{\ubar<\ubar_k\}$, where $\ubar_k\to+\infty$ and $\ubar$ is the analog of the Eddington-Finkelstein advanced time in Kerr\footnote{The value of $\ubar_k$ with respect to $k$ has to be chosen carefully to avoid cancellation by oscillations.}. We find that such a choice simplifies the computations. The difference is however irrelevant in our case and the conclusion is the same. To show this, we notice that only the final integration by parts in \cite[p.19]{sbierskiinextdernier} depends on the exact form $\overline{V}\cap\{\ubarring<\ubarring_k\}$, which allows us to apply most of the results in \cite{sbierskiinextdernier} as a black box, assuming that a Lipschitz extension exists. We then reprove the final integration by parts argument in our case, which is basically just a difference in notations compared to what is done in \cite{sbierskiinextdernier}. The same contradiction as in \cite{sbierskiinextdernier} then applies, concluding the proof of the Lipschitz inextendibility.
 \end{itemize}
\subsection{Overview of the article}
In Section \ref{section:preli}, we review the non-integrable geometric framework of \cite{KSwaveeq}. Next, in Section \ref{section:choiceofgauges}, we present the different gauges that we use, and the Kerr values of the associated geometric quantities. In Section \ref{section:mainthm}, we state our initial data assumptions and the precise version of our main theorem, which we decompose into five main intermediate results. In Section \ref{section:regionun}, we prove existence and precise control of region $\un$ in the principal temporal gauge. In Section \ref{section:IIdoublenull}, we slightly extend the stability estimates of \cite{stabC0} concerning region $\deux\cup\trois$ in the double null gauge. Next, in Section \ref{section:gauge}, we construct the auxiliary non-integrable frame in region $\deux$ required for the instability estimates, which we obtain in Section \ref{section:teukolskydeux} by analyzing the Teukolsky equation relative to this frame. Finally, in Section \ref{section:regionIII}, we obtain the curvature blow-up estimates in region $\trois$ and the Lipschitz inextendibility across the Cauchy horizon.

\subsection{Acknowledgments} I am very grateful to Jérémie Szeftel for his continuous support and encouragements, as well as many helpful discussions. This work was partially supported by
ERC-2023 AdG 101141855 BlaHSt.
	
	\section{Non-integrable formalism}\label{section:preli}
	In this section, we review the formalism for non-integrable structures introduced in \cite{formalismkerr,KSwaveeq}.
	\subsection{General formalism for non-integrable structures}
	\subsubsection{Null pairs and horizontal structures}
	We consider a null pair $(e_3,e_4)$ in a smooth spacetime $(\mcm,\g)$, which satisfies
	$$\g(e_3,e_3)=\g(e_4,e_4)=0,\quad\g(e_3,e_4)=-2.$$
	\begin{defi}
		We define the horizontal distribution 
		$$\mch:=(e_3,e_4)^\bot=\left\{X\in T\mcm,\:\g(X,e_3)=\g(X,e_4)=0\right\}.$$ 
	\end{defi}
	The horizontal distribution $\mch$ is a subbundle of the tangent bundle of $\mcm$. In this paper, we will consider a null pair $(e_3,e_4)$ such that the corresponding horizontal distribution $\mch$ is non-integrable. Given an orientation of $\mcm$ and the corresponding spacetime volume form $\in$, we define the induced volume form on $\mch$ by 
	$$\in(X,Y):=\frac{1}{2}\in(X,Y,e_3,e_4),\quad X,Y\in \mch.$$
Given any $X\in T\mcm$, the orthogonal projection ${}^{(h)}\!X$ on the horizontal distribution is
	$${}^{(h)}\!X=X+\frac{1}{2}\g(X,e_3)e_4+\frac{1}{2}\g(X,e_4)e_3.$$
	\begin{defi}
		For $X,Y\in\mch$, we define
		$$\chi(X,Y)=\g(\D_Xe_4,Y),\quad\chibar(X,Y)=\g(\D_Xe_3,Y),$$
		where $\D$ denotes the Levi-Civita connection of $\g$. We also define the induced metric $\gamma$ on $\mch$ as
		$$\gamma(X,Y)=\g(X,Y).$$
	\end{defi}
	Notice that $\chi$ and $\chibar$ are symmetric if and only if the horizontal distribution $\mch$ is integrable.	In that case, $\chi$ and $\chibar$ correspond to the null second fundamental forms of the integral surfaces of $\mch$. 
	\begin{defi}
		We say that a $k$-tensor $U$ on $\mcm$ is horizontal if for any  $X_1,\cdots, X_k\in T\mcm$, 
		$$U(X_1,\cdots,X_k)=U({}^{(h)}\!X_1,\cdots,{}^{(h)}\!X_K).$$
	\end{defi}
	Note that $\chi,\chibar$ and $\gamma$ are horizontal 2-tensors, once we extend them by
	$$\chi(X,Y)=\chi({}^{(h)}\!X,{}^{(h)}\!Y),\quad X,Y\in T\mcm,$$
	and similarly for $\chibar,\gamma$.  
\begin{defi}
	The trace of a horizontal 2-tensor $U$ is the frame-independant scalar defined by
	$$tr (U):=\delta^{ab}U_{ab}.$$
	Similarly, we define the anti-trace of $U$ to be
	$${}^{(a)}tr (U):=\in^{ab}U_{ab}.$$
\end{defi}
	For a horizontal 2-tensor $U$, we have the following decomposition
$$U_{ab}=\hat{U}_{ab}+\frac{1}{2}tr U\delta_{ab}+\frac{1}{2}\atr U\in_{ab},$$
where $\hat{U}_{ab}$ is the symmetric traceless part of $U_{ab}$.
\begin{defi}
	We introduce the notations
	$$tr\chi:=tr(\chi),\quad tr\chibar:=tr(\chibar),\quad\atrchi:={}^{(a)}tr(\chi),\quad\atrchibar:={}^{(a)}tr(\chibar),$$
	and the decompositions
	$$\chi_{ab}=\chihat_{ab}+\frac12\delta_{ab}tr\chi+\frac12\in_{ab}\atrchi,\quad\chibar_{ab}=\wh{\chibar}_{ab}+\frac12\delta_{ab}tr\chibar+\frac12\in_{ab}\atrchibar.$$
\end{defi}
\begin{rem}
	In all the paper, we extend the Einstein summation convention by stating that when two equal horizontal indices appear in an expression, they are summed with respect to $1,2$. For instance,
	$$\xi_b u_{ba}=\sum_{b=1,2}\xi_b u_{ba}.$$
\end{rem}
\begin{defi}[Hodge dual]
	We define the left dual of a horizontal 1-form $\xi$ and a horizontal 2-tensor $u$ by
	$$\hodge{\xi}_a=\in_{ab}\xi_b,\quad \hodge{u}_{ab}=\in_{ac}u_{cb}.$$
	Note that we have the following identities
	$$\hodge{(\hodge{\xi})}=-\xi,\quad\hodge{(\hodge{u})}=-u.$$
	Moreover, if $u$ is symmetric traceless, then $\hodge{u}$ is also symmetric traceless.
\end{defi}
\begin{defi}
	We denote $\fraks_0(\R)$ the set of pairs of real functions on $\mcm$, $\fraks_1(\R)$ the set of real horizontal 1-forms on $\mcm$, and $\fraks_2(\R)$ the set of real symmetric traceless horizontal 2-tensors on $\mcm$.
\end{defi}

\begin{defi}
	Given $\xi,\eta\in\fraks_1(\R)$, we define 
	$$\xi\cdot\eta=\delta^{ab}\xi_a\eta_b,\quad \xi\wedge\eta=\in^{ab}\xi_a\eta_b=\xi\cdot\hodge{\eta},\quad(\xi\hot\eta)_{ab}=\xi_a\eta_b+\xi_b\eta_a-\delta_{ab}(\xi\cdot\eta).$$
	Given $\xi\in\fraks_1(\R)$ and $u,v\in\fraks_2(\R)$, we define 
	$$(\xi\cdot u)_a=\xi_b u_{ba},\quad u\cdot v=\delta^{ac}\delta^{bd}u_{ab}v_{cd}.$$
	We also define 
	$$|\xi|^2=\xi\cdot\xi,\quad |u|^2=u\cdot u.$$
\end{defi}
\begin{lem}
	The following formulas hold true:
\begin{itemize}
	\item Given $\xi,\eta\in\fraks_1(\R)$, we have
	\begin{equation*}
		\begin{gathered}
					\hodge{\xi}\cdot\eta=-\xi\cdot\hodge{\eta},\quad \hodge{\xi}\cdot\hodge{\eta}=\xi\cdot\eta,\quad \hodge{\xi}\wedge\eta=-\xi\wedge\hodge{\eta},\quad\hodge{\xi}\wedge\hodge{\eta}=\xi\wedge\eta,\\
			\hodge{\xi}\hot\eta=\xi\hot\hodge{\eta},\quad\hodge{(\xi\hot\eta)}=\hodge{\xi}\hot\eta,\quad\hodge{\xi}\hot\hodge{\eta}=-\xi\hot\eta.
		\end{gathered}
	\end{equation*}
	\item Given $\xi\in\fraks_1(\R)$, $u\in\fraks_2(\R)$, we have
		$$\hodge{(\xi\cdot u)}=\xi\cdot\hodge{u}=-\hodge{\xi}\cdot u,\quad\hodge{\xi}\cdot\hodge{u}=\xi\cdot u.$$
	\item Given $u,v\in\fraks_2(\R)$,
	$$\hodge{u}\cdot v=-u\cdot\hodge{v},\quad\hodge{u}\cdot\hodge{v}=u\cdot v,\quad \hodge{u}\wedge v=-u\wedge\hodge{v},\quad\hodge{u}\wedge\hodge{v}=u\wedge v.$$
\end{itemize}
\end{lem}
\begin{proof}
	See \cite[Lemma 2.19]{KSwaveeq}.
\end{proof}
	\subsubsection{Horizontal covariant derivative}\label{section:horizontalcovariantderivative}
	We define the horizontal covariant operator $\nabla$ as the orthogonal projection of the spacetime Levi-Civita connection $\D$ on $\mch$,
	$$\nabla_XY:={}^{(h)}\!\left(\D_X Y\right)=\D_X Y-\frac{1}{2}\chibar(X,Y)e_4-\frac{1}{2}\chi(X,Y)e_3,$$
	for $X,Y\in\mch$. Note that in the integrable case, $\nabla$ coincides with the Levi-Civita connection of the metric induced on the integral surfaces of $\mch$. We also define, for $X\in\mch$,
	\begin{align*}
		&\nabla_3X:={}^{(h)}\!(\D_3X)=\D_3X-\frac{1}{2}\g(X,\D_3e_3)e_4-\frac{1}{2}\g(X,\D_3e_4)e_3,\\
		&\nabla_4X:={}^{(h)}\!(\D_4X)=\D_4X-\frac{1}{2}\g(X,\D_4e_4)e_3-\frac{1}{2}\g(X,\D_4e_3)e_4.
	\end{align*}
	\begin{defi}
		Let $U$ be a horizontal $k$-tensor. We define its horizontal derivative as
		\begin{align*}
			\nabla_ZU(X_1,\cdots,X_k)=&Z(U(X_1,\cdots,X_k))-U(\nabla_ZX_1,\cdots,X_k)\\
			&-\cdots-U(X_1,\cdots,\nabla_ZX_k),
		\end{align*}
		for $X_1,\cdots, X_k,Z\in\mch$. We also define 
		\begin{align*}
			&\nabla_3U(X_1,\cdots,X_k)=e_3(U(X_1,\cdots,X_k))-U(\nabla_3X_1,\cdots,X_k)\\
			&\quad\quad\quad\quad\quad\quad\quad\quad\quad-\cdots-U(X_1,\cdots,\nabla_3X_k),\\
			&\nabla_4U(X_1,\cdots,X_k)=e_4(U(X_1,\cdots,X_k))-U(\nabla_4X_1,\cdots,X_k)\\
			&\quad\quad\quad\quad\quad\quad\quad\quad\quad-\cdots-U(X_1,\cdots,\nabla_4X_k).
		\end{align*}
	\end{defi}
	The following result follows easily from the definitions, and the identity $\D\g=0$.
	\begin{prop}\label{prop:nablagammahzero}
		The operators $\nabla$, $\nabla_3$ and $\nabla_4$ take horizontal tensors into horizontal tensors. Moreover, we have 
		$$\nabla\gamma=\nabla_3\gamma=\nabla_4\gamma=0.$$
	\end{prop}
\begin{proof}
	See \cite[Proposition 2.23]{KSwaveeq}.
\end{proof}

	\subsubsection{Horizontal Hodge operators}
	\begin{defi}
We define, for a horizontal 1-form $\omega$,
$$ \diver\omega=\delta^{ab}\nabla_b\omega_a,\quad \mathrm{curl}\:\omega=\in^{ab}\nabla_a\omega_b,\quad\nabla\widehat{\otimes}\omega_{ab}=\nabla_a\omega_b+\nabla_b\omega_a-\delta_{ab}\diver\omega.$$
	\end{defi}

\begin{lem}\label{lem:2140waveq}
	Let $f\in\fraks_1(\R)$ that we extend to a spacetime 1-form by $f_3=f_4=0$. Then,
	$$\D^\mu f_\mu=\nabla^a f_a+(\eta+\etabar)\cdot f.$$
\end{lem}
\begin{proof}
	See \cite[Lemma 2.40]{KSwaveeq}.
\end{proof}
	\begin{defi}
We define the following frame-independent Hodge-type operators:
\begin{itemize}
	\item $\shl{\mcd}{6.6}_1$ takes $\fraks_1(\R)$ into $\fraks_0(\R)$: $\shl{\mcd}{6.6}_1\xi=(\diver\xi,\curl\xi)$.
	\item $\shl{\mcd}{6.6}_2$ takes $\fraks_2(\R)$ into $\fraks_1(\R)$: $(\shl{\mcd}{6.6}_2\xi)_a=\nabla^b\xi_{ab}$.
	\item $\shl{\mcd}{6.6}^*_1$ takes $\fraks_0(\R)$ into $\fraks_1(\R)$: $(\shl{\mcd}{6.6}^*_1(f,f_*))_a=-\nabla_a f+\in_{ab}\nabla_b f_*$.
	\item $\shl{\mcd}{6.6}^*_2$ takes $\fraks_1(\R)$ into $\fraks_2(\R)$: $\shl{\mcd}{6.6}^*_2\xi=-\frac12\nabla\hot\xi$.
\end{itemize} 
	\end{defi}

	\noindent\textbf{Elliptic estimates in the integrable case}

	\begin{prop}\label{prop:ellipticestimates}
		Let $(S,\gamma)$ be a compact 2-manifold with Gauss curvature $K$. 
		\begin{itemize}
			\item The following identity holds for vector fields $f$ on $S$:
			$$\int_S\left(|\nabla f|^2+K|f|^2\right)=\int_S(|\diver f|^2+|\curl f|^2)=\int_S|\shl{\mcd}{6.6}_1 f|^2.$$
			\item The following identity holds for symmetric, traceless, 2-tensors on $S$:
			$$\int_S\left(|\nabla f|^2+2K|f|^2\right)=2\int_S|\diver f|^2=2\int_S|\shl{\mcd}{6.6}_2 f|^2.$$
			\item The following identity holds for pairs of functions $(f,f_*)$ on $S$:
$$\int_S\left(|\nabla f|^2+|\nabla f_*|^2\right)=\int_S|-\nabla f+\hodge\:{\nabla} f_*|^2=\int_S|\shl{\mcd}{6.6}_1^*(f,f_*)|^2.$$
			\item The following identity holds for vector fields $f$ on $S$:
$$\int_S\left(|\nabla f|^2-K|f|^2\right)=2\int_S|\shl{\mcd}{6.6}_2^* f|^2.$$
		\end{itemize}
	\end{prop}
	\begin{proof}
		See \cite[Chapter 2]{CK93}.
	\end{proof}

	\subsubsection{Gauss equation}
	\begin{prop}\label{prop:gaussequation}
		The following identities hold true:
		\begin{enumerate}
			\item For a scalar $\psi$,
			$$[\nabla_a,\nabla_b]\psi=\frac12\left(\atrchi\nabla_3\psi+\atrchibar\nabla_4\psi\right)\in_{ab}.$$
			\item For $\psi\in\fraks_k(\R)$ with $k=1,2$,
			$$[\nabla_a,\nabla_b]\psi=\left(\frac12(\atrchi\nabla_3\psi+\atrchibar\nabla_4\psi)+k{}^{(h)}K\hodge{\psi}\right)\in_{ab},$$
			where, denoting $\mathbf{R}$ the Riemann curvature tensor of $(\mcm,\g)$,
			\begin{align}
				{}^{(h)}K:=-\frac14tr\chi tr\chibar-\frac14\atrchi\atrchibar+\frac12\wh{\chi}\cdot\wh{\chibar}-\frac14\mathbf{R}_{3434}.
			\end{align}
		\end{enumerate}
	\end{prop}
	\begin{proof}
		See \cite[Proposition 2.43]{KSwaveeq}.
	\end{proof}
\subsubsection{Horizontal Lie derivatives}\label{section:horizontalliederivative}
\begin{defi}
	Given vector fields $X,Y$, we define the horizontal Lie derivative $\lie_X Y$ as the orthogonal projection of the spacetime Lie derivative $\mcl_X Y$ on the horizontal distribution $\mch$:
	$$\lie_X Y=\mcl_X Y+\frac12\g(\mcl_XY, e_3)e_4+\frac12\g(\mcl_XY, e_4)e_3.$$
	Given a horizontal covariant $k$-tensor $U$, the horizontal Lie derivative $\lie_X U$ is defined by
	$$(\lie_X U)(Y_1,\cdots, Y_k):=X(U(Y_1,\cdots, Y_k))-U(\lie_X Y_1,Y_2,\cdots,Y_k)-\cdots-U(Y_1,\cdots,Y_{k-1},\mcl_X Y_k).$$
	Equivalently, for horizontal indices $A=a_1\cdots a_k$,
	$$\lie_X U_{A}=\nabla_X U_A+\D_{a_1}X^bU_{ba_2\cdots a_k}+\cdots+\D_{a_k}X^bU_{a_1\cdots a_{k-1}b}.$$
\end{defi}

	\subsection{Horizontal structures and Einstein equations}
	\subsubsection{Ricci coefficients and curvature components}\label{section:ricciandcurvdef}
	In what follows, we denote by $e_a$, $a=1,2$ an orthonormal basis of the horizontal distribution $\mch$. We define the \textit{Ricci coefficients} as the following horizontal tensors,
	\begin{alignat*}{2}
		\chi_{ab}&=\g(\dgras_a e_4,e_b),\quad& \chibar_{ab}&=\g(\dgras_a e_3,e_b), \\
		\eta_a&=\frac{1}{2}\g(\dgras_3 e_4,e_a),\quad&\underline{\eta}_a&=\frac{1}{2}\g(\dgras_4 e_3,e_a),\\
		\omega&=\frac{1}{4}\g(\dgras_4 e_4,e_3),\quad&\underline{\omega}&=\frac{1}{4}\g(\dgras_3 e_3,e_4),\\
		\xi_a&=\frac{1}{2}\g(\dgras_4 e_4,e_a),\quad&\underline{\xi}_a&=\frac{1}{2}\g(\dgras_3 e_3,e_a),\\
		\zeta_a&=\frac{1}{2}\g(\dgras_ae_4,e_3).
	\end{alignat*}
	In what follows, we denote $\mathbf{R}$ the Riemann curvature tensor. We define the \textit{curvature components} as the following horizontal tensors,
\begin{alignat*}{2}
	\alpha_{ab}&=\mathbf{R}_{4a4b},\quad&\underline{\alpha}_{ab}&=\mathbf{R}_{3a3b},\\
	\beta_a&=\frac{1}{2}\mathbf{R}_{a434},\quad& \underline{\beta}_a&=\frac{1}{2}\mathbf{R}_{a334},\\
	\rho&=\frac{1}{4}\mathbf{R}_{4343},\quad&\hodge{\rho}&=\frac{1}{4}\hodge{}\:\mathbf{R}_{4343}.
\end{alignat*}
	where the tensor $\hodge{}\:\mathbf{R}$ is the Hodge dual of the curvature tensor $\mathbf{R}$, defined by 
	$$\hodge{}\:\mathbf{R}_{\alpha\beta\mu\nu}=\frac{1}{2}\in_{\mu\nu}\!\!{}^{\rho\sigma}\mathbf{R}_{\alpha\beta\rho\sigma}.$$
	Note that defining $\varrho=-\rho\gamma+\hodge{\rho}\in,$
	we have, $\varrho_{ab}=\mathbf{R}_{a3b4}$.
	\subsubsection{Null structure equations}\label{section:nullstructureeq}
	For a vacuum spacetime $(\mcm,\g)$, i.e. such that \eqref{eq:EVE} is satisfied,	with a null pair $(e_3,e_4)$, the Ricci coefficients defined in Section \ref{section:ricciandcurvdef} satisfy the following \emph{null structure equations}:
	\begin{align*}
		\nabla_3tr\chibar&=-|\wh{\chibar}|^2-\frac{1}{2}\left(tr\chibar^2-\atrchibar^2\right)+2\diver\xibar-2\omegabar tr\chibar+2\xibar\cdot (\eta+\etabar-2\zeta),\\
		\nabla_3\atrchibar&=-tr\chibar\atrchibar+2\curl \xibar-2\omegabar\atrchibar+2\xibar\wedge (-\eta+\etabar+2\zeta),\\
		\nabla_3\wh{\chibar}&=-tr\chibar\wh{\chibar}+\nabla\hot\xibar-2\omegabar \wh{\chibar}+\xibar\hot(\eta+\etabar-2\zeta)-\alphabar,
	\end{align*}
	\begin{align*}
		\nabla_3tr\chi&=-\wh{\chibar}\cdot\wh{\chi}-\frac{1}{2}tr\chibar tr\chi+\frac{1}{2}\atrchibar \atrchi+2\diver\eta+2\omegabar tr\chi+2(\xi\cdot\xibar+|\eta|^2)+2\rho,\\
		\nabla_3\atrchi&=-\wh{\chibar}\wedge\chihat-\frac{1}{2}\left(\atrchibar tr\chi+tr\chibar\atrchi\right)+2\curl\eta+2\omegabar\atrchi+2\xibar\wedge\xi-2\hodge{\rho},\\
		\nabla_3\chihat&=-\frac{1}{2}(tr\chi\wh{\chibar}+tr\chibar\chihat)-\frac{1}{2}(\hodge{\chihat}\atrchibar-\hodge{\wh{\chibar}}\atrchi)+\nabla\hot\eta+2\omegabar\wh{\chi}+\xibar\hot\xi+\eta\hot\eta,
	\end{align*}
	\begin{align*}
		\nabla_4 tr\chibar&= -\chihat\cdot\wh{\chibar}-\frac{1}{2}tr\chi tr\chibar+\frac{1}{2}\atrchi\atrchibar+2\diver\etabar+2\omega tr\chibar+2(\xi\cdot\xibar+|\etabar|^2)+2\rho,\\
		\nabla_4\atrchibar&= -\chihat\wedge\wh{\chibar}-\frac{1}{2}\left(\atrchi tr\chibar+tr\chi\atrchibar\right)+2\curl\etabar+2\omega\atrchibar+2\xi\wedge\xibar+2\hodge{\rho},\\
		\nabla_4\wh{\chibar}&=-\frac{1}{2}\left(tr\chibar\wh{\chi}+tr\chi\wh{\chibar}\right)-\frac{1}{2}(\hodge{\wh{\chibar}}\atrchi-\hodge{\chihat}\atrchibar)+\nabla\hot\etabar+2\omega\wh{\chibar}+\xi\hot\xibar+\etabar\hot\etabar,
	\end{align*}
	\begin{align*}
		\nabla_4tr\chi&=-|\chihat|^2-\frac{1}{2}(tr\chi^2-\atrchi^2)+2\diver\xi-2\omega tr\chi+2\xi\cdot(\etabar+\eta+2\zeta)\\
		\nabla_4\atrchi&=-tr\chi\atrchi+2\curl\xi-2\omega\atrchi+2\xi\wedge(-\etabar+\eta-2\zeta),\\
		\nabla_4\chihat&=-tr\chi\chihat+\nabla\hot\xi-2\omega\chihat+\xi\hot(\etabar+\eta+2\zeta)-\alpha.
	\end{align*}
	Also,
	\begin{align*}
		\nabla_3\zeta+2\nabla\omegabar&=-\wh{\chibar}\cdot(\zeta+\eta)-\frac{1}{2}tr\chibar(\zeta+\eta)-\frac{1}{2}\atrchibar(\hodge{\zeta}+\hodge{\eta})+2\omegabar(\zeta-\eta)\\
		&\quad +\chihat\cdot\xibar+\frac{1}{2}tr\chi\xibar+\frac{1}{2}\atrchi\hodge{\xibar}+2\omega\xibar-\betabar,\\
		\nabla_4\zeta-2\nabla\omega&=\chihat\cdot(-\zeta+\etabar)+\frac{1}{2}tr\chi(-\zeta+\etabar)+\frac{1}{2}\atrchi(\hodge{\etabar}-\hodge{\zeta})+2\omega(\zeta+\etabar)\\
		&\quad-\wh{\chibar}\cdot\xi-\frac{1}{2}tr\chibar\xi-\frac{1}{2}\atrchibar\hodge{\xi}-2\omegabar\xi-\beta,\\
		\nabla_3\etabar-\nabla_4\xibar&=-\wh{\chibar}\cdot(\etabar-\eta)-\frac{1}{2}tr\chibar(\etabar-\eta)+\frac{1}{2}\atrchibar(\hodge{\etabar}-\hodge{\eta})-4\omega\xibar+\betabar,\\
		\nabla_4\eta-\nabla_3\xi&=-\chihat\cdot(\eta-\etabar)-\frac{1}{2}tr\chi(\eta-\etabar)+\frac{1}{2}\atrchi(\hodge{\eta}-\hodge{\etabar})-4\omegabar\xi-\beta,\\
		\nabla_3\omega+\nabla_4\omegabar&=4\omega\omegabar+\xi\cdot\xibar+(\eta-\etabar)\cdot\zeta-\eta\cdot\etabar+\rho,
	\end{align*}
	and
	\begin{align*}
		\diver\chihat+\zeta\cdot\chihat&=\frac{1}{2}\nabla tr\chi+\frac{1}{2}tr\chi\zeta-\frac{1}{2}\hodge{}\:\nabla\atrchi-\frac{1}{2}\atrchi\hodge{\zeta}-\atrchi\hodge{\eta}-\atrchibar\hodge{\xi}-\beta,\\
		\diver\wh{\chibar}-\zeta\cdot\wh{\chibar}&=\frac{1}{2}\nabla tr\chibar-\frac{1}{2}tr\chibar\zeta-\frac{1}{2}\hodge{}\:\nabla\atrchibar+\frac{1}{2}\atrchibar\hodge{\zeta}-\atrchibar \hodge{\etabar}-\atrchi\hodge{\xibar}+\betabar,\\
		\curl\zeta&=-\frac{1}{2}\chihat\wedge\wh{\chibar}+\frac{1}{4}\left(tr\chi\atrchibar-tr\chibar\atrchi\right)+\omega\atrchibar-\omegabar\atrchi+\hodge{\rho}.
	\end{align*}
See \cite[Proposition 2.54]{KSwaveeq}.
	\subsubsection{Null Bianchi identities}\label{section:bianchii}
	For a vacuum spacetime $(\mcm,\g)$, the curvature components defined in Section \ref{section:ricciandcurvdef} satisfy the following \emph{null Bianchi identities}:
	\begin{align*}
		\nabla_3\alpha+\frac{1}{2}(tr\chibar\alpha+\atrchibar\hodge{\alpha})-4\omegabar\alpha&=\nabla\hot\beta+(\zeta+4\eta)\hot\beta-3(\rho\wh{\chi}+\hodge{\rho}\hodge{\chihat}),\\
		\nabla_4\beta+2(tr\chi\beta-\atrchi\hodge{\beta})+2\omega\beta&=\alpha\cdot(2\zeta+\etabar)+3(\xi\rho+\hodge{\xi}\hodge{\rho}),\\
		\nabla_3\beta+(tr\chibar\beta+\atrchibar\hodge{\beta})-2\omegabar\beta&=\nabla\rho+\hodge{}\:\nabla\hodge{\rho}+2\betabar\cdot{\chihat}+3(\rho\eta+\hodge{\rho}\hodge{\eta})+\alpha\cdot\xibar,\\
		\nabla_4\rho+\frac{3}{2}(tr\chi\rho+\atrchi\hodge{\rho})&=\diver\beta+(2\etabar+\zeta)\cdot\beta-2\xi\cdot\betabar-\frac{1}{2}\wh{\chibar}\cdot\alpha,\\
		\nabla_4\hodge{\rho}+\frac{3}{2}(tr\chi\hodge{\rho}-\atrchi\rho)&=-\curl\beta-(2\etabar+\zeta)\cdot\hodge{\beta}-2\xi\cdot\hodge{\betabar}+\frac{1}{2}\wh{\chibar}\cdot\hodge{\alpha},\\
		\nabla_3\rho+\frac{3}{2}(tr\chibar\rho-\atrchi\hodge{\rho})&=-\diver\betabar-(2\eta-\zeta)\cdot\betabar+2\xibar\cdot\beta-\frac{1}{2}\wh{\chi}\cdot\alphabar,\\
		\nabla_3\hodge{\rho}+\frac{3}{2}(tr\chibar\hodge{\rho}+\atrchibar\rho)&=-\curl\betabar-(2\etabar-\zeta)\cdot\hodge{\betabar}-2\xibar\cdot\hodge{\beta}-\frac{1}{2}\wh{\chi}\cdot\hodge{\alphabar},\\
		\nabla_4\betabar+(tr\chi\betabar+\atrchibar\hodge{\betabar})-2\omega\betabar&=-\nabla\rho+\hodge{}\:\nabla\hodge{\rho}+2\beta\cdot\wh{\chihat}-3(\rho\etabar-\hodge{\rho}\hodge{\etabar})-\alphabar\cdot\xi,\\
		\nabla_3\betabar+2(tr\chibar\betabar-\atrchibar\hodge{\betabar})+2\omegabar\betabar&=\alphabar\cdot(2\zeta-\eta)-3(\xibar\rho-\hodge{\xibar}\hodge{\rho}),\\
		\nabla_4\alphabar+\frac{1}{2}(tr\chi\alphabar+\atrchi\hodge{\alphabar})-4\omega\alphabar&=-\nabla\hot\betabar+(\zeta-4\etabar)\hot\betabar-3(\rho\wh{\chibar}-\hodge{\rho}\hodge{\wh{\chibar}}).
	\end{align*}
See \cite[Proposition 2.55]{KSwaveeq}.

\subsection{Main equations in complex notations}
\subsubsection{Complex notations}\label{section:complexnotations}
We define the complexified version of horizontal tensors on $\mcm$.
\begin{defi}
	For $k=0,1,2$, we denote by $\fraks_k(\C)$ the set of complex anti-self dual horizontal $k$-tensors on $\mcm$:
	\begin{itemize}
		\item $a+ib\in\fraks_0(\C)$ is a complex scalar function if $(a,b)\in\fraks_0(\R)$.
		\item $F=f+i\hodge{f}$ is a complex anti-self dual horizontal 1-form if $f\in\fraks_1(\R)$.
		\item $U=u+i\hodge{u}$ is a complex anti-self dual symmetric traceless horizontal 2-tensor if $f\in\fraks_1(\R)$.
	\end{itemize}
	Note that $F\in\fraks_1(\C)$ and $U\in\fraks_2(\C)$ are indeed anti-self dual, namely
	$$\hodge{F}=-iF,\quad \hodge{U}=-iU.$$
\end{defi} 
\begin{defi}\label{defi:complexhodgeversionn}
	We define the complexified version of the horizontal covariant derivative by
	$$\mcd=\nabla+i\hodge{}\:\nabla,\quad\overline{\mcd}=\nabla-i\hodge{}\:\nabla.$$
	More precisely, we have:
	\begin{itemize}
		\item For $a+ib\in\fraks_0(\C)$,
		$$\mcd(a+ib)=(\nabla+i\hodge{}\:\nabla)(a+ib),\quad\overline{\mcd}(a+ib)=(\nabla-i\hodge{}\:\nabla)(a+ib).$$
		\item For $F=f+i\hodge{f}\in\fraks_1(\C)$,
		\begin{equation*}
			\begin{gathered}
				\mcd\cdot F=(\nabla+i\hodge{}\:\nabla)\cdot(f+i\hodge{f})=0,\\
				\divc F=(\nabla-i\hodge{}\:\nabla)\cdot(f+i\hodge{f}),\quad\mcd\hot F=(\nabla+i\hodge{}\:\nabla)\hot (f+i\hodge{f}).
			\end{gathered}
		\end{equation*}
		\item For $U=u+i\hodge{u}\in\fraks_2(\C)$, 
		\begin{align*}
			\mcd\cdot U=(\nabla+i\hodge{}\:\nabla)\cdot(u+i\hodge{u})=0,\quad\divc U=(\nabla-i\hodge{}\:\nabla)\cdot(u+i\hodge{u}).
		\end{align*}
		
	\end{itemize}
\end{defi}
\begin{lem}
	The following holds:
	\begin{itemize}
		\item If $f,h\in\fraks_1(\R)$, 
		\begin{align*}
			f\cdot h+i\hodge{f}\cdot h&=\frac12(f+i\hodge{f})\cdot(\overline{h+i\hodge{h}})\in\fraks_0(\C),\\
			f\hot +i\hodge{(f\hot h)}&=\frac12(f+i\hodge{h})\hot({f+i\hodge{h}})\in\fraks_2(\C).
		\end{align*}
	\item If $f\in\fraks_1(\R)$ and $u\in\fraks_2(\R)$, 
	$$u\cdot +i\hodge{u\cdot f}=\frac12(u+i\hodge{u})\cdot(\overline{f+i\hodge{f}})\in\fraks_1(\C).$$
	\item If $u,v\in\fraks_2(\R)$, 
	$$u\cdot v+i\hodge{u\cdot v}=\frac12(u+i\hodge{u})\cdot(\overline{v+i\hodge{v}})\in\fraks_0(\C).$$
	\item If $(a,b)\in\fraks_0(\R)$,
	$$\nabla a-\hodge\:{\nabla}b+i(\hodge\:{\nabla}a+\nabla b)=\mcd(a+ib)\in\fraks_1(\C).$$
	\item If $f\in\fraks_1(\R)$,
	\begin{align*}
		\diver f+i\curl f&=\frac12\divc(f+i\hodge{f})\in\fraks_0(\C),\\
		\nabla\hot f+i\hodge{(\nabla\hot f)}&=\frac12\mcd\hot(f+i\hodge{f})\in\fraks_2(\C).
	\end{align*}
	\item If $u\in\fraks_2(\C)$, 
	$$\diver u+i\hodge{(\diver u)}=\frac12\divc(u+i\hodge{u})\in\fraks_1(\C).$$
	\end{itemize}
\end{lem}
\begin{proof}
	See \cite[Lemma 2.79]{KSwaveeq}.
\end{proof}
\begin{lem}\label{lem:monamitoitcool}
	Let $E=e+i\hodge{e},F=f+i\hodge{f}\in\fraks_1(\C)$ and $U\in\fraks_2(\C)$. Then we have
	\begin{align*}
		E\hot(\overline{F}\cdot U)+F\hot(\overline{E}\cdot U)=2(E\cdot\overline{F}+\overline{E}\cdot F)U,\quad E\hot(\overline{F}\cdot U)=4(e\cdot f-i e\wedge f)U.
	\end{align*}
\end{lem}
\begin{proof}
	See \cite[Lemma 2.80]{KSwaveeq}.
\end{proof}
\subsubsection{Complex Leibniz identities and horizontal integration by parts}
\begin{lem}\label{lem:leibnizchilll}
	Let $F\in\fraks_1(\C)$, $U\in\fraks_2(\C)$. Then
	\begin{align*}
		U\cdot\overline{\mcd}F=U(\divc F),\quad F\hot(\divc U)=2(F\cdot\overline{\mcd})U=4F\cdot\nabla U.
	\end{align*}
\end{lem}
\begin{proof}
	See \cite[Lemma 2.81]{KSwaveeq}.
\end{proof}
\begin{lem}\label{prop:cestpourteukansatz}
	We have, for $F,H\in\fraks_1(\C)$, the identities
	\begin{align*}
		\mcd(\overline{H}\cdot F)=\overline{H}\cdot(\mcd\hot F)+(\mcd\cdot\overline{H}) F,\quad\quad \divc( F\hot H)=2((\divc F)H+(\divc H)F).
	\end{align*}
\end{lem}
\begin{proof}
	We denote $F=f+i\hodge{f}$, $H=h+i\hodge{h}$. We have $\overline{H}\cdot F=2(f\cdot h+i\hodge{f}\cdot h)$. Thus
	\begin{align*}
		\mcd(\overline{H}\cdot F)&=2\nabla(f\cdot h)-2\hodge\:{\nabla}(\hodge{f}\cdot h)+2i\hodge{(\nabla(f\cdot h)-\hodge\:{\nabla}(\hodge{f}\cdot h))}.
	\end{align*}
	Moreover, by the decomposition $\nabla_af_b=\frac{1}{2}((\nabla\hot f)_{ab}+\delta_{ab}\diver f+\in_{ab}\curl f)$, we find 
	\begin{align*}
		\nabla_a(f\cdot h)&=\nabla_a f_b h_b+f_b\nabla_a h_b=\frac{1}{2}(\nabla\hot f)_{ab}h_b+\frac{1}{2}(\diver f) h_a+\frac{1}{2}(\curl f) \hodge{h}_a+[f\leftrightarrow h].
	\end{align*}
where	$[f\leftrightarrow h]$ means that we add the same quantity where $f$ and $h$ are exchanged. This implies
	\begin{align*}
		\nabla(f\cdot h)&=\frac{1}{2}(\nabla\hot f)\cdot h+\frac{1}{2}(\diver f) h+\frac{1}{2}(\curl f) \hodge{h}+[f\leftrightarrow h],\\
		\nabla(\hodge{f}\cdot h)&=\frac{1}{2}\hodge{(\nabla\hot f)}\cdot h+\frac{1}{2}(\curl f) h-\frac{1}{2}(\diver f) \hodge{h}+\frac{1}{2}(\nabla\hot h)\cdot \hodge{f}+\frac{1}{2}(\diver h) \hodge{f}-\frac{1}{2}(\curl h) f.
	\end{align*}
	We deduce, using $\hodge(\xi\cdot U)=\xi\cdot(\hodge{U})=-\hodge{\xi}\cdot U$ for $\xi\in\fraks_1(\mathbb{R})$, $U\in\fraks_2(\mathbb{R})$, 
	$$\hodge\:{\nabla}(\hodge{f}\cdot h)=-\frac{1}{2}(\nabla\hot f)\cdot h+\frac{1}{2}(\curl f) \hodge{h}+\frac{1}{2}(\diver f){h}+\frac{1}{2}(\nabla\hot h)\cdot{f}-\frac{1}{2}(\diver h) {f}-\frac{1}{2}(\curl h) \hodge{f}.$$
	Thus, 
	\begin{align*}
		2\nabla(f\cdot h)-2\hodge\:{\nabla}(\hodge{f}\cdot h)&=2(\nabla\hot f)\cdot h+2(\diver h)f+2(\curl h)\hodge{f}.
	\end{align*}
	This implies
	\begin{align*}
		\mcd(\overline{H}\cdot F)=2(\nabla\hot f)\cdot h+2i\hodge{((\nabla\hot f)\cdot h)}+2(\diver h)F+2(\curl h)(\hodge{f}-i{f})=\mcd\hot F\cdot\overline{H}+(\mcd\cdot\overline{H})F,
	\end{align*}
which concludes the proof of the first stated identity. The second identity is proven similarly.
\end{proof}

\begin{lem}\label{lem:divcfcdotnablaU}
	Let $U=u+i\hodge{u}\in\fraks_k(\mathbb{C})$, $k=1,2$, and $F=f+i\hodge{f}\in\fraks_1(\C)$. We have
\begin{align*}
	\divc(f\cdot\nabla U)=f\cdot\nabla(\divc U)+\nabla f\cdot \nabla U+k{}^{(h)}\!K \overline{F}\cdot U+\frac{i}{2}\overline{F}\cdot\left(\atrchi\nabla_3 U+\atrchibar\nabla_4 U\right).
\end{align*}
Also, for $k=2$,
	\begin{align*}
		\divc(F\cdot\nabla U)=&\frac12((\divc F)\divc U+(\divc\divc U)F),
	\end{align*}
and for $k=1$, 
	\begin{align*}
		\mcd\hot(f\cdot\nabla U)=&f\cdot\nabla(\mcd\hot U)+\nabla f\hot \nabla U-{}^{(h)}\!K {F}\hot U-\frac{i}{2}F\hot\left(\atrchi\nabla_3 U+\atrchibar\nabla_4 U\right).
	\end{align*}
\end{lem}
\begin{proof}
For the first identity, we use for $k=2$ (the proof for $k=1$ is similar),
$$\divc(f\cdot\nabla U)=2(\diver(f\cdot\nabla u)+i\hodge{\diver(f\cdot\nabla u)}),$$
and we compute, 
\begin{align*}
	\diver(f\cdot\nabla u)_c&=\nabla_a(f_b\nabla_b u_{ac})=f\cdot\nabla\diver u_c+\nabla_a f_b\nabla_b u_{ac}+f_b[\nabla_a,\nabla_b]u_{ac}
\end{align*}
and we conclude by using Proposition \ref{prop:gaussequation}. The last stated identity is proven by a similar computation. The second identity is a direct consequence of Lemmas \ref{lem:leibnizchilll} and \ref{prop:cestpourteukansatz}.
\end{proof}

\begin{lem}\label{lem:nouveauleibniz}
	We have, for $F=f+i\hodge{f}\in\fraks_1(\C)$ and $U\in\fraks_2(\C)$,
	$$\mcd\hot(\overline{F}\cdot U)=2(\mcd\cdot\overline{F})U+2(\overline{F}\cdot\mcd)U=2(\mcd\cdot\overline{F})U+8f\cdot\nabla U-F\hot(\divc U).$$
\end{lem}
\begin{proof}
	The proof is a simple combination of the identities \cite[(2.74), (2.75)]{KSwaveeq}.
\end{proof}

\begin{lem}\label{lem:lempourbianchipairpreli}
	Let $U\in\fraks_2(\C)$ and $V\in\fraks_1(\C)$. We have the identity
	\begin{align*}
		\frac12\Real\left(\overline{V}\cdot(\divc U)\right)+\frac14\Real\left(\overline{U}\cdot(\mcd\hot V)\right)=\nabla\cdot\Real(\overline{V}\cdot U).
	\end{align*}
Now, let  $U\in\fraks_1(\C)$, $V\in\fraks_0(\C)$. We have the identity
	$$\frac12\Real(\overline{V}(\divc U))+\frac12\Real(\overline{U}\cdot\mcd V)=\nabla\cdot\Real(\overline{V}U).$$
\end{lem}
\begin{proof}
	Both statements are the result of \cite[Lemma 15.6]{KSwaveeq}.
\end{proof}
\subsubsection{Horizontal Laplacian computations in the non-integrable case}
For $U\in\fraks_k(\C)$, $k=0,1,2$, we define the horizontal Laplacian operator $\triangle_k$ by
$$\triangle_kU:=\nabla^a\nabla_a U.$$
\begin{lem}\label{lem:ignocoucou}
	Let $U\in\fraks_2(\mathbb{C})$. We have the formula
	\begin{align}\label{eq:igno}
		\frac{1}{4}\mcd\hot(\divc U)=\triangle_2 U-2{}^{(h)}\!KU-\frac{i}{2}({}^{(a)}\!tr\chi\nabla_3U+{}^{(a)}\!tr\chibar\nabla_4U).
	\end{align}
\end{lem}
\begin{proof}
	See \cite[Lemma 2.82]{KSwaveeq} (which lacks a factor $\frac12$ on the LHS).
\end{proof}
\begin{lem}\label{lem:laplaciens1(C)}
	Let $F\in\mathfrak{s}_1(\C)$. We have the following formula
	$$\frac{1}{4}\divc (\mcd\hot F)=\triangle_1F+{}^{(h)}KF+\frac{i}{2}({}^{(a)}tr\chi\nabla_3F+{}^{(a)}tr\chibar\nabla_4F).$$
\end{lem}
\begin{proof}
	We have the following computation, for $F=f+i\hodge{f}$,
	\begin{align*}
		\frac{1}{4}\divc(\mcd\hot F)=\diver\nabla\hot f+i\hodge{}\left(\diver\nabla\hot f\right).
	\end{align*}
	Moreover, by \cite[Lemma 2.36]{KSwaveeq} combined with Proposition \ref{prop:gaussequation}, we have 
	\begin{align*}
		\diver\nabla\hot f&=\triangle_1f-\frac{1}{2}\in_{ab}[\nabla_a,\nabla_b]\hodge{f}\\
		&=\triangle_1f-\frac{1}{2}\left(\frac{1}{2}\left(\atrchi\nabla_3\hodge{f}+\atrchibar\nabla_4\hodge{f}\right)-{}^{(h)}\!Kf\right)\in^{ab}\in_{ab}\\
		&=\triangle_1f-\frac{1}{2}\left(\atrchi\nabla_3\hodge{f}+\atrchibar\nabla_4\hodge{f}\right)+{}^{(h)}\!Kf,
	\end{align*}
	which concludes the proof.
\end{proof}
\begin{lem}\label{lem:unlemsympa}
	Let $F=f+i\hodge{f}\in\fraks_1(\C)$. We have the following formula
	$$\frac{1}{4}\divc(\mcd\hot F)=\frac{1}{2}\mcd(\divc F)+2{}^{(h)}\!KF+i\left(\atrchi\nabla_3F+\atrchibar\nabla_4F\right).$$
\end{lem}
\begin{proof}
	This formula comes from Lemma \ref{lem:laplaciens1(C)} combined with $\triangle_1 F=\triangle_1 f+i\hodge{(\triangle_1f)}$. Indeed, by \cite[(2.21)]{KSwaveeq}, 
	\begin{align*}
		\triangle_1 f&=\nabla\diver f-\hodge{}\:\nabla\curl f-\frac{1}{2}\in_{ab}[\nabla_a,\nabla_b]\hodge{f}\\
		&=\nabla\diver f-\hodge{}\:\nabla\curl f+{}^{(h)}\!Kf-\frac{1}{2}\left(\atrchi\nabla_3\hodge{f}+\atrchibar\nabla_4\hodge{f}\right),
	\end{align*}
	which concludes the proof, as $\divc F=2(\diver f+i\curl f)$.
\end{proof}
\begin{lem}\label{lem:unlemsympascalair}\label{lem:ignocoucoulast}
	Let $h=a+ib$ be a scalar function. We have the following formula
	$$\frac{1}{2}\divc(\mcd h)=\triangle_0 h+\frac{i}{2}\left(\atrchi\nabla_3h+\atrchibar\nabla_4h\right).$$
\end{lem}
\begin{proof}
	Denoting $h=a+ib$, we have $\frac{1}{2}\divc(\mcd h)=\diver f+i\curl f$ where $f=\nabla a-\hodge\:{\nabla} b$. Then we use Proposition \ref{prop:gaussequation} to compute
	\begin{align*}
		\diver f=\triangle_0 a-\frac12(\atrchi\nabla_3+\atrchibar\nabla_4)b,\quad \curl f=\triangle_0 b+\frac12(\atrchi\nabla_3+\atrchibar\nabla_4)a,
	\end{align*}
	which concludes the proof.
\end{proof}

	\subsubsection{Commutation formulas}\label{section:commutationformulas}
\begin{lem}\label{lem:commutnablageneral}
	We define the mixed tensor $\mathbf{B}_{ab\mu\nu}$ as follows : 
	\begin{align*}
		\mathbf{B}_{abc3}&=2\left(-\chibar_{ca}\eta_b+\chibar_{cb}\eta_a-\chi_{ca}\xibar_b+\chi_{cb}\xibar_a\right),\\
		\mathbf{B}_{abc4}&=2\left(-\chi_{ca}\etabar_b+\chi_{cb}\etabar_a-\chibar_{ca}\xi_b+\chibar_{cb}\xi_a\right),\\
		\mathbf{B}_{ab34}&=4(-\xibar_a\xi_b+\xi_a\xibar_b-\eta\etabar_b+\etabar_a\eta_b),\\
		\mathbf{B}_{abcd}&=\chi_{bc}\chibar_{ad}+\chibar_{bc}\chi_{ad}-\chi_{ac}\chibar_{bd}+\chibar_{ac}\chi_{bd}.
	\end{align*}
	Then, for $U_A=U_{a_1\cdots a_k}$ a general horizontal $k$-tensor, we have the commutation identities
	\begin{align*}
		[\nabla_3,\nabla_b]U_A=&-\chibar_{bc}\nabla_c U_A+(\eta_b-\zeta_b)\nabla_3U_A+\xibar_b\nabla_4 U_A+\sum_{i=1}^k\left(-\in_{a_i c}\hodge{\betabar}_b+\frac12\mathbf{B}_{a_ic3b}\right)U_{a_1\cdots c\cdots a_k},\\
		[\nabla_4,\nabla_b]U_A=&-\chi_{bc}\nabla_c U_A+(\etabar_b+\zeta_b)\nabla_4U_A+\xi_b\nabla_3 U_A+\sum_{i=1}^k\left(\in_{a_i c}\hodge{\beta}_b+\frac12\mathbf{B}_{a_ic4b}\right)U_{a_1\cdots c\cdots a_k},\\
		[\nabla_4,\nabla_3]U_A=&2(\etabar_b-\eta_b)\nabla_b U_A+2\omega\nabla_3 U_A-2\omegabar\nabla_4 U_A+\sum_{i=1}^k\left(-\in_{a_i c}\hodge{\rho}+\frac12\mathbf{B}_{a_ic43}\right)U_{a_1\cdots c\cdots a_k}.
	\end{align*}
\end{lem}
\begin{proof}
	See \cite[Lemma 2.56]{KSwaveeq}.
\end{proof}

\begin{prop}\label{prop:commnabla}
Let $u\in\fraks_k(\R)$ with $k=0,1,2$. Then for $a=1,2$ and any set of indices $A=b_1\cdots b_k$ we have
\begin{align*}
	[\nabla_3,\nabla_a]u_A&=-\frac{1}{2}(tr\chibar\nabla_a u+\atrchibar\hodge{}\:\nabla_au)_A+(\eta_a-\zeta_a)\nabla_3u_A+\xibar_a\nabla_4u_A+\wh{\chibar}_{ac}\nabla_c u_A+B_k^{(3)}[U]_{aA},\\
		[\nabla_4,\nabla_a]u_A&=-\frac{1}{2}(tr\chi\nabla_a u+\atrchi\hodge{}\:\nabla_au)_A+(\etabar_a+\zeta_a)\nabla_4u_A+\xi_a\nabla_3u_A+\wh{\chi}_{ac}\nabla_c u_A+B_k^{(4)}[U]_{aA},
\end{align*}
	where $B^{(3)}_0[u]_a=B^{(4)}_0[u]_a=0$  for $k=0$,
	\begin{align*}
		B^{(3)}_1[u]_{ab}=&-\hodge{\betabar}_a\hodge{u}_b-\frac12(tr\chibar(\eta_b u_a-\delta_{ab}(\eta\cdot u))+\atrchibar(\eta_b\hodge{u}_a-\in_{ab}(\eta\cdot u)))\\
		&-\xibar_b\chi_{ac}u_c+\chi_{ab}\xibar\cdot u-\eta_b\wh{\chibar}_{ac}u_c+\wh{\chibar}_{ab}\eta\cdot u,\\
		B^{(4)}_1[u]_{ab}=&\hodge{\beta}_a\hodge{u}_b-\frac12(tr\chi(\etabar_b u_a-\delta_{ab}(\etabar\cdot u))+\atrchi(\etabar_b\hodge{u}_a-\in_{ab}(\etabar\cdot u)))\\
		&-\xi_b\chibar_{ac}u_c+\chibar_{ab}\xi\cdot u-\etabar_b\wh{\chi}_{ac}u_c+\wh{\chi}_{ab}\etabar\cdot u,
	\end{align*}
	for $k=1$, and
		\begin{align*}
		B^{(3)}_2[u]_{abc}=&-2\hodge{\betabar}_a\hodge{u}_b-\frac12tr\chibar(\eta_b u_{ac}+\eta_c u_{ab}-\delta_{ab}(\eta\cdot u)_c-\delta_{ac}(\eta\cdot u)_b)\\
		&-\frac12\atrchibar(\eta_b\hodge{u}_{ac}+\eta_cu_{ab}-\in_{ab}(\eta\cdot u)_c-\in_{ac}(\eta\cdot u)_b)\\
		&-\xibar_{[b}\chi_{ad}u_{dc]}+\chi_{a[b}(\xibar\cdot u)_{c]}-\eta_{[b}\wh{\chibar}_{ad}u_{dc]}+\wh{\chibar}_{a[b}(\eta\cdot u)_{c]},
\end{align*}
\begin{align*}
		B^{(4)}_2[u]_{abc}=&2\hodge{\beta}_a\hodge{u}_b-\frac12tr\chi(\etabar_b u_{ac}+\etabar_c u_{ab}-\delta_{ab}(\etabar\cdot u)_c-\delta_{ac}(\etabar\cdot u)_b)\\
		&-\frac12\atrchi(\etabar_b\hodge{u}_{ac}+\etabar_cu_{ab}-\in_{ab}(\etabar\cdot u)_c-\in_{ac}(\etabar\cdot u)_b)\\
		&-\xi_{[b}\chibar_{ad}u_{dc]}+\chibar_{a[b}(\xi\cdot u)_{c]}-\etabar_{[b}\wh{\chi}_{ad}u_{dc]}+\wh{\chi}_{a[b}(\etabar\cdot u)_{c]},
	\end{align*}
	for $k=2$. We extend the operators $B_k^{(3,4)}$ to $\fraks_k(\C)$ by defining $$B_k^{(3,4)}[u+i\hodge{u}]=B_k^{(3,4)}[u]+iB_k^{(3,4)}[\hodge{u}].$$
\end{prop}
\begin{proof}
See \cite[Lemma 2.57]{KSwaveeq}.
\end{proof}
\begin{prop}\label{prop:commnabladivc}
	Let $U\in\fraks_k(\mathbb{C})$, $k=1,2$. We have
	\begin{align}
		[\nabla_4,\divc]U=&-\frac{1}{2}\overline{trX}\divc U+(\overline{\Hbar+Z})\cdot\nabla_4 U+\overline{\Xi}\cdot\nabla_3 U-\wh{\chi}\cdot\overline{\mcd}U+D_4^{(k)}[U],\label{eq:commnab4divc}\\
		[\nabla_3,\divc]U=&-\frac{1}{2}\overline{tr\Xbar}\divc U+(\overline{H-Z})\cdot\nabla_3 U+\overline{\Xibar}\cdot\nabla_4 U-\wh{\chibar}\cdot\overline{\mcd}U+D_3^{(k)}[U],\label{eq:commnab3divc}
	\end{align}
	where
	\begin{align*}
	D_4^{(k)}[U]=&\frac{k}{2}\overline{trX}\overline{\Hbar}\cdot U+k\overline{B}\cdot U+\overline{\Xi}\cdot\chibar\cdot U+\Hbar\cdot\chihat\cdot U,\nn\\
	D_3^{(k)}[U]=&\frac{k}{2}\overline{tr\Xbar}\overline{H}\cdot U-k\overline{\Bbar}\cdot U+\overline{\Xibar}\cdot\chi\cdot U+H\cdot\wh{\chibar}\cdot U\nn,
\end{align*}
where the terms of the type $A\cdot B\cdot C$ denote arbitrary metric contractions between $A,B$ and $C$.
\end{prop}
\begin{proof}
	See \cite[Lemma 4.7]{KSwaveeq}.
\end{proof}
\begin{prop}\label{prop:commnab34mcdhot}
	Let $F\in\fraks_1(\C)$. We have
	\begin{align}
		[\nabla_4,\mcd\hot]F&=-\frac{1}{2}trX\mcd\hot F+(\Hbar+Z)\hot\nabla_4 F+\Xi\hot\nabla_3 F-\frac12\wh{X}\cdot\overline{\mcd}F+E_4[F],\label{eq:commnab4mcdhot}\\
		[\nabla_3,\mcd\hot]F&=-\frac{1}{2}tr\Xbar\mcd\hot F+(H-Z)\hot\nabla_3 F+\Xibar\hot\nabla_4 F-\frac12\wh{\Xbar}\cdot\overline{\mcd}F+E_3[F],\label{eq:commnab3mcdhot}
	\end{align}
	where
	\begin{align*}
		E_4[F]&=-\frac{1}{2}trX\Hbar\hot F-B\hot F-\Xi\hot(\chibar\cdot F)+\frac12\wh{X}(\overline{\Hbar}\cdot F)+\wh{\Xbar}(\xi\cdot F),\\
		E_3[F]&=-\frac{1}{2}tr\Xbar H\hot F+\Bbar\hot F-\Xibar\hot(\chi\cdot F)+\frac12\wh{\Xbar}(\overline{H}\cdot F)+\wh{X}(\xibar\cdot F).
	\end{align*}
	\begin{proof}
		See \cite[Lemma 4.7]{KSwaveeq}.
	\end{proof}
\end{prop}
\begin{prop}\label{prop:commnab34mcd}
	Let $h\in\fraks_0(\C)$ be a scalar function. Then we have  
	\begin{align*}
		[\nabla_4,\mcd]h&=-\frac{1}{2}trX\mcd h+(\Hbar+Z)\nabla_4h-\frac{1}{2}\wh{X}\cdot\overline{\mcd}h+\Xi\nabla_3h,\\
		[\nabla_3,\mcd]h&=-\frac{1}{2}tr\Xbar\mcd h+(H-Z)\nabla_4h-\frac{1}{2}\wh{\Xbar}\cdot\overline{\mcd}h+\Xibar\nabla_4h.
	\end{align*}
\end{prop}
\begin{proof}
		See \cite[Lemma 4.7]{KSwaveeq}.
\end{proof}
\begin{lem}\label{lem:comm34}
	Let ${U}\in s_k(\C)$, $k\in\{0,1,2\}$. We have 
	$$[\nabla_3,\nabla_4]{U}=-2\omega\nabla_3{U}+2\omegabar\nabla_4{U}+2(\eta-\etabar)\cdot\nabla{U}+C_k[{U}]+\err_{34}[{U}],$$
	where $C_k[{U}]$ is the following $k$-dependent $0$-order operator,
	\begin{equation}\label{eq:ckdeu}
		\begin{aligned}
			&C_0[{U}]=0,\quad C_1[{U}]=\Real(\overline{\Hbar}\cdot {U})H-\Real(\overline{H}\cdot {U})\Hbar-2i\hodge{\rho}{{U}}\\
			&C_2[{U}]=H\hot (\overline{\Hbar}\cdot {U})-\Hbar\hot (\overline{H}\cdot {U})-4i\hodge{\rho}{U},
		\end{aligned}
	\end{equation}
	and where the error term takes the schematic form 
	$$\err_{34}[{U}]=\Xi\cdot\Xibar\cdot{U}.$$
\end{lem}
\begin{proof}
	This is the complexified version of \cite[Lemma 2.56]{KSwaveeq}.
\end{proof}

\subsubsection{Main equations in complex form}\label{section:ricciandcurvcomplex}
\begin{defi}\label{defi:defcomplexquantities}
	We define the following complex anti-self dual tensors: 
	$$A=\alpha+i\:\hodge{\alpha},\quad B=\beta+i\:\hodge{\beta},\quad  P=\rho+i\:\hodge{\rho},\quad\underline{B}=\underline{\beta}+i\:\hodge{\underline{\beta}},\quad \underline{A}=\underline{\alpha}+i\:\hodge{\underline{\alpha}},$$
	and 
	\begin{align*}
		&X=\chi+i\:\hodge{\chi},\quad\underline{X}=\underline{\chi}+i\:\hodge{\underline{\chi}},\quad H=\eta+i\:\hodge{\eta},\quad\underline{H}=\underline{\eta}+i\:\hodge{\underline{\eta}},\quad Z=\zeta+i\:\hodge{\zeta},\\
		&\Xi=\xi+i\:\hodge{\xi},\quad\underline{\Xi}=\underline{\xi}+i\:\hodge{\underline{\xi}}.
	\end{align*}
	Note that we have the identities
	$$trX=tr\chi-i\atrchi,\quad \wh{X}=\wh{\chi}+i\:\hodge{\chihat},\quad tr\Xbar=tr\chibar-i\atrchibar,\quad \wh{\Xbar}=\wh{\chibar}+i\:\hodge{\wh{\chibar}}.$$
\end{defi}
The null structure equations satisfied by the Ricci coefficients can be rewritten in a more compact form using the complex notations.
\begin{prop}\label{prop:nullstructurecomplex}
	We have in a vacuum spacetime,
	\begin{align*}
		\nabla_3 tr\Xbar+\frac{1}{2}(tr\Xbar)^2+2\omegabar tr\Xbar&=\mcd\cdot\overline{\Xibar}+\Xibar\cdot\overline{\Hbar}+\overline{\Xibar}\cdot(H-2Z)-\frac{1}{2}\wh{\Xbar}\cdot\overline{\wh{\Xbar}},\\
		\nabla_3\wh{\Xbar}+\Real(tr\Xbar)\wh{\Xbar}+2\omegabar\wh{\Xbar}&=\frac{1}{2}\mcd\hot\Xibar+\frac{1}{2}\Xibar\hot(H+\Hbar-2Z)-\Abar,
	\end{align*}
	\begin{align*}
		\nabla_3 trX+\frac{1}{2}tr\Xbar trX-2\omegabar trX&=\mcd\cdot\overline{H}+H\cdot\overline{H}+2P+\Xibar\cdot\overline{\Xi}-\frac{1}{2}\wh{\Xbar}\cdot\overline{\wh{X}},\\
		\nabla_3\wh{X}+\frac{1}{2}tr\Xbar\wh{X}-2\omegabar\wh{X}&=\frac{1}{2}\mcd\hot H+\frac{1}{2}H\hot H-\frac{1}{2}\overline{tr X}\wh{\Xbar}+\frac{1}{2}\Xibar\cdot\Xi,
	\end{align*}
	\begin{align*}
		\nabla_4 tr\Xbar+\frac{1}{2}trXtr\Xbar-2\omega tr\Xbar&=\mcd\cdot\overline{\Hbar}+\Hbar\cdot\overline{\Hbar}+2\overline{P}+\Xi\cdot\overline{\Xibar}-\frac{1}{2}\wh{X}\cdot\overline{\wh{\Xbar}},\\
		\nabla_4\wh{\Xbar}+\frac{1}{2}trX \wh{\Xbar}-2\omega\wh{\Xbar}&=\frac{1}{2}\mcd\hot\overline{\Hbar}+\frac{1}{2}\Hbar\hot\Hbar-\frac{1}{2}\overline{tr\Xbar}\wh{X}+\frac{1}{4}\Xi\hot\Xibar,
	\end{align*}
	\begin{align*}
		\nabla_4 trX+\frac{1}{2}(trX)^2+2\omega trX&=\mcd\cdot\overline{\Xi}+\Xi\cdot\overline{H}+\overline{\Xi}\cdot(\Hbar+2Z)-\frac{1}{2}\wh{X}\cdot\overline{\wh{X}},\\
		\nabla_4\wh{X}+\Real(trX)\wh{X}+2\omega\wh{X}&=\frac{1}{2}\mcd\hot\Xi+\frac{1}{2}\Xi\hot(H+\Hbar+2Z)-A.
	\end{align*}
	Also, 
	\begin{align*}
		\nabla_3Z+\frac12tr\Xbar(Z+H)-2\omegabar(Z-H)&=-2\mcd\omegabar-\frac12\wh{\Xbar}\cdot(\overline{Z+H})\\
		&\quad+\frac12trX\Xibar+2\omega\Xibar-\Bbar+\frac12\overline{\Xibar}\cdot\wh{X},\\
		\nabla_4Z+\frac12trX(Z-\Hbar)-2\omega(Z+\Hbar)&=2\mcd\omega+\frac12\wh{X}\cdot(\overline{-Z+\Hbar})\\
		&\quad-\frac12tr\Xbar\Xi-2\omegabar\Xi-B-\frac12\overline{\Xi}\cdot\wh{\Xbar},\\
		\nabla_3\Hbar-\nabla_4\Xibar&=-\frac12\overline{tr\Xbar}(\Hbar-H)-\frac12\wh{\Xbar}\cdot(\overline{\Hbar-H})-4\omega\Xibar+\Bbar,\\
		\nabla_4 H-\nabla_3\Xi&=-\frac12\overline{trX}(H-\Hbar)-\frac12\wh{X}\cdot(\overline{H-\Hbar})-4\omegabar\Xi-B,
	\end{align*}
and 
$$\nabla_3\omega+\nabla_4\omegabar-4\omega\omegabar-\xi\cdot\xibar-(\eta-\etabar)\cdot\xi+\eta\cdot\etabar=\rho.$$
Also,
\begin{align*}
	\frac12\divc\wh{X}+\frac12\wh{X}\cdot\overline{Z}&=\frac12\mcd\overline{trX}+\frac12\overline{trX}Z-i\Imag(trX)H-i\Imag(tr\Xbar)\Xi-B,\\
	\frac12\divc\wh{\Xbar}-\frac12\wh{\Xbar}\cdot\overline{Z}&=\frac12\mcd\overline{tr\Xbar}-\frac12\overline{tr\Xbar}Z-i\Imag(tr\Xbar)\Hbar-i\Imag(trX)\Xibar+\Bbar,
\end{align*}
and
$$\curl\zeta=-\frac12\chihat\wedge\wh{\chibar}+\frac14(tr\chi\atrchibar-tr\chibar\atrchi)+\omega\atrchibar-\omegabar\atrchi+\hodge{\rho}.$$
\end{prop}
The complex notations also allow us to rewrite the Bianchi identities as follows.
\begin{prop}\label{prop:bianchicomplex}
	We have in a vacuum spacetime,
	\begin{align*}
		\nabla_3 A+\frac{1}{2} tr \underline{X} A-4 \underline{\omega} A & =\frac{1}{2}\mathcal{D} \widehat{\otimes} B+\frac{1}{2}(Z+4 H) \widehat{\otimes} B-3 \overline{P} \widehat{X}, \\
		\nabla_4 B+2 \overline{tr X} B+2 \omega B & =\frac{1}{2} \overline{\mathcal{D}} \cdot A+\frac{1}{2} A \cdot(\overline{2 Z+\underline{H}})+3 \overline{P} \Xi, \\
		\nabla_3 B+tr \underline{X} B-2 \underline{\omega} B & =\mathcal{D} \overline{P}+\underline{\overline{B}} \cdot \widehat{X}+3 \overline{P} H+\frac{1}{2} A \cdot \overline{\underline{\Xi}}, \\
		\nabla_4 P+\frac{3}{2} tr X P & =\frac{1}{2} \mathcal{D} \cdot \overline{B}+\frac{1}{2}(2 \underline{H}+Z) \cdot \overline{B}-\overline{\Xi} \cdot \underline{B}-\frac{1}{4} \underline{\widehat{X}} \cdot \overline{A}, \\
		\nabla_3 P+\frac{3}{2} \overline{tr \underline{X}} P & =-\frac{1}{2} \overline{\mathcal{D}} \cdot \underline{B}-\frac{1}{2}(\overline{2 H-Z}) \cdot \underline{B}+\underline{\Xi} \cdot \overline{B}-\frac{1}{4} \overline{\widehat{X}} \cdot \underline{A}, \\
		\nabla_4 \underline{B}+trX \underline{B}-2 \omega \underline{B} & =-\mathcal{D} P+\overline{B} \cdot \underline{\widehat{X}}-3 P \underline{H}-\frac{1}{2} \underline{A} \cdot \overline{\Xi}, \\
		\nabla_3 \underline{B}+2 \overline{tr \underline{X}} \underline{B}+2 \underline{\omega} \underline{B} & =-\frac{1}{2} \overline{\mathcal{D}} \cdot \underline{A}-\frac{1}{2} \underline{A} \cdot(\overline{-2 Z+H})-3 P \underline{\Xi}, \\
		\nabla_4 \underline{A}+\frac{1}{2} tr X \underline{A}-4 \omega \underline{A} & =-\frac{1}{2} \mathcal{D} \widehat{\otimes} \underline{B}+\frac{1}{2}(Z-4 \underline{H}) \hot\underline{B}-3 P \wh{\underline{X}} .
	\end{align*}
	
\end{prop}

\subsection{Teukolsky equations in the non-linear setting}
\begin{prop}\label{prop:teuk}
	In a vacuum spactime, for any null pair $(e_3,e_4)$ with associated horizontal distribution, the horizontal tensor $A$ satisfies 
	\begin{align}\label{eq:teukA}
		\mcl(A)=\err[\mcl(A)],
	\end{align}
	where the operator $\mcl$ is defined by 
	\begin{equation}\label{eq:teukop}
		\begin{aligned}
			\mcl(A):=&\nabla_4\nabla_3 A-\frac{1}{4}\mcd\hot\parentheses{\overline{\mcd}\cdot A}+\parentheses{\frac{1}{2}tr X+2\overline{tr X}+2\omega}\nabla_3 A+\parentheses{\frac{1}{2}tr \underline{X}-4\omegabar} \nabla_4A\\
			&-\parentheses{2Z+\Hbar+\overline{2Z+\Hbar}+4H}\cdot\nabla A- H\hot\parentheses{(\overline{2 Z+\underline{H}})\cdot A}+VA
		\end{aligned}
	\end{equation}
	with potential 
	\begin{equation}	\label{eq:teukpotential}
		\begin{aligned}
			V:=\overline{tr X}tr\Xbar-&2\omegabar trX+2\omega tr\Xbar-8\omegabar(\omega+\overline{tr X})\\
			&-\mcd\cdot\overline{Z}-2Z\cdot\overline{Z}-2\Real(Z\cdot\overline{\Hbar})-2\overline{P}-4\nabla_4\omegabar,
		\end{aligned}
	\end{equation}
	and where the error term is 
	\begin{align}
		&\err[\mcl(A)]\nn:=\\
		&2\parentheses{\Xi \widehat{\otimes} \nabla_3 B+\nabla_3\Xi\hot B}+\frac{3}{2}\parentheses{tr \underline{X} B-2 \underline{\omega} B-\underline{\overline{B}} \cdot \widehat{X}-\frac{1}{2} A \cdot \overline{\underline{\Xi}}}\hot\Xi\nn\\
		&+\frac{1}{2}\parentheses{-B \widehat{\otimes} B-\frac{1}{2}tr \Xbar\Xi \hot B-\frac{1}{2}(\divc B)\wh{X}+\frac{1}{2}\widehat{X}(\overline{\Hbar} \cdot B)+\frac{1}{2}\underline{\widehat{X}}(\overline{\Xi} \cdot B)+\frac{1}{2}\Xi\hot(\overline{\wh{\Xbar}}\cdot B)}\nn\\
		&-\frac{1}{2}\parentheses{\Xi\cdot\overline{\Xibar}-\frac{1}{2}\wh{X}\cdot\overline{\wh{\Xbar}}}A-3\parentheses{\frac{1}{2} \overline{\mathcal{D}} \cdot {B}+\frac{1}{2}(2 \overline{\underline{H}+Z}) \cdot {B}-{\Xi} \cdot \overline{\underline{B}}}\wh{X}+\frac{3}{4}\parentheses{\overline{\wh{\Xbar}}\cdot A}\wh{X}\nn\\
		&+\frac{1}{2}\parentheses{-2\wh{X}\cdot(\overline{H}-\overline{\Hbar})-16\omegabar\Xi-5B+\frac{1}{2}\wh{X}\cdot(-\overline{Z}+\overline{\Hbar})-\frac{1}{2}tr\Xbar\Xi-2\omegabar\Xi-\frac{1}{2}\overline{\Xi}\wh{\Xbar}}\hot B\nn\\
		&-\parentheses{\overline{\mcd}\cdot\wh{X}+\wh{X}\cdot\overline{Z}+2i\Imag(tr\Xbar)\Xi+2B}\hot B.\label{eq:teukerror}
	\end{align}
\end{prop}
\begin{proof}
This is the analog to the Teukolsky equation stated in \cite[Proposition 5.1]{KSwaveeq} with conformal derivatives (which we do not use in the present work), for completeness we provide a derivation of the equation above in Appendix \ref{appendix:teukderivation}.
\end{proof}
\begin{rem}\label{rem:boundteukerr}We make the following observations:
	\begin{itemize}
		\item The error term $\err[\mcl(A)]$ is bounded by a quadratic expression in $B,\wh{X},\wh{\Xbar},\Xi,\Xibar,A$ with factors $H,\Hbar,\omegabar,$ $Z,tr\Xbar,$ where the only derivatives that appear are $\divc B$, $\divc\wh{X},\nabla_3B,\nabla_3\Xi$.
		\item By Lemma \ref{lem:monamitoitcool}, denoting $V_0':=V-4(\eta\cdot(2\zeta+\etabar)-i\eta\wedge(2\zeta+\etabar))$, then the zero-order term of the Teukolsky operator \eqref{eq:teukop} rewrites
		$$- H\hot\parentheses{(\overline{2 Z+\underline{H}})\cdot A}+VA=V_0' A.$$
	\end{itemize}
	
\end{rem}
\subsection{Schematic notations and conventions}\label{section:conventions}
\noindent\textbf{Conventions.} Our conventions are as follows:
\begin{itemize}
	\item We use the Einstein summation convention and the usual notation to raise and lower  indices with the metric. Greek letters $\alpha,\beta,\mu,\nu,\ldots$ go from 1 to 4 while latin letters $a,b,c,d,\ldots$ and $A,B,C,D,\ldots$ go from 1 to 2, except $i,j,k$ which go from 1 to 3.
	\item We use respectively the notations $A\lesssim B$ or $A=O(B)$ whenever $A\leq C B$ where $C>0$ is a constant which depends only on the black hole parameters $(a,M)$. We also denote $A\lesssim_Q B$ or $A=O_Q(B)$ when $A\leq C B$ where $C>0$ depends on $(a,M)$ and some quantity $Q$. We write $A\sim B$ when $A\lesssim B$ and $B\lesssim A$, and $A\sim_Q B$ when $A\lesssim_Q B$ and $B\lesssim_Q A$. 
	\item For any function $f$ and open set or causal hypersurface $\Sigma\subset\mcm$, we denote $\int_\Sigma f=\int_\Sigma f\mathrm{vol}_\Sigma$ where $\mathrm{vol}_\Sigma$ is the volume form on $\Sigma$ induced by the ambiant spacetime $(\mcm,\g)$.
	\item By ‘LHS’ and ‘RHS’ we mean respectively ‘left-hand side’ and ‘right-hand side’.
	\item If $P$ is an operator acting on some tensorial quantity $U$, for any norm $\|\cdot\|$ and integer $k\geq 0$ we use the notation
	$$\|P^{\leq k}U\|=\sum_{i=0}^k\|P^iU\|.$$
	\item For any tensor $U$, we use the square bracket notation to denote anti-symmetrization with respect to specific indices $b_i,b_j$:
	\begin{align*}
		U_{b_1\cdots b_{i-1}[b_i b_{i+1}\cdots b_{j-1} b_j] b_{j+1}\cdots b_k }:=U_{b_1\cdots b_k }-U_{b_1\cdots b_{i-1}b_j b_{i+1}\cdots b_{j-1} b_i b_{j+1}\cdots b_k }.
	\end{align*}
	\item In the context of a summation with respect to $i$, for any indices $A_1,\ldots,A_k,B$, at rank $i$ we denote $A_1\cdots B\cdots A_k$ the indices $A_1\cdots A_k$ where $A_i$ is replaced with $B$. For instance,
	$$\sum_{i=1}^3U_{A_1\cdots B\cdots A_3}=U_{BA_2A_3}+U_{A_1BA_3}+U_{A_1A_2B}.$$
\end{itemize}
\noindent\textbf{Schematic notations.}
We use the schematic notation introduced in Section 3.2 in \cite{stabC0} which we recall here: We will write the \textit{schematic equations} with the symbol $=_s$ with the following conventions:
\begin{itemize}
	\item We keep the exact constants on the LHS of $=_s$ but we do not keep track of the exact constants on the RHS of $=_s$.
	\item Any metric contraction between tensors $U,V$ (which could be horizontal or $S(u,\ubar)$-tangent in a double null gauge) is denoted on the RHS of $=_s$ as $UV$ or $U\cdot V$.
	\item We use brackets as follows: for any quantities $U,V,W$, we denote $U(V,W)$ the sum of all terms of the form $UV$ or $UW$.
	\item There may be extra terms on the RHS of $=_s$ which are not present in the original equation.
	\item We suppress the rank of tensors in the RHS of $=_s$.
\end{itemize}
Notice that as a consequence of Proposition \ref{prop:nablagammahzero}, we can take covariant derivatives of the schematic notations: if $U,V$ are some horizontal tensors (with respect to any null pair $(e_3,e_4)$) such that $U=_sV$, then $\nabla_3 U=_s\nabla_3V$, $\nabla_4U=_s\nabla_4V$, and $\nabla U=_s\nabla V$. Note that some care will have to be taken when taking horizontal Lie derivatives of schematic equations\footnote{Some extra terms involving the Lie derivative of the metric will appear, see Section \ref{section:schematicnotations} in the double null gauge.}.
\\\\
\noindent\textbf{Reduced schematic notations.} Similarly, we will use the reduced schematic notation introduced in Section 7.1 of \cite{stabC0} : We will write the \textit{reduced schematic equations} with the symbol $=_{rs}$ with the following conventions:
\begin{itemize}
	\item All the conventions for the schematic equations are also used for the reduced schematic equations.
	\item All \emph{factors} on the RHS of the symbol $=_{rs}$ which are bounded by a constant which only depends on $a,M$ in $L^\infty$ in the region considered will not be written.
\end{itemize}
Note that reduced schematic equations cannot be differentiated.
\section{Choice of gauges}\label{section:choiceofgauges}

\subsection{Coordinates and principal frames in the Kerr spacetime}

\subsubsection{Boyer-Lindquist and Eddington-Finkelstein coordinates in Kerr}\label{section:BLEFKerr}
We first consider the Kerr metric in standard Boyer-Lindquist (B-L) coordinates $(t,r,\theta,\phi)$,
$$\g_{a,M}=\frac{a^2\sin^2\theta-\Delta}{|q|^2}\dee t^2-\frac{4aMr}{|q|^2}\sin^2\theta\dee t\dee\phi+\frac{|q|^2}{\Delta}\dee r^2+|q|^2\dee\theta^2+\frac{(r^2+a^2)^2-a^2\sin^2\theta\Delta}{|q|^2}\sin^2\theta\dee\phi^2,$$
where $a\in\R\backslash\{0\}$, $M>0$ are the black hole parameters, and where 
\begin{align*}
	\Delta=r^2+a^2-2Mr,\quad\quad q=r+ia\cos\theta,
\end{align*}
with $i$ being the imaginary unit. We also define the scalar
$$R^2=r^2+a^2+\frac{2Ma^2r\sin^2\theta}{|q|^2}=\frac{(r^2+a^2)^2-a^2\sin^2\theta\Delta}{|q|^2}.$$
We recall that $\g_{a,M}$ is a solution of the Einstein vacuum equation \eqref{eq:EVE}, and that in the subextremal setting, i.e. $|a|<M$, the scalar $\Delta=(r-r_-)(r-r_+)$ has two positive distincts roots
\begin{align}\label{eq:rplusoumoins}
	r_-:=M-\sqrt{M^2-a^2},\quad r_+:=M-\sqrt{M^2-a^2},\quad r_-<r_+.
\end{align}
\textbf{In this article, we only consider the subextremal setting such that $0<|a|<M$.} The Kerr black hole interior is the region $\{r_-\leq r\leq r_+\}$, where the scalar $\Delta$ satisfies $$\Delta\leq 0,$$ and $\Delta$ only vanishes at $r=r_{\pm}$. We choose the time orientation on the Kerr black hole interior such that $-\partial_r$ is future directed, and define the tortoise coordinate $r^*$ on $(r_-,r_+)$ by 
\begin{align}\label{eq:defderstarvraiment}
	\frac{\dee r^*}{\dee r}=\frac{r^2+a^2}{\Delta},\quad\quad r^*(M)=0.
\end{align}
Notice that $r^*\to -\infty$ as $r\to r_+$ and $r^*\to +\infty$ as $r\to r_-$. More precisely we have
\begin{align}\label{eq:rstarexpre}
	r^*=r+\frac{1}{2\kappa_+}\log|r-r_+|+\frac{1}{2\kappa_-}\log|r-r_-|-M-\frac{1}{2\kappa_+}\log|M-r_+|-\frac{1}{2\kappa_-}\log|M-r_-|,
\end{align}
where the surface gravities of the event and Cauchy horizon $\kappa_+$ and $\kappa_-$ are defined as
\begin{align}\label{eq:surfacegravities}
	\kappa_+:=\frac{r_+-r_-}{4Mr_+}>0,\quad\quad\kappa_-:=\frac{r_--r_+}{4Mr_-}<0.
\end{align}
Note that \eqref{eq:rstarexpre} implies the following identity for $r_-< r< r_+$,
\begin{align}\label{eq:deltaenexp}
	-\Delta=e^{-2|\kappa_-|r^*}e^{h_-(r)},
\end{align}
where the function $h_-(r)$ admits a finite limit as $r\to r_-$. 

\noindent\textbf{Eddington-Finkelstein coordinates in Kerr.} We define the advanced and retarded times
\begin{align}\label{eq:defuubarEFkerr}
	\ubar:=r^*+t,\quad\quad u:=r^*-t.
\end{align}
We also define the scalar $r_{mod}$ by
$$\frac{\dee r_{mod}}{\dee r}=\frac{a}{\Delta},\quad\quad r_{mod}(M)=0,$$
which can be expressed more precisely as
\begin{align}\label{eq:rmodexpre}
	r_{mod}=\frac{a}{r_+-r_-}\log\left|\frac{r-r_+}{r-r_-}\right|,
\end{align}
as well as the renormalized angular coordinates
$$\phi_+=\phi+r_{mod},\quad\quad \phi_-=\phi-r_{mod}.$$
In the present paper we are interested in a result localized to a neighborhood of timelike infinity $i_+$, so for now we restrict the analysis to the region 
\begin{align}\label{eq:restrictanalyiai}
	\mcm=\{\ubar> 1\}\cap\{u<-1\}.
\end{align}
Then, the B-L coordinate singularities of the Kerr metric at $r=r_{\pm}$ can be removed by considering the more adapted Eddington-Finkelstein (E-F) coordinates systems 
$$(r,\ubar,\theta,\phi_+),\quad\quad (r,u,\theta,\phi_-).$$
More precisely,  we can define the \textit{Cauchy horizon}\footnote{Actually, $\ch$ is the right part of the full bifurcate Cauchy horizon $\ch\cup\mathcal{CH}'_+=\{r=r_-\}$ but since we rescrit the analysis to $\mcm$ as in \eqref{eq:restrictanalyiai}, we can call $\ch$ the Cauchy horizon without risk of confusion. Similar remarks apply to the event horizon $\mch_+$.}
$$\ch=\mcm\cap \{r=r_-\}$$
by using the \emph{outgoing} E-F coordinates $(r,u,\theta,\phi_-)$ in which the metric is smooth at $\mcm\cap\{r=r_-\}$. Similarly, we can define the \textit{event horizon}
$$\eh=\mcm\cap\{r=r_+\}$$
by using the \emph{ingoing} E-F coordinates $(r,\ubar,\theta,\phi_+)$ in which the metric is smooth at $\mcm\cap\{r=r_+\}$. We also have
$$\ch=\mcm\cap\{\ubar=+\infty\},\quad \mch_+=\mcm\cap\{u=-\infty\}.$$
\noindent\textbf{Spacetime volume form.} In coordinates $(u,\ubar,\theta,\phi_{\pm})$, the spacetime volume form is
\begin{align}\label{eq:kerrvolumeformexact}
	\mathrm{vol}_{\g_{a,M}}=\frac{|q|^2|\Delta|}{2(r^2+a^2)}\sin\theta\dee\theta\dee\phi_{\pm}\dee u\dee\ubar.
\end{align}

\noindent\textbf{Metric induced on the spheres $S(r,\ubar)$.} The Riemannian metric induced by $\g_{a,M}$ on the spheres $S(r,\ubar)$ of constant $r$ and $\ubar$ is
\begin{align}\label{eq:metriquesphereI}
	\gslash_{a,M}=|q|^2\dee\theta^2+\frac{(r^2+a^2)^2-a^2\sin^2\theta\Delta}{|q|^2}\sin^2\theta\dee\phi_+^2,
\end{align}
in coordinates $(\theta,\phi_+)$. We denote
\begin{align}\label{eq:defJplusJmoins}
	\Jplus=\sin\theta\cos\phi_+,\quad\quad \Jmoins=\sin\theta\sin\phi_+.
\end{align}
Note that the coordinate system $(r,\ubar,x^1_{(1)}=\theta,x^2_{(1)}=\phi_+)$ is regular away from the poles, for instance on the set
$$\mcu^{(1)}=\{\pi/4<\theta<3\pi/4\},$$
while the coordinate system $(r,\ubar,x^1_{(2)}=x^1_{(3)}=\Jplus,x^2_{(2)}=x^2_{(3)}=\Jmoins)$ is regular on $\mcu^{(2)}\cup\mcu^{(3)}$, where
$$\mcu^{(2)}=\{0\leq\theta<\pi/3\},\quad\quad\mcu^{(3)}=\{2\pi/3<\theta\leq\pi\}.$$
 The metric $\gslash_{a,M}$ takes the following form in the $x^1,x^2$ coordinate system (see \cite[Lem. 2.46]{KS21}),
\begin{align}
	\gslash_{a,M}=|q|^2\Bigg[&\left(\frac{1-(x^2)^2}{1-|x|^2}+\frac{a^2(x^2)^2}{|q|^2}\left(1+\frac{2Mr}{|q|^2}\right)\right)(\dee x^1)^2\nn\\
	&+\left(\frac{2 x^1 x^2}{1-|x|^2}-\frac{2a x^1 x^2}{|q|^2}\left(1+\frac{2Mr}{|q|^2}\right)\right)\dee x^1\dee x^2\label{eq:metricsphx1x2}\\
	&+\left(\frac{1-(x^1)^2}{1-|x|^2}+\frac{a^2(x^1)^2}{|q|^2}\left(1+\frac{2Mr}{|q|^2}\right)\right)(\dee x^2)^2\Bigg].\nn
\end{align}
\noindent\textbf{Kerr metric in ingoing EF coordinates.} The Kerr metric is given in the $(r,\ubar,\theta,\phi_+)$ coordinate system by
\begin{align}\label{eq:kerrmetricEF}
	\g_{a,M}=&-\left(1-\frac{2M}{|q|^2}\right)\dee \ubar^2+2\dee r\dee\ubar-2a\sin^2\theta\dee r\dee\phi_+-\frac{4Mra\sin^2\theta}{|q|^2}\dee\ubar\dee\phi_++\gslash_{a,M},
\end{align}
and in the $(r,\ubar,x^1,x^2)$ coordinate system by
\begin{equation}\label{eq:kerrmetricx1x2}
	\begin{aligned}
		\g_{a,M}=&-\left(1-\frac{2M}{|q|^2}\right)\dee \ubar^2+2\dee r\dee\ubar+2a\dee r\left(x^2\dee x^1-x^1\dee x^2\right)\\
		&-\frac{4Mra}{|q|^2}\dee\ubar\left(-x^2\dee x^1+x^1\dee x^2\right)+\gslash_{a,M}.
	\end{aligned}
\end{equation}

\subsubsection{Principal null frames}
In the Kerr black hole interior, the \textit{ingoing} principal null pair is defined by,
\begin{align}
	e_4=\frac{r^2+a^2}{|q|^2}\partial_t+\frac{\Delta}{|q|^2}\partial_r+\frac{a}{|q|^2}\partial_\phi,\quad\quad e_3=\frac{r^2+a^2}{\Delta}\partial_t-\partial_r+\frac{a}{\Delta}\partial_\phi,\label{eq:principalingoinkerr}
\end{align}
 in B-L coordinates, while the \textit{outgoing} principal null pair is given by
\begin{align}
	e_4'=-\frac{|q|^2}{\Delta}e_4=-\frac{r^2+a^2}{\Delta}\partial_t-\partial_r-\frac{a}{\Delta}\partial_\phi,\quad e_3'=-\frac{\Delta}{|q|^2}e_3=-\frac{r^2+a^2}{|q|^2}\partial_t+\frac{\Delta}{|q|^2}\partial_r-\frac{a}{|q|^2}\partial_\phi.\label{eq:principaloutgoinkerr}
\end{align}
\begin{rem}
	Notice that the normalization of the outgoing principal null pair differs by a factor $-1$ from the one used in many references studying the stability of the exterior of Kerr spacetime, for example \cite{KSwaveeq}. This is because $\Delta\leq 0$ in the black hole interior which implies that the pair as defined in \eqref{eq:principaloutgoinkerr} is future-directed.
\end{rem}
Both the ingoing and outgoing principal null pairs are completed with the following horizontal frame (which is defined everywhere but on the axis $\sin\theta=0$),
\begin{align}\label{eq:horizontalframekerr}
	e_1=\frac{1}{|q|}\partial_\theta,\quad e_2=\frac{a\sin\theta}{|q|}\partial_t+\frac{1}{|q|\sin\theta}\partial_\phi.
\end{align}
We define the \textit{canonical complex 1-form} $\frakJ$ as follows
\begin{align}\label{eq:deffrakJkerr}
	\frakJ_1=\frac{i\sin\theta}{|q|},\quad\quad \frakJ_2=\frac{\sin\theta}{|q|},
\end{align}
with respect to the horizontal frame \eqref{eq:horizontalframekerr}. Note that $\frakJ$ is smooth including on the axis and satisfies $\hodge{\frakJ}=-i\frak{J}$. Moreover, $\frakJ=j+i\hodge{j}$ where
$$j_1=0,\quad\quad j_2=\frac{\sin\theta}{|q|}.$$
\noindent\textbf{Ricci and curvature coefficients in the ingoing principal frame.} Defined with respect to the ingoing principal pair \eqref{eq:principalingoinkerr}, the Ricci coefficients as defined in Section \ref{section:ricciandcurvcomplex} are
\begin{align}
	\wh{X}=\wh{\Xbar}=\Xi=\Xibar=\omegabar=0,\quad H=Z=\frac{aq}{|q|^2}\frakJ,\quad\Hbar=-\frac{a\qbar}{|q|^2}\frakJ,\quad trX=\frac{2\Delta\qbar}{|q|^4},\quad tr\Xbar=-\frac{2}{\qbar},\label{eq:tracesxxbarkerr}
\end{align}
and
$$\omega=-\frac12\partial_r\left(\frac{\Delta}{|q|^2}\right)=\frac{M-r}{|q|^2}+\frac{r\Delta}{|q|^4}.$$
\begin{rem}\label{rem:redshiftt}
	Note that for $\delta_-$ small, we have $\omega\gtrsim 1$ in the region $\{r_-\leq r\leq r_-(1+\delta_-)\}$. This is a manifestation of the blueshift effect at $\ch$ and will be used to control solutions of the Teukolsky equation in the \textit{ingoing} principal frame near $\ch$, similarly as in \cite{spin+2}. 
\end{rem}
The curvature coefficients defined with respect to the ingoing principal pair \eqref{eq:principalingoinkerr} are
\begin{align}\label{eq:curvingoingKerr}
	A=B=\Bbar=\Abar=0,\quad\quad P=-\frac{2M}{q^3}.
\end{align}
\begin{rem}
	These identities show that the RHS \eqref{eq:teukerror} of the Teukolsky equation in the principal frame is indeed an error term as it is quadratic with respect to quantities which vanish in exact Kerr.
\end{rem}
\noindent\textbf{Ricci and curvature coefficients in the outgoing principal frame.} Defined with respect to the outgoing principal pair \eqref{eq:principalingoinkerr} of Kerr spacetime, the Ricci coefficients defined in Sections \ref{section:ricciandcurvdef}, \ref{section:ricciandcurvcomplex} write
\begin{align}
	\wh{X}'=\wh{\Xbar}'=\Xi'=\Xibar'=\omega'=0,\quad H'=\frac{aq}{|q|^2}\frakJ,\quad\Hbar'=-Z'=-\frac{a\qbar}{|q|^2}\frakJ,\quad trX'=-\frac{2}{q},\quad tr\Xbar'=\frac{2\Delta q}{|q|^4},\label{eq:tracesxxbarkerrprim}
\end{align}
as well as 
$$\omegabar'=-\frac12\partial_r\left(\frac{\Delta}{|q|^2}\right)=\frac{M-r}{|q|^2}+\frac{r\Delta}{|q|^4}.$$
Note that the curvature components defined with respect to the outgoing principal pair \eqref{eq:principaloutgoinkerr} also satisfy \eqref{eq:curvingoingKerr}.

\noindent\textbf{Other canonical 1-forms $\frakJ_\pm$.} We define the 1-forms $\frakJ_\pm\in\fraks_1(\C)$ as follows
\begin{align}
	\frakJ_\pm=j_\pm+ i\hodge{j_\pm},
\end{align}
where the real 1-forms $j_\pm$ are given by
\begin{align*}
	&(j_+)_1=\frac{1}{|q|}\cos\theta\cos\phi_+,\quad(j_+)_2=-\frac{1}{|q|}\sin\phi_+,\\
	&(j_-)_1=\frac{1}{|q|}\cos\theta\sin\phi_+,\quad(j_-)_2=\frac{1}{|q|}\cos\phi_+.
\end{align*}
Recall the scalars $J^{(\pm)}$ defined in \eqref{eq:defJplusJmoins}. The horizontal derivatives of $J^{(\pm)}$, $\cos\theta$, $\ubar$, $r$, can be expressed as follows with $\frakJ_\pm$ and $\frakJ$,
\begin{align}\label{eq:followfromxixi}
	\mcd J^{(\pm)}=\frakJ_\pm,\quad\mcd\cos\theta=i\frakJ,\quad\mcd\ubar=a\frakJ,\quad\mcd r=0.
\end{align}
\noindent\textbf{Expression of coordinate vector fields with respect to the ingoing principal null frame.} The $(r,\ubar,\theta,\phi_+)$ and $(r,\ubar,x^1,x^2)$ coordinate vector fields can be expressed as follows with respect the the B-L coordinate vector fields $\partial_t,\partial_\phi$ and $\frakJ,\frakJ_\pm$:
\begin{itemize}
	\item For both coordinate systems, 
	\begin{align}\label{eq:dmuemukerr1}
		\partial_r=-e_3,\quad\quad\partial_\ubar=\partial_t.
	\end{align}
\item In the $(r,\ubar,\theta,\phi_+)$ coordinate system,
\begin{align}\label{eq:dmuemukerr2}
	\partial_\theta=\frac{|q|^2\Imag(\mathfrak{J})^b}{\sin\theta}e_b,\quad\partial_{\phi_+}=\partial_\phi,
\end{align}
\item In the $(r,\ubar,x^1,x^2)$ coordinate system,
\begin{align}\label{eq:dmuemukerr3}
	\partial_{x^1}=\frac{|q|^2}{\cos\theta}\Imag(\frakJ_-^b)e_b+a x^2\partial_t,\quad\partial_{x^2}=-\frac{|q|^2}{\cos\theta}\Imag(\frakJ_+^b)e_b-a x^1\partial_t.
\end{align}
\end{itemize}
Note that $\partial_t$ and $\partial_\phi$ can be expressed with respect to the principal frame and $\frakJ$, see \eqref{eq:defTZdansKerr}. The expressions above easily follow from \eqref{eq:followfromxixi} combined with 
the scalar product computations
\begin{equation}
	\begin{gathered}
		\hodge{j_-}\cdot j_+=-\hodge{j_+}\cdot j_-=\frac{\cos\theta}{|q|^2},\\
		 \hodge{j_-}\cdot j=-\frac{\sin\theta\cos\theta\sin\phi_+}{|q|^2},\quad\hodge{j_+}\cdot j=-\frac{\sin\theta\cos\theta\cos\phi_+}{|q|^2}.
	\end{gathered}
\end{equation}
We also compute the remaining scalar products:
\begin{equation}
	\begin{gathered}
		j_-\cdot{j_-}=\frac{\cos^2\theta\sin^2\phi_++\cos^2\theta_+}{|q|^2},\quad j_+\cdot{j_+}=\frac{\cos^2\theta\cos^2\phi_++\sin^2\theta_+}{|q|^2},\\
		j_-\cdot j_+=-\frac{\sin^2\theta\sin\phi_+\cos\phi_+}{|q|^2},\quad j\cdot j_+=-\frac{\sin\theta\sin\phi_+}{|q|^2},\quad j\cdot j_-=\frac{\sin\theta\cos\phi_+}{|q|^2}.
	\end{gathered}
\end{equation}
\subsection{Ingoing principal temporal gauge}\label{section:onintroPTgauge}
\subsubsection{Main equations}

We recall \cite[Def. 2.83]{KS21}.
\begin{defi}\label{defi:PTframedef}
	An ingoing PT structure $\{(e_3,e_4,\mch),r,\theta,\frakJ\}$ in a Lorentzian spacetime $(\mcm,\g)$ consists of a null pair $(e_3,e_4)$, the induced horizontal structure $\mch=(e_4,e_4)^\bot$, functions $(r,\theta)$ and a horizontal 1-form $\frakJ$ such that the following holds true:
\begin{itemize}
	\item $e_3$ is geodesic: $\D_3 e_3=0$.
	\item We have $e_3(r)=-1$, $e_3(\theta)=0$, $\nabla_3(\qbar\frakJ)=0$, where $q=r+ia\cos\theta$.
	\item Recalling the definition of $H$ in Definition \ref{defi:defcomplexquantities}, we have
	$$H=\frac{aq}{|q|^2}\frakJ.$$
\end{itemize}
An extended ingoing PT structure possesses in addition a function $\ubar$ such that $e_3(\ubar)=0$.
\end{defi}
We also recall \cite[Definition 2.84, Lemma 2.85]{KS21}, which prove that PT frames can be locally extended.
\begin{defi}\label{defi:PTinitial}
	An ingoing PT initial data set consists of a hypersurface $\Sigma$ transversal to $e_3$ together with a null pair $(e_3,e_4)$, the induced horizontal structure $\mch$, scalar functions $(r,\theta)$, and a horizontal 1-form $\frakJ$, all defined on $\Sigma$.
\end{defi}
\begin{lem}\label{lem:locextPTframe}
	Any ingoing PT initial data set as defined in Definition \ref{defi:PTinitial} can be locally extended to an ingoing PT structure.
\end{lem}
In our setting, we will  have a function $\phi_+$ and complex 1-forms $\frakJ_\pm\in\fraks_1(\C)$ such that
\begin{align}\label{eq:easlmitjfofgu}
	e_3(\phi_+)=0,\quad\quad\nabla_3(\qbar\frakJ_\pm)=0.
\end{align}
Similarly as \eqref{eq:defJplusJmoins} in Kerr, we define
$$\Jplus=\sin\theta\cos\phi_+,\quad\quad \Jmoins=\sin\theta\sin\phi_+.$$
\subsubsection{Kerr values and linearized quantities}\label{section:PTlinearized}
We consider an extended ingoing PT structure as defined in Definition \ref{defi:PTframedef}. We first define the Kerr values of the PT quantities which do not vanish in exact Kerr.
\begin{defi}\label{def:kerrvalues}
	We define the following functions and horizontal 1-forms with respect to $(r,\theta,\frakJ)$,
	\begin{alignat*}{4}
		trX_\mck & := \frac{2\qbar\Delta}{|q|^4}, \quad  & tr\Xbar_\mck & := -\frac{2}{\qbar},\quad& Z_\mck&:=\frac{aq}{|q|^2}\mathfrak{J},\\ \Hbar_\mck&:=-\frac{a\qbar}{|q|^2}\mathfrak{J}  ,\quad&
		\omega_\mck&:=-\frac12\partial_r\left(\frac{\Delta}{|q|^2}\right),\quad & P_\mck&:=-\frac{2m}{q^3},
	\end{alignat*}
	as well as
	\begin{alignat*}{2}
		(e_4(r))_\mck & := \frac{\Delta}{|q|^2}, \quad  & (\mcd\cos\theta)_\mck & := i\frakJ, \\
		(\mcd\ubar)_\mck&:=a\frakJ,\quad & (e_4(\ubar))_\mck&:=\frac{2(r^2+a^2)}{|q|^2},\\
		(\divc \frakJ)_\mck&:=\frac{4i(r^2+a^2)\cos\theta}{|q|^4},\quad & (\nabla_4\frakJ)_\mck&:=-\frac{\Delta\qbar}{|q|^4}\frak,
	\end{alignat*}
and 
	\begin{alignat*}{2}
	(\divc\frakJ_\pm)_\mck & :=-\frac{4}{r^2}J^{(\pm)}\mp\frac{4ia\cos^2\theta}{|q|^4}J^{(\mp)}, \quad  & (\nabla_4\frakJ_\pm)_\mck & :=-\frac{\Delta \qbar}{|q|^4}\frakJ_\pm\mp\frac{2a}{|q|^2}\frakJ_\mp, \\
	(\mcd J^{(\pm)})_\mck&:=\frakJ_\pm,\quad & (e_4(J^{(\pm)}))_\mck&:=\mp\frac{2a}{|q|^2}J^{(\mp)}.
\end{alignat*}
\end{defi}
Now, we define the linearized quantities of the PT structure by subtracting the Kerr values.
\begin{defi}\label{def:linearizedquantities}
	We define the following linearized quantities,
	\begin{alignat*}{4}
		\widecheck{trX} & := trX-trX_\mck, \quad  & \widecheck{tr\Xbar} & :=tr\Xbar-tr\Xbar_\mck , \quad &
		\widecheck{Z}&:=Z-Z_\mck,\\
		\widecheck{\Hbar}&:=\Hbar-\Hbar_\mck ,\quad&
		\widecheck{\omega}&:=\omega-\omega_\mck,\quad & \widecheck{P}&:=P-P_\mck,
	\end{alignat*}
	as well as
	\begin{alignat*}{2}
		\widecheck{e_4(r)} & := e_4(r)-(e_4(r))_\mck, \quad  & \widecheck{\mcd\cos\theta} & := \mcd\cos\theta-(\mcd\cos\theta)_\mck, \\
		\widecheck{\mcd\ubar}&:=\mcd\ubar-(\mcd\ubar)_\mck,\quad & \widecheck{e_4(\ubar)}&:=e_4(\ubar)-(e_4(\ubar))_\mck,\\
		\widecheck{\divc \frakJ}&:=\divc \frakJ-(\divc \frakJ)_\mck,\quad & \widecheck{\nabla_4\frakJ}&:=\nabla_4\frakJ-(\nabla_4\frakJ)_\mck,
	\end{alignat*}
and
	\begin{alignat*}{2}
	\widecheck{\divc\frakJ_\pm} & :={\divc\frakJ_\pm}-({\divc\frakJ_\pm})_\mck, \quad  & \widecheck{\nabla_4\frakJ_\pm} & := \nabla_4\frakJ_\pm-(\nabla_4\frakJ_\pm)_\mck, \\
	\widecheck{\mcd J^{(\pm)}}&:=\mcd J^{(\pm)}-(\mcd J^{(\pm)})_\mck,\quad & \widecheck{e_4(J^{(\pm)})}&:=e_4(J^{(\pm)})-(e_4(J^{(\pm)}))_\mck.
\end{alignat*}
\end{defi}
\begin{rem}
	Note that the quantities $\wh{X},\wh{\Xbar},\Xi,A,B,\Bbar,\Abar,\mcd r,e_4(\cos\theta),\mcd\hot\frakJ, \mcd\hot\frakJ_\pm$ vanish in exact Kerr so that they do not need to be linearized. Also, by definition of the PT frame,
	\begin{equation}\label{eq:idbasePT}
		\begin{aligned}
			&\Xibar=0,\quad\omegabar=0,\quad H=H_\mck=\frac{aq}{|q|^2}\frakJ,\\
			&e_3(r)=-1,\quad e_3(\theta)=0,\quad e_3(\ubar)=0,\quad e_3(\phi_+)=0.
		\end{aligned}
	\end{equation}
\end{rem}
\subsubsection{Linearized equations for ingoing PT structures}
We consider an extended PT structure equipped with a function $\phi_+$ and horizontal 1-forms $\frakJ_\pm$ as in \eqref{eq:easlmitjfofgu}.
\begin{defi}\label{def:O(1)mck}
	We denote with $O(1)_\mck$:
	\begin{itemize}
		\item any function of $(r,\theta)$ which satisfies for any $k\geq 0$ and $r\in[r_-,r_+]$,
		$$|(\partial_r,\partial_{\cos\theta})^{\leq k}O(1)_\mck|\lesssim_k 1,$$
		\item and any horizontal 1-form of the form $O(1)'_\mck\frakJ$, or $O(1)'_\mck\frakJ_\pm$, where $O(1)'_\mck$ is a function of $(r,\theta)$ which satisfies for any $k\geq 0$ and $r\in[r_-,r_+]$,
		$$|(\partial_r,\partial_{\cos\theta})^{\leq k}O(1)'_\mck|\lesssim 1.$$
	\end{itemize}
\end{defi}
In what follows, we denote the following sets of linearized quantities:
\begin{equation*}
	\begin{aligned}
		\widecheck{\Gamma}:=\big\{&\Xi,\widecheck{\omega},\widecheck{trX},\wh{X},\widecheck{Z},\widecheck{\Hbar},\widecheck{tr\Xbar},\nabla r,\widecheck{e_4(r)},\widecheck{e_4(\ubar)},e_4(\cos\theta),\widecheck{\nabla_4\mathfrak{J}},\widecheck{\nabla\cos\theta},\wh{\Xbar},\widecheck{\nabla\ubar},\widecheck{\divc\mathfrak{J}},\mcd\hot\mathfrak{J}\\
		&\widecheck{\divc\frakJ_\pm},\widecheck{\nabla_4\frakJ_\pm},\widecheck{\mcd J^{(\pm)}},\widecheck{e_4(J^{(\pm)})}\big\},\\
		\widecheck{R}:=\big\{&A,\Abar,B,\Bbar,\widecheck{P}\big\}.
	\end{aligned}
\end{equation*}
\begin{prop}\label{prop:linearizednullstructure}
	The linearized equations in a PT structure take the form
	\begin{alignat*}{3}
		\nabla_3\widecheck{tr\Xbar}&=_sO(1)_\mck\Gammacheck+\Gammacheck^2,\quad&\nabla_3\wh{\Xbar}&=_s\Rcheck+O(1)_\mck\Gammacheck+\Gammacheck^2,\\
		\nabla_3\widecheck{Z}&=_s\Rcheck+O(1)_\mck\Gammacheck+\Gammacheck^2,\quad&\nabla_3\widecheck{\Hbar}&=_s\Rcheck+O(1)_\mck\Gammacheck+\Gammacheck^2,
	\end{alignat*}
	as well as
	\begin{alignat*}{3}
		\nabla_3\widecheck{trX}&=_s\Rcheck+O(1)_\mck\Gammacheck+\Gammacheck^2,\quad&\nabla_3\wh{X}&=_sO(1)_\mck\Gammacheck+\Gammacheck^2,\\
		\nabla_3\widecheck{\omega}&=_s\Rcheck+O(1)_\mck\Gammacheck+\Gammacheck^2,\quad&\nabla_3\Xi&=_s\Rcheck+O(1)_\mck\Gammacheck+\Gammacheck^2.
	\end{alignat*}
	Also, we have
	\begin{align*}
		\nabla_3\widecheck{\mcd\cos\theta}=_sO(1)_\mck\Gammacheck+\Gammacheck^2,\quad\nabla_3\mcd r=_sO(1)_\mck\Gammacheck+\Gammacheck^2,\quad\nabla_3\widecheck{\mcd\ubar}=_sO(1)_\mck\Gammacheck+\Gammacheck^2,
	\end{align*}
	and 
	\begin{align*}
		e_3(e_4(\cos\theta))=_sO(1)_\mck\Gammacheck+\Gammacheck^2,\quad e_3(\widecheck{e_4(\ubar)})=_sO(1)_\mck\Gammacheck+\Gammacheck^2,\quad e_3(\widecheck{e_4(r)})=_sO(1)_\mck\Gammacheck+\Gammacheck^2.
	\end{align*}
	Finally, we have
	\begin{align*}
		\nabla_3\mcd\hot\frakJ=_sO(1)_\mck(\Gammacheck,\Rcheck)+\Gammacheck^2,\quad\nabla_3\widecheck{\divc\frakJ}=_sO(1)_\mck(\Gammacheck,\Rcheck)+\Gammacheck^2,\quad\nabla_3\widecheck{\nabla_4\frakJ}=_sO(1)_\mck(\Gammacheck,\Rcheck)+\Gammacheck^2.
	\end{align*}
\end{prop}
\begin{proof}
The proof is the same as the one of Proposition 9.27 in \cite{KS21}, except that here we do not need to keep track of the powers of $r$ and instead we notice that non-constant factors in the equations above can simply be put in the form $O(1)_\mck$.
\end{proof}
We can also rewrite the Bianchi identities.
\begin{prop}\label{prop:biaenchidansptgauge}

	The Bianchi identities can be rewritten in the following schematic form,
	\begin{align*}
		\nabla_3A-\frac12\mcd\hot B&=_sO(1)_\mck(\Gammacheck,\Rcheck)+\Gammacheck\cdot\Rcheck,\quad\quad\quad\quad\nabla_4B-\frac12\divc A=_sO(1)_\mck(\Gammacheck,\Rcheck)+\Gammacheck\cdot\Rcheck,\\
		\nabla_3B-\mcd\overline{\widecheck{P}}&=_sO(1)_\mck(\Gammacheck,\Rcheck)+\Gammacheck\cdot\Rcheck,\quad\quad\quad\quad \nabla_4\widecheck{P}-\frac12\mcd\cdot\overline{B}=_sO(1)_\mck(\Gammacheck,\Rcheck)+\Gammacheck\cdot\Rcheck,\\
		\nabla_3\widecheck{P}+\frac12\divc\Bbar&=_sO(1)_\mck(\Gammacheck,\Rcheck)+\Gammacheck\cdot\Rcheck,\quad\quad\quad\quad\quad\:\:\nabla_4\Bbar+\mcd\widecheck{P}=_sO(1)_\mck(\Gammacheck,\Rcheck)+\Gammacheck\cdot\Rcheck,\\
		\nabla_3\Bbar+\frac12\divc\Abar&=_sO(1)_\mck(\Gammacheck,\Rcheck)+\Gammacheck\cdot\Rcheck,\quad\quad\quad\quad\nabla_4\Abar+\frac12\mcd\hot\Bbar=_sO(1)_\mck(\Gammacheck,\Rcheck)+\Gammacheck\cdot\Rcheck.
	\end{align*}
\end{prop}
\begin{proof}
	By Proposition \ref{prop:bianchicomplex}, the only equations above which should be checked are the ones which involve $\widecheck{P}$. For those equations, the non-zero exact Kerr terms vanish and the remaining terms are $O(1)_\mck$ quantities multiplied by one of the following linearized quantity,
	$$\widecheck{e_4(r)},e_4(\cos\theta),\nabla r,\widecheck{\mcd\cos\theta},\widecheck{trX},\widecheck{tr\Xbar},\widecheck{\Hbar}\in\Gammacheck,\quad\widecheck{P}\in\Rcheck,$$
	which concludes the proof.
\end{proof}
\begin{prop}\label{prop:eqpourderjplusomoins}
	The linearized derivatives of $J^{(\pm)}$ and $\frakJ_\pm$ satisfy the following equations,
\begin{align*}
			\nabla_3\mcd\hot\frakJ_\pm&=_sO(1)_\mck(\Gammacheck,\Rcheck)+\Gammacheck^2,\\
	\nabla_3\widecheck{\divc\frakJ_\pm}&=_sO(1)_\mck(\Gammacheck,\Rcheck)+\Gammacheck^2,\\
	\nabla_3\widecheck{\nabla_4\frakJ_\pm}&=_sO(1)_\mck(\Gammacheck,\Rcheck)+\Gammacheck^2,
\end{align*}
and 
\begin{align*}
	\nabla_3\widecheck{\mcd J^{(\pm)}}=O(1)_\mck(\Gammacheck,\Rcheck)+\Gammacheck^2,\\
	\nabla_3\widecheck{\nabla_4 J^{(\pm)}}=O(1)_\mck(\Gammacheck,\Rcheck)+\Gammacheck^2.
\end{align*}
\end{prop}
\begin{proof}
	The proof is the same as the ones of Lemmas 7.4 and 7.5 in \cite{KS21} except that once again we do not need to keep track of the powers of $r$, and instead we notice that non-constant factors in the equations above take the form $O(1)_\mck$.
\end{proof}
\subsection{Double null foliations}\label{section:doublenullgrandesection}
We briefly present the notion of spacetimes covered by double null foliations. We only cover the main properties that we will use in the present paper, and we refer to Chapter 3 in \cite{klnico} for more details.
\subsubsection{General definitions and properties}\label{section:DNspacetimegeneral}
We consider a smooth spacetime $(\mcm,\g)$ with two smooth optical functions $\uring,\ubarring$, that is
$$\g^{\alpha\beta}\partial_\alpha \uring\partial_\beta \uring=\g^{\alpha\beta}\partial_\alpha \ubarring\partial_\beta\ubarring=0,$$
We assume that $\mcm$ is foliated by the level sets of $\uring$ and $\ubarring$, and we require that $\uring,\ubarring$ increase towards the future. We denote $C_{\uring}$ and $\underline{C}_{\ubarring}$ the level sets of $\uring,\ubarring$, which are null hypersurfaces, and $S(\uring,\ubarring)$ the surfaces
$$S(\uring,\ubarring)=C_{\uring}\cap \underline{C}_{\ubarring},$$
which we assume are spacelike 2-spheres. We now introduce the geodesic vectorfields
$$ \underline{L}:=-\D\ubarring,\quad L:=-\D \uring,$$
where $\D$ is the Levi-Civita connection, and the lapse function $\Omega>0$ by
$$\g(L,\underline{L})=-\frac{1}{2\Omega^2}.$$
We also define the renormalized null pair $(\ering_3,\ering_4)$ as follows,
$$\ering_3=2\Omega^2 \underline{L},\quad \ering_4=2 L,$$
which satisfies $$\g(\ering_3,\ering_3)=\g(\ering_4,\ering_4)=0,\quad\g(\ering_3,\ering_4)=-2.$$
\noindent\textbf{Coordinates and metric.}\label{section:DNdefandproperties} As in \cite{stabC0}, we will consider a system of local coordinates $\theta^A$, $A=1,2$, on $S(\uring,\ubarring)$ such that
$$\ering_3(\theta^A)=0.$$
Then, in coordinates $(\uring,\ubarring,\theta^1,\theta^2)$, the metric $\g$ takes the following form,
\begin{align}\label{eq:metricdoublenull}
	\g=-2\Omega^2\left(\dee \uring\otimes\dee\ubarring+\dee\ubarring\otimes\dee \uring\right)+\gamma_{AB}\left(\dee\theta^A-b^A\dee\ubarring\right)\otimes\left(\dee\theta^B-b^B\dee\ubarring\right),
\end{align} where:
\begin{itemize}
	\item $b$ is a smooth vector field tangent to the spheres $S(\uring,\ubarring)$,
	\item  $\gamma$ is the Riemannian metric induced by $\g$ on the spheres $S(\uring,\ubarring)$.
\end{itemize}
Note that in such double-null coordinates, the spacetime volume form writes as follows,
\begin{align}\label{eq:volumeendoublenull}
	\vol=2\Omega^2\mathrm{vol}_\gamma\dee \uring\dee \ubarring,
\end{align}
where $\mathrm{vol}_\gamma$ is the volume form on $S(\uring,\ubarring)$ induced by $\gamma$,
$$\mathrm{vol}_\gamma=\sqrt{\det\gamma_{AB}}\dee\theta^1\dee\theta^2.$$
We denote $\partial_{\uring},\partial_{\ubarring},\partial_{\theta^A}$ the coordinate vector fields with respect to the coordinates $(\uring,\ubarring,\theta^A)$. Then we have
\begin{align}\label{eq:DuDubardoublenull}
	\ering_3=\partial_{\uring},\quad \ering_4=\Omega^{-2}\left(\partial_{\ubarring}+b^A\partial_{\theta^A}\right),
\end{align}
Note that the horizontal distribution $\mathring{\mch}:=(\ering_3,\ering_4)^\bot$ induced by $(\ering_3,\ering_4)$ satisfies
$$\mathring{\mch}=TS(\uring,\ubarring),$$
and is thus integrable. We will consider tensor fields $U$ on $\mcu$ which are $S(\uring,\ubarring)$-tangent, namely which are of the form
$$U=U^{B_1\cdots B_s}_{A_1\cdots A_r}\dee\theta^{A_1}\otimes\cdots\otimes\dee\theta^{A_r}\otimes\partial_{\theta^{B_1}}\otimes\cdots\otimes\partial_{\theta^{B_s}}.$$
Then, we respectively denote $\nabring_3,\nabring_4,\nabring$ the projections to $S(\uring,\ubarring)$ of the covariant derivatives $\D_{\ering_3}$, $\D_{\ering_4}$, and the Levi-Civita connection induced by $\gamma$ on $S(\uring,\ubarring)$, which all act on $S(\uring,\ubarring)$ tensors. More precisely, denoting the Christoffel symbols on $(S(\uring,\ubarring),\gamma)$ as follows,
\begin{align}\label{eq:christodoublenulspheres}
	\mathring{\Gamma}_{AB}^C:=\frac12\gamma^{CF}\left(\partial_{\theta^D}\gamma_{EF}+\partial_{\theta^E}\gamma_{DF}-\partial_{\theta^F}\gamma_{DE}\right),
\end{align}
and denoting
$$\mathring{\chi}_{AB}=\g(\D_{\partial_{\theta^A}}\ering_4,\partial_{\theta^B}),\quad \mathring{\chibar}_{AB}=\g(\D_{\partial_{\theta^A}}\ering_3,\partial_{\theta^B}),$$
the null second fundamental forms defined as in Section \ref{section:ricciandcurvdef} with respect to the null pair $(\ering_3,\ering_4)$ with associated horizontal distribution $\mathring{\mch}=TS(\uring,\ubarring)$, we get that for any $S(\uring,\ubarring)$-tangent $r$-tensor $U$ and for any for $B,A_1,\cdots,A_r=1,2$,
\begin{align}
	\nabring_BU_{A_1\cdots A_k}&=\partial_{\theta^B}(U_{A_1\cdots A_k})-\sum_{i=1}^k\mathring{\Gamma}_{A_iB}^CU_{A_1\cdots C\cdots A_k},\label{eq:nabdoublenul}\\
	\nabring_3U_{A_1\cdots A_k}&=\partial_{\uring}(U_{A_1\cdots A_k})-\sum_{i=1}^k\gamma^{BC}\mathring{\chibar}_{A_iC}U_{A_1\cdots B\cdots A_k},\label{eq:nab3doublenul}\\
	\nabring_4U_{A_1\cdots A_k}&=\Omega^{-2}\left(\partial_{\ubarring}+b^C\partial_{\theta^C}\right)(U_{A_1\cdots A_k})\nn\\
	&\quad-\sum_{i=1}^k\left(\gamma^{BC}\mathring{\chi}_{A_iC}-\Omega^{-2}\partial_{\theta^{A_i}}b^B\right)U_{A_1\cdots B\cdots A_k},\label{eq:nab4doublenul}
\end{align}
(recall that the notation $A_1\cdots B\cdots A_k$ at rank $i$ denotes that $A_i$ is replaced with $B$). Moreover,
$$\nabring_3\gamma=\nabring_4\gamma=\nabring\gamma=0.$$
\noindent\textbf{Ricci and curvature coefficients in the double null frame.} We denote
\begin{align*}
	\mathring{\chi},\quad\mathring{\chibar},\quad\mathring{\eta},\quad\mathring{\etabar},\quad\mathring{\omega},\quad\mathring{\omegabar},\quad\mathring{\xi},\quad\mathring{\xibar},\quad\mathring{\zeta},
\end{align*}
and
$$\mathring{\alpha},\quad\mathring{\alphabar},\quad\mathring{\beta},\quad\mathring{\betabar},\quad\mathring{\rho},\quad,\mathring{\hodge{\rho}}$$
the Ricci and curvature coefficients defined in Section \ref{section:ricciandcurvdef} with respect to the double null pair $(\ering_3,\ering_4)$ with associated horizontal distribution $\mathring{\mch}=TS(\uring,\ubarring)$, as well as the symmetric traceless quantities
$$\mathring{\chihat}_{AB}=\mathring{\chi}_{AB}-\frac12\mathring{tr\chi}\gamma_{AB},\quad\mathring{\wh{\chibar}}_{AB}=\mathring{\wh{\chibar}}_{AB}-\frac12\mathring{tr\chibar}\gamma_{AB},$$
where $\mathring{tr\chi}=\gamma^{AB}\mathring{\chi}_{AB}$, $\mathring{tr\chibar}=\gamma^{AB}\mathring{\wh{\chibar}}_{AB}$. We denote $\mathring{K}$ the Gauss curvature of the spheres $S(\uring,\ubarring)$,
\begin{align}\label{eq:gausscurvintrinsic}
	\mathring{K}=\frac12\gamma^{BC}\left(\partial_{\theta^A}\mathring{\Gamma}_{BC}^A-\partial_{\theta^C}\mathring{\Gamma}_{BA}^A+\mathring{\Gamma}_{AD}^A\mathring{\Gamma}_{BC}^D-\mathring{\Gamma}_{CD}^A\mathring{\Gamma}_{BA}^D\right),
\end{align}
and, following \cite{stabC0}, we denote by $\hodge{\mathring{K}}$ the quantity
 \begin{align*}
 	\hodge{\mathring{K}}=\mathring{\hodge{\rho}}+\frac12\mathring{\wh{\chibar}}\wedge\mathring{\chihat}.
 \end{align*}
We also define the mass aspect functions $\mu,\underline{\mu}$ and the scalar ${\barred{\omegabar}}$ by
\begin{align}
	\mu=-\mathring{\diver}\mathring{\eta}+\mathring{K},\quad\underline{\mu}=-\mathring{\diver}\mathring{\etabar}+\mathring{K},\quad{\barred{\omegabar}}=\mathring{\triangle}\mathring{\omegabar}+\frac12\mathring{\diver}\mathring{\betabar},
\end{align}
with Laplace, divergence and curl operators on $S(\uring,\ubarring)$ defined as
\begin{align}\label{eq:laplacienSdoublenulll}
	\mathring{\triangle}U:=\nabring^A\nabring_AU,\quad\mathring{\diver}\:U_{A_1\cdots A_{r-1}}=\nabring^B U_{B A_1\cdots A_{r-1}},\quad \mathring{\curl}\:U_{A_1\cdots A_{r-1}}=\in^{BC}{\nabring}_BU_{C A_1\cdots A_{r-1}},
\end{align}
for any $S(\uring,\ubarring)$-tangent $r$-tensor $U$, where $\in$ is the volume form of $(S(\uring,\ubarring),\gamma)$. Note that the operators defined above coincide with operators defined in Section \ref{section:horizontalcovariantderivative} in the special case of a double null foliation.
\subsubsection{Main equations in the double null gauge}\label{section:equationsDNnonlinearize}
\noindent\textbf{Preliminary identities.} In a double null gauge, we have the identities
\begin{align}
	\mathring{\omega}=0,\quad\mathring{\xi}=\mathring{\xibar}=0,\quad\mathring{\atrchi}=\mathring{\atrchibar}=0,\label{eq:trucsnulendoublenul}\\
	\mathring{\omegabar}=-\ering_3(\log\Omega),\quad\mathring{\nabla}\log\Omega=\frac12(\mathring{\eta}+\mathring{\etabar}),\quad \mathring{\eta}=\mathring{\zeta},\label{eq:omegabarendoublenul}
\end{align}
as well as
\begin{align}
	\partial_u b^A=2\Omega^2(\mathring{\eta}^A-\mathring{\etabar}^A)=4\Omega^2(\mathring{\zeta}^A-\nabring^A\log\Omega),\label{eq:parubA}\\
	\partial_u\gamma_{AB}=2\mathring{\chibar}_{AB},\quad(\partial_\ubar+b^C\partial_{\theta^C})\gamma_{AB}+\gamma_{AC}\partial_{\theta^B}b^C+\gamma_{BC}\partial_{\theta^A}b^C=2\Omega^2\mathring{\chi}_{AB}.\label{eq:pargammadn}
\end{align}
see for example \cite[Prop. 2.3,2.4]{stabC0}. We also have the following identities for $\mathring{K},\hodge{\mathring{K}}$,
\begin{align}\label{eq:KKstarDNdeff}
	\mathring{K}=-\mathring{\rho}+\frac12\mathring{\wh{\chi}}\cdot\mathring{\wh{\chibar}}-\frac14\mathring{tr\chi} \mathring{tr\chibar},\quad \mathring{\hodge{K}}=\hodge{\mathring{\rho}}+\frac12\mathring{\wh{\chibar}}\wedge\mathring{\chihat}.
\end{align}
\noindent\textbf{Null structure and Bianchi equations.} If $\g$ is a solution of the Einstein vacuum equations \eqref{eq:EVE}, then the null structure equations of Section \ref{section:nullstructureeq} rewrite as follows in a double null gauge,
\begin{alignat*}{3}
	\nabring_4\mathring{tr\chi}&=-|\mathring{\chihat}|^2-\frac12\mathring{tr\chi}^2,\quad&\nabring_4\mathring{\chihat}+\mathring{tr\chi}\mathring{\chihat}&=-\mathring{\alpha},\\
	\nabring_3\mathring{tr\chibar}+2\mathring{\omegabar}\mathring{tr\chibar}&=-|\mathring{\wh{\chibar}}|^2-\frac12\mathring{tr\chibar}^2,\quad&\nabring_3\mathring{\wh{\chibar}}+\mathring{tr\chibar}\mathring{\wh{\chibar}}+2\mathring{\omegabar}\wh{\chibar}&=-\mathring{\alphabar},\\
	\nabring_4\mathring{tr\chibar}+\mathring{tr\chi}\mathring{tr\chibar}&=-2\mathring{K}+2\mathring{\diver}\mathring{\etabar}+2|\mathring{\etabar}|^2,\quad &\nabring_3\mathring{tr\chi}+\mathring{tr\chibar}\mathring{tr\chi}&=2\mathring{\omegabar}\mathring{tr\chi}-2\mathring{K}+2\mathring{\diver}\mathring{\eta}+2|\mathring{\eta}|^2,\\
	\nabring_4\mathring{\wh{\chibar}}+\frac12\mathring{tr\chi}\mathring{\wh{\chibar}}&=\nabring\hot\mathring{\etabar}-\frac12\mathring{tr\chibar}\mathring{\chihat}+\mathring{\etabar}\hot\mathring{\etabar},\quad & \nabring_3\mathring{\wh{\chi}}+\frac12\mathring{tr\chibar}\mathring{\wh{\chi}}&=\nabring\hot\mathring{\eta}+2\mathring{\omegabar}\mathring{\wh{\chi}}-\frac12\mathring{tr\chi}\mathring{\wh{\chibar}}+\mathring{\eta}\hot\mathring{\eta},
\end{alignat*}
as well as 
\begin{align*}
	\nabring_4\mathring{\eta}=-\mathring{\chi}\cdot(&\mathring{\eta}-\mathring{\etabar})-\mathring{\beta},\quad  \nabring_3\mathring{\etabar}=-\mathring{\chibar}\cdot(\mathring{\etabar}-\mathring{\eta})+\mathring{\betabar},\quad\nabring_4\mathring{\omegabar}=\mathring{\zeta}\cdot(\mathring{\eta}-\mathring{\etabar})-\mathring{\eta}\cdot\mathring{\etabar}+\mathring{\rho}.
\end{align*}
We also have the following constraint equations,
\begin{equation}\label{eq:constraintDNouioui}
	\begin{aligned}
		\mathring{\diver}\mathring{\wh{\chi}}=\frac12\nabring \mathring{tr\chi}-\zeta\cdot(\mathring{\wh{\chi}}-&\mathring{tr\chi}\gamma)-\mathring{\beta},\quad\mathring{\diver}\mathring{\wh{\chibar}}=\frac12\nabring \mathring{tr\chibar}+\zeta\cdot(\mathring{\wh{\chibar}}-\mathring{tr\chibar}\gamma)+\mathring{\betabar},\\
		&\mathring{\curl}\:\mathring{\eta}=-\mathring{\curl}\:\mathring{\etabar}=\mathring{\hodge{K}}.
	\end{aligned}
\end{equation}
See \cite[(3.7)]{stabC0}, the Bianchi equations of Section \ref{section:bianchii} rewrite with respect to $\mathring{K},\mathring{\hodge{K}}$ as follows,
\begin{align*}
	\nabring_3\mathring{\beta}+\mathring{tr\chibar}\mathring{\beta}=&-\nabring\mathring{K}+\hodge{\:\nabring\hodge{\mathring{K}}}+2\mathring{\omegabar}\mathring{\beta}+2{\mathring{\wh{\chi}}}\cdot\mathring{\betabar}-3(\mathring{\eta}\mathring{K}-\hodge{\mathring{\eta}}\hodge{\mathring{K}})+\frac12(\nabring(\mathring{\wh{\chi}}\cdot\mathring{\wh{\chibar}})+\hodge{\:\nabring(\mathring{\wh{\chi}}\wedge\mathring{\wh{\chibar}})})\\
	&+\frac32(\mathring{\eta}\mathring{\wh{\chi}}\cdot\mathring{\wh{\chibar}}+\hodge{\mathring{\eta}}\mathring{\wh{\chi}}\wedge\mathring{\wh{\chibar}})-\frac14(\mathring{tr\chibar}\nabring \mathring{tr\chi}+\mathring{tr\chi}\nabring \mathring{tr\chibar})-\frac34\mathring{\eta}\mathring{tr\chi}\mathring{tr\chibar},\\
	\nabring_4\hodge{\mathring{K}}+\frac32\mathring{tr\chi}\hodge{\mathring{K}}=&-\mathring{\diver}\hodge{\mathring{\beta}}-\mathring{\zeta}\wedge\mathring{\beta}-2\mathring{\etabar}\wedge\mathring{\beta}-\frac12\mathring{\wh{\chi}}\wedge(\nabring\hot\mathring{\etabar})-\frac12\mathring{\wh{\chi}}\wedge(\mathring{\etabar}\hot\mathring{\etabar}),\\
	\nabring_4\mathring{K}+\mathring{tr\chi}\mathring{K}=&-\mathring{\diver}\mathring{\beta}-\mathring{\zeta}\cdot\mathring{\beta}-2\mathring{\etabar}\cdot\mathring{\beta}+\frac12\mathring{\wh{\chi}}\cdot(\nabring\hot\mathring{\etabar})+\frac12\mathring{\wh{\chi}}\cdot(\mathring{\etabar}\hot\mathring{\etabar})-\frac12\mathring{tr\chi}\mathring{\diver}\mathring{\etabar}-\frac12\mathring{tr\chi}|\mathring{\etabar}|^2,\\
	\nabring_3\mathring{K}+\mathring{tr\chibar}\mathring{K}=&\:\mathring{\diver}\mathring{\betabar}-\mathring{\zeta}\cdot\mathring{\betabar}+2\mathring{\eta}\cdot\mathring{\betabar}+\frac12\mathring{\wh{\chibar}}\cdot(\nabring\hot\mathring{\eta})+\frac12\mathring{\wh{\chibar}}\cdot(\mathring{\eta}\hot\mathring{\eta})-\frac12\mathring{tr\chibar}\mathring{\diver}\mathring{\eta}-\frac12\mathring{tr\chibar}|\mathring{\eta}|^2,\\
	\nabring_3\hodge{\mathring{K}}+\frac32\mathring{tr\chibar}\hodge{\mathring{K}}=&-\mathring{\diver}\hodge{\mathring{\betabar}}+\mathring{\zeta}\wedge\mathring{\betabar}-2\mathring{\eta}\wedge\mathring{\betabar}+\frac12\mathring{\wh{\chibar}}\wedge(\nabring\hot\mathring{\eta})+\frac12\mathring{\wh{\chibar}}\wedge(\mathring{\eta}\hot\mathring{\eta}),\\
	\nabring_4\mathring{\betabar}+\mathring{tr\chi}\mathring{\betabar}=&\:\nabring\mathring{K}+\hodge{\:\nabring\hodge{\mathring{K}}}+2{\mathring{\wh{\chibar}}}\cdot\mathring{\beta}+3(\mathring{\etabar}\mathring{K}+\hodge{\mathring{\etabar}}\hodge{\mathring{K}})-\frac12(\nabring(\mathring{\wh{\chibar}}\cdot\mathring{\wh{\chi}})-\hodge{\:\nabring(\mathring{\wh{\chi}}\wedge\mathring{\wh{\chibar}})})\\
	&-\frac32(\mathring{\etabar}\mathring{\wh{\chi}}\cdot\mathring{\wh{\chibar}}-\hodge{\mathring{\etabar}}\mathring{\wh{\chi}}\wedge\mathring{\wh{\chibar}})+\frac14(\mathring{tr\chibar}\nabring \mathring{tr\chi}+\mathring{tr\chi}\nabring \mathring{tr\chibar})+\frac34\mathring{\etabar}\mathring{tr\chi}\mathring{tr\chibar}.
\end{align*} 
\noindent\textbf{Some derivative and commutation identities in the double null gauge.} Recall the convention on integrals stated in Section \ref{section:conventions} (which implies $\intS f:=\intS f\mathrm{vol}_\gamma$).
\begin{lem}\label{lem:derintSf}
	Let $f$ be a scalar function. We have the identities
	$$\frac{\partial}{\partial\ubar}\left(\intS f \right)=\intS \Omega^2\left(\ering_4(f)+\mathring{tr\chi} f\right),\quad\frac{\partial}{\partial u}\left(\intS f \right)=\intS \left(\ering_3(f)+\mathring{tr\chibar} f\right).$$
\end{lem}
\begin{proof}
	See \cite{klnico}.
\end{proof}
\begin{cor}\label{prop:dernorme2}
	Let $X$ be a $S(\uring,\ubarring)$-tangent tensor. We have:
	\begin{align*}
		&\frac{\partial}{\partial u}\left(\intS|X|_{{\gamma}}^2\right)=2\intS\Big(\langle X,\nabring_3X\rangle_\gamma+\frac{1}{2}\mathring{tr\chibar}|X|^2_\gamma\Big),\\
		&\frac{\partial}{\partial \ubar}\left(\intS|X|_{{\gamma}}^2\right)=2\intS\Omega^2\Big(\langle X,\nabring_4X\rangle_\gamma+\frac{1}{2}\mathring{tr\chi}|X|^2_\gamma\Big).
	\end{align*}
\end{cor}

\begin{prop}\label{prop:commnablaboudnenulll}
	We have the following commutation formulas, for $\phi$ a $S(\uring,\ubarring)$-tangent $r$-tensor,
	\begin{align*}
		[\nabring_4,\nabring_B]\phi_{A_1\cdots A_r}=&(\mathring\etabar_B+\mathring\eta_B)\nabring_4\phi_{A_1\cdots A_r}-\mathring\chi_B^C\nabring_C\phi_{A_1\cdots A_r}\\
		&+\sum_{i=1}^r(\mathring\chi_{A_i B}\mathring\etabar^C-\mathring\chi_B^C\mathring\etabar_{A_i}+\in_{A_i}^C\hodge{\mathring\beta}_B)\phi_{A_1\cdots C\cdots A_r},
	\end{align*}
	\begin{align*}
		[\nabring_3,\nabring_B]\phi_{A_1\cdots A_r}=&-\mathring\chibar_B^C\nabring_C\phi_{A_1\cdots A_r}+\sum_{i=1}^r(\mathring\chibar_{A_i B}\mathring\etabar^C-\mathring\chibar_B^C\mathring\etabar_{A_i}+\in_{A_i}^C\hodge{\mathring\betabar}_B)\phi_{A_1\cdots C\cdots A_r},
	\end{align*}
	\begin{align*}
		[\nabring_3,\nabring_4]\phi_{A_1\cdots A_r}=& 2\mathring\omegabar\nabring_4\phi_{A_1\cdots A_r}+(\mathring\eta-\mathring\etabar)\cdot\nabring\phi_{A_1\cdots A_r}\\
		&+2\sum_{i=1}^r(\mathring\etabar_{A_i}\mathring\eta^C-\mathring\eta_{A_i}\etabar^C+\in_{A_i}^C\mathring{\hodge{\rho}})\phi_{A_1\cdots C\cdots A_r},
	\end{align*}
	$$[\nabring_B,\nabring_C]\phi_{A_1\cdots A_r}=\mathring{K}\sum_{i=1}^r\left(\gamma_{A_i B}U_{A_1\cdots C\cdots A_r}-\gamma_{A_i C}U_{A_1\cdots B\cdots A_r}\right).$$
\end{prop}
\begin{proof}
	See for example \cite[Prop. 7.1]{stabC0}.
\end{proof}
\begin{rem}
	Using the identity \eqref{eq:omegabarendoublenul} we also have
	\begin{align}\label{eq:bonnermk}
		[\Omega^2\nabring_4,\nabring_B]\phi_{A_1\cdots A_r}=\Omega^2\left(-\mathring\chi_B^C\nabring_C\phi_{A_1\cdots A_r}+\sum_{i=1}^r(\mathring\chi_{A_i B}\mathring\etabar^C-\chi_B^C\mathring\etabar_{A_i}+\in_{A_i}^C\hodge{\mathring\beta}_B)\phi_{A_1\cdots C\cdots A_r}\right).
	\end{align}
\end{rem}
\subsubsection{Kerr values for double null coordinates and linearized quantities}\label{section:linearizationdoublenull}
For the results stated in this subsection and many additional results, we refer to Section 2.4 and Appendix A in \cite{stabC0}.

\noindent\textbf{Kerr metric in double null coordinates.} We consider the Isreal-Pretorius coordinates\footnote{Note that we write $r'_*$ instead of $r_*$ as in \cite{isrealpretorius} (and \cite{stabC0}) to avoid confusion with the quantity $r^*$ defined in \eqref{eq:rstarexpre} in the context of the Eddington-Finkelstein coordinates.} $$(t,r'_*,\theta_*,\phi)$$ of Kerr spacetime constructed in \cite{isrealpretorius}, see also Appendix A in \cite{stabC0} for a thorough investigation of these coordinates inside Kerr black holes. We recall that $r'_*(r,\theta)$, $\theta_*(r,\theta)$ only depend on $r,\theta$, and that we have the identities
\begin{align}\label{eq:derparrstar}
	\frac{\partial r'_*}{\partial r}=\frac{\sqrt{(r^2+a^2)^2-a^2\sin^2\theta_*\Delta}}{\Delta},&\quad \frac{\partial r'_*}{\partial \theta}=a\sqrt{\sin^2\theta_*-\sin^2\theta},	\\
	\frac{\partial \theta_*}{\partial r}=\frac{1}{G\sqrt{(r^2+a^2)^2-a^2\sin^2\theta_*}},&\quad \frac{\partial \theta_*}{\partial \theta}=\frac{-1}{aG\sqrt{\sin^2\theta_*-\sin^2\theta}},\label{eq:derparthetasta}
\end{align}
where $G$ is defined in \cite[(A.19)]{stabC0} and satisfies the following estimate, 
\begin{align}\label{eq:bornepourGaa}
	-\frac{C_{M,a}}{\sqrt{\sin^2\theta_*-\sin^2\theta}}\leq G\leq -\frac{\sin(2\theta)}{a\sin(2\theta_*)\sqrt{\sin^2\theta_*-\sin^2\theta}},
\end{align}
for some constant $C_{M,a}>0$ which only depends on the black hole parameters $(a,M)$, see \cite[Prop. A.3]{stabC0}. Moreover, see \cite[Prop. A.5]{stabC0}, the partial derivatives of $(r,\theta)$ with respect to $(r_*,\theta_*)$ are given by 
\begin{equation}\label{eq:derpardanslautresens}
	\begin{gathered}
		\frac{\partial r}{\partial r_*}=\frac{\Delta\sqrt{(r^2+a^2)^2-a^2\sin^2\theta_*\Delta}}{(r^2+a^2)^2	-a^2\sin^2\theta\Delta},\quad \frac{\partial \theta}{\partial r_*}=\frac{a\Delta\sqrt{\sin^2\theta_*-\sin^2\theta}}{(r^2+a^2)^2	-a^2\sin^2\theta\Delta},\\
		\frac{\partial r}{\partial \theta_*}=\frac{a^2\Delta(\sin^2\theta_*-\sin^2\theta)G\sqrt{(r^2+a^2)^2-a^2\sin^2\theta_*\Delta}}{(r^2+a^2)^2	-a^2\sin^2\theta\Delta},\\
		\frac{\partial \theta}{\partial \theta_*}=-\frac{aG\sqrt{\sin^2\theta_*-\sin^2\theta}((r^2+a^2)^2-a^2\sin^2\theta_*\Delta)}{(r^2+a^2)^2	-a^2\sin^2\theta\Delta}.
	\end{gathered}
\end{equation}

We then define
\begin{align}\label{eq:uubarDNdefikerr}
	\mathring{u}=r'_*-t,\quad\mathring{\ubar}= r'_*+t.
\end{align}
\underline{Notice that our values of $\mathring{u},\mathring{\ubar}$ in \eqref{eq:uubarDNdefikerr} correspond to twice the ones defined in \cite[(A.11)]{stabC0}.}

\begin{lem}\label{lem:ubarssontpareils}
	We have in the Kerr black hole interior the bounds
	$$|\ubar-\mathring{\ubar}|\lesssim 1,\quad\quad |u-\mathring{u}|\lesssim 1.$$
\end{lem}
\begin{proof}
	By \eqref{eq:uubarDNdefikerr} and \eqref{eq:defuubarEFkerr}, it is enough to prove the bound $|r'_*-r^*|\lesssim 1$. Moreover, considering $r'_*-r^*$ as a function of $(r,\theta)$ we have
	$$\left|\frac{\partial}{\partial r}(r'_*-r^*)\right|=\left|\frac{\sqrt{(r^2+a^2)^2-a^2\sin^2\theta_*\Delta}-(r^2+a^2)}{\Delta}\right|\lesssim 1.$$
	Thus, integrating the bound above from $r=M$ (where the bound $|r'_*-r^*|\lesssim 1$ clearly holds because $r'_*-r^*$ is a continuous function on the sphere) to any $r\in(r_-,r_+)$, we conclude the proof.
\end{proof}
Next, we define the function $h(r'_*,\theta_*)$ which solves the transport equation
$$\frac{\partial h}{\partial r'_*}(r'_*,\theta_*)=-\frac{2aMr}{|q|^2R^2}$$
for any fixed $\theta_*\in(0,\pi)$, with initial data $h(r'_*=0,\theta_*)=0$, and then
$$\phi_*=:\phi-h(r'_*,\theta_*).$$
Then, defining $\theta^1=\theta_*,\theta^2=\phi_*$, the Kerr metric in coordinates $(\mathring{u},\mathring{\ubar},\theta^A)$ takes the double null form
$$\g_{a,M}=-2\Omega^2_\mck(\dee \mathring{u}\otimes\dee\mathring{\ubar}+\dee\mathring{\ubar}\otimes\dee \mathring{u})+(\gamma_\mck)_{AB}(\dee\theta^A-b_\mck^A\dee\mathring{\ubar})\otimes(\dee\theta^B-b_\mck^B\dee\mathring{\ubar}),$$
where 
\begin{align}
	\Omega^2_\mck=-\frac{\Delta}{4R^2},\quad b_\mck=\frac{2aMr}{|q|^2 R^2}\partial_{\phi_*},\nn\\
	(\gamma_\mck)_{\phi_* \phi_*}=R^2\sin^2\theta,\quad (\gamma_\mck)_{\theta_* \theta_*}=\frac{\ell^2}{R^2}+\left(\frac{\partial h}{\partial\theta_*}\right)^2R^2\sin^2\theta,\label{eq:bDNdanskerr}\\
	(\gamma_\mck)_{\theta_*\phi_*}=(\gamma_\mck)_{\phi_*\theta_*}=\left(\frac{\partial h}{\partial\theta_*}\right)^2R^2\sin^2\theta,\nn
\end{align}
where $\ell$ is defined as $\ell=-Ga\sqrt{\sin\theta^2_*-\sin^2\theta}\sqrt{(r^2+a^2)^2-a^2\sin^2\theta_*\Delta}$ with $G(r'_*,\theta_*)$ being the quantity defined above. Note in particular that the scalar appearing in the expression of $b_\mck$ satisfies the identity
\begin{align}\label{eq:identitybelow}
	\frac{2Mar}{|q|^2 R^2}=\frac{2Mar}{(r^2+a^2)^2-a^2\sin^2\theta\Delta}.
\end{align} 
Also, in coordinates $\theta_*,\phi_*$,
\begin{align}\label{eq:detgammaKsim}
	\det(\gamma_\mck)_{AB}=\ell^2\sin^2\theta\sim\sin^2\theta,
\end{align}
where we used $\ell^2\sim 1$, which can be deduced from Propositions A.3, A.1 in \cite{stabC0} (see also \cite[Prop. A.4, Eq. (A.21)]{stabC0}). Moreover, for any $C_R\in\R$, we have the following estimate in $\{\mathring{u}+\mathring{\ubar}\geq C_R\}$\footnote{Comparing with \cite[Prop. A.15]{stabC0}, note that we have $\frac{r_+-r_-}{r_-^2+a^2}=2|\kappa_-|$ which is coherent because our values of $\mathring{u},\mathring{\ubar}$ here are twice the ones in \cite{stabC0}.},
\begin{align}\label{eq:omegamcksim}
	\Omega_\mck^2\sim e^{-|\kappa_-|(\mathring{u}+\mathring{\ubar})}.
\end{align}
where the implicit constants above depend on $a,M$ but not on $C_R$ for, say, $C_R\geq1$.

\noindent\textbf{Renormalized Kerr double null coordinates regular on $\ch$.} We recall from \cite[Section A.5.2]{stabC0} the following renormalized coordinates :
\begin{itemize}
	\item We define $\mathring{\ubar}_\ch=f(\ubarring)$ where $f:\R\rightarrow(-\infty,0)$ is a smooth and strictly increasing function such that $f(\mathring{\ubar})=\mathring{\ubar}$ for $\mathring{\ubar}\leq -1$, $f\to 0$ as $\mathring{\ubar}\to+\infty$ and there exists $\mathring{\ubar}_-\geq 1$ such that $f'(\mathring{\ubar})=e^{-|\kappa_-|\mathring{\ubar}}$ for $\mathring{\ubar}\geq\mathring{\ubar}_-$. 
	\item We define $\theta_{*,\ch}$ and $\phi_{*,\ch}$ as follows (recalling $\mathring{\ubar}=r'_*+t$ and \underline{not} $(r'_*+t)/2$)
\begin{align}\label{eq:DNthetaCH}
	\phi_{*,\ch}=\phi_*-\frac{2aMr}{|q|^2R^2}\mathring{\ubar}\:\:(\mathrm{mod}\:2\pi),\quad\quad\theta_{*,\ch}=\theta_*.
\end{align}
\end{itemize}
Then, $(\mathring{u},\mathring{\ubar}_\ch,\theta_*,\phi_{*,\ch})$ is a regular local coordinate system in $\{\mathring{u}+\mathring{\ubar} \geq C_R\}$ up to $\ch$ which corresponds to $\ch=\{\mathring{\ubar}_\ch=0\}$.

\noindent\textbf{Bounds for double null Ricci and curvature coefficients in Kerr.} Defining the following set of derivatives in Kerr,
$$\df_\mck:=\{\Omega^2_\mck(\nabring_4)_\mck,(\nabring_3)_\mck,\nabring_\mck\},$$
we have the following bounds in $\{\uring+\ubarring\geq C_R\}\cap\{\uring\leq-1\}$
\begin{equation}\label{eq:bornesdanskerr}
		\begin{aligned}
		&|\df_\mck^{\leq N}\log\Omega_\mck|{\lesssim} 1,\quad |\df_\mck^{\leq N}\mathring{K}_\mck|{\lesssim} 1,\quad |\df_\mck^{\leq N}\mathring{\hodge{K}}_\mck|{\lesssim} 1,\quad|\df_\mck^{\leq N}b_\mck|{\lesssim} 1,\\
		& |\df_\mck^{\leq N}\mathring{\chi}_\mck|\lesssim 1,\quad |\df_\mck^{\leq N}\mathring{\chibar}_\mck|\lesssim\Omega_\mck^2,\quad |\df_\mck^{\leq N}(\mathring{\eta}_\mck,\mathring{\etabar}_\mck)|\lesssim 1\quad|\df^{\leq N}\mathring{\omegabar}_\mck|\lesssim 1,\\
		& |\df_\mck^{\leq N}\mathring{\beta}_\mck|{\lesssim} 1,\quad |\df_\mck^{\leq N}\mathring{\betabar}_\mck|{\lesssim} \Omega_\mck^2,\quad|\df_\mck^{\leq N}\mathring{\rho}_\mck|{\lesssim} 1,\quad |\df_\mck^{\leq N}\hodge{\mathring{\rho}}_\mck|{\lesssim} 1, \\
		&|\df^{\leq N}\mathring{\alphabar}_\mck|\lesssim\Omega^2_\mck,\quad |\df^{\leq N}\mathring{\alpha}_\mck|\lesssim\Omega^{-2}_\mck,
	\end{aligned}
\end{equation}
where the implicit constants in the bounds above depend on $N,a,M$, see Proposition \ref{prop:boundtoutkerr}.

\begin{rem}
	We note that the bounds \eqref{eq:bornesdanskerr} are not sharp in terms of powers of $\Omega^2_\mck$ for $\mathring{\alphabar}_\mck$ and $\mathring{\alpha}_\mck$ because of the method of proof which relies on the null structure equations expressing $\mathring{\alpha}$ and $\mathring{\alphabar}$ in terms of $\nabring_4\mathring{\wh{\chi}}$ and $\nabring_3\mathring{\wh{\chibar}}$ in Kerr. Indeed, these components (and their derivatives) should be bounded by $\Omega^4_\mck$ and $1$ respectively. However the rough bounds in \eqref{eq:bornesdanskerr} will be enough for our purpose so we prefer to avoid the technical computations required to obtain the corresponding sharp bounds.
\end{rem}
\noindent\textbf{Linearization procedure in the double null gauge.} We follow the linearization procedure in \cite[Section A.4]{stabC0}. Being given a spacetime in the double null gauge as described in Section \ref{section:DNspacetimegeneral} with coordinates $(\uring,\ubarring,\theta^A)$, and with $\mcu=\{\mathring{u}+\mathring{\ubar}\geq C_R\}\cap\{\mathring{\ubar}<\mathring{\ubar}_f\}\cap\{\mathring{u}<\mathring{u}_f\}$, we denote
$$\mathcal{\mcf}:(\mathring{u},\mathring{\ubar},\theta^A)\in\mcu\longmapsto(\mathring{u},\mathring{\ubar},\theta^A)\in\mcu_\mck $$
the identification of the $(\mathring{u},\mathring{\ubar},\theta^A)$ coordinates of $\mcu$ and $\mcu_\mck$ which is the analog region in Kerr, see \cite[Section A.4]{stabC0} for more details. Now, let $U_\mck$ be a tensor field (which includes scalars) defined on the background Kerr spacetime, for example $U_\mck=\gamma_\mck,tr\mathring{\chi}_\mck,\mathring{\alpha}_\mck,(\ering_4)_\mck\ldots$ Then $U_\mck$ can be pulled back to $\mcu$ by the diffeomorphism $\mathcal{F}$, which yields a tensor field $\mathcal{F}^* U_\mck$ on $\mcu$, which we still denote $U_\mck$. Now, if $U$ is a metric, Ricci, or curvature coefficient as defined in Section \ref{section:DNspacetimegeneral}, we define the linearized $S(\mathring{u},\mathring{\ubar})$-tangent tensor
$$\widecheck{U}:=U-U_\mck.$$
\subsubsection{Schematic notations specific to the double null gauge}\label{section:specificdoublenull}
We consider a double null foliation as defined in Section \ref{section:DNdefandproperties}. We follow \cite{stabC0} and introduce the following notations for linearized metric components and some subsets of the linearized Ricci coefficients:
$$\gcheck:=\{\gammacheck,\widecheck{\gamma^{-1}},\widecheck{\log\Omega},\Omega^{-2}\widecheck{\Omega^2}\},\quad \psicheck=\{\widecheck{\mathring{\eta}},\widecheck{\mathring{\etabar}}\},\quad \psicheck_H=\{\widecheck{\mathring{tr\chi}},\widecheck{\mathring{\chihat}}\},\quad\psicheck_{\Hbar}=\{\widecheck{\mathring{tr\chi}},\widecheck{\mathring{\hat{\chibar}}}\},$$
as well as the analog unchecked notations $\psi,\psi_H,\psi_{\Hbar}$ defined in the obvious way.
\subsection{Frame transformations}
\subsubsection{Frame transformations and change of frame formulas}\label{section:frametransfoandformulas}
We recall \cite[Lemma 3.1]{GCMKS}.
\begin{lem}
	Given a null frame $(e_3,e_4,e_1,e_2)$, a general transformation from the null frame $(e_3,e_4,e_1,e_2)$ to another null frame $(e'_3,e'_4,e'_1,e'_2)$ can be written as follows,
	\begin{align}
		&e_4'=\lambda \left(e_4+f^be_b+\frac{1}{4}|f|^2e_3\right),\nn\\
		&{e_a'}=\left(\delta_{ab}+\frac{1}{2}\fbar_a f^b\right)e_b+\frac{1}{2}\fbar_a e_4+\left(\frac{1}{2}f_a+\frac{1}{8}|f|^2\fbar_a\right)e_3\label{eq:frametransfo}\\
		&e_3'=\lambda^{-1}\left(\left(1+\frac{1}{2}f\cdot\fbar+\frac{1}{16}|f|^2|\fbar|^2\right)e_3+\left(\fbar^b+\frac{1}{4}|\fbar|^2f^b\right)e_b+\frac{1}{4}|\fbar|^2e_4\right),\nn
	\end{align}
where $\lambda$ is a scalar, $f$ and $\fbar$ are horizontal 1-forms (with respect to the null pair $e_3,e_4$). We call $f,\fbar,\lambda$ the transition coefficients of the change of frame.
\end{lem}
\begin{rem}\label{rem:phistarpullback}
	Note that, given a null pair $(e_3,e_4)$ and $(e_3',e_4')$ as in \eqref{eq:frametransfo}, then the map
	$$\Phi_p: X\in \mch_p\longmapsto X+\frac12 \fbar(X)f^be_b+\frac12\fbar(X) e_4+\left(\frac12 f(X)+\frac18|f|^2\fbar(X)\right)e_3\in\mch'_p$$
	is an isometry between the horizontal structures $\mch_p,\mch'_p$ respectively associated to $(e_3,e_4)$ and $(e_3',e_4')$, for any $p\in\mcm$. Now, if $V$ is a $\mch'$-horizontal tensor, we denote $\Phi^* V$ (which is the pullback of $V$) the $\mch$-horizontal tensor defined by 
	$$\Phi^* V(X_1,\cdots, X_k)=V(\Phi(X_1),\cdots,\Phi(X_k)),$$
	for any $X_1,\cdots, X_k\in \mch$. Conversely, if $U$ is a $\mch$-horizontal tensor, we denote $\Phi_* U$ (the pushforward of $U$) the $\mch'$-horizontal tensor which is defined by
	$$\Phi_* U(Y_1,\cdots,Y_k)=U(\Phi^{-1}(Y_1),\cdots \Phi^{-1}(Y_k)),$$
	for any $Y_1,\cdots,Y_k\in\mch'$. Also note that for $k=0,1,2$ the pullback map $\Phi^*$ and the pushforward map $\Phi_*$ take $\fraks'_k(\C)$ to $\fraks_k(\C)$ and $\fraks_k(\C)$ to $\fraks'_k(\C)$ respectively.
\end{rem}
\subsubsection*{Change of frame formulas}
We classify the Ricci coefficients according to their signatures $0,\pm 1,\pm 2$ (in the sense of \cite{CK93}),
\begin{align}\label{eq:gammasignature}
	\Gamma_{-2}=\{\xibar\},\quad\Gamma_{-1}=\{\omegabar,\chibar\},\quad\Gamma_0=\{\eta,\etabar,\zeta\},\quad\Gamma_{+1}=\{\omega,\chi\},\quad\Gamma_{+2}=\{\xi\}.
\end{align}
For $p=0,\pm 1,\pm 2$, we define schematic lower-order terms by
\begin{align}\label{eq:schematicG}
	\mathcal{G}_{[p]}=_s\sum_{p'=-2}^2\underset{i_1-i_2+p'=p}{\sum_{3\leq i_1+i_2\leq 12}} f^{i_1}\fbar^{i_2}\Gamma_{p'}.
\end{align}
Similarly, we classify the Ricci coefficients according to their signatures $0,\pm 1,\pm 2$,
\begin{align}\label{eq:rsignature}
	R_{-2}=\{\alphabar\},\quad R_{-1}=\{\betabar\},\quad R_0=\{\rho,\hodge{\rho}\},\quad R_{+1}=\{\beta\},\quad R_{+2}=\{\alpha\},
\end{align}
and for $p=0,\pm 1,\pm 2$, we define schematic lower-order terms by
\begin{align}\label{eq:schematicR}
	\mathcal{R}_{[p]}=_s\sum_{p'=-2}^2\underset{i_1-i_2+p'=p}{\sum_{3\leq i_1+i_2\leq 12}} f^{i_1}\fbar^{i_2}R_{p'}.
\end{align}
\begin{prop}\label{prop:ellfollows}
	Under the frame transformation \eqref{eq:frametransfo}, recalling the notation introduced in Remark \ref{rem:phistarpullback}, we have the following change of frame formulas for the Ricci coefficients,
	\begin{align}
			\lambda^{-2}\Phi^*\xi'=&\xi+\frac12\nabla_{\lambda^{-1}e_4'}f+\frac14(tr\chi f-\atrchi\hodge{f})+\omega f+\frac12 (f\cdot\hat{\chi})+\frac14|f|^2\eta+\frac12(f\cdot\zeta)f\nn\\	&-\frac14|f|^2\etabar-\frac14|f|^2\omegabar f+\frac14|f|^2(f\cdot\chibar)+\frac18|f|^4\xibar+\lambda^{-2}\left(\frac12(f\cdot\Phi^*\xi')\fbar+\frac12(f\cdot\fbar)\Phi^*\xi'\right),\label{eq:changexi}\\
			\lambda^{-1}\omega'=&\omega-\frac12\lambda^{-1}e_4'(\log\lambda)+\frac12f\cdot(\zeta-\etabar)-\frac14|f|^2\omegabar-\frac18tr\chibar|f|^2-\frac14 f\cdot(f\cdot\hat{\chibar})-\frac18|f|^2(f\cdot\xibar)\nn\\
			&+\frac12\lambda^{-2}\fbar\cdot\Phi^*\xi',\label{eq:changeomega}
	\end{align}
\begin{align}
	\lambda^2\Phi^*\xibar'=&\xibar+\frac12\lambda\nabla'_3\fbar'+\omegabar\fbar+\frac14tr\chibar\fbar-\frac14\atrchibar\hodge{\fbar}+\frac12\fbar\cdot\wh{\chibar}-\frac12(\fbar\cdot\zeta)\fbar+\frac14|\fbar|^2\etabar\label{eq:changexibar}\\
	&-\frac14|\fbar|^2\Phi^*\eta'+\mathcal{G}_{[-2]},\nn\\
	\lambda\omegabar'=&\omegabar+\frac12\lambda e_3'(\log\lambda)-\frac12\fbar\cdot(\zeta+\eta)+f\cdot\fbar\omegabar-\frac14|\fbar|^2\omega+\frac12 f\cdot\xibar+\frac18(f\cdot\fbar)tr\chibar\label{eq:changeomegabar}\\
	&+\frac18(\fbar\wedge f)\atrchibar-\frac18|\fbar|^2tr\chi-\frac14\lambda\fbar\cdot\nabla'_3f+\frac12(f\cdot\fbar)(\fbar\cdot\Phi^*\eta')-\frac14|\fbar|^2(f\cdot\Phi^*\eta')+\mathcal{G}_{[-1]},\nn
\end{align}
\begin{align}
	\lambda^{-1}tr\chi'=&tr\chi+\diver'f+f\cdot(\eta+\zeta)+\fbar\cdot\xi+\frac{1}{4}\fbar\cdot\left(ftr\chi-\hodge{f}\atrchi\right)+(f\cdot\fbar)\omega\label{eq:changetrchi}\\
	&-\omegabar|\fbar|^2-\frac{1}{4}|f|^2tr\chibar-\frac{1}{4}(f\cdot\fbar)\lambda^{-1}tr\chi'+\frac{1}{4}(\fbar\wedge f)\lambda^{-1}\atrchi'+\mathcal{G}_{[+1]},\nn\\
	\lambda^{-1}\atrchi'=&\atrchi+\curl' f+f\wedge(\eta+\zeta)+\fbar\wedge\xi+\frac14(\fbar\wedge ftr\chi+(f\cdot\fbar)\atrchi)\label{eq:changeatrchi}\\
	&+\omega f\wedge\fbar-\frac14|f|^2\atrchibar-\frac14(f\cdot\fbar)\lambda^{-1}\atrchi'+\frac14\lambda^{-1}(f\wedge\fbar)tr\chi'+\mathcal{G}_{[+1]},\nn\\
	\lambda^{-1}\Phi^*\chihat'=&\chihat+\nabla'\hot f+f\hot(\eta+\zeta)+\fbar\hot\xi+\frac14\fbar\hot(f tr\chi-\hodge{f}\atrchi)+\omega f\hot\fbar-\omegabar f\hot f\label{eq:changechihat}\\
	&-\frac14|f|^2\atrchibar+\frac14(f\hot\fbar)\lambda^{-1}tr\chi'+\frac14\hodge{f}\hot\fbar\lambda^{-1}\atrchi'+\frac12\fbar\hot(f\cdot\lambda^{-1}\Phi^*\hat{\chi}')+\mathcal{G}_{[+1]},\nn
\end{align}
\begin{align}
	\lambda tr\chibar'=&tr\chibar+\diver'\fbar+\fbar\cdot(\etabar-\zeta)+\frac12(f\cdot\fbar)tr\chibar+f\cdot\xibar-|\fbar|^2\omega+(f\cdot\fbar)\omegabar\label{eq:changetrchibar}\\
	&-\frac14\lambda^{-1}tr\chi'+\mathcal{G}_{[-1]},\nn\\
	\lambda\atrchibar'=&\atrchibar+\curl' \fbar+\fbar\wedge(\etabar+\zeta)+\frac12(f\cdot\fbar)\atrchibar+f\wedge\xibar+(f\wedge\fbar)\omegabar\label{eq:changeatrchibar}\\
	&-\frac14|\fbar|^2\lambda^{-1}\atrchi'+\mathcal{G}_{[-1]},\nn\\
	\lambda^{-1}\Phi^*\wh{\chibar}'=&\wh{\chibar}+\nabla'\hot \fbar+\fbar\hot(\etabar-\zeta)+f\hot\xibar+\frac12(f\hot\fbar)tr\chibar-\omega \fbar\hot\fbar+\omegabar \fbar\hot f\label{eq:changechibarhat}\\
	&-\frac14|\fbar|^2\lambda^{-1}\Phi^*\wh{\chi}'+\mathcal{G}_{[-1]},\nn
\end{align}
\begin{align}
	\Phi^*\zeta'=&\zeta-\nabla'\log\lambda-\frac14tr\chibar f+\frac14\atrchibar\hodge{f}+\omega\fbar-\omegabar f+\frac14\fbar tr\chi-\frac12\wh{\chibar}\cdot f+\frac12 (f\cdot\zeta)\fbar\nn\\
	&-\frac12(f\cdot\etabar)\fbar+\frac14\fbar(f\cdot\eta)+\frac14\fbar(f\cdot\zeta)+\frac14\hodge{\fbar}(f\wedge\eta)+\frac14\hodge{\fbar}(f\wedge\zeta)+\frac14\fbar\diver'f\nn\\
	&+\frac14\hodge{\fbar}\curl'f+\frac12\lambda^{-1}\fbar\cdot\Phi^*\chihat'-\frac{1}{16}(f\cdot\fbar)\fbar\lambda^{-1}tr\chi'+\frac{1}{16}(\fbar\wedge f)\fbar\lambda^{-1}\atrchi'\nn\\
	&-\frac{1}{16}\hodge{\fbar}(f\cdot\fbar)\lambda^{-1}\atrchi'+\frac{1}{16}\hodge{\fbar}\lambda^{-1}(f\wedge\fbar)tr\chi'+\mathcal{G}_{[0]},\label{eq:changezeta}
\end{align}
\begin{align}
	\Phi^*\eta'=&\eta+\frac12\lambda\nabla'_3f+\frac14\fbar tr\chi-\frac14\hodge{\fbar}\atrchi-\omegabar f+\frac12(f\cdot\fbar)\eta+\frac12\fbar\cdot\chihat+\frac12f(\fbar\cdot\zeta)\nn\\
	&-(\fbar\cdot f)\Phi^*\eta'+\frac12\fbar(f\cdot\Phi^*\eta')+\mathcal{G}_{[0]},\label{eq:changeeta}\\
	\Phi^*\etabar'=&\etabar+\frac12\nabla_{\lambda^{-1}e_4'}\fbar+\frac14tr\chibar f-\frac14\atrchibar\hodge{f}-\omega\fbar+\frac12 f\cdot\hat{\chibar}+\frac12(f\cdot\eta)\fbar-\frac14(f\cdot\zeta)\fbar\nn\\
	&-\frac14|\fbar|^2\lambda^{-2}\Phi^*\xi'+\mathcal{G}_{[0]},\label{eq:changeetabar}
\end{align}
where the notation $\mathcal{G}_{[p]}$, $p=0,\pm1,\pm2$ for lower-order terms has been introduced in \eqref{eq:schematicG}. We also have the following change of frame formulas for the curvature components,
\begin{align}
	\lambda^{-2}\Phi^*\alpha'=&\alpha+f\hot\beta-\hodge{f}\hot\hodge{\beta}+\left(f\hot f-\frac12\hodge{f}\hot\hodge{f}\right)\rho+\frac32(f\hot\hodge{f})\hodge{\rho}+\mathcal{R}_{[+2]},\label{eq:changealpha}\\
	\lambda^2\Phi^*\alphabar'=&\alphabar-\fbar\hot\betabar+\hodge{\fbar}\hot\hodge{\betabar}+\left(\fbar\hot\fbar-\frac12\hodge{\fbar}\hot\hodge{\fbar}\right)\rho+\frac32(\fbar\hot\hodge{\fbar})\hodge{\rho}+\mathcal{R}_{[-2]},\label{eq:changealphabar}
\end{align}
\begin{align}
	\lambda^{-1}\Phi^*\beta'=&\beta+\frac32(f\rho+\hodge{f}\hodge{\rho})+\frac12\alpha\cdot\fbar+\mcr_{[+1]},\label{eq:changebeta}\\
	\lambda\Phi^*\betabar'=&\betabar-\frac32(\fbar\rho+\hodge{\fbar}\hodge{\rho})-\frac12\alphabar\cdot f+\mcr_{[-1]},\label{eq:changebetabar}
\end{align}
\begin{align}
	\rho'=&\rho+\fbar\cdot\beta-f\cdot\betabar+\frac32\rho(f\cdot\fbar)-\frac32\hodge{\rho}(f\wedge\fbar)+\mathcal{R}_{[0]},\label{eq:changerho}\\
	\hodge{\rho}'=&\hodge{\rho}-\fbar\cdot\hodge{\beta}-f\cdot\hodge{\betabar}+\frac32\hodge{\rho}(f\cdot\fbar)+\frac32\rho(f\wedge\fbar)+\mathcal{R}_{[0]},\label{eq:changerhostar}
\end{align}
where the notation $\mathcal{R}_{[p]}$, $p=0,\pm1,\pm2$ for lower-order terms has been introduced in \eqref{eq:schematicR}.
\end{prop}
\begin{proof}
		This follows from \cite[Prop. 3.3]{GCMKS} and the precise computation of the lower-order terms $l.o.t.$ which take the schematic forms $\mathcal{G}_{[p]}$ and $\mcr_{[p]}$ with appropriate $p$.
\end{proof}

\subsubsection{Covariant derivatives and frame transformations}\label{section:derivativechangeofframe}
In this subsection we assume that we are given a null pair $(e_3,e_4)$ with associated horizontal structure $\mch$ and another null pair defined as in \eqref{eq:frametransfo} with coefficients $\lambda,f,\fbar$ (which are respectively a scalar function and $\mch$-horizontal 1-forms). We denote $(e_a)_{a=1,2}$ a $\mch$-horizontal frame and the associated $\mch'$-horizontal frame $(e_a')_{a=1,2}$ defined by $e_a'=\Phi(e_a)$, such that \eqref{eq:frametransfo} holds. 
	\begin{prop}\label{prop:changebasisdergeneral}
Let $U$ be a $\mch$-horizontal $k$-tensor. Then for $a,b_1,\ldots,b_k=1,2$ we have
	\begin{align}
		\nabla'_{4} (\Phi_*U)_{b_1\cdots b_k}&={\lambda}\left(\nabla_4U_{b_1\cdots b_k}+f\cdot\nabla U_{b_1\cdots b_k}+\frac{1}{4}|f|^2\nabla_3 U_{b_1\cdots b_k}\right)+\sum_{i=1}^k\hat{\mct}^{(4)}_{b_ic}U_{b_1\cdots c\cdots b_k},\\
		\nabla'_{a} (\Phi_*U)_{b_1\cdots b_k}&=\nabla_aU_{b_1\cdots b_k}+\frac12\fbar_a f\cdot\nabla U_{b_1\cdots b_k}+\frac12\fbar_a\nabla_4U_{b_1\cdots b_k}\nn\\
		&\quad+\left(\frac12f_a+\frac18|f|^2\fbar_a\right)\nabla_3U_{b_1\cdots b_k}+\sum_{i=1}^k\hat{\mct}_{ab_ic}U_{b_1\cdots c\cdots b_k},\label{eq:guiliguili}\\
		\nabla'_{3} (\Phi_*U)_{b_1\cdots b_k}&={\lambda}^{-1}\Bigg(\left(1+f\cdot\fbar+\frac{1}{16}|f|^2|\fbar|^2\right)\nabla_3U_{b_1\cdots b_k}+\left(\fbar+\frac14|\fbar|^2f\right)\cdot\nabla U_{b_1\cdots b_k}\nn\\
		&\quad\quad\quad\quad+\frac{1}{4}|\fbar|^2\nabla_4 U_{b_1\cdots b_k}\Bigg)+\sum_{i=1}^k\hat{\mct}^{(3)}_{b_ic}U_{b_1\cdots c\cdots b_k},\label{eq:diffdesnabtiens}
	\end{align}
	where the $\mch$-horizontal tensors $\hat{\mct}^{(4)}_{bc}$, $\hat{\mct}_{abc}$,  $\hat{\mct}^{(3)}_{bc}$ are given by 
	\begin{align*}
		\hat{\mct}^{(4)}_{bc}&={\lambda}\left(\mct^{(4)}_{bc}+f^a \mct_{abc}+\frac14|f|^2\mct^{(3)}_{bc}\right),\\
		\hat{\mct}_{abc}&=\mct_{abc}+\frac12\fbar_af^d\mct_{dbc}+\frac12\fbar_a\mct^{(4)}_{bc}+\left(\frac12 f_a+\frac18|f|^2\fbar_a\right)\mct^{(3)}_{bc},\\
		\hat{\mct}^{(3)}_{bc}&={\lambda}^{-1}\left(\left(1+f\cdot\fbar+\frac{1}{16}|f|^2|\fbar|^2\right)\mct^{(3)}_{bc}+\left(\fbar^a+\frac14|\fbar|^2f^a\right)\mct_{abc}+\frac{1}{4}|\fbar|^2\mct^{(4)}_{bc}\right),
	\end{align*}
with $\mct^{(4)}_{bc}$, $\mct_{abc}$,  $\mct^{(3)}_{bc}$ as in Propositions \ref{prop:diffchristo}, \ref{prop:diffchristo4}, \ref{prop:diffchristo3},
\end{prop}
\begin{proof}
	These formulas are direct consequences of Propositions \ref{prop:diffchristo}, \ref{prop:diffchristo4}, \ref{prop:diffchristo3}.
\end{proof}

\subsection{Approximate symetries}
\subsubsection{Killing vector fields $\T,\Z$ in Kerr spacetime}
In Boyer-Lindquist coordinates introduced in Section \ref{section:BLEFKerr}, we define the following regular vector fields in Kerr
$$\T=\partial_t,\quad\quad\Z=\partial_\phi,$$
which are Killing vector fields: denoting by $\mcl$ the spacetime Lie derivative, we have
$$\mcl_{\T}\g_{a,M}=\mcl_{\Z}\g_{a,M}=0.$$
They can be expressed with respect to the ingoing principal frame \eqref{eq:principalingoinkerr} as follows:
\begin{align}\label{eq:defTZdansKerr}
		\T:=\frac12\left(e_4+\frac{\Delta}{|q|^2}e_3-2a\Real(\mathfrak{J})^be_b\right),\quad \Z:=\frac12\left(2(r^2+a^2)\Real(\frakJ)^be_b-a\sin^2\theta e_4-\frac{a\sin^2\theta\Delta}{|q|^2}e_3\right),
\end{align}
where we recall the definition \eqref{eq:deffrakJkerr} of the horizontal 1-form $\frakJ$.
\begin{lem}\label{lem:transfokerrIun}
	Let $F_\mck(r,\theta)$ be the following function,
	$$F_\mck(r,\theta):=\frac{-a^2+(r^2+a^2)\sin^{-2}\theta\left(1-\sqrt{1-\frac{a^2\sin^2\theta\Delta}{(r^2+a^2)^2}}\right)}{\sqrt{(r^2+a^2)^2-a^2\sin^2\theta\Delta}}.$$
	Note that $F_\mck(r,\theta)$ is analytic with respect to $r$ and $\cos\theta$. Then the map
	$$\Theta_\mck:X\in \mch\longmapsto X-a\Real(\frakJ)(X)\T+F_\mck(r,\theta)\Real(\frakJ)(X)\Z\in TS(r,\ubar)$$
	is an isometry between the horizontal distribution $\mch$ and $TS(r,\ubar)$ in Kerr.
\end{lem}
\begin{proof}
	Recalling \eqref{eq:metriquesphereI} we get that the vector fields
	$$\barre{e}_1:=\frac{1}{|q|}\partial_\theta,\quad\quad\barre{e}_2:=\frac{|q|}{\sin\theta\sqrt{(r^2+a^2)^2-a^2\sin^2\theta\Delta}}\partial_\phi$$
	form a frame of $TS(r,\ubar)$ which is smooth everywhere but on the axis. Thus, it is only left to check that the map $\Theta_\mck$ as defined above satisfies
	$$\Theta_\mck(e_b)=\barre{e}_b,\quad b=1,2,$$
	with $(e_1,e_2)$ the horizontal frame defined in \eqref{eq:horizontalframekerr}. For $b=1$ this is trivial using $\Real(\frakJ)_1=0$, and for $b=2$ we compute
\begin{align*}
	\Theta_\mck(e_2)&=\frac{a\sin\theta}{|q|}\partial_t+\frac{1}{|q|\sin\theta}\partial_\phi-\frac{a\sin\theta}{|q|}\partial_t+F_\mck(r,\theta)\frac{\sin\theta}{|q|}\partial_\phi\\
	&=\frac{1}{|q|\sin\theta}\left(1+\frac{-a^2\sin^2\theta+(r^2+a^2)\left(1-\sqrt{1-\frac{a^2\sin^2\theta\Delta}{(r^2+a^2)^2}}\right)}{\sqrt{(r^2+a^2)^2-a^2\sin^2\theta\Delta}}\right)\partial_\phi\\
	&=\frac{1}{|q|\sin\theta}\left(1+\frac{|q|^2-\sqrt{(r^2+a^2)^2-a^2\sin^2\theta\Delta}}{\sqrt{(r^2+a^2)^2-a^2\sin^2\theta\Delta}}\right)\partial_\phi=\barre{e}_2,
\end{align*}
which concludes the proof.
\end{proof}

We can also express $\T,\Z$ in Kerr with respect to the double null coordinate vector fields. More precisely, in double null coordinates $(\uring,\ubarring,\theta_*,\phi_*)$ introduced in Section \ref{section:linearizationdoublenull} we have
\begin{align}\label{eq:TenDNlolkerr}
	\T=\frac12\left(\partial_{\ubarring}-\partial_{\uring}\right),\quad\quad\Z=\partial_{\phi_*}.
\end{align}
These expressions will be used to define natural approximate Killing vector fields in perturbations of Kerr spacetime.
\subsubsection{Approximate symmetries in perturbations of Kerr}

\noindent\textbf{Definition of $\T,\Z$ with respect to a PT structure.} Being given an ingoing PT structure $(e_3,e_4,\mch,r,\theta,\frakJ)$ as defined in Section \ref{section:onintroPTgauge}, recalling \eqref{eq:defTZdansKerr} we define $\T,\Z$ by
\begin{align}\label{eq:defTZdansPT}
	\T:=\frac12\left(e_4+\frac{\Delta}{|q|^2}e_3-2a\Real(\mathfrak{J})^be_b\right),\quad \Z:=\frac12\left(2(r^2+a^2)\Real(\frakJ)^be_b-a\sin^2\theta e_4-\frac{a\sin^2\theta\Delta}{|q|^2}e_3\right).
\end{align}
These vector fields will be used for the control of region $\un$ as follows :
\begin{itemize}
	\item First, we will use $\T,\Z$ to define an isometry between the PT horizontal distribution $\mch$ and the tangent space of appropriate spheres $S(r,\ubar)$ in region $\un$, which is a perturbation of the map $\Theta_\mck$ defined in Lemma \ref{lem:transfokerrIun} (see Lemma \ref{lem:transformationIspheres}).
	\item Second, the good commutation properties of $\lieT,\lie_{\Z}$ with other covariant derivative operators will be used to control derivatives of the linearized Ricci and curvature coefficients, see Lemma \ref{lem:commdflieTZI}.
\end{itemize}

\noindent\textbf{Definition of $\T$ with respect to double null coordinates.} Let us consider a double null foliation as described in Section \ref{section:doublenullgrandesection}. Recalling the identity \eqref{eq:TenDNlolkerr} in Kerr, we will define similarly the approximate symmetry
\begin{align}\label{eq:T}
	\T:=\frac12(\partial_{\ubarring}-\partial_{\uring})
\end{align}
in perturbations of Kerr (we will not use the fact that $\partial_{\phi_*}$ is an approximate symmetry in the double null gauge, hence we do not define an analog vector field $\Z$ here). Equivalently,
\begin{align}\label{eq:expreT}
	\T=\frac12(\Omega^2 \ering_4-\ering_3-b^A\partial_{\theta^A})
\end{align}
with $\ering_3,\ering_4$ as in \eqref{eq:DuDubardoublenull}. We already note that for any $S(\uring,\ubarring)$-tangent tensor $U_\mck$ defined on the background Kerr spacetime, recalling the definition of the horizontal Lie derivative in Section \ref{section:horizontalliederivative}, we have
$$\lieT U=\T(\widecheck{U}^{B_1\cdots B_s}_{A_1\cdots A_r})\dee\theta^{A_1}\otimes\cdots\otimes\dee\theta^{A_r}\otimes\partial_{\theta^{B_1}}\otimes\cdots\otimes\partial_{\theta^{B_s}},$$
or equivalently $\lieT U_\mck=0$, because in our context, $(U_\mck)^{B_1\cdots B_s}_{A_1\cdots A_r}$ will be a function of $u+\ubar=2r'_*$ and $\theta_*$ only.
\begin{lem}\label{lem:lienablaT}
	Let $U$ be a $S(\uring,\ubarring)$-tangent $k$-tensor. We have 
	$$\lieT U_{A_1\cdots A_k}=\nabring_\T U_{A_1\cdots A_k}+\frac12\sum_{i=1}^k\gamma^{BC}\left(\Omega^2\mathring{\chi}_{A_iC}-\mathring{\chibar}_{A_iC}-\nabring_{A_i}b_C\right)U_{A_1\cdots B\cdots A_k}.$$
\end{lem}
\begin{proof}
Using that $\T=\frac12(\partial_{\ubarring}-\partial_{\uring})$ commutes with any angular $\partial_{\theta^A}$ derivative we get 
\begin{align*}
	\lieT U_{A_1\cdots A_k}=\T(U_{A_1\cdots A_k})&=\nabring_\T U_{A_1\cdots A_k}+\sum_{i=1}^k\gamma^{BC}\g(\D_\T \partial_{\theta^{A_i}},\partial_{\theta^C}) U_{A_1\cdots B\cdots A_k},
\end{align*}
and we conclude the proof using $\g(\D_\T \partial_{\theta^{A_i}},\partial_{\theta^C})=\g(\D_{A_i}\T ,\partial_{\theta^C})$ combined with \eqref{eq:expreT}.
\end{proof}

\section{The main theorem}\label{section:mainthm}

\subsection{Main ansatz for solutions of the Teukolsky equation}\label{section:ansatzdef}
In this section, we define the main ansatz for the curvature component $A$ defined with respect to a frame which will be a perturbation of the ingoing principal null frame \eqref{eq:principalingoinkerr} in the Kerr black hole interior. We refer the reader to Section \ref{section:assumptionpricelaw} for the explanation and context of the definitions and results which are stated below without further comments. For $|m|\leq 2$, we define the following function of $r$,
\begin{equation}\label{eq:dedfAm(r)}
	\begin{aligned}
		A_m(r)=\frac13\Big[&3\Delta^2+(r-M)(4(a^2-M^2)+6\Delta)iam\\
		&-(2\Delta+6(a^2-M^2)+4(r-M)^2)a^2m^2-4(r-M)ia^3m^3+2a^4m^4\Big].
	\end{aligned}
\end{equation}
Notice that, for any $|m|\leq 2$, we have the identity
\begin{align}\label{eq:amdermoins}
	A_m(r_-)=\frac23am(am+i(r_+-r_-))(M^2+a^2(m^2-1)),
\end{align}
which implies $A_0(r_-)=0$ and $A_m(r_-)\neq 0$ for $m=\pm 1,\pm 2$. We assume initially that in some PT gauge, there are constants $Q_m\in\C$ for $|m|\leq 2$ such that the curvature component $A$ behaves like the ansatz
$$\Psi:=\frac{1}{\ubar^6}\sum_{|m|\leq 2}\frac{Q_m A_m(r)}{\qbar^2}\mcd\hot(\mcd(Y_{m,2}(\cos\theta)e^{im\phi_+})),$$
in a way which will be precised in the following subsection. Note that, recalling the Teukolsky operator $\mcl$ in \eqref{eq:teukop}, from Proposition \ref{prop:teukansatzkerr} we get $\mcl( A_m(r)\qbar^{-2}\mcd\hot(\mcd(Y_{m,2}(\cos\theta)e^{im\phi_+})))=0$ in Kerr. This implies that the ansatz $\Psi$ is an approximate solution of the tensorial Teukolsky equation in Kerr in the sense that
$$\mcl(\Psi)=O(\ubar^{-7}).$$
\begin{rem}
	Lemma \ref{lem:DhotDYm2} in Kerr shows that the horizontal tensors $\mcd\hot(\mcd(Y_{m,2}(\cos\theta)e^{im\phi_+}))$ should be interpreted as the tensorial analog of the standard $\ell=2$, spin $+2$ spherical harmonics. Similarly, $\Psi$ as defined above should be interpreted as the tensorial analog of the RHS in \eqref{eq:expectedpricelaw}.
\end{rem}
\subsection{Initial data assumptions}\label{section:initialdatavrai}
Let $(a,M)$ be subextremal Kerr black hole parameters satisfying $0<|a|<M$. We define the initial hypersurface
$$\mathcal{A}:=(1,+\infty)_\ubar\times\mathbb{S}^2,$$
and we initialize the scalar function $r$ on $\mathcal{A}$ as follows
\begin{align}\label{eq:rmcainit}
	r|_\mca=r_\mca,\quad r_\mca:=r_+(1-\delta_+),
\end{align}
for some given $\delta_+>0$. Recall from Section \ref{section:assumptionpricelaw} that in the context of the global analysis of a Kerr black hole perturbation, the value of $\delta_+$ comes from the analysis in the exterior region which slightly extends into the black hole region, up to the hypersurface $\mca$ for some sufficiently small $\delta_+>0$.  We assume that $\mca$ is equipped with a smooth Riemannian metric $g$ and a smooth symmetric 2-tensor $k$ which satisfy the constraint equations
$$R_g+(tr k)^2-|k|^2=0,\quad \diver k-\overline{\nabla} tr k=0,$$
where $R_g$ is the scalar curvature of $g$, $\overline{\nabla}$ is the covariant derivative of $(\mca,g)$, and the divergence, trace, and tensor norm above are taken with respect to $g$. Using the Choquet-Bruhat theorem \cite{YCB52}, we get that $g$ can be locally extended as a Lorentzian metric $\g$ satisfying the Einstein vacuum equation \eqref{eq:EVE} on a neighborhood of $\mca$, such that the metric induced by $\g$ on $\mca$ is $g$, in a unique way up to diffeomorphism. 

On such a local extension, we assume the following:
\begin{enumerate}
	\item We assume that there is an extended ingoing PT structure $(e_3,e_4,\mch,r,\theta,\frakJ,\ubar)$ (see Section \ref{section:onintroPTgauge}) with associated horizontal distribution $\mch$, and coordinates $(r,\ubar,\theta,\phi_+)$ such that $r|_\mca,\ubar|_\mca$ coincide with the ones defined on $\mca$, along with two horizontal 1-forms $\frakJ_\pm$ and such that, denoting $q=r+ia\cos\theta$,
$$e_3(r)=-1,\quad e_3(\ubar)=e_3(\theta)=e_3(\phi_+)=0,\quad\nabla_3(\qbar\frakJ)=\nabla_3(\qbar\frakJ_\pm)=0.$$
	Here, $\theta,\phi_+$ are local coordinates on the spheres $S(r,\ubar)$ of constant $r,\ubar$ such that defining
	$$\Jmoins=\sin\theta\sin\phi_+,\quad\Jplus=\sin\theta\cos\phi_+,$$
	the usual coordinates patches
	\begin{align*}
	(x^1_{(1)}=\theta,\: x^2_{(1)}=\phi_+)\quad\text{on}\quad\mcu^{(1)}&:=\{\pi/4<\theta<3\pi/4\},\\
	(x^1_{(2)}=\Jplus,\: x^2_{(2)}=\Jmoins)\quad\text{on}\quad\mcu^{(2)}&:=\{0\leq\theta<\pi/3\},\\
	(x^1_{(3)}=\Jplus,\: x^2_{(3)}=\Jmoins)\quad\text{on}\quad\mcu^{(3)}&:=\{2\pi/3<\theta\leq\pi\},
	\end{align*}
	cover the whole sphere.
	\item\label{item:deltaintro} We assume that $\mca$ settles down to Kerr$(a,M)$ in the following sense: there are small real numbers $\varepsilon>0$, $\delta>0$, and a large integer $N_0$ such that denoting $\df_{\mathrm{init}}$ the following set of derivatives,
	$$\df_{\mathrm{init}}=\{\nabla_3,\nabla_4,\nabla\},$$
	we assume that the Ricci and curvature coefficients as defined in Section \ref{section:ricciandcurvcomplex} with respect to $e_3,e_4,\mch$ satisfy the following initial bounds on $\mca$:
	\begin{equation}
		\begin{gathered}
			\left|\df_{\mathrm{init}}^{\leq N_0}\left(trX-\frac{2\qbar\Delta}{|q|^4}\right)\right|\leq\frac{\varepsilon}{\ubar^{3+\delta}},\quad \left|\df_{\mathrm{init}}^{\leq N_0}\left(tr\Xbar-\frac{2}{\qbar}\right)\right|\leq\frac{\varepsilon}{\ubar^{3+\delta}},\\
			|\df_{\mathrm{init}}^{\leq N_0}(\wh{X},\wh{\Xbar})|\leq\frac{\varepsilon}{\ubar^{3+\delta}},\quad |\df_{\mathrm{init}}^{\leq N_0}(\Xi,\Xibar)|\leq\frac{\varepsilon}{\ubar^{3+\delta}},\\
			\left|\df_{\mathrm{init}}^{\leq N_0}\left(Z-\frac{aq}{|q|^2}\mathfrak{J}\right)\right|\leq\frac{\varepsilon}{\ubar^{3+\delta}},\quad \left|\df_{\mathrm{init}}^{\leq N_0}\left(H-\frac{aq}{|q|^2}\mathfrak{J}\right)\right|\leq\frac{\varepsilon}{\ubar^{3+\delta}},\quad \left|\df_{\mathrm{init}}^{\leq N_0}\left(\Hbar-\frac{a\qbar}{|q|^2}\mathfrak{J}\right)\right|\leq\frac{\varepsilon}{\ubar^{3+\delta}},\\
			\left|\df_{\mathrm{init}}^{\leq N_0}\left(\omega+\frac12\partial_r\left(\frac{\Delta}{|q|^2}\right)\right)\right|\leq\frac{\varepsilon}{\ubar^{3+\delta}},\quad \left|\df_{\mathrm{init}}^{\leq N_0}\omegabar\right|\leq\frac{\varepsilon}{\ubar^{3+\delta}},\label{eq:assumptionricci}
		\end{gathered}
	\end{equation}
	
as well as
\begin{align}
	|\df_{\mathrm{init}}^{\leq N_0}(A,\Abar,B,\Bbar)|\leq\frac{\varepsilon}{\ubar^{3+\delta}},\quad \left|\df_{\mathrm{init}}^{\leq N_0}\left(P+\frac{2m}{q^3}\right)\right|\leq\frac{\varepsilon}{\ubar^{3+\delta}},\label{eq:assumptioncurvature}
\end{align}
and the derivative of the coordinates $r,\ubar,\theta$ and of the horizontal 1-form $\frakJ$ satisfy
\begin{equation}
	\begin{gathered}
		\left|\df_{\mathrm{init}}^{\leq N_0}\left(e_4(r)-\frac{\Delta}{|q|^2}\right)\right|\leq\frac{\varepsilon}{\ubar^{3+\delta}},\quad\left|\df_{\mathrm{init}}^{\leq N_0}\left(e_4(\ubar)-\frac{2(r^2+a^2)}{|q|^2}\right)\right|\leq\frac{\varepsilon}{\ubar^{3+\delta}},\\
		 |\df_{\mathrm{init}}^{\leq N_0}e_4(\cos\theta)|\leq\frac{\varepsilon}{\ubar^{3+\delta}},\quad\left|\df_{\mathrm{init}}^{\leq N_0}\left(\nabla_4\frakJ+\frac{\Delta\qbar}{|q|^4}\frakJ\right)\right|\leq\frac{\varepsilon}{\ubar^{3+\delta}}, \\
		|\df_{\mathrm{init}}^{\leq N_0}\mcd r|\leq\frac{\varepsilon}{\ubar^{3+\delta}},\quad\left|\df_{\mathrm{init}}^{\leq N_0}\left(\mcd\cos\theta- i\frakJ\right)\right|\leq\frac{\varepsilon}{\ubar^{3+\delta}},\quad\left|\df_{\mathrm{init}}^{\leq N_0}\left(\mcd\ubar-a\frakJ\right)\right|\leq\frac{\varepsilon}{\ubar^{3+\delta}},\\
		\left|\df_{\mathrm{init}}^{\leq N_0}\left(\divc \frakJ-\frac{4i(r^2+a^2)\cos\theta}{|q|^4}\right)\right|\leq\frac{\varepsilon}{\ubar^{3+\delta}},\quad  |\df_{\mathrm{init}}^{\leq N_0}\mcd\hot\frakJ|\leq\frac{\varepsilon}{\ubar^{3+\delta}}.\label{eq:assumptiondercoord}
	\end{gathered}
\end{equation}
Moreover, we assume that the derivatives of $\Jplus,\Jmoins$ can be controlled as follows on $\mca$:
\begin{equation}\label{eq:initderx1X2EHEH}
	\begin{gathered}
			 \left|\df_{\mathrm{init}}^{\leq N_0}\left(\mcd \Jplus -\frakJ_+\right)\right|\leq\frac{\varepsilon}{\ubar^{3+\delta}},\quad \left|\df_{\mathrm{init}}^{\leq N_0}\left(\mcd \Jmoins -\frakJ_-\right)\right|\leq\frac{\varepsilon}{\ubar^{3+\delta}},\\
		\left|\df_{\mathrm{init}}^{\leq N_0}\left(e_4(\Jplus) +\frac{2a}{|q|^2} \Jmoins\right)\right|\leq\frac{\varepsilon}{\ubar^{3+\delta}},\quad \left|\df_{\mathrm{init}}^{\leq N_0}\left(e_4(\Jmoins) -\frac{2a}{|q|^2} \Jplus\right)\right|\leq\frac{\varepsilon}{\ubar^{3+\delta}}.
	\end{gathered}
\end{equation}
Finally, we assume the following estimates for the derivatives of $\frakJ_\pm$ on $\mca$:
\begin{equation}\label{eq:initderderivefrakJpm}
	\begin{gathered}
		\left|\df_{\mathrm{init}}^{\leq N_0}\left(\divc\frakJ_++\frac{4\Jplus}{r^2}+\frac{4ia(r^2+a^2)\cos\theta}{|q|^4}\Jmoins\right)\right|\leq\frac{\varepsilon}{\ubar^{3+\delta}},\\
		\left|\df_{\mathrm{init}}^{\leq N_0}\left(\divc\frakJ_-+\frac{4\Jmoins}{r^2}-\frac{4ia(r^2+a^2)\cos\theta}{|q|^4}\Jplus\right)\right|\leq\frac{\varepsilon}{\ubar^{3+\delta}},\\
		\left|\df_{\mathrm{init}}^{\leq N_0}\mcd\hot\frakJ_\pm\right|\leq\frac{\varepsilon}{\ubar^{3+\delta}},\quad	\left|\df_{\mathrm{init}}^{\leq N_0}\left(\nabla_4\frakJ_\pm+\frac{\Delta\qbar}{|q|^4}\frakJ_\pm\pm\frac{2a}{|q|^2}\frakJ_\mp\right)\right|\leq\frac{\varepsilon}{\ubar^{3+\delta}}.
	\end{gathered}
\end{equation}
	\item Defining
	$$j=\Real(\frakJ),\quad\quad j_\pm=\Real(\frakJ_\pm),$$
	 we assume the following identities on $\mca$ for scalar products between $j,j_\pm$:
	\begin{equation}\label{eq:inithypscalarproducts}
		\begin{gathered}
			j\cdot j=\frac{\sin^2\theta}{|q|^2},\quad\hodge{j_-}\cdot j_+=-\hodge{j_+}\cdot j_-=\frac{\cos\theta}{|q|^2},\\
			\hodge{j_-}\cdot j=-\frac{\sin\theta\cos\theta\sin\phi_+}{|q|^2},\quad\hodge{j_+}\cdot j=-\frac{\sin\theta\cos\theta\cos\phi_+}{|q|^2},\\
			j_-\cdot{j_-}=\frac{\cos^2\theta\sin^2\phi_++\cos^2\theta_+}{|q|^2},\quad j_+\cdot{j_+}=\frac{\cos^2\theta\cos^2\phi_++\sin^2\theta_+}{|q|^2},\\
			j_-\cdot j_+=-\frac{\sin^2\theta\sin\phi_+\cos\phi_+}{|q|^2},\quad j\cdot j_+=-\frac{\sin\theta\sin\phi_+}{|q|^2},\quad j\cdot j_-=\frac{\sin\theta\cos\phi_+}{|q|^2}.
		\end{gathered}
	\end{equation}

	\item We assume that there are five complex constants $(Q_m)_{|m|\leq 2}\in\C^5$ such that, denoting by $\Psi$ the following ansatz,
	\begin{align}\label{eq:definitansatzun}
		\Psi:=\frac{1}{\ubar^6}\sum_{|m|\leq 2}\frac{Q_mA_m(r)}{\qbar^2}\mcd\hot(\mcd(Y_{m,2}(\cos\theta)e^{im\phi_+})),
	\end{align}
	the curvature component $A$ satisfies the following version of Price's law on $\mca$\footnote{Actually it is natural to expect that the stronger bound $	\|\df_{\mathrm{init}}^{\leq 4}(A-\Psi)\|_{L^2(S(r_\mca,\ubar))}\leq\varepsilon\ubar^{-6-\delta}$ holds on $\mca$, but the extra smallness coming from the $\varepsilon$ factor will not be used in the present paper.}:
	\begin{align}\label{eq:initialhyponmca}
		\|\df_{\mathrm{init}}^{\leq 4}(A-\Psi)\|_{L^2(S(r_\mca,\ubar))}\leq\ubar^{-6-\delta}.
	\end{align}
\end{enumerate}

\begin{rem}\label{rem:initialPTset}
	We remark the following:
	\begin{itemize}
		\item 	As explained in introduction, in practice we expect that one obtains such initial data by choosing $\delta_+>0$ sufficiently small and defining $\mca$ as the level set $\{r=r_+(1-\delta_+)\}$ of a foliation $r,\ubar,\theta,\phi_+$ just inside the black hole region of a perturbation of Kerr spacetime as in \cite{KS21}. Indeed, we expect that a careful analysis of the perturbed exterior region (which would also naturally cover a small part of the black hole region) yields such a foliation and appropriate structure $(e_3,e_4,\mch)$, which satisfy the estimates written above.
		\item From the linearized analysis we expect that $Q_m\neq 0$ generically, and the bounds\footnote{In particular, $\varepsilon$ can not be chosen to be small with respect to $(Q_m)_{|m|\leq 2}$.}
		$$|Q_m|\lesssim \varepsilon^2,\quad\text{for}\:\: |m|\leq 2.$$
		However these bounds will not be needed in the present paper. 
		\item Assumption \eqref{eq:inithypscalarproducts} could be replaced by approximate estimates with error terms bounded by $\varepsilon\ubar^{-3-\delta}$, but the analysis in the exterior region in \cite{KS21} actually provides equalities, since $\frakJ,\frakJ_\pm$ are initialized on a sphere $S^*$ such that \eqref{eq:inithypscalarproducts} holds, and $\frakJ,\frakJ_\pm,\theta,\ubar,\phi_+$ are then transported up to $\mca$ in a way that these scalar products identities remain true. 
		\item Notice that the bounds for $H,\Xibar,\omegabar$ in \eqref{eq:assumptionricci} are automatically satisfied as $(e_3,e_4,\mch,r,\theta,\frakJ)$ is assumed to be a PT structure. Actually, if assumptions \eqref{eq:assumptionricci}--\eqref{eq:initialhyponmca} hold with respect to a \emph{non-necessarily PT} null pair $(e_3,e_4)$ with horizontal distribution $\mch$, then by performing a change of frame of the type \eqref{eq:frametransfo} one can recover a PT structure which also satisfies \eqref{eq:assumptionricci}--\eqref{eq:initialhyponmca}\footnote{ In particular concerning the Price's law-type estimate \eqref{eq:initialhyponmca}, this relies on the fact that $A$ is gauge invariant at the linear level by the transformation formula \eqref{eq:changealpha}.
}
	\end{itemize}
\end{rem}

\subsection{Spacetime regions}\label{section:regions}
In this subsection, we present the successive subregions which will be constructed throughout the proof. \textbf{We refer to Figure \ref{fig:regionsNL} for an illustration of the different regions considered.}
\subsubsection{Region $\un$}
Using a bootstrap argument in the PT gauge initialized on $\mca=\{r=r_\mca\}\cap\{\ubar>1\}$, we will prove that for any constants $\delta_->0$ and suitable $w_f<-1$, the initial geometric quantities (namely the metric, the frame and horizontal distribution) on $\mca$ can be extended to the region
\begin{align}\label{eq:ecouteregIdefigg}
	\un:=\{(r,\ubar,\theta,\phi_+)\in(r_-(1+\delta_-),r_{\mca}]\times(1,+\infty)\times\mathbb{S}^2\:/\:w(r,\ubar)<w_f\},
\end{align}
where the scalar $w(r,\ubar)$ has spacelike level sets and is defined by
\begin{align}\label{eq:defwregIunee}
	w(r,\ubar)=2r^*-\ubar-r
\end{align}
with $r^*(r)$ defined as in \eqref{eq:defderstarvraiment}. Choosing a constant $w_f(a,M,\delta_\pm,\delta,(Q_m)_{|m|\leq 2})\ll -1$ negative enough, this extension is such that the initial estimates on $\mca$ are propagated to $\un$ (provided $\varepsilon$ is small enough). The small $\delta_->0$ depends on the large constant $C_R$, which only depends on $a,M$, chosen in the analysis of regions $\deux$ and $\trois$, see Sections \ref{section:defdeuxettrois} and \ref{section:smallness}.

\subsubsection{Regions $\deux$ and $\trois$}\label{section:defdeuxettrois}
Note that the future spacelike boundary of region $\un$ is $\{r=r_-(1+\delta_-)\}\cup\{w=w_f\}$, so that $\un$ is far away from the (yet to be defined) Cauchy horizon. The regions $\deux$ and $\trois$ that we will construct in the future of region $\un$ bridge this gap, with the Cauchy horizon being part of the future null boundary of region $\trois$. Both regions $\deux$ and $\trois$ are constructed at once using the double null gauge built by Dafermos and Luk \cite{stabC0}. 

More precisely, denoting $\mathring{u},\mathring{\ubar}$ the Kerr values of the double null advanced and retarded time defined in Section \ref{section:linearizationdoublenull} with respect to the coordinates $(r,\ubar,\theta,\phi_+)$ of region $\un$, we define the hypersurface
\begin{align}\label{eq:sigma0mghd}
	\Sigma_0:=\{\mathring{u}+\mathring{\ubar}=C_R\}\subset\un,
\end{align}
where $C_R>1$ is a large enough constant chosen in the proof, which depends only on $a,M$. Note that $\delta_-(a,M)>0$ can then be fixed, depending on the choice of $C_R(a,M)$, to make sure that $\Sigma_0$ is indeed included in region $\un$.

Then, following \cite{stabC0} and using these coordinates $(\mathring{u},\mathring{\ubar},\theta^A)$ we initialize a double null foliation on $\Sigma_0$ which can then be extended in the future of $\Sigma_0$ to the whole region 
$$\deux\cup\trois:=\{(\mathring{u},\mathring{\ubar},\theta^A)\in\mathbb{R}^2\times\mathbb{S}^2\:/\:\mathring{u}<u_f,\:\mathring{u}+\mathring{\ubar}\geq C_R\},$$
for some negative enough constant $u_f<-1$. We also define the subregions $\deux$ and $\trois$ as follows:
\begin{align}
	\deux&:=\{(\mathring{u},\mathring{\ubar},\theta^A)\in\mathbb{R}^2\times\mathbb{S}^2\:/\:\mathring{u}<u_f,\:\mathring{u}+\mathring{\ubar}\geq C_R,\:u+\ubar\leq\ubar^{\gamma}\},\label{eq:defdedeuxx}\\
	\trois&:=\{(\mathring{u},\mathring{\ubar},\theta^A)\in\mathbb{R}^2\times\mathbb{S}^2\:/\:\mathring{u}<u_f,\:\mathring{u}+\mathring{\ubar}\geq C_R,\:u+\ubar\geq\ubar^{\gamma}\},\label{eq:defdetroiss}
\end{align}
where $\gamma>0$ is a small enough constant, depending only on $\delta$. More precisely we will eventually chose $\gamma<\delta/12$, where $\delta>0$ is the small initial polynomial decay parameter introduced in Section \ref{section:initialdatavrai}, see \eqref{eq:assumptionricci}--\eqref{eq:initialhyponmca}. We also denote by $\Gamma$ the boundary between $\deux$ and $\trois$,
\begin{align}\label{eq:defiGammahyp}
	\Gamma:=(\deux\cup\trois)\cap\{\uring+\ubarring=\ubarring^\gamma\}.
\end{align}
 As explained in the introduction, the point of introducing regions $\deux$ and $\trois$ is that in region $\trois$, the scalar $\Omega^2_\mck$ decays (sup-)exponentially with respect to $\mathring{\ubar}$, namely
\begin{align*}
	\Omega^2_\mck(\mathring{u},\mathring{\ubar},\theta^A)\lesssim e^{-|\kappa_-|\mathring{\ubar}^\gamma},\quad\text{in}\:\:\trois.
\end{align*}
This substantially simplifies the analysis, as already observed in \cite{scalarMZ,spin+2}.
\begin{lem}\label{lem:usimubarII}
	We have $|\mathring{u}|\sim\mathring{\ubar}$ in $\deux$ and $\mathring{\ubar}\geq|\mathring{u}|$ in $\deux\cup\trois$, provided $C_R\geq 0$.
\end{lem}
\begin{proof}
	The first point comes from $\mathring{u}+\mathring{\ubar}\leq\mathring{\ubar}^{\gamma}$ in $\deux$ which implies $\mathring{\ubar}\leq |\mathring{u}|+\mathring{\ubar}^\gamma$ thus $\mathring{\ubar}\lesssim|\mathring{u}|$ provided $\mathring{\ubar}\geq 2$, which holds in this context since $\ubarring\geq C_R-u_f$ with $-u_f$ which will eventually be chosen large. Moreover in $\deux\cup\trois$ we also have $\mathring{u}+\mathring{\ubar}\geq C_R$ which implies $|\mathring{u}|=-\mathring{u}\leq\mathring{\ubar}-C_R\leq\mathring{\ubar}$ concluding the proof of the first point, while also proving the second point.
\end{proof}

\subsection{Main norms}
\subsubsection{Main norm in region $\un$ in the PT gauge}
Recalling the linearized quantities in the PT gauge defined Section \ref{section:PTlinearized}, we define the following norm in region $\un$:
\begin{align}
	\mcn_{\un}:=\sup_{\un}\Big|\ubar^{3+\delta/2}\df^{\leq N_1}&\Big(\Xi,\widecheck{\omega},\widecheck{trX},\wh{X},\widecheck{Z},\widecheck{\Hbar},\widecheck{tr\Xbar},\wh{\Xbar},\widecheck{e_4(r)},\widecheck{e_4(\ubar)},e_4(\cos\theta),\nabla r,\widecheck{\nabla\cos\theta},\widecheck{\nabla\ubar},\nn\\
	&\widecheck{\nabla_4\mathfrak{J}},\widecheck{\nabla_4\frakJ_{\pm}},\widecheck{\nabla_4 J^{(\pm)}},\widecheck{\divc\mathfrak{J}},\mcd\hot\mathfrak{J},\widecheck{\divc\mathfrak{J}_\pm},\mcd\hot\mathfrak{J}_\pm,\widecheck{\mcd J ^{(\pm)}},A,B,\widecheck{P},\Bbar,\Abar\Big)\Big|,\label{eq:normeun}
\end{align}
where $\df=\{\nabla_3,\nabla_4,\nabla\}$, and where $N_1:=N_0-3$, with $N_0$ the initial number of derivatives controlled on $\mca$, see \eqref{eq:assumptionricci}--\eqref{eq:initderderivefrakJpm}.

\subsubsection{Main norms in region $\deux\cup\trois$ in the double null gauge}
Recalling the linearized quantities in the double null gauge defined in Sections \ref{section:linearizationdoublenull} and \ref{section:specificdoublenull}, we define the following norm in region $\deux$:
\begin{equation}	\label{eq:normedeux}
	\begin{aligned}
		\mcn_{\deux}:=&\sup_{\deux}\Big|{\ubarring}^{2+\delta/3}\mathring{\df}^{\leq N_{max}}\Big(\gcheck,\bcheck,\psicheck_{\Hbar},\widecheck{\mathring{\omegabar}},\psicheck,\widecheck{\mathring{K}},\widecheck{\hodge{\mathring{K}}},\widecheck{\mathring{\betabar}},\widecheck{\mathring{\alphabar}}\Big)\Big|\\
		&+\sup_{\deux}\Big|\Omega^2{\ubarring}^{2+\delta/3}\mathring{\df}^{\leq N_{max}}\Big(\psicheck_H,\widecheck{\mathring{\beta}}\Big)\Big|+\sup_{\deux}\Big|\Omega^4{\ubarring}^{2+\delta/3}\mathring{\df}^{\leq N_{max}}\widecheck{\mathring{\alpha}}\Big|
	\end{aligned}
\end{equation}
where $\mathring{\df}=\{\nabring_3,\Omega^2\nabring_4,\nabring\}$, and where $N_{max}$ is the integer part of $(N_0-12)/5-1$. We also define the following norm in region $\trois$ which only involves angular derivatives:
\begin{equation}\label{eq:normetrois}
	\begin{aligned}
		\mcn_{\trois}:=&\sup_{\trois}\Big||\uring|^{2+\delta/3}\nabring^{\leq N_0-12}\Big(\gcheck,\psicheck_{\Hbar},\widecheck{\mathring{\omegabar}},\psicheck,\widecheck{\mathring{K}},\widecheck{\hodge{\mathring{K}}},\widecheck{\mathring{\betabar}}\Big)\Big|+\sup_{\trois}\Big|\Omega^2\ubarring^{2+\delta/3}\nabring^{\leq N_0-12}\psicheck_H\Big|\\
		&+\sup_{\trois}\Big|\Omega^2|\uring|^{2+\delta/3}\nabring^{\leq N_0-12}\widecheck{\mathring{\beta}}\Big|+\sup_{\trois}\Big|\ubarring^{2+\delta/3}\nabring^{\leq N_0-12}\bcheck\Big|.
	\end{aligned}
\end{equation}
\subsection{Main theorem}
\subsubsection{Smallness constants}\label{section:smallness}
We present an exhaustive list of the different constants used in this paper and the relations between them.

\noindent\textbf{Constants involved in the initial data assumptions (see Section \ref{section:initialdatavrai}).}
\begin{itemize}
	\item $(a,M)$ are the black hole parameters for which the initial data on $\mca$ approaches the Kerr spacetime. We consider the subextremal setting with $0<|a|<M$.
	\item $\delta_+>0$ is a given constant, such that $r_+(1-\delta_+)>r_-$, which indicates the level set of $r$ that corresponds to the initial hypersurface $\mca$.
	\item $\delta>0$ is involved in the initial polynomial decay in $\ubar$ of Kerr perturbations on $\mca$, see \eqref{eq:assumptionricci}--\eqref{eq:initialhyponmca}.
	\item The constants $(Q_m)_{|m|\leq 2}\in\C^5$ appear in the initial leading-order term $\Psi$ of the curvature component $A$ on $\mca$, see \eqref{eq:definitansatzun}. 
	\item $\varepsilon>0$ is the size of the initial data (or more precisely, the size of the initial perturbation with respect to Kerr($a,M$) on $\mca$), see \eqref{eq:assumptionricci}--\eqref{eq:initderderivefrakJpm}. It will be chosen small enough with respect to the black hole parameters $(a,M)$ and the parameter $\delta_+$. 
\end{itemize}
Recall from Section \ref{section:conventions} that in all the paper, \emph{except when explicitly mentioned}, the implicit constants in the bounds denoted by $\lesssim$ depend only on $(a,M)$, and not on $\varepsilon,\delta_+,\delta$ nor on the auxiliary smallness constants which we present now.

\noindent\textbf{Auxiliary smallness constants introduced during the proof of the main theorem.}
\begin{itemize}
	\item The constant $C_R\geq 1$ depends only $(a,M)$. It will be chosen large enough such that by \eqref{eq:omegamcksim},
	$$\sup_{\uring+\ubarring\geq C_R}\Omega^2_\mck(\uring,\ubarring,\theta^A)\lesssim e^{-|\kappa_-|C_R}$$
	is small enough with respect to the black hole parameters $(a,M)$.
	\item The small constant $\delta_->0$ is chosen such that, denoting $r^*(\mathring{u}+\mathring{\ubar},\theta_*)$ the dependence of $r^*$ on $\mathring{u}+\mathring{\ubar}=2 r'_*$ and $\theta_*$ (see Section \ref{section:linearizationdoublenull}), we have
	\begin{align}\label{eq:conditionCRR}
		r^*(r_-(1+\delta_-))>\sup_{\theta_*} r^*(C_R,\theta_*).
	\end{align}
	This condition ensures that the spacelike hypersurface $\Sigma_0=\{\mathring{u}+\mathring{\ubar}=C_R\}$ is fully included in region $\un$, see Figure \ref{fig:regionsNL}. Thus $\delta_-$ depends on $C_R$ and on $a,M$. Note that once $C_R$ and hence $\delta_-$ are fixed, they only depend on $a,M$.
	\item The constant $w_f$ defines a part of the future boundary of region $\un$, see Figure \ref{fig:regionsNL}. We choose it as follows: for any fixed $\delta_->0$, recalling the definition \eqref{eq:defwregIunee} of $w$ and the definition \eqref{eq:ecouteregIdefigg} of region $\un$, we get the following bound in $\un$:
	$$\ubar\geq 2r^*-r-w\geq 2r^*_\mca-r_--w_f,$$
	where $r^*_\mca=r^*(r_\mca)=r^*(r_+(1-\delta_+))$. This implies the following bound in $\un$,
	$$\ubar^{-3-\delta}\leq\ubar^{-3-\delta/2}(2r^*_\mca-r_--w_f)^{-\delta/2}.$$
	Now, the natural polynomial upper bounds that we will obtain in region $\un$ will be of the type
	$$C(a,M,\delta_\pm)\ubar^{-3-\delta},$$
	where the constants $C(a,M,\delta_\pm)$ depend on $\delta_\pm$ on top of $(a,M)$. We will thus choose $w_f\ll-1$ negative enough depending on $a,M,\delta_\pm,\delta,$ such that 
	\begin{align*}
		C(a,M,\delta_\pm)(2r^*_\mca-r_--w_f)^{-\delta/2}\leq 1,
	\end{align*}
	namely such that we get rid of the dependence in $\delta_\pm$ in the bounds in $\un$.
	\item The constant $u_f$ defines the left part of the future boundary of region $\deux\cap\trois$, see Figure \ref{fig:regionsNL}. We chose $u_f\ll -1$ negative enough such that, to begin with, the sphere $S(u_f,C_R-u_f)$ is included in region $\un$, negative enough such that the results of \cite{stabC0} can be applied, and negative enough with respect to $a,M,Q_{m}$ such that Proposition \ref{prop:explosionintegrale} applies. This yields a choice $u_f\ll -1$ which depends on $a,M,\delta_+,\delta,(Q_m)_{|m|\leq 2}$. 
	\item The constants $p$ and $N$ will appear respectively in the analysis of the Bianchi, null structure, and Teukolsky equation in the PT gauge in region $\un$, and in the analysis of the Teukolsky equation in the auxiliary non-integrable frame in region $\deux$. They are both chosen large depending only on $a,M,\delta_+$.
	
\end{itemize}
\subsubsection{Statement of the main theorem}
We are finally ready to state our main result.
\begin{thm}[Main theorem, precise version]\label{thm:mainthm}
	Let $(\mca,g,k)$ be an initial data set for the Einstein vacuum equations satisfying the assumptions of Section \ref{section:initialdatavrai} with $N_0$ large enough\footnote{Inspecting the proof, we find that $N_0\geq 70$ is enough, although it is likely far from optimal.}.
Then, for $\varepsilon>0$ small enough depending on $(a,M,\delta_+)$, there exist subregions $\un\cup\deux\cup\trois\subset\mcm$ inside the maximal globally hyperbolic development $(\mcm,\g)$ of $(\mca,g,k)$ such that:
\begin{enumerate}
	\item $\un$ is isometric to $\{(r,\ubar,\theta,\phi_+)\in(r_-(1+\delta_-),r_{\mca}]\times(1,+\infty)\times\mathbb{S}^2\:/\:w(r,\ubar)<w_f\}$, where $w(r,\ubar)$ is defined in \eqref{eq:defwregIunee}, for some negative enough constant $w_f(a,M,\delta_+,\delta)\ll -1$ and some small enough constant $\delta_-(a,M)>0$. Moreover $\un$ is equipped with an extended ingoing PT structure (see Definition \ref{defi:PTframedef}) such that, recalling \eqref{eq:normeun},
	$$\mcn_{\un}\lesssim\varepsilon,$$
	and the PT outgoing extremal curvature component $A$ satisfies in particular, in $\un$,
	$$\left|A-\frac{1}{\ubar^6}\sum_{|m|\leq 2}\frac{Q_mA_m(r)}{\qbar^2}\mcd\hot(\mcd(Y_{m,2}(\cos\theta)e^{im\phi_+}))\right|\lesssim \frac{1}{\ubar^{6+\delta/2}},$$
	where the implicit constant above depends on $a,M,\delta_\pm,w_f,(Q_m)_{|m|\leq 2})$.
	\item $\deux\cup\trois$ is the MGHD of the spacelike hypersurface $\Sigma_0\cap\{\mathring{u}<u_f\}\subset\un$, where $\Sigma_0$ and $\uring$ are defined in \eqref{eq:sigma0mghd}, for some negative enough constant $u_f(a,M,\delta_+,\delta,(Q_m)_{|m|\leq 2})\ll -1$. Moreover, $\deux\cup\trois$ is equipped with a double null foliation $(\uring,\ubarring)$, is isometric to $$\{(\mathring{u},\mathring{\ubar},\theta^A)\in\mathbb{R}^2\times\mathbb{S}^2\:/\:\mathring{u}<u_f,\:\mathring{u}+\mathring{\ubar}\geq C_R\},$$
	for some large enough constant $C_R(a,M)\gg 1$, and satisfies, recalling \eqref{eq:normedeux} and \eqref{eq:normetrois},
	$$\mcn_{\deux}+\mcn_{\trois}\lesssim\varepsilon.$$
	\item Moreover, assuming that $(Q_m)_{|m|\leq 2}$ satisfies
	$$(Q_{-2},Q_{-1},Q_{+1},Q_{+2})\neq (0,0,0,0),$$
	then the double null extremal curvature component $\mathring{\alpha}$ blows up as $\ubarring\to+\infty$ in the sense of the result of Corollary \ref{cor:curvintblowup} (see also Proposition \ref{prop:curvintpertblowup} and \eqref{eq:enoughforldeux}). In that case, the future boundary of the MGHD $(\mcm,\g)$ contains a non-trivial piece of Cauchy horizon $\ch$, corresponding to $\{\mathring{\ubar}=+\infty\}$, across which the spacetime metric is Lipschitz-inextendible in the following sense: there is no $C^{0,1}_{loc}$-extension $\hat{\iota}:\mcm\rightarrow\hat{\mcm}$ with the property that for $t_0>0$ there is an affinely parametrised, future directed and future inextendible timelike geodesic $\tau:(-t_0,0)\rightarrow\mcm$ for $t_0>0$ with $\lim_{s\to 0}\mathring{\ubar}(\tau(s))=+\infty$, $\lim_{s\to 0}\mathring{u}(\tau(s))<u_f$, and such that $\lim_{s\to 0}(\hat{\iota}\circ\tau)(s)\in\hat{\mcm}$ exists. 
\end{enumerate}
\end{thm}
\begin{rem} We remark the following:
	\begin{itemize}
		\item The Cauchy horizon $\ch$ is the Dafermos-Luk Cauchy horizon constructed in \cite{stabC0}, where their initial hypersurface $\Sigma_0$ corresponds in this context to the one defined in \eqref{eq:sigma0mghd}. $\ch$ can be defined more precisely using the Dafermos-Luk $C^0$ coordinates $(\uring,\ubarring_\ch,\theta^A_\ch)$, for which it corresponds to the level set $\ch=\{\ubarring_\ch=0\}$ (see Section \ref{section:C0extension}). It is also in this coordinate system that the metric $\g$ extends continuously to $\ch$ and remains $C^0$ close to Kerr, as proven in \cite{stabC0}.
		\item The curvature blow-up in coordinates stated in the rough version of the main theorem (Theorem \ref{thm:roughversion}) is a consequence of the stated blow-up asymptotic behavior for $$\mathring{\alpha}_{AB}=\mathbf{R}\left(\ering_4,\partial_{\theta^A_\ch},\ering_4,\partial_{\theta^B_\ch}\right),$$ where in the $C^0$ coordinates we have the expression 
		$$\ering_4=\Omega^{-2}_\ch\left(\partial_{\ubarring_\ch}+b^A_\ch\partial_{\theta^A_\ch}\right),$$
		where both $\Omega^{-2}_\ch$ and $b^A_\ch$ extend continuously to $\ch$ as proven in \cite{stabC0}.
		\item Independently of Price's law type assumption and of the blow-up and inextendibility results, this theorem also completes the exterior stability results in \cite{KS21} and the $C^0$ interior stability results in \cite{stabC0}, implying a global stability statement for slowly rotating Kerr black holes, up to (a non-trivial piece of) the Cauchy horizon. We bridge the gap by providing the quantitative change of gauge estimates required to initialize the double null foliation used in \cite{stabC0} on $\Sigma_0$\footnote{Note that the result in \cite{stabC0} is stated for $\Sigma_0$ which can be arbitrarily close to $\mch_+$. However in our proof we choose $\Sigma_0$ close to $\ch$, so that we must also construct the spacetime and obtain precise estimates in the no-shift region $\un$.} from the geometric framework of \cite{KS21} coming from the exterior, so that both results can be used as a black box\footnote{The polynomial decay proven in \cite{KS21} corresponds to $\ubar^{-1-\delta}$ on $\mca$ but it is expected that $\ubar^{-3-\delta}$ decay holds for sufficiently fast decaying initial data. Here we assume initial $\ubar^{-3-\delta}$ decay on $\mca$ but all the stability results proven in $\un$ remain true replacing $3+\delta$ with $1+\delta$ with exact same proof, showing that the initial data assumptions in \cite{stabC0} are satisfied for some appropriate choice of $\Sigma_0$.}.
		
	\end{itemize}
\end{rem}

The proof of Theorem \ref{thm:mainthm} is a consequence of the main intermediate results stated in the following subsection.

\subsection{Main intermediate results and proof of the main theorem}\label{section:proofmainthm}
\begin{thm}\label{thm:regionI}
	We consider initial data for the Einstein vacuum equation as described in Section \ref{section:initialdatavrai}. Then, for any $\delta_->0$ such that $r_-(1+\delta_-)<r_+(1-\delta_+)$, provided $\varepsilon>0$ is small enough with respect to $(a,M,\delta_\pm)$, there exists $w_f(a,M,\delta,\delta_+,\delta_-)\ll -1$ such that the MGHD of the initial data $(\mca,g,k)$ contains the region $\un$ isometric to 
	$$\{(r,\ubar,\theta,\phi_+)\in(r_-(1+\delta_-),r_{\mca}]\times(1,+\infty)\times\mathbb{S}^2\:/\:w(r,\ubar)<w_f\},$$
	(recall \eqref{eq:rmcainit} and \eqref{eq:defwregIunee}), such that $\mca\cap\{w(r_\mca,\ubar)<w_f\}$ coincides with the level set $\{r=r_\mca\}$ in $\un$. Moreover $\un$ is equipped with an extended PT structure, a function $\phi_+$, and 1-forms $\frakJ_\pm\in\fraks_1(\C)$ such that 
	$$\mcn_{\un}\lesssim\varepsilon.$$
\end{thm}
\begin{thm}\label{thm:AregionI}
	In region $\un$ given by Theorem \ref{thm:regionI}, the PT curvature component $A$ satisfies
$$\left|\df^{\leq 1}\left(A-\frac{1}{\ubar^6}\sum_{|m|\leq 2}\frac{Q_mA_m(r)}{\qbar^2}\mcd\hot(\mcd(Y_{m,2}(\cos\theta)e^{im\phi_+}))\right)\right|\lesssim \frac{1}{\ubar^{6+\delta/2}},$$
where we recall $\df=\{\nabla_3,\nabla_4,\nabla\}$ in the PT gauge, and the implicit constant above depends on $a$, $M$, $\delta_\pm$, $w_f$, $(Q_m)_{|m|\leq 2}$.
\end{thm}
\begin{thm}\label{thm:regionII}
	We consider the setting of Theorem \ref{thm:regionI} for some $\delta_->0$, and the double null foliation initia data $(\Sigma_0,\hat{g},\hat{k},\mathring{u}|_{\Sigma_0},\mathring{\ubar}|_{\Sigma_0},\theta^A|_{\Sigma_0})$ given by the hypersurface $\Sigma_0\subset\un$ defined in \eqref{eq:sigma0mghd} for some $C_R\in\R$ such that \eqref{eq:conditionCRR} holds, where $\hat{g}$ and $\hat{k}$ are respectively the Riemannian metric on $\Sigma_0$ and the second fundamental form on $\Sigma_0$ induced by the spacetime metric $\g$ in $\un$, and where $\mathring{u}|_{\Sigma_0},\mathring{\ubar}|_{\Sigma_0},\theta^A|_{\Sigma_0}$ are defined in Section \ref{section:defdeuxettrois}. Then, provided $\varepsilon>0$ is small enough with respect to $a,M,C_R$ and $u_f(a,M,C_R,w_f)\ll -1$ is negative enough, the MGHD of the restriction of the initial data to $\Sigma_0\cap\{\mathring{u}<u_f\}$ can be endowed with a double null foliation $(\uring,\ubarring)$ with coordinates $\theta^A$, $A=1,2$ on the spheres $S(\uring,\ubarring)$ so that it is isometric to
$$\deux\cup\trois:=\{(\mathring{u},\mathring{\ubar},\theta^A)\in\mathbb{R}^2\times\mathbb{S}^2\:/\:\mathring{u}<u_f,\:\mathring{u}+\mathring{\ubar}\geq C_R\},$$
and such that $$\mcn_{\deux}+\mcn_{\trois}\underset{a,M,C_R}{\lesssim}\varepsilon.$$
\end{thm}

The proof of the following theorem relies on the auxiliary non-integrable frame constructed in region $\deux$.  
\begin{thm}\label{thm:alphaGamma}
We consider the spacetime region $\deux\cup\trois$ constructed in Theorem \ref{thm:regionII}. Then, provided $C_R(a,M)\gg 1$ is large enough, $\varepsilon(a,M)\ll1$ is small enough, $u_f(a,M)\ll -1$ is negative enough, and $0<\gamma<\delta/12$, we have the following estimate for $\mathring{\alpha}$ in region $\trois$,
\begin{align*}
	\|\mathring{\alpha}-\Omega^{-4}\psi\|_{L^2(S(\mathring{u},\mathring{\ubar}))}\underset{{(Q_m)_{|m|\leq 2}}}{\lesssim} e^{2|\kappa_-|(\mathring{u}+\mathring{\ubar})}\left(\frac{C}{\mathring{\ubar}^{6+\delta/3}}+\frac{1}{\mathring{\ubar}^{6}|u_f|}\right),
\end{align*}
where $\psi\sim 1/\mathring{\ubar}^6$ is the ansatz defined in \eqref{eq:defpsiansatz} and $C$ depends on $u_f,a,M,(Q_m)_{|m|\leq2},\delta_+,\delta$. 
\end{thm}

\begin{thm}\label{thm:inextensibilite}
	We consider the setting of Theorem \ref{thm:alphaGamma}. Provided that $u_f(a,M,(Q_m)_{|m|\leq 2})\ll -1$ is negative enough, and that the following condition is satisfied:
	$$(Q_{-2},Q_{-1},Q_{+1},Q_{+2})\neq (0,0,0,0),$$
	then we have the integral curvature blow-up as $\ubar\to+\infty$ stated in Corollary \ref{cor:curvintblowup} (see also Proposition \ref{prop:curvintpertblowup}), and the Lipschitz inextendibility stated at the end of Theorem \ref{thm:mainthm} holds.
\end{thm}
\textbf{For the locations of the proofs of Theorems \ref{thm:regionI}, \ref{thm:AregionI}, \ref{thm:regionII}, \ref{thm:alphaGamma}, \ref{thm:inextensibilite} in the paper, see Section \ref{section:structurerestpaper}.} We are now ready to prove the main Theorem \ref{thm:mainthm}.
\begin{proof}[\textbf{Proof of Theorem \ref{thm:mainthm}}]
We chose the smallness constants as described in Section \ref{section:smallness}: 
\begin{enumerate}
	\item First, we fix $C_R(a,M)\gg 1$ as in the assumption of Theorem \ref{thm:alphaGamma}. Then, we chose $\delta_-(a,M)>0$ small enough such that \eqref{eq:conditionCRR} holds, which allows us to apply Theorems \ref{thm:regionI}, \ref{thm:regionII} for this choice of constants $(\delta_-,C_R)$ which only depend on $(a,M)$, as long as $\varepsilon>0$ is small enough with respect to $(a,M,\delta_+)$. 
	\item By Theorems \ref{thm:regionI} this yields the value $w_f$ defining part of the future boundary of region $\un$, which depends only on $(a,M,\delta_+,\delta,(Q_m)_{|m|\leq2})$.
	\item By Theorem \ref{thm:regionII} this yields the value of $u_f$ defining the left part of the future boundary of $\deux\cup\trois$, which depends only on $(a,M,\delta_+,\delta,(Q_m)_{|m|\leq2})$. Choosing $u_f(a,M,\delta_+,\delta,(Q_m)_{|m|\leq2})$ even more negative if necessary, we can also apply Theorem \ref{thm:inextensibilite}.
\end{enumerate}
Now that the auxiliary smallness constants $C_R,\delta_-,w_f,u_f$ are fixed, provided $\varepsilon>0$ is small enough with respect to $(a,M,\delta_+)$, so that we can apply Theorems \ref{thm:regionI}, \ref{thm:regionII}, \ref{thm:alphaGamma}, then all the statements in Theorem \ref{thm:mainthm} follow from the successive applications of Theorems \ref{thm:regionI}, \ref{thm:AregionI}, \ref{thm:regionII}, \ref{thm:alphaGamma}, and \ref{thm:inextensibilite}.
\end{proof}
\subsection{Structure of the rest of the paper}\label{section:structurerestpaper}
The rest of the paper is dedicated to the proofs of Theorems \ref{thm:regionI} to \ref{thm:inextensibilite}. In Section \ref{section:regionun}, we prove the existence and control of region $\un$ in the PT gauge, which proves Theorems \ref{thm:regionI} and \ref{thm:AregionI}. In Section \ref{section:IIdoublenull}, we prove the existence and control of region $\deux\cup\trois$ in the double null gauge, proving Theorem \ref{thm:regionII}. In Section \ref{section:gauge}, we prove the existence and control in region $\deux$ of an auxiliary non-integrable frame which is a perturbation of the Kerr ingoing principal null frame. In Section \ref{section:teukolskydeux}, we analyse the Teukolsky equation with respect to this non-integrable frame in region $\deux$. In Section \ref{section:regionIII}, using the results of the two previous sections, we finish the proof of Theorem \ref{thm:alphaGamma}, and we also prove integral curvature blow-up results and the Lipschitz inextendibility across $\ch$, hence proving Theorem \ref{thm:inextensibilite}.

\section{Control of region $\un$ in ingoing principal temporal gauge}\label{section:regionun}

In all of Section \ref{section:regionun}, we assume that the assumptions of Theorem \ref{thm:regionI} are satisfied, and we fix a parameter $\delta_->0$ such that $r_-(1+\delta_-)<r_+(1-\delta_+)$.
\subsection{Bootstrap assumptions in $\un$ and first consequences}\label{section:prooftheo1}
In this subsection we prove existence and control of an ingoing PT structure in all of region $\un$ as defined in \eqref{eq:ecouteregIdefigg}, for some negative enough parameter $w_f(a,M,\delta_\pm,\delta)\ll -1$, which we fix for now and which will be chosen later in the proof.
\subsubsection{Schematic linearized quantities and bootstrap assumptions}\label{section:linearizedPTun}

In any ingoing PT gauge as described in Section \ref{section:onintroPTgauge}, also equipped with a function $\phi_+$ and 1-forms $\frakJ_\pm\in\fraks_1(\C)$ such that \eqref{eq:easlmitjfofgu} holds, recalling Definition \ref{def:linearizedquantities}, we define the following sets of linearized quantities,
\begin{align}\label{eq:defgammacheckvrai}
	\widecheck{\Gamma}:=\bigg\{&\Xi,\widecheck{\omega},\widecheck{trX},\wh{X},\widecheck{Z},\widecheck{\Hbar},\widecheck{tr\Xbar},\wh{\Xbar},\widecheck{e_4(r)},\widecheck{e_4(\ubar)},e_4(\cos\theta),\nabla r,\widecheck{\nabla\cos\theta},\nn\\
	&\widecheck{\nabla\ubar},\widecheck{\nabla_4\mathfrak{J}},\widecheck{\nabla_4\frakJ_\pm},\widecheck{\divc\mathfrak{J}},\mcd\hot\mathfrak{J},\widecheck{\divc\mathfrak{J}_\pm},\mcd\hot\mathfrak{J}_\pm,\widecheck{\nabla_4 J^{(\pm)}},\widecheck{\mcd J ^{(\pm)}}\bigg\},
\end{align}
$$\widecheck{R}:=\big\{A,B,\widecheck{P},\Bbar,\Abar\big\}.$$
We also define, for any quantity $\widecheck{x}\in(\Gammacheck,\Rcheck)$, the non-linearized quantity $x=x_\mck+\widecheck{x}$. We define the function
\begin{align}\label{eq:wbardefun}
	\wbar(r,\ubar):=\ubar-r,
\end{align}
as well as
\begin{align}\label{eq:wbarzeroinitial}
	\wbar_0:=2r^*_\mca-2r_\mca-w_f
\end{align}
which corresponds to the value of $\wbar$ defined in \eqref{eq:wbardefun} at the intersection  $\mca\cap\{w=w_f\}$ with $w$ defined in \eqref{eq:defwregIunee}. Next, recalling the definition \eqref{eq:ecouteregIdefigg} of region $\un$, we define, for $\wbar_*>\wbar_0$, the bootstrap region
\begin{align}
	\un[\wbar_*]&:=\un\cap\{\wbar<\wbar_*\}=\mathcal{B}[\wbar_*]\times\mathbb{S}^2,\label{eq:defmcbwbarstar}\\ \mathcal{B}[\wbar_*]&:=\{(r,\ubar)\in(r_-(1+\delta_-),r_{\mca}]\times(1,+\infty)\:/\:w(r,\ubar)<w_f,\:\wbar(r,\ubar)<\wbar_*\}.\nn
\end{align}
\begin{defi}\label{defi:bootstrapdansunI}
	We say that the bootstrap assumptions in $\un$ hold for some $\wbar_*>\wbar_0$ if there exists a smooth extension of $(\mca,g,k)$ which is isometric to $(\un[\wbar_*],\g)$, with $\un[\wbar_*]$ defined as in \eqref{eq:defmcbwbarstar}, such that:
	\begin{itemize}
		\item The Lorentzian metric $\g$ satisfies the Einstein vacuum equations \eqref{eq:EVE}.
		\item $\mca\cap\{w(r_\mca,\ubar)<w_f\}$ coincides with the level set $\{r=r_\mca\}$ in $\un[\wbar_*]$.
		\item The metric and second fundamental form induced by $\g$ on $\mca$ are respectively $g$ and $k$.
		\item $(\un[\wbar_*],\g)$ is equipped with: an extended ingoing PT structure, a function $\phi_+$, and two complex 1-forms $\frakJ_\pm$ such that $e_3(\phi_+)=0$, $\nabla_3(\qbar\frakJ_\pm)=0$, which extend the initial data $(e_3,e_4,\mch,r,\theta,\ubar,\phi_+,\frakJ,\frakJ_\pm)|_{\mca}$ defined in Section \ref{section:initialdatavrai} and which satisfy the estimate\footnote{We assume that $N_0$ is even.}
		\begin{align}\label{eq:BA}
			\sup_{\un[\wbar_*]}|\df^{\leq N_0/2}(\Gammacheck,\Rcheck)|\leq {\sqrt{\varepsilon}},
		\end{align}
		where we defined the following set of derivatives,
		\begin{align}\label{eq:setofderivatives}
			\df=\{\nabla_3,\nabla_4,\nabla\}.
		\end{align}
	\end{itemize}
\end{defi}
As a consequence of the local well-posedness of ingoing PT gauges \cite[Lemma 2.85]{KS21} (more precisely, as a consequence of the local uniqueness), if the bootstrap assumptions hold for some $\wbar_*>\wbar_0$, then by the initial data assumptions in Section \ref{section:initialdatavrai} we actually have 
\begin{align}\label{eq:initialhypoth}
	|{\df}^{\leq N_0}(\Gammacheck,\Rcheck)|\lesssim \varepsilon/\ubar^{3+\delta},\quad\text{on }\mathcal{A}\cap\{\wbar<\wbar_*\}.
\end{align}

\textbf{In Sections  \ref{subsection:preleminaryestI} to \ref{section:sectionalasuite6}, we fix $\wbar_*>\wbar_0$ and we assume that the bootstrap assumptions hold for $\wbar_*$. We will then improve these bootstrap assumptions and show by a standard continuity argument that they in fact hold for $\wbar_*=+\infty$.}

\subsubsection{Control of coordinate systems}\label{subsection:preleminaryestI}
Note that we have in $\un[\wbar_*]$,
\begin{align}\label{eq:coordotranspoPT}
	e_3(r)=-1,\quad e_3(\theta)=e_3(\ubar)=e_3(\phi_+)=e_3(\Jplus)=e_3(\Jmoins)=0,
\end{align}
where we recall the notations
\begin{align}\label{eq:jplusoumoins}
	\Jplus=\sin\theta\cos\phi_+,\quad\Jmoins=\sin\theta\sin\phi_+.
\end{align}
\begin{rem}\label{rem:propagatedjjj}
	The conditions $\nabla_3(\qbar(\frakJ,\frakJ_\pm))=0$ ensure that the initial conditions \eqref{eq:inithypscalarproducts} on $\mca$ are propagated in $\un[\wbar_*]$, see \cite[Lemma 2.64]{KS21}\footnote{Lemma 2.64 in \cite{KS21} does not deal with every scalar products between $j,j_+,j_-$ and their Hodge duals, however the proof is the same for all possible products.}.
\end{rem}	
Now, we have that $\un[\wbar_*]$ is covered by three coordinate patches $\mcu^{(1)}$, $\mcu^{(2)}$, $\mcu^{(3)}$, where:
\begin{align*}
	\mcu^{(1)}&:=\un[\wbar_*]\cap\{\pi/4<\theta<3\pi/4\},\\
	\mcu^{(2)}&:=\un[\wbar_*]\cap\{0\leq\theta<\pi/3\},\\
	\mcu^{(3)}&:=\un[\wbar_*]\cap\{2\pi/3<\theta\leq\pi\},
\end{align*}
and where the coordinates that we use on each $\mcu^{(i)}$, $i=1,2,3$, are $(r,\ubar,x^1_{(i)},x^2_{(i)})$ with 
\begin{align*}
	(x^1_{(1)}:=\theta, \: x^2_{(1)}:=\phi_+),\quad (x^1_{(2)}=x^1_{(3)}=\Jplus,\: x^2_{(2)}=x^2_{(3)}=\Jmoins).
\end{align*}

\begin{rem}\label{rem:productderun}
	By the Leibniz rule and the bootstrap assumption \eqref{eq:BA}, we have the reduced schematic equation for $k\leq N_0$,
	$$\df^{k}(\Gammacheck\cdot\Gammacheck)=_s\sum_{i_1+i_2= k}\df^{i_1}\Gammacheck\df^{i_2}\Gammacheck=_{rs}\df^{\leq k}\Gammacheck,$$
	where in the last step we used that either $i_1\leq N_0/2$ or $i_2\leq N_0/2$ so that one of the terms is bounded by the bootstrap assumption. This observation is used extensively in the present section to bound the derivatives of nonlinear terms.
\end{rem}
To obtain Sobolev and elliptic estimates on the spheres $S(r,\ubar)$, we need to control the metric $\g$ in coordinates $(r,\ubar,x^1_{(i)},x^2_{(i)})$.
\begin{lem}\label{lem:lemmainmcuiee}
For $i=1,2,3$, $j=1,2$, and $k\leq N_0$, we have in $\mcu^{(i)}$ the bounds 
\begin{align*}
	|\df^{\leq k}(e_4(x^j_{(i)})-e_4(x^j_{(i)})_\mck)|+|\df^{\leq k}(\mcd(x^j_{(i)})-\mcd(x^j_{(i)})_\mck)|\lesssim|\df^{\leq k}\Gammacheck|,
\end{align*}
	where we recall that the Kerr values $e_4(x^j_{(i)})_\mck$, $\mcd(x^j_{(i)})_\mck$ for $i=2,3$ correspond to $e_4(J^{(\pm)})_\mck,(\mcd J^{(\pm)})_\mck$ as defined in Definition \ref{def:kerrvalues}, and where, for $i=1$,
\begin{align*}
	&e_4(\phi_+)_\mck=\frac{2a}{|q|^2},\quad(\mcd\phi_+)_\mck=\sin^{-2}\theta\left(\Jplus\frakJ_--\Jmoins\frakJ_+\right),\\
	&e_4(\theta)_\mck=0,\quad (\mcd\theta)_\mck=-\frac{i}{\sin\theta}\frakJ.
\end{align*}
\end{lem}
\begin{proof}
For $i=2,3$, this is immediate by the definition of $\Gammacheck$. For $i=1$, this is a consequence of
\begin{equation*}
	\begin{gathered}
			e_4(\phi_+)=\sin^{-2}\theta(\Jplus e_4(\Jplus)-\Jmoins e_4(\Jmoins)),\quad \mcd\phi_+=\sin^{-2}\theta(\Jplus \mcd\Jplus-\Jmoins\mcd\Jmoins),\\
			e_4(\theta)=-\frac{1}{\sin\theta}e_4(\cos\theta),\quad \mcd\theta=-\frac{1}{\sin\theta}\mcd\cos\theta,
	\end{gathered}
\end{equation*}
combined with the observation in Remark \ref{rem:productderun}, and the estimate $\sin\theta\sim 1$ in $\mcu^{(1)}$.
\end{proof}
\begin{defi}
	We denote by $\err$ any quantity in $\un[\wbar_*]$ such that for $k\leq N_0$, 
		$$|\df^{\leq k}\err|\lesssim|\df^{\leq k}\Gammacheck|.$$
	In particular by the bootstrap assumptions \eqref{eq:BA} this implies $|\df^{\leq N_0/2}\err|\lesssim\sqrt{\varepsilon}.$
\end{defi}
\begin{prop}\label{prop:diffmetricdansIee}
	We have in $\mcu^{(1)}$, in the coordinates $(r,\ubar,\theta,\phi_+)$,
	$$\g=\g_{a,M}+\err\cdot\left(\dee\ubar,\dee r,\dee\theta,\dee\phi_+\right)^2,$$
	where $\g_{a,M}$ is given in coordinates $(r,\ubar,\theta,\phi_+)$ by \eqref{eq:kerrmetricEF}. Also, for $i=2,3$, we have in $\mcu^{(i)}$,
		$$\g=\g_{a,M}+\err\cdot\left(\dee\ubar,\dee r,\dee x^1_{(i)},\dee x^2_{(i)}\right)^2,$$
		where $\g_{a,M}$ is given in coordinates $(r,\ubar,x^1_{(i)},x^2_{(i)})$ by \eqref{eq:kerrmetricx1x2}.
\end{prop}
\begin{proof}
	We use the computation for the inverse metric
	$$\g^{\mu\nu}=\g(\D x^\mu,\D x^\nu)=-\frac12 e_3(x^\nu)e_4(x^\mu)-\frac12 e_3(x^\mu)e_4(x^\nu)+\nabla x^\nu\cdot\nabla x^\mu$$
	in the three coordinate systems $(x^\mu)$. Then, using
\begin{align*}
	X(r)=X(r)_\mck+\err,\quad X(\ubar)=X(\ubar)_\mck+\err
\end{align*}
for $X\in\{e_3,e_4,\nabla\}$ by Lemma \ref{lem:lemmainmcuiee}, we get
$$\g^{\mu\nu}=\g^{\mu\nu}_{a,M}+\err$$
in $\mcu^{(i)}$ for $i=1,2,3$, where $\g^{\mu\nu}_{a,M}$ is the inverse of the Kerr metric in coordinates $(r,\ubar,x^1_{(i)},x^2_{(i)})$, see \eqref{eq:kerrmetricEF}, \eqref{eq:kerrmetricx1x2}. As all the coefficients of the Kerr metric inverse $\g^{\mu\nu}_{a,M}$ in coordinates $(r,\ubar,x^1_{(i)},x^2_{(i)})$ are bounded by some constant which depends on $(a,M)$, this implies
$$\g_{\mu\nu}=(\g_{a,M})_{\mu\nu}+\err$$
in coordinates $(r,\ubar,x^1_{(i)},x^2_{(i)})$ in $\mcu^{(i)}$, hence the conclusion.
\end{proof}
We define the following approximate Killing vector fields by (recall \eqref{eq:defTZdansKerr}),
\begin{align}\label{eq:defTdansI}
\T:=\frac12\left(e_4+\frac{\Delta}{|q|^2}e_3-2a\Real(\mathfrak{J})^be_b\right),\quad \Z:=\frac12\left(2(r^2+a^2)\Real(\frakJ)^be_b-a\sin^2\theta e_4-\frac{a\sin^2\theta\Delta}{|q|^2}e_3\right).
\end{align}
We express the $(r,\ubar,x^1_{(i)},x^2_{(i)})$ coordinate vector fields with respect to the PT frame as follows.
\begin{prop}\label{prop:expredmuavecenu}
The $(r,\ubar,\theta,x^1,x^2)$-coordinate vector fields can be expressed as follows with respect to the PT frame,
	\begin{align*}
		\partial_r=-e_3,\qquad \partial_\ubar=\T+\err\cdot\df,
	\end{align*}
	Moreover, in $\mcu^{(1)}$ we have
	\begin{align*}
		\partial_\theta=\frac{|q|^2\Imag(\mathfrak{J})^b}{\sin\theta}e_b+\err\cdot\df,\qquad\partial_{\phi_+}=\Z+\err\cdot\df,
	\end{align*}
	and for $i=2,3$ we have in $\mcu^{(i)}$,
	\begin{align*}
		\partial_{x^1_{(i)}}=\frac{|q|^2}{\cos\theta}\Imag(\frakJ_-^b)e_b+a x^2_{(i)}\T+\err\cdot\df,\qquad	\partial_{x^2_{(i)}}=-\frac{|q|^2}{\cos\theta}\Imag(\frakJ_+^b)e_b-a x^1_{(i)}\T+\err\cdot\df.
	\end{align*}
\end{prop}
\begin{proof}
We clearly have $e_3=-\partial_r$ by \eqref{eq:coordotranspoPT}, and the bootstrap assumptions \eqref{eq:BA} imply
\begin{align*}
	\T(\ubar)&=1+\err,\quad\T(r),\T(\theta),\T(\phi_+)=\err,\\
	\frac{|q|^2\Imag(\mathfrak{J})^b}{\sin\theta}e_b(\theta)&=1+\err,\quad \frac{|q|^2\Imag(\mathfrak{J})^b}{\sin\theta}e_b(r),\frac{|q|^2\Imag(\mathfrak{J})^b}{\sin\theta}e_b(\ubar),\frac{|q|^2\Imag(\mathfrak{J})^b}{\sin\theta}e_b(\phi_+) =\err,\\
	\Z(\phi_+)&=1+\err,\quad\Z(r),\Z(\theta),\Z(\ubar)=\err,
\end{align*}
which concludes the proof in $\mcu^{(1)}$. We omit the proofs in $\mcu^{(2)}$ and $\mcu^{(3)}$, that are very similar.
\end{proof}
\subsubsection{Isometry between $\mch$ and $TS(r,\ubar)$}\label{section:isometrteri}
We define the surfaces $S(r,\ubar)$ as the intersections of the corresponding level sets of $r$ and $\ubar$.
\begin{defi}
	In what follows, we denote $\gslash$ the metric induced on the spheres $S(r,\ubar)$. We also denote $\gslash_\mck$ the metric induced on the topological spheres $S(r,\ubar)$ by the exact Kerr metric $\g_{a,M}$, pulled back to $\mck$ by the coordinates $(r, \ubar,x^j_{(i)})$, $j=1,2$, see \eqref{eq:metriquesphereI} and \eqref{eq:metricsphx1x2}. We also denote $\nablaslash$ the induced Levi-Civita connection on the spheres $S(r,\ubar)$.
\end{defi}
Now, we introduce a transformation which relates horizontal tensors and tensors tangential to the spheres $S(r,\ubar)$.
\begin{rem}\label{rem:transfoIKerrr}
	Recall from Lemma \ref{lem:transfokerrIun} the smooth function $F_\mck(r,\theta)$, and the isometry $\Theta_\mck$	between $\mch$ and $TS(r,\ubar)$ in Kerr. The lemma below shows that, up to error terms, this result can be extended to perturbations of Kerr as considered here.
\end{rem}
\begin{lem}\label{lem:transformationIspheres}
	There exist horizontal 1-forms $\lambda_1,\lambda_2$, and a linear map $\widecheck{\Theta}:\mch\longrightarrow TS(r,\ubar)$, which satisfy
	$$(\lambda_1,\lambda_2,\widecheck{\Theta})=\err,$$
such that the map
	\begin{align}\label{eq:retransfoPhiI}
		\Theta:X\in \mch\longmapsto X-a\Real(\mathfrak{J})(X)\T+F_\mck(r,\theta)\Real(\frakJ)(X)\Z+\lambda_1(X)e_3+\lambda_2(X)e_4+\widecheck{\Theta}(X),
	\end{align}
	is an isometry between $\mch$ and $(TS(r,\ubar),\gslash)$.
\end{lem}
\begin{proof}
	With $\T,\Z$ defined as in \eqref{eq:defTdansI} we compute, using the bootstrap assumption \eqref{eq:BA},
\begin{align}\label{eq:Tappliquearetubar}
	\T(r)=_s\err,\quad \T(\ubar)-1=_s\err,\quad\Z(r)=_s\Gammacheck,\quad\Z(\ubar)=_s\err,
\end{align}
Thus for a given horizontal frame $(e_1,e_2)$ and any horizontal 1-forms $\lambda_1,\lambda_2$, denoting
$$e'_b=e_b-a\Real(\mathfrak{J})_b\T+F_\mck(r,\theta)\Real(\frakJ)_b\Z+(\lambda_1)_be_3+(\lambda_2)_be_4$$
we have
\begin{align*}
	e_b'(r)&=\nabla_b r+\err-(\lambda_1)_b+(\lambda_2)_b\left(\frac{\Delta}{|q|^2}+\widecheck{e_4(r)}\right),\\
	e_b'(\ubar)&=\nabla_b\ubar-a\Real(\frakJ)\T(\ubar)+\err+(\lambda_2)_b\left(\frac{2(r^2+a^2)}{|q|^2}+\widecheck{e_4(\ubar)}\right)\\
	&=\widecheck{\nabla_b \ubar}+\err+(\lambda_2)_b\left(\frac{2(r^2+a^2)}{|q|^2}+\widecheck{e_4(\ubar)}\right).
\end{align*}
This implies that, provided $\lambda_1$ and $\lambda_2$ satisfy the following system,
\begin{align*}
	\err+\lambda_2\left(\frac{2(r^2+a^2)}{|q|^2}+\widecheck{e_4(\ubar)}\right)=0,\qquad\err-\lambda_1+\lambda_2\left(\frac{\Delta}{|q|^2}+\widecheck{e_4(r)}\right)=0,
\end{align*}
then $e'_b\in TS(r,\ubar)$. Solving the system above, we get $\lambda_1=\err$, $\lambda_2=\err$, and $e_b'\in TS(r,\ubar)$. Moreover, using Remark \ref{rem:propagatedjjj} we compute, for $b,c=1,2$,
\begin{align*}
	\gslash(e_b',e_c')=\delta_{bc}+a(1+\sin^2\theta F_\mck)\Real(\frakJ)_b\left((\lambda_1)_c+\frac{\Delta}{|q|^2}(\lambda_2)_c\right)-2(\lambda_1)_c(\lambda_2)_b+[b\leftrightarrow c]
\end{align*}
which rewrites schematically as
\begin{align}\label{eq:whereweuseddd}
	\widecheck{U}_{bc}:=\gslash(e_b',e_c')-\delta_{bc}&=_s(\lambda_1,\lambda_2)\cdot(\lambda_1,\lambda_2,O(1)_\mck)=\err.
\end{align}
Now we modify $e_b'$ to ensure that the transformation is an isometry, which is done by adding the error term $\widecheck{\Theta}$. First, we denote 
$$\Theta':X\in\longmapsto X-a\Real(\mathfrak{J})(X)\T+F_\mck(r,\theta)\Real(\frakJ)(X)\Z+\lambda_1(X)e_3+\lambda_2(X)e_4\in TS(r,\ubar)$$
the transformation such that $\Theta'(e_b)=e_b'$. We will now prove the existence of a  horizontal symmetric tensor $\Mcheck_{ab}$ such that, defining 
\begin{align}\label{eq:lia}
	\Theta:X\in H\longmapsto \Theta(X):=\Theta'(X)+\Theta'(\widecheck{M}_a^bX^ae_b)\in TS(r,\ubar),
\end{align}
the transformation $\Theta$ is an isometry. This translates into a fixed point problem for $\Mcheck$. Indeed, recalling $e_b'$ and $e_b$ defined above, we get that $\Theta$ is an isometry if and only if for $b,c=1,2$,
$$\gslash(e'_b+\Mcheck_{ab} e'_a,e'_c+\Mcheck_{ac} e'_a)=\delta_{bc}.$$
Moreover, we can compute the LHS above:
\begin{align*}
	\gslash(e'_b+\Mcheck_{ab} e'_a,e'_c+\Mcheck_{ac} e'_a)&=\gslash(e'_b,e'_c)+\Mcheck_{ab}\gslash(e'_a,e'_c)+\Mcheck_{ac}\gslash(e'_a,e'_b)+\Mcheck_{ab}\Mcheck_{cd}\gslash(e'_a,e'_d)\\
	&=\delta_{bc}+\widecheck{U}_{bc}+2\widecheck{M}_{bc}+\widecheck{M}_{ab}\widecheck{U}_{ac}+\widecheck{M}_{dc}\widecheck{U}_{bd}+\widecheck{M}_{ab}\widecheck{M}_{ac}+\widecheck{M}_{ab}\widecheck{M}_{dc}\widecheck{U}_{ad},
\end{align*}
where we used \eqref{eq:whereweuseddd} and the symmetry of $\Mcheck$. Thus, we get that $\Theta$ as defined above is an isometry if and only if $\Mcheck$ satisfies a fixed point problem which writes schematically
\begin{align}\label{eq:fixedpointcafaisait}
	\Mcheck=_s\widecheck{U}+\widecheck{U}\cdot\widecheck{M}+\widecheck{M}^2+\widecheck{M}^2\cdot\widecheck{U}.
\end{align}
By a standard fixed-point argument, from the bootstrap assumptions \eqref{eq:BA} and $\widecheck{U}=\err$ by \eqref{eq:whereweuseddd}, for $\varepsilon(a,M)$ small enough we get that there exists $\Mcheck$ such that \eqref{eq:fixedpointcafaisait} holds, which satisfies for $k\leq N_0$,
\begin{align}\label{eq:borneMcheck}
	|\df^{\leq k}\Mcheck|\lesssim|\df^{\leq k}\Gammacheck|.
\end{align}
Finally, defining $\widecheck{\Theta}(X):=\Theta'(\widecheck{M}_a^bX^ae_b)$ for $X$ horizontal yields $|\df^{\leq k}\widecheck{\Theta}|\lesssim|\df^{\leq k}\Mcheck|\lesssim|\df^{\leq k}\Gammacheck|$, and defining $\Theta$ as in \eqref{eq:lia} we get that $\Theta$ is an isometry, which concludes the proof.
\end{proof}
The following definition allows to pass from horizontal to $S(r,\ubar)$-tangent tensors, and conversely.
\begin{defi}\label{defi:verscncersdefi}
	If $U$ is a horizontal $k$-tensor in $\un$, we denote here $U^S$ the $S(r,\ubar)$-tangent $k$-tensor defined by, for $Y_1,\cdots, Y_k\in TS(r,\ubar)$,
	\begin{align}\label{eq:transformationIdeux}
		U^S(Y_1,\cdots, Y_k)=U(\Theta^{-1}(Y_1),\cdots,\Theta^{-1}(Y_k)).
	\end{align}
Conversely, if $V$ is a $S(r,\ubar)$-tangent $k$-tensor in $\un$, we denote $V^H$ the horizontal tensor such that $(V^H)^S=V$, namely for $X_1,\cdots,X_k\in\mch$,
\begin{align}\label{eq:conversetransfoI}
	V^H(X_1,\cdots, X_k)=V(\Theta(X_1),\cdots,\Theta(X_k)).
\end{align}
\end{defi}
\begin{prop}\label{prop:comparnormestransfoI}
	For any horizontal $k$-tensor $U$ in $\un$, we have 
	$$|U|= |U^S|.$$
\end{prop}
\begin{proof}
	This is a direct consequence of the fact that $\Theta$ is an isometry.
\end{proof}
\begin{prop}\label{prop:premderchgtframeI}
	Let $U$ be a horizontal $k$-tensor. Then denoting $(e_1,e_2)$ a horizontal frame and $\barre{e}_a=\Theta(e_a)$ the associated frame of $TS(r,\ubar)$, we have the following identities:
\begin{align*}
	\nablaslash_{e_3} U^S_{a_1\cdots a_k}&=\nabla_3 U_{a_1\cdots a_k}+\sum_{i=1}^kU_{a_1\cdots b\cdots a_k}\mct^{(3)}_{a_ib},\\
	\nablaslash_{e_4} U^S_{a_1\cdots a_k}&=\nabla_4 U_{a_1\cdots a_k}+\sum_{i=1}^kU_{a_1\cdots b\cdots a_k}\mct^{(4)}_{a_ib},\\
	\nablaslash_{e_c} U^S_{a_1\cdots a_k}&=\nabla_c U_{a_1\cdots a_k}+\sum_{i=1}^kU_{a_1\cdots b\cdots a_k}{\mct}^{(0)}_{a_ibc},
\end{align*}	
where $c=1,2$ for some horizontal tensors $\mct^{(3)}$, $\mct^{(4)}$, $\mct^{(0)}$ which satisfy	
	$$|\df^{\leq N_0/2-1}(\mct^{(3)}, \mct^{(4)}, \mct^{(0)})|\lesssim 1.$$
Moreover, denoting $\nablaslash_c=\nablaslash_{\barre{e}_c}$ we also have
$$\nablaslash_c U^S_{a_1\cdots a_k}=\nabla_c U_{a_1\cdots a_k}+F_{cd}\nabla_d U_{a_1\cdots a_k}+\lambda_c\nabla_3 U_{a_1\cdots a_k}+\lambdabar_c\nabla_4 U_{a_1\cdots a_k}+\sum_{i=1}^kU_{a_1\cdots b\cdots a_k}{\mct}_{a_ibc},$$
for some horizontal tensors $F_{cd},\lambda_c,\lambdabar_c,{\mct}_{abc}$ which satisfy $\barre{e}_b=e_b+F_{bc}e_c+\lambda_b e_3+\lambdabar_b e_4$, and
$$|\df^{\leq N_0/2-1}(F,\lambda,\lambdabar,{\mct})|\lesssim 1.$$
\end{prop}
\begin{proof}
First of all we notice that the transformation $\Theta$ defined in \eqref{eq:retransfoPhiI} can be rewritten 
\begin{align}\label{eq:rnencoreunchanger}
	\barre{e}_b:=\Theta(e_b)=e_b+F_{bc}e_c+\lambda_b e_3+\lambdabar_b e_4,
\end{align}
where, see \eqref{eq:lia},
\begin{align*}
	F_{bc}&=\Mcheck_{bc}+\left(F_\mck(r,\theta)(r^2+a^2)+a^2\right)(\Real(\frakJ)_b+\Mcheck_{bd}\Real(\frakJ)_d)\Real(\frakJ)_c,\\
	\lambda_b&=(\lambda_1)_b+\Mcheck_{bc}(\lambda_1)_c-\frac{a\Delta}{2|q|^2}\left({\sin^2\theta}F_\mck(r,\theta)+1\right)(\Real(\frakJ)_b+\Mcheck_{bd}\Real(\frakJ)_d)\Real(\frakJ)_c,\\
	\lambdabar_b&=(\lambda_2)_b+\Mcheck_{bc}(\lambda_2)_c-\frac{a}{2}\left(\sin^2\theta F_\mck(r,\theta)+1\right)(\Real(\frakJ)_b+\Mcheck_{bd}\Real(\frakJ)_d)\Real(\frakJ)_c.
\end{align*}
Note that by $\delta_{bc}=\g(\barre{e}_a,\barre{e}_b)$ we get the constraint
\begin{align}\label{eq:cosntraiet}
	F_{bc}+F_{cb}+F_{cd}F_{bd}-2(\lambda_b\lambdabar_c+\lambda_c\lambdabar_b)=0.
\end{align}
Now we compute for $\mu=1,2,3,4$ and $a_1,\cdots,a_k=1,2$, using the definition of the tensor $U^S_{a_1\cdots a_k}=U_{a_1\cdots a_k}$ and the definition of the induced connection $\nablaslash$,
\begin{align*}
	\nablaslash_{e_\mu}U^S_{a_1\cdots a_k}&=e_\mu(U^S_{a_1\cdots a_k})-\sum_{i=1}^kU^S_{a_1\cdots b\cdots a_k}\g(\nablaslash_{e_\mu}\barre{e}_{a_i},\barre{e}_b)\\
	&=e_\mu(U_{a_1\cdots a_k})-\sum_{i=1}^kU_{a_1\cdots b\cdots a_k}\g(\D_{e_\mu}\barre{e}_{a_i},\barre{e}_b)\\
	&=\nabla_\mu U_{a_1\cdots a_k}+\sum_{i=1}^kU_{a_1\cdots b\cdots a_k}\left(\g(\D_{e_\mu}{e}_{a_i},{e}_b)-\g(\D_{e_\mu}\barre{e}_{a_i},\barre{e}_b)\right).
\end{align*}
Moreover, using \eqref{eq:rnencoreunchanger} and the constraint \eqref{eq:cosntraiet}, we compute\footnote{This computation is very similar to the proofs of Propositions \ref{prop:diffchristo}, \ref{prop:diffchristo3}, \ref{prop:diffchristo4}, except that here we consider a more general isometric transformation.} 
\begin{align*}
	\g(\D_{e_\mu}{e}_{a},{e}_b)-\g(\D_{e_\mu}\barre{e}_{a},\barre{e}_b)=&\frac12\left(\nabla_\mu F_{[ab]}+F_{[bd}\nabla_\mu F_{a]d}-2\lambda_{[b}\nabla_\mu\lambdabar_{a]}-2\lambdabar_{[b}\nabla_\mu\lambda_{a]}\right)\\
	&+\lambda_{[a}\g(\D_\mu e_3,e_{b]})+\lambdabar_{[a}\g(\D_\mu e_4,e_{b]})+F_{[bd}\lambda_{a]}\g(\D_\mu e_3,e_d)\\
	&+F_{[bd}\lambdabar_{a]}\g(\D_\mu e_4,e_d)+\lambdabar_{[a}\lambda_{b]}\g(\D_\mu e_4,e_3)
\end{align*}
Now, choosing $\mu=3,4,1,2$ and using the definitions of the Ricci coefficients in Section \ref{section:ricciandcurvdef} we get that the identities stated in the Proposition \ref{prop:premderchgtframeI} holds with
\begin{align*}
	\mct^{(3)}_{ab}=&\frac12\left(\nabla_\mu F_{[ab]}+F_{[bd}\nabla_\mu F_{a]d}-2\lambda_{[b}\nabla_\mu\lambdabar_{a]}-2\lambdabar_{[b}\nabla_\mu\lambda_{a]}\right)+2\lambdabar_{[a}\eta_{b]}+2F_{[bd}\lambdabar_{a]}\eta_d,\\
	\mct^{(4)}_{ab}=&\frac12\left(\nabla_\mu F_{[ab]}+F_{[bd}\nabla_\mu F_{a]d}-2\lambda_{[b}\nabla_\mu\lambdabar_{a]}-2\lambdabar_{[b}\nabla_\mu\lambda_{a]}\right)+2\lambda_{[a}\etabar_{b]}+2\lambdabar_{[a}\xi_{b]}+2F_{[bd}\lambda_{a]}\etabar_d,\\
	\mct^{(0)}_{abc}=&\frac12\left(\nabla_c F_{[ab]}+F_{[bd}\nabla_c F_{a]d}-2\lambda_{[b}\nabla_c\lambdabar_{a]}-2\lambdabar_{[b}\nabla_c\lambda_{a]}\right)\\
	&+\lambda_{[a}\chibar_{cb]})+\lambdabar_{[a}\chi_{cb]})+F_{[bd}\lambda_{a]}\chibar_{cd}+F_{[bd}\lambdabar_{a]}\chi_{cd}+\lambdabar_{[a}\lambda_{b]}\zeta_c,
\end{align*}
and $\mct_{abc}=\mct^{(0)}_{abc}+F_{cd}\mct^{(0)}_{abd}+\lambda_c\mct^{(3)}_{ab}+\lambdabar_c\mct^{(4)}_{ab}$. Finally, the bounds 
$$|\df^{\leq N_0/2}(\mct^{(3)}, \mct^{(4)}, \mct^{(0)},F,\lambda,\lambdabar,{\mct})|\lesssim 1$$
holds by the expressions for $\mct^{(3)}, \mct^{(4)}, \mct^{(0)},F,\lambda,\lambdabar,{\mct}$ above combined with the bootstrap assumption \eqref{eq:BA} and the bound for $\Mcheck,\lambda_1,\lambda_2$ in Lemma \ref{lem:transformationIspheres} and \eqref{eq:borneMcheck}.
\end{proof}
\begin{cor}\label{cor:corchillvoila}
	Let $U$ be a horizontal $k$-tensor. Then we have
	$$\nablaslash^2U^S=_{rs}(\df^{\leq2} U)^S.$$
\end{cor}
\begin{proof}
	This directly follows from applying twice the expression for $\nablaslash_c$ in Proposition \ref{prop:premderchgtframeI}.
\end{proof}

\subsubsection{Sobolev embedding and elliptic estimates on the spheres $S(r,\ubar)$}\label{section:sobolevdansUnoklm}

\begin{lem}\label{lem:gslashcheckIborne}
	For $i=1,2,3$ we have in $\mcu^{(i)}$,
	$$\gslash=\gslash_\mck+\err\cdot\left(\dee x^1_{(i)},\dee x^1_{(i)}\right)^2,$$
	where $\gslash_\mck$ is the metric on $S(r,\ubar)$ induced by $\g_{a,M}$ in coordinates $(x^1_{(i)},x^2_{(i)})$, see \eqref{eq:metriquesphereI}, \eqref{eq:metricsphx1x2}.
\end{lem}
\begin{proof}
This is a direct consequence of Proposition \ref{prop:diffmetricdansIee}.
\end{proof}

We can now state a $L^2-L^\infty$ Sobolev embedding result on the spheres $S(r,\ubar)$.
\begin{prop}\label{prop:sobolevspheriqueI}
	Let $k\in\mathbb{N}$. For any $(r,\ubar)\in\mcb[\wbar_*]$ and any $k$-tensor $U$ tangent to $S(r,\ubar)$, we have the following $L^\infty-L^2$ Sobolev estimate on $S(r,\ubar)$,
	$$\|U\|_{L^\infty(S(r,\ubar))}\lesssim_k \|\nablaslash^{\leq 2}U\|_{L^2(S(r,\ubar))}.$$
\end{prop}
\begin{proof}
	This can be proven similarly as \cite[Prop. 5.5]{stabC0}. We recall the definition of the isoperimetric constant for a Riemannian 2-manifold $(S,g)$,
	\begin{align}\label{eq:constanteisoperimetrique}
		\mathbf{I}(S,g):=\sup_{V\in\mathcal{O}_1(S)}\frac{\min(\mathrm{Area}(V,g),\mathrm{Area}(V^c,g))}{(\mathrm{Perimeter}(\partial U,g))^2},
	\end{align}
	where $\mathcal{O}_1(S)$ is the set of open sets $V\subset S$ such that $\partial V$ is $C^1$. Moreover, combining Lemmas 5.1 and 5.2 in \cite{christo} we get that for any $k\geq 0$, there is a constant $C_k>0$ such that for any closed Riemannian 2-manifold $(S,g)$, and any covariant $k$-tensor $U$ on $S$,
	\begin{align}\label{eq:sobobrut0}
		&\|U\|_{L^\infty(S)}\leq\\
		&C_k\max(\mathbf{I}(S,g),1)\left(\mathrm{Area}(S,g)^{\frac12}\|\nablaslash^2U\|_{L^2(S,g)}+2\|\nablaslash U\|_{L^2(S,g)}+\mathrm{Area}(S,g)^{-\frac12}\|U\|_{L^2(S,g)}\right).\nn
	\end{align}
It is easy to see that Lemma \ref{lem:gslashcheckIborne} and the expression \eqref{eq:metriquesphereI} for $\gslash_\mck$ imply
\begin{equation*}
	\begin{gathered}
			\left|\mathrm{Area}(S(r,\ubar),\gslash)-\mathrm{Area}(S(r,\ubar),\gslash_\mck)\right|\lesssim\sqrt{\varepsilon},\quad \mathrm{Area}(S(r,\ubar),\gslash_\mck)\sim 1,\\
		\mathrm{Perimeter}(\partial V,\gslash)\sim \mathrm{Perimeter}(\partial V,g_{\mathbb{S}^2}),\\
		\mathrm{Area}(V,\gslash)\lesssim \mathrm{Area}(V,g_{\mathbb{S}^2}),
	\end{gathered}
\end{equation*}
for any $V\in\mco_1(S(r,\ubar))$. From the definition \eqref{eq:constanteisoperimetrique} of $\mathbf{I}(S(r,\ubar),\gslash)$, we thus get 
$$\mathbf{I}(S(r,\ubar),\gslash)\lesssim \mathbf{I}(\mathbb{S}^2,g_{\mathbb{S}^2})\lesssim 1,$$
which concludes the proof in regard of \eqref{eq:sobobrut0}.
\end{proof}
\begin{cor}\label{cor:sobospheriqueI}
	Let $k\in\mathbb{N}$. For any $(r,\ubar)\in\mcb[\wbar_*]$ and any $k$-tensor $U$ tangent to $S(r,\ubar)$, we have, using the notation $\shl{\triangle}{7.3}:=\nablaslash^i\nablaslash_i$,
	$$\|U\|_{L^\infty(S(r,\ubar))}\lesssim_k \|\shl{\triangle}{7.3}^{\leq 1}U\|_{L^2(S(r,\ubar))}.$$
\end{cor}
\begin{proof}
	For any $k$-tensor $U$ tangent to $S(r,\ubar)$, by standard Bochner identities and integration by parts we have
\begin{align}
	\intS |\nablaslash^2U|^2\lesssim\intS|\shl{\triangle}{7.3}U|^2+\|\nablaslash^{\leq 1}\shl{K}{7.7}\|_{L^\infty(S(r,\ubar))}\|\nablaslash^{\leq 1}U\|^2_{L^2(S(r,\ubar))},\label{eq:BochnerI}
\end{align}
where $\shl{K}{7.7}$ is the Gauss curvature of $S(r,\ubar)$. Thus, by Lemma \ref{lem:gslashcheckIborne} and \eqref{eq:retransfoPhiI}, we get
\begin{align}\label{eq:bornenabKslashe}
	|\nablaslash^{\leq 1}\shl{K}{7.7}|\lesssim |\df^{\leq 1}\shl{K}{7.7} |\lesssim 1.
\end{align}
By \eqref{eq:BochnerI}, this yields
\begin{align}\label{eq:usingun1}
	\|\nablaslash^{\leq 2}U\|_{L^2(S(r,\ubar))}&\lesssim	\|\shl{\triangle}{7.3}^{\leq 1}U\|_{L^2(S(r,\ubar))}+\|\nablaslash U\|_{L^2(S(r,\ubar))}.
\end{align} 
Combining the estimate above with the following integration by parts,
\begin{align}\label{eq:usingun2}
	\|\nablaslash U\|_{L^2(S(r,\ubar))}=\int_{S(r,\ubar)} \nablaslash_j U\cdot\nablaslash^j U=\int_{S(r,\ubar)}(-\shl{\triangle}{7.3}) U\cdot U\lesssim\int_{S(r,\ubar)} \left(|\shl{\triangle}{7.3} U|^2+|U|^2\right),
\end{align}
we conclude the proof by applying Proposition \ref{prop:sobolevspheriqueI}.
\end{proof}
\begin{cor}\label{cor:sobolevtwiste}
	Let $U$ be a horizontal tensor. For any $(r,\ubar)\in\mcb[\wbar_*]$, we have
	$$\|U\|_{L^\infty(S(r,\ubar))}\lesssim \|\df^{\leq 2}U\|_{L^2(S(r,\ubar))}.$$
\end{cor}
\begin{proof}
	Applying successively Propositions \ref{prop:comparnormestransfoI}, \ref{prop:sobolevspheriqueI} and Corollary \ref{cor:corchillvoila} we get
	$$\|U\|_{L^\infty(S(r,\ubar))}\lesssim \|U^S\|_{L^\infty(S(r,\ubar))}\lesssim \|\nablaslash^{\leq 2}U^S\|_{L^2(S(r,\ubar))}\lesssim \|\df^{\leq 2}U\|_{L^2(S(r,\ubar))},$$	
	which concludes the proof.
\end{proof}
In the following result, we prove elliptic estimates on the spheres $S(r,\ubar)$ that will be used to control Bianchi pairs.

\begin{lem}\label{lem:ellipticIun}
	Let $U\in\barre{\fraks}_p(\C)$, $p=0,1,2$, namely $U$ is a complex scalar or a complex antiself-dual $S(r,\ubar)$-tangent 1-form or symetric traceless 2-tensor, with respect to $\gslash$. We denote
	$$\shl{\mcd}{6.6}_0=\shl{\mcd}{6.6},\quad \shl{\mcd}{6.6}_1=\shl{\overline{\mcd}}{7.2}\cdot,\quad \shl{\mcd}{6.6}_2=\shl{\overline{\mcd}}{7.2}\cdot$$
	the spherical complex Hodge operators, where $\shl{{\mcd}}{6.6}=\nablaslash+i\hodge{}\:\nablaslash$. Then we have, for any $k\geq 1$ such that $\max(2k,2k-1)\leq N_0$ and any $(r,\ubar)\in\mcb[\wbar_*]$,
\begin{align*}
	\int_{S(r,\ubar)}|\nablaslash^{2k}U|^2&\lesssim \int_{S(r,\ubar)}\left(|\shl{\triangle}{7.3}^{\leq k}U|^2+|\shl{\mcd}{6.6}_p^{\leq 1}\shl{\triangle}{7.3}^{\leq k-1}U|^2\right)+\sum_{i_1+i_2\leq 2(k-1)}\int_{S(r,\ubar)}|\df^{i_1+2}(\Gammacheck,\Rcheck)|^2|\nablaslash^{i_2}U|^2,\\
	\int_{S(r,\ubar)}|\nablaslash^{2k-1}U|^2&\lesssim \int_{S(r,\ubar)}|\shl{\mcd}{6.6}_p^{\leq 1}\shl{\triangle}{7.3}^{\leq k-1}U|^2+\sum_{i_1+i_2\leq 2k-3}\int_{S(r,\ubar)}|\df^{i_1+2}(\Gammacheck,\Rcheck)|^2|\nablaslash^{i_2}U|^2.
\end{align*}
\end{lem}
\begin{proof}
	We prove the stated estimates by induction on $k$. Both estimates hold at rank $k=1$ by combining \eqref{eq:usingun1} and \eqref{eq:usingun2} for the first estimate, and by Proposition \ref{prop:ellipticestimates}, combined with the bound $|\shl{K}{7.7}|\lesssim 1$ already proven in \eqref{eq:bornenabKslashe}, for the second bound. Note in particular that for $k=1$, the error terms with $\Gammacheck,\Rcheck$ on the RHS above are not present. Next, assuming that they hold at rank $k$, we prove that they hold at rank $k+1$ where $k$ is such that $\max(2k+2,2k+1)\leq N_0$, namely $\max(2k,2k-1)\leq N_0-2$. We begin by proving the first estimate at rank $k+1$. We have, using the case $k=1$,
	\begin{align*}
		\int_{S(r,\ubar)}|\nablaslash^{2(k+1)}U|^2&\lesssim \int_{S(r,\ubar)}|\shl{\triangle}{7.3}^{\leq 1}\nablaslash^{2k}U|^2.
	\end{align*}
	Moreover, we have the schematic commutation identity
	$$[\shl{\triangle}{7.3},\nablaslash^m]U=_s\sum_{i_1+i_2\leq m}\nablaslash^{i_1}\shl{K}{7.7}\nablaslash^{i_2}U,$$
	which is proven by induction on $m$ starting from $[\nablaslash,\shl{\triangle}{7.3}]U=_s\shl{K}{7.7}\nablaslash U+U\nablaslash\shl{K}{7.7}$. Using this with $m=2k$ implies
	\begin{align*}
		\int_{S(r,\ubar)}|\nablaslash^{2(k+1)}U|^2&\lesssim \int_{S(r,\ubar)}|\nablaslash^{2k}\shl{\triangle}{7.3}^{\leq 1}U|^2+\sum_{i_1+i_2\leq 2k}\int_{S(r,\ubar)}|\nablaslash^{i_1}\shl{K}{7.7}\nablaslash^{i_2}U|^2\\
		&\lesssim \int_{S(r,\ubar)}\left(|\shl{\triangle}{7.3}^{\leq k+1}U|^2+|\shl{\mcd}{6.6}_p^{\leq 1}\shl{\triangle}{7.3}^{\leq k}U|^2\right)+\sum_{i_1+i_2\leq 2k}\int_{S(r,\ubar)}|\nablaslash^{i_1}\shl{K}{7.7}\nablaslash^{i_2}U|^2
	\end{align*}
	where we used the stated estimate at rank $k$. Moreover, combining the definition of the Gauss curvature $\shl{K}{7.7}$ for $\gslash$ and the expression for $\gslash$ in  Lemma \ref{lem:gslashcheckIborne}, we get for $j\leq N_0-2$,
	$$|\nablaslash^j\shl{K}{7.7}|\lesssim 1+|\df^{\leq j+2}(\Gammacheck,\Rcheck)|,$$
	which concludes the proof of the first estimate at rank $k+1$, using the induction assumption to deal with the bounded term above. The proof of the second estimate is similar, which concludes the proof.
\end{proof}

\subsubsection{Additional estimates on hypersurfaces and approximate symmetries}\label{section:encorrreunne}

\noindent\textbf{Volume form estimates}

\begin{lem}\label{lem:volumeformI}
	For $i=1,2,3$ there exists functions $f_i\sim 1$, $h_i\sim 1$ such that on $\mcu^{(i)}$, the spacetime volume form $\vol$ and the volum forms $\mathrm{vol}_{\wbar_0}$ induced on $\Sigma_{\wbar_0}=\{\wbar=\wbar_0\}$ satisfy
$$\vol=f_i\mathrm{vol}_{\wbar}\dee\wbar,\quad \mathrm{vol}_{\wbar}=h_i \dee x^1_{(i)}\dee x^2_{(i)}\dee r.$$
\end{lem}
\begin{proof}
 In coordinates $(r,x^1_{(i)},x^2_{(i)})$ the coordinate vector fields on $\Sigma_{\wbar}\cap\mcu^{(i)}$ are
$$\bar{\partial}_r=\partial_r+\partial_\ubar,\quad \bar{\partial}_{x^1_{(i)}}={\partial}_{x^1_{(i)}},\quad \bar{\partial}_{x^2_{(i)}}={\partial}_{x^2_{(i)}}.$$
Using Proposition \ref{prop:diffmetricdansIee} and the bootstrap assumptions \eqref{eq:BA}, we deduce that the metric $\bar{\g}^{(i)}$ induced on $\Sigma_{\wbar}\cap\mcu^{(i)}$ satisfies, in the coordinate system $(y^j)=(r,x^1_{(i)},x^2_{(i)})$, 
$$|\bar{\g}^{(i)}_{jk}-(\bar{\g}^{(i)}_\mck)_{jk}|\lesssim\sqrt{\varepsilon}\implies |\det\bar{\g}^{(i)}_{jk}-\det(\bar{\g}^{(i)}_\mck)_{jk}|\lesssim\sqrt{\varepsilon} .$$	
Together with $\det(\bar{\g}^{(i)}_\mck)_{jk}\sim 1$\footnote{This follows from an explicit computation or, alternatively, from the fact that in Kerr spacetime, $(r,x^1_{(i)},x^2_{(i)})$ is a global smooth coordinate system on the pre-compact spacelike hypersurface $\un\cap\{\wbar=\wbar_0\}\cap\{u<-1\}\cap{\mcu^{(i)}}$.} we get $\mathrm{vol}_{\wbar}=h_i \dee x^1_{(i)}\dee x^2_{(i)}\dee r$ with $h_i=\sqrt{\det\bar{\g}^{(i)}_{jk}}\sim 1$. By similar arguments, we get $\vol=c_i \dee x^1_{(i)}\dee x^2_{(i)}\dee r\dee\wbar$ with $c_i\sim 1$, which concludes the proof by defining $f_i=c_i/h_i\sim 1$.
\end{proof}
\noindent\textbf{Pointwise estimates from estimates on hypersurfaces $\{\wbar=cst\}$}
\begin{lem}\label{lem:transpoXX}
	Let $\phi$ be a horizontal tensor in $\un[\wbar_*]$. We have the estimate
	$$|\phi|^2(r,\ubar,\theta,\phi_+)\lesssim \|\phi\|^2_{L^\infty(S(r_\mca,\ubar-r+r_\mca))}+\int_{\un[\wbar_*]\cap\{\wbar=\ubar-r\}} |\df^{\leq 3}\phi|^2.$$
\end{lem}
\begin{proof}
	We denote $X:=\lambda e_3+\mu e_4,$
	where
	$$\mu:=-\left(\frac{2(r^2+a^2)}{|q|^2}+\widecheck{e_4(\ubar)}\right)^{-1},\quad \lambda:= -\mu\left(\frac{2(r^2+a^2)-\Delta}{|q|^2}+\widecheck{e_4(\ubar)}-\widecheck{e_4(r)}\right),$$
	are such that  $X(\wbar)=0, X(r)=-1$. We also denote 
	$$\varphi_X^{s}(r,\ubar,\theta,\phi_+)=(r-s,\ubar-s,\theta_s,(\phi_+)_s)$$
	the flow of $X$. Note that $X$ is tangent to $\{\wbar=cst\}$ and that $\varphi_X^{s}$ remains in a constant $\wbar$ hypersurface. We can now compute, by the Cauchy-Schwarz inequality, 
	\begin{align*}
		\left|\frac{\dee}{\dee s}|\phi|^2(\varphi_X^{s}(r,\ubar,\theta,\phi_+))\right|=\left|X(|\phi|^2)(\varphi_X^{s}(r,\ubar,\theta,\phi_+))\right|\lesssim |\phi||\nabla_X\phi|(\varphi_X^{s}(r,\ubar,\theta,\phi_+)).
	\end{align*}
	Integrating this bound between $0$ and $s=r_\mca-r$, we get
	\begin{align*}
		&|\phi|^2(r,\ubar,\theta,\phi_+)\\ &\lesssim|\phi|^2(r_\mca,\ubar+r_\mca-r,\theta_{-s_f},(\phi_{+})_{-s_f})+\int_0^s |\phi||\nabla_X\phi|(r+s',\ubar+s',\theta_{-s'},(\phi_+)_{-s'})\dee s'\\
		&\lesssim \|\phi\|^2_{L^\infty(S(r_\mca,\ubar-r+r_\mca))}\\
		&\quad+\left(\int_0^s |\nabla_X\phi|^2(r+s',\ubar+s',\theta_{-s'},(\phi_+)_{-s'})\dee s'\right)^{1/2}\left(\int_0^s |\phi|^2(r+s',\ubar+s',\theta_{-s'},(\phi_+)_{-s'})\dee s'\right)^{1/2}.
	\end{align*} 
	Moreover, by Corollary \ref{cor:sobolevtwiste} we have 
	\begin{align*}
		\int_0^s |\nabla_X^{\leq 1}\phi|^2(r+s',\ubar+s',\theta_{-s'},(\phi_+)_{-s'})\dee s'\lesssim \int_0^s \|\df^{\leq 3}\phi\|^2_{L^2(S(r+s',\ubar+s'))}\dee s',
	\end{align*}
	where we used $|\df^{\leq 2}(\lambda,\mu)|\lesssim 1$ which comes from the bootstrap assumption \eqref{eq:BA}. This yields
	\begin{align*}
		|\phi|^2(r,\ubar,\theta,\phi_+)\lesssim \|\phi\|^2_{L^\infty(S(r_\mca,\ubar-r+r_\mca))}+\int_0^s \|\df^{\leq 3}\phi\|^2_{L^2(S(r+s',\ubar+s'))}\dee s',
	\end{align*} 
which concludes the proof by Lemma \ref{lem:volumeformI}.
\end{proof}

\noindent\textbf{Identities involving approximate Killing fields $\T,\Z$}
\begin{lem}\label{lem:relatnablalieTZ}
	Recall Definition \ref{def:O(1)mck}, the definition \eqref{eq:defTdansI} of the vector fields $\T,\Z$, and the definition of the horizontal Lie derivative $\lie$ in Section \ref{section:horizontalliederivative}. For a horizontal covariant $k$-tensor $U$, we have
\begin{align*}
	\nabla_\T U_{b_1\cdots b_k}&=\lieT U_{b_1\cdots b_k}+\frac{2aMr\cos\theta}{|q|^4}\sum_{j=1}^k\in_{b_jc}U_{b_1\cdots c\cdots b_k}+O(1)_\mck\Gammacheck\cdot U,\\
	\nabla_\Z U_{b_1\cdots b_k}&=\lie_\Z U_{b_1\cdots b_k}-\frac{\cos\theta((r^2+a^2)^2-a^2\sin^2\theta\Delta)}{|q|^4}\sum_{j=1}^k\in_{b_jc}U_{b_1\cdots c\cdots b_k}+O(1)_\mck\Gammacheck\cdot U.
\end{align*}
\end{lem}
\begin{proof}
	See \cite[Lemma 4.14]{KSwaveeq}.
\end{proof}
The following identities are a consequence of the fact that $\T$ and $\Z$ are approximate Killing field, namely that they are Killing fields in exact Kerr.
\begin{lem}\label{lem:commdflieTZI}
	We have the following schematic identities for the commutators between $\lieT$ and 
	$\df$,
\begin{align*}
	[\nabla,(\lieT,\lie_\Z)]U=_sO(1)_\mck\Gammacheck\cdot\df^{\leq 1}U+O(1)_\mck\df^{\leq 1}\Gammacheck\cdot U,\\
	[\nabla_3,(\lieT,\lie_\Z)]U=_sO(1)_\mck\Gammacheck\cdot\df^{\leq 1}U+O(1)_\mck\df^{\leq 1}\Gammacheck\cdot U,\\
	[\nabla_4,(\lieT,\lie_\Z)]U=_sO(1)_\mck\Gammacheck\cdot\df^{\leq 1}U+O(1)_\mck\df^{\leq 1}\Gammacheck\cdot U.
\end{align*}
\end{lem}
\begin{proof}
	The proof is the same as the one of \cite[Lem. C.2]{KSwaveeq}. The only difference is that in the present paper we do not keep track of the powers of $r$, and we notice that the factors in front of the quadratic terms $\Gammacheck\cdot\df^{\leq 1}U$, $\Gammacheck\cdot\df^{\leq 1}U$ take the form $O(1)_\mck$.
\end{proof}
\subsection{Estimates for the PT Ricci and curvature coefficients in region $\un$}

\subsubsection{Commuted equations in region $\un$}\label{section:sectionalasuite3}

\noindent\textbf{Generic non-homogeneous terms for the linearized null structure equations.}
From Propositions \ref{prop:linearizednullstructure} and \ref{prop:eqpourderjplusomoins}, potentially adding terms on the RHS, we get the following schematic equations satisfied by terms belonging in $\Gammacheck$ as defined in \eqref{eq:defgammacheckvrai},
\begin{align}\label{eq:nullstructureschemphi}
	\forall\phi\in\widecheck{\Gamma},\quad\nabla_3\phi=_sO(1)_\mck(\widecheck{R},\widecheck{\Gamma})+\Gammacheck^2.
\end{align}

\noindent\textbf{Commuted equations.} Recall from \eqref{eq:setofderivatives} the set of derivatives $\df=\{\nabla_3,\nabla_4,\nabla\}$.
\begin{lem}\label{lem:derO(1)mck}
	Recalling Definition \ref{def:O(1)mck}, we have the schematic equation
	$$\df(O(1)_\mck)=_sO(1)_\mck+O(1)_\mck\Gammacheck.$$
\end{lem}
\begin{proof}
This follows immediatly from Definition \ref{def:O(1)mck} for the notation $O(1)_\mck$ and the fact that $\df (r,\cos\theta,\frakJ)=O(1)_\mck+\Gammacheck$.
\end{proof}
We now write the commuted null structure equations satisfied by $\df$ derivatives of $\Gammacheck$. 
\begin{prop}\label{prop:commlinnullstructure}
	The following reduced schematic commuted equations hold for $k\leq N_0$,
	\begin{align*}
		\forall\phi\in\widecheck{\Gamma},\quad\nabla_3\df^k\phi=_{rs}\df^{\leq k}(\Gammacheck,\Rcheck).
	\end{align*}
\end{prop}
\begin{proof}
We define a more precise RHS as
$$\mcq^{(k)}_0:=\sum_{1\leq n\leq k+2,\:i_1+\cdots+i_n\leq k}O(1)_\mck\prod_{j=1}^n\df^{\leq i_j}(\Gammacheck,\Rcheck).$$
Using Lemma \ref{lem:derO(1)mck}, and the commutation formulas 
\begin{align*}
	[\nabla_3,\nabla]U&=_s O(1)_\mck\nabla U+O(1)_\mck U+O(1)_\mck(\Gammacheck,\Rcheck)\df^{\leq 1}U,\\
	[\nabla_3,\nabla_4]U&=_s O(1)_\mck\nabla_3 U+O(1)_\mck\nabla U+O(1)_\mck U+O(1)_\mck(\Gammacheck,\Rcheck) \df^{\leq 1}U.
\end{align*}
which follow from Lemma \ref{lem:commutnablageneral} combined with $\eta=\eta_\mck$ in the PT gauge, it is straightforward to prove by induction that for $k\leq N_0$, for $\phi\in\Gammacheck$, we have the schematic equation
\begin{align}\label{eq:HRrankk}
	\nabla_3\df^k\phi=_s\mcq^{(k)}_0,
\end{align}
where the case $k=0$ holds by \eqref{eq:nullstructureschemphi}. Now, using the bootstrap assumption \eqref{eq:BA}, for each term in the sum defining $\mcq_0^{(k)}$, at maximum only one of its factors is a derivative $\df^{\leq k}$ of $(\Gammacheck,\Rcheck)$ of order $k>N_0/2$, and hence all the other terms are bounded by $\sqrt\varepsilon\lesssim 1$. This yields
$$\mcq^{(k)}_0=_{rs}\df^{\leq k}(\Gammacheck,\Rcheck),$$
which concludes the proof by \eqref{eq:HRrankk}.
\end{proof}
We now turn to writing the schematic commuted linearized Bianchi identities. Here, the hyperbolic structure of the Bianchi equations is crucial to not lose derivatives, and must be preserved when commuting with derivatives. Following \cite[Def. 16.11]{KSwaveeq}, we introduce a subset $\df'$ of the $\df$ derivatives, where the general horizontal derivatives $\nabla$ are replaced with horizontal Hodge operators that preserve the hyperbolic structure when commuted with the Bianchi equations. The general $\df$ derivatives will be recovered at the end of Section \ref{section:sectionalasuite5}.
\begin{defi}\label{defi:dfprimhodgeI}
	Let $U\in\fraks_p(\C)$, $p=0,1,2$. For $j\geq 0$, we define the following operators: 
	\begin{itemize}
		\item If $p=0$,
		\begin{align*}
			\barre{\df}^jU&:=(\divc\mcd)^{\frac{j}{2}}U\quad\text{if }j\text{ is even},\\
			\barre{\df}^jU&:=\mcd(\divc\mcd)^{\frac{j-1}{2}}U\quad\text{if }j\text{ is odd}.
		\end{align*} 
		\item If $p=1$, we denote by $\barre{\df}^j$ any of the following derivatives,
		\begin{alignat*}{2}
			\barre{\df}^jU&:=(\mcd\divc)^{\frac{j}{2}}U\quad&\text{or}\quad\barre{\df}^jU&:=(\divc\mcd\hot)^{\frac{j}{2}}U\quad\text{if }j\text{ is even},\\
			\barre{\df}^jU&:=\divc(\mcd\divc)^{\frac{j-1}{2}}U\quad&\text{or}\quad\barre{\df}^jU&:=\mcd\hot(\divc\mcd\hot)^{\frac{j-1}{2}}U\quad\text{if }j\text{ is odd}.
		\end{alignat*} 
	 \item If $p=2$,
 	\begin{align*}
 	\barre{\df}^jU&:=(\mcd\hot\divc)^{\frac{j}{2}}U\quad\text{if }j\text{ is even},\\
 	\barre{\df}^jU&:=\divc(\mcd\hot\divc)^{\frac{j-1}{2}}U\quad\text{if }j\text{ is odd}.
 	\end{align*} 
	\end{itemize}
We also define the following operators $\mcd_p$ acting on $\fraks_p(\C)$ for $p=0,1,2$,
\begin{align}\label{eq:defdesmcdpla}
	\mcd_0=\mcd,\quad\mcd_1=\overline{\mcd}\:\cdot,\quad\mcd_2=\divc.
\end{align}
Finally, we define the following set of derivatives, which can be seen as a subset of $\df$,
 	$$\df':=\{\nabla_3,\nabla_4,\barre{\df}\}.$$

 \end{defi}
The following definition allows us to write in a natural way the commuted Bianchi equations.
\begin{defi}\label{def:horizontalparity}
	Let $\mco_k\in\df'^k$ be any mixed $\df'$ derivative of order $k$, which acts on $\fraks_p(\C)$, $p=1,2$. We denote $\wt\mco_k$ the analog operator which acts on $\fraks_{p-1}(\C)$ defined as follows: let $\barre{\df}^{j}$ be the horizontal derivative operator acting on $\fraks_p(\C)$, which is obtained by removing the factors $\nabla_3,\nabla_4$ of $\mco_k$. Then, we first define $\wt{\barre{\df}^{j}}$ as follows:
	
	\noindent If $p=2$:
	\begin{itemize}
		\item If $\barre{\df}^j=(\mcd\hot\divc)^{\frac{j}{2}}$, then  $\wt{\barre{\df}^{j}}=(\divc\mcd\hot)^{\frac{j}{2}}$.
		\item If $\barre{\df}^j=\divc(\mcd\hot\divc)^{\frac{j-1}{2}}$, then  $\wt{\barre{\df}^{j}}=\mcd\hot(\divc\mcd\hot)^{\frac{j-1}{2}}$.
	\end{itemize}
	If $p=1$, we only consider the cases $\barre{\df}^j=(\mcd\divc)^{\frac{j}{2}}$ or $\barre{\df}^j=\divc(\mcd\divc)^{\frac{j}{2}}$:
\begin{itemize}
	\item If $\barre{\df}^j=(\mcd\divc)^{\frac{j}{2}}$, then  $\wt{\barre{\df}^{j}}=(\divc\mcd)^{\frac{j}{2}}$.
	\item If $\barre{\df}^j=\divc(\mcd\divc)^{\frac{j-1}{2}}$, then  $\wt{\barre{\df}^{j}}=\mcd(\divc\mcd)^{\frac{j-1}{2}}$.
\end{itemize}
Then, we define the operator $\wt{\mco_k}\in\df'^k$ which acts on $\fraks_{p-1}(\C)$ as $\wt{\barre{\df}^{j}}$ where we add $\nabla_3,\nabla_4$ factors in the same place of $\mco_k$ (namely, it corresponds to $\mco_k$ with ${\barre{\df}^{j}}$ replaced with $\wt{\barre{\df}^{j}}$ factor by factor). We also call \textbf{horizontal parity} of $\mco_k$ the parity of the sum of all powers on top of horizontal $\barre{\df}$ derivatives in the factors of $\mco_k$.
\end{defi}

\begin{prop}\label{prop:bianchieqcommutedI}
Let $k\leq N_0$ and $\mco_k\in\df'^k$ which acts on $\fraks_2(\C)$. Then: 
\begin{itemize}
	\item If the horizontal parity of $\mco_k$ is even, 
	\begin{align*}
		\nabla_3(\mco_k A)-\frac12\mcd\hot(\wt{\mco_k}B)=_{rs}\df^{\leq k}(\Gammacheck,\Rcheck),\quad\quad \nabla_4(\wt{\mco_k}B)-\frac12\divc(\mco_k A)=_{rs}\df^{\leq k}(\Gammacheck,\Rcheck).
	\end{align*}
	\item If the horizontal parity of $\mco_k$ is odd, 
	\begin{align*}
		\nabla_3(\mco_k A)-\frac12\divc(\wt{\mco_k}B)=_{rs}\df^{\leq k}(\Gammacheck,\Rcheck),\quad\quad \nabla_4(\wt{\mco_k}B)-\frac12\mcd\hot(\mco_k A)=_{rs}\df^{\leq k}(\Gammacheck,\Rcheck).
	\end{align*}
\end{itemize}
Also, if $\mco_k\in\df'^k$ acts on $\fraks_1(\C)$, then:
\begin{itemize}
	\item If the horizontal parity of $\mco_k$ is even, 
	\begin{align*}
		\nabla_3(\mco_k B)-\mcd(\wt{\mco_k}\overline{\widecheck{P}})=_{rs}\df^{\leq k}(\Gammacheck,\Rcheck),\quad\quad \nabla_4(\wt{\mco_k}\overline{\widecheck{P}})-\frac12\divc({\mco_k B})=_{rs}\df^{\leq k}(\Gammacheck,\Rcheck).
	\end{align*}
	\item If the horizontal parity of $\mco_k$ is odd, 
	\begin{align*}
		\nabla_3(\mco_k B)-\frac12\divc(\wt{\mco_k}\overline{\widecheck{P}})=_{rs}\df^{\leq k}(\Gammacheck,\Rcheck),\quad\quad \nabla_4(\wt{\mco_k}\overline{\widecheck{P}})-\mcd({\mco_k B})=_{rs}\df^{\leq k}(\Gammacheck,\Rcheck),
	\end{align*}
\end{itemize}
and:
	\begin{itemize}
	\item If the horizontal parity of $\mco_k$ is even, 
	\begin{align*}
		\nabla_4(\mco_k \Bbar)+\mcd(\wt{\mco_k}\widecheck{P})=_{rs}\df^{\leq k}(\Gammacheck,\Rcheck),\quad\quad \nabla_3(\wt{\mco_k}\widecheck{P})+\frac12\divc(\mco_k \Bbar)=_{rs}\df^{\leq k}(\Gammacheck,\Rcheck).
	\end{align*}
	\item If the horizontal parity of $\mco_k$ is odd, 
	\begin{align*}
		\nabla_4(\mco_k \Bbar)+\frac12\divc(\wt{\mco_k}\widecheck{P})=_{rs}\df^{\leq k}(\Gammacheck,\Rcheck),\quad\quad \nabla_3(\wt{\mco_k}\widecheck{P})+\mcd(\mco_k \Bbar)=_{rs}\df^{\leq k}(\Gammacheck,\Rcheck).
	\end{align*}
\end{itemize}
Finally, if $\mco_k\in\df'^k$ acts on $\fraks_2(\C)$, then:
	\begin{itemize}
	\item If the horizontal parity of $\mco_k$ is even, 
	\begin{align*}
		\nabla_4(\mco_k \Abar)+\frac12\mcd\hot(\wt{\mco_k}\Bbar)=_{rs}\df^{\leq k}(\Gammacheck,\Rcheck),\quad\quad \nabla_3(\wt{\mco_k}\Bbar)+\frac12\divc(\mco_k \Abar)=_{rs}\df^{\leq k}(\Gammacheck,\Rcheck).
	\end{align*}
	\item If the horizontal parity of $\mco_k$ is odd, 
	\begin{align*}
		\nabla_4(\mco_k \Abar)+\frac12\divc(\wt{\mco_k}\Bbar)=_{rs}\df^{\leq k}(\Gammacheck,\Rcheck),\quad\quad \nabla_3(\wt{\mco_k}\Bbar)+\frac12\mcd\hot(\mco_k \Abar)=_{rs}\df^{\leq k}(\Gammacheck,\Rcheck).
	\end{align*}
\end{itemize}
\end{prop}
\begin{proof}
	This is similar to \cite[Prop. 16.19]{KSwaveeq} except that the commutator terms are simpler to bound here. Let us present the argument for the first Bianchi pair $(A,B)$. The proof is similar to that of Proposition \ref{prop:commlinnullstructure}, the only difference being that we do not commute horizontal $\barre{\df}$ derivatives with the terms $\mcd\hot B$ and $\divc A$. Namely, we respectively commute the operator $\mathcal{O}_k$ with the derivative $\nabla_3$, and the operator $\wt{\mathcal{O}_k}$ with $\nabla_4$, which produces commutator terms similar as $\mcq_0^{(k)}$ in the proof of Proposition \ref{prop:commlinnullstructure}, which are shown to be bounded by $\df^{\leq k}(\Gammacheck,\Rcheck)$ using the bootstrap assumption \eqref{eq:BA}. To deal with the terms with additional horizontal derivatives $\mcd\hot B$ and $\divc A$, we use the definition of $\barre{\df}^j,\wt{\barre{\df}^j}$ to get 
$$\mathcal{O}_k\left(\mcd\hot B\right)-\mcd\hot(\wt{\mathcal{O}_k}B)=_{rs}\df^{\leq k}(\Gammacheck,\Rcheck),\quad \wt{\mathcal{O}_k}\left(\divc A\right)-\divc(\mathcal{O}_kA)=_{rs}\df^{\leq k}(\Gammacheck,\Rcheck),$$
if the horizontal parity of $\mco_k$ is even, and
$$\mathcal{O}_k\left(\mcd\hot B\right)-\divc(\wt{\mathcal{O}_k}B)=_{rs}\df^{\leq k}(\Gammacheck,\Rcheck),\quad \wt{\mathcal{O}_k}\left(\divc A\right)-\mcd\hot(\mathcal{O}_kA)=_{rs}\df^{\leq k}(\Gammacheck,\Rcheck),$$
if the horizontal parity of $\mco_k$ is odd, where the terms $\df^{\leq k}(\Gammacheck,\Rcheck)$ on the RHS above are generated only by the commutators between the $\nabla_3$ and $\nabla_4$ derivatives which appear in $\mathcal{O}_k$ and respectively $\mcd\hot$ and $\divc$. This concludes the proof in the case of the Bianchi pair $(A,B)$. The proof is similar for the other Bianchi pairs, relying on Definitions \ref{defi:dfprimhodgeI} and \ref{def:horizontalparity}.
\end{proof}
\subsubsection{General transport estimates in region $\un$}\label{section:sectionalasuite1}
Recall the scalar functions $w,\wbar$ defined in \eqref{eq:defwregIunee} and \eqref{eq:wbardefun}, and also \eqref{eq:wbarzeroinitial},
$$\wbar:=\ubar-r,\quad w=2r^*-\ubar-r,\quad \wbar_0:=2r^*_\mca-2r_\mca-w_f.$$
In what follows, for $\wbar_0\leq\wbar_1\leq\wbar_2\leq\wbar_*$ we denote
\begin{align}\label{eq:gjhhjhjhj}
	\un[\wbar_1,\wbar_2]:=\un[\wbar_*]\cap\{\wbar_1\leq\wbar\leq\wbar_2\}.
\end{align}
\begin{lem}\label{lem:transpoIderectione3}
	Let $\phi$ be a horizontal tensor in $\un[\wbar_*]$ and $p\geq 1$. We have the following transport identity in the $e_3$ direction in $\un[\wbar_*]$ for any $\wbar_0\leq\wbar_1\leq\wbar_2\leq\wbar_*$,
\begin{align*}
	&\int_{\un\cap\{\wbar=\wbar_2\}}\frac{r^p|\phi|^2}{\sqrt{-\g(\D\wbar,\D\wbar)}}+\iint_{\un[\wbar_1,\wbar_2]}pr^{p-1}|\phi|^2\\
	&+\int_{\{w=w_f\}\cap\un[\wbar_1,\wbar_2]}\frac{r^p|\phi|^2(1-2\Delta^{-1}(r^2+a^2))}{\sqrt{-\g(\D w,\D w)}}+\int_{\{r=r_-(1+\delta_-)\}\cap\un[\wbar_1,\wbar_2]}\frac{r^p|\phi|^2}{\sqrt{-\g(\D r,\D r)}}\\
	&=\int_{\un\cap\{\wbar=\wbar_1\}}\frac{r^p|\phi|^2}{\sqrt{-\g(\D\wbar,\D\wbar)}}+\int_{\mca\cap\un[\wbar_1,\wbar_2]}\frac{r^p|\phi|^2}{\sqrt{-\g(\D r,\D r)}}\\
	&\quad+2\iint_{\un[\wbar_1,\wbar_2]}r^p\left(\Real(\overline{\phi}\cdot\nabla_3\phi)+\frac12tr\chibar|\phi|^2\right).
\end{align*}
\end{lem}
\begin{proof}
	We compute, using $e_3(r)=-1$,
\begin{align}
	\frac12\D_\mu\left(r^p|\phi|^2e_3\right)^\mu+\frac12 pr^{p-1}|\phi|^2&=\frac12  e_3(r^p|\phi|^2)+\frac12 pr^{p-1}|\phi|^2+\frac12r^p|\phi|^2\D_\mu e_3^\mu\nn\\
	&=r^p\left(\Real(\overline{\phi}\cdot\nabla_3\phi)+\frac12tr\chibar|\phi|^2\right),\label{eq:firdtiddidi}
\end{align}
where we used the identity $\D_\mu e_3^\mu=tr\chibar-2\omegabar=tr\chibar$ because $\omegabar=0$ in the PT gauge. Integrating this on $\un[\wbar_1,\wbar_2]$ we get 
\begin{align}
	\frac12\iint_{\un[\wbar_1,\wbar_2]}\D_\mu\left(r^p|\phi|^2e_3\right)^\mu+\frac12\iint_{\un[\wbar_1,\wbar_2]}pr^{p-1}|\phi|^2=\iint_{\un[\wbar_1,\wbar_2]}r^p\left(\Real(\overline{\phi}\cdot\nabla_3\phi)+\frac12tr\chibar|\phi|^2\right).\label{eq:zzheyhey}
\end{align}
We now compute the first term on the LHS above by Stokes theorem. Note that the boundary of $\un[\wbar_1,\wbar_2]$ is given by
\begin{align*}
	\partial\left(\un[\wbar_1,\wbar_2]\right)=&\{\wbar=\wbar_1\}\cup\{\wbar=\wbar_2\}\cup\left(\{w=w_f\}\cap\{\wbar_1\leq\wbar\leq\wbar_2\}\right)\\
	&\cup\left(\{r=r_-(1+\delta_-)\}\cap\{\wbar_1\leq\wbar\leq\wbar_2\}\right)\cup\left(\mca\cap\{\wbar_1\leq\wbar\leq\wbar_2\}\right),
\end{align*}
where both upper boundaries $\{w=w_f\}\cap\{\wbar_1\leq\wbar\leq\wbar_2\}$ and $\{r=r_-(1+\delta_-)\}\cap\{\wbar_1\leq\wbar\leq\wbar_2\}$ may be empty or non-empty\footnote{Denoting $\wbar_{f}=2r^*(r_-(1+\delta_-))-2r_-(1+\delta_-)-w_f$ the value of $\wbar$ at the intersection $\{w=w_f\}\cap\{r=r_-(1+\delta_-)\}$, then: If $\wbar_{f}<\wbar_1$ then $\{w=w_f\}\cap\{\wbar_1\leq\wbar\leq\wbar_2\}=\varnothing$, if $\wbar_2<\wbar_{f}$ then $\{r=r_-(1+\delta_-)\}\cap\{\wbar_1\leq\wbar\leq\wbar_2\}=\varnothing$, and if $\wbar_1\leq\wbar_{f}\leq\wbar_2$ then $\{w=w_f\}\cap\{\wbar_1\leq\wbar\leq\wbar_2\}\neq\varnothing$ and $\{r=r_-(1+\delta_-)\}\cap\{\wbar_1\leq\wbar\leq\wbar_2\}\neq\varnothing$.}. Also, for the derivatives of $r,w,\wbar$ we have
\begin{align}
	&e_3(\wbar)= 1,\quad e_4(\wbar)=\frac{2(r^2+a^2)-\Delta}{|q|^2}+O(\sqrt{\varepsilon}),\quad \nabla_b(\wbar)=a\Real(\frakJ)_b+O(\sqrt{\varepsilon}),\nn\\
	&e_3(r)=-1,\quad e_4(r)=\frac{\Delta}{|q|^2}+O(\sqrt{\varepsilon}),\quad \nabla_b(r)=O(\sqrt{\varepsilon}),\label{eq:zcoucou2}\\
	&e_3(w)= 1-\frac{2(r^2+a^2)}{\Delta},\quad e_4(w)=\frac{-\Delta}{|q|^2}+O(\sqrt{\varepsilon}),\quad \nabla_b(w)=-a\Real(\frakJ)_b+O(\sqrt{\varepsilon}),\nn
\end{align}
where we used the boostrap assumption \eqref{eq:BA}, which implies 
\begin{align}
	\g(\D\wbar,\D\wbar)=-e_3(\wbar)e_4(\wbar)+|\nabla\wbar|^2&=e_4(r)-e_4(\ubar)+|\nabla r-\nabla\ubar|^2\nn\\
	&=\frac{\Delta-(r^2+a^2+|q|^2)}{|q|^2}+O(\sqrt{\varepsilon})\sim -1,\label{eq:gDwbarDwbarunI}
\end{align}
for $\varepsilon(a,M)>0$ small enough, and similarly
\begin{align}
	\g(\D w,\D w)=\frac{\Delta-(r^2+a^2+|q|^2)}{|q|^2}+O(\sqrt{\varepsilon})\sim -1.\label{eq:gDwDwunI}
\end{align}
Note that using \eqref{eq:BA} and $\ubar\sim\wbar\geq\wbar_0=2r^*_\mca-2r_\mca-w_f$ by \eqref{eq:wbarzeroinitial}, we also have
\begin{align}
	\g(\D r,\D r)=-e_3(r)e_4(r)+|\nabla r|^2=\frac{\Delta}{|q|^2}+O(\sqrt{\varepsilon})<\frac{\Delta}{2|q|^2}<0\label{eq:gDrDrunI}
\end{align}
for $\varepsilon(a,M,\delta_\pm)$ small enough. Stokes theorem and \eqref{eq:zcoucou2} thus yields
\begin{align*}
	\iint_{\un[\wbar_1,\wbar_2]}\D_\mu\left(r^p|\phi|^2e_3\right)^\mu=&\int_{\un\cap\{\wbar=\wbar_2\}}\frac{r^p|\phi|^2}{\sqrt{-\g(\D\wbar,\D\wbar)}}-\int_{\un\cap\{\wbar=\wbar_1\}}\frac{r^p|\phi|^2}{\sqrt{-\g(\D\wbar,\D\wbar)}}\\
	&-\int_{\mca\cap\un[\wbar_1,\wbar_2]}\frac{r^p|\phi|^2}{\sqrt{-\g(\D r,\D r)}}+\int_{\{r=r_-(1+\delta_-)\}\cap\un[\wbar_1,\wbar_2]}\frac{r^p|\phi|^2}{\sqrt{-\g(\D r,\D r)}}\\
	&+\int_{\{w=w_f\}\cap\un[\wbar_1,\wbar_2]}\frac{r^p|\phi|^2(1-2\Delta^{-1}(r^2+a^2))}{\sqrt{-\g(\D w,\D w)}}.
\end{align*}
where we use $\D\wbar(\wbar)=\g(\D\wbar,\D\wbar)<0$, and similarly $\D r(r)<0$, $\D w(w)<0$, which implies that $\D\wbar$ is inwards pointing at $\{\wbar=\wbar_2\}$ and outwards pointing at $\{\wbar=\wbar_1\}$, that $\D r$ is inwards pointing on $\mathcal{A}$ and outwards pointing at $\{r=r_-(1+\delta_-)\}$, and that $\D w$ is inwards pointing on $\{w=w_f\}\cap\{\wbar\leq\wbar_*\}$. Combining this with \eqref{eq:zzheyhey} concludes the proof.
\end{proof}

\begin{lem}\label{lem:transpoIderectione4}
	Let $\phi$ be a horizontal tensor in $\un[\wbar_*]$ and $p\geq 1$. We have the following transport identity in the $e_4$ direction in $\un[\wbar_*]$ for any $\wbar_0\leq\wbar_1\leq\wbar_2\leq\wbar_*$,
\begin{align*}
	&\int_{\un\cap\{\wbar=\wbar_2\}}\frac{r^p|\phi|^2\left(2(r^2+a^2)-\Delta+O\left(\sqrt{\varepsilon}\right)\right)}{|q|^2\sqrt{-\g(\D\wbar,\D\wbar)}}+\iint_{\un[\wbar_1,\wbar_2]}\left(\frac{-\Delta}{|q|^2}+O\left(\sqrt{\varepsilon}\right)\right)pr^{p-1}|\phi|^2\\
	&+\int_{\{w=w_f\}\cap\un[\wbar_1,\wbar_2]}\frac{r^p|\phi|^2\left(-\Delta+O\left(\sqrt{\varepsilon}\right)\right)}{|q|^2\sqrt{-\g(\D w,\D w)}}+\int_{\{r=r_-(1+\delta_-)\}\cap\un[\wbar_1,\wbar_2]}\frac{r^p|\phi|^2\left(-\Delta+O\left(\sqrt{\varepsilon}\right)\right)}{|q|^2\sqrt{-\g(\D r,\D r)}}\\
	&=\int_{\un\cap\{\wbar=\wbar_1\}}\frac{r^p|\phi|^2\left(2(r^2+a^2)-\Delta+O\left(\sqrt{\varepsilon}\right)\right)}{|q|^2\sqrt{-\g(\D\wbar,\D\wbar)}}+\int_{\mca\cap\un[\wbar_1,\wbar_2]}\frac{r^p|\phi|^2\left(-\Delta+O\left(\sqrt{\varepsilon}\right)\right)}{|q|^2\sqrt{-\g(\D r,\D r)}}\\
	&\quad+2\iint_{\un[\wbar_1,\wbar_2]}r^p\left(\Real(\overline{\phi}\cdot\nabla_4\phi)+\frac12(tr\chi-2\omega)|\phi|^2\right).
\end{align*}
\end{lem}
\begin{proof}
	The proof is the same as the one of Lemma \ref{lem:transpoIderectione3}, integrating instead the identity
		\begin{align*}
			\frac12\D_\mu\left(r^p|\phi|^2e_4\right)^\mu-\frac12\left(\frac{\Delta}{|q|^2}+\widecheck{e_4(r)}\right) pr^{p-1}|\phi|^2=r^p\left(\Real(\overline{\phi}\cdot\nabla_4\phi)+\frac12(tr\chi-2\omega)|\phi|^2\right).
		\end{align*}
	and using the estimates for the $e_4$ derivatives of $\wbar,r,w$ in \eqref{eq:zcoucou2}.
\end{proof}
\begin{prop}\label{prop:transportestimates}
There exists $p_0(a,M)\gg1$ which depends only on $a,M$, and a constant $C_{\delta_\pm}>0$ which depend, only on $a,M,\delta_\pm$ such that for any horizontal tensor $\phi$, for any $\wbar_0\leq\wbar_1\leq\wbar_2\leq\wbar_*$, and for any $p\geq p_0$, we have the following transport estimate in the $e_3$ direction,
	\begin{align*}
		&\int_{\un\cap\{\wbar=\wbar_2\}}r^p|\phi|^2+\iint_{\un[\wbar_1,\wbar_2]}pr^{p-1}|\phi|^2\lesssim\\
		&\int_{\un\cap\{\wbar=\wbar_1\}}r^p|\phi|^2+\int_{\mca\cap\un[\wbar_1,\wbar_2]}C_{\delta_\pm}r^p|\phi|^2+\iint_{\un[\wbar_1,\wbar_2]}r^p|\nabla_3\phi|^2.
	\end{align*}
\end{prop}
\begin{proof}
By Lemma \ref{lem:transpoIderectione3}, dropping the non-negative terms on $\{w=w_f\}\cup\{r=r_-(1+\delta_-)\}$ and using the estimates 
$$\g(\D\wbar,\D\wbar)\sim -1,\quad \g(\D w,\D w)\sim -1,\quad 1-2\Delta^{-1}(^2+a^2)>0,\quad |tr\chibar|\lesssim 1,$$
which hold by \eqref{eq:gDwbarDwbarunI}, \eqref{eq:gDwDwunI}, \eqref{eq:BA} as well as \eqref{eq:gDrDrunI} which yields
\begin{align}\label{eq:againehetyty}
	\left(\sqrt{-\g(\D r,\D r)}\right)^{-1}\leq C_{\delta_\pm},
\end{align}
where $C_{\delta_\pm}>0$ depends only on $a,M,\delta_\pm$, we get
	\begin{align*}
	&\int_{\un\cap\{\wbar=\wbar_2\}}r^p|\phi|^2+\iint_{\un\cap\{\wbar_1\leq\wbar\leq\wbar_2\}}pr^{p-1}|\phi|^2\lesssim\\
	&\int_{\un\cap\{\wbar=\wbar_1\}}r^p|\phi|^2+\int_{\mca\cap\{\wbar_1\leq\wbar\leq\wbar_2\}}C_{\delta_\pm}r^p|\phi|^2+\iint_{\un\cap\{\wbar_1\leq\wbar\leq\wbar_2\}}r^p|\nabla_3\phi|^2+\iint_{\un\cap\{\wbar_1\leq\wbar\leq\wbar_2\}}r^{p}|\phi|^2,
\end{align*}
This concludes the proof by choosing $p\geq p_0(a,M)$ large enough (depending only on $a,M$) such that the last term on the RHS above is absorbed in the second term on the LHS. 
\end{proof}

\subsubsection{Transport estimates for Ricci coefficients}\label{section:sectionalasuite4}
From Propositions \ref{prop:commlinnullstructure} and \ref{prop:transportestimates}, and recalling the notations introduced in these propositions, we deduce the following result.

\begin{prop}\label{prop:controlricci}
	For any $k\leq N_0$, $p\geq p_0(a,M)$, $\wbar_0\leq\wbar_1\leq\wbar_2\leq\wbar_*$, we have
	\begin{align*}
		&\sum_{\phi\in\widecheck{\Gamma}}\left(\int_{\un\cap\{\wbar=\wbar_2\}}r^p|\df^{\leq k}\phi|^2+\iint_{\un\cap\{\wbar_1\leq\wbar\leq\wbar_2\}}pr^{p-1}|\df^{\leq k}\phi|^2\right)\lesssim\\
		&\sum_{\phi\in\widecheck{\Gamma}}\int_{\un\cap\{\wbar=\wbar_1\}}r^p|\df^{\leq k}\phi|^2+\int_{\mca\cap\{\wbar_1\leq\wbar\leq\wbar_2\}}C_{\delta_\pm}r^p|{\df}^{\leq k}(\widecheck{\Gamma},\widecheck{R})|^2+\iint_{\un\cap\{\wbar_1\leq\wbar\leq\wbar_2\}}r^p|\df^{\leq k}\widecheck{R}|^2.
	\end{align*}
\end{prop}
\begin{proof}
By Propositions \ref{prop:commlinnullstructure} and \ref{prop:transportestimates}, summing over all $\phi\in\Gammacheck$, we get
	\begin{align}
	&\sum_{\phi\in\widecheck{\Gamma}}\left(\int_{\un\cap\{\wbar=\wbar_2\}}r^p|\df^{\leq k}\phi|^2+\iint_{\un\cap\{\wbar_1\leq\wbar\leq\wbar_2\}}pr^{p-1}|\df^{\leq k}\phi|^2\right)\lesssim\nn\\
	&\sum_{\phi\in\widecheck{\Gamma}}\int_{\un\cap\{\wbar=\wbar_1\}}r^p|\df^{\leq k}\phi|^2+\int_{\mca\cap\{\wbar_1\leq\wbar\leq\wbar_2\}}C_{\delta_\pm}r^p|{\df}^{\leq k}(\widecheck{\Gamma},\widecheck{R})|^2+\iint_{\un\cap\{\wbar\leq\wbar_*\}}r^p|\df^{\leq k}(\Gammacheck,\Rcheck)|^2.\nn
\end{align}	
Thus choosing $p_0(a,M)$ larger if necessary such that the last term with $\df^{\leq k}\Gammacheck$ on the RHS above is absorbed in the second term in the LHS, we conclude the proof.
\end{proof}
\subsubsection{Energy estimates for Bianchi pairs}\label{section:sectionalasuite5}
\begin{lem}\label{lem:lempourbianchipair}
	Let $U\in\fraks_2(\C)$ and $V\in\fraks_1(\C)$. We have the identity
	\begin{align*}
		\frac12\Real\left(\overline{V}\cdot(\divc U)\right)+\frac14\Real\left(\overline{U}\cdot(\mcd\hot V)\right)=\D^\mu\Real(\overline{V}\cdot U)_\mu-(\eta_\mck+\etabar)\cdot\Real(\overline{V}\cdot U),
	\end{align*}
where we extended the horizontal 1-form $\Real(\overline{V}\cdot U)_c=\Real(\overline{V}_bU_{bc})$ to a spacetime 1-form by 
$$\Real(\overline{V}\cdot U)_3=\Real(\overline{V}\cdot U)_4=0.$$
\end{lem}
\begin{proof}
	This a direct consequence of Lemmas \ref{lem:lempourbianchipairpreli} and \ref{lem:2140waveq} and the PT condition $\eta=\eta_\mck$.
\end{proof}
\begin{prop}\label{prop:bianchipair1estimee}
	Let $U\in\fraks_2(\C)$ and $V\in\fraks_1(\C)$ in $\un[\wbar_*]$ be such that
	\begin{align}\label{eq:equationbianchipair1}
		\nabla_3 U-\frac12\mcd\hot V=F,\quad\nabla_4V-\frac12\divc U=G.
	\end{align}
	Then for $p\geq p_0(a,M,\delta_+,\delta_-)$ large enough and $\wbar_0\leq\wbar_1\leq\wbar_2\leq\wbar_*$, we have the estimate
	\begin{align*}
	&\int_{\un\cap\{\wbar=\wbar_2\}}r^p(|U|^2+|V|^2)+\iint_{\un\cap\{\wbar_1\leq\wbar\leq\wbar_2\}}pr^{p-1}(|U|^2+|V|^2)\\
	&\lesssim C'_{\delta_\pm}\Bigg(\int_{\un\cap\{\wbar=\wbar_1\}}r^p(|U|^2+|V|^2)+\int_{\mathcal{A}\cap\{\wbar_1\leq\wbar\leq\wbar_2\}}r^p(|U|^2+|V|^2)\\
	&\quad\quad+\iint_{\un\cap\{\wbar_1\leq\wbar\leq\wbar_2\}}r^p(|F|^2+|G|^2)\Bigg),
\end{align*}
where $C'_{\delta_\pm}$ is a constant which only depends on $a,M,\delta_\pm$.
\end{prop}

\begin{proof}
Combining Lemmas \ref{lem:transpoIderectione3} and \ref{lem:transpoIderectione4}, recalling $\un[\wbar_1,\wbar_2]=\un[\wbar_*]\cap\{\wbar_1\leq\wbar\leq\wbar_2\}$ we get 
	\begin{align}
	&\int_{\un\cap\{\wbar=\wbar_2\}}\frac{r^p}{\sqrt{-\g(\D\wbar,\D\wbar)}}\left({\frac12|U|^2}+\frac{|V|^2(2(r^2+a^2)-\Delta+O(\sqrt{\varepsilon}))}{|q|^2}\right)\nn\\
	&+\iint_{\un[\wbar_1,\wbar_2]}pr^{p-1}\left(\frac12|U|^2+\left(\frac{-\Delta}{|q|^2}+O(\sqrt{\varepsilon})\right)|V|^2\right)\nn\\
	&+\int_{\{w=w_f\}\cap\un[\wbar_1,\wbar_2]}\frac{r^p}{\sqrt{-\g(\D w,\D w)}}\left(\frac12|U|^2(1-2\Delta^{-1}(r^2+a^2))+\frac{|V|^2(-\Delta+O(\sqrt{\varepsilon}))}{|q|^2}\right)\nn\\
	&+\int_{\{r=r_-(1+\delta_-)\}\cap\un[\wbar_1,\wbar_2]}\frac{r^p}{\sqrt{-\g(\D r,\D r)}}\left(\frac12|U|^2+\frac{|V|^2(-\Delta+O(\sqrt{\varepsilon}))}{|q|^2}\right)\nn\\
	&=\int_{\un\cap\{\wbar=\wbar_1\}}\frac{r^p}{\sqrt{-\g(\D\wbar,\D\wbar)}}\left({\frac12|U|^2}+\frac{|V|^2(2(r^2+a^2)-\Delta+O(\sqrt{\varepsilon}))}{|q|^2}\right)\nn\\
	&\quad+\int_{\mathcal{A}\cap\un[\wbar_1,\wbar_2]}\frac{r^p}{\sqrt{-\g(\D r,\D r)}}\left(\frac12|U|^2+\frac{|V|^2(-\Delta+O(\sqrt{\varepsilon}))}{|q|^2}\right)\nn\\
	&\quad+2\iint_{\un[\wbar_1,\wbar_2]}r^p\left(\frac12\Real(\overline{U}\cdot\nabla_3U)+\frac14tr\chibar|U|^2+\Real(\overline{V}\cdot\nabla_4V)+\frac12(tr\chi-2\omega)|V|^2\right).\label{eq:zazazaRHS}
\end{align} 
We first compute the last term on the RHS of \eqref{eq:zazazaRHS}. By Lemma \ref{lem:lempourbianchipair}, and \eqref{eq:equationbianchipair1},
\begin{align*}
	\frac12\Real(\overline{U}\cdot\nabla_3U)+\Real(\overline{V}\cdot\nabla_4V)&=\frac14\Real\left(\overline{U}\cdot(\mcd\hot V)\right)+\frac12\Real\left(\overline{V}\cdot(\divc U)\right)+\Real\left(\frac12\overline{U}\cdot F+\overline{V}\cdot G\right)\\
	&=\D^\mu\Real(\overline{V}\cdot U)_\mu-(\eta_\mck+\etabar)\cdot\Real(\overline{V}\cdot U)+\Real\left(\frac12\overline{U}\cdot F+\overline{V}\cdot G\right).
\end{align*}
Moreover, we have 
\begin{align*}
	\iint_{\un[\wbar_1,\wbar_2]}r^p\D^\mu\Real(\overline{V}\cdot U)_\mu&=\iint_{\un[\wbar_1,\wbar_2]}\left(\D^\mu\left(r^p\Real(\overline{V}\cdot U)\right)_\mu-pr^{p-1}\nabla r\cdot \Real(\overline{V}\cdot U)\right)\\
	&=\iint_{\un[\wbar_1,\wbar_2]}\D^\mu\left(r^p\Real(\overline{V}\cdot U)\right)_\mu+\iint_{\un[\wbar_1,\wbar_2]}pr^{p-1}O(\sqrt{\varepsilon})\cdot \Real(\overline{V}\cdot U),
\end{align*}
where we used the convention $\Real(\overline{V}\cdot U)_3=\Real(\overline{V}\cdot U)_4=0$ and $\nabla r=O(\sqrt{\varepsilon})$ by the bootstrap assumption \eqref{eq:BA}. Moreover, by Stokes theorem, we have
\begin{align*}
	\iint_{\un[\wbar_1,\wbar_2]}\D^\mu\left(r^p\Real(\overline{V}\cdot U)\right)_\mu&=B_{\{\wbar=\wbar_2\}}-B_{\{\wbar=\wbar_1\}}+B_{\{w=w_f\}}+B_{\{r=r_-(1+\delta_-)\}}+B_{\mathcal{A}}
\end{align*}
where the boundary terms are given by 
\begin{align*}
	B_{\{\wbar=\wbar_b\}}&:=\int_{\un\cap\{\wbar=\wbar_b\}}\frac{r^p}{\sqrt{-\g(\D\wbar,\D\wbar)}}\Real(\overline{V}\cdot U)\cdot\nabla\wbar,\\
	B_{\{w=w_f\}}&:=\int_{\{w=w_f\}\cap\un[\wbar_1,\wbar_2]}\frac{r^p}{\sqrt{-\g(\D w,\D w)}}\Real(\overline{V}\cdot U)\cdot\nabla w,\\
	B_{\{r=r_-(1+\delta_-)\}}&:=\int_{\{r=r_-(1+\delta_-)\}\cap\un[\wbar_1,\wbar_2]}\frac{r^p}{\sqrt{-\g(\D r,\D r)}}\Real(\overline{V}\cdot U)\cdot\nabla r,\\
	B_{\mathcal{A}}&:=\int_{\mathcal{A}\cap\un[\wbar_1,\wbar_2]}\frac{r^p}{\sqrt{-\g(\D r,\D r)}}\Real(\overline{V}\cdot U)\cdot\nabla r.
\end{align*}
 Thus, we deduce
 \begin{align}
 	&\iint_{\un[\wbar_1,\wbar_2]}r^p\left(\frac12\Real(\overline{U}\cdot\nabla_3U)+\Real(\overline{V}\cdot\nabla_4V)\right)=\nn\\
 	&\iint_{\un[\wbar_1,\wbar_2]}\left(\Real\left(\frac12\overline{U}\cdot F+\overline{V}\cdot G\right)-(\eta_\mck+\etabar)\cdot\Real(\overline{V}\cdot U)\right)\label{eq:deducedoklm}\\
 	&+\iint_{\un[\wbar_1,\wbar_2]}pr^{p-1}O(\sqrt{\varepsilon})\cdot \Real(\overline{V}\cdot U)+B_{\{\wbar=\wbar_2\}}-B_{\{\wbar=\wbar_1\}}+B_{\{w=w_f\}}+B_{\{r=r_-(1+\delta_-)\}}+B_{\mathcal{A}}.\nn
 \end{align}
By the bootstrap assumption \eqref{eq:BA}, for $b=1,2$ we now write the bounds
\begin{align}
	|B_{\{\wbar=\wbar_b\}}|&\leq\int_{\un\cap\{\wbar=\wbar_b\}}\frac{r^p}{\sqrt{-\g(\D\wbar,\D\wbar)}}|V||U||\nabla\wbar|\nn\\
	&\leq \int_{\un\cap\{\wbar=\wbar_b\}}\frac{r^p}{\sqrt{-\g(\D\wbar,\D\wbar)}}|V||U|\left(\frac{a\sin\theta}{|q|}+O(\sqrt{\varepsilon})\right)\label{eq:boundaryIun1}\\
	&\leq \int_{\un\cap\{\wbar=\wbar_b\}}\frac{r^p}{\sqrt{-\g(\D\wbar,\D\wbar)}}\left(\frac{a^2\sin^2\theta}{4(r^2+a^2)}|U|^2+\frac{r^2+a^2}{|q|^2}|V|^2+O(\sqrt{\varepsilon})\left(|U|^2+|V|^2\right)\right),\nn
\end{align}
where we used the bound $xy\leq \frac{x^2}{4}+y^2$, as well as
\begin{align}
	&|B_{\{w=w_f\}}|\nn\\
	&\leq  \int_{\{w=w_f\}\cap\un[\wbar_1,\wbar_2]}\frac{r^p}{\sqrt{-\g(\D w,\D w)}}|V||U||\nabla w|\nn\\
	&\leq  \int_{\{w=w_f\}\cap\un[\wbar_1,\wbar_2]}\frac{r^p}{\sqrt{-\g(\D w,\D w)}}|V||U|\left(\frac{a\sin\theta}{|q|}+O(\sqrt{\varepsilon})\right)\label{eq:boundaryIun2}\\
	&\leq \frac12\int_{\{w=w_f\}\cap\un[\wbar_1,\wbar_2]}\frac{r^p}{\sqrt{-\g(\D w,\D w)}}\left(\frac{r^2+a^2}{-\Delta}|U|^2+\frac{-\Delta}{|q|^2}\frac{a^2\sin^2\theta}{r^2+a^2}|V|^2+O(\sqrt{\varepsilon})\left(|U|^2+|V|^2\right)\right),\nn
\end{align}
and
\begin{align*}
	|B_{\{r=r_-(1+\delta_-)\}}|&\lesssim \int_{\{r=r_-(1+\delta_-)\}\cap\un[\wbar_1,\wbar_2]}\frac{r^p\sqrt{\varepsilon}}{\sqrt{-\g(\D r,\D r)}}(|U|^2+|V|^2),\\
	|B_{\mathcal{A}}|&\lesssim\int_{\mathcal{A}\cap\un[\wbar_1,\wbar_2]}\left|\frac{r^p}{\sqrt{-\g(\D r,\D r)}}\Real(\overline{V}\cdot U)\cdot\nabla r\right|\lesssim\int_{\mathcal{A}\cap\un[\wbar_1,\wbar_2]}C_{\delta_\pm}r^p\left(|U|^2+|V|^2\right),
\end{align*}
by \eqref{eq:againehetyty}. Combining the bounds \eqref{eq:boundaryIun1} and \eqref{eq:boundaryIun2} with \eqref{eq:deducedoklm} and \eqref{eq:zazazaRHS}, we deduce
	\begin{align}
	&\int_{\un\cap\{\wbar=\wbar_2\}}\frac{r^p}{\sqrt{-\g(\D\wbar,\D\wbar)}}\left(\frac{|q|^2+O(\sqrt{\varepsilon})}{4(r^2+a^2)}|U|^2+\frac{-\Delta+O(\sqrt{\varepsilon})}{|q|^2}|V|^2\right)\nn\\
	&+\frac12\iint_{\un[\wbar_1,\wbar_2]}pr^{p-1}\left(\frac12|U|^2+\frac{-\Delta+O(\sqrt{\varepsilon})}{|q|^2}|V|^2\right)\nn\\
	&\leq\int_{\un\cap\{\wbar=\wbar_1\}}\frac{r^p}{\sqrt{-\g(\D\wbar,\D\wbar)}}\left(\frac{|q|^2+O(\sqrt{\varepsilon})}{4(r^2+a^2)}|U|^2+\frac{-\Delta+O(\sqrt{\varepsilon})}{|q|^2}|V|^2\right)\nn\\
	&+\frac12\int_{\mathcal{A}\cap\un[\wbar_1,\wbar_2]}\frac{r^p}{\sqrt{-\g(\D r,\D r)}}\left(\frac12|U|^2+\frac{|V|^2(-\Delta+O(\sqrt{\varepsilon}))}{|q|^2}\right)\nn\\
	&\quad+\iint_{\un[\wbar_1,\wbar_2]}r^p\Bigg(\frac14tr\chibar|U|^2+\frac12(tr\chi-2\omega)|V|^2+pr^{p-1}O(\sqrt{\varepsilon})\cdot \Real(\overline{V}\cdot U)\nn\\
	&\quad\quad\quad\quad\quad\quad\quad\quad\quad\quad+\Real\left(\frac12\overline{U}\cdot F+\overline{V}\cdot G\right)-(\eta_\mck+\etabar)\cdot\Real(\overline{V}\cdot U)\Bigg)+B_{\mathcal{A}},\label{eq:tropgrossevraiment}
\end{align}
where we dropped the terms on $\{w=w_f\}$ and $\{r=r_-(1+\delta_-)\}$ on the LHS which write
\begin{align*}
	\int_{\{w=w_f\}\cap\un[\wbar_1,\wbar_2]}\frac{r^p}{\sqrt{-\g(\D w,\D w)}}\left(\frac{1+O(\sqrt{\varepsilon})}{4}|U|^2+\frac{-\Delta+O(\sqrt{\varepsilon})}{2(r^2+a^2)}|V|^2\right),\\
	\int_{\{r=r_-(1+\delta_-)\}\cap\un[\wbar_1,\wbar_2]}\frac{r^p}{\sqrt{-\g(\D r,\D r)}}\left(\frac12\left(1+O(\sqrt{\varepsilon})\right)|U|^2+\frac{|V|^2(-\Delta+O(\sqrt{\varepsilon}))}{|q|^2}\right),
\end{align*}
because they are non-negative for $\varepsilon(a,M,\delta_\pm)$ small enough such that 
$1+O(\sqrt{\varepsilon})>1/2>0$ and $-\Delta+O(\sqrt{\varepsilon})>-\Delta/2>0$. Note that \eqref{eq:tropgrossevraiment} combined with \eqref{eq:gDwbarDwbarunI}, \eqref{eq:againehetyty} then implies
	\begin{align*}
	&\int_{\un\cap\{\wbar=\wbar_2\}}r^p(|U|^2+|\Delta||V|^2)+\iint_{\un[\wbar_1,\wbar_2]}pr^{p-1}(|U|^2+|\Delta||V|^2)\lesssim\\
	&\int_{\un\cap\{\wbar=\wbar_1\}}r^p(|U|^2+|\Delta||V|^2)+\int_{\mca\cap\un[\wbar_1,\wbar_2]}C_{\delta_\pm}r^p(|U|^2+|V|^2)\\
	&+\iint_{\un[\wbar_1,\wbar_2]}r^p(|F|^2+|G|^2+|U|^2+|V|^2).
\end{align*}
Thus, choosing $p(a,M,\delta_+,\delta_-)$ sufficiently large such that the bulk term with $|U|^2+|V|^2$ on the RHS above is absorbed in the bulk term on the LHS above\footnote{Note that here $p$ necessarily depends on $\delta_\pm$ because of the factor $\Delta$ in front of $p r^{p-1}|V|^2$ in the LHS.}, we get
	\begin{align*}
	&\int_{\un\cap\{\wbar=\wbar_2\}}r^p(|U|^2+|\Delta||V|^2)+\iint_{\un[\wbar_1,\wbar_2]}pr^{p-1}(|U|^2+|\Delta||V|^2)\lesssim\\
	&\int_{\un\cap\{\wbar=\wbar_1\}}r^p(|U|^2+|\Delta||V|^2)+\int_{\mca\cap\un[\wbar_1,\wbar_2]}C_{\delta_\pm}r^p(|U|^2+|V|^2)+\iint_{\un[\wbar_1,\wbar_2]}r^p(|F|^2+|G|^2).
\end{align*}
Finally we conclude the proof by using 
$$|U|^2+|V|^2\lesssim C'_{\delta_\pm}(|U|^2+|\Delta V|^2),$$
where $C'_{\delta_\pm}:=\sup_{r_-(1+\delta_-)\leq r\leq r_+(1-\delta_+)}|\Delta^{-1}|$.
\end{proof}
We continue this section with the following results, which are hyperbolic estimates analog to the ones in Proposition \ref{prop:bianchipair1estimee}, but for other types of Bianchi pairs.

\begin{prop}\label{prop:bianchipairautresestimee}
	We assume that $U$ and $V$ satisfy one of the following assumptions :
\begin{enumerate}
	\item $U\in\fraks_2(\C)$ and $V\in\fraks_1(\C)$ satisfy in $\un[\wbar_*]$, for some fixed sign $\epsilon\in\{\pm 1\}$,
	\begin{align}\label{eq:equationbianchipair11}
		\nabla_3 U+\epsilon\frac12\mcd\hot V=F,\quad\nabla_4V+\epsilon\frac12\divc U=G.
	\end{align}
	\item $U\in\fraks_1(\C)$ and $V\in\fraks_2(\C)$ satisfy in $\un[\wbar_*]$, for some fixed sign $\epsilon\in\{\pm 1\}$,
	\begin{align}\label{eq:equationbianchipair12}
	\nabla_3 U+\epsilon\frac12\divc V=F,\quad\nabla_4V+\epsilon\frac12\mcd\hot U=G.
	\end{align}
	\item $U\in\fraks_1(\C)$ and $V\in\fraks_0(\C)$ satisfy in $\un[\wbar_*]$, for some fixed sign $\epsilon\in\{\pm 1\}$,
	\begin{align}\label{eq:equationbianchipair13}
	\nabla_3 U+\epsilon\mcd V=F,\quad\nabla_4V+\epsilon\frac12\divc U=G.
	\end{align}
	\item $U\in\fraks_0(\C)$ and $V\in\fraks_1(\C)$ satisfy in $\un[\wbar_*]$, for some fixed sign $\epsilon\in\{\pm 1\}$,
	\begin{align}\label{eq:equationbianchipair14}
		\nabla_3 U+\epsilon\frac12\divc V=F,\quad\nabla_4V+\epsilon \mcd U=G.
	\end{align}
\end{enumerate}
	Then for $p\geq p_0(a,M,\delta_+,\delta_-)$ large enough and $\wbar_0\leq\wbar_1\leq\wbar_2\leq\wbar_*$, we have the estimate
\begin{align*}
	&\int_{\un\cap\{\wbar=\wbar_2\}}r^p(|U|^2+|V|^2)+\iint_{\un\cap\{\wbar_1\leq\wbar\leq\wbar_2\}}pr^{p-1}(|U|^2+|V|^2)\\
	&\lesssim C'_{\delta_\pm}\Bigg(\int_{\un\cap\{\wbar=\wbar_1\}}r^p(|U|^2+|V|^2)+\int_{\mathcal{A}\cap\{\wbar_1\leq\wbar\leq\wbar_2\}}r^p(|U|^2+|V|^2)\\
	&\quad\quad+\iint_{\un\cap\{\wbar_1\leq\wbar\leq\wbar_2\}}r^p(|F|^2+|G|^2)\Bigg),
\end{align*}
where $C'_{\delta_\pm}$ is a constant which only depends on $a,M,\delta_\pm$.
\end{prop}
\begin{proof}
	The proof is basically the same as the one of Proposition Proposition \ref{prop:bianchipair1estimee}. The only slight difference is that to deal with the horizontal derivative terms in the case where $U$ or $V$ is a function in $\fraks_0(\C)$, instead of Lemma \ref{lem:lempourbianchipair} one must use the identity
	$$\frac12\Real(\overline{V}(\divc U))+\frac12\Real(\overline{U}\cdot\mcd V)=\D^\mu\Real(\overline{V}U)_\mu-(\eta_\mck+\etabar)\cdot\Real(\overline{V}U),$$
	for $U\in\fraks_1(\C)$, $V\in\fraks_0(\C)$, which is a direct consequence of Lemmas \ref{lem:lempourbianchipairpreli} and \ref{lem:2140waveq}.
\end{proof}
Before obtaing integral bounds for general $\df$ derivatives of curvature components $\Rcheck$, we first prove the result for less general  $\df'=\{\nabla_3,\nabla_4,\barre{\df}\}$ derivatives defined in Definition \ref{defi:dfprimhodgeI}.
\begin{prop}\label{prop:controlcurvatureprim}
	For $k\leq N_0$, $p\geq p_0(a,M,\delta_+,\delta_-)$ large enough, and  $\wbar_0\leq\wbar_1\leq\wbar_2\leq\wbar_*$, we have the estimate
	\begin{align*}
		&\sum_{\phi\in\widecheck{R}}\left(\int_{\un\cap\{\wbar=\wbar_2\}}r^p|\df'^{\leq k}\phi|^2+\iint_{\un\cap\{\wbar_1\leq\wbar\leq\wbar_2\}}pr^{p-1}|\df'^{\leq k}\phi|^2\right)\lesssim\\
		&C'_{\delta_\pm}\Bigg(\sum_{\phi\in\widecheck{R}}\int_{\un\cap\{\wbar=\wbar_1\}}r^p|\df'^{\leq k}\phi|^2+\int_{\mca\cap\{\wbar_1\leq\wbar\leq\wbar_2\}}r^p|{\df}^{\leq k}(\widecheck{\Gamma},\widecheck{R})|^2+\iint_{\un\cap\{\wbar_1\leq\wbar\leq\wbar_2\}}r^p|\df^{\leq k}(\Gammacheck,\Rcheck)|^2\Bigg),
	\end{align*}
where $C'_{\delta_\pm}$ is a constant which only depends on $a,M,\delta_\pm$.
\end{prop}

\begin{proof}
This is a direct consequence of Propositions \ref{prop:bianchieqcommutedI} and \ref{prop:bianchipairautresestimee}\footnote{Note that Proposition \ref{prop:bianchieqcommutedI} ensures that for any $\df'^k$ derivative of the Bianchi pairs, one of the assumptions of Proposition \ref{prop:bianchipairautresestimee} is satisfied.}.
\end{proof}
\noindent\textbf{Recovering integrated bounds for $\df^{\leq k}$ derivatives of $\Rcheck$}
\begin{defi}\label{defi:pouretrebienahah}
	For $0\leq j\leq N_0$, we denote
	$$\wt{\df}_j=\{P\in(\nabla,\df',\lieT,\lie_\Z)^{\leq N_0}\:/\:\text{there are }\leq j\text{ operators }\nabla\text{ in the factors of }P\},$$
	namely $\wt{\df}_j$ is the set of mixed products of $(\nabla,\df',\lieT,\lie_\Z)^{\leq N_0}$ where $\nabla$ appears $\leq j$ times.
\end{defi}
\begin{lem}\label{lem:cletechnique}
	Let $0\leq j\leq N_0$ and $\phi\in(\df',\lieT,\lie_\Z)^{N_0-j}\Rcheck$. Then recalling Definition \ref{defi:verscncersdefi} we have
	$$\nabla^j \phi=(\nablaslash^j \phi^S)^H+S_j[\phi],$$
	where $S_j[\phi]$ satisfies
	$$|S_j[\phi]|\lesssim |\wt{\df}_{j-1}\Rcheck|+|\df^{\leq N_0}\Gammacheck|+\sqrt\varepsilon|\df^{\leq N_0}\Rcheck|.$$
\end{lem}
\begin{proof}
	Let $U$ be a general horizontal tensor for now. Then by Proposition \ref{prop:premderchgtframeI}, noticing that \eqref{eq:retransfoPhiI} implies
	$$F_{bc}e_c+\lambda_b e_3+\lambdabar_b e_4=\barre{e}_b-e_b=-a j_b \T+F_\mck(r,\theta)j_b\Z+(\lambda_1,\lambda_2,\widecheck{\Theta})\df,$$
	we get for any set of indices $A$,
	\begin{align*}
		\nablaslash_b U^S_A=_s\nabla_b U_A-a j_b\nabla_\T U_A+j_b F_\mck(r,\theta)\nabla_\Z U_A+\mct\cdot U+(\lambda_1,\lambda_2,\widecheck{\Theta})\df U.
	\end{align*}
Hence, using Lemma \ref{lem:relatnablalieTZ} and recalling \eqref{eq:conversetransfoI}, we infer
\begin{align}\label{eq:iteratingesqredcc}
	\nabla U=(\nablaslash U^S)^H+S[U],
\end{align}
where
\begin{align}\label{eq:defidesdeUun}
	S[U]=_sa j\otimes\lieT U-F_\mck(r,\theta)j\otimes \lie_\Z U+O(1)_\mck\Gammacheck^{\leq 1}\cdot U+\mct\cdot U+(\lambda_1,\lambda_2,\widecheck{\Theta})\df U.
\end{align}
Then, iterating the formula \eqref{eq:iteratingesqredcc} $j$ times, for $j\geq 1$ we obtain the recurrence formula
\begin{align}\label{eq:recurrenceformuladerrrun}
	\nabla^j U&=(\nablaslash^j U^S)+S_j[U],\\
	S_j[U]&=S[\nabla^{j-1}U-S_{j-1}[U]]+\nabla (S_{j-1}[U]),\nn
\end{align}
where by convention $S_{0}[U]=0$. Now, for $\phi\in(\df',\lieT,\lie_\Z)^{N_0-j}\Rcheck$ we get
$$\nabla^j\phi=(\nablaslash^j \phi^S)+S_j[\phi]$$
where the induction formula for $S_j$ in \eqref{eq:recurrenceformuladerrrun} and the bootstrap assumption shows that $S_j[\phi]$ is bounded by a number $\leq j-1$ of $\nabla$ derivatives of $\Rcheck$, plus an error term coming from the derivatives of the term $$O(1)_\mck\Gammacheck\cdot U+(\lambda_1,\lambda_2,\widecheck{\Theta})\df U$$
in \eqref{eq:defidesdeUun} which is thus bounded by $\sqrt\varepsilon|\df^{\leq N_0}(\Rcheck,\Gammacheck)|$, and also some $\df^{\leq N_0}$ derivatives of $\Gammacheck$ because of the factors $j,O(1)_\mck$ in \eqref{eq:defidesdeUun}. In other words, recalling Definition \ref{defi:pouretrebienahah} we have
$$|S_j[\phi]|\lesssim |\wt{\df}_{j-1}\Rcheck|+|\df^{\leq N_0}\Gammacheck|+\sqrt\varepsilon|\df^{\leq N_0}\Rcheck|,$$
which concludes the proof.
\end{proof}

\begin{lem}\label{lem:cctcooloui}
We have for $(r,\ubar)\in\mcb[\wbar_*]$,
\begin{align}\label{eq:estcequeholds}
	\sum_{\phi\in\Rcheck}\int_{S(r,\ubar)}|\df^{\leq N_0}\phi|^2\lesssim &\int_{S(r,\ubar)}|(\df',\lieT,\lie_\Z)^{\leq N_0}\Rcheck|^2+\int_{S(r,\ubar)}|\df^{\leq N_0}\Gammacheck|^2.
\end{align}
\end{lem}
\begin{proof}
	We prove by induction on $0\leq i\leq N_0$ the bound
\begin{align}
	\sum_{\phi\in\Rcheck}\int_{S(r,\ubar)}|\wt{\df}_i\phi|^2\lesssim&\int_{S(r,\ubar)}|(\df',\lieT,\lie_\Z)^{\leq N_0}\Rcheck|^2+\int_{S(r,\ubar)}|\df^{\leq N_0}\Gammacheck|^2+\varepsilon\int_{S(r,\ubar)}|\df^{\leq N_0}\Rcheck|^2.\label{eq:horriblle}
\end{align}
It clearly holds for $i=0$ where the two last terms on the RHS above are not needed. Let us assume that it holds at rank $i-1\leq N_0-1$ and we prove that it holds at rank $i$. Let $\phi\in(\df',\lieT,\lie_\Z)^{\leq N_0-i}\Rcheck$. Then, by Lemma \ref{lem:cletechnique} we have 
\begin{align}
	\int_{S(r,\ubar)}|\nabla^i\phi|^2\lesssim& \int_{S(r,\ubar)}|\nablaslash^i\phi^S|^2+\int_{S(r,\ubar)}|\wt{\df}_{i-1}\Rcheck|^2+\int_{S(r,\ubar)}|\df^{\leq N_0}\Gammacheck|^2+\varepsilon\int_{S(r,\ubar)}|\df^{\leq N_0}\Rcheck|^2\nn\\
	\lesssim &\int_{S(r,\ubar)}|\nablaslash^i\phi^S|^2+\int_{S(r,\ubar)}|(\df',\lieT,\lie_\Z)^{\leq N_0}\Rcheck|^2+\int_{S(r,\ubar)}|\df^{\leq N_0}\Gammacheck|^2+\varepsilon\int_{S(r,\ubar)}|\df^{\leq N_0}\Rcheck|^2,\label{eq:encorerandocmob}
\end{align}
where we used the induction assumption \eqref{eq:horriblle} at rank $j-1$ to bound the term $\int_{S(r,\ubar)}|\wt{\df}_{i-1}\Rcheck|^2$. It is only left to bound the term $\int_{S(r,\ubar)}|\nablaslash^i\phi^S|^2$, which we do by using the elliptic estimates on the spheres of Lemma \ref{lem:ellipticIun}, which yield
\begin{align}
	\int_{S(r,\ubar)}|\nablaslash^{i}\phi^S|^2&\lesssim \int_{S(r,\ubar)}\left(|\shl{\triangle}{7.3}^{\leq i/2}\phi^S|^2+|\shl{\mcd}{6.6}_p^{\leq 1}\shl{\triangle}{7.3}^{\leq i/2-1}\phi^S|^2\right)+\sum_{i_1+i_2\leq i-2}\int_{S(r,\ubar)}|\df^{i_1+2}(\Gammacheck,\Rcheck)|^2|\nablaslash^{i_2}\phi^S|^2,\nn\\
	\int_{S(r,\ubar)}|\nablaslash^{i}\phi^S|^2&\lesssim \int_{S(r,\ubar)}|\shl{\mcd}{6.6}_p^{\leq 1}\shl{\triangle}{7.3}^{\leq \frac{i-1}{2}}\phi^S|^2+\sum_{i_1+i_2\leq i-2}\int_{S(r,\ubar)}|\df^{i_1+2}(\Gammacheck,\Rcheck)|^2|\nablaslash^{i_2}\phi^S|^2,\label{eq:nowwedealyy}
\end{align}
where the first bound above corresponds to the case where $i$ is even and the second bound to the case where $i$ is odd. In any case, using $i\leq N_0-2$, the bootstrap assumption and Lemma \ref{lem:cletechnique} to express $\nablaslash^{i_2}\phi^S$ with $\df^{i_2}\phi^S$ we get
\begin{align}\label{eq:randocombinees}
	\sum_{i_1+i_2\leq i-2}\int_{S(r,\ubar)}|\df^{i_1+2}(\Gammacheck,\Rcheck)|^2|\nablaslash^{i_2}\phi^S|^2\lesssim \int_{S(r,\ubar)}|\df^{\leq N_0}\Gammacheck|^2+\varepsilon\int_{S(r,\ubar)}|\df^{\leq N_0}\Rcheck|^2
\end{align}
where we treated both cases: if $i_1+2\geq N_0/2$ we use the boostrap assumption for the term $\nablaslash^{i_2}\phi^S$, and if $i_2\geq N_0/2$ we use the bootstrap assumption for the term $\df^{i_1+2}(\Gammacheck,\Rcheck)$. Now we bound the first terms on the RHS of \eqref{eq:nowwedealyy}. Using Lemma \ref{lem:cletechnique} again and evaluating on the components $a_1a_1\cdots a_ma_m A$ for any components $A$ we get for $m\leq i/2$,
$$(\shl{\triangle}{7.3}^m \phi^S)^H=\triangle^m \phi+O\left(|\wt{\df}_{2m-1}\Rcheck|+|\df^{\leq N_0}\Gammacheck|+\varepsilon|\df^{\leq N_0}\Rcheck|\right),$$
and similarly recalling \eqref{eq:defdesmcdpla}, treating each case $p=0,1,2$, for $m\leq \frac{i-1}{2}$,
$$(\shl{\mcd}{6.6}_p\shl{\triangle}{7.3}^m \phi^S)^H=\mcd_p\triangle^m \phi+O\left(|\wt{\df}_{2m}\Rcheck|+|\df^{\leq N_0}\Gammacheck|+\varepsilon|\df^{\leq N_0}\Rcheck|\right),$$
where we notice that by Lemmas \ref{lem:unlemsympa}, \ref{lem:unlemsympascalair}, \ref{lem:laplaciens1(C)}, \ref{lem:ignocoucou}, we have 
$$\triangle^m \phi=_{rs}\df'^{\leq N_0}\Rcheck+\df^{\leq N_0}\Gammacheck,\quad \mcd_p\triangle^m \phi=_{rs}\df'^{\leq N_0}\Rcheck+\df^{\leq N_0}\Gammacheck.$$
Thus, combining \eqref{eq:nowwedealyy}, \eqref{eq:randocombinees} with \eqref{eq:encorerandocmob} and using the induction assumption at rank $i-1$ again we get
\begin{align*}
	\sum_{\phi\in\Rcheck}\int_{S(r,\ubar)}|\nabla^i(\df',\lieT,\lie_\Z)^{\leq N_0-i}\phi|^2\lesssim&\int_{S(r,\ubar)}\left(|(\df',\lieT,\lie_\Z)^{\leq N_0}\Rcheck|^2+|\df^{\leq N_0}\Gammacheck|^2+\varepsilon|\df^{\leq N_0}\Rcheck|^2\right).
\end{align*}
Commuting $\nabla^i$ with $(\df',\lieT,\lie_\Z)^{\leq N_0-i}$ which creates lower-order terms  by Lemma \ref{lem:commutnablageneral} and Proposition \ref{prop:gaussequation}, that are treated by the induction assumption, we deduce that \eqref{eq:horriblle} holds at rank $i$, concluding the induction. Finally, using \eqref{eq:horriblle} at rank $N_0$ we get
\begin{align*}
	\sum_{\phi\in\Rcheck}\int_{S(r,\ubar)}|\df^{\leq N_0}\phi|^2\lesssim &\int_{S(r,\ubar)}|(\df',\lieT,\lie_\Z)^{\leq N_0}\Rcheck|^2+\int_{S(r,\ubar)}|\df^{\leq N_0}\Gammacheck|^2+\varepsilon\int_{S(r,\ubar)}|\df^{\leq N_0}\Rcheck|^2,
\end{align*}
hence the proof choosing $\varepsilon(a,M)>0$ small enough such that the last term on the RHS above is absorbed in the LHS.
\end{proof}
In the next result, we prove that we can control additional $\lie_\T,\lie_\Z$ derivatives of $\Rcheck$.
\begin{prop}\label{prop:controlcurvatureprimlieTZ}
	For $k\leq N_0$, $p\geq p_0(a,M,\delta_+,\delta_-)$ large enough, and  $\wbar_0\leq\wbar_1\leq\wbar_2\leq\wbar_*$,
	\begin{align*}
		&\sum_{\phi\in\widecheck{R}}\left(\int_{\un\cap\{\wbar=\wbar_2\}}r^p|(\df',\lieT,\lie_\Z)^{\leq k}\phi|^2+\iint_{\un\cap\{\wbar_1\leq\wbar\leq\wbar_2\}}pr^{p-1}|(\df',\lieT,\lie_\Z)^{\leq k}\phi|^2\right)\lesssim\\
		&C'_{\delta_\pm}\Bigg(\sum_{\phi\in\widecheck{R}}\int_{\un\cap\{\wbar=\wbar_1\}}r^p|\df^{\leq k}\phi|^2+\int_{\mca\cap\{\wbar_1\leq\wbar\leq\wbar_2\}}r^p|{\df}^{\leq k}(\widecheck{\Gamma},\widecheck{R})|^2+\iint_{\un\cap\{\wbar_1\leq\wbar\leq\wbar_2\}}r^p|\df^{\leq k}(\Gammacheck,\Rcheck)|^2\Bigg),
	\end{align*}
where $C'_{\delta_\pm}$ is a constant which only depends on $a,M,\delta_\pm$.
\end{prop}
\begin{proof}
This directly follows from commuting the identities of Proposition \ref{prop:bianchieqcommutedI} with $(\lieT,\lie_\Z)$. Indeed, by Lemma \ref{lem:commdflieTZI} the commutator terms are are quadratic error terms which can be treated easily, and hence we conclude using Proposition \ref{prop:bianchipairautresestimee}.
\end{proof}
Finally, we prove that we recover  the following control of $\df$ derivatives for $\Rcheck$.
\begin{prop}\label{prop:controlcurvatureprimdernier}
	For $k\leq N_0$, $p\geq p_0(a,M,\delta_+,\delta_-)$ large enough, and  $\wbar_0\leq\wbar_1\leq\wbar_2\leq\wbar_*$,
	\begin{align*}
		&\sum_{\phi\in\widecheck{R}}\left(\int_{\un\cap\{\wbar=\wbar_2\}}r^p|\df^{\leq k}\phi|^2+\iint_{\un\cap\{\wbar_1\leq\wbar\leq\wbar_2\}}pr^{p-1}|\df^{\leq k}\phi|^2\right)\lesssim\\
		&C'_{\delta_\pm}\Bigg(\int_{\un\cap\{\wbar=\wbar_1\}}r^p|\df^{\leq k}(\Gammacheck,\Rcheck)|^2+\int_{\mca\cap\{\wbar_1\leq\wbar\leq\wbar_2\}}r^p|{\df}^{\leq k}(\widecheck{\Gamma},\widecheck{R})|^2+\iint_{\un\cap\{\wbar_1\leq\wbar\leq\wbar_2\}}r^p|\df^{\leq k}(\Gammacheck,\Rcheck)|^2\Bigg).
	\end{align*}
	where $C'_{\delta_\pm}$ is a constant which only depends on $a,M,\delta_\pm$.
\end{prop}
\begin{proof}
By Lemma \ref{lem:cctcooloui} we have
 	\begin{align*}
 	&\sum_{\phi\in\widecheck{R}}\left(\int_{\un\cap\{\wbar=\wbar_2\}}r^p|\df^{\leq k}\phi|^2+\iint_{\un\cap\{\wbar_1\leq\wbar\leq\wbar_2\}}pr^{p-1}|\df^{\leq k}\phi|^2\right)\lesssim\\
 	&\int_{\un\cap\{\wbar=\wbar_2\}}r^p|((\df',\lieT,\lie_\Z)^{\leq N_0}\Rcheck,\df^{\leq N_0}\Gammacheck)|^2+\iint_{\un\cap\{\wbar_1\leq\wbar\leq\wbar_2\}}pr^{p-1}|((\df',\lieT,\lie_\Z)^{\leq N_0}\Rcheck,\df^{\leq N_0}\Gammacheck)|^2
 \end{align*}
 Thus by Propositions \ref{prop:controlcurvatureprimlieTZ} and \ref{prop:controlricci} we conclude the proof.
\end{proof}
\subsubsection{Recovering the bootstrap assumptions}\label{section:sectionalasuite6}
\noindent\textbf{Integral hypersurface and bulk estimates.}
\begin{prop}\label{prop:recoverinteg}
For $p\geq p_0(a,M,\delta_+,\delta_-)$ large enough, and  $\wbar_0\leq\wbar_1\leq\wbar_2\leq\wbar_*$, we have
	\begin{align*}
		&\sum_{\phi\in\Gammacheck\cup\Rcheck}\Bigg(\int_{\un\cap\{\wbar=\wbar_2\}}r^p|\df^{\leq N_0}\phi|^2+\iint_{\un\cap\{\wbar_1\leq\wbar\leq\wbar_2\}}pr^{p-1}|\df^{\leq N_0}\phi|^2\Bigg)\lesssim\\
		&C_{\delta_\pm}\left(\sum_{\phi\in\Gammacheck\cup\Rcheck}\int_{\un\cap\{\wbar=\wbar_1\}}r^p|\df^{\leq N_0}\phi|^2+\int_{\mca\cap\{\wbar_1\leq\wbar\leq\wbar_2\}}r^p|{\df}^{\leq k}(\widecheck{\Gamma},\widecheck{R})|^2\right),
	\end{align*}
	where $C_{\delta_\pm}$ is a constant which only depends on $a,M,\delta_\pm$.
\end{prop}
\begin{proof}
	By combining Propositions \ref{prop:controlricci} and \ref{prop:controlcurvatureprimdernier} we have
	\begin{align*}
	&\sum_{\phi\in\Gammacheck\cup\Rcheck}\Bigg(\int_{\un\cap\{\wbar=\wbar_2\}}r^p|\df^{\leq N_0}\phi|^2+\iint_{\un\cap\{\wbar_1\leq\wbar\leq\wbar_2\}}pr^{p-1}|\df^{\leq N_0}\phi|^2\Bigg)\lesssim\\
	&C_{\delta_\pm}\sum_{\phi\in\Gammacheck\cup\Rcheck}\left(\int_{\un\cap\{\wbar=\wbar_1\}}r^p|\df^{\leq N_0}\phi|^2+\int_{\mca\cap\{\wbar_1\leq\wbar\leq\wbar_2\}}r^p|{\df}^{\leq k}\phi|^2+\iint_{\un\cap\{\wbar_1\leq\wbar\leq\wbar_2\}}r^{p}|\df^{\leq N_0}\phi|^2\right).
\end{align*}
To conclude the proof, it is thus left to choose $p\geq p_0(a,M,\delta_+,\delta_-)$ large enough such that last term on the RHS above is absorbed in the second term on the LHS.
\end{proof}

\begin{prop}\label{prop:proploleee}
	We have for $\wbar_0\leq\wbar'\leq\wbar_*$,
	$$\sum_{\phi\in\Gammacheck\cup\Rcheck}\int_{\un\cap\{\wbar=\wbar'\}}|\df^{\leq N_0}\phi|^2\lesssim C_{\delta_\pm} \varepsilon^2.$$
\end{prop}
\begin{proof}
	Let us define
	\begin{align}\label{eq:jppenaimarre}
		F(\wbar'):=\sum_{\phi\in\Gammacheck\cup\Rcheck}\int_{\un\cap\{\wbar=\wbar'\}}|\df^{\leq N_0}\phi|^2.
	\end{align}
		Let $p=p_0(a,M,\delta_+,\delta_-)$ be large enough as in Proposition \ref{prop:recoverinteg}. Notice that we have $r_+^p\geq r^p\geq r_-^p$, hence there is a constant $C_{\delta_\pm}$ which depends on $a,M,\delta_\pm$ such that
	\begin{align*}
		&\sum_{\phi\in\Gammacheck\cup\Rcheck}\Bigg(\int_{\un\cap\{\wbar=\wbar_2\}}|\df^{\leq N_0}\phi|^2+\iint_{\un\cap\{\wbar_1\leq\wbar\leq\wbar_2\}}|\df^{\leq N_0}\phi|^2\Bigg)\lesssim\\
		&C_{\delta_\pm}\left(\sum_{\phi\in\Gammacheck\cup\Rcheck}\int_{\un\cap\{\wbar=\wbar_1\}}|\df^{\leq N_0}\phi|^2+\int_{\mca\cap\{\wbar_1\leq\wbar\leq\wbar_2\}}|{\df}^{\leq N_0}(\widecheck{\Gamma},\widecheck{R})|^2\right).
	\end{align*}
	Moreover, by the initial estimates \eqref{eq:initialhypoth} on $\mca$ we have
	$$\int_{\mca\cap\{\wbar_1\leq\wbar\leq\wbar_2\}}|{\df}^{\leq N_0}(\widecheck{\Gamma},\widecheck{R})|^2\lesssim \varepsilon^2\int_{\wbar_1}^{\wbar_2}\wbar^{-6-2\delta}\dee\wbar.$$
	Thus, by Lemma \ref{lem:volumeformI} this implies, for any $\wbar_0\leq\wbar_1\leq\wbar_2\leq\wbar_*$,
\begin{align}\label{eq:jemenfousahah}
	F(\wbar_2)+\int_{\wbar_1}^{\wbar_2}F(\wbar)\dee\wbar\lesssim C_{\delta_\pm}F(\wbar_1)+C_{\delta_\pm}\varepsilon^2\int_{\wbar_1}^{\wbar_2}\wbar^{-6-2\delta}\dee\wbar.
\end{align}	
In particular for $\wbar_1=\wbar_0$ and $\wbar_2=\wbar\in[\wbar_0,\wbar*]$, using $F(\wbar_0)=0$, this yields $F(\wbar)\lesssim C_{\delta_\pm}\varepsilon^2$, and hence concludes the proof.
\end{proof}
\noindent\textbf{Pointwise estimates.} We are now ready to improve the bootstrap assumptions \eqref{eq:BA}.
\begin{prop}\label{prop:bootstrapameliore}
	Assume that the bootstrap assumption is satisfied for $\wbar_*$ in the sense of Definition \ref{defi:bootstrapdansunI}. Then we have the following estimate in $\un[\wbar_*]$, 
	$$\sup_{\un[\wbar_*]}\left(|\df^{\leq N_0-3}(\Gammacheck,\Rcheck)|\right)\lesssim_{\delta_\pm} \varepsilon.$$
\end{prop}
\begin{proof}
Let $\phi\in(\Gammacheck,\Rcheck)$. Then by Lemma \ref{lem:transpoXX} we have in $\un[\wbar_*]$
	\begin{align*}
		|\df^{\leq N_0-3}\phi|^2(r,\ubar,\theta,\phi)\lesssim \|\df^{\leq N_0-3}\phi\|^2_{L^\infty(S(r_\mca,\ubar-r+r_\mca))}+\int_{\un[\wbar_*]\cap\{\wbar=\ubar-r\}} |\df^{\leq N_0}\phi|^2\lesssim C_{\delta_\pm}\varepsilon^2
	\end{align*}
	where in the last step we used  \eqref{eq:initialhypoth} and Proposition \ref{prop:proploleee}, which concludes the proof.
\end{proof}

\subsubsection{Decay estimates in region $\un$ (proof of Theorem \ref{thm:regionI})}\label{section:preuvepremiertheo}
Before proving Theorem \ref{thm:regionI}, we recall from \cite{scalarMZ} the following result which allows us to deduce polynomial energy decay from integrated energy decay.
\begin{lem}\label{lem:decay}
	Let $p>1$, $1\leq b<+\infty$, and let $f:[b,+\infty) \rightarrow[0,+\infty)$ be a continuous function. Assume that there are constants $C_0>0, C_1>0, C_2 \geq 0$, $C_3 \geq 0$ such that for any $b \leq x_1<x_2$,
	\begin{align}\label{eq:ineqassuemd}
		f(x_2)+C_1 \int_{x_1}^{x_2}f(x) \mathrm{d} x \leq C_0 f(x_1)+C_2 \int_{x_1}^{x_2} x^{-p} \mathrm{~d} x+C_3 x_1^{-p} .
	\end{align}
	Then for any $x \in [b,+\infty)$,
	$$
	f(x) \leq C x^{-p}
	$$
	where $C$ is a constant that depends only on $b,f(b), C_0, C_1, C_2, C_3$, and $p$. Moreover, for $x\geq 2^pb$ we also have $f(x)\leq C'x^{-p}$ with $C'$ depending only on $f(b), C_0, C_1, C_2, C_3$.
\end{lem}
\begin{proof}
	See \cite[Lemma 3.4]{scalarMZ} (which treats $b=1$, but the proof for $b\geq 1$ is the same). The last statement is a direct consequence of the proof of \cite[Lemma 3.4]{scalarMZ} for $b\geq 1$.
\end{proof}
\begin{proof}[\textbf{Proof of Theorem \ref{thm:regionI}}]
The proof is a standard bootstrap argument with respect to $\wbar_*$. We denote $\mathcal{I}$ the set of all $\wbar_*>\wbar_0$ such that the bootstrap assumption holds for $\wbar_*$ in the sense of Definition \ref{defi:bootstrapdansunI}. We then prove that $\mathcal{I}=(\wbar_0,+\infty)$ by proving that $\mathcal{I}$ is non-empty, closed, and open, for $\varepsilon(a,M,\delta_\pm)>0$ small enough. This is standard so we omit the details. Roughly speaking, $\mci$ is non-empty by the Choquet-Bruhat theorem \cite{YCB52} and continuity, $\mci$ is closed by construction, and $\mci$ is open for $\varepsilon(a,M,\delta_\pm)>0$ small enough by Proposition \ref{prop:bootstrapameliore}, the Choquet-Bruhat theorem \cite{YCB52}, and continuity, since we choose $N_0\geq 6$ which yields $N_0/2\leq N_0-3$. This proves that the spacetime extends in all of region $\un$ where it satisfies \eqref{eq:BA}. To conclude the proof of Theorem \ref{thm:regionI}, it is only left to upgrade this bound to $\varepsilon\ubar^{-3-\delta/2}$ polynomial decay. This follows from \eqref{eq:jemenfousahah} which implies, together with Lemma \ref{lem:decay} with $p=6+2\delta$, $b=\wbar_0$, $f=F$, and where $C_{0,1,2,3}$ only depend on $a,M,\delta_+,\delta_-$ in this context\footnote{Note that here by \eqref{eq:wbarzeroinitial} we have $\wbar_0=2r^*_\mca-2r_\mca-w_f\geq 1$ for $w_f(a,M,\delta_\pm)$ negative enough, and also $f(b)=F(\wbar_0)=0$.},
\begin{align}\label{eq:bonbaneocorejamaisrecite}
	F(\wbar)\leq C'_{\delta_\pm}\varepsilon^2\wbar^{-6-2\delta},
\end{align}
for $\wbar$ large enough, where $C'_{\delta_\pm}$ only depends on $a,M,\delta_\pm$. Thus choosing $w_f(a,M,\delta,\delta_+,\delta_-)\ll -1$ negative enough, we get $F(\wbar)\leq \varepsilon^2\wbar^{-6-\delta}$ by \eqref{eq:bonbaneocorejamaisrecite}, and recalling \eqref{eq:jppenaimarre} we conclude the proof that $|\df^{\leq N_1-3}(\Gammacheck,\Rcheck)|\lesssim\varepsilon\ubar^{-3-\delta/2}$ exactly like in the proof of Proposition \ref{prop:bootstrapameliore}.
\end{proof}
\begin{rem}\label{rem:defN111}
	\textbf{In the rest of Section \ref{section:regionun}, we consider a fixed $\delta_->0$ (which will be chosen small enough later) and associated negative enough $w_f(a,M,\delta,\delta_+,\delta_-)\ll -1$ and small enough $\varepsilon(a,M,\delta_\pm)>0$ given by Theorem \ref{thm:regionI}. We summarize below for future reference the bounds already proven in region $\un$: denoting $N_1:=N_0-3$,}
\begin{align}\label{eq:controlI}
		\sup_{\un}\Bigg(\ubar^{3+\delta/2}\sum_{\phi\in\Gammacheck\cup\Rcheck}|\df^{\leq N_1}\phi|\Bigg)\lesssim{\varepsilon}.
\end{align} 
\end{rem}

\subsection{Analysis of Teukolsky equation in $\un$ (proof of Theorem \ref{thm:AregionI})}\label{section:teukolskydansIun}
We now analyse the Teukolsky equation by an energy method and obtain the precise asymptotics of the PT curvature component $A$ in $\un$.
\subsubsection{Preliminaries for the energy method in $\un$}

\noindent\textbf{Integration by parts identities in $\un$}

\begin{prop}\label{prop:Iippnab3}
	Let ${U}$ be a horizontal tensor and $p>1$. We have the identities
	\begin{align*}
		\Real\left(r^p\overline{{\nabla_3}{{U}}}\cdot\nabla_4{\nabla_3}{{U}}\right)&=\frac{1}{2}\Big[{r^p}(2\omega-tr\chi)-e_4(r^p)\Big]\left|{{\nabla_3}{{U}}}\right|^2+\D_\mu\left(\frac{r^p}{2}\left|{{\nabla_3}{{U}}}\right|^2e_4^\mu\right),\\
		\Real\left(r^p\overline{{\nabla_4}{{U}}}\cdot\nabla_3{\nabla_4}{{U}}\right)&=-\frac{1}{2}\Big[{r^p}tr\chibar+e_3(r^p)\Big]\left|{{\nabla_4}{{U}}}\right|^2+\D_\mu\left(\frac{r^p}{2}\left|{{\nabla_4}{{U}}}\right|^2e_3^\mu\right).
	\end{align*}
\end{prop}
\begin{proof}
	For the first identity we compute 
	\begin{align*}
		\Real\left(r^p\overline{{\nabla_3}{{U}}}\cdot\nabla_4{\nabla_3}{{U}}\right)&=\frac{r^p}{2}\nabla_4\left(\left|{{\nabla_3}{{U}}}\right|^2\right)=-\frac{1}{2}\Big[{r^p}\D_\mu e_4^\mu+e_4(r^p)\Big]\left|{{\nabla_3}{{U}}}\right|^2+\D_\mu\left(\frac{r^p}{2}\left|{{\nabla_3}{{U}}}\right|^2e_4^\mu\right),\nn
	\end{align*}
	and we conclude using $\D_\mu e_4^\mu=tr\chi-2\omega$. The proof of the second identity is similar, exchanging the roles of $e_3$ and $e_4$ and using $\omegabar=0$.
\end{proof}

\begin{prop}\label{prop:Iipphoriz}
	Let ${U}$ be a horizontal $k$-tensor and $p>1$. Then, for any vector field $X$,
	\begin{align*}
		\Real\left({r^p}\overline{\nabla_X{U}}\cdot(-\triangle_k{U})\right)=&-\frac{1}{2}\left({r^p}\D_\mu X^\mu+\nabla_X({r^p})\right)|\nabla{U}|^2-\D^\mu({r^p}\nu^{(X)}_\mu[U])\\
		&+\D_\mu\left(\frac{r^p}{2}|{\nabla}{U}|^2X^\mu\right)-r^p\Real\left(\overline{[\nabla_X,{\nabla^a}]{U}}\cdot{\nabla}_a{U}\right)\\
		&+{r^p}(\eta+\etabar)\cdot\nu^{(X)}[U]+\nabla(r^p)\cdot \nu^{(X)}[U],
	\end{align*}
	where we define the spacetime 1-form $\nu^{(X)}[U]$ by, for $a=1,2$, 
	$$\nu^{(X)}_a[U]=\Real\left(\overline{\nabla_X{U}}\cdot{\nabla}_a{U}\right),\quad\nu^{(X)}_3[U]=\nu^{(X)}_4[U]=0.$$
\end{prop}
\begin{proof}
	Integrating by parts we get
	\begin{align*}
		-\Real(r^p\overline{\nabla_X{U}}\cdot\triangle_k{U})= -\Real(r^p\overline{\nabla_X{U}}\cdot\nabla^a{\nabla}_a{U})&=-\nabla^a({r^p}\nu^{(X)}[U]_a)+\nabla^a(r^p)\nu^{(X)}[U]_a\\
		&\quad+r^p\overline{\nabla^a\nabla_X{U}}\cdot{\nabla}_a{U}.
	\end{align*}
	Thus by Lemma \ref{lem:2140waveq},
	\begin{align*}
		-\Real(r^p\overline{\nabla_X{U}}\cdot\triangle_k{U})&=-\D^\mu({r^p}\nu^{(X)}_\mu[U])+{r^p}(\eta+\etabar)\cdot\nu^{(X)}[U]\\
		&\quad+\nabla^a(r^p)\nu^{(X)}[U]_a+r^p\Real\left(\overline{\nabla^a\nabla_X{U}}\cdot{\nabla}_a{U}\right).
	\end{align*}
	Next, we deal with the following term,
	\begin{align*}
		r^p\Real\left(\overline{\nabla^a\nabla_X{U}}\cdot{\nabla}_a{U}\right)&=\frac{r^p}{2}\nabla_X(|{\nabla}{U}|^2)-r^p\Real\left(\overline{[\nabla_X,{\nabla^a}]{U}}\cdot{\nabla}_a{U}\right),
	\end{align*}
	and we write
	\begin{align*}
		\frac{r^p}{2}\nabla_X(|{\nabla}{U}|^2)&=\nabla_X\left(\frac{r^p}{2}|{\nabla}{U}|^2\right)-\nabla_X\left(\frac{r^p}{2}\right)|{\nabla}{U}|^2\\
		&=\D_\mu\left(\frac{r^p}{2}|{\nabla}{U}|^2X^\mu\right)-\frac{r^p}{2}|{\nabla}{U}|^2\D_\mu X^\mu-\nabla_X\left(\frac{r^p}{2}\right)|{\nabla}{U}|^2,
	\end{align*}
	which concludes the proof.
\end{proof}

\begin{defi}
	We define the following short-hand notations for $a=1,2$,
	\begin{align}\label{eq:Ishorthandnotation}
		\nu^{(3)}_a[U]:=\nu^{(e_3)}_a[U]=\Real\left(\overline{\nabla_3{U}}\cdot{\nabla}_a{U}\right),\quad \nu^{(4)}_a[U]:=\nu^{(e_4)}_a[U]=\Real\left(\overline{\nabla_4{U}}\cdot{\nabla}_a{U}\right).
	\end{align}
\end{defi}
\noindent\textbf{Energy and Hardy-type inequality in $\un$}
\begin{defi}
	For $U$ a horizontal tensor in $\un$ and $\wbar'\geq \wbar_0$, $p\geq 1$, we define the energies
	\begin{align}
		\mathbf{E}_\un[U](\wbar')&:=\int_{\un\cap\{\wbar=\wbar'\}}\left(|\nabla_3 U|^2+|\nabla_4 U|^2+|\nabla U|^2\right),\label{eq:Idefenerdeg}\\
		\mathbf{E}_\un^{(p)}[U](\wbar')&:=\int_{\un\cap\{\wbar=\wbar'\}}r^p\left(|\nabla_3 U|^2+|\nabla_4 U|^2+|\nabla U|^2\right).\label{eq:Idefenerdeg(p)}
	\end{align}
\end{defi}
We now prove a control of the $L^2(S(r,\ubar))$ norm with $\Eun$ defined above and initial data.
\begin{prop}\label{prop:controlL2avecener}
	Let $U$ be a horizontal tensor and $S(r,\ubar)\subset\un$. Then we have the estimate
	$$\|U\|_{L^2(S(\ubar,r))}^2\lesssim \|U\|^2_{L^2(S(\ubar-r+r_\mca,r_\mca))}+\Eun[U](\ubar-r).$$
\end{prop}

\begin{proof}
We compute in coordinates along the hypersurface $\{\wbar=\wbar_1\}$ with $\wbar_1:=\ubar-r$,
\begin{align*}
	\left|\frac{\dee}{\dee r}\|U\|_{L^2(S(\wbar_1+r,r))}^2\right|&\lesssim\sum_{i=1}^3\left|\frac{\dee}{\dee r}\left(\int_{\mcu^{(i)}_{\mathbb{S}^2}}|U|^2\sqrt{\det\gslash_{x^j_{(i)}}}(\wbar_1+r,r,x^1_{(i)},x^2_{(i)})\dee x^1_{(i)}\dee x^2_{(i)}\right)\right|\nn\\
	&\lesssim\sum_{i=1}^3\left|\int_{\mcu^{(i)}_{\mathbb{S}^2}}\left(\partial_r+\partial_\ubar\right)\left[|U|^2\sqrt{\det\gslash_{x^j_{(i)}}}\right](\wbar_1+r,r,x^1_{(i)},x^2_{(i)})\dee x^1_{(i)}\dee x^2_{(i)}\right|,
\end{align*}
with $\det\gslash_{x^j_{(i)}}$ the determinant of $\gslash$ in coordinates $(x^1_{(i)},x^2_{(i)})$. Using Lemma \ref{lem:gslashcheckIborne} which implies
\begin{align*}
	&\sqrt{\det\gslash_{x^j_{(i)}}}\sim 1,\\
	&|\df\sqrt{\det\gslash_{x^j_{(i)}}}|\lesssim |\df{\det\gslash_{x^j_{(i)}}}|\lesssim \max_{j,k=1,2}|\df (\gslash_\mck)_{x^j_{(i)}x^k_{(i)}}|+\max_{j,k=1,2}|\df\big(\gslash_{x^j_{(i)}x^k_{(i)}}-(\gslash_\mck)_{x^j_{(i)}x^k_{(i)}}\big)|\lesssim 1,
\end{align*}
by Proposition \ref{prop:expredmuavecenu} we thus get 
	\begin{align}
		\left|\frac{\dee}{\dee r}\|U\|_{L^2(S(\wbar_1+r,r))}^2\right|&\lesssim\left|\int_{S(\wbar_1+r,r)} U\cdot \df U\right|+\|U\|_{L^2(S(\wbar_1+r,r))}^2\nn\\
		&\lesssim \int_{S(\wbar_1+r,r)}\left(|\nabla_3 U|^2+|\nabla_4 U|^2+|\nabla U|^2\right)+\|U\|_{L^2(S(\wbar_1+r,r))}^2.\label{eq:derquejutapres}
	\end{align}
	Integrating from $r$ to $r_\mca$, and using Lemma \ref{lem:volumeformI} we deduce 
	\begin{align*}
		\|U\|_{L^2(S(\wbar_1+r,r))}^2\lesssim \|U\|^2_{L^2(S(\wbar_1+r_\mca,r_\mca))}+\Eun[U](\wbar_1)+\int_{r}^{r_\mca}\|U\|_{L^2(S(\wbar_1+r',r'))}^2\dee r'.
	\end{align*}
	Finally, using Grönwall's inequality we conclude the proof.
\end{proof}
Now, we prove a Hardy-type inequality which will allow us to absorb the 0-order terms in the energy method for the Teukolsky equation.
\begin{prop}[Hardy-type inequality in $\un$]\label{prop:hardydansI}
	Let $U$ be a horizontal tensor in $\un$. Then for $p\geq p_0(a,M)$ large enough, we have the following estimate for any $\wbar_1\geq\wbar_0$,
	\begin{align*}
		\int_{\un\cap\{\wbar=\wbar_1\}}r^p|U|^2\lesssim \frac{r_\mca^p}{p}\|U\|^2_{L^2(S(\wbar_1+r_\mca,r_\mca))}+\int_{\un\cap\{\wbar=\wbar_1\}}\frac{r^p}{p}\left(|\nabla_3 U|^2+|\nabla_4 U|^2+|\nabla U|^2\right).
	\end{align*}
\end{prop}
\begin{proof}
	We have the following identity in $\un$,
	\begin{align*}
		&\frac{\dee}{\dee r}\left(r^p\|U\|^2_{L^2(S(\wbar_1+r,r))}\right)=pr^{p-1}\|U\|^2_{L^2(S(\wbar_1+r,r))}+r^p \frac{\dee}{\dee r}\left(\|U\|^2_{L^2(S(\wbar_1+r,r))}\right)\\
		&=pr^{p-1}\|U\|^2_{L^2(S(\wbar_1+r,r))}+r^pO\left(\int_{S(\wbar_1+r,r)}\left(|\nabla_3 U|^2+|\nabla_4 U|^2+|\nabla U|^2+|U|^2\right)\right),
	\end{align*}
	where we used \eqref{eq:derquejutapres} for the second identity. Integrating between $r\leq r_\mca$ and $r_{\mathcal{A}}$ we get 
	\begin{align*}
		\int_r^{r_\mca}pr'^{p-1}\|U\|^2_{L^2(S(\wbar_1+r',r'))}\dee r'=&r_\mca^p\|U\|^2_{L^2(S(\wbar_1+r_\mca,r_\mca))}-r^p\|U\|^2_{L^2(S(\wbar_1+r,r))}\\
		&+\int_r^{r_\mca}r'^pO\left(\int_{S(\wbar_1+r',r')}\left(|\nabla_3 U|^2+|\nabla_4 U|^2+|\nabla U|^2+|U|^2\right)\right)\dee r'.
	\end{align*}
	Thus, choosing $p\geq p_0(a,M)$ large enough so that the term with $|U|^2$ in the last term on the RHS above is absorbed in the LHS, and dropping the non-positive term $-r^p\|U\|^2_{L^2(S(\wbar_1+r,r))}$, we get 
	\begin{align*}
		\int_r^{r_\mca}pr'^{p-1}&\|U\|^2_{L^2(S(\wbar_1+r',r'))}\dee r'\lesssim\\
		&r_\mca^p\|U\|^2_{L^2(S(\wbar_1+r_\mca,r_\mca))}+\int_r^{r_\mca}r'^p\int_{S(\wbar_1+r',r')}\left(|\nabla_3 U|^2+|\nabla_4 U|^2+|\nabla U|^2\right)\dee r'
	\end{align*}
	which implies the statement of the proposition, dividing by $p$ and using Lemma \ref{lem:volumeformI}.
\end{proof}

\subsubsection{Energy and decay estimates for generalized Teukolsky fields in $\un$}

In this section, we assume that ${U}\in\fraks_k(\mathbb{C})$, $k\in\{0,1,2\}$, satisfies a wave equation
\begin{align}\label{eq:Iteukmod}
	\wh{\mcl}({U})=\err[\wh{\mcl}({U})],\quad\text{in}\:\:\un,
\end{align}
where $\wh{\mcl}$ is a Teukolsky-like wave operator, namely such that
\begin{align}\label{eq:IoperateurLhat}
	\wh{\mcl}({U})&=\nabla_4\nabla_3 {U}-\triangle_k{U}+h\nabla_3 {U}+\underline{h}\nabla_4{U}+L_1[U]+L[U],
\end{align}
where $h,\underline{h}$ are functions, $L_1$ is an first-order operator, and $L$ is a $0$-order operator (in a sense which will be clarified later). For $\wbar_0\leq\wbar_1\leq\wbar_2$, we recall the spacetime region $\un[\wbar_1,\wbar_2]:=\un\cap\{\wbar_1\leq\wbar\leq\wbar_2\}$ defined in \eqref{eq:gjhhjhjhj} (here, $\wbar_*=+\infty$). We also define the following hypersurface
$$\Sigma_1:=\un\cap\{r=r_-(1+\delta_-)\}.$$
Finally, we also denote $\wbar_{f}:=2r^*(r_-(1+\delta_-))-2r_-(1+\delta_-)-w_f$ the value of $\wbar$ at the intersection $\{w=w_f\}\cap\{r=r_-(1+\delta_-)\}$.
\begin{prop}\label{prop:Icalculcommunenergy}
	Let $\wh{\mcl}$ be given by \eqref{eq:IoperateurLhat},  ${U}\in\fraks_k(\mathbb{C})$, $k\in\{0,1,2\}$, and $\wbar_f\leq\wbar_1\leq\wbar_2$. We have the following identity, for $p>1$,
	\begin{align*}
		\int_{\un[\wbar_1,\wbar_2]}r^p \Real\left((\overline{\nabla_3{U}+\nabla_4{U}})\cdot\wh{\mcl}({U})\right)=&\int_{\un[\wbar_1,\wbar_2]}\mathbf{B}^{(p)}[U]+\int_{\un\cap\{\wbar=\wbar_2\}}\mathbf{F}_{\wbar}^{(p)}[U]-\int_{\un\cap\{\wbar=\wbar_1\}}\mathbf{F}_{\wbar}^{(p)}[U]\\
		&+\int_{\Sigma_1\cap\{\wbar_1\leq\wbar\leq\wbar_2\}}\mathbf{F}^{(p)}_{\Sigma_1}[U]-\int_{\mathcal{A}\cap\{\wbar_1\leq\wbar\leq\wbar_2\}}\mathbf{F}_{\mathcal{A}}^{(p)}[U],
	\end{align*}
	where the boundary terms satisfy the following estimates : for $a=1,2$, 
	\begin{align}
		&\int_{\un\cap\{\wbar=\wbar_a\}}\mathbf{F}^{(p)}_{\wbar}[U]\sim\int_{\un\cap\{\wbar=\wbar_a\}}r^p \left(|\nabla_3U|^2+|\nabla_4U|^2+|{\nabla} U|^2)\right),\label{eq:Iestimbdwbar}\\
		&\int_{\mathcal{A}\cap\{\wbar_1\leq\wbar\leq\wbar_2\}}\mathbf{F}_{\mathcal{A}}^{(p)}[U]\lesssim C_{\delta_\pm}\int_{\mathcal{A}\cap\{\wbar_1\leq\wbar\leq\wbar_2\}}r^p\left(|\nabla_3U|^2+|\nabla_4U|^2+|{\nabla} U|^2)\right),\label{eq:Iestimbdsigmazero}\\
		&\int_{{\Sigma_1}\cap\{\wbar_1\leq\wbar\leq\wbar_2\}}\mathbf{F}_{{\Sigma_1}}^{(p)}[U]\sim C'_{\delta_\pm} \int_{{\Sigma_1}\cap\{\wbar_1\leq\wbar\leq\wbar_2\}}r^p \left(|\nabla_3U|^2+|\nabla_4U|^2+|{\nabla} U|^2)\right),\label{eq:IestimbdGamma}
	\end{align}
where the constants $C_{\delta_\pm},C'_{\delta_\pm}>0$ only depend on $a,M,\delta_\pm$ and where, recalling \eqref{eq:ckdeu} and \eqref{eq:Ishorthandnotation}, the bulk term is
	\begin{align}
		\mathbf{B}^{(p)}[U]:=&\mathbf{B}_{pr}^{(p)}[U]+\nabla^a(r^p)\Real\left(\overline{\nabla_3U}\cdot\nabla_a U\right)+r^p\Big[-\Real\left(\overline{[\nabla_3,\nabla^a] U}\cdot\nabla_a U\right)\nn\\
		&+\Real\left(\underline{h}\overline{\nabla_3 U}\cdot\nabla_4 U\right)+\Real\left(\overline{\nabla_3 U}\cdot L_1[U]\right)+\Real\left(\overline{\nabla_3 U}\cdot L[U]\right)\Big]+r^p(\eta+\etabar)\cdot \nu^{(3)}[U]\nn\\
		&+\nabla^a(r^p)\Real\left(\overline{\nabla_4U}\cdot\nabla_a U\right)+r^p\Big[-\Real\left(\overline{[\nabla_4,\nabla^a] U}\cdot\nabla_a U\right)\nn\\
		&+\Real\left((h+2\omega)\overline{\nabla_4 U}\cdot\nabla_3 U\right)+\Real\left(\overline{\nabla_4U}\cdot( L_1[U]+2(\eta-\etabar)\cdot\nabla U)\right)\nn\\
		&+\Real\left(\overline{\nabla_4 U}\cdot (L[U]-C_k[U])\right)\Big]+r^p(\eta+\etabar)\cdot \nu^{(4)}[U],\label{eq:bulkIdef}
	\end{align}
	where the principal bulk term is defined as
	\begin{align}
		\mathbf{B}_{pr}^{(p)}[U]:=&\Big(\frac{1}{2}{{r^p}}(2\omega-tr\chi)-\frac{1}{2}e_4({r^p})+{r^p}\Real(h)\Big)\left|{{\nabla_3}{{U}}}\right|^2+\left(r^p \left(\Real(\underline{h})-\frac{1}{2}tr\chibar\right)-\frac{1}{2}e_3({r^p})\right)\left|{{\nabla_4}{{U}}}\right|^2\nn\\
		&-\frac{1}{2}\left(e_4({r^p})+e_3({r^p})+{r^p} (tr\chi -2\omega)+{r^p}tr\chibar\right)|\nabla{U}|^2.\label{eq:bulkprIdef}
	\end{align}
\end{prop}
\begin{proof}
	Using \eqref{eq:IoperateurLhat}, we have 
	\begin{align*}
		r^p \overline{\nabla_3{U}}\cdot\wh{\mcl}({U})=&r^p \overline{\nabla_3{U}}\cdot\Big(\nabla_4\nabla_3 {U}-\triangle_k{U}+h\nabla_3 {U}+\underline{h}\nabla_4{U}+L_1[U]+L[U]\Big).
	\end{align*}
	Using Propositions \ref{prop:Iippnab3} and \ref{prop:Iipphoriz}, and $\D_\mu e_3^\mu=tr\chibar$, $\D_\mu e_4^\mu=tr\chi-2\omega$, we get
	\begin{align}
		\Real\left(r^p \overline{\nabla_3{U}}\cdot\wh{\mcl}({U})\right)=&\Big(\frac{1}{2}{{r^p}}(2\omega-tr\chi)-\frac{1}{2}e_4({r^p})+{r^p}\Real(h)\Big)\left|{{\nabla_3}{{U}}}\right|^2\nn\\
		&-\frac{1}{2}\left({r^p}tr\chibar+ e_3({r^p})\right)|\nabla{U}|^2+\mathbf{B}_3[{U}]+\D_\mu\mathbf{F}_3[{U}]^\mu,\label{eq:Imult3}
	\end{align}
	where
	\begin{align*}
		\mathbf{B}_3[{U}]=&\nabla^a({r^p})\Real\left(\overline{\nabla_3U}\cdot\nabla_a U\right)+{r^p}\Big[-\Real\left(\overline{[\nabla_3,\nabla^a] U}\cdot\nabla_a U\right)+\Real\left(\underline{h}\overline{\nabla_3 U}\cdot\nabla_4 U\right)\\
		&+\Real\left(\overline{\nabla_3 U}\cdot L_1[U]\right)+\Real\left(\overline{\nabla_3 U}\cdot L[U]\right)\Big]+{r^p}(\eta+\etabar)\cdot \nu^{(3)}[U],
	\end{align*}
	and
	\begin{align*}
		\mathbf{F}_3[{U}]^\mu=\frac{{r^p}}{2}\left|{{\nabla_3}{{U}}}\right|^2e_4^\mu+\frac{r^p }{2}|{\nabla}{U}|^2e_3^\mu-{r^p}\nu^{(3)}[U]^\mu.
	\end{align*}
	Next, to compute $r^p \overline{\nabla_4{U}}\cdot\wh{\mcl}({U})$, we use Lemma \ref{lem:comm34} (and $\Xibar=0,\omegabar=0$ here) to get
	\begin{align*}
		[\nabla_4,\nabla_3]{{U}}=2\omega\nabla_3{{U}}+2(\etabar-\eta)\cdot\nabla{{U}}-C_k[U].
	\end{align*}
	This gives the new expression of \eqref{eq:IoperateurLhat}:
	\begin{align*}
		\wh{\mcl}({U})=&\nabla_3\nabla_4{U}-\triangle_k{U}+\parentheses{h+2\omega}\nabla_3 {U}+\underline{h} \nabla_4{U}+L_1[U]+2(\etabar-\eta)\cdot\nabla{{U}}+L[U]-C_k[U].
	\end{align*}
	Combining this with Propositions \ref{prop:Iippnab3} and \ref{prop:Iipphoriz} and $\D_\mu e_3^\mu=tr\chibar$, $\D_\mu e_4^\mu=tr\chi-2\omega$, yields 
	\begin{align}
		\Real\left(r^p \overline{\nabla_4{U}}\cdot\wh{\mcl}({U})\right)=&\left(r^p \left(-\frac{1}{2}tr\chibar+\Real(\underline{h})\right)-\frac{1}{2}e_3({r^p})\right)\left|{{\nabla_4}{{U}}}\right|^2\label{eq:Imult4}\\
		&-\frac{1}{2}\left({r^p} (tr\chi -2\omega) +e_4({r^p})\right)|\nabla{U}|^2+\mathbf{B}_4[{U}]+\D_\mu\mathbf{F}_4[{U}]^\mu,\nn
	\end{align}
	where
	\begin{align*}
		&\mathbf{B}_4[{U}]=\nabla^a({r^p})\Real\left(\overline{\nabla_4U}\cdot\nabla_a U\right)+{r^p}\Big[-\Real\left(\overline{[\nabla_4,\nabla^a] U}\cdot\nabla_a U\right)+\Real\left((h+2\omega)\overline{\nabla_4 U}\cdot\nabla_3 U\right)\\
		&+\Real\left(\overline{\nabla_4U}\cdot(L_1[U]+2(\etabar-\eta)\cdot\nabla{{U}})\right)+\Real\left(\overline{\nabla_4 U}\cdot (L[U]-C_k[U])\right)\Big]+{r^p}(\eta+\etabar)\cdot \nu^{(4)}[U],
	\end{align*}
	and
	\begin{align*}
		\mathbf{F}_4[{U}]^\mu=\frac{{r^p}}{2}\left|{{\nabla_4}{{U}}}\right|^2e_3^\mu+\frac{r^p }{2}|{\nabla}{U}|^2e_4^\mu-{r^p}\nu^{(4)}[U]^\mu.
	\end{align*}
	Combining \eqref{eq:Imult3} and \eqref{eq:Imult4} we get 
	\begin{align}
		\label{eq:Ionrevient}\int_{\un[\wbar_1,\wbar_2]}r^p \Real\left((\overline{\nabla_3{U}+\nabla_4{U}})\cdot\wh{\mcl}({U})\right)=&\int_{\un[\wbar_1,\wbar_2]}\mathbf{B}[{U}]+\int_{\un[\wbar_1,\wbar_2]}\D_\mu\mathbf{F}[{U}]^\mu,
	\end{align}
	where the bulk term is
	\begin{align*}
		\mathbf{B}^{(p)}[{U}]=&\frac{1}{2}\Big({{r^p}}(2\omega-tr\chi)-e_4({r^p})+2{r^p}\Real(h)\Big)\left|{{\nabla_3}{{U}}}\right|^2+\frac{1}{2}\left(r^p \left(2\Real(\underline{h})-tr\chibar)\right)-e_3({r^p})\right)\left|{{\nabla_4}{{U}}}\right|^2\\
		&-\frac{1}{2}\left(e_4({r^p})+e_3({r^p})+{r^p} (tr\chi -2\omega)+{r^p}tr\chibar\right)|\nabla{U}|^2+\mathbf{B}_3[U]+\mathbf{B}_4[U],
	\end{align*}
	and where the flux term is 
	\begin{align}
		\mathbf{F}[{U}]^\mu&=\mathbf{F}_3[{U}]^\mu+\mathbf{F}_4[{U}]^\mu\nn\\
		&=\frac{{r^p}}{2}\left(\left|{{\nabla_3}{{U}}}\right|^2e_4^\mu+\left|{{\nabla_4}{{U}}}\right|^2e_3^\mu+|{\nabla}{U}|^2(e_3^\mu+e_4^\mu)\right)-{r^p}(\nu^{(3)}[U]^\mu+\nu^{(4)}[U]^\mu).\label{eq:Ifluxtermm}
	\end{align}
	Now, by Stokes theorem for manifolds with corners, 
	\begin{align}\label{eq:Istokes}
		\int_{\un[\wbar_1,\wbar_2]}\D_\mu\mathbf{F}[{U}]^\mu=\int_{\partial\un[\wbar_1,\wbar_2]}\mathbf{F}[{U}]_\mu n^\mu,
	\end{align}
	where $n$ is the inwards pointing unit normal vector to the boundary $\partial\un[\wbar_1,\wbar_2]$, where we note that since $\wbar\geq\wbar_f$, the boundary of $\un[\wbar_1,\wbar_2]$ is composed of four spacelike hypersurfaces. More precisely, see Figure \ref{fig:wbarInl}, we have the decomposition
	\begin{align*}
		\partial\un[\wbar_1,\wbar_2]=\left(\un\cap\{\wbar=\wbar_1\}\right)\cup\left(\un\cap\{\wbar=\wbar_2\}\right)\cup \left({\Sigma_1}\cap\{\wbar_1\leq\wbar\leq\wbar_2\}\right)\cup \left(\mathcal{A}\cap\{\wbar_1\leq\wbar\leq\wbar_2\}\right).
	\end{align*}

\begin{figure}[h!]
	\centering
	\includegraphics[scale=0.4]{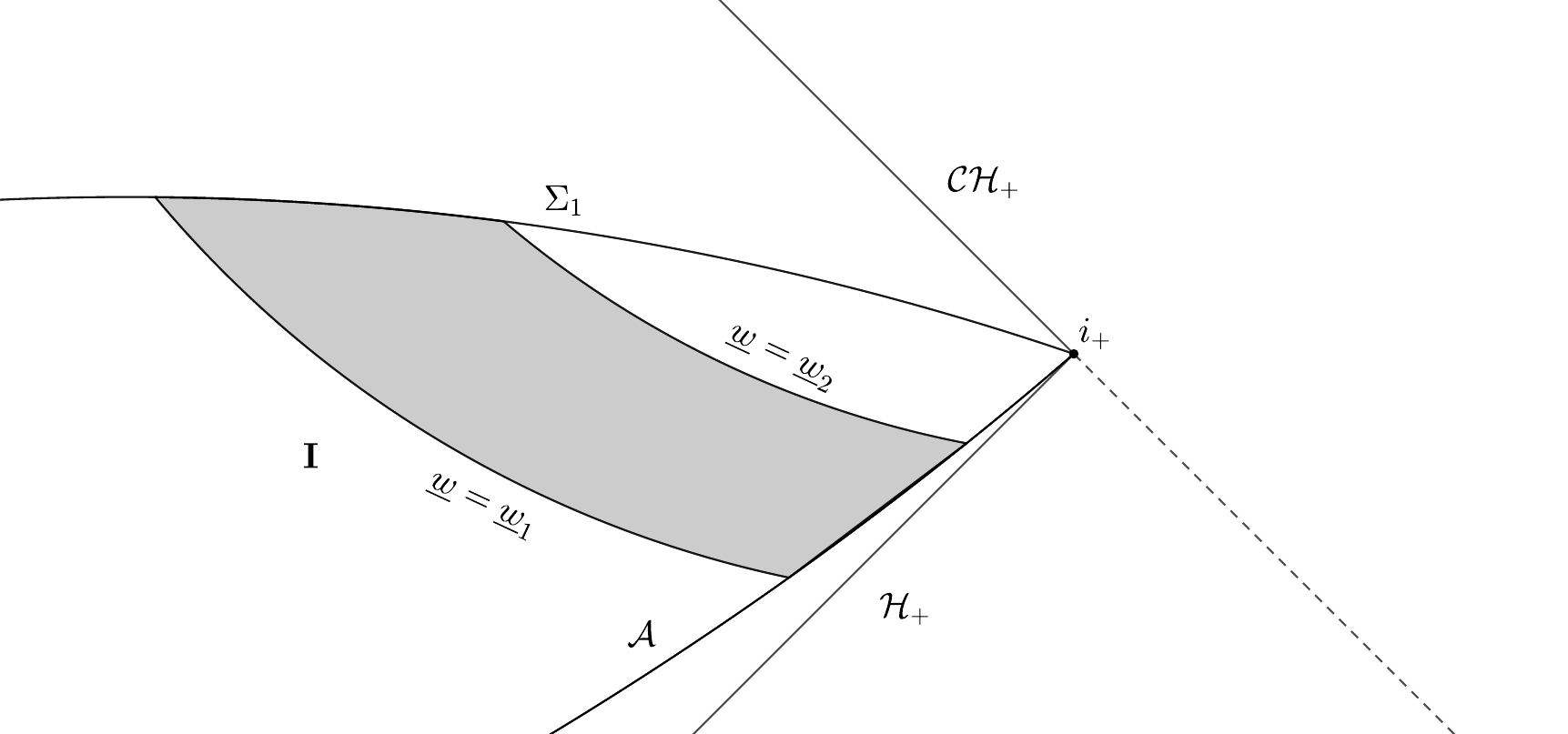}
	\caption{Region $\un[\wbar_1,\wbar_2]$ in grey, for $\wbar_1\geq\wbar_f$.}
	\label{fig:wbarInl}
\end{figure}

	\noindent\textbf{Boundary term on $\{\wbar=\wbar_a\}$, $a=1,2$.} A choice of unit normal vector for the hypersurface $\{\wbar=cst\}$ is given by 
	$$n_{\wbar}:=\frac{\D\wbar}{\sqrt{-\g(\D\wbar,\D\wbar)}},$$
	where we recall $-\g(\D\wbar,\D\wbar)\sim 1$ in $\un$. Also recall that we have $\D\wbar(\wbar)=\g(\D\wbar,\D\wbar)<0,$ which proves that $n_{\wbar}$ is outwards pointing (with respect to $\partial\un[\wbar_1,\wbar_2]$) at $\un\cap\{\wbar=\wbar_1\}$, and inwards pointing at $\un\cap\{\wbar=\wbar_2\}$. This proves that the boundary terms corresponding to $\left(\un\cap\{\wbar=\wbar_1\}\right)\cup\left(\un\cap\{\wbar=\wbar_2\}\right)$ in \eqref{eq:Istokes} is
	$$\mathcal{B}_{\wbar}:=\int_{\un\cap\{\wbar=\wbar_2\}}\mathbf{F}^{(p)}[U]_\mu n_{\wbar}^\mu-\int_{\un\cap\{\wbar=\wbar_1\}}\mathbf{F}^{(p)}[U]_\mu n_{\wbar}^\mu.$$
	Defining $\mathbf{F}_{\wbar}^{(p)}[U]:=\mathbf{F}^{(p)}[U]_\mu n_{\wbar}^\mu$, we deduce from \eqref{eq:Ifluxtermm} the identity
	\begin{align}
		\mathbf{F}_{\wbar}^{(p)}[U]=&\frac{{r^p}}{2}\left(|\nabla_3U|^2\g(e_4,n_{\wbar})+|\nabla_4U|^2\g(e_3,n_{\wbar})+|\nabla U|^2\g(e_3+e_4,n_{\wbar})\right)\nn\\
		&-{r^p}(\nu^{(3)}[U]+\nu^{(4)}[U])^c\g(e_c,n_{\wbar}).\label{eq:Icestuneexpre}
	\end{align}
	Moreover, using \eqref{eq:controlI} we have the following identities,
	\begin{align*}
		\g(e_4,\D\wbar)=e_4(\wbar)=\frac{2(r^2+a^2)-\Delta}{|q|^2}+O(\varepsilon),\quad\g(e_3,\D\wbar)=e_3(\wbar)=1,
	\end{align*}
which implies
	\begin{align}
		&\sqrt{-\g(\D\wbar,\D\wbar)}\frac{{r^p}}{2}\left(|\nabla_3U|^2\g(e_4,n_{\wbar})+|\nabla_4U|^2\g(e_3,n_{\wbar})+|\nabla U|^2\g(e_3+e_4,n_{\wbar})\right)\label{eq:Iprovesabso1}\\
		=&\frac{{r^p}}{2}\Bigg(\left(\frac{2(r^2+a^2)-\Delta}{|q|^2}+O(\varepsilon)\right)|\nabla_3U|^2+|\nabla_4 U|^2+\left(1+\frac{2(r^2+a^2)-\Delta}{|q|^2}+O(\varepsilon)\right)|\nabla U|^2\Bigg)\nn
	\end{align}
	We also have $\g(\D\wbar,e_c)=e_c(\wbar)=a\Real(\mathfrak{J})_c+O(\varepsilon)$, which yields
	\begin{align*}
		r^p(\nu^{(3)}[U]+\nu^{(4)}[U])^c\g(e_c,\D{\wbar})=&r^p a\Real(\mathfrak{J})^c\Real((\overline{\nabla_3 U+\nabla_4 U})\cdot{\nabla_c}U)\\
		&+r^p O\left(\varepsilon(|\nabla_3U|^2+|\nabla_4U|^2+|{\nabla} U|^2)\right).
	\end{align*}
	We now write the bounds:
	\begin{align*}
		\left|r^p a\Real(\mathfrak{J})^c\Real(\overline{\nabla_4 U}\cdot{\nabla_c}U)\right|\leq {r^p}\frac{a\sin\theta}{|q|}|\nabla_4U||{\nabla} U|\leq r^p\left(\frac14|\nabla_4 U|^2+\frac{a^2\sin^2\theta}{|q|^2}|\nabla U|^2\right),
	\end{align*}
	and similarly
	\begin{align*}
		\left|r^p a\Real(\mathfrak{J})^c\Real(\overline{\nabla_3 U}\cdot{\nabla_c}U)\right|&\leq r^p\left(\frac{a^2\sin^2\theta}{|q|^2}|\nabla_3 U|^2+\frac14|\nabla U|^2\right),
	\end{align*}
	which proves, by \eqref{eq:Iprovesabso1} and \eqref{eq:Icestuneexpre},
	\begin{align*}
		\sqrt{-\g(\D\wbar,\D\wbar)}\mathbf{F}_{\wbar}^{(p)}[U]\geq &r^p\left(1-\frac{\Delta}{2|q|^2}\right)|\nabla_3U|^2+\frac14r^p|\nabla_4 U|^2+r^p\left(\frac54-\frac{\Delta}{2|q|^2}\right)|\nabla U|^2\nn\\
		&-\varepsilon{r^p} (|\nabla_3U|^2+|\nabla_4U|^2+|\nabla U|^2).
	\end{align*}
	Provided $\varepsilon(a,M)>0$ is small enough, we thus get 
	$$\mathbf{F}_{\wbar}^{(p)}[U]\gtrsim{r^p}\left(|\nabla_3 U|^2+|\nabla_4U|^2+|\nabla U|^2\right).$$
	Since the reverse bound also clearly holds, we obtain $\mathbf{F}_{\wbar}^{(p)}[U]\sim{r^p}\left(|\nabla_3 U|^2+|\nabla_4U|^2+|\nabla U|^2\right)$, which implies the following estimate for the boundary terms on $\{\wbar=\wbar_a\}$, $a=1,2$:
	\begin{align*}
		\int_{\un\cap\{\wbar=\wbar_a\}}\mathbf{F}^{(p)}[U]_\mu n_{\wbar}^\mu\sim\int_{\un\cap\{\wbar=\wbar_a\}}r^p \left(|\nabla_3U|^2+|\nabla_4U|^2+|{\nabla} U|^2\right).
	\end{align*}
	\noindent\textbf{Boundary term on $\mathcal{A}=\un\cap\{r=r_\mca\}$.} Using the fact that $\D r$ is inwards pointing at $\mathcal{A}$, we get that the boundary term on $\mathcal{A}$ rewrites 
	$$-\int_{\mathcal{A}\cap\{\wbar_1\leq\wbar\leq\wbar_2\}}\mathbf{F}_{\mathcal{A}}^{(p)}[A]=\int_{\mathcal{A}\cap\{\wbar_1\leq\wbar\leq\wbar_2\}}\left(\sqrt{-\g(\D r,\D r)}\right)^{-1}\mathbf{F}^{(p)}[A]_\mu \D r^\mu$$
	hence recalling \eqref{eq:againehetyty}, the following estimate is satisfied,
	\begin{align*}
		\int_{\mathcal{A}\cap\{\wbar_1\leq\wbar\leq\wbar_2\}}\mathbf{F}_{\mathcal{A}}^{(p)}[A]\lesssim C_{\delta_\pm}\int_{\mathcal{A}\cap\{\wbar_1\leq\wbar\leq\wbar_2\}}r^p\left(|\nabla_3U|^2+|\nabla_4U|^2+|{\nabla} U|^2)\right).
	\end{align*}

	\noindent\textbf{Boundary term on ${\Sigma_1}$.} We compute by \eqref{eq:controlI},
	\begin{align*}
		\g(\D r, e_3)=e_3(r)= -1,\quad\g(\D r, e_4)=e_4(r)=\frac{\Delta}{|q|^2}+O(\varepsilon\ubar^{-3-\delta/2})
	\end{align*}
	in $\un$. As $\g(\D r,\D r)<0$,  $\D r$ is outwards pointing at $\Sigma_1$ so that the boundary term on $\Sigma_1$ rewrites $\int_{\Sigma_1\cap\{\wbar_1\leq\wbar\leq\wbar_2\}}\mathbf{F}_{{\Sigma_1}}^{(p)}[U]$, where
	\begin{align*}
		\sqrt{-\g(\D r,\D r)}\mathbf{F}_{{\Sigma_1}}^{(p)}[U]&=\frac{{r^p}}{2}\left(|\nabla_3U|^2\g(e_4,-\D r)+|\nabla_4U|^2\g(e_3,-\D r)+|\nabla U|^2\g(e_3+e_4,-\D r)\right)\nn\\
		&\quad-{r^p}(\nu^{(3)}[U]+\nu^{(4)}[U])^a\g(e_a,-\D r)\\
		&=\frac{{r^p}}{2}\left(\frac{-\Delta}{|q|^2}|\nabla_3U|^2+|\nabla_4U|^2+\left(1-\frac{\Delta}{|q|^2}\right)|\nabla U|^2\right)+r^pO(\varepsilon|\df U|^2),
	\end{align*}
	where we used in particular the bound $|\D r|\lesssim\varepsilon\ubar^{-3-\delta/2}$. As before this implies, provided $w_f(a,M,\delta_\pm)$ is negative enough,
	$$\int_{\Sigma_1\cap\{\wbar_1\leq\wbar\leq\wbar_2\}}\mathbf{F}_{{\Sigma_1}}^{(p)}[U]\sim C'_{\delta_\pm}\int_{\Sigma_1\cap\{\wbar_1\leq\wbar\leq\wbar_2\}}r^p(|\nabla_3U|^2+|\nabla_4 U|^2+|\nabla U|^2),$$
	which concludes the proof of the proposition.
\end{proof}

\begin{prop}\label{prop:Ibulkpos}
	We assume that $\hat{\mcl}$ in \eqref{eq:IoperateurLhat} satisfies the following bounds, 
	\begin{align}
		|h|\lesssim 1,\quad |\hhbar|\lesssim 1,\quad |L_1[U]|\lesssim |\nabla U|,\quad |L[U]|\lesssim |U|.\label{eq:IhypLhat}
	\end{align}
	We also assume that $U$ is a horizontal tensor which decays as follows on $\mathcal{A}$,
	\begin{align}\label{eq:AAhyp}
		\|U\|_{L^2(r_\mca,\ubar)}\lesssim\ubar^{-6-\delta},\quad\text{on }\mathcal{A}.
	\end{align}
	Then, provided $p\geq p_0(a,M,\delta_\pm)\gg 1$ is sufficiently large and $w_f(a,M,\delta_\pm)\ll -1$ is negative enough, for $\wbar_0\leq\wbar_1\leq\wbar_2$ we have the bound
	$$\int_{\un[\wbar_1,\wbar_2]} r^p\left(|\nabla_3 U|^2+|\nabla_4 U|^2+|\nabla U|^2\right)\lesssim \int_{\un[\wbar_1,\wbar_2]}\mathbf{B}^{(p)}[U]+\frac{r_\mca^p}{p}\int_{\wbar_1}^{\wbar_2}\frac{1}{\wbar^{12+2\delta}}\dee\wbar.$$
\end{prop}
\begin{proof}
	Note that we have the estimates
	$$-\frac12 e_3(r^p)=\frac12 p r^{p-1}\gtrsim pr^p,\quad -\frac12 e_4(r^p)=\frac12p\left(\frac{-\Delta}{|q|^2}+O(\varepsilon\ubar^{-3-\delta/2})\right)r^{p-1}\gtrsim |\Delta| pr^{p},$$
	in $\un$ for $w_f(a,M,\delta_\pm)\ll -1$. Thus, using the definition \eqref{eq:bulkprIdef} of the principal bulk term and  the assumed estimates \eqref{eq:IhypLhat} we get for $p\geq p_0(a,M,\delta_+,\delta_-)\gg 1$ sufficiently large
	\begin{align}\label{eq:combibibi1}
		\mathbf{B}^{(p)}_{pr}[U]\gtrsim r^p\left(p|\Delta||\nabla_3 U|^2+p|\nabla_4 U|^2+p|\nabla U|^2\right),
	\end{align}
where we also used \eqref{eq:controlI} to bound the Ricci coefficients. Next, by \eqref{eq:bulkIdef}, using \eqref{eq:IhypLhat}, $\nabla r=O(\varepsilon\ubar^{-3-\delta/2})$, the $L^\infty(\un)$ bounds on $\Gammacheck,\Rcheck$ from \eqref{eq:controlI}, and the commutator identities in Proposition \ref{prop:commnabla} we get
	\begin{align*}
		|\mathbf{B}^{(p)}[U]-\mathbf{B}^{(p)}_{pr}[U]|\lesssim & \frac{\varepsilon p r^p}{\ubar^{3+\delta/2}}  \left(|\nabla_3 U|^2+|\nabla_4 U|^2+|\nabla U|^2\right)+r^p\left(|\nabla_3 U|^2+|\nabla_4 U|^2+|\nabla U|^2+|U|^2\right).
	\end{align*}
	Combining this bound with \eqref{eq:combibibi1} and choosing $w_f(a,M,\delta_\pm)\ll -1$ such that the first term on the RHS above is absorbed in the RHS of \eqref{eq:combibibi1}, and then $p(a,M,\delta_+,\delta_-)\gg 1$ large enough such the second term above is also absorbed in the RHS of \eqref{eq:combibibi1}, we get $r^p\left(|\nabla_3 U|^2+|\nabla_4 U|^2+|\nabla U|^2\right)\lesssim \mathbf{B}^{(p)}[U]+r^p|U|^2$.	Integrating this bound on $\un[\wbar_1,\wbar_2]$ and using Proposition \ref{prop:hardydansI} combined with \eqref{eq:AAhyp} we get 
	\begin{align*}
		\int_{\un[\wbar_1,\wbar_2]} r^p\left(|\nabla_3 U|^2+|\nabla_4 U|^2+|\nabla U|^2\right)\lesssim &\int_{\un[\wbar_1,\wbar_2]}\mathbf{B}^{(p)}[U]+\frac{r_\mca^p}{p}\int_{\wbar_1}^{\wbar_2}\frac{1}{\wbar^{12+2\delta}}\dee\wbar\\
		&+\int_{\un[\wbar_1,\wbar_2]}\frac{r^p}{p}\left(|\nabla_3 U|^2+|\nabla_4 U|^2+|\nabla U|^2\right).
	\end{align*}
	Finally, we conclude the proof by choosing $p$ larger if necessary so that the last term on the RHS above is absorbed in the LHS.
\end{proof}
\begin{prop}\label{prop:Iunenerteukcomm}
	Let ${U}\in\fraks_k(\mathbb{C})$, $k\in\{0,1,2\}$ be such that the inhomogeneous Teukolsky-like equation \eqref{eq:Iteukmod} holds in $\un$, where $\wh{\mcl}$ is as in \eqref{eq:IoperateurLhat}. We assume that \eqref{eq:IhypLhat} holds, and that :
	\begin{itemize}
		\item We have the following initial energy bound on $\mathcal{A}$, for any $\wbar_0\leq\wbar_1\leq\wbar_2$,
		\begin{align}\label{eq:Iinitialenergyhyp}
			\int_{\mathcal{A}\cap\{\wbar_1\leq\wbar\leq\wbar_2\}}\left(|\nabla_3U|^2+|\nabla_4U|^2+|{\nabla} U|^2\right)\lesssim\frac{1}{\wbar_1^{12+2\delta}}+\int_{\wbar_1}^{\wbar_2}\frac{1}{\wbar^{12+2\delta}}\dee\wbar.
		\end{align}
		\item We have the following initial decay on $\mathcal{A}$, for $\ubar\geq 1$,
		\begin{align}\label{eq:L2decayinitialmca}
			\|U\|_{L^2(r_\mca,\ubar)}\lesssim{\ubar^{-6-\delta}},\quad\text{on }\mathcal{A}.
		\end{align}
		\item 	We have the following estimate on $\err[\wh{\mcl}({U})]$, for any $\wbar_f\leq\wbar_1\leq\wbar_2$,
		\begin{align}\label{eq:Ihyperrhat}
			\int_{\un[\wbar_1,\wbar_2]}|\err[\wh{\mcl}({U})]|^2\lesssim C\left(\frac{1}{\wbar_1^{12+2\delta}}+\int_{\wbar_1}^{\wbar_2}\frac{1}{\wbar^{12+2\delta}}\dee\wbar\right),
		\end{align}
		where for some $C>0$.
	\end{itemize}
	Then recalling the definition \eqref{eq:Idefenerdeg(p)} of $\Eun[U]$, we have the following decay in $\un$,
	$$\Eun[U](\wbar)\lesssim\frac{C'}{\wbar^{12+2\delta}},\quad \|U\|_{L^2(S(\ubar,r))}\lesssim\frac{C'}{\ubar^{6+\delta}},$$
	where $C'>0$ depends on $a,M,C,\delta_\pm,w_f$.
\end{prop}
\begin{proof}
	We multiply \eqref{eq:Iteukmod} with $r^p (\overline{\nabla_3{U}+\nabla_4{U}}),$
	where $p(a,M,\delta_+,\delta_-)$ is a large enough integer chosen later in the proof, and we integrate on $\un[\wbar_1,\wbar_2]$ with respect to spacetime volume, and we take the real part. By Proposition \ref{prop:Icalculcommunenergy}, we get 
	\begin{align}\label{eq:Istep1Psi}
		\int_{\un[\wbar_1,\wbar_2]}\mathbf{B}^{(p)}[U]+\int_{\un\cap\{\wbar=\wbar_2\}}\mathbf{F}_{\wbar}^{(p)}[U]\leq&\int_{\un\cap\{\wbar=\wbar_1\}}\mathbf{F}_{\wbar}^{(p)}[U]+\int_{\mathcal{A}\cap\{\wbar_1\leq\wbar\leq\wbar_2\}}\mathbf{F}_{\mathcal{A}}^{(p)}[U]\nn\\
		&+\left|\int_{\un[\wbar_1,\wbar_2]}r^p \Real\left((\overline{\nabla_3{U}+\nabla_4{U}})\cdot\err[\wh{\mcl}({U})]\right)\right|,
	\end{align}
	where we used in particular \eqref{eq:IestimbdGamma} which implies that the boundary term on ${\Sigma_1}$ is non-negative. Next, using Proposition \ref{prop:Ibulkpos}, we get that for $p=p(a,M,\delta_\pm)\gg 1$ and $w_f(a,M,\delta_\pm)\ll -1$, 
	\begin{align*}
		\int_{\un[\wbar_1,\wbar_2]}r^p\left(|\nabla_3 U|^2+|\nabla_4 U|^2+|\nabla U|^2\right)\lesssim \int_{\un[\wbar_1,\wbar_2]}\mathbf{B}^{(p)}[U]+r_{\mathcal{A}}^p\int_{\wbar_1}^{\wbar_2}\frac{1}{\wbar^{12+2\delta}}\dee\wbar.
	\end{align*}
	Using Lemma \ref{lem:volumeformI}, we get
	\begin{align*}
		&\int_{\un[\wbar_1,\wbar_2]}r^p\Big(\left|{{\nabla_3}{U}}\right|^2+\left|{\nabla_4{U}}\right|^2+|{{\nabla}U}|^2\Big)\gtrsim\int_{\wbar_1}^{\wbar_2}\Eun^{(p)}[U](\wbar)\dee\wbar.
	\end{align*}
	Combining the bounds above with the estimates \eqref{eq:Iestimbdwbar}, \eqref{eq:Iestimbdsigmazero}, \eqref{eq:Iinitialenergyhyp} for the boundary terms on $\wbar=\wbar_1,\wbar_2$ and on $\mathcal{A}$, we deduce from \eqref{eq:Istep1Psi} the estimate
	\begin{align}\label{eq:IstepestepU}
		\Eun^{(p)}[U](\wbar_2)+\int_{\wbar_1}^{\wbar_2}\Eun^{(p)}[U](\wbar)\dee\wbar\lesssim & \:\Eun^{(p)}[U](\wbar_1)+C_{\delta_\pm}r_{\mathcal{A}}^p\left(\frac{1}{\wbar_1^{12+4\delta}}+\int_{\wbar_1}^{\wbar_2}\frac{1}{\wbar^{12+4\delta}}\dee\wbar\right)\nn\\
		&+\left|\int_{\un[\wbar_1,\wbar_2]}r^p \Real\left((\overline{\nabla_3{U}+\nabla_4{U}})\cdot\err[\wh{\mcl}({U})]\right)\right|.
	\end{align}
	Now we bound the last term on the RHS above. For any $s>0$, it is bounded by
	\begin{align*}
	s\int_{\un[\wbar_1,\wbar_2]}r^p\left(|\nabla_3{U}|^2+|\nabla_4{U}|^2\right)+s^{-1}\int_{\un[\wbar_1,\wbar_2]}r^p|\err[\wh{\mcl}({U})]|^2,
	\end{align*}
so that choosing $s(a,M)>0$ small enough ensures that the first term above can be absorbed in the second term in the LHS of \eqref{eq:IstepestepU}, and hence we deduce
	$$\Eun[U](\wbar_2)+\int_{\wbar_1}^{\wbar_2}\Eun[U](\wbar)\dee\wbar\lesssim C_{\delta_\pm}\left( \Eun[U](\wbar_1)+\frac{1}{\wbar_1^{12+2\delta}}+\int_{\wbar_1}^{\wbar_2}\frac{1}{\wbar^{12+2\delta}}\dee\wbar\right),$$
	where we used a fixed value $p=p(a,M,\delta_+,\delta_-)$ of $p$, and $r\sim 1$, and where the constant $C_{\delta_\pm}>0$ depends on $a,M,\delta_\pm$. We conclude the proof of the stated energy decay by using Lemma \ref{lem:decay} with $p=12+2\delta$, $f(\wbar)=\Eun[U](\wbar)$, $B=+\infty$, $b=\wbar_f$\footnote{Note that $\Eun[U](\wbar_f)\lesssim C_{\delta_\pm,w_f}$ since  $\un\cap\{\wbar=\wbar_f\}$ is compact. However we could also do the energy method on $\{\wbar\leq\wbar_f\}$ which can be seen to yield $\Eun[U](\wbar_f)\lesssim C_{\delta_\pm}$, and as a consequence the implicit constant in the result of Proposition \ref{prop:Iunenerteukcomm} can be chosen to not depend on $w_f$, but this is not necessary for our purpose.} which yields
	$$\Eun[U](\wbar)\lesssim C_{\delta_\pm,w_f}\wbar^{-12-2\delta}.$$
	Finally, the stated $L^2(S(r,\ubar))$ decay holds by Proposition \ref{prop:controlL2avecener} and the energy decay above, combined with the initial decay \eqref{eq:L2decayinitialmca}.
\end{proof}

\subsubsection{Energy and decay estimates for $A$ in $\un$}\label{section:errunI}
We recall the definition \eqref{eq:definitansatzun} of the ansatz $\Psi$,
$$\Psi=\frac{1}{\ubar^6}\sum_{|m|\leq 2}Q_m\Psi_m,\quad\Psi_m:=\frac{A_m(r)}{\qbar^2}\mcd\hot(\mcd(Y_{m,2}(\cos\theta)e^{im\phi_+})).$$ 
In all of Section \ref{section:errunI}, we denote by $\err$ any error term satisfying for some $k\geq N_0-7$,
\begin{align}\label{eq:asinflemme}
	|\df^{\leq k}\err|\lesssim \varepsilon\ubar^{-3-\delta/2}.
\end{align}
\begin{lem}\label{lem:teukpsimfrakF}
	We denote by $\mathfrak{F}$  the following set of 1-forms,
	$$\mathfrak{F}=\{j,\hodge{j},j_+,\hodge{j_+},j_-,\hodge{j_-}\}.$$
	There exist scalar functions $F_\mck^{j_1,j_2,m}\left(r,\cos\theta,J^{(+)},J^{(-)}\right)$ such that
	\begin{align*}
		\mcl(\Psi_m)=\sum_{j_1,j_2\in\frak{F}}F^{j_1,j_2,m}_\mck\left(r,\cos\theta,J^{(+)},J^{(-)}\right)j_1\hot j_2+\err,\quad|\df^{\leq N_1}F_\mck^{j_1,j_2,m}|\lesssim 1.
	\end{align*}

\end{lem}
\begin{proof}
	See Appendix \ref{appendix:teukpsimfrakF}.
\end{proof}

\begin{prop}\label{prop:teukansatzdansI}
	We have the following decay estimates for the Teukolsky operator $\mcl$ applied to the ansatz $\Psi$,
	\begin{align*}
		|\df^{\leq N_0-7}\mcl(\Psi)|\lesssim_Q\ubar^{-6-\delta},
	\end{align*}
where the notation $\lesssim_Q$ means that the implicit constant depends $a,M$ and on $(Q_m)_{|m|\leq 2}$.
\end{prop}

\begin{proof}
	First of all, using $|\nabla_\mu\ubar|\lesssim 1$ for $\mu=1,2,3,4$, it is clear that any time that a derivative falls on the factor $\ubar^{-6}$, this will yield a term bounded by $\ubar^{-7}\lesssim\ubar^{-6-\delta}$. Hence by \eqref{eq:teukop} we get the following estimate for $k\leq N_0-3$,
	\begin{align}\label{eq:bahoubliepascellela}
		|\df^{\leq k}\mcl(\Psi)|\lesssim_Q\frac{1}{\ubar^{6+\delta}}+\frac{1}{\ubar^6}\sum_{|m|\leq 2}|\df^{\leq k}\mcl(\Psi_m)|,
	\end{align}
Now, the point is that $\mcl(\Psi_m)=0$ in exact Kerr, see Proposition \ref{prop:teukansatzkerr}, so that in perturbed Kerr it is only left to show that $\mcl(\Psi_m)$ can be expressed with $\err$ terms, and then use \eqref{eq:controlI}. By Lemma \ref{lem:teukpsimfrakF},  we have
\begin{align}\label{eq:hlhhkhkhkhlkhlh}
	\mcl(\Psi_m)=\sum_{j_1,j_2\in\frak{F}}F_\mck^{j_1,j_2,m}\left(r,\cos\theta,J^{(+)},J^{(-)}\right)j_1\hot j_2+\err,
\end{align}
where $F_\mck^{j_1,j_2,m}\left(r,\cos\theta,J^{(+)},J^{(-)}\right)$ are regular and explicit complex valued functions which appear in the analog computation of $\mcl(\Psi_m)$ in Kerr, setting $\err=0$, see Appendix \ref{appendix:teukpsimfrakF}. But, using Remark \ref{rem:propagatedjjj} we get that the function
\begin{align*}
	\left|\sum_{j_1,j_2\in\frak{F}}F_\mck^{j_1,j_2,m}\left(r,\cos\theta,J^{(+)},J^{(-)}\right)j_1\hot j_2\right|^2,
\end{align*}
that can be expressed a sum of products of $j_1\cdot j_2=(j_1\cdot j_2)_\mck$ with exact Kerr coefficients, coincides with its Kerr value with respect to $(r,\ubar,\theta,\phi_+)$, which is $|\mcl(\Psi_m)|^2_\mck=0$ by Corollary \ref{cor:teukansatzdansun}. Hence by \eqref{eq:hlhhkhkhkhlkhlh} we deduce $\mcl(\Psi_m)=\err$, which concludes the proof by \eqref{eq:bahoubliepascellela}.\end{proof}

We now present the decay results in $\un$ for the horizontal tensor $$\err[A]:=A-\Psi.$$ Recall the initial assumption \eqref{eq:initialhyponmca} on $\mathcal{A}$ for $\err[A]$, which rewrites
\begin{align*}
	\|\df^{\leq 4}\err[A]\|_{L^2(S(r_\mca,\ubar))}\lesssim\ubar^{-6-\delta},\quad\text{on }\mathcal{A}.
\end{align*}

\begin{prop}\label{prop:leteukdeerrA}
	Recalling the notation introduced in Proposition \ref{prop:teukansatzdansI}, we have in $\un$
	\begin{align*}
		|\df^{\leq N_0-7}\mcl(\err[A])|\lesssim_Q\ubar^{-6-\delta}.
	\end{align*}
\end{prop}
\begin{proof}
	By Proposition \ref{prop:teukansatzdansI}, combined with the Teukolsky equation \eqref{eq:teukA} (and more precisely, using the quadratic structure of the error term \eqref{eq:teukerror}) we get for $k\leq N_0-7$,
	\begin{align*}
		|\df^{\leq k}\mcl(\err[A])|\lesssim |\df^{\leq k}\err[\mcl(A)]|+|\df^{\leq k}\mcl(\Psi)|
		\lesssim_Q |\df^{\leq k+1}(\Gammacheck,\Rcheck)|^2+\ubar^{-6-\delta}\lesssim_Q \ubar^{-6-\delta},
	\end{align*}
	where we used the bounds in \eqref{eq:controlI} in the last step.
\end{proof}
\begin{prop}\label{prop:IundecayA}
	We have, in region $\un$, the following decay estimates for $\err[A]$,
	\begin{align*}
			\Eun[\err[A]](\wbar)\lesssim C(\delta_\pm,w_f,(Q_m)_{|m|\leq 2})\wbar^{-12-2\delta},\quad \|\err[A]\|_{L^2(S(r,\ubar))}\lesssim C(\delta_\pm,w_f,(Q_m)_{|m|\leq 2})\ubar^{-6-\delta},
	\end{align*}
where $C(\delta_\pm,w_f,(Q_m)_{|m|\leq 2})>0$ depends on $a,M,\delta_\pm,w_f,(Q_m)_{|m|\leq 2}$.
\end{prop}
\begin{proof}
	We apply Proposition \ref{prop:Iunenerteukcomm}. For $\hat{\mcl}=\mcl$, $U=\err[A]$, and hence $\err[\hat{\mcl}(U)]=\mcl(\err[A])$ so that \eqref{eq:Iteukmod} holds, we have that : 
	\begin{itemize}
		\item \eqref{eq:Iinitialenergyhyp} and \eqref{eq:L2decayinitialmca} hold by \eqref{eq:initialhyponmca}.
		\item \eqref{eq:Ihyperrhat} holds by Proposition \ref{prop:leteukdeerrA}.
	\end{itemize}
	Moreover, \eqref{eq:IhypLhat} holds by the expression \eqref{eq:teukop} of the Teukolsky operator $\mcl$ and the bounds on $\Gammacheck,\Rcheck$ given by \eqref{eq:controlI}. Note that here we also use Lemma \ref{lem:ignocoucou} to express the horizontal Laplacian term $\frac14\mcd\hot\divc$ in $\mcl$ with $\triangle_2$ and lower-order terms. Hence the proof by Proposition \ref{prop:Iunenerteukcomm}.
\end{proof}
Using a suitable order, from the commutation identities proven in Section \ref{section:commutingteukwith} we are also able to control some $\barre{\df}^{\leq 3}$ derivatives of $\err[A]$ in $\un$, as can be seen in the following result.
\begin{prop}\label{prop:L2lesderrivesdeerrA}
	We have the following decay estimates in $\un$ for $\err[A]$: for any
	\begin{align}
		P\in\mathcal{S}:=\Big\{&(\lieT,\lie_\Z)^{\leq 3},(\lieT,\lie_\Z)^{\leq 2}\divc,(\lieT,\lie_\Z)^{\leq 1}\divc\divc,(\lieT,\lie_\Z)^{\leq 1}\mcd\hot\divc,(\lieT,\lie_\Z)^{\leq 2}\nabla_4,\nn\\
		&(\lieT,\lie_\Z)^{\leq 1}\nabla_4\divc,\divc\mcd\hot\divc,\mcd\divc\divc,\nabla_4\divc\divc, \nabla_4\mcd\hot\divc,(\lieT,\lie_\Z)^{\leq 1}\nabla_4\nabla_4\Big\},\label{eq:deefSensop}
	\end{align}
	we have 
	$$\|P\err[A]\|_{L^2(S(r,\ubar))}\lesssim C(\delta_\pm,w_f,(Q_m)_{|m|\leq 2})\ubar^{-6-\delta},$$
	where $C(\delta_\pm,w_f,(Q_m)_{|m|\leq 2})>0$ depends on $a,M,\delta_\pm,w_f,(Q_m)_{|m|\leq 2}$.
\end{prop}
\begin{proof}
We simplify the notation $\lesssim_Q$ and write instead $\lesssim$ in this proof. We introduce the ordered set of derivatives
		\begin{align*}
		(P_k)_{1\leq k\leq 10}:=\Big(&\divc,\divc\divc,\mcd\hot\divc,\divc\mcd\hot\divc,\mcd\divc\divc,\\
		&\nabla_4,\nabla_4\divc,\nabla_4\divc\divc, \nabla_4\mcd\hot\divc,\nabla_4\nabla_4\Big),
	\end{align*}
and denoting $C=C(\delta_\pm,w_f,(Q_m)_{|m|\leq 2})$ we prove by induction on $1\leq k\leq 10$ the bounds
\begin{align}\label{eq:inductioncoolunUN}
	\Eun[P_k\err[A]](\wbar)\lesssim C\wbar^{-12-2\delta},\quad \|P_k\err[A]\|_{L^2(S(r,\ubar))}\lesssim C\ubar^{-6-\delta}.
\end{align}
\textbf{Case $k=1$:} by Proposition \ref{prop:nouvellescommut} with $U=\err[A]$, $\divc\err[A]$ satisfies 
$$\mcl^{(1)}(\divc\err[A])=\err[\mcl^{(1)}(\divc\err[A])],$$
where $\mcl^{(1)}=\nabla_4\nabla_3-\triangle_2+O(1)\df^{\leq 1}$, and where
\begin{align}\label{eq:errdeiverraborne}
	|\err[\mcl^{(1)}(\divc \err[A])]|\lesssim |\nabla_3\err[A]|+|\nabla_4 \err[A]|+|\nabla \err[A]|+|\err[A]|+\ubar^{-6-\delta},
\end{align}
where the last term on the RHS above corresponds to the error terms $\df^{\leq k}(\Gammacheck,\Rcheck)\cdot\df^{\leq k+1}U+\df^{\leq k}\err[\mcl(U)]$ which we bound by $\ubar^{-6-\delta}$ using \eqref{eq:controlI} and the bound $|\df^{\leq 2}\Psi|\lesssim\ubar^{-6}$, plus the term $\df\err[\mcl(A)]=O(\ubar^{-6-\delta})$ by Proposition \ref{prop:leteukdeerrA}. Thus, we can use Proposition \ref{prop:Iunenerteukcomm} for $\hat{\mcl}=\mcl^{(1)}$, $U=\divc\err[A]$, and hence $\err[\hat{\mcl}(U)]=\err[\mcl^{(1)}(\divc \err[A])]$ so that \eqref{eq:Iteukmod} holds, we have that : 
\begin{itemize}
	\item \eqref{eq:Iinitialenergyhyp} and \eqref{eq:L2decayinitialmca} hold by \eqref{eq:initialhyponmca}.
	\item \eqref{eq:Ihyperrhat} holds by \eqref{eq:errdeiverraborne} combined with Proposition \ref{prop:IundecayA} to control the integral of the terms $|\nabla_3\err[A]|^2+|\nabla_4 \err[A]|^2+|\nabla \err[A]|^2+|A|^2$ .
\end{itemize}
Moreover, \eqref{eq:IhypLhat} holds by the expression of $\mcl^{(1)}$ and the bounds on $\Gammacheck,\Rcheck$ given by \eqref{eq:controlI}. Hence the result for $n=1$ by Proposition \ref{prop:Iunenerteukcomm}.

\textbf{Induction step:} Now we assume that $n\leq 10$ is such that \eqref{eq:inductioncoolunUN} holds for all $k\leq n-1$ and we prove that \eqref{eq:inductioncoolunUN} holds for $k=n$. This follows from the following observation: by Proposition \ref{prop:nouvellescommut} for $U=\err[A]$, $P_{n}\err[A]$ satisfies a Teukolsky-like equation
$$\hat{\mcl}(P_n\err[A])=\err[\hat{\mcl}(P_n\err[A])]$$
where $\hat{\mcl}=\nabla_4\nabla_3-\triangle+O(1)\df^{\leq 1}$, and where
\begin{align}\label{eq:larectassuretkt}
	|\err[\hat{\mcl}(P_n\err[A])]|\lesssim \ubar^{-6-\delta}+\sum_{k=0}^{n-1}\left(|\nabla_3 P_k\err[A]|+|\nabla_4 P_k\err[A]|+|\nabla P_k\err[A]|+|P_k\err[A]|\right),
\end{align}
where the term $\ubar^{-6-\delta}$ comes from the quadratic error terms $\df^{\leq k}(\Gammacheck,\Rcheck)\cdot\df^{\leq k+1}U+\df^{\leq k}\err[\mcl(U)]$  which we bound using \eqref{eq:controlI} and the bound $|\df^{\leq k}\Psi|\lesssim\ubar^{-6}$, plus the term $\df^{\leq k}\err[\mcl(A)]=O(\ubar^{-6-\delta})$ by Proposition \ref{prop:leteukdeerrA}. We can apply Proposition \ref{prop:Iunenerteukcomm} because:
\begin{itemize}
	\item \eqref{eq:Iinitialenergyhyp} and \eqref{eq:L2decayinitialmca} hold for $P_n\err[A]$ by \eqref{eq:initialhyponmca}.
	\item \eqref{eq:Ihyperrhat} holds for $\err[\hat{\mcl}(P_n\err[A])]$ by \eqref{eq:larectassuretkt} combined with the induction assumption for $k\leq n-1$ to control the integral of the terms $|\nabla_3 P_k\err[A]|^2+|\nabla_4 P_k\err[A]|^2+|\nabla P_k\err[A]|^2+|P_k\err[A]|^2$ by \eqref{eq:inductioncoolunUN}.
\end{itemize}
Moreover, \eqref{eq:IhypLhat} holds for $\hat{\mcl}=\nabla_4\nabla_3-\triangle+O(1)\df^{\leq 1}$. Hence we deduce that \eqref{eq:inductioncoolunUN} holds for $k=n$ by Proposition \ref{prop:Iunenerteukcomm}. This concludes the induction. Moreover, we claim that the bounds \eqref{eq:inductioncoolunUN} also hold when adding $(\lieT,\lie_\Z)$ derivatives. Indeed, by commuting the Teukolsky-like equations satisfied by $P_n\err[A]$ for $1\leq n\leq 10$ given by Proposition \ref{prop:nouvellescommut}, and using Lemma \ref{lem:commdflieTZI} we see that the commutator terms with $\lieT,\lie_\Z$ only add quadratic error terms which are bounded by $\ubar^{-6-\delta}$ by \eqref{eq:controlI}. It is thus straightforward to extend the bounds \eqref{eq:inductioncoolunUN} to get that for any $P\in\mathcal{S}$ as defined in \eqref{eq:deefSensop}, we have
$$\|P\err[A]\|_{L^2(S(r,\ubar))}\lesssim_Q C_{\delta_\pm}\ubar^{-6-\delta},$$
which concludes the proof.
\end{proof}

\begin{cor}\label{cor:1dererrAI}
	We have in region $\un$ the following pointwise decay estimates for $\df^{\leq 1}\err[A]$,
	\begin{align}
		|\df^{\leq 1}\err[A]|\lesssim C(\delta_\pm,w_f,(Q_m)_{|m|\leq 2})\ubar^{-6-\delta},
	\end{align}
	where $C(\delta_\pm,w_f,(Q_m)_{|m|\leq 2})>0$ depends on $a,M,\delta_\pm,w_f,(Q_m)_{|m|\leq 2}$.
\end{cor}
\begin{proof}We begin by deriving a Sobolev embedding for horizontal tensors more precise than Proposition \ref{prop:sobolevspheriqueI} (see \eqref{eq:sobolevplusprecis}), which will be needed for technical reasons\footnote{This is necessary because we do not control all $\df^{\leq 2}$ derivatives for $\err[A]$, but only a subset of $\barre{\df}^{\leq 2}$ derivatives.}. Recall from Lemma \ref{lem:cletechnique} and \eqref{eq:recurrenceformuladerrrun} the following identity for any horizontal tensor $U$ and $j\geq 1$:
\begin{align}\label{eq:ecoutetrkql}
	\nabla^jU=(\nablaslash^j U)^S+S_j[U],
\end{align}
where the values of $S_j[U]$ for $j=1,2,3$ are determined as follows:
\begin{align*}
	S_1[U]&=a j\otimes\lieT U_A-F_\mck(r,\theta)j\otimes \lie_\Z U_A+O(1)_\mck\cdot U+O(1)_\mck\Gammacheck\cdot U+\mct\cdot U+(\lambda_1,\lambda_2,\widecheck{\Theta})\df U,\\
	S_2[U]&=S_1[\nabla U-S_1[U]]+\nabla(S_1[u]),\quad S_3[U]=S_1[\nabla^2 U-S_2[U]]+\nabla(S_2[u]).
\end{align*}
More precisely using the bounds in \eqref{eq:controlI}, and using Lemma \ref{lem:commdflieTZI}, we compute 
$$S_2[U]=_{rs}\nabla(\lieT,\lie_\Z)^{\leq 1} U+(\lieT,\lie_\Z)^{\leq 2}U+O(\varepsilon\ubar^{-3-\delta/2})\cdot\df^{\leq 2}U.$$
Thus, by the Sobolev embedding in Corollary \ref{cor:sobospheriqueI},
\begin{align*}
	|U|^2= |U^S|^2&\lesssim \int_{S(r,\ubar)}|\shl{\triangle}{7.3}^{\leq 1}U^S|^2\\
	&\lesssim \int_{S(r,\ubar)}\left(|{\triangle}U|^2+|\nabla(\lieT,\lie_\Z)^{\leq 1} U|^2+|(\lieT,\lie_\Z)^{\leq 2}U|^2+\ubar^{-6-\delta}|\df^{\leq 2}U|^2\right).
\end{align*}
We also have for any horizontal tensor $V\in\fraks_p(\C)$ $p=0,1,2$, by \eqref{eq:ecoutetrkql}, recalling \eqref{eq:defdesmcdpla},
\begin{align}
	\int_{S(r,\ubar)}|\nabla V|^2&\lesssim\int_{S(r,\ubar)}\left(|\nablaslash V^S|^2+|(\lieT,\lie_\Z)^{\leq 1}V|^2+\ubar^{-6-\delta}|\df^{\leq 1}V|^2\right)\nn \\
	&\lesssim \int_{S(r,\ubar)}\left(|\barred{\mcd}_p V^S|^2+|(\lieT,\lie_\Z)^{\leq 1}V|^2+\ubar^{-6-\delta}|\df^{\leq 1}V|^2\right)\nn\\
	&\lesssim \int_{S(r,\ubar)}\left(|{\mcd}_p V|^2+|(\lieT,\lie_\Z)^{\leq 1}V|^2+\ubar^{-6-\delta}|\df^{\leq 1}V|^2\right)\label{eq:nablautileeeu}
\end{align}
where we used the elliptic estimate of Lemma \ref{lem:ellipticIun} in the second step, and \eqref{eq:ecoutetrkql} for $j=1$ again in the last step. We deduce for any $U\in\fraks_p(\C)$, in $\un$,
\begin{align}\label{eq:sobocooltcoolkk}
	|U|^2\lesssim \int_{S(r,\ubar)}\left(|{\triangle}U|^2+|\mcd_p(\lieT,\lie_\Z)^{\leq 1} U|^2+|(\lieT,\lie_\Z)^{\leq 2}U|^2+\ubar^{-6-\delta}|\df^{\leq 2}U|^2\right).
\end{align}
Also notice that by Lemma \ref{lem:relatnablalieTZ} and \eqref{eq:defTdansI} we have 
\begin{align}\label{eq:ggghhhhjpppdesnoms}
	\nabla_3=\frac{|q|^2}{\Delta}\left(2a\Real(\frakJ)\cdot\nabla-\nabla_4+2\lieT+O(1)_\mck\cdot U+O(\varepsilon\ubar^{-3-\delta/2})\cdot\df^{\leq 1}U\right).
\end{align}
Combining this with Lemma \ref{lem:ignocoucou} and \eqref{eq:nablautileeeu} we get that for any $U\in\fraks_2(\C)$, \eqref{eq:sobocooltcoolkk} rewrites 
\begin{align}\label{eq:sobolevplusprecis}
	|U|^2\lesssim C_{\delta_\pm}\int_{S(r,\ubar)}\left(|\mcd\hot\divc U|^2+|(\lieT,\lie_\Z)^{\leq 1}\divc U|^2+|(\lieT,\lie_\Z)^{\leq 2}U|^2+|\nabla_4 U|^2+\ubar^{-6-\delta}|\df^{\leq 2}U|^2\right).
\end{align}
Thus, by \eqref{eq:sobolevplusprecis} and \eqref{eq:controlI} to bound the quadratic error terms, Proposition \ref{prop:L2lesderrivesdeerrA} yields
\begin{align}\label{eq:cmaborne1}
	|\err[A]|\lesssim C(\delta_\pm,w_f,(Q_m)_{|m|\leq 2})\ubar^{-6-\delta}.
\end{align}
Similarly, for $\nabla_4\err[A]$, by \eqref{eq:commnab4divc} and \eqref{eq:commnab4mcdhot} combined with \eqref{eq:sobolevplusprecis} for $U=\nabla_4\err[A]$ we get
\begin{align}
	|\nabla_4\err[A]|^2&\lesssim  C_{\delta_\pm}\int_{S(r,\ubar)}\Big(|\nabla_4^{\leq 1}\mcd\hot^{\leq 1}\divc^{\leq 1} \err[A]|^2+|(\lieT,\lie_\Z)^{\leq 1}\nabla_4^{\leq 1}\divc^{\leq 1} \err[A]|^2\nn\\
	&\quad\quad\quad\quad\quad\quad+|(\lieT,\lie_\Z)^{\leq 2}\nabla_4^{\leq 1} \err[A]|^2+|\nabla_4^2 \err[A]|^2+\ubar^{-6-\delta}|\df^{\leq 3}U|^2\Big)\nn\\
	&\lesssim C(\delta_\pm,w_f,(Q_m)_{|m|\leq 2})\ubar^{-12-2\delta}\label{eq:cmaborne2}
\end{align}
where we used Proposition \ref{prop:L2lesderrivesdeerrA} in the last step. By the same argument we also get 
\begin{align}\label{eq:cmaborne3}
	|(\lieT,\lie_\Z)\err[A]|\lesssim C(\delta_\pm,w_f,(Q_m)_{|m|\leq 2})\ubar^{-6-\delta}.
\end{align}
Now we bound $|\nabla \err[A]|$ which requires a bit more work. By \eqref{eq:ecoutetrkql} with $j=1$ combined with \eqref{eq:controlI} to bound the quadratic error terms we have
\begin{align}\label{eq:goingbacktogoingback}
	|\nabla \err[A]|&\lesssim |\nablaslash \err[A]^S|+|(\lieT,\lie_\Z)^{\leq 1}\err[A]|+\ubar^{-6-\delta}\lesssim |\nablaslash \err[A]^S|+C(\delta_\pm,w_f,(Q_m)_{|m|\leq 2})\ubar^{-6-\delta},
\end{align}
where we used \eqref{eq:cmaborne1} and \eqref{eq:cmaborne3} in the last step above. Moreover, by the elliptic estimates in Lemma \ref{lem:ellipticIun}, and using that $\nablaslash^{\leq 1}\shl{K}{7.3}=O(1)$ we get
\begin{align}
	|\nablaslash \err[A]^S|^2\lesssim& \int_{S(r,\ubar)}|\nablaslash^3\err[A]^S|^2\nn\\
	\lesssim& \int_{S(r,\ubar)} \left(|\shl{\overline{\mcd}}{7.2}\cdot\shl{\triangle}{7.3}\err[A]^S|^2+|\shl{\triangle}{7.3}\err[A]^S|^2+|\shl{\overline{\mcd}}{7.2}\cdot\err[A]^S|^2+|\err[A]|^2\right).\nn
\end{align}
Moreover by \eqref{eq:ecoutetrkql} with $j=1$ and $j=2$ and \eqref{eq:cmaborne1} we have
\begin{align*}
	&\int_{S(r,\ubar)} \left(|\shl{\triangle}{7.3}\err[A]^S|^2+|\shl{\overline{\mcd}}{7.2}\cdot\err[A]^S|^2+|\err[A]|^2\right)\lesssim \\
	&C_{\delta_\pm}\int_{S(r,\ubar)}\Big(|\mcd\hot\divc \err[A]|^2+|(\lieT,\lie_\Z)^{\leq 1}\divc \err[A]|^2+|(\lieT,\lie_\Z)^{\leq 2}\err[A]|^2+|\nabla_4 \err[A]|^2\Big)\\
	&+\ubar^{-12-2\delta}\lesssim C(\delta_\pm,w_f,(Q_m)_{|m|\leq 2})\ubar^{-12-2\delta}
\end{align*}
by Proposition \ref{prop:L2lesderrivesdeerrA}. Thus it is only left to write that by \eqref{eq:ecoutetrkql} for $j=3$ we have 
\begin{align*}
	\int_{S(r,\ubar)} |\shl{\overline{\mcd}}{7.2}\cdot\shl{\triangle}{7.3}\err[A]^S|^2\lesssim\int_{S(r,\ubar)}\Bigg(&|\divc\triangle_2\err[A]|^2+|\nabla^{\leq 2}(\lieT,\lie_\Z)^{\leq 1}\err[A]|^2\\
	&+ |\nabla^{\leq 1}(\lieT,\lie_\Z)^{\leq 2}\err[A]|^2+|(\lieT,\lie_\Z)^{\leq 3}\err[A]|^2\Bigg)+\ubar^{-12-2\delta}.
\end{align*}
Moreover using \eqref{eq:ecoutetrkql} with $j=1,2$ and Proposition \ref{prop:L2lesderrivesdeerrA} as before we get
\begin{align*}
	&\int_{S(r,\ubar)}|\nabla^{\leq 2}(\lieT,\lie_\Z)^{\leq 1}\err[A]|^2
	+ |\nabla^{\leq 1}(\lieT,\lie_\Z)^{\leq 2}\err[A]|^2+|(\lieT,\lie_\Z)^{\leq 3}\err[A]|^2\Bigg)\\
	&\lesssim\sum_{P\in\mathcal{S}}\|P\err[A]\|_{L^2(S(r,\ubar))}^2\lesssim C(\delta_\pm,w_f,(Q_m)_{|m|\leq 2})\ubar^{-12-2\delta}.
\end{align*}
Finally, using Lemma \ref{lem:ignocoucou} again and \eqref{eq:ggghhhhjpppdesnoms} we have 
\begin{align*}
	\int_{S(r,\ubar)}|\divc\triangle_2\err[A]|^2\lesssim &C_{\delta_\pm}\int_{S(r,\ubar)}\Big(|\divc\mcd\hot\divc\err[A]|^2+|(\nabla_4,\lieT)^{\leq 1}\divc A|^2+|\nabla^{\leq 2}A|^2\Big)\\
	&\lesssim C(\delta_\pm,w_f,(Q_m)_{|m|\leq 2})\ubar^{-12-2\delta},
\end{align*}
where we used using \eqref{eq:ecoutetrkql} with $j=1,2$ as before, to bound the last term on the RHS above and Proposition \ref{prop:L2lesderrivesdeerrA}. Going back to \eqref{eq:goingbacktogoingback}, this concludes the proof that 
\begin{align}
	|\nabla\err[A]|\lesssim C(\delta_\pm,w_f,(Q_m)_{|m|\leq 2})\ubar^{-6-\delta}.
\end{align}
Finally, we deduce that $|\nabla_3\err[A]|\lesssim C(\delta_\pm,w_f,(Q_m)_{|m|\leq 2})\ubar^{-6-\delta}$ by the identity \eqref{eq:ggghhhhjpppdesnoms} combined with \eqref{eq:cmaborne1}, \eqref{eq:cmaborne2}, \eqref{eq:cmaborne3} and the bound above. This yields, in $\un$,
$$		|\df^{\leq 1}\err[A]|\lesssim_Q C(\delta_\pm,w_f,(Q_m)_{|m|\leq 2})\ubar^{-6-\delta},$$
which concludes the proof.
\end{proof}
\begin{rem}
The statement of Corollary \ref{cor:1dererrAI} concludes the proof of Theorem \ref{thm:AregionI}.
\end{rem}
\begin{defi}\label{defi:sigmazeroII}
	We define, in region $\un$, the coordinates $(\mathring{u},\mathring{\ubar},\theta_*,\phi_*)$ as the Kerr values of the Pretorius-Israel double null coordinates of Kerr spacetime defined in Section \ref{section:linearizationdoublenull}, expressed with respect to the PT coordinates $(r,\ubar,\theta,\phi_+)$. Also, denoting $C_R(a,M,\delta_-)$ any large constant such that \eqref{eq:conditionCRR} holds, which will be chosen later in the proof\footnote{Actually, in the proof of the main theorem (Theorem \ref{thm:mainthm}), we first choose $C_R$ large enough depending only on $a,M$ and then we choose $\delta_-(a,M,C_R)$ small enough such that \eqref{eq:conditionCRR} holds, see Section \ref{section:proofmainthm}.}, we define the hypersurface
	$$\Sigma_0:=\{\mathring{u}+\mathring{\ubar}=C_R\}\subset\un.$$
\end{defi}

From Corollary \ref{cor:1dererrAI}, we deduce the following pointwise and energy decay along $\Sigma_0$,
\begin{align}
	|A-\Psi|&\lesssim C(\delta_\pm,w_f,(Q_m)_{|m|\leq 2})\mathring{\ubar}^{-6-\delta},\label{eq:Aunsig00}\\
	\int_{\Sigma_0\cap\{\mathring{\ubar}_1\leq\mathring{\ubar}\leq\mathring{\ubar}_2\}}|\df^{\leq 1}(A-\Psi)|^2&\lesssim C(\delta_\pm,w_f,(Q_m)_{|m|\leq 2})\int_{\mathring{\ubar}_1}^{\mathring{\ubar}_2}\frac{\dee\mathring{\ubar}}{\mathring{\ubar}^{6+\delta}},\label{eq:Aenerunsig00}
\end{align}
for any $1\lesssim\mathring{\ubar}_1\leq\mathring{\ubar}_2$, where we used the estimate $\ubar\sim\mathring{\ubar}$, which holds because $|\ubar-\ubarring|\lesssim1$ by Lemma \ref{lem:ubarssontpareils}.

\section{Control of region $\deux\cup\trois$ in double null gauge}\label{section:IIdoublenull}
In this section, we start from region $\un$ already constructed in Section \ref{section:regionun}, as well as the estimates proven therein. We will consider initial data (constructed in Section \ref{section:initialdatasigmazero}) on the hypersurface $\Sigma_0$ for $C_R\gg 1$ (see Definition \ref{defi:sigmazeroII}) and we construct regions $\deux$ and $\trois$, which extend up to the Cauchy horizon $\ch$. The spacetime is constructed in the double null gauge, relying on the work \cite{stabC0} of Dafermos and Luk on the $C^0$ stability of the Kerr Cauchy horizon. On top of the estimates proven in \cite{stabC0}, we prove estimates for more derivatives, with more decay (which is possible because we have more initial decay on $\Sigma_0$ for the linearized quantities in the double null gauge), and we also deduce pointwise decay from the integrated estimates.
\begin{rem}\label{rem:CRfixee}
	In Section \ref{section:IIdoublenull}, we consider the constant $C_R$ to be fixed. Thus, in Section \ref{section:IIdoublenull} only, the implicit constants in the bounds $\lesssim$ depend on $a,M$, and $C_R$ (except at the end where we recover bounds which do not depend on $C_R$, sacrificing some small polynomial decay).
\end{rem}
\subsection{Setting up the double null foliation on $\Sigma_0$}\label{section:initialdatasigmazero}

Recall from Definition \ref{defi:sigmazeroII} that $\Sigma_0=\{\mathring{u}+\mathring{\ubar}=C_R\}\subset\un$. We follow \cite[Section A.8]{stabC0} in the following definition:
\begin{defi}\label{defi:initialdatasig0}
	We denote $\hat{g}$ the metric induced by $(\un,\g)$ on $\Sigma_0$. We also denote $\Sigma_\mck$ the analog hypersurface in exact Kerr, and $\hat{g}_\mck$ the induced metric on $\Sigma_\mck$ by the Kerr metric. Finally, we respectively define $\hat{k}$ and $\hat{k}_\mck$ to be the second fundamental forms of $\Sigma_0$ in $(\un,\g)$, and of $\Sigma_\mck$ in $(\un_\mck,\g_\mck)$. 
\end{defi}
From \cite[Section A.8]{stabC0}, in local coordinates $\ubarring,\theta^1,\theta^2$ where $\theta^1,\theta^2$ is local coordinate system on the topological spheres
$\mathring{S}(\mathring{u}_0,\mathring{\ubar}_0):=\{\mathring{u}=\mathring{u}_0\}\cap\{\mathring{\ubar}=\mathring{\ubar}_0\}$, $\hat{g}_\mck$ takes the form
\begin{equation}\label{eq:ghatdanskerr}
	\begin{gathered}
		\hat{g}_\mck=\Phi_\mck\dee\ubarring\otimes\dee\ubarring+(w_\mck)_A\left(\dee\ubarring\otimes\dee\theta^A+\dee\theta^A\otimes\dee\ubarring\right)+(\gamma_\mck)_{AB}\dee\theta^A\otimes\dee\theta^B,\\
		\Phi_\mck:=(-\gamma_\mck)_{AB}(b_\mck)^A(b_\mck)^B+4\Omega_\mck^2,\quad (w_\mck)_A:=-(\gamma_\mck)_{AB}(b_\mck)^B.
	\end{gathered}
\end{equation}

In view of \cite[Section 4.1]{stabC0}, there exists a natural diffeomorphism $\Phi:\Sigma_0\longrightarrow\Sigma_\mck$ such that $\hat{g}-\Phi_* \hat{g}_\mck$ and $\hat{k}-\Phi_*\hat{k}_\mck$ are small. Denoting $\theta^1=\theta_*,\theta^2=\phi_*$, this diffeomorphism is given by
$$\Phi : x\in\Sigma_0\longmapsto (\mathring{u}(x), C_R-\mathring{u}(x),\theta^A(x))\in\Sigma_\mck,$$
namely $\Phi$ is the identification of $\Sigma_0$ and $\Sigma_\mck$ with respect to the coordinates $(\mathring{u},\theta^A)$, which is a well defined local coordinate system in $\Sigma_0$ (because it $\mathring{u},\theta^A$ are defined as their Kerr values with respect to the coordinate system $r,\ubar,\theta,\phi_+$ in $\un$). \textbf{From now on, we omit the pullback map $\Phi_*$ in the notations $\Phi_*\hat{g}_\mck$, $\Phi_*\hat{k}_\mck$.}

\subsubsection{Preliminary estimates on the coordinates $(\mathring{u},\mathring{\ubar},\theta^A)$ in $\un$}\label{section:premsecdoublenulll}
As $\mathring{u},\mathring{\ubar}$ are defined as their Kerr values with respect to $(r,\ubar,\theta)$ we have in $\un$\footnote{Note that in the future of $\Sigma_0$, once these have been constructed we will later use the notations $\uring,\ubarring$ to denote the double null advanced and retarded time, which do not coincide with these Kerr values with respect to $(r,\ubar,\theta)$.},
$$\mathring{u}+\mathring{\ubar}=2r'_*(r,\theta),$$
where the function $r'_*(r,\theta)$ is defined in Section \ref{section:linearizationdoublenull}. This implies that $r'_*,\theta_*$ satisfy, in $\un$,
\begin{align}\label{eq:derparrstar22}
	\frac{\partial r'_*}{\partial r}=\frac{\sqrt{(r^2+a^2)^2-a^2\sin^2\theta_*\Delta}}{\Delta},&\quad \frac{\partial r'_*}{\partial \theta}=a\sqrt{\sin^2\theta_*-\sin^2\theta},	\\
	\frac{\partial \theta_*}{\partial r}=\frac{1}{G\sqrt{(r^2+a^2)^2-a^2\sin^2\theta_*}},&\quad \frac{\partial \theta_*}{\partial \theta}=\frac{-1}{aG\sqrt{\sin^2\theta_*-\sin^2\theta}},\label{eq:derparthetastar}
\end{align}
where $G$ is defined in \cite[(A.19)]{stabC0} and satisfies \eqref{eq:bornepourGaa} in $\un$. We also have the identities for the derivatives of $\phi_*=\phi_*(\phi_+,r,\theta)$ in $\un$, 
\begin{align}\label{eq:derparphistar}
	\frac{\partial \phi_*}{\partial \phi_+}=1,\quad\frac{\partial \phi_*}{\partial r}=-\frac{a}{\Delta}-\frac{\partial r'_*}{\partial r}\frac{\partial h}{\partial r'_*}-\frac{\partial \theta_*}{\partial r}\frac{\partial h}{\partial \theta_*},\quad\frac{\partial \phi_*}{\partial\theta}=-\frac{\partial r'_*}{\partial \theta}\frac{\partial h}{\partial r'_*}-\frac{\partial \theta_*}{\partial \theta}\frac{\partial h}{\partial \theta_*},
\end{align}
where we used the Kerr value $\phi_*=\phi_+-r_{mod}-h(r'_*,\theta_*)$ where $r_{mod}'(r)=a/\Delta$ and where recalling Section \ref{section:linearizationdoublenull}, the function $h$ is defined via the transport equation
\begin{align}\label{eq:definitiondehlatusais}
	\frac{\partial h(r'_*,\theta_*)}{\partial r'_*}=-\frac{2Mar}{|q|^2(r^2+a^2)+2Ma^2r\sin^2\theta}=-\frac{2Mar}{(r^2+a^2)^2-a^2\sin^2\theta\Delta}
\end{align}
for any fixed $\theta_*$, and $h(0,\theta_*)=0$. 
\begin{prop}\label{prop:derparubarprim}
	We have, in Kerr, the following identities for the derivatives of $\ubarring$,
	\begin{align*}
		e_3(\mathring{\ubar})_\mck&=\frac{r^2+a^2-\sqrt{(r^2+a^2)^2-a^2\sin^2\theta_*\Delta}}{\Delta}=\frac12\frac{a^2\sin^2\theta_*}{r^2+a^2}+O(\Delta),\\
		e_4(\mathring{\ubar})_\mck&=\frac{r^2+a^2+\sqrt{(r^2+a^2)^2-a^2\sin^2\theta_*\Delta}}{|q|^2},\\
		(\nabla\mathring{\ubar})_\mck&=\frac{a\Imag(\mathfrak{J})\sqrt{\sin^2\theta_*-\sin^2\theta}}{\sin\theta}+a\Real(\mathfrak{J}),
	\end{align*}
the following identities for the derivatives of $\mathring{u}$,
\begin{align*}
	e_3(\mathring{u})_\mck&=-\frac{r^2+a^2+\sqrt{(r^2+a^2)^2-a^2\sin^2\theta_*\Delta}}{\Delta},\\
	e_4(\mathring{u})_\mck&=\frac{r^2+a^2-\sqrt{(r^2+a^2)^2-a^2\sin^2\theta_*\Delta}}{|q|^2}=O(\Delta),\\
	(\nabla\mathring{u})_\mck&=\frac{a\Imag(\mathfrak{J})\sqrt{\sin^2\theta_*-\sin^2\theta}}{\sin\theta}-a\Real(\mathfrak{J}),
\end{align*}
the following identities for the derivatives of $\theta_*$,
\begin{align*}
	e_3(\theta_*)_\mck&=-\frac{1}{G\sqrt{(r^2+a^2)^2-a^2\sin^2\theta_*\Delta}},\\
	e_4(\theta_*)_\mck&=\frac{\Delta}{|q|^2G\sqrt{(r^2+a^2)^2-a^2\sin^2\theta_*\Delta}},\\
	(\nabla\theta_*)_\mck&=-\frac{1}{\sin\theta a G\sqrt{\sin^2\theta_*-\sin^2\theta}}\Imag(\mathfrak{J}),
\end{align*}
and the following identities for the derivatives of $\phi_*$,
\begin{align*}
	e_3(\phi_*)_\mck&=\frac{a}{\Delta}+\frac{\partial r'_*}{\partial r}\frac{\partial h}{\partial r'_*}+\frac{\partial \theta_*}{\partial r}\frac{\partial h}{\partial \theta_*},\\
	e_4(\phi_*)_\mck&=\frac{a}{|q|^2}-\frac{\Delta}{|q|^2}\left(\frac{\partial r'_*}{\partial r}\frac{\partial h}{\partial r'_*}+\frac{\partial \theta_*}{\partial r}\frac{\partial h}{\partial \theta_*}\right),\\
	(\nabla\phi_*)_\mck&=\sin^{-2}\theta\left(\Jplus\Real(\frakJ_-)-\Jmoins\Real(\frakJ_+)\right)-\frac{1}{\sin\theta}\Imag(\mathfrak{J})_b\left(\frac{\partial r'_*}{\partial \theta}\frac{\partial h}{\partial r'_*}+\frac{\partial \theta_*}{\partial \theta}\frac{\partial h}{\partial \theta_*}\right).
\end{align*}
\end{prop}
\begin{proof}
	These identities are straightforward using \eqref{eq:derparrstar22}, \eqref{eq:derparthetastar}, \eqref{eq:derparphistar} and the Kerr values of $e_\mu(r,\theta,\phi_+)$, $\mu=1,2,3,4$.
\end{proof}

\begin{rem}\label{rem:eqregthetathetastar}
	Just like in Kerr, we have the following estimates in $\un$,
	\begin{align}\label{eq:eqregthetathetastar}
		\sin\theta\sim\sin\theta_*,\quad\cos\theta\sim\cos\theta_*.
	\end{align}
	This comes from the fact that $\theta_*\in[\theta,\pi/2)$ for $\theta\in(0,\pi/2)$ and $\theta_*=\pi-\theta_*(r,\pi-\theta)$ for $\theta\in(\pi/2,\pi)$ (see \cite[Section A.1]{stabC0}), and the estimate $\sin\theta_*\lesssim\sin\theta$, $\cos\theta\lesssim \cos\theta_*$ for $\theta\in(0,\pi/2)$ (see \cite[Prop. A.1]{stabC0}).
\end{rem}
Now we introduce coordinate patches which cover the whole spheres $\mathring{S}(\mathring{u}_0,\mathring{\ubar}_0)$. 
\begin{defi}\label{defi:defthetadansmcvi}
To control the metric and second fundamental form induced on $\Sigma_0$, it is convenient introduce the following local coordinate systems on the spheres $\mathring{S}(\mathring{u},\mathring{\ubar})$ as follows : 
\begin{itemize}
	\item On $\mathcal{V}_1:=\un\cap\{\pi/4<\theta_*<3\pi/4\}$ we define $$(\theta^{1}_{(1)},\theta^2_{(1)}):=\left(\theta_*,\phi_*\right).$$
	\item On $\mathcal{V}_2:=\un\cap\{0\leq\theta_*<\pi/3\}$ and $\mathcal{V}_3:=\un\cap\{2\pi/3<\theta_*\leq\pi\}$ we define for $i=2,3$, $$(\theta^1_{(i)},\theta^2_{(i)}):=\left(\sin\theta_*\cos\phi_*,\sin\theta_*\sin\phi_*\right).$$
\end{itemize}

\end{defi}
\begin{prop}\label{prop:derstereoII}
	Recalling Proposition \ref{prop:derparubarprim}, we have the following identities in Kerr for the derivatives of $\theta^A_{(i)}$, $A=1,2$, $i=2,3$:
	\begin{align*}
		e_3(\sin\theta_*\cos\phi_*)&=\cos\theta_*\cos\phi_* e_3(\theta_*)_\mck-\sin\theta_*\sin\phi_* e_3(\phi_*)_\mck,\\ e_4(\sin\theta_*\cos\phi_*)&=\cos\theta_*\cos\phi_* e_4(\theta_*)_\mck-\sin\theta_*\sin\phi_* e_4(\phi_*)_\mck,\\
		e_3(\sin\theta_*\sin\phi_*)&=\cos\theta_*\sin\phi_* e_3(\theta_*)_\mck+\sin\theta_*\cos\phi_* e_3(\phi_*)_\mck,\\ e_4(\sin\theta_*\sin\phi_*)&=\cos\theta_*\sin\phi_* e_4(\theta_*)_\mck+\sin\theta_*\cos\phi_* e_4(\phi_*)_\mck,
	\end{align*}
as well as
\begin{align*}
	\nabla(\sin\theta_*\cos\phi_*)=&-\frac{\cos\theta_*\cos\phi_*}{a\cos\theta G\sqrt{\sin^2\theta_*-\sin^2\theta}}\left(\cos\phi_+\Real(\frakJ_+)+\sin\phi_+\Real(\frakJ_-)\right)-\sin\theta_*\sin\phi_* (\nabla\phi_*)_\mck,\\
	\nabla(\sin\theta_*\sin\phi_*)=&-\frac{\cos\theta_*\sin\phi_*}{a\cos\theta G\sqrt{\sin^2\theta_*-\sin^2\theta}}\left(\cos\phi_+\Real(\frakJ_+)+\sin\phi_+\Real(\frakJ_-)\right)+\sin\theta_*\cos\phi_* (\nabla\phi_*)_\mck.
\end{align*}
\end{prop}
\begin{proof}
	This is a direct consequence of Proposition \ref{prop:derparubarprim}, where we also use the identity
	$$\cos\theta\nabla\cos\theta+J^{(+)}\nabla J^{(+)}+J^{(-)}\nabla J^{(-)}=\frac12\nabla(\cos^2\theta+(J^{(+)})^2+(J^{(-)})^2)=\frac12\nabla (1)=0$$
	for the $\nabla$ derivatives.
\end{proof}
\begin{rem}
	The identity used in the proof above generalizes as follows: for any vector field $X$ we have in $\un$
\begin{align}\label{eq:switchsympadisdonc}
	\cos\theta X(\cos\theta)+J^{(+)} X(J^{(+)})+J^{(-)} X(J^{(-)})=0.
\end{align}
\end{rem}
\begin{defi}
	For $i=1,2,3$, we denote for convenience $(x'^\nu_{(i)})$, $\nu=1,2,3,4$ the coordinate system $(\mathring{u},\mathring{\ubar},\theta^1_{(i)},\theta^2_{(i)})$ in region $\mcv_i$. We also denote for $i=1,2$ and $\nu=1,2,3,4$,
	$$\widecheck{e_{4}(x'^\nu_{(i)})}-e_{4}(x'^\nu_{(i)})_\mck,\quad \widecheck{\nabla x'^\nu_{(i)}}=\nabla x'^\nu_{(i)}-(\widecheck{\nabla x'^\nu_{(i)}})_\mck,$$
	where $e_{4}(x'^\nu_{(i)})_\mck,(\widecheck{\nabla x'^\nu_{(i)}})_\mck$ are defined in Propositions \ref{prop:derparubarprim} and \ref{prop:derstereoII}.
\end{defi}
 We are now ready to state and prove the main result of this section.
\begin{prop}\label{prop:diffderiveescoordsDN}
	We have in $\mcv_i$, for $i=1,2,3$, and $\nu=1,2,3,4$,
	$$|\df^{\leq N_0-3}(\widecheck{e_{4}(x'^\nu_{(i)})},\widecheck{\nabla x'^\nu_{(i)}})|\lesssim \varepsilon\ubarring^{-3-\delta/2},$$
and ${e_{3}(x'^\nu_{(i)})}=e_{3}(x'^\nu_{(i)})_\mck$ (where $e_{3}(x'^\nu_{(i)})_\mck$ is defined in Propositions \ref{prop:derparubarprim} and \ref{prop:derstereoII}).
\end{prop}
\begin{proof}
For any $x\in(u,\ubar,\theta_*,\phi_*)$ and $X\in\{e_3,e_4,e_b\}$,
\begin{align*}
	X(x)&=X(r)\frac{\partial x}{\partial r}+X(\ubar)\frac{\partial x}{\partial \ubar}+X(\theta)\frac{\partial x}{\partial \theta}+X(\phi_+)\frac{\partial x}{\partial \phi_+}\\
	&=X(r)\left(\frac{\partial x}{\partial r}\right)_\mck+X(\ubar)\left(\frac{\partial x}{\partial \ubar}\right)_\mck+X(\theta)\left(\frac{\partial x}{\partial \theta}\right)_\mck+X(\phi_+)\left(\frac{\partial x}{\partial \phi_+}\right)_\mck,
\end{align*} 
where the Kerr values $\frac{\partial x}{\partial r},\frac{\partial x}{\partial \ubar},\frac{\partial x}{\partial \theta},\frac{\partial x}{\partial \phi_+}$ can be deduced from \eqref{eq:derparrstar}, \eqref{eq:derparthetastar}, \eqref{eq:derparphistar}.
Using this combined with the fact that the $e_3$ derivatives of $r,\ubar,\theta,\phi_+$ coincide with their Kerr values we get in $\un$,
$${e_{3}(x'^\nu_{(i)})}=e_{3}(x'^\nu_{(i)})_\mck$$
as well as the following identities for the derivatives of $\mathring{\ubar}$,
	\begin{align*}
	{e_4(\mathring{\ubar})}-{e_4(\mathring{\ubar})}_\mck&=\frac{\sqrt{(r^2+a^2)^2-a^2\sin^2\theta_*\Delta}}{\Delta}\widecheck{e_4(r)}+\widecheck{e_4(\ubar)}-ae_4(\cos\theta)\frac{\sqrt{\sin^2\theta_*-\sin^2\theta}}{\sin\theta},\\
	\nabla\mathring{\ubar}-(\nabla\mathring{\ubar})_\mck&=\frac{\sqrt{(r^2+a^2)^2-a^2\sin^2\theta_*\Delta}}{\Delta}\nabla r+\widecheck{\nabla\ubar}-a\sqrt{\sin^2\theta_*-\sin^2\theta}\frac{\widecheck{\nabla\cos\theta}}{\sin\theta},
\end{align*}
the following identities for the derivatives of $\mathring{u}$,
	\begin{align*}
	{e_4(\mathring{u})}-{e_4(\mathring{u})}_\mck&=\frac{\sqrt{(r^2+a^2)^2-a^2\sin^2\theta_*\Delta}}{\Delta}\widecheck{e_4(r)}-\widecheck{e_4(\ubar)}-ae_4(\cos\theta)\frac{\sqrt{\sin^2\theta_*-\sin^2\theta}}{\sin\theta},\\
	\nabla\mathring{u}-(\nabla\mathring{u})_\mck&=\frac{\sqrt{(r^2+a^2)^2-a^2\sin^2\theta_*\Delta}}{\Delta}\nabla r-\widecheck{\nabla\ubar}-a\sqrt{\sin^2\theta_*-\sin^2\theta}\frac{\widecheck{\nabla\cos\theta}}{\sin\theta},
\end{align*}
the following identities for the derivatives of $\theta_*$,
\begin{align*}
	e_4(\theta_*)-e_4(\theta_*)_\mck&=\frac{e_4(\cos\theta)}{a\sin\theta G\sqrt{\sin^2\theta_*-\sin^2\theta}}+\frac{\widecheck{e_4(r)}}{G\sqrt{(r^2+a^2)^2-a^2\sin^2\theta_*\Delta}},\\
	\nabla\theta_*-(\nabla\theta_*)_\mck&=\frac{\widecheck{\nabla\cos\theta}}{a\sin\theta G\sqrt{\sin^2\theta_*-\sin^2\theta}}+\frac{\nabla r}{G\sqrt{(r^2+a^2)^2-a^2\sin^2\theta_*\Delta}},
\end{align*}
the following identities for the derivatives of $\phi_*$,
\begin{align*}
	e_4(\phi_*)-e_4(\phi_*)_\mck&=\frac{\cos\phi_+\widecheck{e_4(\Jplus)}-\sin\phi_+\widecheck{e_4(\Jmoins)}}{\sin\theta}-\frac{a}{\Delta}\widecheck{e_4(r)}\\
	\quad&-\frac{\partial h}{\partial r'_*}\left(\widecheck{e_4(r)}\frac{\partial r'_*}{\partial r}-\frac{e_4(\cos\theta)}{\sin\theta}\frac{\partial \theta_*}{\partial\theta}\right)-\frac{\partial h}{\partial \theta_*}(e_4(\theta_*)-e_4(\theta_*)_\mck),\\
	\nabla\phi_*-(\nabla\phi_*)_\mck&=\frac{\cos\phi_+\widecheck{\nabla \Jplus}-\sin\phi_+\widecheck{\nabla\Jmoins}}{\sin\theta}-\frac{a}{\Delta}\nabla r\\
	\quad&-\frac{\partial h}{\partial r'_*}\left(\nabla r\frac{\partial r'_*}{\partial r}-\frac{\widecheck{\nabla\cos\theta}}{\sin\theta}\frac{\partial \theta_*}{\partial\theta}\right)-\frac{\partial h}{\partial \theta_*}(\nabla\theta_*-(\nabla\theta_*)_\mck),
\end{align*}
the following identities for the derivatives of $\sin\theta_*\cos\phi_*$ and $\sin\theta_*\cos\phi_*$,
\begin{equation}\label{eq:gathereddddwsh}
	\begin{gathered}
		e_4(\sin\theta_*\cos\phi_*)-e_4(\sin\theta_*\cos\phi_*)_\mck=\widecheck{e_4(\theta_*)}\cos\theta_*\cos\phi_*-\widecheck{e_4(\phi_*)}\sin\theta_*\sin\phi_*,\\	\nabla(\sin\theta_*\cos\phi_*)-\nabla(\sin\theta_*\cos\phi_*)_\mck=\widecheck{\nabla\theta_*}\cos\theta_*\cos\phi_*-\widecheck{\nabla\phi_*}\sin\theta_*\sin\phi_*,\\	e_4(\sin\theta_*\sin\phi_*)-e_4(\sin\theta_*\sin\phi_*)_\mck=\widecheck{e_4(\theta_*)}\cos\theta_*\sin\phi_*+\widecheck{e_4(\phi_*)}\sin\theta_*\cos\phi_*,\\	\nabla(\sin\theta_*\sin\phi_*)-\nabla(\sin\theta_*\sin\phi_*)_\mck=\widecheck{e_4(\theta_*)}\cos\theta_*\sin\phi_*+\widecheck{e_4(\phi_*)}\sin\theta_*\cos\phi_*.
	\end{gathered}
\end{equation}
The stated estimates thus directly follow from the bounds \eqref{eq:controlI} for $\Gammacheck$, keeping in mind Remark \ref{rem:eqregthetathetastar} which gives $\sin\theta\sim\sin\theta_*\sim 1$ in $\mcv_1$ and $\cos\theta\sim\cos\theta_*\sim 1$ in $\mcv_2\cup\mcv_3$, and \eqref{eq:switchsympadisdonc}.
\end{proof}
\subsubsection{Estimates for $\hat{g}-\hat{g}_\mck$ and $\hat{k}-\hat{k}_\mck$ on $\Sigma_0$}

\begin{defi}\label{def:normalSig0}
	Recalling $r'_*=(\mathring{u}+\mathring{\ubar})/2$, we define on $\Sigma_0$ the following vector field,
	$$N:=\frac{\D r'_*}{\sqrt{-\g(\D r'_*,\D r'_*)}},$$
	which is the future directed unit normal to $\Sigma_0$.
\end{defi}
Note that $N$ is well-defined. Indeed, we have	$\D r'_*=-\frac12 e_3(r'_*) e_4-\frac12 e_4(r'_*)e_3+\nabla^b r'_* e_b$ where, by Proposition \ref{prop:derparubarprim},
\begin{equation*}
	\begin{gathered}
		e_3(r'_*)=-\frac{\sqrt{(r^2+a^2)^2-a^2\sin^2\theta_*\Delta}}{\Delta},\quad e_4(r'_*)=\frac{\sqrt{(r^2+a^2)^2-a^2\sin^2\theta_*\Delta}}{|q|^2}+O(\varepsilon\ubar^{-3-\delta/2}),\\
	\nabla r'_*=\frac{a\Imag(\mathfrak{J})\sqrt{\sin^2\theta_*-\sin^2\theta}}{\sin\theta}+O(\varepsilon\ubar^{-3-\delta/2}).
	\end{gathered}
\end{equation*}
This implies 
\begin{align*}
	\g(\D r'_*,\D r'_*)&=-e_3(r'_*)e_4(r'_*)+|\nabla r'_*|^2\\
	&=\frac{(r^2+a^2)^2-a^2\sin^2\theta\Delta}{\Delta|q|^2}+O(\varepsilon\ubar^{-3-\delta/2})<-C(a,M,\delta_\pm),
\end{align*}
where $C(a,M,\delta_\pm)>0$ depends only on $a,M,\delta_\pm$, hence the fact that $N$ is well defined. Also note that from the estimates above we have
\begin{align}\label{eq:expreNnormalsig0}
	N=&\frac{\sqrt{(r^2+a^2)^2-a^2\sin^2\theta_*\Delta}}{2\Delta}e_4-\left(\frac{\sqrt{(r^2+a^2)^2-a^2\sin^2\theta_*\Delta}}{2|q|^2}+\err\right)e_3\nn\\
	&+\left(\frac{a\sqrt{\sin^2\theta_*-\sin^2\theta}}{\sin\theta}\Imag(\mathfrak{J})^b+\err^b\right)e_b,
\end{align}
where $|\df^{\leq N_1}\err|\lesssim\varepsilon\ubar^{-3-\delta/2}$, where we recall $N_1=N_0-3$.

\begin{defi}We introduce the following notations:
\begin{itemize}
	\item For $i=1,2,3$ we denote $\bar{x}^j_{(i)}$, $j=1,2,3$ the coordinates $$(\bar{x}^j_{(i)})=(\mathring{\ubar},\theta^1_{(i)}, \theta^2_{(i)})\quad\text{ on }\quad \Sigma_0\cap\mcv_i,$$ where $\theta^A_{(i)}$ is defined in Definition \ref{defi:defthetadansmcvi}. The coordinate vector fields with respect to these coordinates (which are tangent to $\Sigma_0$) are denoted as $\bar{\partial}_j^{(i)}$, $j=1,2,3$. 
	\item We also denote $\hat{\nabla}$ the Levi-Civita connection of $(\Sigma_0,\hat{g})$.
\end{itemize}

\end{defi}
\textbf{We use the convention that on $\Sigma_0$, the norms of tensors tangential to $\Sigma_0$ are taken with respect to the Riemannian metric $\hat{g}_\mck$}, which we recall is given by \eqref{eq:ghatdanskerr}. 

In the following result, we denote $\err$ any function such that $|\df^{\leq N_1}\err|\lesssim\varepsilon\ubar^{-3-\delta/2}$ and for $i=1,2,3$, we denote $\err^{(i)}$ any function such that $|\df^{\leq N_1}\err^{(i)}|\lesssim\varepsilon\ubar^{-3-\delta/2}$ in $\mcv_i$. 
\begin{lem}\label{lem:expredmuDN}
	Recalling \eqref{eq:defTdansI} have the following identities on $\Sigma_0$, for $i=1,2,3$,
	\begin{align*}
		\bar{\partial}_1^{(i)}=\T+\err\cdot\df.
	\end{align*}
Also, for $i=1$,
\begin{align*}
	\bar{\partial}_2^{(1)}=F_3 e_3+F_4e_4+\left(\frac{F}{\sin\theta}\Imag(\frakJ)^b+\frac{\partial h}{\partial \theta_*}(r^2+a^2)\Real(\frakJ)^b\right)e_b+\err^{(1)}\cdot\df
\end{align*}
where
\begin{align*}
	F_3&=\frac12\left(\frac{\partial h}{\partial\theta_*}\frac{a\sin^2\theta\Delta}{|q|^2}-\frac{a^2\Delta(\sin^2\theta_*-\sin^2\theta)G\sqrt{(r^2+a^2)^2-a^2\sin^2\theta_*\Delta}}{(r^2+a^2)^2-a^2\sin^2\theta\Delta}\right),\\
	F_4&=\frac{|q|^2}{2\Delta}\left(\frac{\partial h}{\partial\theta_*}\frac{a\sin^2\theta\Delta}{|q|^2}+\frac{a^2\Delta(\sin^2\theta_*-\sin^2\theta)G\sqrt{(r^2+a^2)^2-a^2\sin^2\theta_*\Delta}}{(r^2+a^2)^2-a^2\sin^2\theta\Delta}\right),\\
	F&=-\frac{a\sqrt{\sin^2\theta_*-\sin^2\theta}G|q|^2({(r^2+a^2)^2-a^2\sin^2\theta_*\Delta})}{(r^2+a^2)^2-a^2\sin^2\theta\Delta},
\end{align*}
and
\begin{align*}
	\bar{\partial}_3^{(1)}&=\Z+\err^{(1)}\cdot\df.
\end{align*}
Finally, for $i=2,3$,
\begin{align*}
	\bar{\partial}_2^{(i)}=&\left(\frac{\cos\phi_*}{\cos\theta_*}F_3+\frac{a\sin\theta^2\sin\phi_*\Delta}{|q|^2\sin\theta_*}\right)e_3+\left(\frac{\cos\phi_*}{\cos\theta_*}F_4+\frac{a\sin\theta^2\sin\phi_*}{\sin\theta_*}\right)e_4\\
	&+\Bigg(\frac{\cos\phi_*}{\cos\theta_*}\left(\frac{\partial h}{\partial \theta_*}(r^2+a^2)\Real(\frakJ)^b+F\frac{\cos\phi_+\Real(\frakJ_+)^b+\sin\phi_+\Real(\frakJ_-)^b}{\cos\theta}\right)\\
	&\quad\quad+\frac{\sin\phi_*\sin\theta(r^2+a^2)}{\sin\theta_*\cos\theta}(\cos\phi_+\Imag(\frakJ_+)^b+\sin\phi_+\Imag(\frakJ_-)^b)\Bigg)e_b+\err^{(i)}\cdot\df
\end{align*}
and 
\begin{align*}
	\bar{\partial}_3^{(i)}=&\left(\frac{\sin\phi_*}{\cos\theta_*}F_3-\frac{a\sin\theta^2\cos\phi_*\Delta}{|q|^2\sin\theta_*}\right)e_3+\left(\frac{\sin\phi_*}{\cos\theta_*}F_4-\frac{a\sin\theta^2\cos\phi_*}{\sin\theta_*}\right)e_4\\
	&+\Bigg(\frac{\sin\phi_*}{\cos\theta_*}\left(\frac{\partial h}{\partial \theta_*}(r^2+a^2)\Real(\frakJ)^b+F\frac{\cos\phi_+\Real(\frakJ_+)^b+\sin\phi_+\Real(\frakJ_-)^b}{\cos\theta}\right)\\
	&\quad\quad-\frac{\cos\phi_*\sin\theta(r^2+a^2)}{\sin\theta_*\cos\theta}({\cos\phi_+\Imag(\frakJ_+)^b+\sin\phi_+\Imag(\frakJ_-)^b})\Bigg)e_b+\err^{(i)}\cdot\df.
\end{align*}
\end{lem}
\begin{proof}
	We denote $E_j^{(i)}$ the vector fields such that the identities stated above rewrite
	$$\bar{\partial}_j^{(i)}=E_j^{(i)}+\err^{(i)}\cdot\df,$$
	which can be seen to coincide with $\bar{\partial}_j^{(i)}$ in exact Kerr. Moreover, by Proposition \ref{prop:diffderiveescoordsDN} we can compute in each $\mcv_i$,
	$$E_j^{(i)}(\bar{x}^k_{(i)})=\delta_{kj}+\err^{(i)}.$$ 
	Thus we get $E_j^{(i)}=\bar{\partial}_j^{(i)}+\err^{(i)}\cdot(\bar{\partial}_1^{(i)},\bar{\partial}_2^{(i)},\bar{\partial}_3^{(i)})$, and inversing the system concludes the proof.
\end{proof}
\begin{prop}\label{prop:bornegchecksig0}
	We have the following estimate on $\Sigma_0$,
	$$|\hat{\nabla}_\mck^{\leq N_1}(\hat{g}-\hat{g}_\mck)|\lesssim\varepsilon\mathring{\ubar}^{-3-\delta/2}.$$
\end{prop}
\begin{proof}
	By Lemma \ref{lem:expredmuDN} we compute, in coordinates $\bar{x}^j_{(i)}$,
	$$\hat{g}_{jk}=\g(\bar{\partial}_j^{(i)},\bar{\partial}_k^{(i)})=(\hat{g}_\mck)_{jk}+\err^{(i)},$$
	where as before $|\df^{\leq N_1}\err^{(i)}|\lesssim\varepsilon\ubar^{-3-\delta/2}$ in $\mcv_i$. Using Lemma \ref{lem:expredmuDN} we deduce
	\begin{align}\label{eq:wantedentens}
		|(\bar{\partial}_1^{(i)},\bar{\partial}_2^{(i)},\bar{\partial}_3^{(i)})^{\leq N_1}(\hat{g}_{jk}-(\hat{g}_\mck)_{jk})|\lesssim |\df^{\leq k}\err^{(i)}|\lesssim\varepsilon\ubar^{-3-\delta/2}
	\end{align}
on $\Sigma_0\cap\mcv_i$. Now we deduce from the coordinate bound above the stated tensorial bound for $\hat{\nabla}^m_\mck(\hat{g}-\hat{g}_\mck)$. First, note that for $U$ a $\Sigma_0$-tangent $N$-tensor field, we have for $k,j_1,\ldots, j_N=1,2,3$,
\begin{align}\label{eq:identitynabSigzero}
	(\hat{\nabla}_\mck)_k U_{j_1\cdots j_N}=\bar{\partial}_kU_{j_1\cdots j_N}-\sum_{m=1}^N(\hat{\Gamma}_\mck)_{j_m k}^l U_{j_1\cdots l \cdots j_N}, 
\end{align}
where the Christoffel symbol $(\hat{\Gamma}_\mck)_{j_m k}^l$ are given by $(\hat{\Gamma}_\mck)_{jk}^l=\frac12\hat{g}_\mck^{ln}\left(\bar{\partial_j} (\hat{g}_\mck)_{nk}+\bar{\partial}_k(\hat{g}_\mck)_{nj}-\bar{\partial}_n(\hat{g}_\mck)_{jk}\right)$, which thus satisfy $|(\bar{\partial}_1^{(i)},\bar{\partial}_2^{(i)},\bar{\partial}_3^{(i)})^{\leq k}(\hat{\Gamma}_\mck)_{jk}^l|\lesssim_k 1$ in $\Sigma_0\cap\mcv_i$ for any $k\geq 0$. Iterating the identity \eqref{eq:identitynabSigzero} we infer
$$|\hat{\nabla}_\mck^{\leq N_1}(\hat{g}-\hat{g}_\mck)|\lesssim \max_{i=1,2,3}\sup_{\mcv_i}\max_{j,k=1,2}|(\bar{\partial}_1^{(i)},\bar{\partial}_2^{(i)},\bar{\partial}_3^{(i)})^{\leq N_1}(\hat{g}_{jk}-(\hat{g}_\mck)_{jk})|\lesssim \varepsilon\mathring{\ubar}^{-3-\delta},$$
where we used \eqref{eq:wantedentens} combined with Lemma \ref{lem:ubarssontpareils}. This concludes the proof.
\end{proof}
We now obtain the same result for the second fundamental form $\hat{k}$.
\begin{prop}\label{prop:bornekchecksig0}
	We have the estimate
	$$|\hat{\nabla}_\mck^{\leq N_1-1}(\hat{k}-\hat{k}_\mck)|\lesssim\varepsilon\mathring{\ubar}^{-3-\delta/2},$$
	where we recall from Definition \ref{defi:initialdatasig0} that $\hat{k}$ is the second fundamental form of $\Sigma_0$ in $(\un,\g)$,
	$$\hat{k}_{ij}:=-\g(\D_{\bar{\partial}_i}N,\bar{\partial}_j),$$
	where $N$ is the unit future directed vector normal to $\Sigma_0$, see Definition \ref{def:normalSig0}.
\end{prop}
\begin{proof}
The proof is very similar to the one of Proposition \ref{prop:bornegchecksig0}, using additionally the expression  \eqref{eq:expreNnormalsig0} for the normal $N$.
\end{proof}
The following result is a simple corollary of Propositions \ref{prop:bornegchecksig0} and \ref{prop:bornekchecksig0}.
\begin{cor}\label{cor:hypDLsigma0}
	For any $C_R\in\mathbb{R}$, there exists $u_f(a,M,\delta,C_R)\ll -1$ negative enough such that we have the following estimates along $\Sigma_0'=\Sigma_0\cap\{\ubar< C_R-u_f\}$,
\begin{align*}
	\|\mathring{\ubar}^{5/2+\delta/3}\hat{\nabla}_\mck^{\leq N_1}(\hat{g}-\hat{g}_\mck)\|_{L^2(\Sigma_0',\hat{g}_\mck)}+\|\mathring{\ubar}^{5/2+\delta/3}\hat{\nabla}_\mck^{\leq N_1-1}(\hat{k}-\hat{k}_\mck)\|_{L^2(\Sigma_0',\hat{g}_\mck)}&\lesssim\varepsilon,\\
	\|\mathring{\ubar}^{3+\delta/2}\hat{\nabla}_\mck^{\leq N_1}(\hat{g}-\hat{g}_\mck)\|_{L^\infty(\Sigma_0',\hat{g}_\mck)}+\|\mathring{\ubar}^{3+\delta/2}\hat{\nabla}_\mck^{\leq N_1-1}(\hat{k}-\hat{k}_\mck)\|_{L^\infty(\Sigma_0',\hat{g}_\mck)}&\lesssim\varepsilon.
\end{align*}
\end{cor}
\begin{proof}
	We have, by Proposition \ref{prop:bornegchecksig0},
	\begin{align*}
		\|\mathring{\ubar}^{5/2+\delta/3}\hat{\nabla}_\mck^{\leq N_1}(\hat{g}-\hat{g}_\mck)\|_{L^2(\Sigma_0\cap\{\ubar< C_R-u_f\},\hat{g}_\mck)}^2&\lesssim\int_{C_R-u'_f}^{+\infty}\mathring{\ubar}^{5+2\delta/3}\left(\int_{\mathring{S}(C_R-\mathring{\ubar},\mathring{\ubar})}|\hat{\nabla}_\mck^{\leq N_1}(\hat{g}-\hat{g}_\mck)|^2\right)\dee\mathring{\ubar}\\
		&\lesssim \varepsilon^2\int_{C_R-u'_f}^{+\infty}\mathring{\ubar}^{-1-\delta/3}\dee\mathring{\ubar}\lesssim\varepsilon^2\delta^{-1}(C_R-u_f)^{-\delta/3}\lesssim\varepsilon^2
	\end{align*}
for $u_f(a,M,\delta,C_R)\ll -1$. The $L^\infty(\Sigma_0,\hat{g}_\mck)$ bound for $\hat{g}-\hat{g}_\mck$ are direct consequences of Proposition \ref{prop:bornegchecksig0}. The bounds for $\hat{k}-\hat{k}_\mck$ are proven similarly, using Proposition \ref{prop:bornekchecksig0}.
\end{proof}
\begin{rem}\label{rem:alsoholdsDL}
	Note that Corollary \ref{cor:hypDLsigma0} implies, provided $N_1\geq 5$, 
	\begin{align*}
		\|\mathring{\ubar}^{1/2+\delta}\hat{\nabla}_\mck^{\leq 5}(\hat{g}-\hat{g}_\mck)\|_{L^2(\Sigma_0,\hat{g}_\mck)}+\|\mathring{\ubar}^{1/2+\delta}\hat{\nabla}_\mck^{\leq 4}(\hat{k}-\hat{k}_\mck)\|_{L^2(\Sigma_0,\hat{g}_\mck)}&\lesssim\varepsilon,\\
		\|\mathring{\ubar}^{1/2+\delta}\hat{\nabla}_\mck^{\leq 3}(\hat{g}-\hat{g}_\mck)\|_{L^\infty(\Sigma_0,\hat{g}_\mck)}+\|\mathring{\ubar}^{1/2+\delta}\hat{\nabla}_\mck^{\leq 2}(\hat{k}-\hat{k}_\mck)\|_{L^\infty(\Sigma_0,\hat{g}_\mck)}&\lesssim\varepsilon.
	\end{align*}
These estimates are  exactly the initial data assumptions of \cite{stabC0}. \textbf{Thus we can apply the main theorem in \cite{stabC0}, and any estimate in that paper also holds in the context of the present paper, for example the Sobolev and elliptic estimates proven in \cite[Section 5]{stabC0}}. However for our purpose we need a version of the main theorem in \cite{stabC0} with more decay and derivatives, which will be the main goal of the following subsection.
\end{rem}

\subsection{Dafermos-Luk estimates with more decay and angular derivatives}
\label{section:DLmoredecaynabring}
Before proving Theorem \ref{thm:regionII} starting from the initial data estimates stated in Corollary \ref{cor:hypDLsigma0}, we introduce and recall some notations along with a few remarks.
\subsubsection{Conventions and schematic notations in double null gauge}\label{section:schematicnotations}

\noindent\textbf{General notations.} From now on, and in the rest of the paper \textbf{we change notations for convenience and we denote $u,\ubar$ the double null advanced and retarded time which we construct in region $\deux\cup\trois$}, instead of $\mathring{u}$ and $\mathring{\ubar}$ as in region $\un$. There is no danger of confusion with the PT coordinate $\ubar$ of region $\un$ because we will not mention the double null coordinates $\mathring{u}$, $\mathring{\ubar}$ in region $\un$ anymore, as they will only be used in regions $\deux$ and $\trois$. Also, in Section \ref{section:IIdoublenull} all geometric quantities are now linearized with respect to the double null $(u,\ubar,\theta^1,\theta^2)$ differential structure, see Section \ref{section:linearizationdoublenull}. 

Moreover, following Section \ref{section:DNspacetimegeneral} we recall the notations for the double null pair \eqref{eq:DuDubardoublenull}:
$$\ering_3=-2\Omega^2\D\ubar=\partial_u,\quad \ering_4=-2\D u=\Omega^{-2}\left(\partial_\ubar+b^A\partial_{\theta^A}\right).$$
We also recall that we denote $\nabring_3,\:\nabring_4,\:\nabring_A,\:A=1,2$ the covariant derivative acting on $S(u,\ubar)$-tangent tensors in the directions $\ering_3$, $\ering_4$, $\partial_{\theta^A}$, as defined in \eqref{eq:nabdoublenul}, \eqref{eq:nab3doublenul}, \eqref{eq:nab4doublenul}. Finally, recall from Section \ref{section:DNspacetimegeneral} the notations $\mathring{\chi},\mathring{\chibar},\mathring{\eta},\ldots$ and $\mathring{\alpha},\mathring{\beta},\ldots$ for the Ricci and curvature coefficients in the double null gauge, as well as the metric coefficients $\Omega^2,\gamma,b$ in \eqref{eq:metricdoublenull}, and the scalars $\mathring{K},\hodge{\mathring{K}}$ in \eqref{eq:KKstarDNdeff}.

\noindent\textbf{Schematic notations.} We also recall from Section \ref{section:conventions} that the schematic notations $=_s$, $=_{rs}$ that we use coincide in the double null gauge with the Dafermos-Luk schematic and reduced schematic notations $=_S$, $=_{RS}$ as introduced respectively in \cite[Section 3.2]{stabC0} and \cite[Section 7.1]{stabC0}.

We note that although the schematic $=_s$ equations can be differentiated when taking covariant derivatives, when taking $\lieT$ derivatives of such schematic equations, there are additional terms which come from the fact that schematic equations hide the metric contractions in the notation, and $\T$ is only an approximate symetry (thus $\lieT\gamma\neq 0$ in general). For example, 
$$A=_s BC\implies \lieT A=_s (\lieT B)C+B(\lieT C)+BC(\lieT\gamma),$$
where $\gamma$ above can denote either the metric $\gamma$ on the spheres $S(u,\ubar)$ or the inverse metric $\gamma^{-1}$. We note however that these additional $\lieT\gamma$ terms will be easy to deal with because they will not change the global schematic structure of the null structure and Bianchi equations for $\lieT$ derivatives of the linearized double null quantities.

\noindent\textbf{Additional notations.} Following \cite[Section 3.2]{stabC0} we recall the following notations:
\begin{equation*}
	\begin{gathered}
			\psi\in\{\mathring\eta,\mathring\etabar\},\quad\psi_H\in\{\mathring{tr\chi},\mathring\chihat\},\quad\psi_{\Hbar}\in\{\mathring{tr\chibar},\mathring{\wh{\chibar}}\},\quad\psicheck\in\{\widecheck{\mathring\eta},\widecheck{\mathring\etabar}\},\\
		\widecheck{\psi}_H\in\{\widecheck{\mathring{tr\chi}},\widecheck{\mathring\chihat}\},\quad\widecheck{\psi}_{\Hbar}\in\{\widecheck{\mathring{tr\chibar}},\widecheck{\mathring{\wh{\chibar}}}\},\quad\gcheck\in\left\{\gammacheck,\widecheck{\gamma^{-1}},\frac{\Omega^2-\Omega_\mck^2}{\Omega^2},\log\Omega-\log\Omega_\mck\right\},
	\end{gathered}
\end{equation*}
and for any tensor $U$ and any operator $P$ (for example $P=\nabring,\lieT...$), we define schematically
$$P^i U^j=_s\sum_{i_1+\cdots+i_j=i}P^{i_1}\psi\cdots P^{i_j}U.$$

\noindent\textbf{Energies for controlling the solutions.} In Section \ref{section:DLmoredecaynabring}, we will prove a slightly different version of the Dafermos-Luk Theorem \cite[Thm. 4.24]{stabC0}, in the sense that we control more angular derivatives, using the initial decay assumption in Corollary \ref{cor:hypDLsigma0}. More precisely, let
 \begin{align}
 	N_2:= N_1-2,
 \end{align}
and for some auxiliary constant $N>0$ chosen later in the proof, we define the energies\footnote{The $L^p L^q L^r$ norms are precisely defined in \cite[Section 4.6]{stabC0}.},
\begin{align}
	\mcn_{int}^{(0)}:=\sum_{i\leq N_2}\Big(&\||u|^{\frac{5}{2}+\frac{\delta}{3}}\varpi^N\Omega_\mck\nabring^i(\psicheck_{\Hbar},\widecheck{\omegabar})\|_{L^2_u L^2_\ubar L^2_S}^2+\|\ubar^{\frac{5}{2}+\frac{\delta}{3}}\varpi^N\Omega_\mck^3\nabring^i\psicheck_{H}\|_{L^2_u L^2_\ubar L^2_S}^2\nn\\
	&+\|\ubar^{\frac{5}{2}+\frac{\delta}{3}}\varpi^N\Omega_\mck(\nabring^i(\psicheck,\gcheck,\bcheck),\nabring^{\min(i,N_2-1)}\widecheck{\mathring{K}})\|_{L^2_u L^2_\ubar L^2_S}^2\Big),\label{eq:mcnintnab}
\end{align}
as well as
\begin{align}
	\mcn_{hyp}^{(0)}:=&\sum_{i\leq N_2-1}\Big(\||u|^{\frac{5}{2}+\frac{\delta}{3}}\varpi^N\nabring^i(\psicheck_{\Hbar},\widecheck{\omegabar})\|^2_{L^2_u L^\infty_\ubar L^2_S}+\|\ubar^{\frac{5}{2}+\frac{\delta}{3}}\varpi^N\Omega_\mck\nabring^i\etacheck\|^2_{L^2_u L^\infty_\ubar L^2_S}\nn\\
	&+\||u|^{\frac{5}{2}+\frac{\delta}{3}}\varpi^N\Omega_\mck(\nabring^i\etabarcheck,\nabring^{\min(i,N_2-2)}\widecheck{\mathring{K}})\|^2_{L^\infty_\ubar L^2_u  L^2_S}+\|\ubar^{\frac{5}{2}+\frac{\delta}{3}}\varpi^N\Omega_\mck^2\nabring^i\psicheck_H\|^2_{L^2_\ubar L^\infty_u L^2_S}\nn\\
	&+ \|\ubar^{\frac{5}{2}+\frac{\delta}{3}}\varpi^N\Omega_\mck(\nabring^i\etacheck,\nabring^i\etabarcheck,\nabring^{\min(i,N_2-2)}\widecheck{\mathring{K}})\|^2_{L^2_\ubar L^\infty_u  L^2_S}\Big)\nn\\
	&+\||u|^{\frac{5}{2}+\frac{\delta}{3}}\varpi^N\nabring^{N_2}(\psicheck_{\Hbar},\widecheck{\omegabar})\|^2_{L^\infty_\ubar L^2_u  L^2_S}+\|\ubar^{\frac{5}{2}+\frac{\delta}{3}}\varpi^N\Omega_\mck^2\nabring^{N_2}\psicheck_H\|^2_{ L^\infty_u L^2_\ubar L^2_S}\nn\\
	&+\||u|^{\frac{5}{2}+\frac{\delta}{3}}\varpi^N\Omega_\mck(\nabring^{N_2}\etabarcheck,\nabring^{N_2-1}\widecheck{\mathring{K}})\|^2_{L^\infty_u L^2_\ubar  L^2_S}+\|\ubar^{\frac{5}{2}+\frac{\delta}{3}}\varpi^N\Omega_\mck(\nabring^{N_2}\etacheck,\nabring^{N_2-1}\widecheck{\mathring{K}})\|^2_{L^\infty_\ubar L^2_u L^2_S}\nn\\
	&+\sum_{i\leq N_2}\Big(\|\ubar^{\frac{5}{2}+\frac{\delta}{3}}\varpi^N\Omega_\mck\nabring^i\gcheck\|_{L^2_\ubar L^\infty_u L^2_S}^2+\||u|^{\frac{5}{2}+\frac{\delta}{3}}\varpi^N\Omega_\mck\nabring^i\gcheck\|_{L^\infty_\ubar L^2_u  L^2_S}^2+\|\ubar^{\frac{5}{2}+\frac{\delta}{3}}\varpi^N\nabring^i\bcheck\|_{L^2_\ubar L^\infty_u  L^2_S}^2\Big),\label{eq:mcnhypnab}
\end{align}
and
\begin{align}
	\mcn_{sph}^{(0)}:=&\|\nabring^{\leq N_2}\gcheck\|^2_{ L^\infty_\ubar L^\infty_u L^2_S}+\|\nabring^{\leq N_2-1}(\psicheck_{\Hbar},\widecheck{\omegabar},\psicheck,\bcheck)\|^2_{L^\infty_\ubar L^\infty_u L^2_S}\nn\\
	&+\|\Omega_\mck^2\nabring^{\leq N_2-1}\psicheck_H\|^2_{L^\infty_\ubar L^\infty_u L^2_S}+\|\nabring^{\leq N_2-2}\widecheck{\mathring{K}}\|^2_{L^\infty_\ubar L^\infty_u L^2_S}\nn\\
	&+\|\nabring^{\leq N_2-1}\psicheck_{\Hbar}\|^2_{L^1_u L^\infty_\ubar L^2_S}+\|\Omega_\mck^2\nabring^{\leq N_2-1}\psicheck_H\|^2_{L^1_\ubar L^\infty_u L^2_S}.\label{eq:mcnsphnab}
\end{align}
Then, one of the conclusions of the main theorem in \cite{stabC0} implies, in the case $N_2=3$, 
\begin{align}\label{eq:conDLgener}
	\mcn_{int}^{(0)}+\mcn_{hyp}^{(0)}+\mcn_{sph}^{(0)}\lesssim\varepsilon
\end{align}
in $\deux\cup\trois$ for powers ${\frac{5}{2}+\frac{\delta}{3}}$ on top of the polynomial $|u|,\ubar$ weights replaced with $1/2+\delta$. In this section, our conclusion will be that \eqref{eq:conDLgener} holds with $\mcn_{int}^{(0)},\mcn_{hyp}^{(0)},\mcn_{sph}^{(0)}$ defined as in \eqref{eq:mcnintnab}, \eqref{eq:mcnhypnab}, \eqref{eq:mcnsphnab}, with $N_2=N_1-2$. The proof of this result is a straightforward adaptation of the proof in \cite{stabC0} so we will omit the details. We will only highlight the slight differences.

\subsubsection{Initial data energy in the double null gauge}

We define the initial data on $\Sigma_0$ in the double null gauge similarly as in \cite[Section 4.5]{stabC0}, as well as the following data energy $\mcd$, with respect to any local extension of $(\Sigma_0,\hat{g},\hat{k})$ in the double null gauge,
\begin{align*}
	\mcd:=&\sum_{\gcheck,\psicheck, \psicheck_{H},\psicheck_{\Hbar}}\left\|\ubar^{\frac{5}{2}+\frac{\delta}{3}}\left(\nabring^{\leq N_2}(\gcheck,\psicheck, \psicheck_{H},\psicheck_{\Hbar},\widecheck{\mathring\omegabar},\bcheck,\nabring_4\widecheck{\log\Omega}),\nabring^{\leq N_2-1}\widecheck{\mathring{K}}\right)\right\|_{L^2_uL^2(S(u,-u+C_R))}^2\nn\\
	&+\sum_{\gcheck,\psicheck, \psicheck_{H},\psicheck_{\Hbar}}\left\|\ubar^{3+\delta/2}\left(\nabring^{\leq N_2}(\gcheck,\psicheck, \psicheck_{H},\psicheck_{\Hbar},\widecheck{\mathring\omegabar},\bcheck,\nabring_4\widecheck{\log\Omega}),\nabring^{\leq N_2-1}\widecheck{\mathring{K}}\right)\right\|_{L^\infty_uL^2(S(u,-u+C_R))}^2.
\end{align*}
The metric $\hat{g}$ takes the following form on $\Sigma_0$ in coordinates $\ubar,\theta^A$:
\begin{align}\label{eq:followingformonsig0}
	\hat{g}=\Phi\dee\ubar\otimes\dee\ubar+w_A(\dee\ubar\otimes\dee\theta^A+\dee\theta^A\otimes\dee\ubar)+(\gamma_\Sigma)_{AB}\dee\theta^A\otimes\dee\theta^B,
\end{align}
for some $\Phi,w,\gamma_\Sigma$ which are close to their Kerr values by Corollary \ref{cor:hypDLsigma0}. Moreover, denoting 
$$n:=\frac{1}{\sqrt{\Phi-|w|_\gamma^2}}\left(\frac{\partial}{\partial\ubar}-(\gamma^{-1}_\Sigma)^{CD}w_C\frac{\partial}{\partial\theta^D}\right)$$
the unit normal to the spheres $\mathring{S}(C_R-\ubar,\ubar)$ in $\Sigma_0$ (see Propositions 4.9 and 4.11 in \cite{stabC0}), we have the following expressions on $\Sigma_0$,
\begin{equation}\label{eq:pourrelatergeompasgeom}
	\begin{gathered}
		\gamma_{AB}|_{\Sigma_0}=(\gamma_\Sigma)_{AB},\quad\Omega|_{\Sigma_0}=\frac12\sqrt{\Phi-|w|_\gamma^2},\quad b^A|_{\Sigma_0}=-\gamma^{AB}w_B,\\
		\mathring\chibar_{AB}|_{\Sigma_0}=\Omega\hat{k}_{AB}-\Omega\hat{g}\left(\hat{\nabla}_A n,\frac{\partial}{\partial\theta^B}\right),\quad\mathring\chi_{AB}|_{\Sigma_0}=\Omega^{-1}\hat{k}_{AB}-\Omega^{-1}\hat{g}\left(\hat{\nabla}_A n,\frac{\partial}{\partial\theta^B}\right),\\
		\mathring\zeta_A|_{\Sigma_0}=\mathring\eta_A|_{\Sigma_0}=\frac{\partial}{\partial\theta^A}\log\Omega-\hat{k}\left(\frac{\partial}{\partial\theta^A},n\right),\quad \mathring\etabar_A|_{\Sigma_0}=\frac{\partial}{\partial\theta^A}\log\Omega+\hat{k}\left(\frac{\partial}{\partial\theta^A},n\right),\\
		\mathring\omegabar|_{\Sigma_0}=-\Omega k(n,n)+\Omega n(\log\Omega),\quad e_4(\log\Omega)=-\Omega^{-1} k(n,n)+\Omega^{-1} n(\log\Omega).
	\end{gathered}
\end{equation}
From the identities above and Corollary \ref{cor:hypDLsigma0} we easily deduce the higher-order generalization of Proposition 4.13 of \cite{stabC0} summarized in the following result.
\begin{prop}\label{prop:doublenullDborne}
	We have the following estimate for $\mcd$,
	$$\mcd\lesssim\varepsilon^2.$$
\end{prop}
\subsubsection{Bootstrap assumptions and main result on the control of $\nabring$ derivatives}

The proof of \eqref{eq:conDLgener} uses a bootstrap, see \cite[Section 4.8]{stabC0}\footnote{After using the main theorem of \cite{stabC0}, the equations become linear in the highest-order $\nabring$ derivatives, which can thus be controlled by induction using the same estimates as in \cite{stabC0}. However we find it convenient to do the bootstrap argument to control all derivatives at once.}.  For $\ubar_f> C_R-u_f$, defining
  \begin{equation}
  	\begin{gathered}
		\mcw_{\ubar_f}:=\{(u,\ubar)\:/\:-u_f+C_R<-u+C_R\leq\ubar<\ubar_f\},\quad\mcu_{\ubar_f}:=\mcw_{\ubar_f}\times\mathbb{S}^2,
  	\end{gathered}
  \end{equation}
 we say that the bootstrap assumption holds for $\ubar_f$ if $(\mcu_{\ubar_f},\g)$ is a smooth solution to the Einstein vacuum equations such that the metric $\g$ takes the double null form \eqref{eq:metricdoublenull} and achieves the prescribed initial data $(\hat{g},\hat{k})$ on $\Sigma_0=\{(u,\ubar)\:/\:u+\ubar=C_R\}\times\mathbb{S}^2\subset\mcu_{\ubar_f}$ (see Definition \ref{defi:initialdatasig0}), and such that recalling \eqref{eq:mcnsphnab}, 
\begin{align}\label{eq:BAII}
	\mcn_{sph}^{(0)}\leq\varepsilon,\quad\text{in}\:\:\mcu_{\ubar_f}.
\end{align}
We will prove the following bootstrap theorem, which is an adaptation of \cite[Thm. 4.27]{stabC0}.

\begin{thm}\label{thm:bootstrapII}
	For every Kerr black hole parameters $a,M$ such that $0<|a|<M$ and every $\delta>0$ and $C_R\in(-\infty,+\infty)$, there exists $\varepsilon_0=\varepsilon_0(a,M,C_R)>0$ and $u_f=u_f(a,M,\delta,C_R)\leq -1$ such that if the bootstrap assumption holds for $\ubar_f$ with $\varepsilon\leq\varepsilon_0$, then we have the estimate
$$\mcn_{int}^{(0)}+\mcn_{hyp}^{(0)}+\mcn_{sph}^{(0)}\leq C\varepsilon^2$$
in $\mcu_{\ubar_f}$ for some $C>0$ which depends only on $a,M$ and $C_R$.
\end{thm}
Sections \ref{section:schematicequationsangular} to \ref{section:recoveringbootstrapII} deal with the preliminary estimates needed for the proof of Theorem \ref{thm:bootstrapII}, so in these sections we assume the bootstrap assumption \ref{eq:BAII}. See the end of Section \ref{section:recoveringbootstrapII} for the proof of Theorem \ref{thm:bootstrapII}. As previously mentioned, we omit the details of the proof and focus on the slight differences with the proof of Theorem 4.27 in \cite{stabC0}. More precisely, these differences (which are basically differences in notations) are treated in the following sections: 
\begin{itemize}
	\item In Section \ref{section:schematicequationsangular}, we write the schematic equations satisfied by $\nabring^k$ derivatives of the linearized metric, Ricci and curvature coefficients, with $k$ arbitrary large, and then the reduced schematic equations implied by the bootstrap assumption \eqref{eq:BAII}.
	\item In Remark \ref{rem:transpodoublenullII}, we note that the transport estimates in Section 6 of \cite{stabC0} also hold replacing the polynomial weights $\ubar^{\frac12+\delta},|u|^{\frac12+\delta}$ by $\ubar^{\frac{5}{2}+\frac{\delta}{3}},|u|^{\frac{5}{2}+\frac{\delta}{3}}$, with same proofs.
	\item In Section \ref{section:recoveringbootstrapII} we conclude the proof of the bootstrap theorem, still following \cite{stabC0}.
\end{itemize}

\subsubsection{Schematic equations satisfied by angular derivatives of linearized quantities}\label{section:schematicequationsangular}
\textbf{From now on, we drop the ring notation for the double null linearized quantities.} We define the following generalizations of the inhomogeneous terms defined in \cite[Section 7.3]{stabC0}, 
\begin{align*}
	\mathcal{F}_1=&\Omega_\mck^2\sum_{{i_1+i_2\leq N_2}}\left(1+\nabring^{i_1}(\gcheck,\psicheck)\right)\Big(\nabring^{i_2}(\gcheck,\psicheck,\psicheck_{\Hbar},\widecheck{\omegabar})+\Omega_\mck^2\nabring^{i_2}\psicheck_H+\nabring^{i_2-1}\Kcheck\Big)\\
	&+\Omega_\mck^2\sum_{{i_1+i_2+i_3\leq N_2}}(1+\nabring^{i_1}\gcheck+\nabring^{\min(i_1,N_2-1)}\psicheck)\nabring^{i_2}(\psicheck_{\Hbar},\widecheck{\omegabar})\nabring^{i_3}\psicheck_H,\\
	\mcf_2=&\sum_{{i_1+i_2\leq N_2}}(1+\nabring^{i_1}(\gcheck,\psicheck))(\Omega_\mck^{-2}\nabring^{i_2}\bcheck+\nabring^{i_2}(\gcheck,\psicheck,\psicheck_H)+\nabring^{i_2-1}\Kcheck)+\sum_{i_1+i_2\leq N_2-1}\nabring^{i_1}\Kcheck\nabring^{i_2}\psicheck_{H},\\
	\mathcal{F}_3=&\sum_{{i_1+i_2\leq N_2}}\left(1+\nabring^{i_1}(\gcheck,\psicheck)\right)\left(\Omega_\mck^2\nabring^{i_2}(\gcheck,\psicheck)+\nabring^{i_2}(\psicheck_{\Hbar},\widecheck{\omegabar})+\Omega^2_\mck\nabring^{i_2-1}\Kcheck\right)+\sum_{i_1+i_2\leq N_2-1}\nabring^{i_1}\Kcheck\nabring^{i_2}\psicheck_{\Hbar},
\end{align*}
\begin{align*}
	\mcf_4=&\sum_{{i_1+i_2\leq N_2}}(1+\nabring^{i_1}\gcheck+\nabring^{\min(i_2,N_2-1)}\psicheck)(\nabring^{i_2}(\gcheck,\bcheck,\psicheck,\psicheck_{\Hbar},\widecheck{\omegabar})+\Omega_\mck^2\nabring^{i_2}\psicheck_H+\nabring^{i_2-1}\Kcheck),\\
	&+\sum_{{i_1+i_2+i_3\leq N_2}}(1+\nabring^{i_1}\gcheck+\nabring^{\min(i_1,N_2-1)}\psicheck)\nabring^{i_2}(\psicheck_{\Hbar},\widecheck{\omegabar})(\nabring^{i_3}\psicheck_H+\Omega_\mck^{-2}\nabring^{i_3}\bcheck),
\end{align*}
as well as
\begin{align*}
	\mcf_1'=&\Omega_\mck^2\nabring^{N_2}(\gcheck,\etacheck)+\Omega^2_\mck\nabring^{N_2-1}\Kcheck+\Omega_\mck^2\sum_{{i_1+i_2\leq N_2-1}}\big(1+\nabring^{i_1}(\gcheck,\psicheck)\big)\big(\nabring^{i_2}(\gcheck,\psicheck,\psicheck_{\Hbar},\widecheck{\omegabar})+\Omega_\mck^2\nabring^{i_2}\psicheck_H\big)\\
	&+\Omega_\mck^2\sum_{{i_1+i_2+i_3\leq N_2-1}}(1+\nabring^{i_1}(\gcheck,\psicheck))\nabring^{i_2}(\psicheck_{\Hbar},\widecheck{\omegabar})\nabring^{i_3}\psicheck_H,\\
	\mcf_2'=&\sum_{{i_1+i_2\leq N_2}}(1+\nabring^{i_1}\gcheck+\nabring^{\min(i_1,N_2-1)}\psicheck)(\Omega_\mck^{-2}\nabring^{i_2}\bcheck+\nabring^{\min(i_2,N_2-1)}\psicheck+\nabring^{i_2}(\gcheck,\psicheck_H)),\\
	\mathcal{F}_3'=&\sum_{{i_1+i_2\leq N_2}}\left(1+\nabring^{i_1}\gcheck+\nabring^{\min(i_1,N_2-1)}\psicheck\right)\left(\Omega_\mck^2\nabring^{i_2}\gcheck+\Omega_\mck^2\nabring^{\min(i_2,N_2-1)}\psicheck+\nabring^{i_2}(\psicheck_{\Hbar},\widecheck{\omegabar})\right)\\
	\mcf_4'=&\nabring^{N_2}(\gcheck,\bcheck,\etabarcheck)+\nabring^{N_2-1}\Kcheck+\sum_{{i_1+i_2\leq N_2-1}}(1+\nabring^{i_1}(\gcheck,\psicheck))(\nabring^{i_2}(\gcheck,\bcheck,\psicheck,\psicheck_{\Hbar},\widecheck{\omegabar})+\Omega_\mck^2\nabring^{i_2}\psicheck_H),\\
	&+\sum_{{i_1+i_2+i_3\leq N_2-1}}(1+\nabring^{i_1}\gcheck+\nabring^{\min(i_1,N_2-1)}\psicheck)\nabring^{i_2}(\psicheck_{\Hbar},\widecheck{\omegabar})(\nabring^{i_3}\psicheck_H+\Omega^{-2}_\mck\nabring^{i_3}\bcheck).
\end{align*}
which are slightly modified inhomogeneous terms.\footnote{Compared to the homogeneous terms defined in Section 7.3 of \cite{stabC0} which correspond to $N_2=3$, we have simpler expressions for $\mcf_1$ and $\mcf_4$ since there is no analog of the quadratic terms $\nabring^2\gcheck(\nabring\Kcheck,\nabring^2\psicheck)$ above. This is because we control a large number $N_2$ of angular derivative in the bootstrap assumption \eqref{eq:BAII} which implies that the $N_2> 3$ analog of these quadratic terms, which had to be dealt with specifically in the case $N_2=3$, become linear by the bootstrap assumption in the case where $N_2$ is larger than 7 (which holds here). The inhomogeneous terms $\mcf_i,\mcf_i'$ could be simplified further but this is not necessary for our purpose.} We also introduce a specific inhomogeneous term which will appear in the equation for $\bcheck$ ,
\begin{align*}
	\mcf_b:=&\sum_{{i_1+i_2\leq N_2}}(1+\nabring^{i_1}(\gcheck,\psicheck))(\Omega^2_\mck\nabring^{i_2}(\gcheck,\psicheck,\bcheck))+\sum_{{i_1+i_2+i_3\leq N_2}}(1+\nabring^{i_1}(\gcheck,\psicheck))\nabring^{i_2}\psicheck_{\Hbar}\nabring^{i_3}\bcheck.
\end{align*}
We now write down the reduced schematic equations satisfied by an arbitrary number of $\nabring$ derivatives of the linearized metric, Ricci and curvature coefficients.
\begin{prop}\label{prop:DNmetricnabla}
	For $i\leq N_2$, we have the following reduced schematic equations for the linearized metric components,
	\begin{align*}
		\nabring_3\nabring^i\gammacheck=_{rs}\mathcal{F}_3',\quad\nabring_3\nabring^i\logomegacheck=_{rs}\mathcal{F}_3',\quad\nabring_3\nabring^i\bcheck=_{rs}\mcf_b.
	\end{align*}
\end{prop}
\begin{proof}
	See Section \ref{section:proofsschemIInbabla}.
\end{proof}
\begin{prop}\label{prop:DNKchecknabla}
For $i\leq N_2-2$, $\Kcheck$ satisfies
$$\nabring^i\Kcheck=_{rs}\sum_{i_1\leq i+2}\nabring^i\gcheck.$$
\end{prop}
\begin{proof}
This follows from the definition \eqref{eq:gausscurvintrinsic} of the Gauss curvature $K$ and the bootstrap assumption \eqref{eq:BAII}.
\end{proof}
Now we write the reduced schematic equations satisfies by the differences of Ricci coefficients.
\begin{prop}\label{prop:DNriccinabla}
	For $i\leq N_2-1$, we have the following commuted null structure equations,
	\begin{alignat*}{3}
		\nabring_3\nabring^i\etabarcheck&=_{rs}\mcf_3',\quad&\nabring_3\nabring^i\widecheck{\underline{\mu}}&=_{rs}\mcf_3,\\
		\nabring_4\nabring^i\etacheck&=_{rs}\mcf_2',\quad&\nabring_4\nabring^i\widecheck{\mu}&=_{rs}\mcf_2\\
		\nabring_3\nabring^i(\Omega_\mck^2\psicheck_H)&=_{rs}\mcf_1',\quad&\nabring_4\nabring^i(\psicheck_{\Hbar},\widecheck{\omegabar})&=_{rs}\mcf_4'.
	\end{alignat*}
Moreover, $\nabring^{N_2-2}\widecheck{\barred{\omegabar}}$ satisfies the following reduced schematic equation,
$$	\nabring_4\nabring^{N_2-2}\widecheck{\barred{\omegabar}}=_{rs}\mcf_4.$$
\end{prop}
\begin{proof}
	See Section \ref{section:proofsschemIInbabla}.
\end{proof}
We now write the additional reduced schematic equations for $\widecheck{tr\chi}$ and $\widecheck{tr\chibar}$, which are used to control the top order derivatives of $\psicheck_H$ and $\psicheck_{\Hbar}$.
\begin{prop}\label{prop:DNadditionalriccinabla}
	 We have the following reduced schematic equations
	\begin{align*}
		\nabring_3(\Omega_\mck^{-2}\nabring^{N_2}\widecheck{tr\chibar})=_{rs}\Omega_\mck^{-2}\sum_{{i_1+i_2+i_3\leq N_2}}(1+\nabring^{i_1}\gcheck+\nabring^{i_1-1}\psicheck)(\nabring^{i_2}\psicheck_{\Hbar}+\Omega_\mck^2)(\nabring^{i_3}(\psicheck_{\Hbar},\widecheck{\omegabar})+\Omega_\mck^2\nabring^{i_3}\gcheck),
	\end{align*}
	as well as
	\begin{align*}
	\nabring_4\nabring^{N_2}\widecheck{tr\chi}=_{rs}\sum_{{i_1+i_2+i_3\leq N_2}}(1+\nabring^{i_1}\gcheck+\nabring^{i_1-1}\psicheck)(\nabring^{i_2}\psicheck_H+1)(\nabring^{i_3}(\psicheck_H,\gcheck)+\Omega_\mck^{-2}\nabring^{i_3}\bcheck).
	\end{align*}
\end{prop}
\begin{proof}
	See Section \ref{section:proofsschemIInbabla}.
\end{proof}
We now write the reduced schematic Bianchi equations. 
\begin{prop}\label{prop:DNbianchinabla}
	We have the following reduced schematic equations,
	\begin{align*}
		\nabring_3\nabring^{N_2-1}_{A_1\cdots A_{N_2-1}}(\Omega\Omega_\mck\widecheck{\beta})_C+\nabring_C(\Omega\Omega_\mck\nabring^{N_2-1}_{A_1\cdots A_{N_2-1}}\widecheck{K})-\in_C^D\nabring_D(\Omega\Omega_\mck\nabring^{N_2-1}_{A_1\cdots A_{N_2-1}}\hodge{\widecheck{K}})&=_{rs}\mcf_1,\\
		\nabring_4\nabring^{N_2-1}_{A_1\cdots A_{N_2-1}}\hodge{\widecheck{K}}+\nabring^C\nabring^{N_2-1}_{A_1\cdots A_{N_2-1}}(\in_C^D\widecheck{\beta}_D)&=_{rs}\mcf_2,\\
		\nabring_4\nabring^{N_2-1}_{A_1\cdots A_{N_2-1}}{\widecheck{K}}+\nabring^C\nabring^{N_2-1}_{A_1\cdots A_{N_2-1}}\widecheck{\beta}_C&=_{rs}\mcf_2,
	\end{align*}
	 for $A_1,\ldots,A_{N_2-1}=1,2$, as well as 
	\begin{align*}
		\nabring_4\nabring^{N_2-1}_{A_1\cdots A_{N_2-1}}\widecheck{\betabar}_C+\nabring_C\nabring^{N_2-1}_{A_1\cdots A_{N_2-1}}\widecheck{K}-\in_C^D\nabring_D\nabring^{N_2-1}_{A_1\cdots A_{N_2-1}}\hodge{\widecheck{K}}&=_{rs}\mcf_4,\\
		\nabring_3\nabring^{N_2-1}_{A_1\cdots A_{N_2-1}}\hodge{\widecheck{K}}+\nabring^C\nabring^{N_2-1}_{A_1\cdots A_{N_2-1}}(\in_C^D\widecheck{\betabar}_D)&=_{rs}\mcf_3,\\
		\nabring_3\nabring^{N_2-1}_{A_1\cdots A_{N_2-1}}{\widecheck{K}}+\nabring^C\nabring^{N_2-1}_{A_1\cdots A_{N_2-1}}\widecheck{\betabar}_C&=_{rs}\mcf_3.
	\end{align*}
\end{prop}
\begin{proof}
	See Section \ref{section:proofsschemIInbabla}.
\end{proof}

\subsubsection{Recovering the bootstrap assumptions}\label{section:recoveringbootstrapII}

Before explaining how to recover the bootstrap assumptions, we do a quick remark on the general transport estimates used in \cite[Section 6]{stabC0}.
\begin{rem}\label{rem:transpodoublenullII}
	An straightforward adaptation shows that the transport estimates written in Propositions 6.1 to 6.7 in Section 6 of \cite{stabC0} also hold replacing the polynomial power $1/2+\delta$ on top of the weights $\ubar$ and $|u|$ by ${\frac{5}{2}+\frac{\delta}{3}}$. Therefore, in what follows we will refer to these results in our context with polynomial power ${\frac{5}{2}+\frac{\delta}{3}}$.
\end{rem}
We now define error terms, which are adaptations of the ones defined in \cite[Section 8]{stabC0},
\begin{align*}
	&\mathcal{E}_1:=\sum_{{i_1+i_2\leq N_2}}\left(1+|\nabring^{i_1}(\gcheck,\psicheck)|+|\nabring^{i_1-1}\Kcheck|\right)\\
	&\quad\quad\quad\quad\times\left(\Omega_\mck|\nabring^{i_2}(\psicheck_{\Hbar},\widecheck{\omegabar},\bcheck)|+\Omega_\mck^3|\nabring^{i_2}(\psicheck,\gcheck,\psicheck_H)|+\Omega_\mck^3|\nabring^{i_2-1}\Kcheck|\right),\\
	&\mathcal{E}_2:=\sum_{{i_1+i_2\leq N_2}}\Omega_\mck\left(1+|\nabring^{i_1}(\gcheck,\psicheck)|+|\nabring^{i_1-1}\Kcheck\right)\left(|\nabring^{i_2}(\psicheck,\gcheck)|+|\nabring^{i_2-1}\Kcheck|\right),
\end{align*}
as well as
\begin{align*}
	\mct_1:=&\sum_{i_1+i_2\leq N_2}\left(1+|\nabring^{i_1}\gcheck|+|\nabring^{\min(i_1,N_2-1)}\psicheck|\right)|\nabring^{i_2}(\psicheck_{\Hbar},\widecheck{\omegabar})|\\
	&\times\left(\sum_{i_3+i_4+i_5\leq N_2}\left(1+|\nabring^{i_3}\gcheck|+|\nabring^{\min(i_3,N_2-1)}\psicheck|\right)\left(|\nabring^{i_4}\psicheck_H|,\Omega_\mck^{-2}|\nabring^{i_4}\bcheck|\right)|\nabring^{i_5}(\psicheck_{\Hbar},\widecheck{\omegabar})|\right),\\
	\mct_2:=&\sum_{i_1+i_2\leq N_2}\left(1+|\nabring^{i_1}\gcheck|+|\nabring^{\min(i_1,N_2-1)}\psicheck|\right)(|\nabring^{i_2}\psicheck_H|,\Omega_\mck^{-2}|\nabring^{i_2}\bcheck|)\\
	&\times\left(\sum_{i_3+i_4+i_5\leq N_2}\left(1+|\nabring^{i_3}\gcheck|+|\nabring^{\min(i_3,N_2-1)}\psicheck|\right)\left(|\nabring^{i_4}\psicheck_H|,\Omega_\mck^{-2}|\nabring^{i_4}\bcheck|\right)|\nabring^{i_5}(\psicheck_{\Hbar},\widecheck{\omegabar})|\right).
\end{align*}
We also define the following subsets of $\mcn_{int}$ and $\mcn_{hyp}$,
	\begin{align*}
	\mcn_{int,1}^{(0)}:=&\sum_{i\leq N_2}\Big(\||u|^{\frac{5}{2}+\frac{\delta}{3}}\varpi^N\Omega_\mck\nabring^i(\psicheck_{\Hbar},\widecheck{\omegabar})\|^2_{L^2_u L^2_\ubar L^2_S}+\|\ubar^{\frac{5}{2}+\frac{\delta}{3}}\varpi^N\Omega^3_\mck\nabring^i\psicheck_{H}\|^2_{L^2_u L^2_\ubar L^2_S}\\
	&\quad\quad+\|\ubar^{\frac{5}{2}+\frac{\delta}{3}}\varpi^N\Omega_\mck^2\nabring^{\min(i,N_2-1)}\Kcheck\|^2_{L^2_u L^2_\ubar L^2_S}+\|\ubar^{\frac{5}{2}+\frac{\delta}{3}}\varpi^N\Omega_\mck\nabring^i\bcheck\|^2_{L^2_u L^2_\ubar L^2_S}\\
	&\quad\quad+\|\ubar^{\frac{5}{2}+\frac{\delta}{3}}\varpi^N\Omega_\mck^2\nabring^i(\psicheck,\gcheck)\|^2_{L^2_u L^2_\ubar L^2_S}\Big),
\end{align*}
as well as
\begin{align*}
	\mcn_{int,2}^{(0)}:=\sum_{i\leq N_2}\Big(\|\ubar^{\frac{5}{2}+\frac{\delta}{3}}\varpi^N\Omega_\mck\nabring^i(\psicheck,\gcheck)\|^2_{L^2_u L^2_\ubar L^2_S}+\|\ubar^{\frac{5}{2}+\frac{\delta}{3}}\varpi^N\Omega_\mck\nabring^{\min(i,N_2-1)}\Kcheck\|^2_{L^2_u L^2_\ubar L^2_S}\Big),
\end{align*}
and 
	\begin{align*}
	\mcn_{hyp,1}^{(0)}:=&\sum_{i\leq N_2-1}\Big(\||u|^{\frac{5}{2}+\frac{\delta}{3}}\varpi^N\nabring^i\psicheck_{\Hbar}\|^2_{L^2_u L^\infty_\ubar L^2_S}+\|\ubar^{\frac{5}{2}+\frac{\delta}{3}}\varpi^N\Omega^2_\mck\nabring^i\psicheck_{H}\|^2_{L^2_\ubar L^\infty_u L^2_S}\Big)\\
	&\quad\quad+\||u|^{\frac{5}{2}+\frac{\delta}{3}}\varpi^N\nabring^{N_2}\psicheck_{\Hbar}\|^2_{L^\infty_\ubar L^2_u  L^2_S}+\|\ubar^{\frac{5}{2}+\frac{\delta}{3}}\varpi^N\Omega^2_\mck\nabring^{N_2}\psicheck_{H}\|^2_{ L^\infty_u L^2_\ubar L^2_S}\\
	&\quad\quad+\sum_{i\leq N_2}\|\ubar^{\frac{5}{2}+\frac{\delta}{3}}\varpi^N\nabring^i \bcheck\|^2_{L^2_\ubar L^\infty_u L^2_S}.
\end{align*}

\begin{prop}\label{prop:controlerrorIInabla}
	We have the following estimates for the error terms,
	\begin{equation*}
		\begin{gathered}
			\|\ubar^{\frac{5}{2}+\frac{\delta}{3}}\varpi^N\mathcal{E}_1\|_{L^2_u L^2_\ubar L^2_S}^2\lesssim\mcn_{int,1}^{(0)}+\varepsilon\mcn_{int,2}^{(0)}+N^{-1}\mcn_{hyp,1}^{(0)},\quad\|\ubar^{\frac{5}{2}+\frac{\delta}{3}}\varpi^N\mathcal{E}_2\|_{L^2_u L^2_\ubar L^2_S}^2\lesssim \mcn_{int,2}^{(0)},\\
			\||u|^{5+\frac{2\delta}{3}}\varpi^{2N}\Omega_\mck^2\mathcal{T}_1\|_{L^1_u L^1_\ubar L^1_S}\lesssim(\mcn_{hyp,1}^{(0)})^{3/2},\quad\|\ubar^{5+\frac{2\delta}{3}}\varpi^{2N}\Omega_\mck^4\mathcal{T}_2\|_{L^1_u L^1_\ubar L^1_S}\lesssim(\mcn_{hyp,1}^{(0)})^{3/2}.
		\end{gathered}
	\end{equation*}
\end{prop}
\begin{proof}
	The proof of this Proposition is the same as the ones of Propositions 12.2, 12.3, 12.4, 12.5 in \cite{stabC0}.
\end{proof}

\begin{prop}\label{prop:controlenergyIInabla}
	The following estimate holds,
	\begin{align*}
		\mcn_{int}^{(0)}+\mcn_{hyp}^{(0)}+N\mcn_{int,1}^{(0)}\lesssim& N2^{2N}\mcd+\|\ubar^{\frac{5}{2}+\frac{\delta}{3}}\varpi^N\mathcal{E}_1\|_{L^2_u L^2_\ubar L^2_S}^2+N^{-1}\|\ubar^{\frac{5}{2}+\frac{\delta}{3}}\varpi^N\mathcal{E}_2\|_{L^2_u L^2_\ubar L^2_S}^2\\
		&+\||u|^{5+\frac{2\delta}{3}}\varpi^{2N}\Omega_\mck^2\mathcal{T}_1\|_{L^1_u L^1_\ubar L^1_S}+\|\ubar^{5+\frac{2\delta}{3}}\varpi^{2N}\Omega_\mck^4\mathcal{T}_2\|_{L^1_u L^1_\ubar L^1_S}.
	\end{align*}
\end{prop}
\begin{proof}
	The proof of this estimate is the same as the one of Proposition 8.1 in \cite{stabC0}. Namely, we use the reduced schematic equations proven in Propositions \ref{prop:DNmetricnabla}, \ref{prop:DNriccinabla} to prove the bounds involving metric components and $\nabring^{\leq N_2-1}$ derivatives of the Ricci coefficients via the transport estimates adapted to our context (see Remark \ref{rem:transpodoublenullII}), exactly like in Section 9 of \cite{stabC0}. Then we use Proposition \ref{prop:DNbianchinabla} to control the top order $\nabring^{N_2-1}$ derivatives of the curvature coefficients via energy estimates, exactly like in Section 10 of \cite{stabC0}. Next, by using Proposition \ref{prop:DNadditionalriccinabla}, we control the top order derivatives $\nabring^{N_2}(\widecheck{tr\chi},\widecheck{tr\chibar})$. Finally, we prove the estimates for the top order derivatives of all the Ricci coefficients by elliptic estimates\footnote{Note that the elliptic estimates in Propositions 5.7 and 5.8 in \cite{stabC0} hold replacing $i\leq 3$ with $i\leq N_2$, as in our context in the bootstrap assumption \eqref{eq:BAII}  we control up to $\nabring^{N_2-2}$ derivatives of the Gauss curvature $K$.}: the top order derivative of $\etacheck$, $\etabarcheck$ are controlled by the lower-order derivatives of $\mucheck,\underline{\mucheck}$ (and other derivatives of the linearized quantities which we already control), the top order derivative of $\widecheck{\chihat},\widecheck{\underline{\chihat}},\widecheck{\omegabar}$ is controlled by the top order derivatives of $(\widecheck{tr\chi},\widecheck{tr\chibar},\gcheck,\widecheck{\barred{\omegabar}})$ and some lower-order derivatives. The precise order of control is the same as what is done in Section 11 of \cite{stabC0}.
\end{proof}
\begin{cor}\label{cor:apresilresteplusqueimprove}
	We have the estimate
	$$\mcn_{int}^{(0)}+\mcn_{hyp}^{(0)}\lesssim\varepsilon^2.$$
\end{cor}
\begin{proof}
	The proof is a corollary of Propositions \ref{prop:controlerrorIInabla} and \ref{prop:controlenergyIInabla}, see Proposition 12.1 in \cite{stabC0}.
\end{proof}

\noindent\textbf{End of the proof of Theorem \ref{thm:bootstrapII} and consequences}\label{section:endofproofBtII}

\begin{proof}[Proof of Theorem \ref{thm:bootstrapII}]
By Corollary \ref{cor:apresilresteplusqueimprove}, it is only left to improve the bootstrap assumption by proving $\mcn_{sph}^{(0)}\lesssim\varepsilon^2$. This is done exactly as in Section 13 of \cite{stabC0}, by using the Cauchy-Schwarz inequality to control the terms with norm $L^1_\ubar L^\infty_u L^2_S$ and $L^1_u L^\infty_\ubar L^2_S$ appearing in the definition of $\mcn_{sph}^{(0)}$, and by integrating the transport equations satisfied by $$\nabring^{\leq N_2}\gcheck,\nabring^{\leq N_2-1}(\etabarcheck,\bcheck),\nabring^{\leq N_2-1}\psicheck_H, \nabring^{\leq N_2-1}\etacheck, \nabring^{\leq N_2-1}(\psicheck_{\Hbar},\widecheck{\omegabar})$$ to bound the $L^\infty_u L^\infty_\ubar L^2_S$ norms appearing in $\mcn_{sph}^{(0)}$ (here we must use the fact that the RHS on the transport equations satisfied by these quantities takes the form $\mcf_i'$, and $\mcf_b$ for $\bcheck$). 
\end{proof}

\begin{thm}\label{thm:controlangulairedoublenull}
The vacuum spacetime with initial data $(\Sigma_0,\hat{g},\hat{k})$ (see Definitions \ref{defi:sigmazeroII} and \ref{defi:initialdatasig0}) extends in double null gauge \eqref{eq:metricdoublenull} in all of region
$$\deux\cup\trois=\{(u,\ubar)\:/\:\ubar\in[-u+C_R,+\infty),\: u\in[-\ubar+C_R,u_f)\}\times\mathbb{S}^2,$$
where it satisfies the estimate
	$$\mcn_{int}^{(0)}+\mcn_{hyp}^{(0)}+\mcn_{sph}^{(0)}\lesssim\varepsilon^2.$$
\end{thm}
\begin{proof}
	Using the initial data estimates in Corollary \ref{cor:hypDLsigma0} (or, equivalently, Proposition \ref{prop:doublenullDborne}), we proceed exactly as in Section 14  of \cite{stabC0} (which deals with the propagation of higher regularity, which is needed in the bootstrap argument which uses a local existence theorem in the smooth category) and Section 15  of \cite{stabC0} (which completes the bootstrap argument), using Theorem \ref{thm:bootstrapII}.
\end{proof}
\subsubsection{$L^2(S(u,\ubar))$ and pointwise decay estimates for angular derivatives}

Now that we have extended the spacetime up to the Cauchy horizon $\ch$ and that we have the integral estimate in Theorem \ref{thm:controlangulairedoublenull}, we prove $L^2(S(u,\ubar))$ and pointwise polynomial decay estimates. This is done by integrating again the reduced schematic equations for the angular derivatives of the linearized quantities of Section \ref{section:schematicequationsangular}, while precisely estimating some inverse-polynomial integrals. These integrals appeared in the proof of Theorem \ref{thm:controlangulairedoublenull} to recover the bootstrap assumptions, but so far we only used the fact that they are bounded, while they actually decay at a polynomial rate. We first recall the Sobolev embedding \cite[Prop. 5.5]{stabC0} (recall Remark \ref{rem:alsoholdsDL}).

\begin{prop}\label{prop:sobolevdoublenullII}
	Let $\phi$ be a $S(u,\ubar)$-tangent tensor field. We have in $\deux\cup\trois$,
	$$\|\phi\|_{L^\infty(S(u,\ubar))}\lesssim\|\nabring^{\leq 2}\phi\|_{L^2(S(u,\ubar))}.$$
\end{prop}
We can reformulate this Sobolev estimate using the Laplace operator on the spheres $S(u,\ubar)$, which is defined in \eqref{eq:laplacienSdoublenulll}.
	\begin{cor}\label{cor:sobolaplace}
	Let $\phi$ be a $S(u,\ubar)$-tangent tensor field. We have in $\deux\cup\trois$,
$$\|\phi\|_{L^\infty(S(u,\ubar))}\lesssim\|\mathring{\triangle}^{\leq 1}\phi\|_{L^2(S(u,\ubar))}.$$
\end{cor}
\begin{proof}
	By Proposition \ref{prop:sobolevdoublenullII} we have $\|\phi\|_{L^\infty(S(u,\ubar))}\lesssim\|{\nabring}^{\leq 2}\phi\|_{L^2(S(u,\ubar))}$. Moreover, by standard Bochner identities for tensors,
\begin{align*}
	\intS |\nabring^2\phi|^2&=\intS\left(|\mathring{\triangle}\phi|^2+\nabring_B\phi\cdot[\nabring^B,\mathring{\triangle}]\phi+\nabring^C\left(\nabring_B\phi\cdot\nabring_C\nabring^B\phi-\nabring_C\phi\cdot\mathring{\triangle}\phi\right)\right)\\
	&=\intS\left(|\mathring{\triangle}\phi|^2+\nabring_B\phi\cdot[\nabring^B,\mathring{\triangle}]\phi\right)\lesssim \intS \left(|\mathring{\triangle}\phi|^2+|\nabring \phi|^2+|\phi|^2\right),
\end{align*}
where we used the estimate $|[\nabring,\mathring{\triangle}]\phi|\lesssim |\nabring^{\leq 1} K||\nabring^{\leq 1}\phi|\lesssim |\nabring^{\leq 1}\phi|$ because the Gauss curvature $K$ of $S(u,\ubar)$ satisfies $|\nabring^{\leq 1} K|\lesssim 1$ by $\mcn_{sph}^{(0)}\lesssim\varepsilon^2$. Finally, we integrate by parts again to get
	\begin{align*}
		\intS |\nabring\phi|^2=\intS \nabring^A\phi\cdot\nabring_A\phi=\intS(-\mathring{\triangle})\phi\cdot\phi\lesssim\intS \left(|\mathring{\triangle}\phi|^2+|\phi|^2\right),
	\end{align*}
	which concludes the proof.
\end{proof}
\begin{rem}
	In the proof of Corollary \ref{cor:sobolaplace}, we have also proven
\begin{align}\label{eq:thebound222}
	\intS |\nabring^{\leq 2}\phi|^2\lesssim\intS \left(|\mathring{\triangle}\phi|^2+|\phi|^2\right).
\end{align}
\end{rem}
Now we recall \cite[Prop. 13.2]{stabC0}.
\begin{prop}\label{prop:13.2DL}
	Let $\phi$ be a $S(u,\ubar)$-tangent tensor field. We have in $\deux\cup\trois$,
	\begin{align*}
		\Ldeux{\phi}&\lesssim \|\phi\|_{L^2(S(u,-u+C_R))}+\int_{-u+C_R}^\ubar\|\Omega_\mck^2\nabring_4\phi\|_{L^2(S(u,\ubar'))},\\
		\Ldeux{\phi}&\lesssim \|\phi\|_{L^2(S(-\ubar+C_R,\ubar))}+\int_{-\ubar+C_R}^u\|\nabring_3\phi\|_{L^2(S(u,\ubar'))}.
	\end{align*}
Moreover, integrating instead from $\Gamma$ defined in \eqref{eq:defiGammahyp} instead of from $\Sigma_0$, we also have
	\begin{align*}
	\Ldeux{\phi}&\lesssim \|\phi\|_{L^2(S(u,\ubar_\Gamma(u)))}+\int_{\ubar_\Gamma(u)}^\ubar\|\Omega_\mck^2\nabring_4\phi\|_{L^2(S(u,\ubar'))},\\
	\Ldeux{\phi}&\lesssim \|\phi\|_{L^2(S(u_\Gamma(\ubar),\ubar))}+\int_{u_\Gamma(\ubar)}^u\|\nabring_3\phi\|_{L^2(S(u,\ubar'))},
\end{align*}
where $u_\Gamma(\ubar),\ubar_\Gamma(u)$ are such that $(u_\Gamma(\ubar),\ubar),(u,\ubar_\Gamma(u))\in\Gamma$.
\end{prop}
The following lemma will be useful to deduce polynomial decay estimates in $\ubar$.
\begin{lem}\label{lem:intdecaymieux}
	We have, for $|u_f|$ chosen large enough with respect to $a,M,C_R,\delta$ and $C_R\geq 1$, for any $u,\ubar$ such that 
	$u+\ubar\geq C_R$ and $u\leq u_f$,
	$$\int_{-\ubar+C_R}^u \frac{e^{-|\kappa_-|(u'+\ubar)}}{|u'|^{5+\frac{2\delta}{3}}}\dee u'\lesssim \frac{1}{\ubar^{5+\frac{2\delta}{3}}}.$$
\end{lem}
\begin{proof}
Integrating by parts, we compute
\begin{align*}
	&\int_{-\ubar+C_R}^u \frac{e^{-|\kappa_-|(u'+\ubar)}}{|u'|^{5+\frac{2\delta}{3}}}\dee u'\\
	&=e^{-|\kappa_-|\ubar}\Bigg(\frac{1}{|\kappa_-|}\left(\frac{e^{|\kappa_-|(\ubar-C_R)}}{(\ubar-C_R)^{5+\frac{2\delta}{3}}}-\frac{e^{-|\kappa_-|u}}{|u|^{5+\frac{2\delta}{3}}}\right)+\frac{5+\frac{2\delta}{3}}{|\kappa_-|}\int_{-\ubar+C_R}^u \frac{e^{-|\kappa_-|u'}}{|u'|^{6+\frac{2\delta}{3}}}\dee u'\Bigg)\\
	&\leq \frac{e^{-|\kappa_-|C_R}}{|\kappa_-|(\ubar-C_R)^{5+\frac{2\delta}{3}}}+\frac{5+\frac{2\delta}{3}}{|u_f||\kappa_-|}\int_{-\ubar+C_R}^u \frac{e^{-|\kappa_-|(u'+\ubar)}}{|u'|^{5+\frac{2\delta}{3}}}\dee u'.
\end{align*}
Thus, choosing $|u_f|$ large enough so that the last term on the RHS above is absorbed on the LHS, we conclude the proof of the stated estimate.
\end{proof}

\begin{prop}\label{prop:L2decaydoublenullangular}
	For $N_2'=N_2-2$, we have the following decay estimates in $\deux\cup\trois$,
	\begin{equation}\label{eq:IIlabelfirstbound}
		\begin{aligned}
					\|\nabring^{\leq N_2'}\gcheck\|_{L^2(S(u,\ubar))}+\|\nabring^{\leq N_2'-1}(\psicheck_{\Hbar},\widecheck{\omegabar},\psicheck)\|_{L^2(S(u,\ubar))}+\|\nabring^{\leq N_2'-2}\Kcheck\|_{L^2(S(u,\ubar))}&\lesssim\frac{\varepsilon}{|u|^{2+\frac{\delta}{3}}},\\
			\|\nabring^{\leq N_2'-1}(\Omega_\mck^2\psicheck_H)\|_{L^2(S(u,\ubar))}+\|\nabring^{\leq N_2'-1}\bcheck\|_{L^2(S(u,\ubar))}&\lesssim\frac{\varepsilon}{\ubar^{2+\frac{\delta}{3}}}.
		\end{aligned}
	\end{equation}
\end{prop}
\begin{proof}
	According to Propositions \ref{prop:DNmetricnabla}, \ref{prop:DNriccinabla}, Proposition \ref{prop:13.2DL}, and the $L^\infty_uL^2_S$ estimates given by Proposition \ref{prop:doublenullDborne} we have the following bounds\footnote{Here we also use \cite[Prop. 9.1, 9.2]{stabC0} generalized to $N_2\geq 3$ which shows that the $\nabring$ derivatives of $\gammacheck,\logomegacheck$ control the ones of $\gcheck$.},
	
	\begin{align*}
		\Ldeux{\nabring^{\leq N_2'}\gcheck}+\Ldeux{\nabring^{\leq N_2'-1}\etabarcheck}&\lesssim \frac{\varepsilon}{\ubar^{2+\frac{\delta}{3}}}+\int_{-\ubar+C_R}^u \|{}^{(N_2')}\!\mcf_3'\|_{L^2(S(u',\ubar))}\dee u',\\
		\Ldeux{\nabring^{\leq N_2'-1}\bcheck}&\lesssim \frac{\varepsilon}{\ubar^{2+\frac{\delta}{3}}}+\int_{-\ubar+C_R}^u \|{}^{(N_2'-1)}\!\mcf_b\|_{L^2(S(u',\ubar))}\dee u',\\
		\Ldeux{\nabring^{\leq N_2'-1}(\Omega^2_\mck\psicheck_H)}&\lesssim \frac{\varepsilon}{\ubar^{2+\frac{\delta}{3}}}+\int_{-\ubar+C_R}^u \|{}^{(N_2')}\!\mcf_1'\|_{L^2(S(u',\ubar))}\dee u',\\
		\Ldeux{\nabring^{\leq N_2'-1}\etacheck}&\lesssim \frac{\varepsilon}{|u|^{2+\frac{\delta}{3}}}+\int_{-u+C_R}^\ubar \|\Omega_\mck^2{}^{(N_2')}\!\mcf_2'\|_{L^2(S(u,\ubar'))}\dee\ubar',\\
		\Ldeux{\nabring^{\leq N_2'-1}(\psicheck_{\Hbar},\widecheck{\omegabar})}&\lesssim \frac{\varepsilon}{|u|^{2+\frac{\delta}{3}}}+\int_{-u+C_R}^\ubar \|\Omega_\mck^2{}^{(N_2')}\!\mcf_4'\|_{L^2(S(u,\ubar'))}\dee \ubar',
	\end{align*}   
where each of the ${}^{(N_2')}\!\mcf_i'$ correspond to the $\mcf'_i$ as defined in Section \ref{section:schematicequationsangular} with $N_2$ replaced by $N_2'$, and similarly ${}^{(N_2'-1)}\!\mcf_b$ corresponds to $\mcf_b$ with $N_2$ replaced by $N_2'-1$. Note that using Grönwall's inequality, we actually have the following more precise bound for $(\psicheck_{\Hbar},\widecheck{\omegabar})$,
	\begin{align*}
		\Ldeux{\nabring^{\leq N_2'-1}(\psicheck_{\Hbar},\widecheck{\omegabar})}&\lesssim \frac{\varepsilon}{|u|^{2+\frac{\delta}{3}}}+\int_{-u+C_R}^\ubar \|\Omega_\mck^2{}^{(N_2')}\!\mcf_4''\|_{L^2(S(u,\ubar'))}\dee \ubar',
	\end{align*}
	where ${}^{(N_2')}\!\mcf_4''$ is the part of ${}^{(N_2')}\!\mcf_4'$ which does not have any $\psicheck_{\Hbar},\widecheck{\omegabar}$ factor, namely
	\begin{align*}
		{}^{(N_2')}\!\mcf_4'':=\nabring^{N_2'}(\gcheck,\etabarcheck)+\nabring^{N_2'-1}\Kcheck+\sum_{{i_1+i_2\leq N_2'-1}}(1+\nabring^{i_1}(\gcheck,\psicheck))(\nabring^{i_2}(\gcheck,\bcheck,\psicheck)+\Omega_\mck^2\nabring^{i_2}\psicheck_H).
	\end{align*} 
Here we used that $N_2'\leq N_2-2$ so that by the bound $\mcn_{sph}^{(0)}\lesssim\varepsilon^2$ and the Sobolev embedding, the factors involving $\nabring$ derivatives of $\gcheck,\psicheck$ in front of $\nabring^{i_2}(\psicheck_{\Hbar},\widecheck{\omegabar})$ in ${}^{(N_2')}\mcf_4'$ are bounded in $L^\infty$. Similarly, for $\psicheck_H$, we have 
	\begin{align*}
		\Ldeux{\nabring^{\leq N_2'-1}(\Omega^2_\mck\psicheck_H)}&\lesssim \frac{\varepsilon}{\ubar^{2+\frac{\delta}{3}}}+\int_{-\ubar+C_R}^u \|{}^{(N_2')}\!\mcf_1''\|_{L^2(S(u',\ubar))}\dee u',
	\end{align*}
	where
	\begin{align*}
		{}^{(N_2')}\!\mcf_1''=&\Omega_\mck^2\left(\nabring^{N_2'}(\gcheck,\etacheck)+\nabring^{N_2'-1}\Kcheck+\sum_{{i_1+i_2\leq N_2'-1}}\big(1+\nabring^{i_1}\gcheck\big)\nabring^{i_2}(\gcheck,\psicheck,\psicheck_{\Hbar},\widecheck{\omegabar})\right),
	\end{align*}
and for $\bcheck$ we have 
$$\Ldeux{\nabring^{\leq N_2'-1}\bcheck}\lesssim \frac{\varepsilon}{\ubar^{2+\frac{\delta}{3}}}+\int_{-\ubar+C_R}^u \|{}^{(N_2'-1)}\!\mcf_b'\|_{L^2(S(u',\ubar))}\dee u'$$
where 
$${}^{(N_2'-1)}\!\mcf_b'=\sum_{{i_1+i_2\leq N_2'-1}}(1+\nabring^{i_1}(\gcheck,\psicheck))\Omega^2_\mck\nabring^{i_2}(\gcheck,\psicheck).$$
Now, using a Cauchy-Schwarz inequality we get the following bound: 
	\begin{align}
		\int_{-\ubar+C_R}^u \|{}^{(N_2')}\!\mcf'_3\|_{L^2(S(u',\ubar))}\dee u'&=\int_{-\ubar+C_R}^u \left(\int_{S(u',\ubar)} |{}^{(N_2')}\!\mcf'_3|^2\right)^{1/2}\dee u'\nn\\
		&\lesssim \left(\int_{-\ubar+C_R}^u \frac{1}{|u'|^{5+\frac{2\delta}{3}}} \dee u'\right)^{1/2}\left(\int_{-\ubar+C_R}^u |u'|^{5+\frac{2\delta}{3}}\int_{S(u',\ubar)} |{}^{(N_2')}\!\mcf'_3|^2 \dee u'\right)^{1/2}\nn\\
		&\lesssim \frac{1}{|u|^{2+\frac{\delta}{3}}}\left(\int_{-\ubar+C_R}^u |u'|^{5+\frac{2\delta}{3}}\int_{S(u',\ubar)} |{}^{(N_2')}\!\mcf'_3|^2 \dee u'\right)^{1/2},\nn
	\end{align}
	where we used
	\begin{align}\label{eq:usedussusu}
		\int_{-\ubar+C_R}^u \frac{1}{|u'|^{5+\frac{2\delta}{3}}} \dee u'=\frac{1}{4+\delta/2}\left(\frac{1}{|u|^{4+\delta/2}}-\frac{1}{(\ubar-C_R)^{4+\delta/2}}\right).
	\end{align}
	We also have, using $\Omega_\mck^2\sim e^{-|\kappa_-|(u+\ubar)}$ (recall \eqref{eq:omegamcksim}), for $\mcf={}^{(N_2')}\!\mcf'_1,{}^{(N_2'-1)}\!\mcf_b'$, 
	\begin{align}
		\int_{-\ubar+C_R}^u \|\mcf\|_{L^2(S(u',\ubar))}\dee u'&\lesssim \left(\int_{-\ubar+C_R}^u \frac{e^{-|\kappa_-|(u'+\ubar)}}{|u'|^{5+\frac{2\delta}{3}}} \dee u'\right)^{1/2}\left(\int_{-\ubar+C_R}^u |u'|^{5+\frac{2\delta}{3}}\int_{S(u',\ubar)} |\Omega_\mck^{-1}\mcf|^2 \dee u'\right)^{1/2}\nn\\
		&\lesssim \frac{1}{\ubar^{2+\frac{\delta}{3}}}\left(\int_{-\ubar+C_R}^u |u'|^{5+\frac{2\delta}{3}}\int_{S(u',\ubar)} |\Omega_\mck^{-1}\mcf|^2 \dee u'\right)^{1/2},\label{eq:bRRRouloublou}
	\end{align}
	where we used Lemma \ref{lem:intdecaymieux} (and dropped some $\ubar^{-1/2}$ additional decay). Next, we also have
	\begin{align}
		\int_{-u+C_R}^\ubar\|\Omega_\mck^2{}^{(N_2')}\!\mcf'_2&\|_{L^2(S(u,\ubar'))}\dee \ubar'\nn\\
		&\lesssim \left(\int_{-u+C_R}^\ubar \frac{1}{\ubar'^{5+\frac{2\delta}{3}}} \dee \ubar'\right)^{1/2}\left(\int_{-u+C_R}^\ubar \ubar'^{5+\frac{2\delta}{3}}\int_{S(u',\ubar)} |\Omega_\mck^2{}^{(N_2')}\!\mcf'_2|^2 \dee \ubar'\right)^{1/2}\nn\\
		&\lesssim \frac{1}{|u|^{2+\frac{\delta}{3}}}\left(\int_{-u+C_R}^\ubar \ubar'^{5+\frac{2\delta}{3}}\int_{S(u',\ubar)} |\Omega_\mck^2{}^{(N_2')}\!\mcf'_2|^2 \dee \ubar'\right)^{1/2},
	\end{align}
	where we used $\ubar^{-5-\delta/2}\lesssim |u|^{-5-\delta/2}$ and \eqref{eq:usedussusu}. Finally, we also have
	\begin{align}
		\int_{-u+C_R}^\ubar\|\Omega_\mck^2{}^{(N_2')}\!\mcf''_4&\|_{L^2(S(u,\ubar'))}\dee \ubar'\nn\\
		&\lesssim \left(\int_{-u+C_R}^\ubar \frac{\Omega_\mck}{|u|^{5+\frac{2\delta}{3}}} \dee \ubar'\right)^{1/2}\left(\int_{-u+C_R}^\ubar |u|^{5+\frac{2\delta}{3}}\int_{S(u',\ubar)} |\Omega_\mck{}^{(N_2')}\!\mcf''_4|^2 \dee \ubar'\right)^{1/2}\nn\\
		&\lesssim \frac{1}{|u|^{2+\frac{\delta}{3}}}\left(\int_{-u+C_R}^\ubar |u|^{5+\frac{2\delta}{3}}\int_{S(u',\ubar)} |\Omega_\mck{}^{(N_2')}\!\mcf''_4|^2 \dee \ubar'\right)^{1/2}.
	\end{align}
	
	This implies, using also Proposition \ref{prop:DNKchecknabla},
	
	\begin{align}
		&\|\nabring^{\leq N_2'}\gcheck\|_{L^2(S(u,\ubar))}+\|\nabring^{\leq N_2'-1}(\psicheck_{\Hbar},\widecheck{\omegabar},\psicheck)\|^2_{L^2(S(u,\ubar))}+\|\nabring^{\leq N_2'-2}\Kcheck\|^2_{ L^2(S(u,\ubar))}\nn\\
		&\lesssim \frac{1}{|u|^{2+\frac{\delta}{3}}}\Bigg[\varepsilon+\left(\int_{-\ubar+C_R}^u |u'|^{5+\frac{2\delta}{3}}\int_{S(u',\ubar)} |{}^{(N_2')}\!\mcf'_3|^2 \dee u'\right)^{1/2}\label{eq:IIthirddd}\\
		&+\left(\int_{-u+C_R}^\ubar \ubar'^{5+\frac{2\delta}{3}}\int_{S(u',\ubar)} |\Omega_\mck^2{}^{(N_2')}\!\mcf'_2|^2 \dee \ubar'\right)^{1/2}+\left(\int_{-u+C_R}^\ubar |u|^{5+\frac{2\delta}{3}}\int_{S(u',\ubar)} |\Omega_\mck{}^{(N_2')}\!\mcf''_4|^2 \dee \ubar'\right)^{1/2}\Bigg].\nn
	\end{align}
	Using the estimate for $\mcn_{sph}^{(0)}$ in Theorem \ref{thm:controlangulairedoublenull} combined with the Sobolev embedding in Proposition \ref{prop:sobolevdoublenullII}, we get that all the factors $1+\nabring^{i_1}(\gcheck,\psicheck)$ in ${}^{(N_2')}\!\mcf'_1$, ${}^{(N_2')}\!\mcf'_4$ and ${}^{(N_2')}\!\mcf'_b$, and $1+\nabring^{i_1}\gcheck+\nabring^{\min(i_1,N_2'-1)}\psicheck$ in ${}^{(N_2')}\!\mcf'_2$, ${}^{(N_2')}\!\mcf'_3$, ${}^{(N_2')}\!\mcf'_4$ and ${}^{(N_2'-1)}\!\mcf'_b$, are bounded\footnote{This is where we lose 2 derivatives which is why we have to introduce $N_2'=N_2-2$, and why we control $\bcheck$ with one less derivative compared to the other linearized quantities.}. Thus by Theorem \ref{thm:controlangulairedoublenull} we deduce
	\begin{align*}
		&\int_{-\ubar+C_R}^u |u'|^{5+\frac{2\delta}{3}}\int_{S(u',\ubar)} |{}^{(N_2')}\!\mcf'_3|^2 \dee u'\\
		&\lesssim\||u|^{\frac{5}{2}+\frac{\delta}{3}}\Omega_\mck\nabring^{\leq N_2}\gcheck\|^2_{L^\infty_\ubar L^2_u L^2_S}+\|\ubar^{\frac{5}{2}+\frac{\delta}{3}}\Omega_\mck\nabring^{\leq N_2-1}\etacheck\|^2_{ L^2_u L^\infty_\ubar L^2_S}+\||u|^{\frac{5}{2}+\frac{\delta}{3}}\Omega_\mck\nabring^{\leq N_2-1}\etabarcheck\|^2_{ L^\infty_\ubar L^2_u L^2_S}\\
		&\quad+\||u|^{\frac{5}{2}+\frac{\delta}{3}}\nabring^{\leq N_2}(\psicheck_{\Hbar},\widecheck{\omegabar})\|^2_{L^\infty_\ubar L^2_u  L^2_S}\lesssim \mcn_{hyp}^{(0)}\lesssim\varepsilon^2,
	\end{align*}
and, similarly,
	\begin{align*}
		&\int_{-u+C_R}^\ubar \ubar'^{5+\frac{2\delta}{3}}\int_{S(u',\ubar)} |\Omega_\mck^2{}^{(N_2')}\!\mcf'_2|^2 \dee \ubar'\lesssim \mcn_{hyp}^{(0)}\lesssim\varepsilon^2,\\
		&\int_{-u+C_R}^\ubar |u|^{5+\frac{2\delta}{3}}\int_{S(u',\ubar)} |\Omega_\mck{}^{(N_2')}\!\mcf''_4|^2 \dee \ubar'\lesssim \mcn_{hyp}^{(0)}\lesssim\varepsilon^2.
	\end{align*}
By \eqref{eq:IIthirddd}, this concludes the proof the first line of \eqref{eq:IIlabelfirstbound}. Similarly, we get the bound
	\begin{align*}
		\int_{-\ubar+C_R}^u& |u'|^{5+\frac{2\delta}{3}}\int_{S(u',\ubar)} |\Omega_\mck^{-1}\left({}^{(N_2')}\!\mcf''_1,{}^{(N_2'-1)}\!\mcf'_b\right)|^2 \dee u'\\
		\lesssim& \||u|^{\frac{5}{2}+\frac{\delta}{3}}\Omega_\mck\nabring^{\leq N_2}\gcheck\|^2_{L^\infty_\ubar L^2_u L^2_S}+\|\ubar^{\frac{5}{2}+\frac{\delta}{3}}\Omega_\mck\nabring^{ N_2}\etacheck\|^2_{L^\infty_\ubar L^2_u L^2_S }\\
		&+\||u|^{\frac{5}{2}+\frac{\delta}{3}}\Omega_\mck\nabring^{\leq N_2-1}\Kcheck\|^2_{L^\infty_\ubar L^2_u L^2_S }+\||u|^{\frac{5}{2}+\frac{\delta}{3}}\Omega_\mck\nabring^{\leq N_2-1}\etabarcheck\|^2_{ L^\infty_\ubar  L^2_u L^2_S}\\
		&+\|\ubar^{\frac{5}{2}+\frac{\delta}{3}}\Omega_\mck\nabring^{\leq N_2-1}\etacheck\|^2_{ L^2_u  L^\infty_\ubar L^2_S}+\||u|^{\frac{5}{2}+\frac{\delta}{3}}\nabring^{\leq N_2}(\psicheck_{\Hbar},\widecheck{\omegabar})\|^2_{L^\infty_\ubar L^2_u  L^2_S}\lesssim\mcn_{hyp}^{(0)}\lesssim\varepsilon^2,
	\end{align*}
	which, by \eqref{eq:bRRRouloublou}, yields that
	$$\|\nabring^{\leq N_2'-1}(\Omega_\mck^2\psicheck_H)\|^2_{ L^2(S(u,\ubar))}+\|\nabring^{\leq N_2'-1}\bcheck\|_{L^2(S(u,\ubar))}\lesssim\varepsilon/\ubar^{2+\frac{\delta}{3}},$$
	and concludes the proof of \eqref{eq:IIlabelfirstbound}.
\end{proof}

Using the Sobolev embedding, and the schematic constraint equations \eqref{eq:constraintDNouioui} which rewrite
\begin{align}\label{eq:constraintbetabatabar}
	\beta=_s\nabring \psi_H+\psi\psi_H,\quad\betabar=_s\nabring \psi_{\Hbar}+\psi\psi_{\Hbar},\quad \hodge{K}=_s\nabring\psi,
\end{align}
we obtain pointwise decay estimates for the angular derivatives of the linearized quantities.
\begin{prop}\label{prop:decaypointwiseDNpourIII}
	We have the following decay estimates in the double null frame, for $N_2''=N_2-7$, in region $\deux\cup\trois$,
	\begin{align*}
		|\nabring^{\leq N_2''}(\gcheck,\psicheck_{\Hbar},\widecheck{\omegabar},\psicheck,\Kcheck,\hodge{\Kcheck},\betabarcheck)|&\lesssim \frac{\varepsilon}{|u|^{2+\frac{\delta}{3}}},\\
		|\nabring^{\leq N_2''}\psicheck_H|\lesssim \frac{\varepsilon\Omega^{-2}_\mck}{\ubar^{2+\frac{\delta}{3}}},\quad |\nabring^{\leq N_2''}\betacheck|&\lesssim \frac{\varepsilon\Omega^{-2}_\mck}{|u|^{2+\frac{\delta}{3}}},\quad|\nabring^{\leq N_2''}\bcheck|\lesssim \frac{\varepsilon}{\ubar^{2+\frac{\delta}{3}}}.
	\end{align*}
\end{prop}
\begin{proof}
	These bounds follow easily from the $L^2(S(u,\ubar))$ bounds proven in Proposition \ref{prop:L2decaydoublenullangular} combined with the Sobolev embedding of Proposition \ref{prop:sobolevdoublenullII}, where the constraint equations \eqref{eq:constraintbetabatabar} also have to be used for $\widecheck{\beta},\betabarcheck,\widecheck{\hodge{K}}$. Also, to get the bound for $\psicheck_H$ from the $L^2(S(u,\ubar))$ bound in \eqref{eq:IIlabelfirstbound}, we commute the scalar $\Omega_\mck^{-2}$ with the $\nabring^{\leq N_2''}$ derivatives. Using $|\nabring^{\leq N_2''}\Omega_\mck^{-2}|\lesssim\Omega_\mck^{-2}$ by the bound $|\nabring^{\leq N_2''}(\gcheck,\psicheck)|\lesssim 1$, we get that these commutations only introduce lower order terms in $\psicheck_{H}$ which can controlled by induction in $L^\infty(S(u,\ubar))$ by $\Omega_\mck^{-2}\varepsilon/\ubar^{2+\frac{\delta}{3}}$.
\end{proof}

\begin{rem} We remark the following:
	\begin{itemize}
		\item We could prove more precise decay with respect to $\ubar$ for $\betacheck$, using the Bianchi identity in the $\nabring_3$ direction for $\betacheck$, instead of the constraint equations \eqref{eq:constraintbetabatabar}.
		\item Actually, see Section \ref{section:regionIII}, the only bounds for angular derivatives of the double null linearized quantities that we will use in region $\trois$ are
		\begin{align}
			&|\nabring^{\leq 1}\beta|\lesssim\Omega^{-2},\quad |\psicheck_H|\lesssim\frac{\Omega^{-2}}{\ubar^{2+\frac{\delta}{3}}},\quad |(\psicheck_{\Hbar},\widecheck{\omegabar})|\lesssim\frac{1}{|u|^{2+\frac{\delta}{3}}},\quad|\nabring^{\leq 1}\psi|+|K|+|\hodge{K}|\lesssim 1,\nn\\
			&|\betabar|\lesssim 1,\quad |(\rho,\hodge{\rho})|\lesssim\Omega^{-2},\quad|\nabring^{\leq 1}\gcheck|\lesssim \frac{1}{|u|^{2+\frac{\delta}{3}}},\quad |\bcheck|\lesssim\frac{1}{\ubar^{2+\frac{\delta}{3}}}.\label{eq:onlyused}
		\end{align}
		which are easily deduced from the proposition above\footnote{For the bound on $(\rho,\hodge{\rho})$, we use the definitions of $K,\hodge{K}$ which imply $(\rho,\hodge{\rho})=_s(K,\hodge{K})+\psi_{\Hbar}\psi_H$ together with the previous bounds on $\psi_H,\psi_{\Hbar},K,\hodge{K}$.}.
	\end{itemize}
	
\end{rem}
\subsubsection{The Dafermos-Luk $C^0$ extension and additional estimates}\label{section:C0extension}
In $\deux\cup\trois$, we can choose local coordinates $(u,\ubar,\theta^1=\theta_*,\theta^2=\phi_*)$, where $(\theta_*,\phi_*)$ are local coordinates on $S(u,\ubar)$ corresponding to the pullback to $\deux\cup\trois$ of the analog $(\theta_*,\phi_*)$ coordinates in Kerr as defined in Section \ref{section:linearizationdoublenull} (see also Section 4.3 in \cite{stabC0}). The Dafermos-Luk coordinates for which the spacetime is $C^0$ extensible across $\ch$ are
\begin{align}\label{eq:C0coordinates}
	(u,\ubar_\ch,\theta^A_\ch)
\end{align}
for $A=1,2$, which are defined in \cite[Section 16]{stabC0} as follows:
\begin{itemize}
	\item $\ubar_\ch$ is defined by $\ubar_\ch=\ubar$ for $\ubar\leq -1$ and there exists $\ubar_-\geq -1$ such that\footnote{Recall here that compared to the ones defined in \cite{stabC0}, our values of $u$ and $\ubar$ differ by a factor 2, see \eqref{eq:uubarDNdefikerr}.}
	$$\frac{\dee\ubar_\ch}{\dee\ubar}=e^{-|\kappa_-|\ubar},$$
	for $\ubar\geq\ubar_-$, and which satisfies $\ubar_\ch\to 0$ as $\ubar\to+\infty$.
	\item For $A=1,2$, $\theta^A_\ch$ is defined by $\theta^A_\ch=\theta^A$ on $S(u_f,\ubar_0)$ for some $\ubar_0$ such that $\frac{4Mar_-\ubar_0}{r_-^2+a^2}\in2\pi\mathbb{Z}$, and by
	\begin{align}\label{eq:thecomebacktientien}
		\left(\partial_\ubar\theta^A_\ch+b^B\partial_{\theta^B}\theta^A_\ch\right)(u_f,\ubar,\theta^1,\theta^2)=0
	\end{align}
	on $\{u=u_f\}$, and then
	$$\partial_u\theta^A_\ch=0$$
	in $\deux\cup\trois$. Actually two sets of coordinates $(\theta^1_{(i),\ch},\theta^2_{(i),\ch})$, $i=1,2$ must be constructed via the stereographic projection, to cover the whole sphere $\mathbb{S}^2$. More precisely, defining 
	\begin{align}
		(\theta^1_{(1)},\theta^2_{(1)})&=(\cot\frac{\theta_*}{2}\cos\phi_*,\cot\frac{\theta_*}{2}\sin\phi_*),\quad\text{on }\:\mcv_1=\mathbb{S}^2\backslash\{\theta_*=0\},\label{eq:defstereocoord}\\
		(\theta^1_{(2)},\theta^2_{(2)})&=(\tan\frac{\theta_*}{2}\cos\phi_*,\tan\frac{\theta_*}{2}\sin\phi_*),\quad\text{on }\:\mcv_2=\mathbb{S}^2\backslash\{\theta_*=\pi\},\nn
	\end{align}
and the precompact sets $\mcv'_1\subset\mcv''_1\subset\mcv_1$, $\mcv'_2\subset\mcv''_2\subset\mcv_2$ by
\begin{align*}
	\mcv'_1&:=\{(\theta^1_{(1)})^2+(\theta^2_{(1)})^2<2\},\quad \mcv''_1:=\{(\theta^1_{(1)})^2+(\theta^2_{(1)})^2<3\},\\
	\mcv'_2&:=\{(\theta^1_{(2)})^2+(\theta^2_{(2)})^2<2\},\quad \mcv''_2:=\{(\theta^1_{(2)})^2+(\theta^2_{(2)})^2<3\},
\end{align*}
it is proven in Lemmas 16.8 and 16.10 of \cite{stabC0} that, defining $\theta^A_{\ch,(i)}$ with respect to $\theta^A_{(i)}$ on $\mcv_i$ by \eqref{eq:thecomebacktientien}, the definition is coherent on the intersection $\mcv_1\cap\mcv_2$, and that for any $(\theta^1_{(i)},\theta^2_{(i)})\in\mcv'_i$, we have $(\theta^1_{(i)},\theta^2_{(i)})\in\mcv''_i$ and 
 \begin{align}\label{eq:oulalaclutch}
 	\sup_{A=1,2}|\theta^A_{\ch,(i)}-\theta^A_{\ch,(i),\mck}|\lesssim\varepsilon.
 \end{align}
\end{itemize}
Then, it is proven in \cite[Theo. 16.14]{stabC0} that in these coordinates, the spacetime metric extends continuously to the Cauchy horizon $\ch=\{\ubar_\ch=0\}$. Moreover, in these coordinates the metric takes the form
\begin{align}\label{eq:metricformC0}
	\g=-2&\Omega^2_\ch\left(\dee u\otimes\dee\ubar_\ch+\dee\ubar_\ch\otimes\dee u\right)\\
	&+(\gamma_\ch)_{AB}\left(\dee\theta^A_\ch-(b_\ch)^A\dee\ubar_\ch\right)\otimes\left(\dee\theta^B_\ch-(b_\ch)^B\dee\ubar_\ch\right),\nn
\end{align}
where
\begin{align}\label{eq:quantitesC0}
	&(\gamma_\ch)_{AB}=\gamma_{A'B'}\left(\left(\partial_\theta\theta_\ch\right)^{-1}\right)_A^{\:\:\:A'}\left(\left(\partial_\theta\theta_\ch\right)^{-1}\right)_B^{\:\:\:B'},\\
	\Omega^2_\ch&=\Omega^2e^{|\kappa_-|\ubar},\quad (b_\ch)^A=e^{|\kappa_-|\ubar}\left(\partial_\ubar\theta^A_\ch+b^B\partial_{\theta^B}\theta^A_\ch\right),\nn
\end{align}
are such that $\gamma_\ch,\Omega^2_\ch,b^A_\ch$ extend continuously to $\ch$ in $(u,\ubar_\ch,\theta^A_\ch)$ coordinates.

\textbf{We now prove some technical estimates that will be needed in Section \ref{section:regionIII} to apply Sbierski's theorem to deduce the Lipschitz inextendibility from curvature blow-up.} We begin with assumption \cite[(3.7)]{sbierskiinextdernier}. Note that from the identities above and the definitions of $\ubar_\ch,\theta^A_\ch$, we have, in the coordinates $(u,\ubar_\ch,\theta^A_\ch)$, the identity 
$$\partial_{\ubar_\ch}=e^{|\kappa_-|\ubar}\left(\partial_\ubar+b^B\partial_{\theta^B}\right)-b^A_\ch\partial_{\theta^A_\ch}.$$
Indeed, the vector field on the RHS above vanishes when applied to $u,\theta^A_\ch$, $A=1,2$, and yields $1$ when applied to $\ubar_\ch$. This implies in particular
\begin{align}\label{eq:e4estC0}
	\ering_4=\Omega^{-2}_\ch\left(\partial_{\ubar_\ch}+b^A_\ch\partial_{\theta^A_\ch}\right).
\end{align}
Also note that the coordinate vector field $\partial_{u_\ch}$ with respect to $u$ in the coordinates $(u,\ubar_\ch,\theta^A_\ch)$ remains unchanged, $ \partial_{u_\ch}=\partial_u=\ering_3$. Sbierski's work \cite{sbierskiinextdernier} uses the $C^0$ coordinates \eqref{eq:C0coordinates}, and hence the metric form \eqref{eq:metricformC0} and the double null quantities \eqref{eq:quantitesC0}. By \eqref{eq:e4estC0} and the identity $\ering_3=\partial_{u_\ch}$ we get that the Ricci coefficients $\omegabar,\eta,\etabar,\chibar,\chi$ defined with respect to the double null frame
$$\partial_{u_\ch},\quad\Omega^{-2}_\ch\left(\partial_{\ubar_\ch}+b^A_\ch\partial_{\theta^A_\ch}\right),$$
in \cite[Section 2.2]{sbierskiinextdernier}, coincide with the Ricci coefficients $\mathring{\omegabar},\mathring{\eta},\mathring{\etabar},\mathring{\chibar},\mathring{\chi}$ defined with respect to the double null frame $\ering_\mu$ in our context. Thus, using \eqref{eq:onlyused}, we get for any $u_0<u_1<u_f$
\begin{align}\label{eq:hypsbierski1}
	\sup_{\deux\cup\trois}\left(|\mathring{\omegabar}|+|\mathring{\eta}|+|\mathring{\etabar}|+|\mathring{\chibar}|\right)\lesssim 1,\quad \int_{-C}^0\sup_{u_0<u<u_1}\|\mathring{\chi}\|_{L^\infty(S(u,\ubar_\ch))}\dee\ubar_\ch\lesssim 1,
\end{align}
for any $C>0$, where for the second bound above we computed 
\begin{align*}
	\int_{-C}^0\sup_{u_0<u<u_1}\|\mathring{\chi}\|_{L^\infty(S(u,\ubar))}\dee\ubar_\ch&=\int_{C'}^\infty e^{-|\kappa_-|\ubar}\sup_{u_0<u<u_1}\|\mathring{\chi}\|_{L^\infty(S(u,\ubar))}\dee\ubar\\
	&\lesssim\int_{C'}^\infty\left(e^{-|\kappa_-|\ubar}+\frac{e^{|\kappa_-|u_1}}{\ubar^{2+\frac{\delta}{3}}}\right)\dee\ubar\lesssim 1,
\end{align*}
for some $C'>0$, where we used \eqref{eq:onlyused} and $e^{|\kappa_-|u_1}\leq 1$ because $u_1<0$. This proves assumption \cite[(3.7)]{sbierskiinextdernier}. Now, we prove assumption \cite[(3.15)]{sbierskiinextdernier}. This assumption writes as follows: for some $C>0$, and for any $u_0<u_1<u_f$, and $i=1,2$, $A,B=1,2$,
\begin{align}\label{eq:hypsbierski2}
	\int_{-C}^0\sup_{u_0<u<u_1}\|\partial_{\theta^B_{(i),\ch}}(b_\ch)^A_{(i)}\|_{L^\infty(S(u,\ubar_\ch)\cap\mcv'_i)}\dee\ubar_\ch\lesssim_{u_0,u_1} 1.
\end{align}

\noindent\textbf{Proof of \eqref{eq:hypsbierski2}.} To prove \eqref{eq:hypsbierski2}, we use that $(b_\ch)^A_{(i)}$ vanishes on $\{u=u_f\}$, and that $\theta^A_\ch$ is transported along $\partial_u$. By \eqref{eq:quantitesC0} we compute, omitting the subscripts $(i)$,
\begin{align*}
	\partial_u \partial_{\theta^B_{\ch}}(b_\ch)^A&=\partial_{\theta^B_{\ch}}\left(e^{|\kappa_-|\ubar}\partial_u\left(\partial_\ubar\theta^A_\ch+b^B\partial_{\theta^B}\theta^A_\ch\right)\right)=e^{|\kappa_-|\ubar}\partial_{\theta^B_{\ch}}\left(\partial_ub^C\partial_{\theta^C}\theta^A_\ch\right)\\
	&=2e^{|\kappa_-|\ubar}\partial_{\theta^B_{\ch}}\left(\Omega^2(\eta^C-\etabar^C)\partial_{\theta^C}\theta^A_\ch\right)\\
	&=2e^{|\kappa_-|\ubar}\left(\left((\partial_{\theta^B_{\ch}}\Omega^2)(\eta^C-\etabar^C)+\Omega^2\partial_{\theta^B_{\ch}}(\eta^C-\etabar^C)\right)\partial_{\theta^C}\theta^A_\ch\right)\\
	&\quad +2e^{|\kappa_-|\ubar}\Omega^2(\eta^C-\etabar^C)\partial_{\theta^B_{\ch}}\left(\partial_{\theta^C}\theta^A_\ch\right),
\end{align*}
where we used \eqref{eq:parubA}. Moreover, we have
\begin{align}\label{eq:transitionouioui}
	\partial_{\theta^B_{\ch}}=\left(\left(\partial_\theta\theta_\ch\right)^{-1}\right)_B^{\:\:\:C}\partial_{\theta^C},
\end{align}
where the matrix $\left(\left(\partial_\theta\theta_\ch\right)^{-1}\right)_B^{\:\:\:C}$ has bounded components in $\mcv'_i$ (see \cite[Prop. 16.11]{stabC0}). Thus by the bounds on $\nabring\psi$ and $\nabring\log\Omega$ in \eqref{eq:onlyused}, we get in $\mcv'_i$
\begin{align*}
	|(\partial_{\theta^B_{\ch}}\Omega^2)(\eta^C-\etabar^C)|\lesssim \Omega^2|\partial_{\theta^B_{\ch}}\log\Omega|\lesssim \Omega^2,\quad|\Omega^2\partial_{\theta^B_{\ch}}(\eta^C-\etabar^C)|\lesssim \Omega^2.
\end{align*}
Using $\Omega^2\sim e^{-|\kappa_-|(u+\ubar)}$ we deduce, for $(\theta^1_{(i)},\theta^2_{(i)})\in\mcv'_i$ and $u>u_0$,
\begin{align}\label{eq:equatandhence}
	|\partial_u \partial_{\theta^B_{\ch}}(b_\ch)^A|(u,\ubar,\theta^1_{(i)},\theta^2_{(i)})\lesssim_{u_0} 1+|\partial_{\theta^B_{\ch}}\left(\partial_{\theta^C}\theta^A_\ch\right)|. 
\end{align}
Thus, by \eqref{eq:transitionouioui} and the boundedness of $\left(\left(\partial_\theta\theta_\ch\right)^{-1}\right)_B^{\:\:\:C}$, it is only left to prove the bound
\begin{align}\label{eq:provetheboundahah}
	|\partial_{\theta^B}\partial_{\theta^C}\theta^A_\ch|(u,\ubar,\theta^1_{(i)},\theta^2_{(i)})\lesssim 1,
\end{align}
for $(\theta^1_{(i)},\theta^2_{(i)})\in\mcv'_i$. To do this, we proceed as in \cite[Prop. 16.11]{stabC0} where the bound
$$\left|\partial_{\theta^B}(\theta^A_\ch-\theta^A_{\ch,\mck})\right|\lesssim\varepsilon$$
is proven, except that we commute the estimates with one more $\partial_{\theta^C}$ derivative\footnote{Here, compared to the estimates done in \cite{stabC0}, we will use the estimates for $\nabring$ derivatives of order higher than 3, which we proved in Theorem \ref{thm:controlangulairedoublenull}).}. More precisely, similarly as in \cite[Prop. 16.11]{stabC0}, it is enough to prove the bound \eqref{eq:provetheboundahah} on $\{u=u_f\}$. Differentiating \cite[Eq. (16.45)]{stabC0} by $\partial_{\theta^C}$ we get the transport equation along $\{u=u_f\}$
\begin{align*}
	(\partial_\ubar+b^D\partial_{\theta^D})(\partial_{\theta^C}\partial_{\theta^B}\theta^A_\ch-\partial_{\theta^C}\partial_{\theta^B}\theta^A_{\ch,\mck})=I+II+III,
\end{align*}
and hence using the fact that initial data is zero on $S(u_f,\ubar_0)$ and integrating along the flow of $\partial_\ubar+b^C\partial_{\theta^C}$ we get for $(\theta^1_{(i)},\theta^2_{(i)})\in\mcv_{i}'$,
\begin{align}\label{eq:summingest}
	|\partial_{\theta^C}\partial_{\theta^B}\theta^A_\ch-\partial_{\theta^C}\partial_{\theta^B}\theta^A_{\ch,\mck}|(u_f,\ubar,\theta^1_{(i)},\theta^2_{(i)})\lesssim \int_{-u_f+C_R}^\ubar \left(|I|+|II|+|III|\right)\dee\ubar',
\end{align}
and where the terms $I,II,III$ are defined and bounded as follows:
\begin{itemize}
	\item $I:=-\partial_{\theta^C}\partial_{\theta^B}\bcheck^D\partial_{\theta^D}\theta^A_{\ch,\mck}-\partial_{\theta^B}\bcheck^D\partial_{\theta^C}\partial_{\theta^D}\theta^A_{\ch,\mck}+\partial_{\theta^C}\bcheck^D\partial_{\theta^D}\partial_{\theta^B}\theta^A_{\ch,\mck}+\bcheck^D\partial_{\theta^C}\partial_{\theta^D}\partial_{\theta^B}\theta^A_{\ch,\mck}$ satisfies by the Sobolev embedding
	$$\int_{-u_f+C_R}^\ubar|I|\dee\ubar'\lesssim \|\ubar^{\frac{5}{2}+\frac{\delta}{3}}\nabring^{\leq 4}\bcheck\|_{L^\infty_u L^2_\ubar L^2_S}\lesssim\varepsilon,$$
	where we used the bound for $\mcn_{hyp}^{(0)}$ in Theorem \ref{thm:controlangulairedoublenull}, and $|(\partial_{\theta^C})^{\leq 3}\theta^A_{\ch,\mck}|\lesssim 1$ by \eqref{eq:DNthetaCH}.
	\item $II:=-\partial_{\theta^C}\partial_{\theta^B}b^D(\partial_{\theta^D}\theta^A_\ch-\partial_{\theta^D}\theta^A_{\ch,\mck})$ satisfies
	$$\int_{-u_f+C_R}^\ubar|II|\dee\ubar'\lesssim \varepsilon\int_{-u_f+C_R}^\ubar e^{-|\kappa_-|(u_f+\ubar')}\dee\ubar'+\|\ubar^{\frac{5}{2}+\frac{\delta}{3}}\nabring^{\leq 4}\bcheck\|_{L^\infty_u L^2_\ubar L^2_S}\lesssim\varepsilon$$
	where we used the bound for $\mcn_{hyp}^{(0)}$ again and the bound 
	\begin{align}\label{eq:derdebomegacarre}
		|(\partial_{\theta^C})^{\leq 1}\partial_{\theta^B}b_\mck^D|\lesssim\Omega^2,
	\end{align}
which comes from \eqref{eq:bDNdanskerr}, \eqref{eq:identitybelow}\footnote{Also note the computation of $\partial_{\theta_*}r$ in \cite[Prop. A.5]{stabC0} which shows that $\partial_{\theta^B}$ derivatives of $r$ are controlled by $\Omega^2$. This also follows from Proposition A.14 in \cite{stabC0}.}, as well as the bound $|\partial_{\theta^D}\theta^A_\ch-\partial_{\theta^D}\theta^A_{\ch,\mck}|\lesssim\varepsilon$ which is the result of \cite[Prop. 16.11]{stabC0}. 
\item $III:=-\partial_{\theta^B}b^D(\partial_{\theta^C}\partial_{\theta^D}\theta^A_\ch-\partial_{\theta^C}\partial_{\theta^D}\theta^A_{\ch,\mck})+\partial_{\theta^C}b^D(\partial_{\theta^D}\partial_{\theta^B}\theta^A_\ch-\partial_{\theta^D}\partial_{\theta^B}\theta^A_{\ch,\mck})$ which satisfies
\begin{align*}
	&\int_{-u_f+C_R}^\ubar|III|\dee\ubar'\lesssim\\
	& \int_{-u_f+C_R}^\ubar \left(e^{-|\kappa_-|(u_f+\ubar')}\dee\ubar'+\|\nabring^{\leq 3}\bcheck\|_{L^2(S(u_f,\ubar'))}\right)\sup_{S(u_f,\ubar')\cap\mcv''_i}|\partial_{\theta^C}\partial_{\theta^B}\theta^A_\ch-\partial_{\theta^C}\partial_{\theta^B}\theta^A_{\ch,\mck}|,
\end{align*}
where we used \eqref{eq:derdebomegacarre} again.
\end{itemize}
Finally, using $\mcv''_1\backslash\mcv'_1\subset\mcv_2'$ and $\mcv''_2\backslash\mcv'_2\subset\mcv_1'$, summing the estimates \eqref{eq:summingest} for $i=1,2$ and injecting the bounds for $I,II,III$ above we can apply Grönwall's inequality to the function
$$A(\ubar):=\sum_{i=1,2}\sup_{\theta^{1,2}_{(i)}\in\mcv'_i}|\partial_{\theta^C}\partial_{\theta^B}\theta^A_\ch-\partial_{\theta^C}\partial_{\theta^B}\theta^A_{\ch,\mck}|(u_f,\ubar,\theta^1_{(i)},\theta^2_{(i)}),$$
and deduce $A(\ubar)\lesssim\varepsilon$, where we used again the bound for $\bcheck$ in the estimate $\mcn_{hyp}^{(0)}\lesssim\varepsilon$. Combining this estimate with $|\partial_{\theta^C}\partial_{\theta^B}\theta^A_{\ch,\mck}(u_f,\ubar,\theta^1_{(i)},\theta^2_{(i)})|\lesssim 1$ for $\theta^{1,2}_{(i)}\in\mcv'_i$, we deduce that \eqref{eq:provetheboundahah} holds, and hence by \eqref{eq:equatandhence}, for $(\theta^1_{(i)},\theta^2_{(i)})\in\mcv'_i$ and $u>u_0$,
$$	|\partial_u \partial_{\theta^B_{\ch}}(b_\ch)^A|(u,\ubar,\theta^1_{(i)},\theta^2_{(i)})\lesssim_{u_0} 1.$$
Thus, integrating the bound above from $u>u_0$ to $u_f$ we deduce
$$|\partial_{\theta^B_{\ch}}(b_\ch)^A|(u,\ubar,\theta^1_{(i)},\theta^2_{(i)})\lesssim_{u_0} 1+|\partial_{\theta^B_{\ch}}(b_\ch)^A|(u_f,\ubar,\theta^1_{(i)},\theta^2_{(i)})\lesssim 1,$$
where we used that $b^A_\ch|_{\{u=u_f\}}=0$ by \eqref{eq:quantitesC0} and \eqref{eq:thecomebacktientien}. Integrating the bound above with respect to $\dee\ubar_\ch=e^{-|\kappa_-|\ubar}\dee\ubar$, this concludes the proof of \eqref{eq:hypsbierski2}.

{Finally, using our higher decay assumptions, we slightly improve the bound \eqref{eq:oulalaclutch}.} This is done as follows : in the proof of \eqref{eq:oulalaclutch} in \cite{stabC0} (p.178), it is proven that the LHS of \eqref{eq:oulalaclutch} is bounded by the following integral
$$\int_{-u_f+C_R}^\ubar\|\nabring^{\leq 2}\bcheck\|_{L^2(S(u_f,\ubar'))}\dee\ubar',$$
which we bound as follows by Cauchy-Schwarz: 
\begin{align*}
	\int_{-u_f+C_R}^\ubar\|\nabring^{\leq 2}\bcheck\|_{L^2(S(u_f,\ubar'))}\dee\ubar'&\lesssim\|\ubar^{\frac{5}{2}+\frac{\delta}{3}}\nabring^{\leq 2}\bcheck\|_{L^2_\ubar L^\infty_u L^2_S}\left(\int_{-u_f+C_R}^\ubar\ubar'^{-5-2\delta/3}\dee\ubar'\right)^{1/2}\\
	&\lesssim(\mcn_{hyp}^{(0)})^{1/2}|u_f|^{-2-\delta/3}\lesssim \varepsilon|u_f|^{-2-\delta/3}
\end{align*}
by Theorem \ref{thm:controlangulairedoublenull}. We deduce the following improved bound on $\mcv'_i$,
 \begin{align}\label{eq:oulalaclutchbetter}
	\sup_{A=1,2}|\theta^A_{\ch,(i)}-\theta^A_{\ch,(i),\mck}|\lesssim\varepsilon|u_f|^{-2-\delta/3}.
\end{align}

\subsection{Control of $(\Omega^2\nabring_4,\nabring_3,\nabring)^k$ derivatives in $\deux$  (proof of Theorem \ref{thm:regionII})}\label{section:controllietdn}

In this section, we prove decay estimates in $\deux$ for arbitrary $\df^{\leq k}$ derivatives of the linearized quantities, where $\df$ corresponds in this section to the following set of renormalized derivatives,
\begin{align}\label{eq:dfdoublenull}
	\df=\{\Omega^2\nabring_4,\nabring_3,\nabring\}.
\end{align}
\begin{rem}
	 In this subsection we will eventually restrict the analysis to region $\deux$, where $\ubar\sim|u|$, see Lemma \ref{lem:usimubarII}. This estimate is what allows us to prove decay for general renormalized $\df$ derivatives of the linearized quantities. Indeed, in region $\deux$, once some linearized quantities are renormalized by an appropriate $\Omega^2$ factor, they can be proven to all decay at a polynomial rate with respect to $\ubar$. It is thus possible to prove the bounds for $\df$ derivatives by general induction arguments on the number of derivatives\footnote{Note that in Section \ref{section:notethtatt} below, we prove separate $\ubar$ and $|u|$ for $\lieT^{k'}\nabring^k$ derivatives of the linearized quantities in region $\deux\cup\trois$. This is possible thanks to the good commutation relations between $\lieT$ and $\nabring_{3,4,A}$. More precisely, commutations with $\lieT$ only add quadratic error terms, so that quantities with different $\ubar$ and $|u|$ decay are not mixed at the linear level for $\lieT^{k'}\nabring^k$ derivatives.}.
\end{rem}
Before proving the estimates for arbitrary $\df$ derivatives, we begin in the following subsection by controlling $\lieT^{k'}\nabring^k$ derivatives. This is done by commuting the schematic equations satisfied by $\nabring^k$ derivatives of the linearized quantities with $\lieT$ derivatives, which satisfy some good commutation properties as $\T$, as defined in \eqref{eq:T}, is an approximate symmetry.

\subsubsection{Control of $\lieT^{k'}\nabring^k$ derivatives of the linearized quantities}\label{section:notethtatt}
Recall from Section \ref{section:horizontalliederivative} the definition of the horizontal lie derivative $\lieT$, where here in double null coordinates the horizontal distribution is $\mathring{\mch}=TS(u,\ubar)$. In this section, we prove the pointwise control of $\lieT^{k'}\nabring^{ k}$ derivatives of the linearized double null quantities, which will be done by induction. To this end, we derive the schematic equations satisfies by $\lieT^{k'}\nabring^{ k}$ derivatives in Section \ref{section:schematicequationslieT}, \textbf{where we also introduce the energy terms $\mcn_{int}^{(k',k)}$, $\mcn_{hyp}^{(k',k)}$, $\mcn_{sph}^{(k',k)}$, which generalize the ones defined in Section \ref{section:schematicnotations}}. We now introduce the following induction assumption:
 \begin{align*}
 	H(k',k):\quad\begin{cases}
 			&|\lieT^{\leq k'}\nabring^{\leq k}\gcheck|+|\lieT^{\leq k'}\nabring^{\leq k-1}(\psicheck_{\Hbar},\widecheck{\omegabar},\psicheck)|+|\lieT^{\leq k'}\nabring^{\leq k-2}\Kcheck|\lesssim\dfrac{\varepsilon}{|u|^{2+\frac{\delta}{3}}},\\
 		&|\Omega_\mck^2\lieT^{\leq k'}\nabring^{\leq k-1}\psicheck_H|+|\lieT^{\leq k'}\nabring^{\leq k-1}\bcheck|\lesssim\dfrac{\varepsilon}{\ubar^{2+\frac{\delta}{3}}},\\
 		&\mcn_{int}^{(k',k)}+\mcn_{hyp}^{(k',k)}+\mcn_{sph}^{(k',k)}\lesssim\varepsilon^2.
 	\end{cases}\quad\text{in }\deux\cup\trois.
 \end{align*}
\begin{prop}\label{prop:inductionliekprim}
	Let $k'\geq 1,k_0\geq 0$ such that $k_0+k'-1\leq N_2''$. Then, we have the implication
	$$H(k'-1,k_0)\implies H(k',k_0-5).$$
\end{prop}
\begin{proof}
See Section \ref{section:proofinducliekprim} (and the preliminary commutations in Section \ref{section:schematicequationslieT}).
\end{proof}

\begin{rem}\label{rem:seerem}
	We remark the following: 
	\begin{itemize}
		\item We know that $H(0,N_2'')$ holds by Proposition \ref{prop:decaypointwiseDNpourIII} and Theorem \ref{thm:controlangulairedoublenull}. The result above can thus be used to control by induction any number of $\lieT^{k'}\nabring^k$ derivative, provided $N_2$ is large enough. 
		\item We note that here, the number 5 of angular derivatives lost in the induction step to recover one $\lieT$ derivative is not optimal with respect to the loss of derivatives. We chose to lose enough angular derivatives at each step so that many auxiliary terms which appear in the schematic equations satisfied by $\lieT^{k'}\nabring^k$ derivatives can be put in $L^\infty$. This ensures that the corresponding reduced schematic equations can be easily integrated with no additional difficulty to deal with non-linear terms. Using the quadratic structure of the commutators with $\lieT$ derivative, and bounding the terms with lower order $\lieT$ derivatives (which may present some higher order $\nabring$ derivatives) by $H(k'-1,k_0)$, it is then straightforward to prove $H(k',k_0-5)$ by applying the same Dafermos-Luk estimates as for the $\nabring^k$ derivatives in Section \ref{section:DLmoredecaynabring}.
	\end{itemize}
\end{rem}

\begin{cor}\label{cor:n0primprimprim}
	Let $N_2'''$ be the integer part of $N_2''/5$ where $N_2''$ is defined in Proposition \ref{prop:decaypointwiseDNpourIII}. Then for any $k',k\geq 0$ such that $k'+k\leq N_2'''-2$, we have the following estimates in $\deux\cup\trois$,
\begin{align*}
	&|\lieT^{\leq k'}\nabring^{\leq k}(\gcheck,\psicheck_{\Hbar},\widecheck{\omegabar},\psicheck,\Kcheck,\widecheck{\hodge{K}},\betabarcheck)|\lesssim \dfrac{\varepsilon}{|u|^{2+\frac{\delta}{3}}},\quad |\lieT^{\leq k'}\nabring^{\leq k}\betacheck|\lesssim \dfrac{\varepsilon\Omega^{-2}}{|u|^{2+\frac{\delta}{3}}},\\
	&|\lieT^{\leq k'}\nabring^{\leq k}\psicheck_H|\lesssim \dfrac{\varepsilon\Omega^{-2}}{\ubar^{2+\frac{\delta}{3}}},\quad |\lieT^{\leq k'}\nabring^{\leq k}\bcheck|\lesssim \dfrac{\varepsilon}{\ubar^{2+\frac{\delta}{3}}}.
\end{align*}
\end{cor}
\begin{proof}
	As mentioned in Remark \ref{rem:seerem}, this is a simple consequence of Propositions \ref{prop:inductionliekprim} (the induction step) and \ref{prop:decaypointwiseDNpourIII} and Theorem \ref{thm:controlangulairedoublenull} (the fact that $H(0,N_2'')$ holds). Note that we need $k'+k\leq N_2'''-2$ instead of $k'+k\leq N_2'''$ because we state a control of same order of derivative for all linearized quantities, including $\Kcheck$. Also, we recover $\betacheck,\betabarcheck,\widecheck{\hodge{K}}$ using the constraint equations \eqref{eq:constraintbetabatabar}.
\end{proof}
\begin{rem}
	Combining the Bianchi equation for $\nabring_3\wh{{\chibar}}$ which rewrites ${\alphabar}=_s\psi_{\Hbar}^2+\omegabar\psi_{\Hbar}+\nabring_3\wh{{\chibar}}$, Lemma \ref{lem:lienablaT}, \eqref{eq:expreT}, and the schematic null structure equation $\Omega^2\nabring_4\wh{{\chibar}}=_s\Omega^2(\psi_H\psi_{\Hbar}+\nabring\psi+\psi^2)$, we get in $\deux\cup\trois$ the bound
	\begin{align}
		|{\alphabar}|\lesssim |\lieT\wh{{\chibar}}|+|(\Omega^2\psi _H,\psi_{\Hbar},b)||\psi_{\Hbar}|+\Omega^2|(\psi_H\psi_{\Hbar},\nabring\psi,\psi^2)|+|b||\nabring\psi_{\Hbar}|+|\psi_{\Hbar}|^2+|\omegabar||\psi_{\Hbar}|\lesssim 1,\label{eq:atsomepointtkt}
	\end{align}
	where we used Corollary \ref{cor:n0primprimprim} in the last bound above. This bound for ${\alphabar}$ will be used in $\trois$. 
\end{rem}
\subsubsection{Control of the remaining derivatives by induction}
Recall from \eqref{eq:dfdoublenull} the set $\df=\{\Omega^2\nabring_4,\nabring_3,\nabring\}$ of derivatives in the double null gauge.
\begin{thm}\label{thm:controldoublenull}
	Defining $N_{max}=N_2'''-3$, we have the following decay estimates in the double null frame in region $\deux$,
	\begin{align*}
		|\df^{\leq N_{max}}(\gcheck,\bcheck,\psicheck_{\Hbar},\widecheck{\omegabar},\psicheck,\Kcheck,\hodge{\Kcheck},\betabarcheck,\widecheck{\alphabar})|&\lesssim \frac{\varepsilon}{\ubar^{2+\frac{\delta}{3}}},\\
		|\df^{\leq N_{max}}\psicheck_H|\lesssim \frac{\varepsilon\Omega^{-2}}{\ubar^{2+\frac{\delta}{3}}},\quad |\df^{\leq N_{max}}\betacheck|&\lesssim \frac{\varepsilon\Omega^{-2}}{\ubar^{2+\frac{\delta}{3}}},\quad |\df^{\leq N_{max}}\widecheck{\alpha}|\lesssim \frac{\varepsilon\Omega^{-4}}{\ubar^{2+\frac{\delta}{3}}},
	\end{align*}
where we recall from Remark \ref{rem:CRfixee} that the implicit constants above depend only on $a,M,C_R$.
\end{thm}
\begin{proof}
	See Section \ref{section:proofthmdoublenull}. 
\end{proof}
\begin{rem}We remark the following:
	\begin{itemize}
		\item The polynomial decay estimates with respect to $\ubar$ in region $\deux$ above are consistant with the decay estimates in Proposition \ref{prop:decaypointwiseDNpourIII} and Corollary \ref{cor:n0primprimprim} in region $\deux\cup\trois$, because in region $\deux$ we have $\ubar\sim|u|$, see Lemma \ref{lem:usimubarII}.
		\item As a consequence of the identities $K=-\rho+\frac12\wh{\chibar}\cdot\chihat-\frac14tr\chi tr\chibar$ and $ \hodge{K}=\hodge{\rho}+\frac12\hat{\chibar}\wedge\chihat$, we also have the following less precise estimates in $\deux$,
		$$|\df^{\leq N_{max}}(\widecheck{\rho},\widecheck{\hodge{\rho}})|\lesssim\frac{\varepsilon\Omega^{-2}}{\ubar^{2+\frac{\delta}{3}}}.$$
		\item The proof of Theorem \ref{thm:controldoublenull} has two steps. First, we control all the linearized quantities except $\widecheck{\alphabar}$ and $\widecheck{\alpha}$. To this end, we use that $\nabring^i$ derivatives of the renormalized quantities
		$$\widecheck{S}=\big\{\gammacheck,\logomegacheck, \bcheck,\etabarcheck,\etacheck,\Omega_\mck^2\psicheck_H,\psicheck_{\Hbar},\widecheck{\omegabar},\Omega\Omega_\mck\betacheck,\Kcheck,\hodge{\Kcheck},\betabarcheck\big\},$$
		satisfy a closed system of equations in $\nabring_3$ and $\Omega^2\nabring_4$. More precisely, see Section \ref{section:schematicequationsangular}, each quantity $\phi\in\nabring^i\widecheck{S}$ satisfies either a $\nabring_3$ or a $\Omega^2\nabring_4$ schematic equation where the RHS can be expressed as a bounded sum of products of terms in $\nabring^{\leq i+1}\widecheck{S}$. Using the identity
		$$\lieT U =_s \Omega^2\nabring_4 U-\nabring_3U-b\cdot\nabring U+(\Omega^2\chi-\chibar-\nabring b)\cdot U,$$ 
		for any $S(u,\ubar)$-tangent tensor $U$ (see Lemma \ref{lem:lienablaT}), we get that for $\phi\in\nabring^i\widecheck{S}$, the quantity $(\nabring_3,\Omega^2\nabring_4)\phi$ can be expressed as a sum of products of bounded terms in $\lieT^{\leq 1}\nabring^{\leq i+1}\widecheck{S}$. We then proceed by induction (also using the structure of the system satisfied by $\lieT^{k'}\nabring^k$ derivatives in Section \ref{section:schematicequationslieT}) to prove that for $l\geq 1$ the higher order derivatives $(\nabring_3,\Omega^2\nabring_4)^l\phi$ can be expressed with quantities $\lieT^{k'}\nabring^k\widecheck{S}$, and we use Corollary \ref{cor:n0primprimprim} to conclude the stated bounds for $(\nabring_3,\Omega^2\nabring_4)^l\nabring^i\widecheck{S}$. Then we prove similar bounds for the quantities $\widecheck{\alphabar}$ and $\Omega^4\widecheck{\alpha}$ by using the null structure equations for $\nabring_4\chihat$ and $\nabring_3\wh{\chibar}$.
		\item Theorem \ref{thm:regionII} is a consequence of Theorem \ref{thm:controlangulairedoublenull}, Proposition \ref{prop:decaypointwiseDNpourIII}, and Theorem \ref{thm:controldoublenull}.
	\end{itemize}

\end{rem}

\section{Auxiliary non-integrable frame in region $\deux$}\label{section:gauge}

We recall that we denote with a ring, for example $\mathring{tr\chi}$, the quantities defined in Section \ref{section:doublenullgrandesection} with respect to the outgoing double null frame $(\ering_3,\ering_4,\ering_1,\ering_2)$ constructed in Section \ref{section:IIdoublenull}. In this section, we will construct in region $\deux$ a null-frame $({e_3}',{e_4}',{e_a}',a=1,2)$ that we will call \emph{outgoing non-integrable}, in the sense that it is a perturbation of the outgoing principal null frame \eqref{eq:principaloutgoinkerr} in exact Kerr. We will moreover require the gauge conditions ${\Xibar}'=0,\quad{\Xi}'=0,\quad{\omega}'=0$, where the primed quantities are the Ricci and curvature coefficients with respect to the outgoing non-integrable frame. Thus $e_4'$ will be geodesic.\footnote{These gauge conditions will allow us to both simplify the analysis of the Teukolsky equation in region $\deux$, and use the Bianchi identities for $\wh{X}'$ and $B'$ to prove improved decay of these quantities from the improved decay for $A$ implied by the Teukolsky equation.}

There will also be an ingoing normalization of the outgoing non-integrable frame: the \textit{ingoing non-integrable frame}, denoted $(e_3,e_4,e_a, a=1,2)$, which will satisfy $\D_3 e_3=0$, and which will be a perturbation of the Kerr ingoing principal frame \eqref{eq:principalingoinkerr}. The Ricci and curvature components defined with respect to this frame will be denoted without any additional symbol, for example $trX$. \textbf{Note that the notations for the ingoing non-integrable frame in $\deux$ and for the ingoing PT frame in $\un$ coincide. This is consistant with the fact that we will initialize the ingoing non-integrable frame on $\Sigma_0$ so that it coincides with the PT frame there.}

\begin{rem}
	In this section, we denote by $u,\ubar$ the double null advanced and retarded time constructed in region $\deux\cup\trois$ in Section \ref{section:IIdoublenull}.
\end{rem}
\subsection{Construction of the non-integrable frames in $\deux$}
\subsubsection{Null frame transformations and Kerr values}
\textbf{Frame construction.} We will perform a null frame transformation of the type \eqref{eq:frametransfo}, with $(e_\mu)=(\ering_\mu)$ being the double null frame of in $\deux\cup\trois$ constructed in Section \ref{section:IIdoublenull} to construct the non-integrable frames. Before that, the first step is to show that there exists such a transformation in exact Kerr.
\begin{prop}\label{prop:transfodanskerr}
	In region $\{u+\ubar\geq C_R\}\cap\{u<u_f\}$ of the Kerr black hole interior, the outgoing principal frame \eqref{eq:principaloutgoinkerr} is obtained from the Kerr double null frame \eqref{eq:DuDubardoublenull} by a transformation of the type \eqref{eq:frametransfo} with coefficients $(\lambda_\mck,f_\mck,\fbar_\mck)$, where $\lambda_\mck$ is a scalar and $f_\mck,\fbar_\mck$ are $S(u,\ubar)$-tangent 1-forms, which satisfy the estimates
	$$\lambda_\mck\sim 1,\quad |f_\mck|\lesssim 1,\quad |\fbar_\mck|\lesssim\Omega^2_\mck.$$
Also, recalling the derivatives $\df_\mck=\{\Omega^2_\mck(\nabring_4)_\mck,(\nabring_3)_\mck,\nabring_\mck\}$ in Kerr, we have for $N\geq 0$,
\begin{align}\label{rem:deriveesdanskerrr}
	|\df_\mck^{\leq N}\lambda_\mck|\lesssim 1,\quad |\df_\mck^{\leq N}f_\mck|\lesssim 1,\quad |\df_\mck^{\leq N}\fbar_\mck|\lesssim\Omega^2_\mck,
\end{align}
in $\{u+\ubar\geq C_R\}\cap\{u<u_f\}$, where the implicit constants in the bounds above depend only on $a,M,N$ provided $C_R(a,M)$ is large enough.
\end{prop}
\begin{proof}
	As $(e_4')_\mck$ and $(e_3')_\mck$ are null, there exists $\lambda_\mck,\wt{\lambda}_\mck$ and $f_\mck$, $\underline{\wh{g}}_\mck$ such that 
	\begin{align*}
		(e_4')_\mck=\lambda _\mck\left((\ering_4)_\mck+f_\mck^b(\ering_b)_\mck+\frac{1}{4}|f_\mck|^2(\ering_3)_\mck\right),\quad(e_3')_\mck=\wt{\lambda} _\mck\left((\ering_3)_\mck+\underline{\wh{g}}^b_\mck(\ering_b)_\mck+\frac{1}{4}|\underline{\wh{g}}_\mck|^2(\ering_4)_\mck\right).
	\end{align*}
	\noindent\textbf{Step 1 : obtaining the bounds $\lambda_\mck\sim 1$, $|f_\mck|\lesssim 1$, $|\underline{\wh{g}}_\mck|\lesssim \Omega_\mck^2$.} We have, recalling the definition of $r'_*$ in Section \ref{section:doublenullgrandesection} and \eqref{eq:derparrstar},
	\begin{align}
		\lambda_\mck&=\Omega_\mck^2(e_4')_\mck(\ubar)=\Omega^2_\mck(e_4')_\mck(r)\frac{\partial r'_*}{\partial r}+\Omega^2_\mck(e_4')_\mck(t)\nn\\
		&={\Omega^2_\mck}\frac{\sqrt{(r^2+a^2)^2-\Delta a^2\sin^2\theta_*}+r^2+a^2}{-\Delta}\sim 1.\label{eq:lambdamckkerr}
	\end{align}
	This implies
\begin{align}
	|f_\mck|^2=4{\lambda_\mck^{-1}} (e_4')_\mck(u)&=4{\lambda_\mck^{-1}}\left((e_4')_\mck(r'_*)-(e_4')_\mck(t)\right)\nn\\
	&=4{\lambda_\mck^{-1}}\left(\frac{r^2+a^2-\sqrt{(r^2+a^2)^2-a^2\sin^2\theta_*\Delta}}{\Delta}\right)\lesssim 1.\label{eq:pourfcadonnecagge}
\end{align}
Next, we have
	\begin{align}
		\wt{\lambda}_\mck&=(e_3')_\mck(u)=(e_3')_\mck(r)\frac{\partial r_*}{\partial r}-(e_3')_\mck(t)\nn\\
		&=\frac{1}{|q|^2}\left(\sqrt{(r^2+a^2)^2-\Delta a^2\sin^2\theta_*}+r^2+a^2\right)\sim 1.\label{eq:alatoutefinnn}
	\end{align}
	We deduce 
	\begin{align}
		|\underline{\wh{g}}_\mck|^2&=4\wt{\lambda}^{-1}_\mck\Omega^2_\mck(e_3')_\mck(\ubar)=4\wt{\lambda}^{-1}_\mck\Omega^2_\mck(e_3')_\mck(r)\frac{\partial r_*}{\partial r}+4\wt{\lambda}^{-1}_\mck\Omega^2_\mck(e_3')_\mck(t)\nn\\
		&=\frac{4\wt{\lambda}^{-1}_\mck\Omega^2_\mck}{|q|^2}\left(\sqrt{(r^2+a^2)^2-\Delta a^2\sin^2\theta_*}-(r^2+a^2)\right)\nn\\
		&=\frac{4\wt{\lambda}^{-1}_\mck\Omega^2_\mck(r^2+a^2)}{|q|^2}\left(\sqrt{1-\frac{\Delta a^2\sin^2\theta_*}{(r^2+a^2)^2}}-1\right)=-\frac{2\wt{\lambda}^{-1}_\mck\Omega^2_\mck\Delta a^2\sin^2\theta_*}{|q|^2(r^2+a^2)}+O(\Delta^2\Omega^2_\mck)\lesssim\Omega^4_\mck.\label{eq:caonsaitdejafacile}
	\end{align}
Moreover, see  Proposition \ref{prop:annexeptddd} in appendix, we have the bounds
\begin{align}\label{eq:relyingonehoui}
		|\df_\mck^{\leq N}\lambda_\mck|\lesssim 1,\quad |\df_\mck^{\leq N}\wt{\lambda}_\mck|\lesssim 1,\quad|\df_\mck^{\leq N}f_\mck|\lesssim 1,\quad |\df_\mck^{\leq N}\widehat{\gbar}_\mck|\lesssim\Omega^2_\mck.
\end{align}
	\noindent\textbf{Step 2 : proving existence and bound of $\fbar_\mck$.} We now look for $\fbar_\mck$ such that 
	\begin{align}\label{eq:eqlienfbarfbartilde}
		\underline{\wh{g}}_\mck=\frac{\fbar_\mck+\frac{1}{4}|\fbar_\mck|^2f_\mck}{1+\frac{1}{2}f_\mck\cdot\fbar_\mck+\frac{1}{16}|\fbar_\mck|^2|f_\mck|^2}.
	\end{align}
	We have that \eqref{eq:eqlienfbarfbartilde} holds if and only if $\fbar_\mck$ is a fixed point of the map 
	$$\Phi(\fbar_\mck)=\left(1+\frac{1}{2}f_\mck\cdot\fbar_\mck+\frac{1}{16}|\fbar_\mck|^2|f_\mck|^2\right)\underline{\wh{g}}_\mck-\frac{1}{4}|\fbar_\mck|^2f_\mck.$$
	The existence and bounds for $\fbar$ then follow from standard fixed-point arguments choosing $C_R(a,M)$ large enough and relying on the bounds \eqref{eq:relyingonehoui} for $\widehat{\gbar}_\mck$ and $f_\mck$, we omit the details.

	\noindent\textbf{Step 3 : conclusion.} By $\g_\mck((e_3')_\mck,(e_4')_\mck)=-2$ and \eqref{eq:eqlienfbarfbartilde} we obtain 
	$$\wt{\lambda}_\mck=\lambda^{-1}_\mck\left(1+\frac{1}{2}f_\mck\cdot\fbar_\mck+\frac{1}{16}|f_\mck|^2|\fbar_\mck|^2\right),$$
	thus
	$$(e_3')_\mck=\lambda^{-1}_\mck\left(\left(1+\frac{1}{2}f_\mck\cdot\fbar_\mck+\frac{1}{16}|f_\mck|^2|\fbar_\mck|^2\right)(\ering_3)_\mck+\left(\fbar^b_\mck+\frac{1}{4}|\fbar_\mck|^2f^b_\mck\right)(\ering_b)_\mck+\frac{1}{4}|\fbar_\mck|^2(\ering_4)_\mck\right),$$
which ensures that we have a transformation of the type \eqref{eq:frametransfo}, hence the proof.
\end{proof}

\begin{rem}
	We can compute more precisely the components of the $S(u,\ubar)$-tangent 1-forms $f_\mck,\widehat{\gbar}_\mck$. For $f_\mck$ we choose the following coordinates on $S(u,\ubar)$,
	\begin{align}\label{eq:meilleurecoordsphDN}
		\Big(\theta_*,\quad\phi_{*,\ch}':=\phi_*-\frac{4Ma r_-}{(r_-^2+a^2)^2}r'_*\:\:(\mathrm{mod}2\pi)\Big),
	\end{align}
	which is a regular system of coordinates on $S(u,\ubar)$ up to $\ch$\footnote{Note that recalling \eqref{eq:DNthetaCH} we have $\phi_{*,\ch}'=\phi_{*,\ch}-\frac{2Mar_-}{(r_-^2+a^2)^2}u+O(\Delta)$.} (except on the axis $\{\sin\theta_*=0\}$).
	Using $(e_4')_\mck(\theta^A)={\lambda}_\mck f_\mck^A$ and $(e_3')_\mck(\theta^A)=\wt{\lambda}_\mck\widehat{\gbar}_\mck^A$, and \eqref{eq:principaloutgoinkerr}, \eqref{eq:derparrstar}, \eqref{eq:derparthetasta} we get
	\begin{align}
		f_\mck^{\theta_*}=-\lambda_\mck^{-1}\frac{\partial\theta_*}{\partial r}=-\frac{\lambda_\mck^{-1}}{G\sqrt{(r^2+a^2)^2-a^2\sin^2\theta_*\Delta}},\label{eq:fthetamck}
	\end{align}
	and, recalling $\ubar=2r'_*-u$,
	\begin{align}
		\lambda_\mck f_\mck^{\phi_{*,\ch}'}&=e_4'(\phi)_\mck-e_4'(h(r'_*,\theta_*))_\mck-\frac{4Ma r_-}{(r_-^2+a^2)^2}e_4'\left(r'_*\right)_\mck\nn\\
		&=-\frac{a}{\Delta}+\frac{\partial r'_*}{\partial r}\frac{\partial h}{\partial  r'_*}+\frac{\partial \theta_*}{\partial r}\frac{\partial h}{\partial  \theta_*}+\frac{4Ma r_-}{(r_-^2+a^2)^2}\frac{\partial r'_*}{\partial r}\nn\\
		&=\frac{r^2+a^2}{\Delta}\left(\left(\frac{4Ma r_-}{(r_-^2+a^2)^2}-\frac{2aMr}{|q|^2 R^2}\right){\sqrt{1-\frac{a^2\sin^2\theta_*\Delta}{(r^2+a^2)^2}}}-\frac{a}{r^2+a^2}\right)+\frac{\partial \theta_*}{\partial r}\frac{\partial h}{\partial  \theta_*}.\label{eq:fphich}
	\end{align}
	Where, recalling \eqref{eq:identitybelow}, we note the following bound
	$$\left(\frac{4Ma r_-}{(r_-^2+a^2)^2}-\frac{2aMr}{|q|^2 R^2}\right){\sqrt{1-\frac{a^2\sin^2\theta_*\Delta}{(r^2+a^2)^2}}}-\frac{a}{r^2+a^2}=O(\Delta),$$
	which also holds for angular derivatives of the LHS above. Next for $\widehat{\gbar}_\mck$ we use $(\theta_*,\phi_*)$ coordinates, and we get
	\begin{align}
		\widehat{\gbar}_\mck^{\theta_*}=\wt{\lambda}_\mck^{-1}e_3'(\theta_*)_\mck=\frac{\Delta}{|q|^2 G\sqrt{(r^2+a^2)^2-a^2\sin^2\theta_*\Delta}},\label{eq:gbarhattheta}
	\end{align}
and
\begin{align}
	\wt{\lambda}_\mck(\widehat{\gbar}_\mck)^{\phi_{*}}&=e_3'(\phi)_\mck-e_3'(h(r'_*,\theta_*))_\mck=-\frac{a}{|q|}^2+\frac{2Mar}{|q|^2 R^2}\frac{\sqrt{(r^2+a^2)^2-a^2\sin^2\theta_*\Delta}}{|q|^2}-e_3'(\theta_*)_\mck\frac{\partial h}{\partial  \theta_*}\nn\\
	&=\frac{a}{|q|^2}\left(\frac{2Mr\sqrt{(r^2+a^2)^2-a^2\sin^2\theta_*\Delta}}{(r^2+a^2)^2-a^2\sin^2\theta\Delta}-1\right)-\frac{\Delta}{|q|^2}\frac{\partial\theta_*}{\partial r}\frac{\partial h}{\partial  \theta_*},\label{eq:gbarhatphi}
\end{align}
where we note the bound 
$$\frac{2Mr\sqrt{(r^2+a^2)^2-a^2\sin^2\theta_*\Delta}}{(r^2+a^2)^2-a^2\sin^2\theta\Delta}-1=O(\Delta),$$
which also holds for angular derivatives of the LHS above.
\end{rem}
\begin{rem}\label{rem:valeurkerrgbarlambdabar}
	The proof of Proposition \ref{prop:transfodanskerr} also shows that there exists a scalar $\wt{\lambda}_\mck\sim 1$ and a $S(u,\ubar)$-tangent 1-form $\underline{\wh{g}}_\mck$ satisfying $|\underline{\wh{g}}_\mck|\lesssim\Omega^2_\mck$ such that 
	$$(e_3')_\mck=\wt{\lambda} _\mck\left((\ering_3)_\mck+\underline{\wh{g}}^b_\mck(\ering_b)_\mck+\frac{1}{4}|\underline{\wh{g}}_\mck|^2(\ering_4)_\mck\right).$$
	Thus, recalling from \eqref{eq:principalingoinkerr} the renormalized field $(e_3)_\mck=-|q_\mck|^2\Delta_\mck^{-1}(e_3')_\mck$, we have in Kerr,
	$$(e_3)_\mck=\lambdabar_\mck\left(\Omega^{-2}_\mck (\ering_3)_\mck+\gbar_\mck^a(\ering_a)_\mck+\frac14|\gbar|_\mck^2\Omega_\mck^2 (\ering_4)_\mck\right),$$
	where
	\begin{align}\label{eq:enfaitsijutilise}
		\lambdabar_\mck=\frac{|q_\mck|^2\wt{\lambda}_\mck\Omega^2_\mck}{-\Delta_\mck}\sim 1,\quad \gbar_\mck=\Omega^{-2}_\mck\underline{\wh{g}}_\mck+O(1).
	\end{align}
\end{rem}

\subsubsection{Link between PT and double null frame on $\Sigma_0$}

The following result will be used to initialize the non-integrable frames of region $\deux$ on $\Sigma_0$. 
\begin{prop}\label{prop:initf0gbar0ettout}
	We have, on $\Sigma_0$,
	\begin{align}\label{eq:asinyo}
		e_3|_{\Sigma_0}={\lambdabar_0}\left(\Omega^{-2}\mathring{e}_3+\gbar_0^A\partial_{\theta^A}+\frac{1}{4}|\gbar_0|^2\Omega^2\mathring{e}_4\right),\quad -\frac{|q|^2}{\Delta}e_4|_{\Sigma_0}=\lambda_0\left(\ering_4+f_0^A\partial_{\theta^A}+\frac{1}{4}|f_0|^2\ering_3\right),
	\end{align}
	where $(e_3,e_4)$ is the ingoing PT null pair in $\un$ constructed in Section \ref{section:regionun}, $\lambda_0, {\lambdabar_0}$ are scalar functions and $f_0,\gbar_0$ are $S(u,\ubar)$-tangent 1-forms. Moreover, denoting $\overline{\nabla}=\{\nabring_\T,\nabring_A\}$ the derivatives tangential to $\Sigma_0$ and recalling $N_1=N_0-3$ (see Remark \ref{rem:defN111}), we have the estimates
	\begin{align}\label{eq:estimsigma0fclambdac}
		|\overline{\nabla}^{\leq N_1}(\widecheck{\lambda_0},\widecheck{{\lambdabar}}_0,\widecheck{f_0},\widecheck{\gbar_0})|\lesssim\varepsilon\ubar^{-3-\delta/2}, 
	\end{align}
	where the implicit constant above depends on $a,M,C_R$ and where $(\widecheck{\lambda_0},\widecheck{{\lambdabar}}_0,\widecheck{f_0},\widecheck{\gbar_0})$ are defined by
	\begin{align*}
		\widecheck{\lambda_0}=\lambda_0-\lambda_\mck|_{\Sigma_0},\quad \widecheck{\lambdabar_0}=\lambdabar_0-\lambdabar_\mck|_{\Sigma_0},\quad\widecheck{f_0}^A=f_0^A-f_\mck^A|_{\Sigma_0},\quad \widecheck{\gbar_0}^A=\gbar_0^A-\gbar_\mck^A|_{\Sigma_0},
	\end{align*}
with $\lambda_\mck,\lambdabar_\mck,f_\mck,\gbar_\mck$ defined in Proposition \ref{prop:transfodanskerr} and in Remark \ref{rem:valeurkerrgbarlambdabar}.
\end{prop}
\begin{proof} In this proof, the implicit constants in the bounds $\lesssim$ may depend on $a,M,C_R$. As $e_3$ is null and $(\Omega^{-2}\ering_3,\Omega^2\ering_4,\ering_a)$ is a null frame, we can always express $e_3$ as in \eqref{eq:asinyo} on $\Sigma_0$ for some $\lambdabar_0,\gbar_0$. We now compute 
\begin{align}\label{eq:secondidee}
	\lambdabar_0=\Omega^2e_3(u)|_{\Sigma_0},\quad |\gbar_0|^2=4\lambdabar_0^{-1}e_3(\ubar)|_{\Sigma_0},\quad \gbar_0^A=\lambdabar_0^{-1}e_3(\theta^A)|_{\Sigma_0}-\frac14|\gbar_0|^2b^A|_{\Sigma_0}.
\end{align}
We decompose $e_3|_{\Sigma_0}$ into a tangent and orthogonal part with respect to $\Sigma_0$\footnote{Note that here to estimate $e_3(u),e_3(\ubar),e_3(\theta^A)$ on $\Sigma_0$, we cannot invoke Proposition \ref{prop:diffderiveescoordsDN} since in that result we consider the coordinates $x'^\nu$ which are the Kerr values of the Israel-Pretorius coordinates $u,\ubar,\theta^A$ with respect to the PT coordinates in region $\un$, and hence do not coincide with the double null coordinates $u,\ubar,\theta^A$ constructed dynamically in the future of $\Sigma_0$ in Section \ref{section:IIdoublenull}.}:
$$e_3|_{\Sigma_0}=(e_3+\g(e_3,N)N)|_{\Sigma_0}-\g(e_3,N)N|_{\Sigma_0},$$ 
where $N$ is the unit timelike normal to $\Sigma_0$, see Definition \ref{def:normalSig0}. Thus, for any double null coordinate function $x\in\{u,\ubar,\theta^A\}$ we compute 
$$e_3(x)|_{\Sigma_0}=(e_3+\g(e_3,N)N)(x)|_{\Sigma_0}-\g(e_3,N)N(x)|_{\Sigma_0},$$ 
where:
\begin{itemize}
	\item Since $(e_3+\g(e_3,N)N)|_{\Sigma_0}\in T\Sigma_0$, the quantity $(e_3+\g(e_3,N)N)(x)|_{\Sigma_0}$ coincides with $(e_3+\g(e_3,N)N)(x')|_{\Sigma_0}$ where $x'$ is the Kerr value of the Isreal-Pretorius coordinate $x$ with respect to the PT coordinates $r,\ubar^{PT},\theta,\phi_+$ of region $\un$, because $x$ is initialized as $x'$ on $\Sigma_0$. Moreover, we note that we have precise estimates for $(e_3+\g(e_3,N)N)(x')|_{\Sigma_0}$ combining Proposition \ref{prop:diffderiveescoordsDN} and \eqref{eq:expreNnormalsig0}.
	\item We can express the unit timelike normal $N$ with respect to the double null pair $\ering_3,\ering_4$ on $\Sigma_0$ as follows (see for instance from \cite[Eq. (4.8)]{stabC0} and \eqref{eq:DuDubardoublenull}):
	$$N|_{\Sigma_0}=\frac{\sqrt{\Phi-|w|^2_\gamma}}{4}(\Omega^{-2}\ering_3+\ering_4)|_{\Sigma_0},$$
	where we recall the expressions \eqref{eq:followingformonsig0}, \eqref{eq:ghatdanskerr}.
	\item By \eqref{eq:expreNnormalsig0}, we have on $\Sigma_0$,
	$$\g(e_3,N)|_{\Sigma_0}=-\frac{\sqrt{(r^2+a^2)^2-a^2\sin^2\theta_*\Delta}}{\Delta}+\err,$$
	where $|\df^{\leq N_1}\err|\lesssim\varepsilon\ubar^{-3-\delta/2}$.
\end{itemize}
Combining these observations, and noticing that on $\Sigma_0$ the linearization procedures with respect to the PT coordinates $r,\ubar^{PT},\theta,\phi_+$ and with respect to the double null coordinates of region $\un$ coincide (because $u,\ubar,\theta^A$ is initialized on $\Sigma_0$ as their Kerr values with respect to the PT coordinates) so that we can can use the checked notation to denote unambiguously the linearized quantities, we get for $x\in\{u,\ubar,\theta^A\}$,
\begin{align}\label{eq:gplusideedenomstppp}
	|\overline{\nabla}^{\leq N_1}\widecheck{e_3(x)}|_{\Sigma_0}|\lesssim\varepsilon\ubar^{-3-\delta/2},
\end{align}
where we also used the bounds in Proposition \ref{prop:bornegchecksig0} to control $\widecheck{\Phi}, \widecheck{w}, \widecheck{\log\Omega},\bcheck$ on $\Sigma_0$. Also note that for $x=\theta^A$, more precisely the bound \eqref{eq:gplusideedenomstppp} holds for $x=\theta^A_{(i)}$ in $\mcv_i$, $i=1,2,3$, recalling Definition \ref{defi:defthetadansmcvi}. Combining this with the bound 
$$|\overline{\nabla}^{\leq N_1}\widecheck{\Omega^2}|_{\Sigma_0}|\lesssim |\hat{\nabla}_\mck^{\leq N_1}(\hat{g}-\hat{g}_\mck)|\lesssim\varepsilon\ubar^{-3-\delta/2},$$
(which also holds by Proposition \ref{prop:bornegchecksig0}) we get the stated bound for $\widecheck{\lambdabar_0}$. Also, using this result for $\widecheck{\lambdabar_0}$, from the second identity in \eqref{eq:secondidee}, and \eqref{eq:gplusideedenomstppp} with $x=\ubar$, we get $|\overline{\nabla}^{\leq k}\widecheck{|\gbar_0|^2}|\lesssim\varepsilon\ubar^{-3-\delta}$. Combining this with the last identity in \eqref{eq:secondidee} and \eqref{eq:gplusideedenomstppp} once again for $x=\theta^A_{(i)}$ in $\mcv_i$, and the bound $$|\overline{\nabla}^{\leq N_1}\widecheck{b}|_{\Sigma_0}|\lesssim |\hat{\nabla}_\mck^{\leq N_1}(\hat{g}-\hat{g}_\mck)|\lesssim\varepsilon\ubar^{-3-\delta/2},$$
which holds by Proposition \ref{prop:bornegchecksig0}, we deduce finally $|\overline{\nabla}^{\leq N_1}\widecheck{\gbar_0}|\lesssim \varepsilon\ubar^{-3-\delta/2}$. We proceed similarly with $e_4$ to prove the stated bounds on $\widecheck{\lambda_0},\widecheck{f_0}$.
\end{proof}

\subsubsection{Non-linear transport equations for $f,\underline{g},\lambda,\lambdabar$}
\begin{rem}\label{rem:notationemuun}From now on, in the rest of the paper we denote by
$$(e_3^{(\un)},e_4^{(\un)},e_a^{(\un)})$$
the ingoing PT frame of region $\un$, to avoid conflicts of notations with the ingoing non-integrable frame $(e_3,e_4,e_a)$ in region $\deux$, while the outgoing non-integrable frame will be denoted by $(e_3',e_4',e_a')$.
\end{rem}
\begin{rem}\label{rem:depuisla}
	From here and up to Remark \ref{rem:jusquala}, the implicit constants in the bounds $\lesssim$ may depend on $a,M$ and $C_R$. 
\end{rem}
We will first construct the outgoing null vector ${e_4'}$ in $\deux$ by 
\begin{align}\label{eq:e4primdef}
	{e_4'}=\lambda\left(\ering_4+f^A\partial_{\theta^A}+\frac{1}{4}|f|^2\ering_3\right),
\end{align}
where we require that ${e_4'}$ is geodesic. This is equivalent to ${\xi'}=0$, ${\omega'}=0$ in the new frame, which rewrites as non-linear transport equations for $\lambda$, $f$. 
\begin{prop}\label{prop:eqtranspofetlambda}
	Let $f$ be a $S(u,\ubar)$-tangent 1-form, and $\lambda$ a scalar function in $\deux$. Defining
	$$e_4'=\lambda\left(\ering_4+f^A\partial_{\theta^A}+\frac{1}{4}|f|^2\ering_3\right),$$
	then the vector field $e_4'$ is geodesic, namely $\D_{4'}e_4'=0$ in $\deux$, if and only if $f$ and $\lambda$ satisfy the following (non-linear) transport equations,
	\begin{align}\label{eq:transpof}
		&\left(\nabring_4+f\cdot\nabring+\frac{1}{4}|f|^2\nabring_3\right)[f]+\frac{1}{4}\mathring{tr\chi} f+\frac{1}{2}f\cdot\mathring{\chihat}+\frac{1}{4}|f|^2(\mathring{\eta}-\mathring{\etabar})\nn\\
		&\quad+\frac{1}{2}(f\cdot\mathring{\zeta}) f-\frac{1}{4}\mathring{\omegabar}|f|^2f-\frac{1}{8}|f|^2 f\cdot\mathring{\wh{\chibar}}-\frac{1}{8}|f|^2 \mathring{tr\chibar} f=0,
	\end{align}
	and
	\begin{align}\label{eq:lambdaeqprelim}
		-\frac{1}{2}\left(\ering_4+f^A\partial_{\theta^A}+\frac{1}{4}|f|^2\ering_3\right)[\log\lambda]+\frac{1}{2}f\cdot(\mathring{\zeta}-\mathring{\etabar})-\frac{1}{4}|f|^2\mathring{\omegabar}-\frac{1}{4}f\cdot(f\cdot\mathring{\chibar})=0.
	\end{align}
\end{prop}
\begin{proof}
	This a direct consequence of the change of frame formulas \eqref{eq:changexi}, \eqref{eq:changeomega} with $\fbar=0$, and the identity \eqref{eq:trucsnulendoublenul} in the double null gauge which implies $\mathring{\xi}=0,\:\mathring{\omega}=0,\: \mathring{\atrchi}=0$.
\end{proof}
Next, we will construct the vector field $e_3$ in the \emph{ingoing} normalization. We will define 
\begin{align}\label{eq:e3daggerdef}
	e_3=\underline{\lambda}\left(\Omega^{-2}\ering_3+\gbar^A\partial_{\theta^A}+\frac{1}{4}|\gbar|^2\Omega^2\ering_4\right),
\end{align}
such that $e_3$ is geodesic, which rewrites as non-linear transport equations for $\underline{\lambda}$, $\gbar$. 

\begin{defi}\label{def:doublenullingoinggg}
We define the \emph{ingoing} double null pair as
$$(\wh{e}_3=\Omega^{-2}\ering_3,\wh{e}_4=\Omega^2\ering_4),$$
and we denote with a hat the Ricci and curvature coefficients $(\wh{tr\chibar},\wh{\wh{\chibar}},\wh{\zeta},\cdots)$ defined with respect to the ingoing double null frame. Similarly, as in the outgoing doube null frame $(\ering_3,\ering_4)$, we define the schematic notations
$$\wh{\psi}_H=\{\wh{tr\chi},\wh{\chihat}\}=\Omega^2\psi_H,\quad\wh{\psi}_{\Hbar}=\{\wh{tr\chibar},\wh{\wh{\chibar}}\}=\Omega^{-2}\psi_{\Hbar},\quad\wh{\psi}=\{\wh{\eta},\wh{\etabar}\}=\psi,$$
as well as the checked versions which denote the corresponding linearized quantities. Note that the metric components remain unchanged, $\wh{g}=g$. Also, in this gauge we have the identities
$$\wh{\omegabar}=\Omega^{-2}\left(\mathring\omegabar+\frac{1}{2}e_3(\log\Omega^2)\right)=0,\quad \wh{\zeta}=-\wh{\etabar}=-\mathring\etabar,\quad \wh{\omega}=-\wh{\nabla}_4(\log\Omega).$$
\end{defi}
\begin{prop}\label{prop:eqgbarlambdabar}
	Let $\gbar$ be a $S(u,\ubar)$-tangent 1-form, and $\lambdabar$ a scalar function in $\deux$. Defining
	$$e_3=\lambdabar\left(\wh{e}_3+\gbar^A\partial_{\theta^A}+\frac{1}{4}|\gbar|^2\wh{e}_4\right),$$
	then the vector field $e_3$ is geodesic, namely $\D_{e_3}e_3=0$ in $\deux$, if and only if $\gbar$ and $\lambdabar$ satisfy
	\begin{align}
		&\left(\wh{\nabla}_3+\gbar\cdot\wh{\nabla}+\frac{1}{4}|\gbar|^2\wh{\nabla}_4\right)\left[\gbar\right]+\frac{1}{4}\wh{tr\chibar} \gbar+\frac{1}{2}\gbar\cdot\wh{\wh{\chibar}}+\frac{1}{4}|\gbar|^2(\wh{\etabar}-\wh{\eta})\nn\\
		&-\frac{1}{2}(\gbar\cdot\wh{\zeta}) \gbar-\frac{1}{4}\wh{\omega}|\gbar|^2\gbar-\frac{1}{8}|\gbar|^2 \gbar\cdot\wh{\wh{\chi}}-\frac{1}{8}|\gbar|^2 \wh{tr\chi} \gbar=0,\label{eq:transpogbar}
	\end{align}
	where $\wh{\nabla}_\mu=\nabring_{\wh{e}_\mu}$ for $\mu=1,2,3,4$ and
	\begin{align}
		-\frac{1}{2}\left(\wh{e}_3+\gbar^A\partial_{\theta^A}+\frac{1}{4}|\gbar|^2\wh{e}_4\right)[\log\lambdabar]-\frac{1}{2}\gbar\cdot(\wh{\zeta}+\wh{\eta})-\frac{1}{4}|\gbar|^2\wh{\omega}-\frac{1}{4}\gbar\cdot(\gbar\cdot\wh{\chihat})-\frac18|\gbar|^2\wh{tr\chi}=0.\label{eq:transpolambdabar}
	\end{align}
\end{prop}
\begin{proof}
	The proof is the same as the one of Proposition \ref{prop:eqtranspofetlambda}, exchanging $e_4$ by $\wh{e}_3$ and $e_3$ by $\wh{e}_4$. Note that this time we use $\wh{\xibar}=\Omega^{-4}\xibar=0,\: \wh{\omegabar}=0,\: \wh{{}^{(a)}tr\chibar}=\Omega^{-2}\atrchibar=0$, by the identities \eqref{eq:trucsnulendoublenul}, \eqref{eq:omegabarendoublenul}.
\end{proof}

\subsection{Existence and control of $f$ and $\lambda$ in $\deux$}\label{section:controlef}
\subsubsection{Transport estimates in the perturbed outgoing direction}\label{section:transpoestimoutgoing}
In this section, we consider $\ubar_f> C_R-u_f$ and a $S(u,\ubar)$ tangent 1-form $f$ in $$\deux[\ubar_f]:=\deux\cap\{\ubar<\ubar_f\},$$ which satisfies
\begin{align}\label{eq:fbronehypdeux}
	|f|\lesssim 1,\:\text{in}\:\:\deux[\ubar_f],
\end{align}
and we define the vector field 
\begin{align}\label{eq:Ysadefquoi}
	Y:=\Omega^2\left(\ering_4+f^A\partial_{\theta^A}+\frac{1}{4}|f|^2\ering_3\right)=\partial_\ubar + (b^A+\Omega^2 f^A)\partial_{\theta^A} +\frac{1}{4}\Omega^2|f|^2\partial_u.
\end{align}
We denote $\varphi_Y(s,u,\ubar,\theta^A)$ the flow of $Y$ at time $s$ starting at $(u,\ubar,\theta^A)$ at $s=\ubar$, namely $\varphi_Y(s,u,\ubar,\theta^A)$ satisfies the following ODE in $\deux[\ubar_f]$,
\begin{align*}
	\frac{\partial}{\partial s}\varphi_Y(s,u,\ubar,\theta^A)=Y(\varphi_Y(s,u,\ubar,\theta^A)),\quad\varphi_Y(s=\ubar,u,\ubar,\theta^A)=(u,\ubar,\theta^A).
\end{align*}
For $p\in\deux[\ubar_f]$, we also introduce the coordinate functions evaluated along the flow of $Y$,
\begin{align}
	u_{Y}^p(s):=u(\varphi_Y(s,p)),\quad \ubar_{Y}^p(s):=\ubar(\varphi_Y(s,p)),\quad (\theta^A)_{Y}^p(s):=\theta^A(\varphi_Y(s,p)).
\end{align}
\begin{prop}\label{prop:flowofY}
	For any $p\in\deux[\ubar_f]$ and $s\in\mathbb{R}$ such that $\varphi_Y(s,p)$ is well-defined, we have
	\begin{align}\label{eq:equivtransporte4'}
		\ubar_Y^p(s)=s,\quad |u_Y^p(s)-u(p)|\lesssim 1.
	\end{align}
\end{prop}
\begin{proof}
	By definition of the flow $\varphi_Y(s,p)$, we have 
	\begin{align*}
		Y(\varphi_Y(s,p))=\frac{\partial}{\partial s}\varphi_Y(s,u,\ubar,\theta^A)=\frac{\dee}{\dee s}u_{Y}^p(s)\partial_u+\frac{\dee}{\dee s}\ubar_{Y}^p(s)\partial_\ubar+\frac{\dee}{\dee s}(\theta^A)_{Y}^p(s)\partial_{\theta^A}.
	\end{align*}
	Thus, using \eqref{eq:Ysadefquoi}, we get the identities
	\begin{align*}
		\frac{\dee}{\dee s}u_{Y}^p(s)=\frac{1}{4}[\Omega^2|f|^2 ](u_{Y}^p(s),\ubar_{Y}^p(s),(\theta^A)_{Y}^p(s)),\quad\frac{\dee}{\dee s}\ubar_{Y}^p(s)=1.
	\end{align*}
Using the initial condition $\ubar_Y^p(\ubar(p))=\ubar(p)$, we deduce $\ubar_Y^p(s)=s$, which concludes the first part of the proposition, and which implies
	\begin{align}\label{eq:deruY}
		\frac{\dee}{\dee s}u_{Y}^p(s)&=\frac{1}{4}[\Omega^2|f|^2 ](u_{Y}^p(s),s,(\theta^A)_{Y}^p(s)).
	\end{align}
	This yields, using the initial condition $u_Y^p(\ubar(p))=u(p)$, 
	\begin{align}\label{eq:faitepourconclencore}
		|u_Y^p(s)-u(p)|\lesssim\left|\int_{\ubar(p)}^se^{-|\kappa_-|(u_Y^p(s')+s')}\dee s'\right|,
	\end{align}
	where we used the bound $\Omega^2\lesssim e^{-|\kappa_-|(u+\ubar)}$. Moreover, integrating by parts we compute
	\begin{align}
		\int_{\ubar(p)}^se^{-|\kappa_-|(u_Y^p(s')+s')}\dee s'=&\frac{1}{|\kappa_-|}\left(e^{-|\kappa_-|(u(p)+\ubar(p))}-e^{-|\kappa_-|(u_Y^p(s)+s)}\right)\nn\\
		&-\int_{\ubar(p)}^s\left[\frac{\dee}{\dee s}u_{Y}^p\right](s')e^{-|\kappa_-|(u_Y^p(s')+s')}\dee s'.\label{eq:panecess}
	\end{align}
	Thus, we can separate the following two cases : 
	\begin{itemize}
		\item If $s\leq\ubar(p)$, 
		\begin{align*}
			\left|\int_{\ubar(p)}^se^{-|\kappa_-|(u_Y^p(s')+s')}\dee s'\right|=&\int_s^{\ubar(p)}e^{-|\kappa_-|(u_Y^p(s')+s')}\dee s'\\
			=&\frac{1}{|\kappa_-|}\left(e^{-|\kappa_-|(u_Y^p(s)+s)}-e^{-|\kappa_-|(u(p)+\ubar(p))}\right)\\
			&-\int_s^{\ubar(p)}\left[\frac{\dee}{\dee s}u_{Y}^p\right](s')e^{-|\kappa_-|(u_Y^p(s')+s')}\dee s'\lesssim e^{-|\kappa_-|(u_Y^p(s)+s)},
		\end{align*}
		where we used \eqref{eq:panecess} and \eqref{eq:deruY} which implies $\frac{\dee}{\dee s}u_{Y}^p\geq 0$.
		\item If $\ubar(p)\leq s$, 
		\begin{align*}
			\left|\int_{\ubar(p)}^se^{-|\kappa_-|(u_Y^p(s')+s')}\dee s'\right|=&\int_{\ubar(p)}^se^{-|\kappa_-|(u_Y^p(s')+s')}\dee s'\lesssim e^{-|\kappa_-|(u(p)+\ubar(p))},
		\end{align*}
		where we used \eqref{eq:panecess} and $\frac{\dee}{\dee s}u_{Y}^p\geq 0$ again.
	\end{itemize}
	In any case, using $e^{-|\kappa_-|(u+\ubar)}\lesssim 1$ in $\deux$ we conclude the proof of the proposition by \eqref{eq:faitepourconclencore}.
\end{proof}
\begin{rem}
By proposition \ref{prop:flowofY} and the definition of $\Sigma_0=\{u+\ubar=C_R\}$, we deduce that for any $p\in\deux[\ubar_f]$, for some $\ubar'\leq\ubar(p)$ sufficiently close to $C_R-u_f$, 
$$\varphi_Y(\ubar',p)\in\Sigma_0.$$
Indeed, from the identity $\ubar_Y^p(s)=s$ and \eqref{eq:deruY} we get that the flow starting from $p$ directed towards the past remains in $\{u<u_f\}$, hence in $\deux[\ubar_f]$, as long as it does not reach $\Sigma_0$ (which it does in finite time).
\end{rem}
\begin{defi}
The special value of $\ubar'$, which is such that the flow $\varphi_Y(s,p)$ starting from $p$ reaches $\Sigma_0$ at time $s=\ubar'$, will be denoted by
$$\ubar_{Y,\Sigma_0}(p):=\sup\big\{\ubar'\leq\ubar(p)\:\big|\:\varphi_Y(\ubar',p)\in\Sigma_0\big\}.$$
We also introduce the following simplified notations,
$$u_{Y,\Sigma_0}(p):=u_Y^p(\ubar_{Y,\Sigma_0}(p)),\quad \theta^A_{Y,\Sigma_0}(p):=(\theta^A)_Y^p(\ubar_{Y,\Sigma_0}(p)).$$
\end{defi}
Note that we have for any $p=(u,\ubar,\theta^A)\in\deux[\ubar_f]$,
\begin{align}\label{eq:equivinitubaru}
	\ubar_{Y,\Sigma_0}(p)\sim |C_R-\ubar_{Y,\Sigma_0}(p)|=|u_Y^p(\ubar_{Y,\Sigma_0}(p))|\sim |u(p)|\sim \ubar(p),
\end{align}
where we used \eqref{eq:equivtransporte4'} in the third step, and the estimate $|u|\sim\ubar$ (see Lemma \ref{lem:usimubarII}) in $\deux$ in the last step. Now we state a transport estimate along the flow of $Y$.
\begin{prop}\label{prop:transporte4'}
	Let $f,Y$ be as in \eqref{eq:fbronehypdeux} and \eqref{eq:Ysadefquoi}. Then, for any $p=(u,\ubar,\theta^A)\in\deux[\ubar_f]$, for any $S(u,\ubar)$-tangent tensor $U$, 
	\begin{align}\label{eq:inegtransporte4'}
		|U|(p)\lesssim |U|(u_{Y,\Sigma_0}(p),\ubar_{Y,\Sigma_0}(p),\theta^A_{Y,\Sigma_0}(p))+\int_{\ubar_{Y,\Sigma_0}(p)}^{\ubar(p)}|\nabring_Y U|(u_Y^p(\ubar'),\ubar',(\theta^A)_Y^p(\ubar'))\dee\ubar'.
	\end{align}
\end{prop}
\begin{proof}
This directly follows from the integration of the bound
	$$\left|\frac{\dee}{\dee\ubar'}|U|(\varphi_Y(\ubar',p))\right|\leq |\nabring_YU|(u_Y^p(\ubar'),\ubar',(\theta^A)^p_Y(\ubar')),$$
from $\ubar_{Y,\Sigma_0}(p)$ to $\ubar(p)$.
\end{proof}

\subsubsection{Existence and control of $\fc$ and $\lambdacheck$}\label{section:exsicontfc}

In this section, we prove existence and control of a $S(u,\ubar)$-tangent 1-form $f$ which satisfies the non-linear transport equation \eqref{eq:transpof} and appropriate estimates. We then prove existence and control of $\lambda$ which satisfies the linear transport equation \eqref{eq:lambdaeqprelim}, see Theorems \ref{thm:controlfc}, \ref{thm:controllambdacheck}. \textbf{In Section \ref{section:exsicontfc}, we drop the ring notation on top of the double null Ricci and curvature coefficients for conciseness in the notations. Thus in this subsection, $tr\chi,\chihat,\ldots$ denote $\mathring{tr\chi},\mathring{\chihat},\ldots$, and similarly for the analog checked and Kerr quantities. However we keep the ring on top of double null covariant derivatives $\nabring_{3,4,A}$.}

We begin with $\fc$. Let $f_\mck$ be the $f$ coefficient of the frame transformation between the outgoing double null and the principal null frames in Kerr, see Proposition \ref{prop:transfodanskerr}. By Proposition \ref{prop:eqtranspofetlambda}, as $(e_4')_\mck$ is geodesic in Kerr, $f_\mck$ satisfies the following equation,
\begin{align}
	(\nabring&_{V})_\mck f_\mck+\frac14{tr\chi}_\mck f_\mck+\frac{1}{2}f_\mck\cdot_\mck{\chihat}_\mck+\frac{1}{4}|f_\mck|^2_\mck({\eta}_\mck-{\etabar}_\mck)+\frac{1}{2}(f_\mck\cdot_\mck{\zeta}_\mck)f_\mck-\frac{1}{4}{\omegabar}_\mck|f_\mck|^2_\mck f_\mck\nn\\
	&-\frac{1}{8}|f_\mck|^2_\mck f_\mck\cdot_\mck{\wh{\chibar}}_\mck-\frac{1}{8}|f_\mck|^2_\mck {tr\chibar}_\mck f_\mck=0,\label{eq:fkerr}
\end{align}
where $V_\mck:=(\mathring{e}_4)_\mck+f_\mck^A \partial_{\theta^A}+\frac{1}{4}|{f}_\mck|^2_\mck(\mathring{e}_3)_\mck$. Let $f$ be a 1-form, and 
\begin{align}
	\fc:=f-f_\mck.
\end{align}
We define the null vector $Y$ as in \eqref{eq:Ysadefquoi}. Then, subtracting \eqref{eq:fkerr} from \eqref{eq:transpof}, we get that $f$ satisfies \eqref{eq:transpof} if and only if $\fc$ satisfies the following equation,
\begin{align}
	&\Omega^{-2}\nabring_{Y}\widecheck{f}+\frac{1}{4}tr\chi_\mck\widecheck{f}+\frac{1}{2}\chihat_\mck\cdot\fc+\fc\cdot(\nabring_\mck f_\mck)+\frac{1}{2}(\fc\cdot f_\mck)(\nabring_3)_\mck f_\mck+\frac{1}{2}(\fc\cdot f_\mck)(\eta_\mck-\etabar_\mck)\nn\\
	&+\frac{1}{2}(\fc\cdot\zeta_\mck)f_\mck+\frac{1}{2}f_\mck\cdot_\mck\zeta_\mck\fc-\frac{1}{4}\omegabar_\mck|f_\mck|^2_\mck\fc-\frac{1}{2}\omegabar(\fc\cdot f_\mck)f-\frac{1}{4}(\fc\cdot f_\mck)f_\mck\cdot_\mck{\chibar}_\mck-\frac{1}{8}|f|^2\fc\cdot{\chibar}_\mck\nn\\
	&=\frac{1}{4}\widecheck{tr\chi}f_\mck+\frac{1}{4}\widecheck{\chihat}\cdot f_\mck+\frac{1}{2}\chihat_\mck\widecheck{\cdot}f_\mck+\widecheck{\nabring}_4[f_\mck]+f_\mck\widecheck{\cdot}\:\nabring_\mck f_\mck+f\cdot\widecheck{\nabring}[f_\mck]+\frac{1}{4}f_\mck\widecheck{\cdot}f_\mck(\nabring_3)_\mck f_\mck\nn\\
	&\quad+\frac{1}{4}|f|^2\widecheck{\nabring}_3[f_\mck]+\frac{1}{4}f_\mck\widecheck{\cdot}f_\mck(\eta_\mck-\etabar_\mck)+\frac{1}{4}|f|^2(\widecheck{\eta}-\widecheck{\etabar})+\frac{1}{2}(f_\mck\widecheck{\cdot}\zeta_\mck+f\cdot\widecheck{\zeta})f_\mck\nn\\
	&\quad-\frac{1}{4}(\widecheck{\omegabar}|f_\mck|^2_\mck+\omegabar f_\mck\widecheck{\cdot}f_\mck)f-\frac{1}{8}f_\mck\widecheck{\cdot}f_\mck(f_\mck\cdot_\mck\chibar_\mck)-\frac{1}{8}|f|^2(f_\mck\widecheck{\cdot}\chibar_\mck+f\cdot\widecheck{\chibar})+\err[\fc],\label{eq:nlfc}
\end{align}
where $\err[\fc]:=\frac{1}{4}|\fc|^2(\nabring_3)_\mck f_\mck+\frac{1}{4}|\fc|^2(\eta_\mck-\etabar_\mck)-\frac{1}{4}\omegabar|\fc|^2f-\frac{1}{8}|\fc|^2(f_\mck\cdot_\mck\chibar_\mck)$, the symbol $\widecheck{\cdot}$ denotes some $\gcheck$ quantity, and where the operators $\widecheck{\nabring}_3$, $\widecheck{\nabring}_4$, $\widecheck{\nabring}$ are defined at the end of Proposition \ref{prop:linderdoublenull}.
\begin{defi}
	We define the following schematic quantities,
	\begin{align}
		\widecheck{\mcq}&=_s\left((\Omega^2)^{\leq 1}\df^{\leq 1}\gcheck,\psicheck_{\Hbar},\widecheck{\omegabar},\psicheck\right),\\
		\mcq_\mck&=_s\left((\psi_H)_\mck,(\psi_{\Hbar})_\mck,\omegabar_\mck,\psi_\mck,g_\mck,\nabla_\mck f_\mck,(\nabla_3)_\mck f_\mck,f_\mck\right),\label{eq:defmcqindicek}
	\end{align}
namely $\widecheck{\mcq}$ and $\mcq_\mck$ represent respectively all the linearized quantities except $\psicheck_H$, $\bcheck$ and $\fc$, and all the background Kerr quantities, which appear in \eqref{eq:nlfc}. Note that we separate $\psicheck_H$ and $\bcheck$ from the other linearized quantities in $\widecheck{\mcq}$ because these are the ones which present a $\Omega^{-2}$-degenerate bound on the RHS of \eqref{eq:nlfc}, see Theorem \ref{thm:controldoublenull} and \eqref{eq:nlfcsch} below.
\end{defi}
Then, \eqref{eq:nlfc} can be rewritten schematically as follows,
\begin{align}\label{eq:nlfcsch}
	\nabla_{\ering_4+f^A\partial_{\theta^A}+\frac14|f|^2 \ering_3}[\fc]=_s (\mcq_\mck,\widecheck{\mcq},\fc)^{\leq 4}(\psicheck_H,\Omega^{-2}\df^{\leq 1}\bcheck,\widecheck{\mcq},\fc).
\end{align}

\noindent\textbf{Initial data for $f$ and $\lambda$.}  Recall the $S(u,\ubar)$-tangent 1-form $f_0$ and scalar function $\lambda_0$ on $\Sigma_0$ which are defined in Proposition \ref{prop:initf0gbar0ettout}, and the notation introduced in Remark \ref{rem:notationemuun}. We will choose our initial data for $f$ and $\lambda$ such that $e'_4|_{\Sigma_0}$ coincides with $-|q|^2\Delta^{-1}e_4^{(\un)}|_{\Sigma_0}$, namely
\begin{align}\label{eq:initcondfclambdac}
	f|_{\Sigma_0}=f_0,\quad \lambda|_{\Sigma_0}=\lambda_0.
\end{align}
Also recall from \eqref{eq:dfdoublenull} the set $\df=\{\Omega^2\nabring_4,\nabring_3,\nabring\}$ of derivatives.
\begin{thm}\label{thm:controlfc}
	Provided that $\varepsilon(a,M,C_R)>0$ is small enough, and that the constant $\gamma$ defining region $\deux$ in \eqref{eq:defdedeuxx} satisfies $0<\gamma<\delta/12$, there exists a unique solution $\fc$ of \eqref{eq:transpof} in $\deux$ with initial condition given by \eqref{eq:initcondfclambdac} on $\Sigma_0$,  which satisfies moreover
	$$|{\df}^{\leq N_3}\fc|\lesssim\frac{\varepsilon}{\ubar^{2+\delta/4}},\quad\text{in}\:\:\deux,$$
	where $N_3=N_{max}-1$.
\end{thm}
Theorem \ref{thm:controlfc} will be proven by a standard bootstrap argument which relies on the following preliminary results. We notice that for $|f|\lesssim 1$, defining $Y$ as in \eqref{eq:Ysadefquoi}, we have\footnote{Here, we use the identity $2\T=\Omega^2e_4-e_3-b^A\partial_{\theta^A}.$}
\begin{align}
	Y&=\left(1+\frac{\Omega^2|f|^2}{4}\right)\Omega^2\ering_4-\frac{\Omega^2|f|^2}{2}\T+\Omega^2\left(f^A-\frac{1}{4}|f|^2b^A\right)\partial_{\theta^A},\label{eq:recupe4avecdf}\\
	&=\left(1+\frac{\Omega^2|f|^2}{4}\right)\ering_3+2\T+\left(b^A+\Omega^2f^A\right)\partial_{\theta^A}.\label{eq:recupe3avecdf}
\end{align}
This suggests to define the slightly modified set of derivatives
\begin{align}\label{eq:dfprim}
	{\df}'=\{\nabring_Y,\lieT,\nabring\},
\end{align}
which is equivalent to $ {\df}$, the the sense that for any $U$, $| {\df}^{\leq 1}U|\sim|( {\df}')^{\leq 1}U|$. Recall the estimate \eqref{eq:estimsigma0fclambdac} satisfied on $\Sigma_0$ by $\fc_0$ and $\widecheck{\lambda}_0$.
\begin{prop}\label{prop:initdfprimfc}
	Let $\fc$ be a solution of \eqref{eq:nlfcsch} on a neighbourhood of a subset of $\Sigma_0$ included in $\deux$, satisfying the initial condition \eqref{eq:initcondfclambdac}. Then we have, on $\Sigma_0$, 
	\begin{align}\label{eq:initndfprimdfrak}
		|( {\df}')^{\leq N}\fc||_{\Sigma_0}\lesssim\varepsilon\ubar^{-2-\delta/4}.
	\end{align}
\end{prop}
\begin{proof} Note that by \eqref{eq:nlfcsch}, we control $\nabring_{Y}\fc|_{\Sigma_0}$, and as we control appropriately all derivatives of linearized Ricci, curvature, and metric coefficients on $\Sigma_0$ in the double null frame (see Theorem \ref{thm:controldoublenull}), by induction we control higher order $\nabring_Y$ derivatives of $\fc$ on $\Sigma_0$ and we deduce
	$$|(\lieT,\nabring,\nabring_Y)^{\leq N}\fc||_{\Sigma_0}\lesssim\varepsilon\ubar^{-2-\delta/4},$$
	which concludes the proof.
\end{proof}
Before stating and proving the bootstrap proposition (Proposition \ref{prop:bootstrapfcdfrak}) which improves the bootstrap assumptions, we first prove three results that allow us to control $ \df'$ derivatives of solutions of \eqref{eq:nlfcsch} by induction.
\begin{prop}\label{prop:ind1}
	Let $\fc$ be a 1-form tangent to the spheres $S(u,\ubar)$ which satisfies \eqref{eq:nlfcsch}. Then for any integer $k\geq 0$, we have the schematic equation
	\begin{align}
		(\nabring_Y)^{k+1}\fc&=_s F_k^{(1)},\quad\text{where}:\label{eq:ind1}\\
		F_k^{(1)}&=_s\sum_{m=0}^4\sum_{i_1+\cdots+i_{m+2}\leq k,i_1\geq 1}\left(\nabring_Y^{i_1}\fc\df'^{i_2}\Omega^2,\df'^{i_1}(\Omega^2\psicheck_H,\df^{\leq 1}\bcheck),\df'^{i_1}\widecheck{\mcq}\df'^{i_2}\Omega^2\right)\prod_{j=2}^{m+2}\df'^{i_j}(\mcq_\mck,\widecheck{\mcq},\fc).\nn
	\end{align}
\end{prop}
\begin{proof}
By \eqref{eq:nlfcsch} and \eqref{eq:Ysadefquoi} we have $\nabring_Y\fc=_s(\mcq_\mck,\widecheck{\mcq},\fc)^{\leq 4}(\Omega^2\psicheck_H,\df^{\leq 1}\bcheck,\Omega^2\widecheck{\mcq},\Omega^2\fc)$, so that
$$(\nabring_Y)^{k+1}\fc=\nabring_Y^k\left((\mcq_\mck,\widecheck{\mcq},\fc)^{\leq 4}(\Omega^2\psicheck_H,\df^{\leq 1}\bcheck,\Omega^2\widecheck{\mcq},\Omega^2\fc)\right)=_s F_k^{(1)},$$
where we simply used the product rule in the last step.
\end{proof}

\begin{prop}\label{prop:ind2}
	let $\fc$ be a 1-form tangent to the spheres $S(u,\ubar)$ which satisfies \eqref{eq:nlfcsch}. Then for $k,k'\geq 0$, we have the schematic equation
	\begin{align}    
		\nabring_Y(\lieT^{k'}\nabring_Y^k\fc)&=_sF_{k,k'}^{(2)},\quad\text{where}:\label{eq:secondinduction}\\
		F_{k,k'}^{(2)}&=_s\lieT^{k'}F_k^{(1)}+\df'^{\leq k'}(\Omega^2\psicheck_H,\df^{\leq 1}\bcheck,\widecheck{\mcq},\fc)^{\leq k'+2}\df'^{\leq k+k'}\fc.\nn
	\end{align}
\end{prop}
\begin{proof}
We have, using Proposition \ref{prop:ind1},
\begin{align*}
	\nabring_Y(\lieT^{k'}\nabring_Y^k\fc)&=\lieT^{k'}F_k^{(1)}+[\lieT^{k'},\nabring_Y]\nabring_Y^k\fc\\
	&=_s\lieT^{k'}F_k^{(1)}+[\Omega^2\nabring_4,\lieT^{k'}]\nabring_Y^k\fc+[\Omega^2f\cdot\nabring,\lieT^{k'}]\nabring_Y^k\fc+[\Omega^2|f|^2\nabring_3,\lieT^{k'}]\nabring_Y^k\fc.
\end{align*}
Moreover, using \eqref{eq:commlieTschnab4}, see also Proposition \ref{prop:commmmmmm}, we get the schematic identities
\begin{align*}
	[\Omega^2\nabring_4,\lieT^{k'}]\nabring_Y^k\fc&=_s\sum_{i=0}^{k'-1}\lieT^i[\Omega^2\nabring_4,\lieT]\lieT^{k'-1-i}\nabring_Y^k\fc=_s\df'^{\leq k'}(\Omega^2\psicheck_H,\df^{\leq 1}\bcheck,\widecheck{\mcq},\fc)^{\leq k'+2}\df'^{\leq k+k'}\fc,
\end{align*}
and
\begin{align*}
	[\Omega^2f\cdot\nabring,\lieT^{k'}]\nabring_Y^k\fc&=_s\sum_{i_1+i_2+i_3+i_4\leq k',i_4\leq k'-1}\lieT^{i_1}\gamma\lieT^{i_2}\Omega^2\lieT^{i_3}f\lieT^{i_4}\nabring \nabring_Y^k\fc+\Omega^2f[\lieT^{k'},\nabring]\nabring_Y^k\fc\\
	&=_s\df'^{\leq k'}(\Omega^2\psicheck_H,\df^{\leq 1}\bcheck,\widecheck{\mcq},\fc)^{\leq k'+2}\df'^{\leq k+k'}\fc,\\
	[\Omega^2|f|^2\nabring_3,\lieT^{k'}]\nabring_Y^k\fc&=_s\sum_{i_1+i_2\leq k',i_2\leq k'-1}\T^{i_1}(\Omega^2|f|^2)\lieT^{i_2}\nabring_3 \nabring_Y^k\fc+\Omega^2|f|^2[\lieT^{k'},\nabring_3]\nabring_Y^k\fc\\
	&=_s\df'^{\leq k'}(\Omega^2\psicheck_H,\df^{\leq 1}\bcheck,\widecheck{\mcq},\fc)^{\leq k'+2}\df'^{\leq k+k'}\fc,
\end{align*}
which concludes the proof.
\end{proof}
\begin{defi}
	We denote $N_3=N_{max}-1$, with $N_{max}$ defined in Theorem \ref{thm:controldoublenull}.
\end{defi}
In the following proposition, we use the convention $\Omega^2\nabring^{-1}\lieT^{k'+1}\nabring_Y^k\fc=0$.
\begin{prop}\label{prop:ind3}
	Let $\fc$ be a 1-form tangent to the spheres $S(u,\ubar)$ which satisfies \eqref{eq:nlfcsch}. Under the bootstrap assumption $|( \df')^{\leq N_3}\fc|\leq\sqrt{\varepsilon}\ubar^{-2-\delta/4}$, we have the following reduced schematic equation for $k+k'+k''\leq N_3$,
	\begin{align}    
		\nabring_Y(\nabring^{k''}\lieT^{k'}\nabring_Y^k\fc)=_{rs}&\:O(\Omega^2)\nabring^{k''-1}\lieT^{k'+1}\nabring_Y^k\fc+O(\Omega^2)\nabring^{k''}\lieT^{k'}\nabring_Y^k\fc\nn\\
		&+O(\Omega^2)\df'^{\leq k''+k'+k-1}\fc+O(\varepsilon\ubar^{-2-\delta/3}).\label{eq:recform}
	\end{align}
\end{prop}

\begin{proof}
	From Proposition \ref{prop:ind2} we get
\begin{align}\label{eq:lokkatt}
	\nabring_Y(\nabring^{k''}\lieT^{k'}\nabring_Y^k\fc)=_s\nabring^{k''}F_{k,k'}^{(2)}+[\nabring_Y,\nabring^{k''}]\lieT^{k'}\nabring_Y^k\fc.
\end{align}
Moreover by \eqref{eq:secondinduction} we infer
\begin{align*}
	\nabring^{k''}F_{k,k'}^{(2)}&=_{rs}\nabring^{k''}\lieT^{k'}F_k^{(1)}+\nabring^{k''}\left(\df'^{\leq k'}(\Omega^2\psicheck_H,\df^{\leq 1}\bcheck,\widecheck{\mcq},\fc)^{\leq k'+2}\df'^{\leq k+k'}\fc\right)=_{rs}I+II+III+IV,
\end{align*}
where : 
\begin{itemize}
	\item $I$ contains the terms in $\nabring^{k''}\lieT^{k'}F_k^{(1)}$ with highest order derivatives acting on $\fc$. By \eqref{eq:ind1}, the bootstrap assumption, and the bounds in Theorem \ref{thm:controldoublenull}\footnote{In particular, we use the identity $\df^k\Omega^2=\df^k(e^{2\log\Omega})=\Omega^2\df^{\leq k}\log\Omega^{\leq k}=O(\Omega^2)$.}, we have 
	$$I=_{rs}O(\Omega^2)\nabring^{k''}\lieT^{k'}\nabring_Y^k\fc.$$
	\item $II$ contains the terms in $\nabring^{k''}\lieT^{k'}F_k^{(1)}$ which are of order $\leq k+k'+k''-1$ in $\fc$. By \eqref{eq:ind1}, the bootstrap assumption and the bounds in Theorem \ref{thm:controldoublenull}, we get 
	$$II=O(\Omega^2,\sqrt{\varepsilon}\ubar^{-2-\delta/4})\df'^{\leq k+k'+k''-1}\fc=O(\Omega^2)\df'^{\leq k+k'+k''-1}\fc+O(\varepsilon\ubar^{-2-\delta/3}).$$
	\item $III$ contains the terms in $\nabring^{k''}\lieT^{k'}F_k^{(1)}$ which do not involve $\fc$. These terms are $\df'$ derivatives of $\Omega^2\psicheck_H,\widecheck{\mcq}$, see the RHS of \eqref{eq:nlfcsch}. By Theorem \ref{thm:controldoublenull}, these terms satisfy
	$$III=O(\varepsilon\ubar^{-2-\delta/3}).$$
	\item $IV=\nabring^{k''}\left(\df'^{\leq k'}(\Omega^2\psicheck_H,\df^{\leq 1}\bcheck,\widecheck{\mcq},\fc)^{\leq k'+2}\df'^{\leq k+k'}\fc\right)$ is a quadratic error term, which, combining the bootstrap assumption $|( \df')^{\leq N_3}\fc|\leq\sqrt{\varepsilon}\ubar^{-2-\delta/4}$ and the bounds of Theorem \ref{thm:controldoublenull}, is easily bounded by $\varepsilon\ubar^{-4}\lesssim\varepsilon\ubar^{-2-\delta/3}$.
\end{itemize}
Note that to bounds the derivatives of the terms involving $f_\mck$ in the definition \eqref{eq:defmcqindicek} of $\mcq_\mck$, we use \eqref{rem:deriveesdanskerrr}. Now we estimate the commutator term $[\nabring_Y,\nabring^{k''}]\lieT^{k'}\nabring_Y^k\fc$ in \eqref{eq:lokkatt}. We have
\begin{align*}
	[\nabring_Y,\nabring^{k''}]\lieT^{k'}\nabring_Y^k\fc=_s[\Omega^2\nabring_4,\nabring^{k''}]\lieT^{k'}\nabring_Y^k\fc+[\Omega^2 f\cdot\nabring,\nabring^{k''}]\lieT^{k'}\nabring_Y^k\fc+[\Omega^2|f|^2\nabring_3,\nabring^{k''}]\lieT^{k'}\nabring_Y^k\fc.
\end{align*}
Moreover, we have the following computations for $U=\lieT^{k'}\nabring_Y^k\fc$,
\begin{align*}
	[\Omega^2\nabring_4,\nabring^{k''}]U=_s&\sum_{i=0}^{k''-1}\nabring^i[\Omega^2\nabring_4,\nabring]\nabring^{k''-1-i}U\\
	=_s&\sum_{i=0}^{k''-1}\nabring^i\left(\Omega^2(\psi_H\nabring^{k''-i}U+(\psi_H\etabar+\beta)\nabring^{k''-i-1}U)\right)=_sI'+II',
\end{align*}
where we used \eqref{eq:bonnermk} in the second step, and we also have, by Proposition \ref{prop:commnablaboudnenulll}, 
\begin{align*}
	[\Omega^2 f\cdot\nabring,\nabring^{k''}]U&=_s\sum_{i=0}^{k''-1}\nabring^i[\Omega^2 f\cdot\nabring,\nabring]\nabring^{k''-1-i}U\\
	&=_s\sum_{i=0}^{k''-1}\nabring^i\left(\Omega^2 f K\nabring^{k''-i-1}U+\nabring(\Omega^2f)\nabring^{k''-i}U\right)=_sI'+II',
\end{align*}
and 
\begin{align*}
	&[\Omega^2|f|^2\nabring_3,\nabring^{k''}]U=_s\sum_{i=0}^{k''-1}\nabring^i[\Omega^2|f|^2\nabring_3,\nabring]\nabring^{k''-1-i}U\\
	&=_s\sum_{i=0}^{k''-1}\nabring^i\left(\Omega^2|f|^2(\psi_{\Hbar}\nabring^{k''-i}U+(\psi_{\Hbar}\eta+\betabar)\nabring^{k''-1-i}U)+\Omega f\cdot\nabring(\Omega f)\nabring_3\nabring^{k''-1-i}U\right)\\
	&=_sI'+II'+III',
\end{align*}
where:
\begin{itemize}
	\item Once again, $I'$ represents the terms with highest order derivatives in $\fc$ except for the term $\nabring^i\left(\Omega f\cdot\nabring(\Omega f)\nabring_3\nabring^{k''-1-i}\lieT^{k'}\nabring_Y^k\fc\right)$ in the expression of $[\Omega^2|f|^2\nabring_3,\nabring^{k''}]U$ above. We have by Theorem \ref{thm:controldoublenull} the estimate
	$$I'=O(\Omega^2,\varepsilon\ubar^{-2-\delta/3})\nabring^{k''}\lieT^{k'}\nabring_Y^k\fc=O(\Omega^2)\nabring^{k''}\lieT^{k'}\nabring_Y^k\fc+O(\varepsilon\ubar^{-2-\delta/3}).$$
	\item Next, $II'$ represents once again the lower order terms in $\fc$ which satisfy as before
		$$II'=O(\Omega^2,\sqrt{\varepsilon}\ubar^{-2-\delta/4})\df'^{\leq k+k'+k''-1}\fc=O(\Omega^2)\df'^{\leq k+k'+k''-1}\fc+O(\varepsilon\ubar^{-2-\delta/3}).$$
		\item Finally, this time the term $III'$ corresponds to the highest order derivative of $\fc$ which comes from the term $\nabring^i\left(\Omega f\cdot\nabring(\Omega f)\nabring_3\nabring^{k''-1-i}\lieT^{k'}\nabring_Y^k\fc\right)$. Thus, commuting $\nabring^i$ and $\nabring_3$, which produces lower order terms of the type $II$, we get that this term is
		$$III'=_{rs}O(\Omega^2)\nabring_3\nabring^{k''-1}\lieT^{k'}\nabring_Y^k\fc.$$
		Thus, using \eqref{eq:recupe3avecdf} we deduce
		\begin{align*}
			III'&=_{rs}O(\Omega^2)\nabring_Y\nabring^{k''-1}\lieT^{k'}\nabring_Y^k\fc+O(\Omega^2)\nabring^{k''}\lieT^{k'}\nabring_Y^k\fc+O(\Omega^2)\lieT^{\leq 1}\nabring^{k''-1}\lieT^{k'}\nabring_Y^k\fc\\
			&=_{rs}O(\Omega^2)\df^{k+k'+k''-1}\fc+O(\varepsilon\ubar^{-2-\delta/4})+I'+II'+O(\Omega^2)\nabring^{k''-1}\lieT^{k'+1}\nabring_Y^k,
		\end{align*}
where in the last step we used the following identities:
\begin{enumerate}
	\item $\nabring_Y\nabring^{k''-1}\lieT^{k'}\nabring_Y^k\fc=_{rs}\df^{k+k'+k''-1}\fc+O(\varepsilon\ubar^{-2-\delta/3})$ which comes from the previously proven identity
	$$\nabring_Y(\nabring^{k''-1}\lieT^{k'}\nabring_Y^k\fc)=_{rs}I+II+III+I'+II'+III'\quad\text{(rank}\:k''-1\text{)},$$
	at rank $k''-1$, where we treat the term $III'$ at rank $k''-1$ be simply writing 
	$$III'=_{rs}O(\Omega^2)\nabring_3\nabring^{k''-2}\lieT^{k'}\nabring_Y^k\fc=O(\Omega^2)\df^{k+k'+k''-1}\fc.$$
	\item $O(\Omega^2)\lieT^{\leq 1}\nabring^{k''-1}\lieT^{k'}\nabring_Y^k\fc=O(\Omega^2)\nabring^{k''-1}\lieT^{k'+1}\nabring_Y^k\fc+II'$ by commuting $\lieT$ and $\nabring^{k''-1}$, which generates lower order terms of the type $II'$.
\end{enumerate}
\end{itemize}
This concludes the proof of Proposition \ref{prop:ind3}.
\end{proof}

\begin{prop}\label{prop:bootstrapfcdfrak}
We assume that $\fc$ is a $S(u,\ubar)$-tangent 1-form which satisfies \eqref{eq:nlfcsch}, the bootstrap assumption $|( \df')^{\leq N_3}\fc|\leq\sqrt{\varepsilon}\ubar^{-2-\delta/4}$ in $\deux[\ubar_f]$, and the estimate $|( \df')^{\leq N_3}\fc||_{\Sigma_0}\lesssim\varepsilon\ubar^{-2-\delta/4}$ on $\Sigma_0$. Then provided $0<\gamma<\frac{\delta}{12}$, there exists $C(a,M,C_R)>0$ such that, on $\deux[\ubar_f]$,  
	$$|( \df')^{\leq N_3}\fc|\leq C(a,M,C_R)\varepsilon\ubar^{-2-\delta/4}.$$
\end{prop}
\begin{proof}
	Recall that we denote $ \df'=\{\lieT, \nabring, \nabring_Y\}$ and that the implicit constants depend here on $a,M,C_R$, see Remark \ref{rem:depuisla}. We will prove by induction on $N\in\{0,\cdots,N_3\}$ the bound
	\begin{align}\label{eq:inductionhypo}
		|( \df')^{\leq N}\fc|\lesssim\varepsilon\ubar^{-2-\delta/4},\quad\text{in}\:\:\deux[\ubar_f].
	\end{align}
	\noindent\textbf{Step 1.}\textit{ The case $N=0$.} We detail the case $N=0$, to highlight the important points in the estimate, which will be used also in the induction step. Recalling \eqref{eq:nlfcsch}, by the bootstrap assumption, \eqref{eq:Ysadefquoi} and Theorem \ref{thm:controldoublenull} we get in $\deux[\ubar_f]$ the bound
	$$|\nabring_{Y}\fc|\lesssim\varepsilon{\ubar^{-2-\delta/3}}+\Omega^2|\fc|.$$
	Note that the bootstrap assumption also implies $|\fc|+|f|\lesssim 1$. Thus by the transport estimate in Proposition \ref{prop:transporte4'} we get for $p=(u,\ubar,\theta^A)\in\deux[\ubar_f]$,
	\begin{align*}
		|\fc|(p)\lesssim& |\fc|(u_{Y,\Sigma_0}(p),\ubar_{Y,\Sigma_0}(p),\theta^A_{Y,\Sigma_0}(p))+\int_{\ubar_{Y,\Sigma_0}(p)}^\ubar\frac{\varepsilon}{(\ubar')^{2+\delta/3}}\dee\ubar'\\
		&+\int_{\ubar_{Y,\Sigma_0}(p)}^\ubar\Omega^2|\fc|(u_Y^p(\ubar'),\ubar',(\theta^A)^p_Y(\ubar'))\dee\ubar'.
	\end{align*}
	Moreover, we have 
	$$\int_{\ubar_{Y,\Sigma_0}(p)}^\ubar\frac{\varepsilon}{(\ubar')^{2+\delta/3}}\dee\ubar'\lesssim\frac{\varepsilon}{\ubar_{Y,\Sigma_0}(p)^{2+\delta/3}}(\ubar-\ubar_{Y,\Sigma_0}(p)).$$
	Also, as $\Sigma_0=\{u+\ubar=C_R\}$, by Proposition \ref{prop:flowofY} which yields $|u_{Y,\Sigma_0}(p)-u|\lesssim 1$ we get
	\begin{align*}
		\ubar-\ubar_{Y,\Sigma_0}(p)=\ubar-C_R+u_{Y,\Sigma_0}(p)=\ubar+u+O(1)\leq \ubar^\gamma+O(1)\lesssim\ubar^\gamma,
	\end{align*}
where we used $u+\ubar\leq\ubar^\gamma$ in $\deux$, which also implies
	\begin{align}\label{eq:thisisbien}
		\ubar_{Y,\Sigma_0}(p)\sim\ubar.
	\end{align}
	We deduce
	\begin{align}\label{eq:superborneca}
		\int_{\ubar_{Y,\Sigma_0}(p)}^\ubar\frac{\varepsilon}{(\ubar')^{2+\delta/3}}\dee\ubar'\lesssim\frac{\varepsilon}{\ubar^{2+\delta/3-\gamma}}\lesssim\frac{\varepsilon}{\ubar^{2+\delta/4}} 
	\end{align}
	for $0<\gamma<\delta/12$. This yields, using the initial data assumption on $\Sigma_0$,
	$$|\fc|(u,\ubar,\theta^A)\lesssim\frac{\varepsilon}{\ubar^{2+\delta/4}} +\int_{\ubar_{Y,\Sigma_0}(p)}^\ubar\Omega^2|\fc|(u_Y^p(\ubar'),\ubar',(\theta^A)^p_Y(\ubar'))\dee\ubar'.$$
	By Grönwall's inequality to treat the last term on the RHS, we deduce
	\begin{align*}
		|\fc|(u,\ubar,\theta^A)&\lesssim\frac{\varepsilon}{\ubar^{2+\delta/4}}\exp\left(C(a,M,C_R)\int_{\ubar_{Y,\Sigma_0}(p)}^\ubar\Omega^2(u_Y^p(\ubar'),\ubar',(\theta^A)^p_Y(\ubar'))\dee\ubar'\right)\lesssim\frac{\varepsilon}{\ubar^{2+\delta/4}},
	\end{align*}
	where we used the estimate
	$$\Omega^2(u_Y^p(\ubar'),\ubar',(\theta^A)^p_Y(\ubar'))\sim e^{-|\kappa_-|(u_Y^p(\ubar')+\ubar')}\sim e^{-|\kappa_-|(u(p)+\ubar')},$$
	by Proposition \ref{prop:flowofY}, which implies 
	\begin{align}\label{eq:intomegaperturb}
		\int_{\ubar_{Y,\Sigma_0}(p)}^\ubar\Omega^2(u_Y^p(\ubar'),\ubar',(\theta^A)^p_Y(\ubar'))\dee\ubar'\lesssim e^{-|\kappa_-|(u(p)+\ubar_{Y,\Sigma_0}(p))}\lesssim e^{-|\kappa_-|(u_{Y,\Sigma_0}(p)+\ubar_{Y,\Sigma_0}(p))}\lesssim 1,
	\end{align}
	where we used \eqref{eq:equivtransporte4'} again. Hence the result for $N=0$.
	\\
	\noindent\textbf{Step 2.}\textit{ The induction step.} We now assume that\eqref{eq:inductionhypo} holds at rank $N-1$ for some $N\geq 1$. For $k+k'\leq N$ we have $k+k'-1\leq N-1$ thus 
	\begin{align*}
		|\Omega^2( \df')^{\leq k+k'-1}\fc|\lesssim\frac{\varepsilon\Omega^{2}}{\ubar^{2+\delta/4}}.
	\end{align*}
	By Proposition \ref{prop:ind3} at rank $k''=0$, and recalling the convention $\nabring^{-1}=0$, we thus get
	\begin{align*}
		|\nabring_Y(\lieT^{k'}\nabring_Y^k\fc)|\lesssim \Omega^2|\lieT^{k'}\nabring_Y^k\fc|+\frac{\varepsilon}{\ubar^{2+\delta/3}}+\frac{\Omega^2\varepsilon}{\ubar^{2+\delta/4}}.
	\end{align*}
	We now use the transport estimate \ref{eq:inegtransporte4'} which gives, for any $p=(u,\ubar,\theta^A)\in\deux[\ubar_f]$,
	\begin{align*}
		|\lieT^{k'}\nabring_Y^k\fc|(p)&\lesssim |\lieT^{k'}\nabring_Y^k\fc|(u_{Y,\Sigma_0}(p),\ubar_{Y,\Sigma_0}(p),\theta^A_{Y,\Sigma_0}(p))+\varepsilon\int_{\ubar_{Y,\Sigma_0}(p)}^\ubar\left(\frac{1}{(\ubar')^{2+\delta/3}}+\frac{\Omega^2}{(\ubar')^{2+\delta/4}}\right)\dee\ubar'\\
		&\quad+\int^\ubar_{\ubar_{Y,\Sigma_0}(p)}\Omega^2|\lieT^{k'}\nabring_Y^k\fc|(u_Y^p(\ubar'),\ubar',(\theta^A)^p_Y(\ubar'))\dee\ubar'\\
		&\lesssim\frac{\varepsilon}{\ubar^{2+\delta/4}}+\int^\ubar_{\ubar_{Y,\Sigma_0}(p)}\Omega^2|\lieT^{k'}\nabring_Y^k\fc|(u_Y^p(\ubar'),\ubar',(\theta^A)^p_Y(\ubar'))\dee\ubar'.
	\end{align*}
	Note that here we used \eqref{eq:superborneca} as well as the bounds
	\begin{align*}
		\int_{\ubar_{Y,\Sigma_0}(p)}^\ubar\frac{\Omega^2(u_Y^p(\ubar'),\ubar',(\theta^A)^p_Y(\ubar'))}{(\ubar')^{2+\delta/4}}\dee\ubar'\lesssim & \frac{1}{\ubar_{Y,\Sigma_0}(p)^{2+\delta/4}}\int_{\ubar_{Y,\Sigma_0}(p)}^\ubar\Omega^2(u_Y^p(\ubar'),\ubar',(\theta^A)^p_Y(\ubar'))\dee\ubar'\\
		\lesssim& \frac{1}{\ubar_{Y,\Sigma_0}(p)^{2+\delta/4}}\lesssim \frac{1}{\ubar^{2+\delta/4}},
	\end{align*}
	which come from \eqref{eq:intomegaperturb} and \eqref{eq:thisisbien}. Thus, by Grönwall's inequality we obtain
	\begin{align*}
		|\lieT^{k'}\nabring_Y^k\fc|&\lesssim \frac{\varepsilon}{\ubar^{2+\delta/4}}\exp\left(C(a,M,C_R)\int_{\ubar_{Y,\Sigma_0}(p)}^\ubar\Omega^2(u_Y^p(\ubar'),\ubar',(\theta^A)^p_Y(\ubar'))\dee\ubar'\right)\lesssim \frac{\varepsilon}{\ubar^{2+\delta/4}}
	\end{align*}
	by \eqref{eq:intomegaperturb} again. This implies, for $k+k'\leq N$,
	\begin{align}\label{eq:pasletemps}
		|\lieT^{k'}\nabring_Y^k\fc|\lesssim\frac{\varepsilon}{\ubar^{2+\delta/4}}.
	\end{align}
	We now prove by induction on $k''\in\{0,\cdots,N\}$ the following statement
	\begin{itemize}
		\item $(H_{k''})$ : for any $k,k'\geq 0$, and $0\leq k_0''\leq k''$ such that $k+k'+k''_0\leq N$, we have the estimate $$|\nabring^{k_0''}\lieT^{k'}\nabring_Y^{k}\fc|\lesssim\frac{\varepsilon}{\ubar^{2+\delta/4}}.$$
	\end{itemize}
	$(H_0)$ states that for any $k,k'$ such that $k+k'\leq N$, we have $|\lieT^{k'}\nabring_Y^{k}\fc|\lesssim{\varepsilon}{\ubar^{-2-\delta/4}}$. This holds by \eqref{eq:pasletemps}. Now, let $k''\geq 1$ and let us assume that $H_{k''-1}$ holds. Let $k,k'$ and $k_0''\leq k''$ such that $k+k'+k_0''\leq N$. Then by the recurrence formula \eqref{eq:recform} and the induction bound $|\Omega^2( \df')^{\leq k+k'+k''-1}\fc|\lesssim{\varepsilon\Omega^{2}}{\ubar^{-2-\delta/4}}$ we get 
	\begin{align*}
		|\nabring_Y(\nabring^{k''_0}\lieT^{k'}\nabring_Y^{k}\fc)|\lesssim&\frac{\varepsilon}{\ubar^{2+\delta/3}}+\frac{\varepsilon\Omega^2}{\ubar^{2+\delta/4}}+\Omega^2|\nabring^{k''_0-1}\lieT^{k'+1}\nabring_Y^k\fc|+\Omega^2|\nabring^{k''_0}\lieT^{k'}\nabring_Y^{k}\fc|.
	\end{align*}
Moreover, $k''_0-1\leq k''-1$ and $k''-1+k'+1+k=k''+k'+k\leq N$ thus by $(H_{k''-1})$, we have $|\nabring^{k_0''-1}\lieT^{k'+1}\nabring_Y^k\fc|\lesssim{\varepsilon}/{\ubar^{2+\delta/4}}$, which implies
	$$|\nabring_Y(\nabring^{k''_0}\lieT^{k'}\nabring_Y^{k}\fc)|\lesssim \Omega^2|\nabring^{k''_0}\lieT^{k'}\nabring_Y^{k}\fc|+\frac{\varepsilon}{\ubar^{2+\delta/3}}+\frac{\varepsilon\Omega^2}{\ubar^{2+\delta/4}}$$
	and we conclude by the transport inequality of Proposition \ref{prop:transporte4'} and Grönwall's inequality exactly like before, which proves 
	\begin{align}\label{eq:pppinduc}
		|\nabring^{k''_0}\lieT^{k'}\nabring_Y^k\fc|\lesssim\frac{\varepsilon}{\ubar^{2+\delta/4}}.
	\end{align}
	This proves $(H_{k''})$ and concludes the induction in $k''$. Finally, we conclude that \eqref{eq:inductionhypo} holds at rank $N$ by commuting $\nabring,\lieT$ and $\nabring_Y$ which only produces lower-order terms, which concludes the induction in $N$ and hence the proof of Proposition \ref{prop:bootstrapfcdfrak}.
\end{proof}

We are now ready to prove Theorem \ref{thm:controlfc}.
\begin{proof}[\textbf{Proof of Theorem \ref{thm:controlfc}}]
This is a standard bootstrap argument: recalling $\ubar_{\Sigma_0}(u_f)=C_R-u_f$, and defining the interval
	$$\mathcal{I}=\Bigg\{\ubar_f>\ubar_{\Sigma_0}(u_f)\text{ such that:}\begin{array}{l}
		\text{there exists a unique }\fc\text{ solution of \eqref{eq:nlfc} on }\\
		\deux[\ubar_f]\text{ with } \fc|_{\Sigma_0}=\fc_0\text{ and }|(\df')^{\leq N_3}\fc|\leq \sqrt{\varepsilon}\ubar^{-2-\delta/4}
	\end{array}\Bigg\},$$
we prove that $\mci$ is non-empty for $\varepsilon$ small enough by continuity, local existence results for non-linear transport equations in the outgoing direction, and Proposition \ref{prop:initdfprimfc}, closed by construction, open for $\varepsilon$ small enough as a consequence of  Proposition \ref{prop:bootstrapfcdfrak}, and hence $\mci=(\ubar_{\Sigma_0}(u_f),+\infty)$, thereby concluding the proof.
\end{proof}
We now prove existence and control of $\lambdacheck$. 
\begin{thm}\label{thm:controllambdacheck}
	Let $f$ be the solution of \eqref{eq:nlfc} given by Theorem \eqref{thm:controlfc}. Provided $\varepsilon(a,M)>0$ is small enough and $0<\gamma<\delta/12$, there exists a unique solution $\lambdacheck$ of \eqref{eq:lambdaeqprelim} in $\deux$ with initial condition given by \eqref{eq:initcondfclambdac} on $\Sigma_0$,  which satisfies moreover
	$$|\df^{\leq N_3}\lambdacheck|\lesssim\frac{\varepsilon}{\ubar^{2+\delta/4}},\quad\text{in}\:\:\deux.$$
\end{thm}
\begin{proof}
	We only sketch the proof as it is easier than the one of Theorem \ref{thm:controlfc} for $\fc$. The main point is to rewrite the equation for $\lambdacheck$ in a way such that we can use the exact same technics used for the control of $\fc$. We can rewrite equation \eqref{eq:lambdaeqprelim} as
	\begin{align}\label{eq:lambdaeqprelim2}
		\Omega^{-2}Y(\lambda)+h\lambda=0,
	\end{align}
	where we recall the definition \eqref{eq:Ysadefquoi} of $Y$, and where $h$ is the following scalar function
	$$h=-f\cdot(\zeta-\etabar)+\frac{1}{2}|f|^2\omegabar+\frac{1}{2}f\cdot(f\cdot\chibar).$$
	Note that by Theorem \ref{thm:controlfc} and the estimates of Theorem \ref{thm:controldoublenull}, we have in $\deux$,
	$$|\df^{\leq N_3}\widecheck{h}|\lesssim\varepsilon\ubar^{-2-\delta/4},\quad|\df^{\leq N_3}{h}|\lesssim 1.$$
	This implies, recalling Proposition \ref{prop:transfodanskerr} and using $\omega'_\mck=0$ in the Kerr outgoing principal frame,
	\begin{align}\label{eq:edologlambdakerr}
		\left((e_4)_\mck+f_\mck\cdot_\mck\nabring_\mck+\frac{1}{4}|f_\mck|_\mck^2e_3\right)(\lambda_\mck)+h_\mck\lambda_\mck=0.
	\end{align}
	Thus, subtracting \eqref{eq:edologlambdakerr} from \eqref{eq:lambdaeqprelim2}, we get that \eqref{eq:lambdaeqprelim} is equivalent to
	\begin{align}\label{eq:edologlambdacheck}
		Y(\lambdacheck)+\Omega^2h\lambdacheck=\err[\lambdacheck],
	\end{align}
	where, recalling the definition of $\widecheck{\nabla}_4$ in Proposition \ref{prop:linderdoublenull}, and using that $\lambda$ is a scalar,
	\begin{align*}
		\err[\lambdacheck]:=-\Omega^2\left(\widecheck{\nabring}_4[\lambda_\mck]+\widecheck{f}\cdot_\mck(\nabring\lambda)_\mck+f\widecheck{\cdot}(\nabring\lambda)_\mck+\frac{1}{4}(\fc\cdot_\mck f_\mck+f\widecheck{\cdot}f_\mck+f\cdot\fc)(e_3\lambda)_\mck+\widecheck{h}\lambda_\mck\right).
	\end{align*}
	Now, note that using the control of $\fc$ in Theorem \ref{thm:controlfc}, as well as \eqref{eq:nab4diff} and the estimates of Theorem \ref{thm:controldoublenull}, we get the bound
	\begin{align}\label{eq:controlerrlambdacheck}
		|\df^{\leq N_3}\err[\lambdacheck]|\lesssim \varepsilon\ubar^{-2-\delta/3}+\varepsilon\Omega^2\ubar^{-2-\delta/4}.
	\end{align}
Finally, proceeding similarly as in the proof of Theorem \ref{thm:controlfc}, we successively commute the equation \eqref{eq:edologlambdacheck} with $Y^k$, $\T^{k'}$ and $\nabla^{k''}$ and we prove successive control of the derivatives of $\lambdacheck$ by the transport estimate of Proposition \ref{prop:transporte4'}. The proof is roughly the same as the one of Theorem \ref{thm:controlfc}, it is even easier since the equation for $\lambdacheck$ is is linear, and $\lambdacheck$ is a scalar.
\end{proof}
\begin{rem}\label{rem:fixflambda}
\textbf{From now on, we fix the values of $\lambda$ and $f$ to be 
	$$\lambda=\lambda_\mck+\lambdacheck,\quad f=f_\mck+\fc,$$
	where $\lambdacheck$ and $\fc$ are given by Theorems \ref{thm:controllambdacheck} and \ref{thm:controlfc}.}
\end{rem}
\subsection{Existence and control of $\underline{g}$ and $\lambdabar$ in $\deux$\label{section:controlfbar}}
\subsubsection{Transport estimates in the perturbed ingoing direction}\label{section:transpoestperturbingoing}
In this section, we consider $\ubar_f\geq C_R-u_f$ and a $S(u,\ubar)$ tangent 1-form $\gbar$ which satisfies
\begin{align}\label{eq:gbarbronehypdeux}
	|\gbar|\lesssim 1,\:\text{in}\:\:\deux[\ubar_f],
\end{align}
and the vector field 
\begin{align}\label{eq:Ybarsadefquoi}
	\Ybar:=\Omega^2\left(\wh{e}_3+\gbar^A\partial_{\theta^A}+\frac{1}{4}|\gbar|^2\wh{e}_4\right),
\end{align}
where we recall the double null pair $\wh{e}_3=\Omega^{-2}\ering_3,\: \wh{e}_4=\Omega^2 \ering_4$ in the \emph{ingoing} normalization. Note that $\Ybar$ can be rewritten as 
\begin{align*}
	\Ybar=\partial_u + \Omega^2\left(\gbar^A+\frac{1}{4}|\gbar|^2b^A\right)\partial_{\theta^A} +\frac{1}{4}\Omega^2|\gbar|^2\partial_\ubar.
\end{align*}
We denote $\varphi_{\Ybar}(s,u,\ubar,\theta^A)$ the flow of $\Ybar$ at time $s$ starting at $(u,\ubar,\theta^A)$ at $s=u$, namely $\varphi_{\Ybar}(s,u,\ubar,\theta^A)$ satisfies the following ODE in $\deux[\ubar_f]$,
\begin{align*}
	\frac{\partial}{\partial s}\varphi_{\Ybar}(s,u,\ubar,\theta^A)=\Ybar(\varphi_{\Ybar}(s,u,\ubar,\theta^A)),\quad\varphi_{\Ybar}(s=u,u,\ubar,\theta^A)=(u,\ubar,\theta^A).
\end{align*}
For $p\in\deux[\ubar_f]$, we also introduce the coordinate functions along the flow of $\Ybar$,
\begin{align}\label{eq:introcoordflowybar}
	u_{\Ybar}^p(s):=u(\varphi_{\Ybar}(s,p)),\quad \ubar_{\Ybar}^p(s):=\ubar(\varphi_{\Ybar}(s,p)),\quad (\theta^A)_{\Ybar}^p(s):=\theta^A(\varphi_{\Ybar}(s,p)).
\end{align}
\begin{prop}\label{prop:flowofYbar}
	For any $p\in\deux[\ubar_f]$ and $s\in\mathbb{R}$ such that $\varphi_{\Ybar}(s,p)$ is well-defined, we have
	\begin{align}\label{eq:equivtransporte4'bar}
		u_{\Ybar}^p(s)=s,\quad |\ubar_{\Ybar}^p(s)-\ubar(p)|\lesssim 1.
	\end{align}
\end{prop}
\begin{proof}
The proof follws the exact same steps as the one of Proposition \ref{prop:flowofY}, so we omit it.
\end{proof}

\begin{rem}
By proposition \ref{prop:flowofYbar} and the definition of $\Sigma_0=\{u+\ubar=C_R\}$, we deduce that for any $p\in\deux[\ubar_f]$, for some $u'\leq u(p)$ sufficiently negative, 
$$\varphi_{\Ybar}(u',p)\in\Sigma_0.$$
Indeed, from $u_{\Ybar}^p(s)=s$ and the identity
	\begin{align}\label{eq:deruYbar}
	\frac{\dee}{\dee s}\ubar_{\Ybar}^p(s)&=\frac{1}{4}[\Omega^2|\gbar|^2 ](s,\ubar_{\Ybar}^p(s),(\theta^A)_{\Ybar}^p(s)).
\end{align}
 we get that the flow starting from $p$ directed towards the past remains in $\deux[\ubar_f]$, as long as it does not reach $\Sigma_0$ (which it does in finite time).
\end{rem}
\begin{defi}
This special value of $u'$, which is such that the flow $\varphi_{\Ybar}(s,p)$ starting from $p$ reaches $\Sigma_0$ at time $s=u'$ will be denoted 
$$u_{\Ybar,\Sigma_0}(p)=\sup\big\{u'\leq u(p)\:\big|\:\varphi_{\Ybar}(u',p)\in\Sigma_0\big\}.$$
We also introduce the following simplified notations,
$$\ubar_{\Ybar,\Sigma_0}(p):=\ubar_Y^p(u_{\Ybar,\Sigma_0}(p)),\quad \theta^A_{\Ybar,\Sigma_0}(p):=(\theta^A)_{\Ybar}^p(u_{\Ybar,\Sigma_0}(p)).$$
\end{defi}
Note that we have for any $p=(u,\ubar,\theta^A)\in\deux[\ubar_f]$,
\begin{align}\label{eq:equivinitubarubar}
	|u_{\Ybar,\Sigma_0}(p)|\sim |C_R-u_{\Ybar,\Sigma_0}(p)|=|\ubar_Y^p(u_{\Ybar,\Sigma_0}(p))|\sim \ubar(p)\sim |u(p)|,
\end{align}
where we used \eqref{eq:equivtransporte4'bar} in the third step, and Lemma \ref{lem:usimubarII} in the last step. Now we state a transport estimate along the flow of $\Ybar$.
\begin{prop}\label{prop:transporte4'bar}
	Let $\gbar,\Ybar$ be as in \eqref{eq:gbarbronehypdeux} and \eqref{eq:Ybarsadefquoi}. Then, for any $p=(u,\ubar,\theta^A)\in\deux[\ubar_f]$, for any $S(u,\ubar)$-tangent tensor $U$, 
	\begin{align}\label{eq:inegtransporte4'bar}
		|U|(p)\lesssim |U|(u_{\Ybar,\Sigma_0}(p),\ubar_{\Ybar,\Sigma_0}(p),\theta^A_{\Ybar,\Sigma_0}(p))+\int_{u_{\Ybar,\Sigma_0}(p)}^{u(p)}|\nabring_{\Ybar} U|(u',\ubar_{\Ybar}^p(u'),(\theta^A)_{\Ybar}^p(u'))\dee u'.
	\end{align}
\end{prop}
\begin{proof}
	The proof is the same as the one of Proposition \ref{prop:transporte4'}.
\end{proof}

\subsubsection{Existence and control of $\widecheck{\underline{g}}$ and $\widecheck{\underline{\lambda}}$}\label{section:controlgbarlambdabar}

In this section, we prove existence and control of a $S(u,\ubar)$-tangent 1-form $\gbar$ and a scalar $\lambdabar$ which satisfy respectively the equations \eqref{eq:transpogbar}, \eqref{eq:transpolambdabar} and appropriate estimates.

We begin with the control of $\gbar$. Recall the definition of $\gbar_\mck$ from Remark \ref{rem:valeurkerrgbarlambdabar}. By Proposition \ref{prop:eqtranspofetlambda}, as both $(e_3)_\mck$ and $(\wh{e}_3)_\mck$ are geodesic in Kerr, $\gbar_\mck$ satisfies the following equation,
\begin{align}
	&(\nabring_{\underline{V}})_\mck\gbar_\mck+\frac{1}{4}\wh{tr\chibar}_\mck\gbar_\mck+\frac{1}{2}\gbar_\mck\cdot_\mck\wh{\wh{\chibar}}_\mck+\frac{1}{4}|\gbar_\mck|^2_\mck(\wh{\etabar}_\mck-\wh{\eta}_\mck)+\frac{1}{2}(\gbar_\mck\cdot_\mck\wh{\zeta}_\mck)\gbar_\mck\nn\\
	&-\frac{1}{4}\wh{\omega}_\mck|\gbar_\mck|^2_\mck\gbar_\mck-\frac{1}{8}|\gbar_\mck|^2\gbar_\mck\cdot_\mck\wh{\chihat}_\mck-\frac{1}{8}|\gbar_\mck|^2\wh{tr\chi}_\mck\gbar_\mck=0,\label{eq:kerredofbar}
\end{align}
where $\underline{V}_\mck=(\wh{e}_3)_\mck+\gbar^A_\mck\partial_{\theta^A}+\frac{1}{4}|\gbar_\mck|^2_\mck(\wh{e}_4)_\mck$. We define the null vector $\Ybar$ as in \eqref{eq:Ybarsadefquoi}. Then, similarly as in \eqref{eq:nlfc}, subtracting \eqref{eq:kerredofbar} from \eqref{eq:transpogbar} we get that $\gbar$ satisfies \eqref{eq:transpogbar} if and only if $\gbarc$ satisfies a schematic equation of the type
\begin{align}
	\nabla_{\wh{e}_3+\gbar^A\partial_{\theta^A}+\frac14|\gbar|^2\wh{e}_4}[\gbarc]=_s(\mcq_\mck,\widecheck{\mcq},\gbarc)^{\leq 4}(\Omega^{-2}\psicheck_{\Hbar},\Omega^{-2}\gcheck,\widecheck{\mcq},\gbarc)\label{eq:nlgbarc}
\end{align}
where
\begin{align}
	\widecheck{\mcq}&=_s((\Omega^2)^{\leq 1}\df^{\leq 1}\gcheck,\psicheck,\bcheck,\Omega^2\psicheck_H)\\
	\mcq_\mck&=_s((\wh{\psi}_H)_\mck,(\wh{\psi}_{\Hbar})_\mck,\hat{\omega}_\mck, \psi_\mck, g_\mck, (\nabla\gbar)_\mck, (\wh{\nabla}_4\gbar)_\mck, \gbar_\mck).
\end{align}
\begin{rem}
	Note that here we used the identities
	$${\nabring_{\wh{e}_3}}-(\nabring_{\wh{e}_3})_\mck=\widecheck{\Omega^{-2}}(\nabring_3)_\mck+\Omega^{-2}\widecheck{\nabring}_3,\quad{\nabring_{\wh{e}_4}-(\nabring_{\wh{e}_4})_\mck}=\widecheck{\Omega^{2}}(\nabring_4)_\mck+\Omega^{2}\widecheck{\nabring}_4,$$
	with $\widecheck{\nabring}_{3,4}$ defined in Proposition \ref{prop:linderdoublenull}. We also used the identities in Definition \ref{def:doublenullingoinggg} relating the Ricci coefficients defined in the ingoing double null frame $(\wh{e}_\mu)$ to the ones defined in the outgoing double null frame $(\ering_\mu)$.
\end{rem}

\textbf{Initial data for $\gbar$.} Recall the $S(u,\ubar)$-tangent 1-form $\gbar_0$ and the scalar $\lambdabar_0$ on $\Sigma_0$ defined in Proposition \ref{prop:initf0gbar0ettout}. We will choose our initial data for $\gbar$ and $\lambdabar$ such that $e_3|_{\Sigma_0}$ coincides with $e_3^{(\un)}|_{\Sigma_0}$ on $\Sigma_0$, namely such that
\begin{align}\label{eq:initcondgbarlambdabar}
	\gbar|_{\Sigma_0}=\gbar_0,\quad\lambdabar|_{\Sigma_0}=\lambdabar_0.
\end{align}
We will proceed similarly as in Section \ref{section:exsicontfc} to control $\gbarc$. The difference is that the null derivative in \eqref{eq:nlgbarc} is ingoing. This will only slightly modify the analysis, see Remark \ref{rem:flowpaststable}.  The main result of this section if the following control of $\gbarc$.
\begin{thm}\label{thm:controlgbarc}
	Provided $\varepsilon(a,M,C_R)>0$ is small enough and $0<\gamma<\delta/12$, there exists a unique solution $\gbarc$ of \eqref{eq:nlgbarc} in $\deux$ with initial condition given by \eqref{eq:initcondgbarlambdabar} on $\Sigma_0$,  which satisfies moreover
	$$|\df^{\leq N_3}\gbarc|\lesssim\frac{\varepsilon}{\ubar^{2+\delta/4}},\quad\text{in}\:\:\deux.$$
\end{thm}

We start with a few preliminaries before proving Theorem \ref{thm:controlgbarc}. We define the new convenient set of bootstrap derivatives: 
\begin{align}\label{eq:defdfprimprim}
	\df''=\{\lieT, \nabring, \nabring_{\Ybar}\},
\end{align}
with $\Ybar$ defined as in \eqref{eq:Ybarsadefquoi}. Note the following identities, which are analog to \eqref{eq:recupe4avecdf}, \eqref{eq:recupe3avecdf},
\begin{align*}
	\Ybar&=\left(1+\frac{\Omega^2|\gbar|^2}{4}\right)\Omega^2\wh{e}_3+\frac{\Omega^2|\gbar|^2}{2}\T+\Omega^2\left(\gbar^A+\frac{1}{4}|\gbar|^2b^A\right)\partial_{\theta^A}\\
	&=\left(1+\frac{\Omega^2|\gbar|^2}{4}\right)\wh{e}_4+\left(\Omega^2\gbar^A-b^A\right)\partial_{\theta^A}-2\T.
\end{align*}
\begin{prop}\label{prop:fbarinitdfprimfc}
	Let $\gbarc$ be a solution of \eqref{eq:nlgbarc} on a neighbourhood of $\Sigma_0$, such that the initial condition \eqref{eq:initcondgbarlambdabar} is satisfied. Then we have, on $\Sigma_0$, 
	\begin{align}\label{eq:fbarinitndfprimdfrak}
		|(\df'')^{\leq N_3}\gbarc||_{\Sigma_0}\lesssim\varepsilon\ubar^{-2-\delta/4}.
	\end{align}
\end{prop}
\begin{proof} The proof is the same as the one of Proposition \ref{prop:initdfprimfc}.
\end{proof}

\begin{prop}\label{prop:fbarind1}
	Let $\gbarc$ be a 1-form tangent to the spheres $S(u,\ubar)$ which satisfies \eqref{eq:nlgbarc}. Then we have for any integer $k\geq 0$,
\begin{align}
		(\nabring_{\Ybar})^{k+1}\gbarc&=_s G_k^{(1)},\quad\text{where}:\label{eq:fbarind1}\\
		G_k^{(1)}&=_s\sum_{m=0}^4\sum_{i_1+\cdots+i_{m+2}\leq k,i_1\geq 1}\left(\nabla_{\Ybar}^{i_1}\gbarc\df'^{i_2}\Omega^2,\df'^{i_1}(\psicheck_{\Hbar},\df^{\leq 1}\gcheck),\df'^{i_1}\widecheck{\mcq}\df'^{i_2}\Omega^2\right)\prod_{j=2}^{m+2}\df'^{i_j}(\mcq_\mck,\widecheck{\mcq},\gbarc).\nn
	\end{align}
\end{prop}
\begin{proof}
	The proof is the same as the one of Proposition \ref{prop:ind1}.
	\end{proof}
	\begin{prop}\label{prop:fbarind2}
		let $\gbarc$ be a 1-form tangent to the spheres $S(u,\ubar)$ which satisfies \eqref{eq:nlgbarc}. Then we have for $k,k'\geq 0$,
	\begin{align}    
		\nabring_{\Ybar}(\lieT^{k'}\nabring_{\Ybar}^{k}\gbarc)=_s\lieT^{k'}G_k^{(1)}+\df''^{\leq k'}(\psicheck_{\Hbar},\widecheck{\mcq},\gbarc)^{\leq k'+2}\df''^{\leq k+k'}\gbarc.\nn
	\end{align}

	\end{prop}
\begin{proof} 
The proof is the same as the one of Proposition \ref{prop:ind2}.
\end{proof}
	\begin{prop}\label{prop:fbarind3}
Let $\gbarc$ be a 1-form tangent to the spheres $S(u,\ubar)$ which satisfies \eqref{eq:nlgbarc}. Under the bootstrap assumption $|( \df'')^{\leq N_3}\gbarc|\leq\sqrt{\varepsilon}\ubar^{-2-\delta/4}$, we have the following reduced schematic equation for $k,k',k''\geq 0$ such that $k+k'+k''\leq N$,
	\begin{align}    
		\nabring_{\Ybar}(\nabring^{k''}\lieT^{k'}\nabla_{\Ybar}^k\gbarc)=_{rs}&\:O(\Omega^2)\nabla^{k''-1}\lieT^{k'+1}\nabla_{\Ybar}^k\gbarc+O(\Omega^2)\nabla^{k''}\lieT^{k'}\nabla_{\Ybar}^k\gbarc\nn\\
		&+O(\Omega^2)\df''^{\leq k''+k'+k-1}\gbarc+O(\varepsilon\ubar^{-2-\delta/3}).\label{eq:fbarrecform}
	\end{align}
	\end{prop}
\begin{proof}
	The proof is very similar to the one of Proposition \ref{prop:ind3}, so we omit it.
\end{proof}
	\begin{rem}\label{rem:flowpaststable}
		While the equation \eqref{eq:nlfc} for $\fc$ is a transport equation in the outgoing $Y$ direction, the equation \eqref{eq:nlgbarc} for $\gbarc$ is in the \emph{ingoing} $\Ybar$ direction. This only slightly complicates the improvement of the bootstrap assumptions. Indeed, using the identities $\Ybar(\ubar)\geq 0$ and $\Ybar(u)=1$, we get that regions $\mcr$ of the type  
		$$\deux[\ubar_f],\quad\deux[\ubar_f]\cup(\deux\cap\{\ubar_f\leq\ubar<\ubar_f+k\}\cap\{u< u_0\}),$$
	satisfy that $\forall p\in\mcr$, the past-directed integral curve of $\Ybar$ starting from $p$ remains in $\mcr$ up to $\Sigma_0$. This observation allows for a 2-step improvement of the bootstrap assumptions for $\gbarc$, see the proof of Theorem \ref{thm:controlgbarc} below.
	\end{rem}
	\begin{prop}\label{prop:fbarbootstrapfcdfrak}
		Let $\mcr\subset\deux$ which is stable by the past-directed flow of $\Ybar$ up to $\Sigma_0$. We assume that $\gbarc$ is a $S(u,\ubar)$-tangent 1-form which satisfies \eqref{eq:nlgbarc} and the bootstrap assumption $|(\df'')^{\leq N_3}\gbarc|\leq\sqrt{\varepsilon}\ubar^{-2-\delta/4}$ on $\mcr$, and $|(\df'')^{\leq N_3}\gbarc||_{\Sigma_0}\lesssim\varepsilon\ubar^{-2-\delta/4}$ on $\Sigma_0$.  Then provided $0<\gamma<\frac{\delta}{12}$, there is a constant $C(a,M,C_R)>0$ such that, on $\mcr$,  
		$$|(\df'')^{\leq N_3}\gbarc|\leq C(a,M,C_R)\varepsilon\ubar^{-2-\delta/4}.$$
	\end{prop}
	\begin{proof}
		The proof is basically the same as the one of Proposition \ref{prop:bootstrapfcdfrak}, namely we prove by induction on $N\in\{0,\cdots,N_3\}$ the estimate 
		\begin{align}\label{eq:fbarinductionhypo}
			|(\df'')^{\leq N}\gbarc]|\lesssim\varepsilon\ubar^{-2-\delta/4}.
		\end{align}
		\noindent\textbf{Step 1.}\textit{ The case $N=0$.} We detail the case $N=0$. Recalling \eqref{eq:nlgbarc}, by the bootstrap assumption we get in $\mcr$ the bound $|\nabring_{\Ybar}\gbarc|\lesssim {\varepsilon}{\ubar^{-2-\delta/3}}+\Omega^2|\gbarc|$, where we recall \eqref{eq:Ybarsadefquoi}. Note that the bootstrap assumption implies $|\gbarc|+|\gbar|\lesssim 1$. Thus by Proposition \ref{prop:transporte4'bar}, we get for $p=(u,\ubar,\theta^A)\in\mcr$, integrating from $\Sigma_0$ and recalling \eqref{eq:introcoordflowybar},
		\begin{align*}
			|\gbarc|(p)\lesssim& |\gbarc|(u_{\Ybar,\Sigma_0}(p),\ubar_{\Ybar,\Sigma_0}(p),\theta^A_{\Ybar,\Sigma_0}(p))+\int_{u_{\Ybar,\Sigma_0}(p)}^u\frac{\varepsilon}{(\ubar_{\Ybar}^p(u'))^{2+\delta/3}}\dee u'\\
			&+\int_{u_{\Ybar,\Sigma_0}(p)}^u\Omega^2|\gbarc|(u',\ubar_{\Ybar}^p(u'),(\theta^A)^p_{\Ybar}(u'))\dee u'.
		\end{align*}
		Moreover, we have by \eqref{eq:equivtransporte4'bar} the bound 
		$$\int_{u_{\Ybar,\Sigma_0}(p)}^u\frac{\varepsilon}{(\ubar_{\Ybar}^p(u'))^{2+\delta/3}}\dee u'\lesssim\frac{\varepsilon}{\ubar^{2+\delta/3}}(u-u_{\Ybar,\Sigma_0}(p))\lesssim\frac{\varepsilon}{\ubar^{2+\delta/3-\gamma}}\lesssim\frac{\varepsilon}{\ubar^{2+\delta/4}},$$
		for $0<\gamma<\delta/12$ , where we used $\Sigma_0=\{u+\ubar=C_R\}$, \eqref{eq:equivtransporte4'bar} which yields $|\ubar_{\Ybar,\Sigma_0}(p)-\ubar|\lesssim 1$, and $u+\ubar\leq\ubar^\gamma$ in $\deux$, which yields
		\begin{align*}
			u-u_{\Ybar,\Sigma_0}(p)=u-C_R+\ubar_{\Ybar,\Sigma_0}(p)=\ubar+u+O(1)\leq \ubar^\gamma+O(1)\lesssim\ubar^\gamma.
		\end{align*}
Moreover, the initial data assumption on $\Sigma_0$, combined with \eqref{eq:equivtransporte4'bar}, implies
		$$|\gbarc|(u_{\Ybar,\Sigma_0}(p),\ubar_{\Ybar,\Sigma_0}(p),\theta^A_{\Ybar,\Sigma_0}(p))\lesssim \varepsilon\ubar_{\Ybar,\Sigma_0}(p)^{-2-\delta/4}\lesssim\varepsilon\ubar^{-2-\delta/4},$$
		so that we get
	\begin{align*}
		|\gbarc|(p)&\lesssim\frac{\varepsilon}{\ubar^{2+\delta/4}} +\int_{u_{\Ybar,\Sigma_0}(p)}^u\Omega^2|\gbarc|(u',\ubar_{\Ybar}^p(u'),(\theta^A)^p_{\Ybar}(u'))\dee u'\\
		&\lesssim\frac{\varepsilon}{\ubar^{2+\delta/4}}\exp\left(C(a,M,C_R)\int_{u_{\Ybar,\Sigma_0}(p)}^u\Omega^2(u',\ubar_{\Ybar}^p(u'),(\theta^A)^p_{\Ybar}(u'))\dee u'\right)\lesssim\frac{\varepsilon}{\ubar^{2+\delta/4}},
	\end{align*}
		where we used Grönwall's inequality and the estimate
		$$\Omega^2(u',\ubar_{\Ybar}^p(u'),(\theta^A)^p_{\Ybar}(u'))\sim e^{-|\kappa_-|(u'+\ubar_{\Ybar}^p(u'))}\sim e^{-|\kappa_-|(u'+\ubar(p))},$$
		by \eqref{eq:equivtransporte4'bar}, which implies 
		\begin{align}
			\int_{u_{\Ybar,\Sigma_0}(p)}^u\Omega^2(u',\ubar_{\Ybar}^p(u'),(\theta^A)^p_{\Ybar}(u'))\dee u'\lesssim e^{-|\kappa_-|(\ubar(p)+u_{\Ybar,\Sigma_0}(p))}\lesssim  1,\label{eq:instead}
		\end{align}
		where we used \eqref{eq:equivtransporte4'bar} again. Hence the result for $N=0$.
		\\
		\noindent\textbf{Step 2.}\textit{ The induction step.} This is very similar to Step 2 in the proof of Proposition \ref{prop:bootstrapfcdfrak}, relying on Propositions \ref{prop:fbarind2},  \ref{prop:fbarind3}.
	\end{proof}
	\noindent Recalling the set of derivatives $\df=\{\nabring_3,\Omega^2\nabring_4,\nabring\}$, we are now ready to prove Theorem \ref{thm:controlgbarc}. 
	\begin{proof}[\textbf{Proof of Theorem \ref{thm:controlgbarc}}]
		This is again a standard bootstrap argument, the only difference with the proof of Theorem \ref{thm:controlfc} being that here we deal with a transport equation in the ingoing direction. We define the interval 
		$$\mathcal{I}=\Bigg\{{\ubar_f}> \ubar_{\Sigma_0}(u_f)\text{ such that:}\begin{array}{l}
			\text{there exists a unique }\gbarc\text{ solution of \eqref{eq:nlgbarc} on }\\
			\deux[{\ubar_f}]\text{ with } \gbarc|_{\Sigma_0}=\gbarc_0\text{ and }|(\df'')^{\leq N_3}\gbarc|\leq \sqrt{\varepsilon}\ubar^{-2-\delta/4}
		\end{array}\Bigg\},$$
		which is easily seen to be non-empty for $\varepsilon$ small enough, and closed. To prove that it is open, for ${\ubar_f}\in\mathcal{I}$, by Proposition \ref{prop:fbarbootstrapfcdfrak} we have $|(\df'')^{\leq N_3}\gbarc|\lesssim\varepsilon\ubar^{-2-\delta/4}$ in $\deux[{\ubar_f}]$. Then, by continuity and local existence results for transport equations in the ingoing direction, we get $k,k'>0$ such that $\gbarc$ can be extended to $\mcr'=\deux[\ubar_f]\cup(\deux\cap\{\ubar_f\leq\ubar<\ubar_f+k\}\cap\{u< C_R-\ubar_f+k'\})$ where it satisfies $|(\df'')^{\leq N_3}\gbarc|\lesssim\varepsilon\ubar^{-2-\delta/4}$.
We now do a bootstrap argument in the ingoing direction: let
		$$\mathcal{J}=\Bigg\{k''>0\text{ such that:}\begin{array}{l}
			\text{there exists a unique }\gbarc\text{ solution of \eqref{eq:nlgbarc} on }\\
			\deux[{\ubar_f}]\cup(\deux\cap\{\ubar_f\leq\ubar<\ubar_f+k\}\cap\{u< C_R-\ubar_f+k''\}),\\
			\text{ with } \gbarc|_{\Sigma_0}=\gbarc_0\text{ and }|(\df'')^{\leq N_3}\gbarc|\leq \sqrt{\varepsilon}\ubar^{-2-\delta/4}
		\end{array}\Bigg\}.$$
It is easy to check that $\mathcal{J}$ is non-empty and closed, and also open for $\varepsilon$ small enough as a direct consequence of Proposition \ref{prop:fbarbootstrapfcdfrak} (see also Remark \ref{rem:flowpaststable}) and local existence results for transport equations in the ingoing direction. This yields $\mathcal{J}=(0,+\infty)$, thereby proving that $\mci$ is open and hence $\mci=(\ubar_{\Sigma_0}(u_f),+\infty)$, which concludes the proof.
	\end{proof}
We now establish the theorem on the existence and control of $\widecheck{\lambdabar}$. 
	We assume here that $\gbarc$ is the solution of \eqref{eq:nlgbarc} given by Theorem \eqref{thm:controlgbarc}. 
	
	\begin{thm}\label{thm:controllambdabarcheck}
		Provided $\varepsilon>0$ and $\gamma>0$ are small enough, there exists a unique solution $\lambdabarcheck$ of \eqref{eq:transpolambdabar} in $\deux$ with initial condition given by \eqref{eq:initcondgbarlambdabar} on $\Sigma_0$,  which satisfies moreover
		$$|\df^{\leq N_3}\lambdabarcheck|\lesssim\frac{\varepsilon}{\ubar^{2+\delta/4}},\quad\text{in}\:\:\deux.$$
	\end{thm}
\begin{proof}
The proof is similar to the one of Theorem \ref{thm:controllambdacheck}, once the following observations are made : We can rewrite equation \eqref{eq:transpolambdabar} as
\begin{align}\label{eq:lambdabareqprelim2}
	\Omega^{-2}\Ybar(\lambdabar)+\hhbar\lambdabar=0,
\end{align}
where $\hhbar=\gbar\cdot(\wh{\zeta}+\wh{\eta})+\frac{1}{2}|\gbar|^2\wh{\omega}+\frac{1}{2}\gbar\cdot(\gbar\cdot\wh{\chihat})+\frac14\widehat{tr\chi}|\gbar|^2$. Note that by Theorem \ref{thm:controlgbarc} and the estimates in Theorem \ref{thm:controldoublenull}, we have the estimate  $|\df^{\leq N_3}\widecheck{\hhbar}|\lesssim\varepsilon\ubar^{-2-\delta/4},\:|\df^{\leq N_3}{\hhbar}|\lesssim 1,\: \text{in}\:\deux$. From this point, the proof is a consequence of the same transport and integral estimates as in the proof of Theorem \ref{thm:controlgbarc} (except that it is easier because the equation for $\lambdabar$ is linear and $\lambdabar$ is a scalar), hence we omit it for conciseness.
\end{proof}

\begin{rem}\label{rem:fixgbarlambdabar}
	\textbf{From now on, we fix the values of $\lambdabar$ and $\gbar$ to be 
		$$\lambdabar=\lambdabar_\mck+\lambdabarcheck,\quad \gbar=\gbar_\mck+\gbarc,$$
		where $\lambdabarcheck$ and $\gbarc$ are given by Theorems \ref{thm:controllambdabarcheck} and \ref{thm:controlgbarc}.}
\end{rem}

	\subsection{Outgoing and ingoing non-integrable null frames in $\deux$}
Recall that we denote with a ring the double null frame $(\mathring{e}_\mu)$.
	\subsubsection{ Non-integrable null frames}\label{section:defprincipalframes}
	Let $e_4'$ and $e_3$ be defined by \eqref{eq:e4primdef}, \eqref{eq:e3daggerdef}, with $f,\lambda,\gbar,\lambdabar$ defined in Remarks \ref{rem:fixflambda}, \ref{rem:fixgbarlambdabar}. Then we compute 
	\begin{align*}
		\g(e_3,e_4')=\lambda\lambdabar\left(-2\Omega^{-2}+f\cdot\gbar-\frac18\Omega^2|f|^2|\gbar|^2\right)=-2\lambda\lambdabar\Omega^{-2}\left(1-\frac12\Omega^2f\cdot\gbar+\frac{1}{16}\Omega^4|f|^2|\gbar|^2\right).
	\end{align*}
	We now define the scalar function
	\begin{align}\label{eq:omegahat}
		\hat{\Omega}^2:=\Omega^2\lambda^{-1}\lambdabar^{-1}\left(1-\frac12\Omega^2f\cdot\gbar+\frac{1}{16}\Omega^4|f|^2|\gbar|^2\right)^{-1},
	\end{align}
and the following rescaled versions of $e_4'$ and $e_3$, 
	\begin{align}\label{eq:outtoin}
		e_4:=\hat{\Omega}^2e_4',\quad e_3'=\hat{\Omega}^2e_3,
	\end{align}
so that we have by construction
$$\g(e_3,e_4)=\g(e_3',e_4')=-2.$$
	Now, let $e_a$, $a=1,2$, be any orthonormal basis of $\mch=(e_3',e_4')^{\bot}=(e_3,e_4)^{\bot}$.
	\begin{defi}
		In region $\deux$, we respectively call the null frames 
		$$(e_3,e_4,e_1,e_2),\quad (e_3',e_4',e_1'=e_1,e_2'=e_2)$$
		the \emph{ingoing} and \emph{outgoing} non-integrable null frames. 
	\end{defi}
	These frames are perturbations of the Kerr principal null frames, in the sense that they coincide with \eqref{eq:principalingoinkerr}, \eqref{eq:principaloutgoinkerr} in exact Kerr. Moreover we have by construction
	\begin{align}\label{eq:e4prime3geodesic}
		\D_3 e_3=0,\quad\D_{e_4'}e_4'=0,
	\end{align}
	(see Propositions \ref{prop:eqtranspofetlambda}, \ref{prop:eqgbarlambdabar}), which implies $\xi=\xibar=\xi'=\xibar'=0,$ as well as $\omega'=\omegabar=0.$
	
	\begin{rem}
		Notice that by the initial condition \eqref{eq:initcondgbarlambdabar} for $\gbar$ and $\lambdabar$, and \eqref{eq:asinyo}, we have
		\begin{align}\label{eq:coincidesigma0e3prim}
			e_3|_{\Sigma_0}=e_3^{(\un)}|_{\Sigma_0}.
		\end{align}
		Moreover, by the initial condition \eqref{eq:initcondfclambdac} for $\lambda$ and $f$, we also have $e_4'|_{\Sigma_0}=-\frac{|q|^2}{\Delta}e_4^{(\un)}|_{\Sigma_0}$, on $\Sigma_0$. By \eqref{eq:outtoin}, $e_4$ is thus also proportional to $e_4^{(\un)}$, on $\Sigma_0$. Thus, by \eqref{eq:coincidesigma0e3prim}, and using $\g(e_3,e_4)=-2=\g(e_3^{(\un)},e_4^{(\un)})$, we deduce similarly
	\begin{align}\label{eq:coincidesigma0e4prim}
	e_4|_{\Sigma_0}=e_4^{(\un)}|_{\Sigma_0}.
	\end{align}
	\end{rem}

	\subsubsection{Definition of $\fbar$ via a fixed point argument}
	
	By \eqref{eq:omegahat} and \eqref{eq:e3daggerdef}, we have the following identity
	\begin{align}
		e_3'=\hat{\Omega}^2e_3&=\Omega^2\lambda^{-1}\left(1-\frac12\Omega^2f\cdot\gbar+\frac{1}{16}\Omega^4|f|^2|\gbar|^2\right)^{-1}\left(\Omega^{-2}\ering_3+\gbar^A\partial_{\theta^A}+\frac14|\gbar|^2\Omega^2\ering_4\right)\nn\\
		&=\lambda^{-1}\left(1-\frac12f\cdot\widehat{\gbar}+\frac{1}{16}|f|^2|\widehat{\gbar}|^2\right)^{-1}\left(\ering_3+\widehat{\gbar}^A\partial_{\theta^A}+\frac14|\widehat{\gbar}|^2\ering_4\right),\label{eq:ifholds}
	\end{align}
	where
\begin{align}\label{eq:etretatlesfalaises}
	\widehat{\gbar}:=\Omega^2\gbar,\quad|\df^{\leq N_3}\widecheck{\widehat{\gbar}}|\lesssim\frac{\varepsilon\Omega^2}{\ubar^{2+\delta/4}}\quad\text{in}\:\:\deux,
\end{align}
 where we used Theorem \ref{thm:controldoublenull} which implies $|\df^{\leq N_{max}}\widecheck{\Omega^2}|\lesssim\varepsilon\Omega^2\ubar^{-2-\delta/2}$, and Theorem \ref{thm:controlgbarc}.
\begin{rem}\label{rem:jusquala}
	From here we choose $u_f(a,M,\delta,C_R)\ll -1$ negative enough such that the bounds
	\begin{equation}\label{eq:voilauneborne}
		\begin{gathered}
			|\df^{\leq N_3}\widecheck{\widehat{\gbar}}|\lesssim\frac{\varepsilon\Omega^2}{\ubar^{2+\frac{\delta}{5}}},\quad|{\df}^{\leq N_3}(\fc,\lambdacheck,\lambdabarcheck)|\lesssim\frac{\varepsilon}{\ubar^{2+\frac{\delta}{5}}},\\
			|\df^{\leq N_3}(\gcheck,\bcheck,\psicheck_{\Hbar},\widecheck{\mathring\omegabar},\psicheck,\widecheck{\mathring{K}},\hodge{\widecheck{\mathring{K}}},\widecheck{\mathring{\betabar}},\widecheck{\mathring\alphabar})|\lesssim \frac{\varepsilon}{\ubar^{2+\frac{\delta}{5}}},\\
			|\df^{\leq N_{3}}\psicheck_H|\lesssim \frac{\varepsilon\Omega^{-2}}{\ubar^{2+\frac{\delta}{5}}},\quad |\df^{\leq N_{3}}\widecheck{\mathring{\beta}}|\lesssim \frac{\varepsilon\Omega^{-2}}{\ubar^{2+\frac{\delta}{5}}},\quad |\df^{\leq N_{3}}\widecheck{\mathring\alpha}|\lesssim \frac{\varepsilon\Omega^{-4}}{\ubar^{2+\frac{\delta}{5}}},
		\end{gathered}
	\end{equation}
	hold in $\deux$ with implicit constants which depend only on $a,M$ and \underline{not on $C_R$}. This is possible by Theorems \ref{thm:controldoublenull}, \ref{thm:controlfc}, \ref{thm:controllambdacheck}, \ref{thm:controllambdabarcheck}, and the bound above on $\widecheck{\widehat{\gbar}}$. \textbf{As a consequence, from now on the implicit constants in the bounds $\lesssim$ depend only on $a,M$.}
\end{rem}
	The following result will ensure that we can express the non-integrable null frames with the double null frame using a frame transformation of the type \eqref{eq:frametransfo}.
	\begin{lem}\label{lem:mettrebienlatransfo}
		Assume that $\fbar$ is a $S(u,\ubar)$-tangent 1-form which satisfies
		\begin{align}\label{eq:relfbartildefbar}
			\widehat{\gbar}=\frac{\fbar+\frac{1}{4}|\fbar|^2f}{1+\frac{1}{2}f\cdot\fbar+\frac{1}{16}|f|^2|\fbar|^2}.     \end{align}
		Then we have the identity 
		$$e_3'=\lambda^{-1}\left(\left(1+\frac{1}{2}f\cdot\fbar+\frac{1}{16}|f|^2|\fbar|^2\right)\ering_3+\left(\fbar^A+\frac{1}{4}|\fbar|^2f^A\right)\partial_{\theta^A}+\frac{1}{4}|\fbar|^2\ering_4\right).$$
	\end{lem}
	\begin{proof}
This is a straightforward computation starting from \eqref{eq:ifholds}.
	\end{proof}
	Recall the $S(u,\ubar)$-tangent 1-form $\fbar_\mck$ introduced in Proposition \ref{prop:transfodanskerr}.
	\begin{prop}\label{prop:recupfbarcheck}
		For $\varepsilon(a,M)>0$ small enough and $C_R(a,M)>0$ large enough, there exists a $S(u,\ubar)$-tangent 1-form $\widecheck{\fbar}$ in $\deux$ which satisfies \eqref{eq:relfbartildefbar} where $\fbar=\fbar_\mck+\widecheck{\fbar}$ and such that
		\begin{align}\label{eq:voilabornefbarcheckkk}
			|\df^{\leq N_3}\widecheck{{\fbar}}|\lesssim\frac{\varepsilon\Omega^2}{\ubar^{2+\delta/5}}.
		\end{align}
	\end{prop}

	\begin{proof}
		We have \eqref{eq:relfbartildefbar} if and only if 
		$$\fbar+\frac{1}{4}|\fbar|^2f= \widehat{\gbar}\left(1+\frac{1}{2}f\cdot\fbar+\frac{1}{16}|f|^2|\fbar|^2\right).$$
		subtracting the identity \eqref{eq:eqlienfbarfbartilde} in Kerr, which can be rewritten as 
		$$\fbar_\mck+\frac{1}{4}|\fbar_\mck|_\mck^2f_\mck= \widehat{\gbar}_\mck\left(1+\frac{1}{2}f_\mck\cdot_\mck\fbar_\mck+\frac{1}{16}|f_\mck|_\mck^2|\fbar_\mck|_\mck^2\right),$$
		we get that \eqref{eq:relfbartildefbar} holds if and only if 
		$$\widecheck{\fbar}=\Phi[\widecheck{\fbar}],$$
		where:
		\begin{align*}
			\Phi[\widecheck{\fbar}]:=&\widecheck{ \widehat{\gbar}}\left(1+\frac{1}{2}f_\mck\cdot_\mck\fbar_\mck+\frac{1}{16}|f_\mck|_\mck^2|\fbar_\mck|_\mck^2\right)+ \widehat{\gbar}\left(\frac{1}{2}f_\mck\widecheck{\cdot}\fbar_\mck+\frac{1}{2}\fc\cdot\fbar_\mck+\frac{1}{2}f\cdot\widecheck{\fbar}\right)\nn\\
			&+\frac{1}{16} \widehat{\gbar}\Big[|f_\mck|^2_\mck\left(\fbar_\mck\widecheck{\cdot}\fbar_\mck+2\widecheck{\fbar}\cdot\fbar_\mck+|\widecheck{\fbar}|^2\right)+|\fbar|^2\left(f_\mck\widecheck{\cdot}f_\mck+2\widecheck{f}\cdot f_\mck+|\widecheck{f}|^2\right)\Big]\nn\\
			&-\frac{1}{4}\left(\fbar_\mck\widecheck{\cdot}\fbar_\mck+2\widecheck{\fbar}\cdot\fbar_\mck+|\widecheck{\fbar}|^2\right)f_\mck-\frac{1}{4}|\fbar|^2\widecheck{f}.
		\end{align*}
	The existence and bound for $\widecheck{\fbar}$ then follow from standard fixed-points results, choosing $\varepsilon$ small enough, $C_R$ large enough, and relying on the bounds for $\widecheck{ \widehat{\gbar}}$ in \eqref{eq:etretatlesfalaises}.
	\end{proof}
	\begin{rem}\label{rem:remlabel}
		From now on, we fix $\fbar$ to be 
		$$\fbar=\fbar_\mck+\widecheck{\fbar},$$
		where $\widecheck{\fbar}$ is given by Proposition \ref{prop:recupfbarcheck}. Then, $\fbar$ satisfies \eqref{eq:relfbartildefbar}, so that by \eqref{eq:e4primdef} and Lemma \ref{lem:mettrebienlatransfo}, the outgoing non-integrable frame $(e_3',e_4',e_a')$ is obtained from the double null frame $(\mathring{e}_3,\ering_4,\ering_a)$ by a frame transformation of the type \eqref{eq:frametransfo} with coefficients $f,\fbar,\lambda$ which satisfy 
		\begin{align}\label{eq:bornecoeffout}
			|\df^{\leq N_3}\fc|+|\df^{\leq N_3}\widecheck{\lambda}|\lesssim\frac{\varepsilon}{\ubar^{2+\delta/5}},\quad |\df^{\leq N_3}\widecheck{\fbar}|\lesssim\frac{\varepsilon\Omega^2}{\ubar^{2+\delta/5}},\quad\text{in }\:\deux,
		\end{align}
		where we used \eqref{eq:voilauneborne} and Proposition \ref{prop:recupfbarcheck}. By \eqref{eq:outtoin}, we get that the ingoing non-integrable frame $(e_3,e_4,e_a)$ is obtained from the double null frame $(\mathring{e}_3,\ering_4,\ering_a)$ by a frame transformation of the type \eqref{eq:frametransfo} with coefficients
		\begin{align}\label{eq:deflambdahattt}
			f,\fbar,\hat{\lambda}:=\lambda\hat{\Omega}^2=\Omega^2\lambdabar^{-1}\left(1-\frac12\Omega^2f\cdot\gbar+\frac{1}{16}\Omega^4|f|^2|\gbar|^2\right)^{-1},
		\end{align}
	which satisfy 
		\begin{align}\label{eq:bornecoeffin}
			|\df^{\leq N_3}\fc|\lesssim\frac{\varepsilon}{\ubar^{2+\delta/5}},\quad |\df^{\leq N_3}\widecheck{\fbar}|+|\df^{\leq N_3}\widecheck{\hat{\lambda}}|\lesssim\frac{\varepsilon\Omega^2}{\ubar^{2+\delta/5}},\quad\text{in }\:\deux,
		\end{align}
		where we used \eqref{eq:voilauneborne} to control $\widecheck{\hat{\lambda}}$. Note that by $\hat{\lambda}_\mck\sim\Omega^2_\mck$ in Kerr, this implies $\hat{\lambda}\sim\Omega^2$.
	\end{rem}
	\subsection{Ricci and curvature coefficients for the non-integrable frames in $\deux$}
	\subsubsection{Comparing $S(u,\ubar)$-tangent and horizontal tensors}\label{section:comparerintetnonint}
	Recall the notation for the non-integral distribution  $\mch:=(e_3,e_4)^\bot=(e_3',e_4')^\bot$. Following Remark \ref{rem:phistarpullback}, we define the isometry 
	\begin{align}\label{eq:grosPhi}
		\Phi:\begin{cases}
			TS(u,\ubar)\longrightarrow \mch\\X\longmapsto X+\dfrac{1}{2}\fbar(X)f^A\partial_{\theta^A}+\dfrac{1}{2}\fbar(X)\ering_4+\left(\dfrac{1}{2}f(X)+\dfrac{1}{8}|f|^2\fbar(X)\right)\ering_3
		\end{cases}.
	\end{align}
		We will use $\Phi$ to pass from $S(u,\ubar)$-tangent tensors to horizontal tensors and conversely. This will be useful when comparing the Ricci and curvature components defined with respect to the non-integrable frames to the ones defined for the double null frame, see Section \ref{section:ricciandcurvout}.
	\begin{defi}\label{defi:covariantway}
		Let $U$ be a horizontal $k$-tensor in $\deux$ (with respect to $\mch=(e_3,e_4)^\bot$). Still following Remark \ref{rem:phistarpullback}, we define the $S(u,\ubar)$-tangent $k$-tensor $U^S$ by, for $X_1,\cdots,X_k\in TS(u,\ubar)$,
		$$U^S(X_1,\cdots,X_k)=U(\Phi(X_1),\cdots\Phi(X_k)).$$
		Conversely, if $V$ is a $S(u,\ubar)$-tangent $k$-tensor, we define the horizontal $k$-tensor $V^H$ by, for $X_1^H,\cdots,X_k^H\in \mch$,
		$$V^H(X_1^H,\cdots,X_k^H)=V(\Phi^{-1}(X_1^H),\cdots\Phi^{-1}(X_k^H)).$$
	\end{defi}
	\begin{prop}\label{prop:comparisisom}
		We have, for $U$ a horizontal tensor, and $V$ a $S(u,\ubar)$-tangent tensor,
		$$(U^S)^{H}=U,\quad (V^H)^S=V,\quad |U|=|U^S|,\quad |V|=|V^H|.$$
	\end{prop}
	\begin{proof}
		The first two statements are obvious and the last two statements come from the fact that $\Phi$ is an isometry.
	\end{proof}
	In what follows, we denote
	$$\nabla_3,\quad\nabla_4,\quad\nabla\quad(\text{resp.}\quad \nabla_3',\quad\nabla_4',\quad\nabla'),$$
	the covariant derivatives for horizontal tensors in the ingoing (resp. outgoing) non-integrable frame. Recall the scalar function $\hat{\lambda}\sim\Omega^2$ defined in Remark \ref{rem:remlabel}, and recall that we denote $\nabring_{3,4,a}$ the covariant derivatives of $S(u,\ubar)$ tensors in the double null frame $(\ering_3,\ering_4,\ering_a\in TS(u,\ubar))$.
	
	\begin{prop}\label{prop:changebasisder}
		Let $U$ be a horizontal $k$-tensor in $\deux$.
		We have the identities
		\begin{align}
			(\nabla_{4} U)^S_{b_1\cdots b_k}&=\hat{\lambda}\left(\nabring_4U^S+f\cdot\nabring U^S+\frac{1}{4}|f|^2\nabring_3 U^S\right)_{b_1\cdots b_k}+\sum_{i=1}^k\hat{\mct}^{(4)}_{b_ic}U_{b_1\cdots c\cdots b_k}^S,\\
			(\nabla_{a} U)^S_{b_1\cdots b_k}&=\left(\left(\delta_{ab}+\frac12\fbar_a f_b\right)\nabring_bU^S+\frac12\fbar_a\nabring_4U^S+\left(\frac12f_a+\frac18|f|^2\fbar_a\right)\nabring_3U^S\right)_{b_1\cdots b_k}\nn\\
			&\quad+\sum_{i=1}^k\hat{\mct}_{ab_ic}U_{b_1\cdots c\cdots b_k}^S,\quad a=1,2,\\
			(\nabla_{3} U)^S_{b_1\cdots b_k}&=\hat{\lambda}^{-1}\left(\left(1+f\cdot\fbar+\frac{1}{16}|f|^2|\fbar|^2\right)\nabring_3U^S+\left(\fbar+\frac14|\fbar|^2f\right)\cdot\nabring U^S+\frac{1}{4}|\fbar|^2\nabring_4 U^S\right)_{b_1\cdots b_k}\nn\\
			&\quad+\sum_{i=1}^k\hat{\mct}^{(3)}_{b_ic}U_{b_1\cdots c\cdots b_k}^S,
		\end{align}
		where the $S(u,\ubar)$-tangent tensors $\hat{\mct}^{(4)}_{bc}$, $\hat{\mct}_{abc}$,  $\hat{\mct}^{(3)}_{bc}$ are given by 
		\begin{align*}
			\hat{\mct}^{(4)}_{bc}&=\hat{\lambda}\left(\mct^{(4)}_{bc}+f^a \mct_{abc}+\frac14|f|^2\mct^{(3)}_{bc}\right),\\
			\hat{\mct}_{abc}&=\mct_{abc}+\frac12\fbar_af^d\mct_{dbc}+\frac12\fbar_a\mct^{(4)}_{bc}+\left(\frac12 f_a+\frac18|f|^2\fbar_a\right)\mct^{(3)}_{bc},\\
			\hat{\mct}^{(3)}_{bc}&=\hat{\lambda}^{-1}\left(\left(1+f\cdot\fbar+\frac{1}{16}|f|^2|\fbar|^2\right)\mct^{(3)}_{bc}+\left(\fbar^a+\frac14|\fbar|^2f^a\right)\mct_{abc}+\frac{1}{4}|\fbar|^2\mct^{(4)}_{bc}\right),
		\end{align*}
where the tensors $\mct^{(4)}_{bc}$, $\mct_{abc}$,  $\mct^{(3)}_{bc}$ for general coefficients $(\lambda,f,\fbar)$ are defined in Propositions \ref{prop:diffchristo}, \ref{prop:diffchristo4}, \ref{prop:diffchristo3} (which correspond in this context to $S(u,\ubar)$-tangent tensors).
	\end{prop}
	\begin{proof}This is a direct application of Proposition \ref{prop:changebasisdergeneral}, see Remark \ref{rem:remlabel}.
	\end{proof}

\begin{defi}
	We define the following set of renormalized derivatives in $\deux$ acting on horizontal tensors, 
	\begin{align}\label{eq:dfhorizontal}
		\df=\{\nabla,\hat{\lambda}\nabla_3,\nabla_4\}.
	\end{align}
\end{defi}
\begin{rem}We remark the following :
	\begin{itemize}
		\item Note that in \eqref{eq:dfhorizontal}, we normalize $\nabla_3$ with the factor $\hat{\lambda}\sim\Omega^2$ because the $\nabla_3$ derivative presents a $\Omega^{-2}$ degeneracy when applied to linearized quantities (because $(e_3,e_4)$ is a perturbation of the Kerr \emph{ingoing} principal null pair).
		\item Note that we use the same notation $\df$ for derivatives in \eqref{eq:dfhorizontal} acting on horizontal tensors, and derivatives as defined in \eqref{eq:dfdoublenull} acting on $S(u,\ubar)$-tangent tensors. There is no risk on confusion as it will be clear in what follows wether we apply $\df$ derivatives on horizontal or on $S(u,\ubar)$-tangent tensors. Also note that by the frame transformation with coefficients $(f,\fbar,\hat{\lambda})$ in Remark \ref{rem:remlabel}, we get that $(\hat{\lambda}e_3,e_4,e_a)$ can be expressed with respect to $(\ering_3,\Omega^2 \ering_4,\partial_{\theta^A})$ with bounded coefficients, and conversely, see Proposition \ref{prop:855}.
	\end{itemize}
	 
\end{rem}
	\begin{prop}\label{prop:855}
		Let $U$ be a horizontal tensor. We have, schematically, for $k\leq N_3$,
		$$(\df^{\leq k}U)^S=_{rs}\df^{\leq k}U^S,$$
		where on the RHS the derivatives acting on $U^S$ are the ones defined in \eqref{eq:dfdoublenull}.
	\end{prop}
	\begin{proof}
		The proof relies on Proposition \ref{prop:changebasisder}. Let us highlight the case $k=1$. It is a direct consequence of Proposition \ref{prop:changebasisder} and the estimates \eqref{eq:bornecoeffin} combined with \eqref{eq:voilauneborne}, which imply
		$$|\hat{\lambda}\hat{\mct}^{(3)}|+|\hat{\mct}^{(4)}|+|\hat{\mct}|\lesssim 1,$$
		hence $(\df^{\leq 1}U)^S=_{rs}\df^{\leq 1}U^S$. The estimate for general $k\leq N_3$ is proven similarly by induction, relying on the bounds \eqref{eq:bornecoeffin} combined with Theorem \ref{thm:controldoublenull} which imply
		$$|\df^{\leq N_3}(\hat{\lambda}\hat{\mct}^{(3)},\hat{\mct}^{(4)},\hat{\mct})|\lesssim 1,$$
which concludes the proof.
	\end{proof}
	\begin{cor}\label{cor:comparder}
		Let $U$ be a horizontal tensor. We have for $k\leq N_3$,
		$$|\df^{\leq k}U|\lesssim|\df^{\leq k}U^S|.$$
	\end{cor}
	\begin{proof}
		We simply write $|\df^{\leq k}U|\lesssim|(\df^{\leq k}U)^S|\lesssim|\df^{\leq k}U^S|$,
		by Propositions \ref{prop:855} and \ref{prop:comparisisom}.
	\end{proof}

	\subsubsection{Control of non-integrable outgoing Ricci and curvature coefficients}\label{section:ricciandcurvout}
	
	We denote respectively $\xi',\xibar',tr\chi',\atrchi',\chihat',tr\chibar',\atrchibar',\wh{\chibar}',\zeta',\eta',\etabar',\omega',\omegabar'$, and $\alpha',\alphabar',\beta',\betabar',\rho',\hodge{\rho}'$ the Ricci and curvature coefficients as defined in Section \ref{section:ricciandcurvdef} with respect to the outgoing non-integrable frame $(e_\mu')$ (see Section \ref{section:defprincipalframes}). We also denote
	$$\Xi',\Xibar',trX',\wh{X}',tr\Xbar',\wh{\Xbar}',Z',H',\Hbar',\quad\text{and}\quad A',\Abar',B',\Bbar',P',$$
	the corresponding complex Ricci and curvature coefficients defined in Section \ref{section:ricciandcurvcomplex}. Recalling Definition \ref{defi:covariantway}, we define the following linearized quantities: 
	\begin{align*}
		&\widecheck{tr\chi'}=tr\chi'-tr\chi'_\mck(u,\ubar,\theta^A),\quad \widecheck{\atrchi'}=\atrchi'-\atrchi'_\mck(u,\ubar,\theta^A),\\
		&\widecheck{tr\chibar'}=tr\chibar'-tr\chibar'_\mck(u,\ubar,\theta^A),\quad \widecheck{\atrchibar'}=\atrchibar'-\atrchibar'_\mck(u,\ubar,\theta^A),\\
		&\widecheck{\zeta'}=\zeta'-(\zeta'^S_\mck)^H,\quad \widecheck{\eta'}=\eta'-(\eta'^S_\mck)^H,\quad \widecheck{\etabar'}=\etabar'-(\etabar'^S_\mck)^H,\\
		&\widecheck{\omegabar'}=\omegabar'-\omegabar'_\mck(u,\ubar,\theta^A),\quad \widecheck{\rho'}=\rho'-\rho'_\mck(u,\ubar,\theta^A),\quad \widecheck{\hodge{\rho}'}=\hodge{\rho}'-\hodge{\rho}'_\mck(u,\ubar,\theta^A),
	\end{align*}
	where $tr\chi'_\mck,\atrchi'_\mck,tr\chibar'_\mck,\atrchibar'_\mck,\omegabar'_\mck,\rho'_\mck,\hodge{\rho}'_\mck$ are the Kerr values of the corresponding coefficients pulled back to $\deux$ using the double null coordinates $(u,\ubar,\theta^A)$, and where 
	$$\zeta'^S_\mck=(\zeta'^S_\mck)_B(u,\ubar,\theta^A)\dee\theta^B,\quad \eta'^S_\mck=(\eta'^S_\mck)_B(u,\ubar,\theta^A)\dee\theta^B,\quad\etabar'^S_\mck=(\etabar'^S_\mck)_B(u,\ubar,\theta^A)\dee\theta^B,$$
	where $(\zeta'^S_\mck)_B(u,\ubar,\theta^A),(\eta'^S_\mck)_B(u,\ubar,\theta^A),(\etabar'^S_\mck)_B(u,\ubar,\theta^A)$ are the $\theta^B$-coefficients of the exact Kerr $S(u,\ubar)$-tangent 1-forms $\zeta'^S_\mck,\eta'^S_\mck,\etabar'^S_\mck$, pulled back to $\deux$ using the double null coordinates $(u,\ubar,\theta^A)$. We also define in the obvious way the following complex versions of the linearized quantities,
	$$\widecheck{trX'},\quad \widecheck{tr\Xbar'},\quad \widecheck{Z'},\quad\widecheck{H'},\quad\widecheck{\Hbar'}. $$
	Recalling $\df=\{\nabla,\hat{\lambda}\nabla_3,\nabla_4\}$, the goal of this section is to prove the following result.
	\begin{prop}\label{prop:coeffout}
		Defining $N_4=N_3-1$, provided $C_R(a,M)$ is large enough, we have the following identities and bounds in $\deux$ for the outgoing non-integrable Ricci coefficients,
		\begin{align}
			&\Xi'=0,\quad\Xibar'=0,\quad\omega'=0,\label{eq:xiprimxibarprimomegaprimzero}\\
			&|\df^{\leq N_4}\widecheck{trX'}|+|\df^{\leq N_4}\wh{X}'|\lesssim\varepsilon\Omega^{-2}\ubar^{-2-\delta/5},\label{eq:boundtrxchihat}\\
			&|\df^{\leq N_4}\widecheck{tr\Xbar'}|+|\df^{\leq N_4}\wh{\Xbar}'|\lesssim\varepsilon\ubar^{-2-\delta/5},\label{eq:boundtrxbarchibarhat}\\
			&|\df^{\leq N_4}\widecheck{Z'}|+|\df^{\leq N_4}\widecheck{H'}|+|\df^{\leq N_4}\widecheck{\Hbar'}|\lesssim\varepsilon\ubar^{-2-\delta/5},\label{eq:boundzhhbar}\\
			&|\df^{\leq N_4}\widecheck{\omegabar'}|\lesssim\varepsilon\ubar^{-2-\delta/5}.\label{eq:boundomegabarprim}
		\end{align}
		Also, we have the following bounds for ${\atrchi'}$, ${\atrchibar'}$ in $\deux$,
		\begin{align}\label{eq:borneantitraces}
			|{\atrchi'}|\lesssim 1,\quad |{\atrchibar'}|\lesssim \Omega^2.
		\end{align}
		Moreover, we have the following bounds in $\deux$ for the curvature coefficients,
		\begin{align}
			&|\df^{\leq N_4}A'|\lesssim \varepsilon\Omega^{-4}\ubar^{-2-\delta/5},\quad |\df^{\leq N_4}\Abar'|\lesssim \varepsilon\ubar^{-2-\delta/5},\label{eq:boundalphaprim}\\
			&|\df^{\leq N_4}B'|\lesssim \varepsilon\Omega^{-2}\ubar^{-2-\delta/5},\quad |\df^{\leq N_4}\Bbar'|\lesssim \varepsilon\ubar^{-2-\delta/5},\label{eq:boundbbbarprim}\\
			&|\df^{\leq N_4}\widecheck{P}|\lesssim\varepsilon\Omega^{-2}\ubar^{-2-\delta/5}\label{eq:boundP}.
		\end{align}
	\end{prop}

\begin{proof}The identities \eqref{eq:xiprimxibarprimomegaprimzero} follows directly from the fact that $e_4'$ and $e_3=\hat{\Omega}^{-2}e_3'$ are geodesic, see \eqref{eq:e4prime3geodesic}. We now prove the bounds \eqref{eq:boundtrxchihat}--\eqref{eq:boundP}. Before that, we note that the terms $\mathcal{G}_{[p]},\mathcal{R}_{[p]}$ which appear on the RHS of the change of frame formulas are lower-order terms (with respect to powers of $\Omega^2$) which systematically behave better than the other terms which appear in this proof. This can easily be checked by their definitions \eqref{eq:schematicG}, \eqref{eq:schematicR} and the bounds \eqref{eq:voilauneborne} and \eqref{eq:voilabornefbarcheckkk}.
	
	\noindent\textbf{Control of $\widecheck{trX'},\wh{X}'$.} By the change of frame formulas \eqref{eq:changetrchi} and \eqref{eq:changeatrchi} we get
$$trX'=_s\frac{\lambda}{1+\frac14(f\cdot\fbar-i f\wedge\fbar)}\left(\mathring{tr\chi},(f,\Omega^{-2}\fbar)^{\leq 3}\df f,f\psi,f\fbar\mathring{tr\chi},\mathring{\omegabar}\fbar^2,f^2\mathring{tr\chibar},\mathcal{G}_{[+1]}\right),$$
where we used \eqref{eq:bornecoeffout} which implies that provided $C_R(a,M)$ is large enough, the scalar $1+\frac14(f\cdot\fbar-i f\wedge\fbar)\neq 0$ does not vanish in $\deux$, and where we used that by \eqref{eq:frametransfo},
$$\diver' f=\mathring{\diver}f+\frac12\fbar\cdot(f\cdot\nabring f)+\frac12\fbar\cdot\nabring_4 f+\left(\frac12 f+\frac18|f|^2\fbar\right)\cdot\nabring_3 f=_s(f,\Omega^{-2}\fbar)^{\leq 3}\df f,$$
and similarly for $\curl 'f$. Subtracting the analog identity in Kerr we deduce 
$$|\df^{\leq N_4}\widecheck{trX'}|\lesssim{\varepsilon}\ubar^{-2-\delta/5}+|\df^{\leq N_3}\psicheck_H|\lesssim\varepsilon\Omega^{-2}\ubar^{-2-\delta/5}$$
in $\deux$, where we used \eqref{eq:voilauneborne} and \eqref{eq:voilabornefbarcheckkk} in the first step and \eqref{eq:voilauneborne} again in the second step. Note that here, to control the derivatives of Kerr values of $\lambda,f,\fbar$ and of the double null quantities, we also use \eqref{eq:bornesdanskerr} and \eqref{rem:deriveesdanskerrr}, as well as Proposition \ref{prop:linderdoublenull} for the linearization of derivatives. These bounds are also used without explicit reference in the rest of the proof. Next, by Corollary \ref{cor:comparder} and the change of frame formula \eqref{eq:changechihat} we get 
	\begin{align*}
		|\df^{\leq N_4}{\wh{\chi}'}|\lesssim |\df^{\leq N_4}(\wh{\chi}')^S|=|\df^{\leq N_4}((\wh{\chi}')^S-(\wh{\chi}')^S_\mck)|\lesssim\frac{\varepsilon}{\ubar^{2+\delta/5}}+|\df^{\leq N_3}\psicheck_H|+\Omega^2|\df^{\leq N_4}\widecheck{trX'}|\lesssim\frac{\varepsilon\Omega^{-2}}{\ubar^{2+\delta/5}}
	\end{align*}
	where we used the identity $(\wh{\chi}')^S_\mck=0$ in Kerr (see \eqref{eq:tracesxxbarkerrprim}), and where we also absorbed the term $\fbar\hot(f\cdot\wh{\chi}')$ on the RHS of \eqref{eq:changechihat} to the LHS using \eqref{eq:voilabornefbarcheckkk}, concluding the proof of \eqref{eq:boundtrxchihat}. Also, note that the first bound in \eqref{eq:borneantitraces} is a direct consequence of \eqref{eq:changeatrchi} and $\mathring{\atrchi}=0$ (see \eqref{eq:trucsnulendoublenul}) and the bounds \eqref{eq:voilauneborne}, \eqref{eq:voilabornefbarcheckkk}.
	
	\noindent\textbf{Control of $\widecheck{tr\Xbar'},\wh{\Xbar}'$. }The proof of \eqref{eq:boundtrxbarchibarhat} is similar to the one of \eqref{eq:boundtrxchihat}. More precisely using the change of frame formulas \eqref{eq:changeatrchibar}, \eqref{eq:changetrchibar} and \eqref{eq:changechibarhat} we get
	\begin{align*}
		|\df^{\leq N_4}\widecheck{tr\Xbar'}|&\lesssim {\varepsilon\Omega^2}{\ubar^{-2-\delta/5}}+|\df^{\leq N_3}\psicheck_{\Hbar}|+\Omega^4|\df^{\leq N_4}\widecheck{trX'}|\lesssim{\varepsilon}{\ubar^{-2-\delta/5}},\\
		|\df^{\leq N_4}\wh{\Xbar}'|&\lesssim 
		{\varepsilon\Omega^2}{\ubar^{-2-\delta/5}}+|\df^{\leq N_3}\psicheck_{\Hbar}|+\Omega^4|\df^{\leq N_4}\wh{X}'|\lesssim
		{\varepsilon}{\ubar^{-2-\delta/5}},
	\end{align*}
	where we used the bounds \eqref{eq:bornesdanskerr} and \eqref{rem:deriveesdanskerrr} combined with \eqref{eq:boundtrxchihat}. Moreover, the second bound in \eqref{eq:borneantitraces} is a direct consequence of \eqref{eq:changeatrchibar} and $\mathring{\atrchibar}=0$ (see \eqref{eq:trucsnulendoublenul}) and the bounds \eqref{eq:voilauneborne}, \eqref{eq:voilabornefbarcheckkk}.

	\noindent\textbf{Control of $\widecheck{Z'},\widecheck{H'},\widecheck{\Hbar'}$.} Similarly as before, by the change of frame formulas \eqref{eq:changezeta}, \eqref{eq:changeeta}, \eqref{eq:changeetabar} and \eqref{eq:voilauneborne}, \eqref{eq:voilabornefbarcheckkk} we get
	\begin{align*}
		&|\df^{\leq N_4}\widecheck{\zeta'}|\lesssim |\df^{\leq N_4}\widecheck{\zeta'^S}|\lesssim{\varepsilon}{\ubar^{-2-\delta/5}}+\Omega^2|\df^{\leq N_4}(\wh{X}',\widecheck{trX'})|\lesssim{\varepsilon}{\ubar^{-2-\delta/5}},\\
		&|\df^{\leq N_4}\widecheck{\eta'}|\lesssim |\df^{\leq N_4}\widecheck{\eta'^S}|\lesssim
		{\varepsilon}{\ubar^{-2-\delta/5}}+\Omega^2|\df^{\leq N_4}\widecheck{\eta'}|\lesssim{\varepsilon}{\ubar^{-2-\delta/5}},\quad|\df^{\leq N_4}\widecheck{\etabar'}|\lesssim |\df^{\leq N_4}\widecheck{\etabar'^S}|\lesssim
		{\varepsilon}{\ubar^{-2-\delta/5}}.
	\end{align*}
	Note that here we used also the estimate $\lambda\sim 1$, which comes from \eqref{eq:bornecoeffout} and $\lambda_\mck\sim 1$, and which implies
	$$|\widecheck{\lambda^{-1}}|\lesssim|\widecheck{\lambda}|\lesssim\varepsilon\ubar^{-2-\delta/2}.$$
	We also absorbed in the bound for $\widecheck{\eta'}$ to the LHS the terms $\Omega^2|\df^{\leq N_4}\widecheck{\eta'}|$ for $C_R(a,M)$ large enough. This proves \eqref{eq:boundzhhbar}.
	
	\noindent\textbf{Control of $\widecheck{\omegabar'}$.} By the change of frame formula \eqref{eq:changeomegabar} we get similarly as before 
	$$|\df^{\leq N_4}\widecheck{\omegabar'}|\lesssim{\varepsilon\Omega^2}{\ubar^{-2-\delta/5}}+|\df^{\leq N_3}(\widecheck{\mathring{\omegabar}},\lambdacheck)|+\Omega^4|\df^{\leq N_4}\widecheck{\eta'}|\lesssim{\varepsilon}{\ubar^{-2-\delta/5}},$$
	\noindent\textbf{Control of $A',\Abar'$.} By the change of frame formulas \eqref{eq:changealpha}, \eqref{eq:changealphabar} we get 
	\begin{align*}
		|\df^{\leq N_4}A'|&\lesssim|\df^{\leq N_4}\widecheck{A'^S}|\lesssim |\df^{\leq N_4}\widecheck{\mathring{\alpha}}|+{\Omega^{-2}\varepsilon}{\ubar^{-2-\delta/5}}\lesssim{\Omega^{-4}\varepsilon}{\ubar^{-2-\delta/5}},\\
		|\df^{\leq N_4}\Abar'|&\lesssim|\df^{\leq N_4}\widecheck{\Abar'^S}|\lesssim |\df^{\leq N_4}\widecheck{\mathring{\alphabar}}|+{\Omega^{2}\varepsilon}{\ubar^{-2-\delta/5}}\lesssim{\varepsilon}{\ubar^{-2-\delta/5}}.
	\end{align*}
	\noindent\textbf{Control of $B',\Bbar'$.} By the change of frame formulas \eqref{eq:changebeta}, \eqref{eq:changebetabar} we get
	\begin{align*}
		|\df^{\leq N_4}B'|&\lesssim|\df^{\leq N_4}\widecheck{B'^S}|\lesssim |\df^{\leq N_4}(\widecheck{\mathring{\beta}},\widecheck{\mathring{\rho}},\widecheck{\hodge{\mathring{\rho}}})|+\Omega^2|\df^{\leq N_4}\widecheck{\mathring{\alpha}}|+{\varepsilon}{\ubar^{-2-\delta/5}}\lesssim{\Omega^{-2}\varepsilon}{\ubar^{-2-\delta/5}},\\
		|\df^{\leq N_4}\Bbar'|&\lesssim|\df^{\leq N_4}\widecheck{\Bbar'^S}|\lesssim |\df^{\leq N_4}(\widecheck{\mathring{\betabar}},\widecheck{\mathring{\alpha}})|+\Omega^2|\df^{\leq N_4}(\widecheck{\mathring{\rho}},\widecheck{\hodge{\mathring{\rho}}})|+{\Omega^2\varepsilon}{\ubar^{-2-\delta/5}}\lesssim{\varepsilon}{\ubar^{-2-\delta/5}}.
	\end{align*}
	\noindent\textbf{Control of $\widecheck{P'}$.} By the change of frame formulas \eqref{eq:changerho}, \eqref{eq:changerhostar} we have
	\begin{align*}
		|\df^{\leq N_4}\widecheck{\rho'}|&\lesssim |\df^{\leq N_4}\widecheck{\mathring{\rho}}|+{\varepsilon}{\ubar^{-2-\delta/5}}\lesssim {\varepsilon\Omega^{-2}}{\ubar^{-2-\delta/5}},\\
		|\df^{\leq N_4}\widecheck{\hodge{\rho}'}|&\lesssim |\df^{\leq N_4}\widecheck{\hodge{\mathring{\rho}}}|+{\varepsilon}{\ubar^{-2-\delta/5}}\lesssim {\varepsilon\Omega^{-2}}{\ubar^{-2-\delta/5}},
	\end{align*}
	which concludes the proof.
\end{proof}

	\subsubsection{Control of non-integrable ingoing Ricci and curvature coefficients}
	We denote respectively $\xi,\xibar,tr\chi,\atrchi,\chihat,tr\chibar,\atrchibar,\wh{\chibar},\zeta,\eta,\etabar,\omega,\omegabar$ and $\alpha,\alphabar,\beta,\betabar,\rho,\hodge{\rho}$ the Ricci and curvature coefficients defined in Section \ref{section:ricciandcurvdef} with respect to the ingoing non-integrable frame $(e_\mu)$ (see Section \ref{section:defprincipalframes}). We also denote
	$$\Xi,\Xibar,trX,\wh{X},tr\Xbar,\wh{\Xbar},Z,H,\Hbar,\quad\quad A,\Abar,B,\Bbar,P,$$
	the corresponding complex Ricci and curvature coefficients defined in Section \ref{section:ricciandcurvcomplex}. We define the following linearized quantities: 
	\begin{align*}
		&\widecheck{tr\chi}=tr\chi-tr\chi_\mck(u,\ubar,\theta^A),\quad \widecheck{\atrchi}=\atrchi-\atrchi_\mck(u,\ubar,\theta^A),\\
		&\widecheck{tr\chibar}=tr\chibar-tr\chibar_\mck(u,\ubar,\theta^A),\quad \widecheck{\atrchibar}=\atrchibar-\atrchibar_\mck(u,\ubar,\theta^A),\\
		&\widecheck{\zeta}=\zeta-(\zeta^S_\mck)^H,\quad \widecheck{\eta}=\eta-(\eta^S_\mck)^H,\quad \widecheck{\etabar}=\etabar-(\etabar^S_\mck)^H,\\
		&\widecheck{\omegabar}=\omegabar-\omegabar_\mck(u,\ubar,\theta^A),\quad \widecheck{\rho}=\rho-\rho_\mck(u,\ubar,\theta^A),\quad \widecheck{\hodge{\rho}}=\hodge{\rho}-\hodge{\rho}_\mck(u,\ubar,\theta^A),
	\end{align*}
	where $tr\chi_\mck,\atrchi_\mck,tr\chibar_\mck,\atrchibar_\mck,\omegabar_\mck,\rho_\mck,\hodge{\rho}_\mck$ are the exact Kerr values of the corresponding coefficients pulled back to perturbed Kerr using the double null coordinates $(u,\ubar,\theta^A)$, and where 
	$$\zeta^S_\mck=(\zeta^S_\mck)_B(u,\ubar,\theta^A)\dee\theta^B,\quad \eta^S_\mck=(\eta^S_\mck)_B(u,\ubar,\theta^A)\dee\theta^B,\quad\etabar^S_\mck=(\etabar^S_\mck)_B(u,\ubar,\theta^A)\dee\theta^B,$$
	where $(\zeta^S_\mck)_B(u,\ubar,\theta^A),(\eta^S_\mck)_B(u,\ubar,\theta^A),(\etabar^S_\mck)_B(u,\ubar,\theta^A)$ are the $\theta^B$-coefficients of the exact Kerr $S(u,\ubar)$-tangent 1-forms $\zeta^S_\mck,\eta^S_\mck,\etabar^S_\mck$, pulled back to perturbed Kerr using the double null coordinates $(u,\ubar,\theta^A)$. We also define in the obvious way the following complexified quantities $\widecheck{trX},\: \widecheck{tr\Xbar},\: \widecheck{Z},\:\widecheck{H},\:\widecheck{\Hbar}$. Recalling $\df=\{\nabla,\hat{\lambda}\nabla_3,\nabla_4\}$, the goal of this section is to prove the following result.
	\begin{prop}\label{prop:coeffin}
		We have the following identities and bounds in $\deux$ for the ingoing non-integrable Ricci coefficients,
		\begin{align}
			&\Xi=0,\quad\Xibar=0,\quad\omegabar=0,\label{eq:xiprimxibarprimomegaprimzeroin}\\
			&|\df^{\leq N_4}\widecheck{trX}|+|\df^{\leq N_4}\wh{X}|\lesssim\varepsilon\ubar^{-2-\delta/5},\label{eq:boundtrxchihatin}\\
			&|\df^{\leq N_4}\widecheck{tr\Xbar}|+|\df^{\leq N_4}\wh{\Xbar}|\lesssim\varepsilon\Omega^{-2}\ubar^{-2-\delta/5},\label{eq:boundtrxbarchibarhatin}\\
			&|\df^{\leq N_4}\widecheck{Z}|+|\df^{\leq N_4}\widecheck{H}|+|\df^{\leq N_4}\widecheck{\Hbar}|\lesssim\varepsilon\ubar^{-2-\delta/5},\label{eq:boundzhhbarin}\\
			&|\df^{\leq N_4}\widecheck{\omega}|\lesssim\varepsilon\ubar^{-2-\delta/5}.\label{eq:boundomegabarprimin}
		\end{align}
		Also, we have the following bounds for ${\atrchi}$, ${\atrchibar}$ in $\deux$,
		\begin{align}\label{eq:borneantitracesin}
			|{\atrchi}|\lesssim \Omega^2,\quad |{\atrchibar}|\lesssim 1.
		\end{align}
		Finally, we have the following bounds in $\deux$ for the ingoing non-integrable curvature coefficients,
		\begin{align}
			&|\df^{\leq N_4}A|\lesssim \varepsilon\ubar^{-2-\delta/5},\quad |\df^{\leq N_4}\Abar|\lesssim\varepsilon\Omega^{-4} \ubar^{-2-\delta/5},\label{eq:boundalphaprimin}\\
			&|\df^{\leq N_4}B|\lesssim \varepsilon\ubar^{-2-\delta/5},\quad |\df^{\leq N_4}\Bbar|\lesssim \varepsilon\Omega^{-2}\ubar^{-2-\delta/5},\label{eq:boundbbbarprimin}\\
			&|\df^{\leq N_4}\widecheck{P}|\lesssim\varepsilon\Omega^{-2}\ubar^{-2-\delta}\label{eq:boundPin}.
		\end{align}
	\end{prop}
	\begin{proof}
		The identities in \eqref{eq:xiprimxibarprimomegaprimzeroin} follow directly from the fact that $e_3$ and $e_4'=\hat{\Omega}^{-2}e_4$ are geodesic, see \eqref{eq:e4prime3geodesic}. We now prove the bounds \eqref{eq:boundtrxchihatin}--\eqref{eq:boundPin}. To this end, notice that the ingoing non-integrable frame $(e_\mu)$ can be obtained from the outgoing non-integrable frame $(e_\mu')$ through a frame transformation \eqref{eq:frametransfo} with coefficients
		\begin{align}\label{eq:coeffnuls}
			f=0,\quad\fbar=0,\quad\lambda=\hat{\Omega}^2,
		\end{align}
		see \eqref{eq:omegahat} and \eqref{eq:outtoin}. By the controls of $\widecheck{\lambda},\widecheck{\lambdabar},\widecheck{f},\widecheck{\gbar}$ and $\widecheck{\Omega^2}$ in \eqref{eq:voilauneborne} we get
		\begin{align}\label{eq:aswellesomegahat}
			|\df^{\leq N_3}\widecheck{\hat{\Omega}}^2|\lesssim\varepsilon\Omega^2\ubar^{-2-\delta/5}.
		\end{align}
		Thus, by the change of frame formulas \eqref{eq:changexi} to \eqref{eq:changerhostar} with coefficients given by \eqref{eq:coeffnuls}, the bounds \eqref{eq:boundtrxchihatin} to \eqref{eq:boundPin} - except the ones for $Z,\omega$ - are direct consequences of the bounds \eqref{eq:boundtrxbarchibarhat} to \eqref{eq:boundP}. Thus, it is only left to prove the bounds for $\widecheck{Z},\widecheck{\omega}$ in \eqref{eq:boundzhhbarin} and \eqref{eq:boundomegabarprimin}.
		
		\noindent\textbf{Control of $\widecheck{Z}$.} The change of frame formula \eqref{eq:changezeta} with coefficients \eqref{eq:coeffnuls} writes $\zeta={\zeta'}-\hat{\Omega}^{-2}\nabla\hat{\Omega}^2$ which proves the stated bounds on $Z$ by \eqref{eq:aswellesomegahat} and \eqref{eq:boundzhhbar}.
		
		\noindent\textbf{Control of $\widecheck{\omega}$.} By the change of frame formula \eqref{eq:changeomega}, and \eqref{eq:xiprimxibarprimomegaprimzero}, we get
		$$\omega=-\frac{1}{2}e_4'(\hat{\Omega}^2)=-\frac12\Omega^2e_4'(W)-\frac12\Omega^2We_4'(\log\Omega),$$
		where, recalling $\widehat{\gbar}=\Omega^2\gbar$, $W:=\lambda^{-1}\lambdabar^{-1}\left(1-\frac12f\cdot\widehat{\gbar}+\frac{1}{16}|f|^2|\widehat{\gbar}|^2\right)^{-1}\sim 1$ satisfies $|\df^{\leq N_3}\widecheck{W}|\lesssim\varepsilon\ubar^{-2-\delta/5}$ and is such that $\hat{\Omega}^2=W\Omega^2$. Thus by \eqref{eq:voilauneborne} and noticing that 
		$$\Omega^2 e_4'=\lambda\left(\Omega^2 \ering_4+\Omega^2f^A\partial_{\theta^A}+\frac{\Omega^2|f|^2}{4}\ering_3\right)=_s\lambda(\df+f \Omega^2\df+f^2\Omega^2\df),$$
		 we deduce $|\df^{\leq N_4}\widecheck{\omega}|\lesssim\varepsilon\ubar^{-2-\delta/5}$, which proves \eqref{eq:boundomegabarprimin}, and concludes the proof.
	\end{proof}
	\begin{rem}\label{rem:omegapos}
		In particular, provided $C_R(a,M)$ is large enough, by the identity
		$$\omega_\mck=\frac{M-r_\mck}{|q_\mck|^2}+\frac{r_\mck\Delta_\mck}{|q_\mck|^4}\gtrsim 1,\quad \text{in}\:\:\deux,$$
		the bound \eqref{eq:boundomegabarprimin} implies the estimate
		\begin{align}\label{eq:omegapos}
			\omega\gtrsim 1,\quad\text{in}\:\:\deux.
		\end{align}
	\end{rem}
	
	\begin{cor}\label{cor:coeffinsanscheck}
		We have the following bounds in $\deux$,
		$$|trX|\lesssim\Omega^2+\frac{\varepsilon}{\ubar^{2+\delta/5}},\quad |tr\Xbar|\lesssim 1+\frac{\varepsilon\Omega^{-2}}{\ubar^{2+\delta/5}},\quad|\df^{\leq N_4}H|+|\df^{\leq N_4}\Hbar|\lesssim 1.$$
	\end{cor}
	\begin{proof}
		The proof follows from Proposition \ref{prop:coeffin} and the bounds 
		$$|trX_\mck|\lesssim\Omega^2,\quad |tr\Xbar|\lesssim 1,\quad|\df^{\leq N_4}(H_\mck,\Hbar_\mck)|\lesssim|\df^{\leq N_4}(H_\mck^S,\Hbar_\mck^S)|\lesssim|(\df^{\leq N_4}(H_\mck^S,\Hbar_\mck^S))_\mck| \lesssim 1,$$
		where the first two bounds hold by \eqref{eq:tracesxxbarkerr}, and the third bound hold combining Corollary \ref{cor:comparder}, Proposition \ref{prop:linderdoublenull}, and the fact that $H_\mck^S,\Hbar_\mck^S$ are smooth in Kerr up to the Cauchy horizon.
	\end{proof}
	\subsubsection{Boundedness of the non-integrable Gauss curvature}
	Note that although the curvature component $P$ is a quantity with signature $0$ in the sense of \cite{CK93}, from the bound \eqref{eq:boundPin} we expect that $P$ presents a $\Omega^{-2}$ degeneracy at $\ch$. This can be seen from the Bianchi identity for $\nabla_4P$ which presents a term with $A$ (that blows-up like $\Omega^{-4}$) on the RHS. However, similarly as in the double null gauge \cite{stabC0,weakluk} there is a modified quantity constructed from $P,X,\Xbar$ which satisfied a modified Bianchi equation where there is no degenerate terms with factors $A$ on the RHS. As a consequence, we will see that this quantity remains bounded up to $\ch$. We introduce the renormalised quantity
	\begin{align}\label{eq:gausscurvature}
		G:=P-\frac{1}{4}\overline{\wh{X}}\cdot\wh{\Xbar}+\frac{1}{4}trX\overline{tr\Xbar}.
	\end{align}
Recall the scalar ${}^{(h)}K$ (the non-integrable Gauss curvature) defined in Proposition \ref{prop:gaussequation}, which we denote by $K$ to simplify the notations in what follows.
	\begin{lem}\label{lem:GexxpreavecK}
		We have 
		$$G=-K+i\hodge{K},$$
		where
		\begin{align*}
			K&=-\rho+\frac{1}{2}\wh{\chi}\cdot\wh{\chibar}-\frac{1}{4}(tr\chi tr\chibar+\atr\chi\atr\chibar),\\
			\hodge{K}&=\hodge{\rho}+\frac{1}{2}\wh{\chibar}\wedge\wh{\chi}+\frac{1}{4}(tr\chi\atr\chibar-tr\chibar\atr\chi).
		\end{align*}
	\end{lem}
	\begin{proof}
		The proof is a simple computation, relying on the identity $\overline{\wh{X}}\cdot\wh{\Xbar}=2(\chihat\cdot\wh{\chibar}+i\chihat\wedge\wh{\chibar})$.
	\end{proof}
	The following result is a renormalized Bianchi identity for $G$. The result holds in a general null frame, but we will use it in the ingoing non-integrable frame in $\deux$ where $\Xi=\Xibar=0$.
	\begin{prop}
		We have the following renormalized Bianchi equation for $G$,
		\begin{align}
			\nabla_4G+trXG=\frac{1}{2}&\mcd\cdot\overline{B}+\frac{1}{2}(2\Hbar+Z)\cdot\overline{B}-\overline{\Xi}\cdot\Bbar-\frac{1}{8}\wh{\Xbar}\cdot(\overline{\mcd}\hot\overline{\Xi})\nn\\
			&-\frac{1}{8}\wh{\Xbar}\cdot(\overline{\Xi}\hot(\overline{\Hbar}+\overline{H}+2\overline{Z}))-\frac{1}{8}\overline{\wh{X}}\cdot(\mcd\hot\Hbar)-\frac{1}{8}\overline{\wh{X}}\cdot(\Hbar\hot\Hbar)\nn\\
			&-\frac{1}{16}\overline{\wh{X}}\cdot(\Xi\hot\Xibar)+\frac{1}{4}\overline{tr\Xbar}(\mcd\cdot\overline{\Xi}+\Xi\cdot\overline{H}+\overline{\Xi}\cdot(H+2Z))\nn\\
			&+\frac{1}{4}trX(\overline{D}\cdot\Hbar+\overline{\Hbar}\cdot\Hbar+\overline{\Xibar}\cdot\Xi)+\frac{i\atrchi}{4}\parentheses{{tr X}\overline{tr\Xbar}-\wh{\Xbar}\cdot\overline{\wh{X}}}.\label{eq:renbianchiP}
		\end{align}
	\end{prop}
	\begin{proof}
		We compute, using the equations in Propositions \ref{prop:nullstructurecomplex}, \ref{prop:bianchicomplex},
		\begin{align*}
			\nabla_4G&=\nabla_4P-\frac{1}{4}\overline{\nabla_4 \wh{X}}\cdot\wh{\Xbar}-\frac{1}{4}\overline{\wh{X}}\cdot\nabla_4\wh{\Xbar}+\frac{1}{4}{\nabla_4 trX}\overline{tr\Xbar}+\frac{1}{4}{tr X}\overline{\nabla_4tr\Xbar}\\
			&=\frac{1}{2} \mathcal{D} \cdot \overline{B}-\frac{3}{2} tr X P+\frac{1}{2}(2 \underline{H}+Z) \cdot \overline{B}-\overline{\Xi} \cdot \underline{B}-\frac{1}{4} \underline{\widehat{X}} \cdot \overline{A}\\
			&-\frac{1}{4}\Bigg(\overline{-\Real(trX)\wh{X}-2\omega\wh{X}+\frac{1}{2}\mcd\hot{\Xi}+\frac{1}{2}\Xi\hot(\Hbar+H-2Z)-A}\Bigg)\cdot\wh{\Xbar}\\
			&-\frac{1}{4}\Bigg(-\frac{1}{2}tr X\wh{\Xbar}+2\omega\wh{\Xbar}+\frac{1}{2}\mcd\hot \Hbar+\frac{1}{2}\Hbar\hot \Hbar-\frac{1}{2}\overline{tr \Xbar}\wh{X}+\frac{1}{4}\Xibar\hot\Xi\Bigg)\cdot\overline{\wh{X}}\\
			&+\frac{1}{4}\Bigg({-\frac{1}{2}(tr X)^2-2 \omega tr X+\mathcal{D} \cdot \overline{\Xi}+\Xi \cdot \overline{H}+\overline{\Xi} \cdot(H+2 Z)-\frac{1}{2} \widehat{X} \cdot \overline{\widehat{X}}}\Bigg)\overline{tr\Xbar}\\
			&+\frac{1}{4}\Bigg(\overline{-\frac{1}{2} tr X tr \underline{X}+2 \omega tr \underline{X}+\mathcal{D} \cdot \underline{\overline{H}}+\underline{H} \cdot \underline{\overline{H}}+2 \overline{P}+\Xi \cdot \underline{\overline{\Xi}}-\frac{1}{2} \widehat{X} \cdot \overline{\underline{\widehat{X}}}}\Bigg){tr X},
		\end{align*}
		which rewrites, noticing the  crucial cancellation of the terms $\wh{\Xbar}\cdot \overline{A}$,
		\begin{align*}
			\nabla_4G=&\frac{1}{2} \mathcal{D} \cdot \overline{B}+\frac{1}{2}(2 \underline{H}+Z) \cdot \overline{B}-\overline{\Xi} \cdot \underline{B}-\frac{3}{2} tr X P\\
			&-\frac{1}{4}\Bigg(-\Real(trX)-\frac{1}{2}trX\Bigg)\wh{\Xbar}\cdot\overline{\wh{X}}-\frac{1}{4}\Real(tr X){tr X}\overline{tr\Xbar}+\frac{1}{2}trX\parentheses{G-\frac{1}{4}trX\overline{tr\Xbar}}\\
			&-\frac{1}{4}\Bigg(\frac{1}{2}\mcd\hot \Hbar+\frac{1}{2}\Hbar\hot \Hbar+\frac{1}{4}\Xibar\hot\Xi\Bigg)\cdot\overline{\wh{X}}-\frac{1}{4}\Bigg(\overline{\mathcal{D} \cdot \underline{\overline{H}}+\underline{H} \cdot \underline{\overline{H}}+\Xi \cdot \underline{\overline{\Xi}}}\Bigg) {tr{X}}\\
			&-\frac{1}{4}\Bigg(\overline{\frac{1}{2}\mcd\hot{\Xi}+\frac{1}{2}\Xi\hot(\Hbar+H-2Z)}\Bigg)\cdot\wh{\Xbar}+\frac{1}{4}\left({\mathcal{D} \cdot \overline{\Xi}+\Xi \cdot \overline{H}+\overline{\Xi} \cdot(H+2 Z)}\right) \overline{tr{\Xbar}}.
		\end{align*}
		To conclude, note that the last term $-\frac{3}{2} tr X P$ on the first line above combined with the second line above rewrite
		\begin{align*}
-trXG+\frac{i\Imag(trX)}{4}\parentheses{{tr X}\overline{tr\Xbar}-\wh{\Xbar}\cdot\overline{\wh{X}}},
		\end{align*}
		which concludes the proof.
	\end{proof}
	\begin{prop}\label{prop:gausscurvbornee}
		The scalar $G$ is bounded in $\deux$. As a consequence, 
		$$|K|\lesssim 1.$$
	\end{prop}
	\begin{proof}
		By \eqref{eq:renbianchiP}, the identities $\Xi=\Xibar=0$ and the estimate $|\df^{\leq 1}B|\lesssim\ubar^{-2-\delta/5}$ from Proposition \ref{prop:coeffin}, and Corollary \ref{cor:coeffinsanscheck} we get in $\deux$
		$$|\nabla_4G|\lesssim (\Omega^2+\ubar^{-2-\delta/5})|G|+\Omega^2+\ubar^{-2-\delta/5}+|\atrchi| (1+\Omega^{-2}\ubar^{4+2\delta/5}), $$
		where the last term on the RHS corresponds to the bound for the last term on the RHS of \eqref{eq:renbianchiP}, given by Proposition \ref{prop:coeffin} and Corollary \ref{cor:coeffinsanscheck}. Thus, by the bound \eqref{eq:borneantitracesin} for $\atrchi$ we get  $|\nabla_4G|\lesssim (\Omega^2+\ubar^{-2-\delta/5})|G|+\Omega^2+\ubar^{-2-\delta/5}$. By \ref{eq:outtoin}, using $\hat{\Omega}^2\sim\Omega^2$ we deduce
		$$|\nabring_YG|\lesssim \Omega^2|\nabla_{4'}G|\lesssim|\nabla_4G|\lesssim (\Omega^2+\ubar^{-2-\delta/5})|G|+\Omega^2+\ubar^{-2-\delta/5},$$
		where $Y$ is as in \eqref{eq:Ysadefquoi}. We then conclude the proof by the transport estimate of Proposition \ref{prop:transporte4'} combined with Grönwall's inequality, and Proposition \ref{prop:coeffin} which yields $|G|\lesssim 1$ on $\Sigma_0$. 
	\end{proof}
	\subsubsection{More precise bounds for $trX$ and $tr\Xbar$ using Raychaudhuri's equation}
		\begin{prop}\label{prop:conseqraych}
		We have, in $\deux$, the bounds 
		\begin{align}\label{eq:boundtrxprim}
			|trX'|\lesssim 1+\frac{\varepsilon^2\Omega^{-2}}{\ubar^{4+2\delta/5}},\quad |tr\Xbar|\lesssim 1+\frac{\varepsilon^2\Omega^{-2}}{\ubar^{4+2\delta/5}}.
		\end{align}
		\end{prop}
	\begin{proof}
	This relies on the quadratic nature of the RHS of the equations for $\nabla_3tr\Xbar$ and $\nabla_4trX$ in Proposition \ref{prop:nullstructurecomplex} in both the outgoing and ingoing non-integrable frames, which write
	\begin{align*}
		\nabla_{4'} tr X'+\frac{1}{2}(tr X')^2=-\frac{1}{2} \widehat{X}' \cdot \overline{\widehat{X}'},\quad\nabla_3tr\Xbar+\frac{1}{2}tr\Xbar^2=-\frac{1}{2}\wh{\Xbar}\cdot\overline{\wh{\Xbar}}.
	\end{align*}
	Note that we used here, in the non-integrable gauge, the identities $\omega'=0$, $\xi'=0$, $\omegabar=0$, $\xibar=0$ (see \eqref{eq:xiprimxibarprimomegaprimzero}, \eqref{eq:xiprimxibarprimomegaprimzeroin}). Moreover, we have
	$$|trX'|+|tr\Xbar|\lesssim 1\:\text{ on }\Sigma_0,\quad |\widecheck{trX}'|+|\widecheck{tr\Xbar}|+|\wh{\Xbar}|+|\wh{X}'|\lesssim\frac{\varepsilon\Omega^{-2}}{\ubar^{2+\delta/5}}\:\text{ in }\deux,$$
by Propositions \ref{prop:coeffout}, \ref{prop:coeffin}. Let $h,g$ be the solutions in $\deux$ of the transport equations
	$$e_4'(h)=\frac{1}{2}trX',\quad e_3(g)=\frac{1}{2}tr\Xbar, \quad h|_{\Sigma_0}=g|_{\Sigma_0}=0.$$
	By the bounds $|trX'|\lesssim 1+{\varepsilon\Omega^{-2}}{\ubar^{-2-\delta/5}}, \quad |tr\Xbar|\lesssim 1+{\varepsilon\Omega^{-2}}{\ubar^{-2-\delta/5}}$
	in $\deux$ (see Corollary \ref{cor:coeffinsanscheck}) we deduce from the transport estimates of Propositions \ref{prop:transporte4'} and \ref{prop:transporte4'bar}, that
	\begin{align}\label{eq:hbornedansdeux}
		|h|+|g|\lesssim 1,\quad\text{in}\:\:\deux.
	\end{align}
	This yields 
	\begin{align*}
		\nabla_{4'}(e^htrX')=e^h\left(\nabla_{4'}trX'+\frac{1}{2}(trX')^2\right)&=-\frac{1}{2} e^h\widehat{X}' \cdot \overline{\widehat{X}'},\\
		\nabla_{3}(e^gtr\Xbar)=e^g\left(\nabla_{3}tr\Xbar+\frac{1}{2}(tr\Xbar)^2\right)&=-\frac{1}{2} e^g\widehat{\Xbar} \cdot \overline{\widehat{\Xbar}}.
	\end{align*}
	Thus, using \eqref{eq:hbornedansdeux}, we get in $\deux$,
	$$|\nabla_{4'}(e^htrX')|\lesssim\frac{\varepsilon^2\Omega^{-4}}{\ubar^{4+2\delta/5}},\quad |\nabla_{3}(e^gtr\Xbar)|\lesssim\frac{\varepsilon^2\Omega^{-4}}{\ubar^{4+2\delta/5}}.$$
	By Propositions \ref{prop:transporte4'} and \ref{prop:transporte4'bar}, this implies in $\deux$ the bounds
	\begin{align*}
		|e^htrX'|(p)&\lesssim \|trX'\|_{L^\infty(\Sigma_0)}+\int_{\ubar_{Y,\Sigma_0}(p)}^\ubar\frac{\varepsilon^2\Omega^{-2}}{(\ubar')^{4+2\delta/5}}\dee\ubar'\lesssim 1+\frac{\varepsilon^2\Omega^{-2}}{\ubar^{4+2\delta/5}},\\
		|e^gtr\Xbar|(p)&\lesssim \|tr\Xbar\|_{L^\infty(\Sigma_0)}+\int_{u_{\Ybar,\Sigma_0}(p)}^u\frac{\varepsilon^2\Omega^{-2}}{(\ubar_{\Ybar}^p(u'))^{4+2\delta/5}}\dee u'\lesssim 1+\frac{\varepsilon^2\Omega^{-2}}{\ubar^{4+2\delta/5}}.
	\end{align*}
	Here, we used the following computations:
	\begin{itemize}
		\item For the first bound, we used  Proposition \ref{prop:flowofY} to get 
		$$\int_{\ubar_{Y,\Sigma_0}(p)}^\ubar\frac{\varepsilon^2\Omega^{-2}}{(\ubar')^{4+2\delta/5}}\dee\ubar'\lesssim \varepsilon^2e^{|\kappa_-|u}\int_{\ubar_{Y,\Sigma_0}(p)}^\ubar\frac{e^{|\kappa_-|\ubar'}}{(\ubar')^{4+2\delta/5}}\dee\ubar'\lesssim \frac{\varepsilon^2e^{|\kappa_-|(u+\ubar)}}{\ubar^{4+2\delta/5}}\lesssim\frac{\varepsilon^2\Omega^{-2}}{\ubar^{4+2\delta/5}}.$$
		\item For the second bound, we used Proposition \ref{prop:flowofYbar} to get
		$$\int_{u_{\Ybar,\Sigma_0}(p)}^u\frac{\varepsilon^2\Omega^{-2}}{(\ubar_{\Ybar}^p(u'))^{4+2\delta/5}}\dee u'\lesssim \frac{\varepsilon^2e^{|\kappa_-|\ubar}}{\ubar^{4+2\delta/5}}\int_{u_{\Ybar,\Sigma_0}(p)}^u e^{|\kappa_-|u'}\dee u'\lesssim \frac{\varepsilon^2e^{|\kappa_-|(u+\ubar)}}{\ubar^{4+2\delta/5}}\lesssim \frac{\varepsilon^2\Omega^{-2}}{\ubar^{4+2\delta/5}}.$$
	\end{itemize}
	We conclude the proof of the proposition by using \eqref{eq:hbornedansdeux} which implies $|e^{-h}|+|e^{-g}|\lesssim 1$.
\end{proof}
	\begin{rem}
		Recalling the hypersurface $\Gamma$ defined in \eqref{eq:defiGammahyp}, \eqref{eq:boundtrxprim} implies the bound
			\begin{align}\label{eq:wantedd}
			|trX'|\lesssim \frac{\Omega^{-2}}{\ubar^{4+2\delta/5}},\quad \text{on}\:\:\Gamma,
		\end{align}
	as $\Omega^{2}$ decreases exponentially in $\ubar$ on $\Gamma$. Indeed, by definition of $\Gamma=\{u+\ubar=\ubar^\gamma\}$ and by $\Omega^2\sim\Omega^2_\mck$ and \eqref{eq:omegamcksim}, $\Omega^2|_\Gamma\sim e^{-|\kappa_-|\ubar^\gamma}$. This bound will be used as some initial data on $\Gamma$ for the analysis of region $\trois$. The bound for $tr\Xbar$ in $\deux$ in \eqref{eq:boundtrxprim} will be used in the energy method for the Teukolsky equation for $A$.
	\end{rem}

	\subsection{More precise pointwise and energy estimates on $\Sigma_0$}
	
	In this section, we prove some pointwise and energy estimates on $\Sigma_0$, which will be used as initial data for the analysis of the Teukolsky equation for $A$ in region $\deux$ in the ingoing non-integrable frame\footnote{These estimates rely on the one on $\Gamma$ proven in Section \ref{section:regionun} for the PT frame.}. Here, we use the notation $(e_\mu^{(\un)})$ for the ingoing PT frame in region $\un$, and $(e_\mu)$ for the ingoing non-integrable frame constructed in Section \ref{section:defprincipalframes}. We extend this notation to the Ricci and curvature coefficients defined with respect to the PT and non-integrable frames in $\un\cap\deux$, which we denote respectively $trX^{(\un)},\wh{X}^{(\un)},\ldots,A^{(\un)},\Abar^{(\un)}\ldots,$ and $trX,\wh{X},\ldots, A,\Abar,\ldots.$
	\subsubsection{Transformation linking the PT and non-integrable frames in $\un\cap\deux$}
	\begin{figure}[h!]
		\centering
		\includegraphics[scale=0.48]{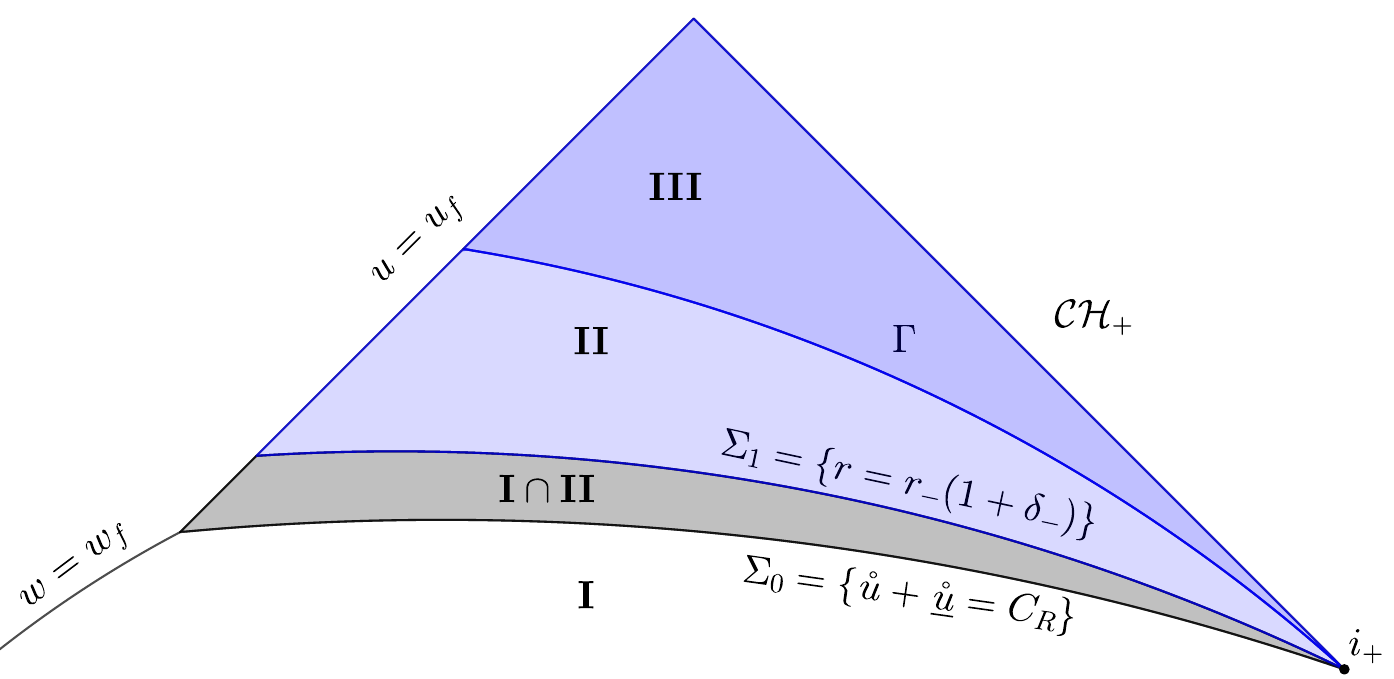}
		\caption{The intersection region $\un\cap\deux$ in grey.}
		\label{fig:IcapIINL}
	\end{figure}

	Recall the definitions of the regions $\un$ and $\deux$ in \eqref{eq:ecouteregIdefigg} and \eqref{eq:defdedeuxx}. Notice that provided \eqref{eq:conditionCRR} is satisfied (which holds in the present context), then $\un\cap\deux$ is an open neighborhood of timelike infinity. Indeed, this condition ensures that $\Sigma_0$, where region $\deux$ is initialized, is strictly in the past of the future boundary $\Sigma_1=\{r=r_-(1+\delta_-)\}$ of region $\un$. As a consequence region $\un$ slightly extends into region $\deux$, see Figure \ref{fig:IcapIINL}.
	\begin{prop}\label{prop:linkPTprincipal}
		In region $\un\cap\deux$, there exists a PT-horizontal 1-form $\hat{f}$ such that the ingoing non-integrable null frame $(e_3,e_4,e_a)$ is given by the frame transformation \eqref{eq:frametransfo} of the \emph{ingoing} PT frame $(e_3^{(\un)},e_4^{(\un)},e_a^{(\un)})$ with coefficients $$\left({f}=\hat{f},\fbar=0,{\lambda}=1\right),$$ where $\hat{f}$ satisfies following estimate, 
		\begin{align}\label{eq:statedestimatefbar}
			|(\nabla_3^{(\un)},\nabla_4^{(\un)},\nabla^{(\un)})^{\leq N_1}\hat{f}|\lesssim\varepsilon\ubar^{-3-\delta/3}\:\text{in}\:\:\un\cap\deux.
		\end{align}
	\end{prop}
	\begin{proof}
		In $\un\cap\deux$, both $e_3$ and $e_3^{(\un)}$ are geodesic by construction. Moreover, $e_3$ and $e_3^{(\un)}$ coincide on $\Sigma_0$ by \eqref{eq:coincidesigma0e3prim}. Thus, by standard uniqueness results for transport equations we deduce 
		$$e_3=e_3^{(\un)},\:\text{in}\:\:\un\cap\deux.$$
		Thus, as $e_4$ is null there exists a PT-horizontal 1-form $\hat{f}$ such that in $\un\cap\deux$\footnote{Note that here, the conformal factor is $\lambda=1$ because $e_3=e_3^{(\un)}$ and $\g(e_3,e_4)=\g(e_3^{(\un)},e_4^{(\un)})=-2.$},
		\begin{align}\label{eq:liene4e4PT}
			e_4=e_4^{(\un)}+\hat{f}^ae_a^{(\un)}+\frac{1}{4}|\hat{f}|^2e_3^{(\un)}.
		\end{align}
		This concludes the first part of the proposition. Next, the change of frame formula \eqref{eq:changexi} gives, using $\xi=0$ in $\un\cap\deux$ in the ingoing non-integrable frame,
		\begin{align*}
			&\nabla_{e_4}\hat{f}+2\omega^{(\un)}\hat\fbar
			+\frac{1}{2}(tr\chi^{(\un)}\hat{f}-\atrchi^{(\un)}\hodge{\hat{f}})=-2\xi^{(\un)}+\err[\hat{f}],\\
			&\err[\hat{f}]:=-\hat{f}\cdot\wh{\chi}^{(\un)}-\frac{1}{2}|\hat{f}|^2\etabar^{(\un)}-(\hat{f}\cdot\zeta^{(\un)})\hat{f}+\frac{1}{2}|\hat{f}|^2\eta^{(\un)}-\frac12|\hat{f}|^2(\hat{f}\cdot\chibar^{(\un)}).\end{align*}
		Thus defining the following complex PT-horizontal 1-form,
		$$\Fhat:=\hat{f}+i\hodge{\hat{f}}\in\fraks_1^{(\un)}(\C),$$
		we get the outgoing transport equation (along the non-integrable null vector field $e_4$)
		\begin{align}
			\nabla_{e_4}\Fhat+2\omega^{(\un)}\Fhat
			+\frac{1}{2}\overline{trX}^{(\un)}\Fhat=-2\Xi^{(\un)}+\err[\Fhat],\quad\err[\Fhat]:=\err[\hat{f}]+i\hodge{\:\err[\hat{f}]},\label{eq:equadiffpourfbarhat}
		\end{align}
	in $\un\cap\deux$.	Moreover, recalling \eqref{eq:coincidesigma0e4prim} on $\Sigma_0$, we have that $\Fhat$ satisfies the initial condition
		\begin{align}\label{eq:condiniteFhat}
			\Fhat|_{\Sigma_0}=0.
		\end{align}
	Also, note that recalling in $\un\cap\deux$ the set of PT derivatives defined in \eqref{eq:setofderivatives}, the error term satisfies in $\un\cap\deux$, 
		$$|\df^{\leq k}\err[\Fhat]|\lesssim \ubar^{-3-\delta/2}|\df^{\leq k}\Fhat|+|\df^{\leq k}\Fhat|^2+l.o.t,$$
		where we used \eqref{eq:controlI}, and where $l.o.t$ denotes terms bounded by cubic or more powers of $\df^{\leq k}\Fhat$. We now prove the decay \eqref{eq:statedestimatefbar} from the non-linear transport equation \eqref{eq:equadiffpourfbarhat} in $\un\cap\deux$ and the initial condition \eqref{eq:condiniteFhat} on $\Sigma_0$. Similarly as what was done in Section \ref{section:controlef}, we only need to improve the bootstrap assumption 
		$$|\df^{\leq k_0}\Fhat|\leq\sqrt{\varepsilon}\ubar^{-3-\delta/2}, \quad\text{in }\un\cap\deux\cap\{\ubar<\ubar_*\}.$$
		This is done by induction on the number $k$ of $\df$ derivatives considered, and it is actually much easier than the analysis for $\fc$ because in region $\un\cap\deux$, we have\footnote{Notice that strictly speaking, $\Omega^2\underset{C_R,\delta_-}{\sim} 1$, but here $\delta_-$ is chosen depending only on $a,M,C_R$ such that \eqref{eq:conditionCRR} holds, which justifies \eqref{eq:omegacarrdansiinterdeux}.}
		\begin{align}\label{eq:omegacarrdansiinterdeux}
			\Omega^2\underset{C_R}{\sim} 1,\quad\text{in}\:\:\un\cap\deux.
		\end{align}
	{Thus, in the remaining part of this proof only, the implicit constants in the bounds $\lesssim $ depend on $a,M$ and $C_R$.} We deal with the control of successive $\df$ derivatives just like in region $\un$, see Section \ref{section:sectionalasuite3}, namely we are able to use rough schematic identities for the commutators between $\df$ derivatives, highly relying on \eqref{eq:omegacarrdansiinterdeux}. For $k=0$, the bootstrap assumption and the bound $|\Xi^{(\un)}|\lesssim\varepsilon\ubar^{-3-\delta/2}$ imply
		$$|\nabla_{e_4}\Fhat|\lesssim |\Fhat|+\varepsilon\ubar^{-3-\delta/2}.$$ 
		Then, integrating along the flow of $e_4$ (relying on Proposition \ref{prop:transporte4'} and \eqref{eq:omegacarrdansiinterdeux}), we get by Grönwall's inequality the bound
		\begin{align}\label{eq:Fhatcontrolkegalzero}
			|\Fhat|\lesssim\varepsilon\ubar^{-3-\delta/2},
		\end{align}
		where we also used the initial condition \eqref{eq:condiniteFhat} on $\Sigma_0$\footnote{Note that contrary to in Section \ref{section:controlef} where the interval of definition of the flow of $e_4'$ is bounded by $\ubar^\gamma$, here in $\un\cap\deux$ it is simply bounded so that we do not lose a $\gamma$ in the exponent of the polynomial bound.}. Next, using Lemma \ref{lem:commutnablageneral} and the bounds \eqref{eq:controlI}, we get the schematic commutator identity $[\df,\nabla_{e_4}]=_{rs}\df^{\leq 1}$,	which holds in $\un\cap\deux$, we get by the bootstrap assumption
		$$\sum_{\partial\in\df}|\nabla_{e_4}\partial\Fhat|\lesssim |\df^{\leq 1}\Fhat|+\varepsilon\ubar^{-3-\delta/2}.$$
		Thus, integrating the bound above and using \eqref{eq:Fhatcontrolkegalzero} and Grönwall's inequality we get  $|\df^{\leq 1}\Fhat|\lesssim\varepsilon\ubar^{-3-\delta/2}$. Proceeding analogously for higher order derivatives\footnote{Here one must use a higher-order schematic commutation identity $[\nabla_{e_4},\df^k]=_{rs}\df^{\leq k}$ in $\un\cap\deux$ which can be proven by induction using Lemma \ref{lem:commutnablageneral} and \eqref{eq:controlI}.}, we deduce the improved bootstrap assumption on $\un\cap\deux\cap\{\ubar\leq\ubar_*\}$,
		$$|\df^{\leq N_1}\Fhat|\lesssim\varepsilon\ubar^{-3-\delta/2}.$$
		This concludes the bootstrap argument, which yields
		$$|(\nabla_3^{(\un)},\nabla_4^{(\un)},\nabla^{(\un)})^{\leq N_1}\hat{f}|\underset{C_R}{\lesssim}\varepsilon\ubar^{-3-\delta/2}$$
		in $\un\cap\deux$, and we get rid of the dependence in $C_R$ in the implicit constant above by choosing $u_f(a,M,\delta,C_R)$ negative enough so that \eqref{eq:statedestimatefbar} holds, hence the proof.
	\end{proof}

	\subsubsection{Sharp pointwise and energy bounds for $A$ on $\Sigma_0$}
We define the following scalar function in $\deux$
	\begin{align}\label{eq:premoccuwbar}
		\wbar:=\ubar-e^{-|\kappa_-|(u+\ubar)}+e^{-|\kappa_-|C_R},
	\end{align}
	which is a variable with spacelike level sets (as will be proven below) satisfying $\wbar\sim\ubar$. To avoid conflict of notations, we denote $\ubar^{(\un)},\wbar^{(\un)}, r^{(\un)}, \theta^{(\un)},\phi_+^{(\un)}$ the quantities defined in Section \ref{section:regionun} with respect to the PT gauge in region $\un$. For convenience, we also denote $\ubar_{(\un)}=\ubar^{(\un)}$. We also define $$\Sigma_0[\wbar_1,\wbar_2]:=\Sigma_0\cap\{\wbar_1\leq\wbar\leq\wbar_2\},$$ and the following energy density,
	$$\mathbf{e}[U]:=|\nabla_3U|^2+|\nabla_4U|^2+|\nabla U|^2,$$
	for any horizontal tensor $U$. We recall from \eqref{eq:Aunsig00} and \eqref{eq:Aenerunsig00} the following pointwise and energy bounds on $\Sigma_0$, 
	\begin{align}
		\left|A^{(\un)}|_{\Sigma_0}-\Psi^{(\un)}|_{\Sigma_0}\right|&\lesssim C(\delta_\pm,w_f,(Q_m)_{|m|\leq2})\ubar^{-6-\delta},\label{pointwisehypfromI}\\
		\int_{\Sigma_0[\wbar_1,\wbar_2]}|\df^{\leq 1}(A^{(\un)}-\Psi^{(\un)})|&\lesssim C(\delta_\pm,w_f,(Q_m)_{|m|\leq2})\int_{\wbar_1}^{\wbar_2}\wbar^{-2(6+\delta)}\dee\wbar,\label{eq:energyhypfromI}
	\end{align}
	where $\Psi^{(\un)}$ is defined as follows in region $\un$ (see \eqref{eq:definitansatzun}),
	\begin{align}\label{eq:ansatzIcapII}
		\Psi^{(\un)}=\frac{1}{\ubar_{(\un)}^6}\sum_{|m|\leq 2} \frac{Q_mA_m(r)}{\qbar^2}
		\mcd^{(\un)}\hot(\mcd^{(\un)}(Y_{m,2}(\cos\theta^{(\un)})e^{im\phi_+^{(\un)}})),
	\end{align}
	and where $A^{(\un)}=\alpha^{(\un)}+i\hodge{\alpha^{(\un)}}$ is the curvature component $A$ defined in Section \ref{section:ricciandcurvdef} with respect to the {ingoing} PT frame $(e_3^{(\un)},e_4^{(\un)},e_a^{(\un)})$	of region $\un$. We now define new coordinates in $\deux\cap\trois$, which are perturbations of ingoing Eddington-Finkelstein coordinates.
\begin{defi} \label{defi:coolcollcool}
	We define the coordinates
		$$(r_\mck,\ubar'_{\mck},\theta_\mck,(\phi_+)_\mck)$$ in region $\deux\cup\trois$ as the Kerr values of $r,\ubar'=r^*+t,\theta,\phi_+$ with respect to the double null coordinates $(u,\ubar,\theta^A)$ in $\deux\cup\trois$. Here, $\ubar'_{\mck}$ is the Kerr value with respect to the double null coordinates $(u,\ubar,\theta^A)$ of the usual Eddington-Finkelstein retarded time $\ubar':=r^*+t$, which coincides with $\ubar^{(\un)}$ in exact Kerr\footnote{Note that we use the prime notation $\ubar'$ to avoid confusions with the double null variable $\ubar$.}. Recalling Definition \ref{defi:covariantway}, we also define the following ansatz in $\deux$,
	\begin{align}\label{eq:defansatzdeux}
			\Psi:=\frac{1}{\ubar_\mck'^6}\sum_{|m|\leq 2}\frac{Q_m A_m(r_\mck)}{\qbar_\mck^2}(\mcd\hot(\mcd(Y_{m,2}(\cos\theta)e^{im\phi_+}))_\mck^S)^{H}.
	\end{align}
where we rely again on the double null coordinates $(u,\ubar,\theta^A)$ to define the Kerr value $$\mcd\hot(\mcd(Y_{m,2}(\cos\theta)e^{im\phi_+}))_\mck^S:=\left[\left(\mcd\hot(\mcd(Y_{m,2}(\cos\theta_\mck)e^{im(\phi_+)_\mck}))\right)^S\right]_\mck$$
of the $S(u,\ubar)$-tangent tensor $\left(\mcd\hot(\mcd(Y_{m,2}(\cos\theta_\mck)e^{im(\phi_+)_\mck}))\right)^S$, and where $q_\mck=r_\mck+ia\cos\theta_\mck$. Note that in Kerr, $\Psi$ and $\Psi^{(\un)}$ coincide, as the PT frame and the ingoing non-integrable frame coincide. Also, we have $\ubar'_\mck\sim\ubar$, because this is the case in exact Kerr, see Lemma \ref{lem:ubarssontpareils}.
\end{defi}
Also recall that we defined the non-integrable frame $(e_3,e_4,e_a)$ in $\deux$ by applying a frame transformation of coefficients $(f,\fbar,\lambda)$ to the double null frame in $\deux$, and that we initialized $(f,\fbar,\lambda)$ such that the ingoing PT and non-integrable frames coincide on $\Sigma_0$, 
	$$(e_3,e_4,e_a)|_{\Sigma_0}=(e_3^{(\un)},e_4^{(\un)},e_a^{(\un)})|_{\Sigma_0},$$
	see \eqref{eq:coincidesigma0e3prim}, \eqref{eq:coincidesigma0e4prim}. By the estimates on the coordinates and their derivatives with respect to the PT frame in $\un$, this was achieved choosing $f,\fbar,\lambda$ such that, on $\Sigma_0$, 
	$$|\df^{\leq N}(\widecheck{f}_0,\widecheck{\fbar}_0,\widecheck{\lambda}_0)|\lesssim\ubar^{-3-\delta/2}.$$
	Now, we want to translate the initial energy decay \eqref{eq:energyhypfromI} into an initial energy decay statement on $\Sigma_0$ for the curvature component $A$ defined with respect to the ingoing non-integrable frame $(e_3,e_3,e_a)$ in $\deux$, and $\Psi$, see \eqref{eq:pointwiseutilisabledansII} and \eqref{eq:energyhyputilisabledansII}.

\begin{defi}
For $U$ a $PT$-horizontal tensor in $\un\cap\deux$, following Remark \ref{rem:phistarpullback} we denote $U^H$ the horizontal tensor defined by $U^H_{ab}=U_{ab}$. More precisely, using Proposition \ref{prop:linkPTprincipal}, we get that the map
$$\mathbf{M}:X\in (e_3^{(\un)},e_4^{(\un)})^\bot\longmapsto X+\frac12 \hat{f}(X)e_3^{(\un)}\in \mch=(e_3,e_4)^\bot$$
is well defined and is an isometry, and we define for $X_1,\cdots,X_k\in \mch$,
$$U^H(X_1,\cdots, X_k)=U(\mathbf{M}^{-1}(X_1),\cdots,\mathbf{M}^{-1}(X_k)).$$
 Conversely, if $V$ is a horizontal tensor in $\un\cap\deux$, we define the $PT$-horizontal tensor $V^{PT}$ by 
$$V^{PT}(X_1,\cdots,X_k):=V(\mathbf{M}(X_1),\cdots,\mathbf{M}(X_k)),$$
for any $PT$-horizontal vectors $X_1,\cdots,X_k$.
\end{defi}
\begin{rem}We use the same notation, namely the supscript ${}^H$, to pullback PT-horizontal (namely $(e_3^{(\un)},e_4^{(\un)})^\bot$-tangent) and $S(u,\ubar)$-tangent tensors to horizontal tensors (see Definition \ref{defi:covariantway}). However, there is no risk of confusion because when we use the supscript ${}^H$, it will be clear wether it is applied to a $PT$-horizontal tensor or to a $S(u,\ubar)$-tangent tensor.
\end{rem}

	\begin{prop}\label{prop:lierAetAun}
		We have the following energy bound on $\Sigma_0$,
		$$\int_{\Sigma_0[\wbar_1,\wbar_2]}\mathbf{e}\left[A-(A^{(\un)})^H\right]\lesssim\int_{\wbar_1}^{\wbar_2}\wbar^{-2(6+2\delta/3)}\dee\wbar.$$
	\end{prop}
	\begin{proof}
By the change of frame formula \eqref{eq:changealpha} we have
$$A^{PT}-A^{(\un)}=\err[A,A^{(\un)}],$$
where the $PT$-horizontal tensor $\err[A,A^{(\un)}]$ satisfies
$$\err[A,A^{(\un)}]=_s\hat{f}\cdot \beta^{(\un)}+\hat{f}^2\rho^{(\un)}+\mcr_{[2]}.$$
By the estimates in \eqref{eq:controlI} and \eqref{eq:statedestimatefbar}, this implies 
$$|\df^{\leq N_1}\err[A,A^{(\un)}]|\lesssim |\df^{\leq N_1}\hat{f}|^2+|\df^{\leq N_1}\Rcheck|^2\lesssim\varepsilon^2\ubar^{-6-2\delta/3}.$$
Moreover, using the computations in Propositions \ref{prop:diffchristo}, \ref{prop:diffchristo4}, \ref{prop:diffchristo3} with $\lambda=1$, $\fbar=0$, and replacing $f$ by $\hat{f}$, by \eqref{eq:statedestimatefbar} for any horizontal tensor $U$ and $k\leq N_1$ we get
\begin{align}\label{eq:bonutilehein}
	|\df^{\leq k}U|\lesssim|\df^{\leq k}U^{PT}|,
\end{align}
in $\un\cap\deux$, and hence we infer
\begin{align}\label{eq:enfaitsicool}
	|\df^{\leq k}(A-(A^{(\un)})^{H})|\lesssim |\df^{\leq k}(A^{PT}-A^{(\un)})|\lesssim |\df^{\leq k}\err[A,A^{(\un)}]|\lesssim\varepsilon^2\ubar^{-6-2\delta/3},
\end{align}
in $\un\cap\deux$. We conclude the proof by integrating this bound for $k=1$ over $\Sigma_0[\wbar_1,\wbar_2]$.
	\end{proof}
	
	\begin{rem}Note that although $A$ and $A^{(\un)}$ coincide on $\Sigma_0$, their derivatives transverse to $\Sigma_0$ do not necessarily coincide.
	\end{rem}
	\begin{lem}\label{lem:ceciestunlem}
		We have the following estimate, in $\un\cap\deux$, 
		\begin{align*}
			&\left|\df^{\leq 1}\left(\left(\big(\mcd\hot\mcd(Y_{m,2}(\cos\theta)e^{im\phi_+})\big)^S_\mck\right)^H-\big[\mcd^{(\un)}\hot\mcd^{(\un)}(Y_{m,2}(\cos\theta^{(\un)})e^{im\phi_+^{(\un)}})\big]^H\right)\right|\\
			&\underset{C_R}{\lesssim} \varepsilon\ubar^{-2-\delta/5}+|\df^{\leq 3}(J^{(+)}-J^{(+)}_\mck,J^{(-)}-J^{(-)}_\mck,J^{(0)}-J^{(0)}_\mck)|,
		\end{align*}
	where
	\begin{align*}
		J^{(+)}&=\sin\theta^{(\un)}\cos\phi_+^{(\un)},\quad J^{(+)}_\mck=\sin\theta_\mck\cos(\phi_+)_\mck,\\
		J^{(-)}&=\sin\theta^{(\un)}\sin\phi_+^{(\un)},\quad J^{(-)}_\mck=\sin\theta_\mck\sin(\phi_+)_\mck,\\
		J^{(0)}&=\cos\theta^{(\un)},\quad J^{(0)}_\mck=\cos\theta_\mck.
	\end{align*}
	\end{lem}
	\begin{proof}
		We first write the bound
		\begin{align}
			&\left|\df^{\leq 1}\left(\left(\big[\mcd\hot\mcd(Y_{m,2}(\cos\theta)e^{im\phi_+})\big]^S_\mck\right)^H-\big[\mcd^{(\un)}\hot\mcd^{(\un)}(Y_{m,2}(\cos\theta^{(\un)})e^{im\phi_+^{(\un)}})\big]^H\right)\right|\nn\\
			&\lesssim \left|\df^{\leq 1}\left(\mcd\hot\mcd(Y_{m,2}(\cos\theta_\mck)e^{im(\phi_+)_\mck})-\left(\big[\mcd\hot\mcd(Y_{m,2}(\cos\theta)e^{im\phi_+})\big]^S_\mck\right)^H\right)\right|\nn\\
			&\quad +\left|\df^{\leq 1}\left(\mcd\hot\mcd(Y_{m,2}(\cos\theta_\mck)e^{im(\phi_+)_\mck})-\big[\mcd^{(\un)}\hot\mcd^{(\un)}(Y_{m,2}(\cos\theta^{(\un)})e^{im\phi_+^{(\un)}})\big]^H\right)\right|\nn\\
			&\lesssim \left|\df^{\leq 1}\left(\big[\mcd\hot\mcd(Y_{m,2}(\cos\theta_\mck)e^{im(\phi_+)_\mck})\big]^S-\big[\mcd\hot\mcd(Y_{m,2}(\cos\theta)e^{im\phi_+})\big]^S_\mck\right)\right|\nn\\
			&\quad +\left|\df^{\leq 1}\left(\mcd\hot\mcd(Y_{m,2}(\cos\theta_\mck)e^{im(\phi_+)_\mck})-\big[\mcd^{(\un)}\hot\mcd^{(\un)}(Y_{m,2}(\cos\theta^{(\un)})e^{im\phi_+^{(\un)}})\big]^H\right)\right|,\label{eq:lastermmmm}
		\end{align}
		where we used Corollary \ref{cor:comparder} to bound the first term in the last step. We begin by bounding the last term in the RHS of \eqref{eq:lastermmmm}. First, denoting $h:=Y_{m,2}(\cos\theta_\mck)e^{im(\phi_+)_\mck}$, for any PT-horizontal frame $(e_1^{(\un)},e_2^{(\un)})$, using \eqref{eq:guiliguili} twice in the context of with frame transformation \eqref{eq:frametransfo} with coefficients given by Proposition \ref{prop:linkPTprincipal} we get
\begin{align*}
	\nabla_a\nabla_bh=_s&\nabla_a^{(\un)}\nabla_b^{(\un)}h+\frac12\hat{f}_a\nabla_3^{(\un)}\nabla^{(\un)}h+\frac12\nabla_ae_3^{(\un)}(h)\hat{f}_b+\hat{\mct}\cdot\nabla^{(\un)}h\\
	&+\frac12 e_3^{(\un)}(h)\left(\nabla^{(\un)}_a\hat{f}_b+\frac12\hat{f}_a\nabla^{(\un)}_3\hat{f}_b+\hat{\mct}\cdot\hat{f}\right).
\end{align*} 
This yields
\begin{align*}
	\nabla_a\nabla_bY_{m,2}(\cos\theta_\mck)e^{im(\phi_+)_\mck}=_{s}&\nabla_a^{(\un)}\nabla_b^{(\un)}Y_{m,2}(\cos\theta^{(\un)})e^{im\phi_+^{(\un)}}+(\df^{\leq 1}{\wh{f}},\wh{f}\wh{f}^{\leq 2}\chibar^{(\un)})\df^{\leq 2}Y_{m,2}(\cos\theta_\mck)e^{im(\phi_+)_\mck}\\
	&+\df^{\leq 2}(Y_{m,2}(\cos\theta_\mck)e^{im(\phi_+)_\mck}-Y_{m,2}(\cos\theta^{(\un)})e^{im\phi_+^{(\un)}}),
\end{align*}
which implies, combined with the expressions \eqref{eq:bcpplustardsisdonc} for the $\ell=2$ spherical harmonics,
\begin{align*}
	&\left|\df^{\leq 1}\left(\mcd\hot\mcd(Y_{m,2}(\cos\theta_\mck)e^{im(\phi_+)_\mck})-\big[\mcd^{(\un)}\hot\mcd^{(\un)}(Y_{m,2}(\cos\theta^{(\un)})e^{im\phi_+^{(\un)}})\big]^H\right)\right|\nn\\
	&\lesssim \varepsilon\ubar^{-3-\delta/3}+|\df^{\leq 3}(J^{(+)}-J^{(+)}_\mck,J^{(-)}-J^{(-)}_\mck,J^{(0)}-J^{(0)}_\mck)|,
\end{align*}
where we also used \eqref{eq:statedestimatefbar}. All is left to do is to bound the first line on the RHS of \eqref{eq:lastermmmm}. By Proposition \ref{prop:changebasisder} we have for any scalar $h$,
\begin{align*}
	(\mcd\hot\mcd h)^S&=_s\mathring{\mcd}\hot(\mcd h)^S+\frac12\fbar\hot (f\cdot\nabring (\mcd h)^S)+\frac12\fbar\hot\nabring_4(\mcd h)^S\nn\\
	&\quad+\left(\frac12f+\frac18|f|^2\fbar\right)\hot\nabring_3(\mcd h)^S+\hat{\mct}\hot (\mcd h)^S,
\end{align*}
where $(\mcd h)^S=\mathring{\mcd} h+\frac12\fbar (f\cdot\nabring h)+\frac12\fbar\nabring_4 h+\left(\frac12f+\frac18|f|^2\fbar\right)\nabring_3 h$. Using this with the function $h=Y_{m,2}(\cos\theta_\mck)e^{im(\phi_+)_\mck}=h_\mck$ we deduce in $\un\cap\deux$ the bound 
\begin{align*}
	&\left|\df^{\leq 1}\left(\big[\mcd\hot\mcd(Y_{m,2}(\cos\theta_\mck)e^{im(\phi_+)_\mck})\big]^S-\big[\mcd\hot\mcd(Y_{m,2}(\cos\theta)e^{im\phi_+})\big]^S_\mck\right)\right|\\
	&\lesssim \left|\df^{\leq 1}\left(\df^{\leq 2}\widecheck{g},\df^{\leq 2}\widecheck{b},\widecheck{\fbar},\widecheck{f},\df^{\leq 1}\widecheck{\chi},\widecheck{\chibar}\right)\right|\lesssim\varepsilon\ubar^{-2-\delta/5},
\end{align*}
by \eqref{eq:voilauneborne}, where we used also \eqref{eq:omegacarrdansiinterdeux}.
	\end{proof}
\begin{prop}\label{eq:assezdinfo}
	For $J\in\{J^{(+)},\Jmoins,J^{(0)}\}$ (as defined in Lemma \ref{lem:ceciestunlem}) we have in $\un\cap\deux$
		\begin{align*}
	\left|\df^{\leq 3}(J-J_\mck,\ubar^{(\un)}-\ubar'_\mck,r^{(\un)}-r_\mck)\right|\underset{C_R}{\lesssim}\varepsilon\ubar^{-2-\delta/5}.
\end{align*}
\end{prop}
\begin{proof}
	By construction we have $e_3^{(\un)}(J)=0$ in $\un$. Moreover we have (see Proposition \ref{prop:linkPTprincipal} and Remark \ref{rem:remlabel}) in $\un\cap\deux$ the identity
	\begin{align*}
		e_3^{(\un)}&=e_3=\hat{\lambda}^{-1}\left(\left(1+\frac12f\cdot\fbar+\frac{1}{16}|f||\fbar|^2\right)\ering_3+\left(\fbar^A+\frac14|\fbar|^2 f^A\right)\partial_{\theta^A}+\frac14|\fbar|^2\ering_4\right)\\
		&=\hat{\lambda}^{-1}\left(\left(1+\frac12f\cdot\fbar+\frac{1}{16}|f||\fbar|^2\right)\partial_u+\left(\fbar^A+\frac14|\fbar|^2 f^A+\frac14|\fbar|^2\Omega^{-2}b^A\right)\partial_{\theta^A}+\frac14|\fbar|^2\Omega^{-2}\partial_\ubar\right).
	\end{align*}
Thus by the identity $(e_3(J))_\mck=0$ and \eqref{eq:voilauneborne} and \eqref{eq:voilabornefbarcheckkk} we get
\begin{align}\label{eq:llliun}
	|\df^{\leq 3}e_3^{(\un)}(J-J_\mck)|\underset{C_R}{\lesssim}\varepsilon\ubar^{-2-\delta/5},\quad (\un\cap\deux).
\end{align}
We argue similarly to obtain the bounds 
\begin{align}\label{eq:lllideux}
	|\df^{\leq 3}e_3^{(\un)}(\ubar^{(\un)}-\ubar'_\mck,r^{(\un)}-r_\mck)|\underset{C_R}{\lesssim}\varepsilon\ubar^{-2-\delta/5},\quad (\un\cap\deux).
\end{align}
Moreover, by the initialization of $u,\ubar,\theta^A$ on $\Sigma_0$ (see Definition \ref{defi:sigmazeroII}), we have the identities
$$(J-J_\mck)|_{\Sigma_0}=(\ubar^{(\un)}-\ubar'_\mck)|_{\Sigma_0}=(r^{(\un)}-r_\mck)|_{\Sigma_0}=0.$$
Thus integrating \eqref{eq:llliun} and \eqref{eq:lllideux} along the flow of $e_3^{(\un)}$ from $\Sigma_0$ in $\un\cap\deux$ (recall $e_3^{(\un)}=-\partial_{r^{(\un)}}$ in coordinates $r^{(\un)},\ubar^{(\un)},\theta^{(\un)},\phi_+^{(\un)}$) we get the bound 
$$\left|(J-J_\mck,\ubar^{(\un)}-\ubar'_\mck,r^{(\un)}-r_\mck)\right|\underset{C_R}{\lesssim}\varepsilon\ubar^{-2-\delta/5}$$
in $\un\cap\deux$, which can be generalized to $\df^{\leq 3}$ derivatives by successive commutations, similarly as in the proof of Proposition \ref{prop:linkPTprincipal} (except that there the equation is linear in the present setting so there is no need for a bootstrap argument), we skip the details.
\end{proof}
	\begin{prop}\label{prop:pointwisepsietpsiun}
		Provided $u_f(a,M,\delta,C_R)$ is negative enough, we have in $\un\cap\deux$, 
		$$|\df^{\leq 1}(\Psi-(\Psi^{(\un)})^H)|\lesssim_Q \ubar^{-6-\delta/2}.$$
	\end{prop}
	\begin{proof}
		We have, by \eqref{eq:ansatzIcapII} and \eqref{eq:defansatzdeux} (see also Lemma \ref{lem:ubarssontpareils}),
		\begin{align*}
			&|\df^{\leq 1}(\Psi-(\Psi^{(\un)})^H)|\lesssim \frac{1}{\ubar^6}\left|\df^{\leq 1}\left(J^{(0)}-J^{(0)}_\mck,\ubar_\mck'-\ubar^{(\un)},r_\mck-r^{(\un)}, \right)\right|\\
			&+\frac{1}{\ubar^6}\left|\df^{\leq 1}\left(\left(\big[\mcd\hot\mcd(Y_{m,2}(\cos\theta)e^{im\phi_+})\big]^S_\mck\right)^H-\big[\mcd^{(\un)}\hot\mcd^{(\un)}(Y_{m,2}(\cos\theta^{(\un)})e^{im\phi_+^{(\un)}})\big]^H\right)\right|\underset{C_R}{\lesssim} \ubar^{-8-\delta/5},
		\end{align*}
		where we used Lemma \ref{lem:ceciestunlem} and Proposition \ref{eq:assezdinfo}. We get rid of the dependence in $C_R$ by choosing $u_f(a,M,\delta,C_R)$ negative enough and sacrificing some power of $\ubar$ in the estimate.
	\end{proof}
	\begin{cor}\label{prop:lierpsiunetpsideux}
		We have the following energy bound on $\Sigma_0$,
		$$\int_{\Sigma_0[\wbar_1,\wbar_2]}\mathbf{e}\left[(\Psi^{(\un)})^H-\Psi\right]\lesssim_Q\int_{\wbar_1}^{\wbar_2}\wbar^{-2(6+\delta/2)}\dee\wbar.$$
	\end{cor}	
	We deduce, from Proposition \ref{prop:lierAetAun}, Corollary \ref{prop:lierpsiunetpsideux} and \ref{prop:pointwisepsietpsiun} combined with \eqref{eq:energyhypfromI} and \eqref{pointwisehypfromI}, the following pointwise and energy bounds on $\Sigma_0$, expressed with respect to the ingoing non-integrable frame of region $\deux$\footnote{	Note that to obtain \eqref{eq:pointwiseutilisabledansII}, we also use $A=A^{(\un)}$ on $\Sigma_0$. Also, to obtain \eqref{eq:energyhyputilisabledansII}, we use 
		$$ \int_{\Sigma_0[\wbar_1,\wbar_2]}\mathbf{e}\left[(A^{(\un)})^H-(\Psi^{(\un)})^H\right]\lesssim\int_{\Sigma_0[\wbar_1,\wbar_2]}|\df^{\leq 1}(A^{(\un)}-\Psi^{(\un)})|^2\lesssim C(\delta_\pm,w_f,(Q_m)_{|m|\leq2})\int_{\wbar_1}^{\wbar_2}\wbar^{-2(6+\delta/2)}\dee\wbar,$$
		which holds by \eqref{eq:bonutilehein}.},
	\begin{align}
		\big|A|_{\Sigma_0}-\Psi|_{\Sigma_0}\big|&\lesssim C(\delta_\pm,w_f,(Q_m)_{|m|\leq2})\ubar^{-6-\delta/2},\label{eq:pointwiseutilisabledansII}\\
		\int_{\Sigma_0[\wbar_1,\wbar_2]}\mathbf{e}\left[A-\Psi\right]& \lesssim C(\delta_\pm,w_f,(Q_m)_{|m|\leq2})\int_{\wbar_1}^{\wbar_2}\wbar^{-2(6+\delta/2)}\dee\wbar.\label{eq:energyhyputilisabledansII}
	\end{align}

	\begin{rem}
		Note that \eqref{eq:pointwiseutilisabledansII}, \eqref{eq:energyhyputilisabledansII} imply in particular the following less precise pointwise and energy bounds for $A$ on $\Sigma_0$,
		\begin{align}\label{eq:initA4plusdelta}
			|A||_{\Sigma_0}\lesssim C(\delta_\pm,w_f,(Q_m)_{|m|\leq2})\ubar^{-4-\delta},\quad\int_{\Sigma_0[\wbar_1,\wbar_2]}\mathbf{e}[A]\lesssim C(\delta_\pm,w_f,(Q_m)_{|m|\leq2})\int_{\wbar_1}^{\wbar_2}\wbar^{-2(4+\delta)}\dee\wbar.
		\end{align}
	\end{rem}
	
	We will also need some non-sharp pointwise and energy bounds for derivatives of $A$ on $\Sigma_0$, which we state here for convenience. We define the following set of operators in region $\deux$,
	\begin{align*}
		\mathcal{S}:=\Big\{&\nabla_4^{\leq 2},\:\nabla_4^{\leq 2}\divc,\:\nabla_4^{\leq 2}\divc\divc,\:\nabla_4^{\leq 1}\mcd\divc\divc,\:\mcd\hot\divc,\:\divc\mcd\hot\divc,\:\divc\mcd\divc\divc,\\
		&(\hat{\lambda}\nabla_3)^{\leq 2},\:(\hat{\lambda}\nabla_3)^{\leq 2}\divc,\:(\hat{\lambda}\nabla_3)^{\leq 2}\divc\divc,\:(\hat{\lambda}\nabla_3)^{\leq 1}\mcd\divc\divc\Big\}.
	\end{align*}
	\begin{prop}\label{prop:enerpointwisefromISA}
		For $S\in\mathcal{S}$, we have the following bounds on $\Sigma_0$,
		\begin{align*}
			\big|(SA)\big||_{\Sigma_0}\lesssim\varepsilon\ubar^{-3-\delta/2},\quad\quad\int_{\Sigma_0[\wbar_1,\wbar_2]}\mathbf{e}\left[SA\right]\lesssim\int_{\wbar_1}^{\wbar_2}\wbar^{-2(3+\delta/2)}\dee\wbar.
		\end{align*}
	\end{prop}
	\begin{proof}
		By \eqref{eq:enfaitsicool} combined with \eqref{eq:bonutilehein} and \eqref{eq:controlI} we have
		\begin{align*}
			|(\df^{\leq 1}SA)|_{\Sigma_0}|\lesssim |(\df^{\leq 5}\alpha)|_{\Sigma_0}|+\varepsilon^2\ubar^{-6-2\delta/3}\lesssim\varepsilon\ubar^{-3-\delta/2},
		\end{align*}
which concludes the proof.
	\end{proof}
	\subsubsection{Bounds for $(\divc)^{\leq 1}(\wh{X},B)$ on $\Sigma_0$}
	
	\begin{prop}\label{prop:recoverboundsBXhatonsigma0}
		We have, on $\Sigma_0$,
		$$|(\divc)^{\leq 1}(\wh{X},B)|\lesssim\varepsilon\ubar^{-3-\delta/3}.$$
	\end{prop}
	\begin{proof}
		By the change of frame formulas \eqref{eq:changechihat}, \eqref{eq:changebeta} for $\chihat$ and $\beta$, and Proposition \ref{prop:linkPTprincipal}, we have in $\un\cap\deux$, 
		$$\wh{\chi}^{PT}=_s\wh{\chi}^{(\un)}+\df^{\leq 1}\hat{f}\cdot(\hat{f}^{\leq 1}\cdot\Gamma^{(\un)})+\mathcal{G}_{[+1]},\quad \beta^{PT}=_s\beta^{(\un)}+\hat{f}\cdot R^{(\un)}+\mathcal{R}_{[+1]}$$
		in $\un\cap\deux$, where $(\Gamma^{(\un)},R^{(\un)})$ denote the PT Ricci and curvature coefficients in $\un$ defined in Section \ref{section:linearizedPTun}. Thus, by \eqref{eq:bonutilehein}, \eqref{eq:statedestimatefbar} and by the bounds on $(\Gammacheck,\Rcheck)$ in \eqref{eq:controlI}, we get
		\begin{align*}
			|\df^{\leq 1}\wh{X}|\lesssim|\df^{\leq 1}\wh{X}^{(\un)}|+|\df^{\leq 2}\hat{f}|+|\df^{\leq 1}\mathcal{G}_{[+1]}|\lesssim \varepsilon\ubar^{-3-\delta/3},\\
			 |\df^{\leq 1}B|\lesssim|\df^{\leq 1}B^{(\un)}|+|\df^{\leq 1}\hat{f}|+|\df^{\leq 1}\mathcal{R}_{[+1]}|\lesssim \varepsilon\ubar^{-3-\delta/3},
		\end{align*}
		in $\un\cap\deux$, which concludes the proof.
	\end{proof}

\section{Analysis of Teukolsky equation in region $\deux$ }
\label{section:teukolskydeux}
The goal of this section is to propagate the precise asymptotics \eqref{eq:pointwiseutilisabledansII} of $A$ from $\Sigma_0$ to region $\deux$. This is done using an adaptation to the non-linear setting of the energy method for the Teukolsky equation introduced in \cite{spin+2} in the linearized setting (recall the discussion for region $\deux$ in Section \ref{section:instabesti}). We recall the horizontal distribution $\mch=(e_3,e_4)^{\bot}$, where $(e_3,e_4,e_a)$ corresponds to the ingoing non-integrable frame defined in Section \ref{section:defprincipalframes}. Also recall the notation $(\ering_3,\ering_4)$ for the double null pair in $\deux\cup\trois$.
\subsection{Preliminaries for the energy method in $\deux$}
\subsubsection{Constant $\wbar$ spacelike hypersurfaces}
\begin{prop}\label{prop:toutsurwbar}
	Recall the scalar function $\wbar=\ubar-{e^{-|\kappa_-|(u+\ubar)}}+e^{-|\kappa_-|C_R}$ defined in \eqref{eq:premoccuwbar}. Then the level sets of $\wbar$ are spacelike, and we have 
	\begin{align}
		-\D\wbar&=\frac{1}{2}|\kappa_-|{e^{-|\kappa_-|(u+\ubar)}}\ering_4+\frac{1}{2}\Omega^{-2}\left(1+|\kappa_-|{e^{-|\kappa_-|(u+\ubar)}}\right)\ering_3,\label{eq:Dwbar}\\
		\g(\D\wbar,\D\wbar)&=-|\kappa_-|{e^{-|\kappa_-|(u+\ubar)}}\Omega^{-2}\left(1+|\kappa_-|{e^{-|\kappa_-|(u+\ubar)}}\right)\sim -1.\label{eq:gDwbarDwbar}
	\end{align}
	Moreover, the volume form $\mu_{\wbar_1}$ induced by $\g$ on the level set $\{\wbar=\wbar_1\}$ in $\deux\cup\trois$ satisfies that there is a function $c>0$ such that $\mu_{\wbar_1}=c\Omega^2\mathrm{vol}_\gamma\dee u$, with $c\sim 1$.
\end{prop}
\begin{proof}
	We have $\partial_{\theta^A}\wbar=0$, $A=1,2$, thus $\D\wbar=-\frac{1}{2}\ering_4(\wbar)\ering_3-\frac{1}{2}\ering_3(\wbar)\ering_4$ with $\ering_3(\wbar)=|\kappa_-|{e^{-|\kappa_-|(u+\ubar)}}$, $\ering_4(\wbar)=\Omega^{-2}\left(1+|\kappa_-|{e^{-|\kappa_-|(u+\ubar)}}\right)$, which yields
	\begin{align*}
		\g(\D\wbar,\D\wbar)=-|\kappa_-|{e^{-|\kappa_-|(u+\ubar)}}\Omega^{-2}\left(1+|\kappa_-|{e^{-|\kappa_-|(u+\ubar)}}\right)\sim -1,
	\end{align*}
	where we used $\Omega^2\sim\Omega_\mck^2\sim{e^{-|\kappa_-|(u+\ubar)}}$. This proves that the level sets of $\wbar$ are spacelike. Next, we prove the statement about the volume form on the level sets of $\wbar$. Let $\wbar_1\geq 1$, and $\Sigma=\{\wbar=\wbar_1\}\cap(\deux\cap\trois)$. Then, by \eqref{eq:metricdoublenull} the Riemannian metric induced on $\Sigma$ is given in coordinates $(u,\theta^1,\theta^2)$ by 
	\begin{align*}
		\g|_{\Sigma}=&\frac{4\Omega^2|\kappa_-|{e^{-|\kappa_-|(u+\ubar)}}}{1+|\kappa_-|{e^{-|\kappa_-|(u+\ubar)}}}\dee u^2\\
		&+\gamma_{AB}\left(\dee\theta^A+\frac{|\kappa_-|{e^{-|\kappa_-|(u+\ubar)}}}{1+|\kappa_-|{e^{-|\kappa_-|(u+\ubar)}}}b^A\dee u\right)\left(\dee\theta^B+\frac{|\kappa_-|{e^{-|\kappa_-|(u+\ubar)}}}{1+|\kappa_-|{e^{-|\kappa_-|(u+\ubar)}}}b^B\dee u\right).
	\end{align*}
	This proves that the determinant of $\g|_\Sigma$ written in coordinates $(u,\theta^1,\theta^2)$ is 
	$$\det(\g|_\Sigma)=\frac{4\Omega^2|\kappa_-|{e^{-|\kappa_-|(u+\ubar)}}}{1+|\kappa_-|{e^{-|\kappa_-|(u+\ubar)}}}\det\gamma,$$
	and we conclude using $\Omega^2\sim {e^{-|\kappa_-|(u+\ubar)}} $.
\end{proof}
\begin{defi}
	For $u<u_f$ and $\wbar_1\geq C_R-u_f$, we denote $\ubar_{\wbar_1}(u)$ the value of $\ubar$ such that 
\begin{align}\label{eq:ehbonjoureuh}
	\wbar(u,\ubar_{\wbar_1}(u))=\wbar_1,\quad i.e.\quad \wbar_1=\ubar_{\wbar_1}(u)-e^{-|\kappa_-|(u+\ubar_{\wbar_1}(u))}+e^{-|\kappa_-|C_R}.
\end{align}
\end{defi}

\subsubsection{Energy and Hardy-type inequalities on $\{\wbar=cst\}$}
The next result will be used to get $L^2(S(u,\ubar))$ decay from \emph{degenerate} energy decay in $\deux$.
\begin{prop}\label{prop:inegwbarcst}
	Let $U$ be a horizontal tensor in $\deux$. Then for any $\wbar_1\geq C_R-u_f$, for $S(u,\ubar)\subset\deux$ such that $\wbar(u,\ubar)=\wbar_1$, we have
	\begin{align*}
		\|U\|_{L^2(S(u,\ubar))}\lesssim&\|U\|_{L^2(S(C_R-\wbar_1,\wbar_1))}\\
		&+\ubar^{\gamma/2}\left(\int_{\deux\cap\{\wbar=\wbar_1\}}\Omega^4\Big(|\nabla_{3}U|^2+|\nabla_{4}U|^2+|\nabla U|^2\Big)\mathrm{vol}_\gamma\dee u\right)^{1/2}.
	\end{align*}
\end{prop}
\begin{rem}
	This result is similar to \cite[Lem 4.6]{spin+2} in the linearized setting, except that here in the non-linear setting we bound the $L^2(S(u,\ubar))$ norm with an $\Omega^2$-degenerate energy - this only costs a factor $\ubar^{\gamma/2}$ in region $\deux$ - which is necessary to control some singular non-linear terms caused by the weak null singularity. Also, note that defining the degenerate energy
	\begin{align}\label{eq:defenerdeg}
		\Ener[U](\wbar_1):=\int_{\deux\cap\{\wbar=\wbar_1\}}\Omega^2\Big(|\nabla_{3}U|^2+|\nabla_{4}U|^2+|\nabla U|^2\Big),
	\end{align}
	where we omit the volume form $\mathrm{vol}_{\{\wbar=\wbar_1\}}\sim\Omega^2\mathrm{vol}_\gamma\dee u$, the result of Proposition \ref{prop:inegwbarcst} rewrites 
	$$\|U\|_{L^2(S(u,\ubar))}\lesssim\|U\|_{L^2(S(C_R-\wbar,\wbar))}+\ubar^{\gamma/2}\Ener[U](\wbar)^{1/2}.$$
\end{rem}
\begin{proof}[Proof of Proposition \ref{prop:inegwbarcst}]
	We consider the function 
	\begin{align}\label{eq:deffdeuuu}
		f(u):=\|U\|_{L^2(S(u,\ubar_{\wbar_1}(u)))}^2=\int_{S(u,\ubar_{\wbar_1}(u))}|U|^2,
	\end{align}
	where as usual we omit the volume form on $S(u,\ubar_{\wbar_1}(u))$, and where $|U|$ is the norm of the horizontal tensor $U$. Differentiating \eqref{eq:ehbonjoureuh} yields
	\begin{align}\label{eq:zaamais}
		\frac{\dee\ubar_{\wbar_1}}{\dee u}(u)=\frac{-|\kappa_-|e^{-|\kappa_-|(u+\ubar_{\wbar_1}(u))}}{1+|\kappa_-|e^{-|\kappa_-|(u+\ubar_{\wbar_1}(u))}}\sim e^{-|\kappa_-|(u+\ubar_{\wbar_1}(u))}.
	\end{align}
	Using Lemma \ref{lem:derintSf}, this gives:
	\begin{align*}
		\frac{\dee f}{\dee u}&=\int_{S(u,\ubar_{\wbar_1})} \left(\ering_3(|U|^2)+\mathring{tr\chibar}|U|^2-\frac{\Omega^2|\kappa_-|e^{-|\kappa_-|(u+\ubar_{\wbar_1}(u))}}{1+|\kappa_-|e^{-|\kappa_-|(u+\ubar_{\wbar_1}(u))}}\left(\ering_4(|U|^2)+\mathring{tr\chi} |U|^2\right)\right)\\
		&=2\int_{S(u,\ubar_{\wbar_1})}\Bigg(\langle U,\nabla_{\ering_3}U\rangle+\frac{1}{2}\mathring{tr\chibar}|U|^2-\frac{\Omega^2|\kappa_-|e^{-|\kappa_-|(u+\ubar_{\wbar_1}(u))}}{1+|\kappa_-|e^{-|\kappa_-|(u+\ubar_{\wbar_1}(u))}}\left(\langle U,\nabla_{\ering_4}U\rangle+\frac{1}{2}\mathring{tr\chi}|U|^2\right)\Bigg)\\
		&=\int_{S(u,\ubar_{\wbar_1})}\Big(\langle U,\nabla_W U\rangle+\mathcal{V}|U|^2\Big)\mathrm{vol}_\gamma,
	\end{align*}
	where we used Proposition \ref{prop:nablagammahzero} in the second step, and where
	\begin{align}\label{eq:VW}
		\mathcal{V}=\mathring{tr\chibar}-\frac{\Omega^2|\kappa_-|e^{-|\kappa_-|(u+\ubar_{\wbar_1}(u))}}{1+|\kappa_-|e^{-|\kappa_-|(u+\ubar_{\wbar_1}(u))}}\mathring{tr\chi},\quad W=2\ering_3-\frac{2\Omega^2|\kappa_-|e^{-|\kappa_-|(u+\ubar_{\wbar_1}(u))}}{1+|\kappa_-|e^{-|\kappa_-|(u+\ubar_{\wbar_1}(u))}}\ering_4.
	\end{align}
	By the Cauchy-Schwarz inequality we obtain:
	$$\left|\frac{\dee f}{\dee u}\right|\lesssim \|U\|_{L^2(S(u,\ubar_{\wbar_1}(u)))}\|\nabla_WU\|_{L^2(S(u,\ubar_{\wbar_1}(u)))}+\|\mathcal{V}\|_{L^\infty(S(u,\ubar_{\wbar_1}(u)))}\|U\|_{L^2(S(u,\ubar_{\wbar_1}(u)))}^2,$$
	which yields
	\begin{align}\label{eq:revientdanslegame}
		\left|\frac{\dee }{\dee u}\|U\|_{L^2(S(u,\ubar_{\wbar_1}(u)))}\right|\lesssim \|\nabla_WU\|_{L^2(S(u,\ubar_{\wbar_1}(u)))}+\|\mathcal{V}\|_{L^\infty(S(u,\ubar_{\wbar_1}(u)))}\|U\|_{L^2(S(u,\ubar_{\wbar_1}(u)))}.
	\end{align}
	Integrating this bound from $\Sigma_0$ to $(u,\ubar)$ we thus obtain 
	\begin{align*}
		\|U\|_{L^2(S(u,\ubar))}\lesssim&\|U\|_{L^2(S(C_R-\wbar_1,\wbar_1))}\\
		&+\int_{C_R-\wbar_1}^u\Big(\|\nabla_WU\|_{L^2(S(u',\ubar_{\wbar_1}(u')))}+\|\mathcal{V}\|_{L^\infty(S(u',\ubar_{\wbar_1}(u')))}\|U\|_{L^2(S(u',\ubar_{\wbar_1}(u')))}\Big)\dee u'.
	\end{align*}
	By Grönwall's inequality, this yields for some constant $C(a,M)>0$,
	\begin{align*}
		\|U\|_{L^2(S(u,\ubar))}\lesssim&\left(\|U\|_{L^2(S(C_R-\wbar_1,\wbar_1))}+\int_{\deux\cap\{\wbar=\wbar_1\}}\|\nabla_WU\|_{L^2(S(u',\ubar_{\wbar_1}(u')))}\dee u'\right)\\
		&\times\exp\left(C(a,M)\int_{C_R-\wbar_1}^u\|\mathcal{V}\|_{L^\infty(S(u',\ubar_{\wbar_1}(u')))}\dee u'\right).
	\end{align*}
	Moreover, we have by definition of $\mathcal{V}$ and \eqref{eq:voilauneborne},
	\begin{align*}
		\int_{C_R-\wbar_1}^u\|\mathcal{V}\|_{L^\infty(S(u',\ubar_{\wbar_1}(u')))}\dee u'\lesssim\int_{C_R-\wbar_1}^u\Bigg(\Omega^2(u,\ubar_{\wbar_1}(u'))+\frac{\varepsilon}{\ubar_{\wbar_1}(u')^{2+\delta/5}}\Bigg)\dee u'\lesssim 1,
	\end{align*}
	where we used Lemma \ref{lem:usimubarII} which implies $\ubar_{\wbar_1}(u')\sim |u'|$ in $\deux$. This yields
	\begin{align}\label{eq:qpresquefini}
		\|U\|_{L^2(S(u,\ubar))}\lesssim\|U\|_{L^2(S(C_R-\wbar_1,\wbar_1))}+\int_{\deux\cap\{\wbar=\wbar_1\}}\|\nabla_WU\|_{L^2(S(u',\ubar_{\wbar_1}(u')))}\dee u'.
	\end{align}
	Next, we express $W$ as defined in \eqref{eq:VW} with respect to the ingoing non-integrable frame $(e_\mu)$. By Remark \ref{rem:remlabel} and \eqref{eq:frametransfo}, and noticing $W\in TS(u,\ubar)^\bot$, we compute
	\begin{align*}
		&\g(W,e_a)=\frac{1}{2}\fbar_a\g(W,\ering_4)+\left(\frac{1}{2}f_a+\frac{1}{4}|f|^2\fbar_a\right)\g(W,\ering_3)=O(\Omega^2),\\
		&\g(W,e_3)=\hat{\lambda}^{-1}\left(1+\frac{1}{2}f\cdot\fbar+\frac{1}{16}|f|^2|\fbar|^2\right)\g(W,\ering_3)+\hat{\lambda}^{-1}\frac{|\fbar|^2}{4}\g(W,\ering_4)=O(\Omega^2),\\
		&\g(W,e_4)=\hat{\lambda}\g(W,\ering_4)+\hat{\lambda}\frac{|f|^2}{4}\g(W,\ering_3)=O(\Omega^2),
	\end{align*}
	where we used $|\fbar|\lesssim\Omega^2$, $|\hat{\lambda}|\sim\Omega^{2}$, and $\g(W,\ering_3)={4\Omega^2|\kappa_-|e^{-|\kappa_-|(u+u)}}/{(1+|\kappa_-|e^{-|\kappa_-|(u+u)})}=O(\Omega^4)$, $\g(W,\ering_4)=-4=O(1)$. This yields the following expression of $W$,
	\begin{align}\label{eq:WWWexpre}
		W=O(\Omega^2)e_3+O(\Omega^2)e_4+O(\Omega^2)^ae_a.
	\end{align}
	Going back to \eqref{eq:qpresquefini} and using the Cauchy-Schwarz inequality we get
	\begin{align*}
		&\|U\|_{L^2(S(u,\ubar))}\\
		&\lesssim\|U\|_{L^2(S(C_R-\wbar_1,\wbar_1))}+\int_{\deux\cap\{\wbar=\wbar_1\}}\left(\int_{S(u,\ubar_{\wbar_1}(u'))}\Omega^4\Big(|\nabla_{3}U|^2+|\nabla_{4}U|^2+|\nabla U|^2\Big)\mathrm{vol}_\gamma\right)^{1/2}\dee u'\\
		&\lesssim\|U\|_{L^2(S(C_R-\wbar_1,\wbar_1))}+\left(\int_{C_R-\wbar_1}^u\dee u'\right)^{1/2}\left(\int_{\deux\cap\{\wbar=\wbar_1\}}\Omega^4\Big(|\nabla_{3}U|^2+|\nabla_{4}U|^2+|\nabla U|^2\Big)\mathrm{vol}_\gamma\dee u'\right)^{1/2}.
	\end{align*}
	By the bound $u+\wbar_1-C_R=u+\ubar+O(1)\lesssim\ubar^\gamma$ in $\deux$, we conclude the proof.
\end{proof}
We continue with a Hardy-type inequality on $\{\wbar=cst\}$ which will allow us to deal with the zero-order terms in the energy method for the Teukolsky equation. We define the following scalar functions of $(u,\ubar)$,
\begin{align}\label{eq:defpoids}
	\Deltahat:=e^{-|\kappa_-|(u+\ubar)},\quad\varpi:=1+\Deltahat,
\end{align}
(see the analog scalar $\varpi$ in \cite[(4.18)]{stabC0}) which satisfy
	\begin{align}\label{eq:weightpointwise}
		\Deltahat\sim\Omega^2,\quad\varpi\sim 1.
	\end{align}
\begin{prop}\label{prop:hardy}
	Let $U$ be a horizontal tensor in $\deux$. Then for $C_R(a,M)$ and $|u_f(a,M,C_R)|$ large enough, we have for $\wbar_1\geq C_R-u_f$,
	\begin{align*}
		\int_{\deux\cap\{\wbar=\wbar_1\}}\Omega^2 |U|^2\lesssim \|U\|_{L^2(S(C_R-\wbar_1,\wbar_1))}^2+\int_{\deux\cap\{\wbar=\wbar_1\}}\Omega^4\left(|\nabla_3U|^2+|\nabla_4U|^2+|\nabla U|^2\right).
	\end{align*}
\end{prop}
\begin{proof}
Recall the function $f(u)=\|U\|_{L^2(S(u,\ubar_{\wbar_1}(u)))}$ defined in \eqref{eq:deffdeuuu}. Then, recalling from Proposition \ref{prop:toutsurwbar} the volume form $\mu_{\wbar_1}\sim\Omega^2\mathrm{vol}_\gamma\dee u$ on $\{\wbar=\wbar_1\}$, and defining $u_\Gamma(\wbar_1)$ as the value of $u$ at $\Gamma\cap\{\wbar=\wbar_1\}$, we get
	\begin{align}
		\int_{\deux\cap\{\wbar=\wbar_1\}}\Omega^2 |U|^2&\lesssim\int_{C_R-\wbar_1}^{u_\Gamma(\wbar_1)}\int_{S(u,\ubar_{\wbar_1}(u))}\Omega^4 |U|^2\mathrm{vol}_\gamma\dee u\lesssim \int_{C_R-\wbar_1}^{u_\Gamma(\wbar_1)}\Deltahat^2(u,\ubar_{\wbar_1}(u))f(u)^2\dee u\nn\\
		&\lesssim e^{-2|\kappa_-|\wbar_1}\int_{C_R-\wbar_1}^{u_\Gamma(\wbar_1)}e^{-2|\kappa_-|u}f(u)^2\dee u,\label{eq:ciao}
	\end{align}
	where in the last step we used $|\wbar_1-\ubar_{\wbar_1}(u)|\lesssim 1$ and the definition of $\Deltahat(u,\ubar)$ to write $\Deltahat^2(u,\ubar_{\wbar_1}(u))\sim e^{-2|\kappa_-|(u+\wbar_1)}$. Now, integrating the identity
	$$\partial_u\left(e^{-2|\kappa_-|u}f(u)^2\right)=-2|\kappa_-|e^{-2|\kappa_-|u}f(u)^2+2e^{-2|\kappa_-|u}f'(u)f(u)$$
	from $C_R-\wbar_1$ to $u_\Gamma(\wbar_1)$, and using $|\kappa_-|\sim 1$, we get for any $s>0$,
	\begin{align}
		\int_{C_R-\wbar_1}^{u_\Gamma(\wbar_1)}e^{-2|\kappa_-|u}f(u)^2\dee u&\lesssim  \int_{C_R-\wbar_1}^{u_\Gamma(\wbar_1)}e^{-2|\kappa_-|u}f'(u)f(u)\dee u-\int_{C_R-\wbar_1}^{u_\Gamma(\wbar_1)}\partial_u\left(e^{-2|\kappa_-|u}f(u)^2\right)\dee u\nn\\
		&\lesssim s\int_{C_R-\wbar_1}^{u_\Gamma(\wbar_1)}e^{-2|\kappa_-|u}f(u)^2\dee u+s^{-1}\int_{C_R-\wbar_1}^{u_\Gamma(\wbar_1)}e^{-2|\kappa_-|u}f'(u)^2\dee u\nn\\
		&\quad -\big(e^{-2|\kappa_-|u_\Gamma(\wbar_1)}h^2(u_\Gamma(\wbar_1))-e^{-2|\kappa_-|(C_R-\wbar_1)}f^2(C_R-\wbar_1)\big)\nn\\
		&\lesssim \int_{C_R-\wbar_1}^{u_\Gamma(\wbar_1)}e^{-2|\kappa_-|u}f'(u)^2\dee u+e^{-2|\kappa_-|(C_R-\wbar_1)}f^2(C_R-\wbar_1),\label{eq:ciaociao}
	\end{align}
	where we chose $s>0$ sufficiently small and drop the non-positive term $-e^{-2|\kappa_-|u_\Gamma(\wbar_1)}f^2(u_\Gamma(\wbar_1))$ in the last step. Now, by \eqref{eq:revientdanslegame} we have $f'(u)^2\lesssim  \|\nabla_WU\|_{L^2(S(u,\ubar_{\wbar_1}(u)))}^2+\|\mathcal{V}\|_{L^\infty(S(u,\ubar_{\wbar_1}(u)))}^2 f(u)^2$, where $\mathcal{V}$ and $W$ are defined in \eqref{eq:VW}. Thus, by \eqref{eq:voilauneborne} and $\ubar\sim\wbar$, 
	\begin{align*}
		\int_{C_R-\wbar_1}^{u_\Gamma(\wbar_1)}e^{-2|\kappa_-|u}f'(u)^2\dee u&\lesssim \int_{C_R-\wbar_1}^{u_\Gamma(\wbar_1)}e^{-2|\kappa_-|u} \|\nabla_WU\|_{L^2(S(u,\ubar_{\wbar_1}(u)))}^2\dee u \\
		&\quad+\int_{C_R-\wbar_1}^{u_\Gamma(\wbar_1)}e^{-2|\kappa_-|u}(\Omega^4+\wbar_1^{-4-2\delta/5})f(u)^2\dee u.
	\end{align*}
	Thus, choosing $C_R(a,M)\gg 1$ and $|u_f(a,M,C_R)|\gg 1$ large enough (hence $\wbar_1\gtrsim C_R-u_f\gg 1$ is large) so that the last term on the RHS above is absorbed in the LHS of \eqref{eq:ciaociao}, we get
	\begin{align*}
		\int_{C_R-\wbar_1}^{u_\Gamma(\wbar_1)}e^{-2|\kappa_-|u}f(u)^2\dee u\lesssim&\int_{C_R-\wbar_1}^{u_\Gamma(\wbar_1)}e^{-2|\kappa_-|u} \|\nabla_WU\|_{L^2(S(u,\ubar_{\wbar_1}(u)))}^2\dee u&+e^{2|\kappa_-|(\wbar_1-C_R)}h^2(C_R-\wbar_1).
	\end{align*}
	By \eqref{eq:ciao} we thus deduce 
	\begin{align*}
		\int_{\deux\cap\{\wbar=\wbar_1\}}\Omega^2 |U|^2&\lesssim \int_{C_R-\wbar_1}^{u_\Gamma(\wbar_1)}e^{-2|\kappa_-|(u+\wbar_1)} \|\nabla_WU\|_{L^2(S(u,\ubar_{\wbar_1}(u)))}^2\dee u+h^2(C_R-\wbar_1)\\
		&\lesssim \|U\|_{L^2(S(C_R-\wbar_1,\wbar_1))}^2+\int_{\deux\cap\{\wbar=\wbar_1\}}|\nabla_W U|^2,
	\end{align*}
	where we used $e^{-2|\kappa_-|(u+\wbar_1)}\sim e^{-2|\kappa_-|(u+\ubar_{\wbar_1}(u))}=\Deltahat^2(u,\ubar_{\wbar_1}(u))$ and the statement on the volume form on $\{\wbar=\wbar_1\}$ in Proposition \ref{prop:toutsurwbar}. Finally, we conclude the proof by the identity \eqref{eq:WWWexpre} which implies $|\nabla_W U|^2\lesssim\Omega^4(|\nabla_3U|^2+|\nabla_4U|^2+|\nabla U|^2)$.
\end{proof}
\subsubsection{Weight functions and integration by parts identities}
Recall from  \eqref{eq:defpoids} the scalars $\Deltahat:=e^{-|\kappa_-|(u+\ubar)},\:\:\varpi:=1+\Deltahat$, and the estimates \eqref{eq:weightpointwise}.
\begin{prop}
	We have the following estimates,
	\begin{align}\label{eq:weightder}
		e_3(\varpi)=e_3(\Deltahat)\sim -1,\quad e_4(\varpi)=e_4(\Deltahat)\sim -\Deltahat,\quad |\nabla(\Deltahat,\varpi)|\lesssim\Deltahat.
	\end{align}
\end{prop}
\begin{proof}
By Remark \ref{rem:remlabel} and \eqref{eq:bornecoeffin}, which imply in particular $\hat{\lambda}\sim\Deltahat$ and $\partial_{\theta^A}(\varpi,\Deltahat)=0$,
	\begin{align*}
		e_3(\varpi,\Deltahat)&=\hat{\lambda}^{-1}\Bigg(\left(1+\frac{1}{2}f\cdot\fbar+\frac{1}{16}|f|^2|\fbar|^2\right)\partial_u+\frac{1}{4}|\fbar|^2\Omega^{-2}\partial_\ubar\bigg)\left[e^{-|\kappa_-|(u+\ubar)}\right]\\
		&=-\hat{\lambda}^{-1}\left(1+\frac{1}{2}f\cdot\fbar+\frac{1}{16}|f|^2|\fbar|^2\right)|\kappa_-|e^{-|\kappa_-|(u+\ubar)}+O(\Omega^2)\sim -1,
	\end{align*}
	as well as 
	\begin{align*}
		e_4(\varpi,\Deltahat)&=\hat{\lambda}\left(\Omega^{-2}\partial_\ubar+\frac{1}{4}|f|^2\partial_u\right)\left[e^{-|\kappa_-|(u+\ubar)}\right]=-\hat{\lambda}\Omega^{-2}|\kappa_-|e^{-|\kappa_-|(u+\ubar)}+O(\Omega^4)\sim -\Deltahat,\\
		e_a(\varpi,\Deltahat)&=\Bigg(\frac{1}{2}\fbar_a\Omega^{-2}\partial_\ubar+\left(\frac{1}{2}f_a+\frac{1}{8}|f|^2\fbar_a\right)\partial_u\Bigg)\left[e^{-|\kappa_-|(u+\ubar)}\right]=O(\Deltahat),
	\end{align*}
	which concludes the proof.
\end{proof}

We continue this section with a few integration by parts results.
\begin{prop}\label{prop:ippnab3}
	Let ${U}$ be a horizontal tensor and $N\geq 1$. We have the identity
	\begin{align}
		\Real\left(\Deltahat \varpi^N\overline{{\nabla_3}{{U}}}\cdot\nabla_4{\nabla_3}{{U}}\right)&=\frac{1}{2}\Big[{\Deltahat \varpi^N}(2\omega-tr\chi)-e_4(\Deltahat \varpi^N)\Big]\left|{{\nabla_3}{{U}}}\right|^2+\D_\mu\left(\frac{\Deltahat \varpi^N}{2}\left|{{\nabla_3}{{U}}}\right|^2e_4^\mu\right),\nn\\
				\Real\left(\Deltahat \varpi^N\overline{{\nabla_4}{{U}}}\cdot\nabla_3{\nabla_4}{{U}}\right)&=-\frac{1}{2}\Big[{\Deltahat \varpi^N}tr\chibar+e_3(\Deltahat \varpi^N)\Big]\left|{{\nabla_4}{{U}}}\right|^2+\D_\mu\left(\frac{\Deltahat \varpi^N}{2}\left|{{\nabla_4}{{U}}}\right|^2e_3^\mu\right).\nn
	\end{align}
\end{prop}
\begin{proof}
The proof is the same as the one of Proposition \ref{prop:Iippnab3}.
\end{proof}
\begin{prop}\label{prop:ipphoriz}
	Let ${U}\in\fraks_k(\C)$. Then, for any vector field $X$ and $N\geq 1$, we have
	\begin{align*}
		\Real\left(\poids\overline{\nabla_X{U}}\cdot(-\triangle_k{U})\right)=&-\frac{1}{2}\left(\poids\D_\mu X^\mu+\nabla_X(\poids)\right)|\nabla{U}|^2-\D^\mu(\poids\nu^{(X)}_\mu[U])\\
		&+\D_\mu\left(\frac{\Deltahat\varpi^N}{2}|{\nabla}{U}|^2X^\mu\right)-\Deltahat\varpi^N\Real\left(\overline{[\nabla_X,{\nabla^a}]{U}}\cdot{\nabla}_a{U}\right)\\
		&+\poids(\eta+\etabar)\cdot\nu^{(X)}[U]+\nabla(\Deltahat\varpi^N)\cdot \nu^{(X)}[U],
	\end{align*}
	where we define the spacetime 1-form $\nu^{(X)}[U]$ by 
	$$\nu^{(X)}_a[U]=\Real\left(\overline{\nabla_X{U}}\cdot{\nabla}_a{U}\right),\quad\nu^{(X)}_3[U]=\nu^{(X)}_4[U]=0.$$
\end{prop}
\begin{proof}
The proof is the same as the one of Proposition \ref{prop:Iipphoriz}.
\end{proof}
\begin{defi}
	We define the following short-hand notations,
	\begin{align}\label{eq:shorthandnotation}
		\nu^{(3)}_a[U]:=\nu^{(e_3)}_a[U]=\Real\left(\overline{\nabla_3{U}}\cdot{\nabla}_a{U}\right),\quad \nu^{(4)}_a[U]:=\nu^{(e_4)}_a[U]=\Real\left(\overline{\nabla_4{U}}\cdot{\nabla}_a{U}\right).
	\end{align}
\end{defi}

\subsection{Energy and decay estimates for generalized Teukolsky equations in $\deux$}\label{section:energyteukgeneralizedsectionpassubsection}
In this section, we implement an energy method for generalized Teukolsky-type equations in $\deux$, which will be satisfied by $A$ and some derivatives of $A$. 
\begin{itemize}
	\item First, in Section \ref{section:energygeneralized}, we introduce the energy method for generalized Teukolsky fields, that is horizontal tensors which satisfy a Teukolsly-like equation.
	\item In Section \ref{section:energyA}, we specify this energy method to the case of the Teukolsky equation \eqref{eq:teukA} for $A$. This energy method will be used to control the energy of $A$ (first, in a non-sharp way) and then eventually to obtain sharp control of $A$.
	\item Finally, in Section \ref{section:decayL2generalizedteukolsky}, from the energy method we prove a $L^2(S(u,\ubar))$ decay result for solutions of generalized Teukolsky fields.
\end{itemize}

\subsubsection{Energy estimates for generalized Teukolsky fields}\label{section:energygeneralized}

\begin{figure}[h!]
	\centering
	\includegraphics[scale=0.49]{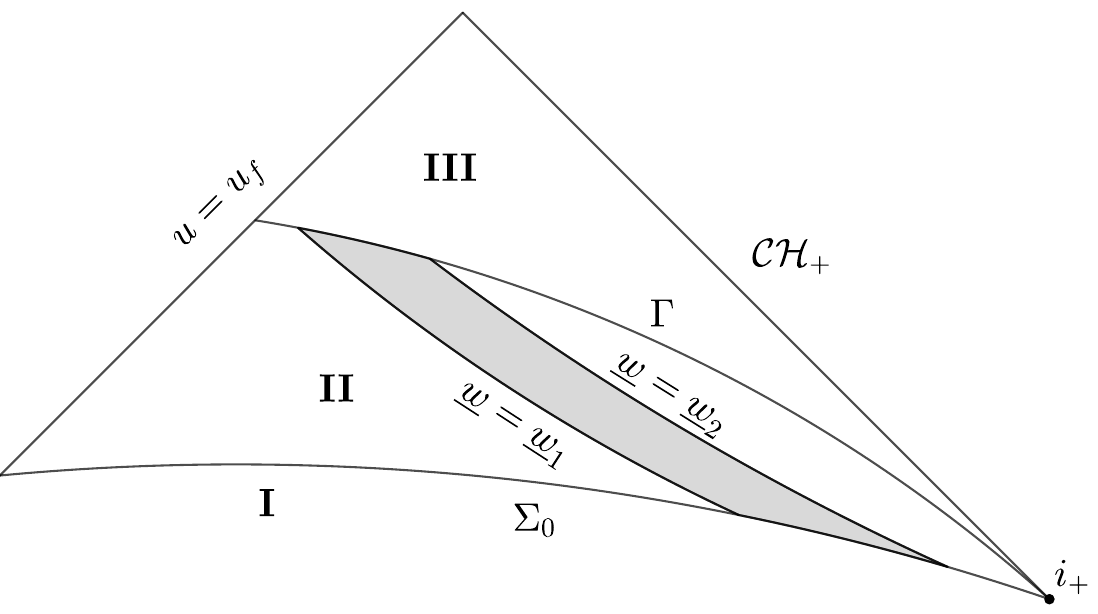}
	\caption{Region $\deux[\wbar_1,\wbar_2]$ in grey, for $\wbar_1\geq\wbar|_{\Gamma\cap\{u=u_f\}}$.}
	\label{fig:wbarIInl}
\end{figure}

In this section, we assume that ${U}\in\fraks_k(\mathbb{C})$, $k\in\{0,1,2\}$, satisfies a wave equation
\begin{align}\label{eq:teukmod}
	\wh{\mcl}({U})=\err[\wh{\mcl}({U})],\quad\text{in}\:\:\deux,
\end{align}
where $\wh{\mcl}$ is a Teukolsky-like wave operator, namely such that
\begin{align}\label{eq:operateurLhat}
	\wh{\mcl}({U})&=\nabla_4\nabla_3 {U}-\triangle_k{U}+h\nabla_3 {U}+\underline{h}\nabla_4{U}+L_1[U]+L[U],
\end{align}
where $h,\underline{h}$ are functions, $L_1$ is a horizontal first-order operator, and $L$ is a $0$-order operator (in a sense which will be clarified later, see \eqref{eq:hypLhat}). For $\wbar_1\leq\wbar_2$, we define the following subregion (see Figure \ref{fig:wbarIInl} for an illustration),
\begin{align}
	\deux[\wbar_1,\wbar_2]:=\deux\cap\{\wbar_1\leq\wbar\leq\wbar_2\}.\label{eq:defIIwbar12}
\end{align}
\begin{prop}\label{prop:calculcommunenergy}
	Let $\wh{\mcl}$ be given by \eqref{eq:operateurLhat},  ${U}\in\fraks_k(\mathbb{C})$, $k\in\{0,1,2\}$, and $\wbar_1\leq\wbar_2$. For $\wbar_1\geq\wbar|_{\Gamma\cap\{u=u_f\}}$ we have the following identity, for any $N>1$,
	\begin{align*}
		\int_{\deux[\wbar_1,\wbar_2]}\Deltahat\varpi^N\Real&\left((\overline{\nabla_3{U}+\nabla_4{U}})\cdot\wh{\mcl}({U})\right)=\\
		&\int_{\deux[\wbar_1,\wbar_2]}\mathbf{B}^{(N)}[U]+\int_{\deux\cap\{\wbar=\wbar_2\}}\mathbf{F}_{\wbar}^{(N)}[U]-\int_{\deux\cap\{\wbar=\wbar_1\}}\mathbf{F}_{\wbar}^{(N)}[U]\\
		&+\int_{\Gamma\cap\{\wbar_1\leq\wbar\leq\wbar_2\}}\mathbf{F}^{(N)}_\Gamma[U]-\int_{\Sigma_0\cap\{\wbar_1\leq\wbar\leq\wbar_2\}}\mathbf{F}_{0}^{(N)}[U],
	\end{align*}
	where the boundary terms satisfy the following estimates : for $i=1,2$, 
	\begin{align}
		&\int_{\deux\cap\{\wbar=\wbar_i\}}\mathbf{F}^{(N)}_{\wbar}[U]\sim\int_{\deux\cap\{\wbar=\wbar_i\}}\Deltahat\varpi^N\left(|\nabla_3U|^2+|\nabla_4U|^2+|{\nabla} U|^2)\right),\label{eq:estimbdwbar}\\
		&\int_{\Sigma_0\cap\{\wbar_1\leq\wbar\leq\wbar_2\}}\mathbf{F}_0^{(N)}[U]\lesssim_{C_R}\int_{\Sigma_0\cap\{\wbar_1\leq\wbar\leq\wbar_2\}}\varpi^N\left(|\nabla_3U|^2+|\nabla_4U|^2+|{\nabla} U|^2)\right),\label{eq:estimbdsigmazero}\\
		&\int_{\Gamma\cap\{\wbar_1\leq\wbar\leq\wbar_2\}}\mathbf{F}_{\Gamma}^{(N)}[U]\geq 0,\label{eq:estimbdGamma}
	\end{align}
	provided $C_R(a,M)\gg 1$ is large enough, and where, recalling \eqref{eq:ckdeu}, \eqref{eq:shorthandnotation}, the bulk term is
	\begin{align*}
		\mathbf{B}^{(N)}[U]:=&\mathbf{B}_{pr}^{(N)}[U]+\nabla^a(\poids)\Real\left(\overline{\nabla_3U}\cdot\nabla_a U\right)+\poids\Big[-\Real\left(\overline{[\nabla_3,\nabla^a] U}\cdot\nabla_a U\right)\\
		&+\Real\left(\underline{h}\overline{\nabla_3 U}\cdot\nabla_4 U\right)+\Real\left(\overline{\nabla_3 U}\cdot L_1[U]\right)+\Real\left(\overline{\nabla_3 U}\cdot L[U]\right)\Big]+\poids(\eta+\etabar)\cdot \nu^{(3)}[U]\\
		&+\nabla^a(\poids)\Real\left(\overline{\nabla_4U}\cdot\nabla_a U\right)+\poids\Big[-\Real\left(\overline{[\nabla_4,\nabla^a] U}\cdot\nabla_a U\right)\\
		&+\Real\left((h+2\omega)\overline{\nabla_4 U}\cdot\nabla_3 U\right)+\Real\left(\overline{\nabla_4U}\cdot( L_1[U]+2(\eta-\etabar)\cdot\nabla U)\right)\\
		&+\Real\left(\overline{\nabla_4 U}\cdot (L[U]-C_k[U])\right)\Big]+\poids(\eta+\etabar)\cdot \nu^{(4)}[U],
	\end{align*}
	where the principal bulk term is defined as
	\begin{align*}
		\mathbf{B}_{pr}^{(N)}[U]=&\Big(\frac{1}{2}{\Deltahat \varpi^N}(2\omega-tr\chi)-\frac{1}{2}e_4(\Deltahat \varpi^N)+\poids\Real(h)\Big)\left|{{\nabla_3}{{U}}}\right|^2\\
		&+\left(\Deltahat\varpi^N\left(-\frac{1}{2}tr\chibar+\Real(\underline{h})\right)-\frac{1}{2}e_3(\Deltahat \varpi^N)\right)\left|{{\nabla_4}{{U}}}\right|^2\\
		&-\frac{1}{2}\left(e_4(\poids)+e_3(\poids)+\poids (tr\chi -2\omega)+\poids tr\chibar\right)|\nabla{U}|^2.
	\end{align*}
\end{prop}
\begin{proof}
	Using \eqref{eq:operateurLhat}, we have 
	\begin{align*}
		\Deltahat\varpi^N\overline{\nabla_3{U}}\cdot\wh{\mcl}({U})=&\Deltahat\varpi^N\overline{\nabla_3{U}}\cdot\Big(\nabla_4\nabla_3 {U}-\triangle_k{U}+h\nabla_3 {U}+\underline{h}\nabla_4{U}+L_1[U]+L[U]\Big).
	\end{align*}
	Using Propositions \ref{prop:ippnab3} and \ref{prop:ipphoriz}, and $\D_\mu e_3^\mu=tr\chibar-2\omegabar=tr\chibar$, we get
	\begin{align}
		\Real\left(\Deltahat\varpi^N\overline{\nabla_3{U}}\cdot\wh{\mcl}({U})\right)=&\Big(\frac{1}{2}{\Deltahat \varpi^N}(2\omega-tr\chi)-\frac{1}{2}e_4(\Deltahat \varpi^N)+\poids\Real(h)\Big)\left|{{\nabla_3}{{U}}}\right|^2\label{eq:mult3}\\
		&-\frac{1}{2}\left(\poids tr\chibar+ e_3(\poids)\right)|\nabla{U}|^2+\mathbf{B}_3[{U}]+\D_\mu\mathbf{F}_3[{U}]^\mu,\nn
	\end{align}
	where
	\begin{align*}
		\mathbf{B}_3[{U}]=&\nabla^a(\poids)\Real\left(\overline{\nabla_3U}\cdot\nabla_a U\right)+\poids\Big[-\Real\left(\overline{[\nabla_3,\nabla^a] U}\cdot\nabla_a U\right)+\Real\left(\underline{h}\overline{\nabla_3 U}\cdot\nabla_4 U\right)\\
		&+\Real\left(\overline{\nabla_3 U}\cdot L_1[U]\right)+\Real\left(\overline{\nabla_3 U}\cdot L[U]\right)\Big]+\poids(\eta+\etabar)\cdot \nu^{(3)}[U],
	\end{align*}
	and
	\begin{align*}
		\mathbf{F}_3[{U}]^\mu=\frac{\Deltahat \varpi^N}{2}\left|{{\nabla_3}{{U}}}\right|^2e_4^\mu+\frac{\Deltahat\varpi^N}{2}|{\nabla}{U}|^2e_3^\mu-\poids\nu^{(3)}[U]^\mu.
	\end{align*}
	Next, to compute $\Deltahat\varpi^N\overline{\nabla_4{U}}\cdot\wh{\mcl}({U})$, we use Lemma \ref{lem:comm34} (and $\omegabar=\Xi=\Xibar=0$ by \eqref{eq:xiprimxibarprimomegaprimzeroin}) which gives $[\nabla_4,\nabla_3]{{U}}=2\omega\nabla_3{{U}}+2(\etabar-\eta)\cdot\nabla{{U}}-C_k[U]$. This gives a new expression of \eqref{eq:operateurLhat}:
	\begin{align*}
		\wh{\mcl}({U})=&\nabla_3\nabla_4{U}-\triangle_k{U}+\parentheses{h+2\omega}\nabla_3 {U}+\parentheses{\underline{h}-2\omegabar} \nabla_4{U}+L_1[U]+2(\etabar-\eta)\cdot\nabla{{U}}+L[U]-C_k[U].
	\end{align*}
	Combining this with Propositions \ref{prop:ippnab3}, \ref{prop:ipphoriz} and $\D_\mu e_4^\mu=tr\chi-2\omega$ yields 
	\begin{align}
		\Real\left(\Deltahat\varpi^N\overline{\nabla_4{U}}\cdot\wh{\mcl}({U})\right)=&-\left(\Deltahat\varpi^N\left(\frac{1}{2}tr\chibar-\Real(\underline{h})\right)+\frac{1}{2}e_3(\Deltahat \varpi^N)\right)\left|{{\nabla_4}{{U}}}\right|^2\label{eq:mult4}\\
		&-\frac{1}{2}\left(\poids (tr\chi -2\omega) +e_4(\poids)\right)|\nabla{U}|^2+\mathbf{B}_4[{U}]+\D_\mu\mathbf{F}_4[{U}]^\mu,\nn
	\end{align}
	where
	\begin{align*}
		&\mathbf{B}_4[{U}]=\nabla^a(\poids)\Real\left(\overline{\nabla_4U}\cdot\nabla_a U\right)+\poids\Big[-\Real\left(\overline{[\nabla_4,\nabla^a] U}\cdot\nabla_a U\right)+\Real\left((h+2\omega)\overline{\nabla_4 U}\cdot\nabla_3 U\right)\\
		&\quad\quad+\Real\left(\overline{\nabla_4U}\cdot(L_1[U]+2(\etabar-\eta)\cdot\nabla{{U}})\right)+\Real\left(\overline{\nabla_4 U}\cdot (L[U]-C_k[U])\right)\Big]+\poids(\eta+\etabar)\cdot \nu^{(4)}[U],
	\end{align*}
	and
	\begin{align*}
		\mathbf{F}_4[{U}]^\mu=\frac{\Deltahat \varpi^N}{2}\left|{{\nabla_4}{{U}}}\right|^2e_3^\mu+\frac{\Deltahat\varpi^N}{2}|{\nabla}{U}|^2e_4^\mu-\poids\nu^{(4)}[U]^\mu.
	\end{align*}
	Combining \eqref{eq:mult3} and \eqref{eq:mult4} we get 
	\begin{align}
		\label{eq:onrevient}\int_{\deux[\wbar_1,\wbar_2]}\Deltahat\varpi^N\Real\left((\overline{\nabla_3{U}+\nabla_4{U}})\cdot\wh{\mcl}({U})\right)=&\int_{\deux[\wbar_1,\wbar_2]}\mathbf{B}^{(N)}[{U}]+\int_{\deux[\wbar_1,\wbar_2]}\D_\mu\mathbf{F}[{U}]^\mu,
	\end{align}
	where the bulk term is
	\begin{align*}
		\mathbf{B}^{(N)}[{U}]=&\Big(\frac{1}{2}{\Deltahat \varpi^N}(2\omega-tr\chi)-\frac{1}{2}e_4(\Deltahat \varpi^N)+\poids\Real(h)\Big)\left|{{\nabla_3}{{U}}}\right|^2\\
		&-\left(\Deltahat\varpi^N\left(\frac{1}{2}tr\chibar-\Real(\underline{h})\right)+\frac{1}{2}e_3(\Deltahat \varpi^N)\right)\left|{{\nabla_4}{{U}}}\right|^2\\
		&-\frac{1}{2}\left(e_4(\poids)+e_3(\poids)+\poids (tr\chi -2\omega)+\poids tr\chibar\right)|\nabla{U}|^2+\mathbf{B}_3[U]+\mathbf{B}_4[U],
	\end{align*}
	and where the flux term is 
	\begin{align}
		\mathbf{F}[{U}]^\mu&=\mathbf{F}_3[{U}]^\mu+\mathbf{F}_4[{U}]^\mu\label{eq:fluxtermm}\\
		&=\frac{\Deltahat \varpi^N}{2}\left(\left|{{\nabla_3}{{U}}}\right|^2e_4^\mu+\left|{{\nabla_4}{{U}}}\right|^2e_3^\mu+|{\nabla}{U}|^2(e_3^\mu+e_4^\mu)\right)-\poids(\nu^{(3)}[U]^\mu+\nu^{(4)}[U]^\mu).\nn
	\end{align}
	Now, by Stokes theorem for manifolds with corners, 
	\begin{align}\label{eq:stokes}
		\int_{\deux[\wbar_1,\wbar_2]}\D_\mu\mathbf{F}[{U}]^\mu=\int_{\partial\deux[\wbar_1,\wbar_2]}\mathbf{F}[{U}]_\mu n^\mu,
	\end{align}
	where $n$ is the inwards pointing unit normal vector to the boundary $\partial\deux[\wbar_1,\wbar_2]$ (note that here the boundary of $\deux[\wbar_1,\wbar_2]$ is composed of four spacelike hypersurfaces). Provided $\wbar_1\geq\wbar|_{\Gamma\cap\{u=u_f\}}$ (see Figure \ref{fig:wbarIInl}), we have the  decomposition
	\begin{align*}
		\partial\deux[\wbar_1,\wbar_2]=\left(\deux\cap\{\wbar=\wbar_1\}\right)\cup\left(\deux\cap\{\wbar=\wbar_2\}\right)\cup \left(\Gamma\cap\{\wbar_1\leq\wbar\leq\wbar_2\}\right)\cup \left(\Sigma_0\cap\{\wbar_1\leq\wbar\leq\wbar_2\}\right).
	\end{align*}
	\noindent\textbf{Boundary term on $\{\wbar=\wbar_a\}$.} A choice of unit vector normal to $\{\wbar=cst\}$ is
	$$n_{\wbar}:=\frac{\D\wbar}{\sqrt{-\g(\D\wbar,\D\wbar)}},$$
	where we recall the computation \eqref{eq:gDwbarDwbar} which rewrites $-\g(\D\wbar,\D\wbar)=|\kappa_-|\Deltahat\Omega^{-2}(1+|\kappa_-|\Deltahat)\sim 1$. Note that we have $\D\wbar(\wbar)=\g(\D\wbar,\D\wbar)<0,$ which proves that $n_{\wbar}$ is outwards pointing (with respect to $\partial\deux[\wbar_1,\wbar_2]$) at $\deux\cap\{\wbar=\wbar_1\}$, and inwards pointing at $\deux\cap\{\wbar=\wbar_2\}$. Thus the boundary term corresponding to $\left(\deux\cap\{\wbar=\wbar_1\}\right)\cup\left(\deux\cap\{\wbar=\wbar_2\}\right)$ in \eqref{eq:stokes} is
	$$\mathcal{B}_{\wbar}=\int_{\deux\cap\{\wbar=\wbar_2\}}\mathbf{F}^{(N)}[U]_\mu n_{\wbar}^\mu-\int_{\deux\cap\{\wbar=\wbar_1\}}\mathbf{F}^{(N)}[U]_\mu n_{\wbar}^\mu.$$
	Defining $\mathbf{F}_{\wbar}^{(N)}[U]:=\mathbf{F}^{(N)}[U]_\mu n_{\wbar}^\mu$, we deduce from \eqref{eq:fluxtermm} the identity
	\begin{align}
		\mathbf{F}_{\wbar}^{(N)}[U]=&\frac{\poids}{2}\left(|\nabla_3U|^2\g(e_4,n_{\wbar})+|\nabla_4U|^2\g(e_3,n_{\wbar})+|\nabla U|^2\g(e_3+e_4,n_{\wbar})\right)\nn\\
		&-\poids(\nu^{(3)}[U]+\nu^{(4)}[U])^a\g(e_a,n_{\wbar}).\label{eq:cestuneexpre}
	\end{align}
	Moreover, using Remark \ref{rem:remlabel} we get the following identities,
	\begin{align*}
		\g(e_4,\D\wbar)=e_4(\wbar)&=\hat{\lambda}\left(\Omega^{-2}(1+|\kappa_-|\Deltahat)+\frac{1}{4}|f|^2|\kappa_-|\Deltahat\right)=\hat{\lambda}\Omega^{-2}+O(\Omega^2),\\
		\g(e_3,\D\wbar)=e_3(\wbar)&=\hat{\lambda}^{-1}\left(\left(1+\frac{1}{2}f\cdot\fbar+\frac{1}{16}|f|^2|\fbar|^2\right)|\kappa_-|\Deltahat+\frac{1}{4}|\fbar|^2\Omega^{-2}(1+|\kappa_-|\Deltahat)\right)\\
		&=\hat{\lambda}^{-1}\left(|\kappa_-|\Deltahat+\frac{1}{4}|\fbar|^2\Omega^{-2}\right)+O(\Omega^2),
	\end{align*}
	which prove 
	\begin{align}
		\sqrt{-\g(\D\wbar,\D\wbar)}&\frac{\poids}{2}\left(|\nabla_3U|^2\g(e_4,n_{\wbar})+|\nabla_4U|^2\g(e_3,n_{\wbar})+|\nabla U|^2\g(e_3+e_4,n_{\wbar})\right)\nn\\
		=&\frac{\poids}{2}\hat{\lambda}\Omega^{-2}|\nabla_3 U|^2+\frac{\poids}{2}\hat{\lambda}^{-1}\left(|\kappa_-|\Deltahat+\frac{1}{4}|\fbar|^2\Omega^{-2}\right)|\nabla_4U|^2\nn\\
		&+\frac{\poids}{2}\left(\hat{\lambda}\Omega^{-2}+\hat{\lambda}^{-1}\left(|\kappa_-|\Deltahat+\frac{1}{4}|\fbar|^2\Omega^{-2}\right)\right)|\nabla U|^2\nn\\
		&+\poids O(\Omega^2(|\nabla_3U|^2+|\nabla_4U|^2+|\nabla U|^2)).\label{eq:provesabso1}
	\end{align}
	We also have
	\begin{align*}
		\g(\D\wbar,e_a)=e_a(\wbar)=\frac{1}{2}\fbar_a\Omega^{-2}(1+|\kappa_-|\Deltahat)+\left(\frac12 f_a+\frac{1}{8}|f|^2\fbar_a\right)|\kappa_-|\Deltahat=\frac{1}{2}\fbar_a\Omega^{-2}+O(\Omega^{2}),
	\end{align*}
	which yields
	\begin{align*}
		-\poids(\nu^{(3)}[U]+\nu^{(4)}[U])^a\g(e_a,\D{\wbar})=&-\frac{1}{2}\Deltahat\varpi^N\Omega^{-2}\Real((\overline{\nabla_3 U+\nabla_4 U})\cdot\fbar^a{\nabla}_aU)\\
		&+\Deltahat\varpi^NO\left(\Omega^{2}(|\nabla_3U|^2+|\nabla_4U|^2+|{\nabla} U|^2)\right).
	\end{align*}
	We now write the bound:
	\begin{align*}
		\left|\frac{1}{2}\poids\Omega^{-2}\Real(\overline{\nabla_4 U}\cdot\fbar^a{\nabla}_a U)\right|&\leq \frac{1}{2}\poids\Omega^{-2}|\nabla_4U||\fbar||{\nabla} U|\nn\\
		&\leq\frac{1}{2}\poids\left(\frac{1}{4}\hat{\lambda}^{-1}\Omega^{-2}|\fbar|^2|{\nabla}_4 U|^2+\Omega^{-2}\hat{\lambda}|{\nabla} U|^2\right),
	\end{align*}
	which proves, by \eqref{eq:provesabso1} and \eqref{eq:cestuneexpre},
	\begin{align*}
		&\sqrt{-\g(\D\wbar,\D\wbar)}\mathbf{F}_{\wbar}^{(N)}[U]\\
		&\geq \frac{\poids}{2}\hat{\lambda}\Omega^{-2}|\nabla_3 U|^2+\frac{\poids}{2}\hat{\lambda}^{-1}|\kappa_-|\Deltahat|\nabla_4U|^2+\frac{\poids}{2}\hat{\lambda}^{-1}\left(|\kappa_-|\Deltahat+\frac{1}{4}|\fbar|^2\Omega^{-2}\right)|\nabla U|^2\nn\\
		&\quad-\left|\frac{1}{2}\poids\Omega^{-2}\Real(\overline{\nabla_3 U}\cdot\fbar^a{\nabla}_a U)\right|+\poids O(\Omega^2(|\nabla_3U|^2+|\nabla_4U|^2+|\nabla U|^2)).
	\end{align*}
	Now, for some sufficiently small $s>0$, we write the bound
	\begin{align*}
		\left|\frac{1}{2}\poids\Omega^{-2}\Real(\overline{\nabla_3 U}\cdot\fbar^a{\nabla}_a U)\right|\leq\frac{1}{2}\poids\left((1-s)\hat{\lambda}\Omega^{-2}|\nabla_3U|^2+\frac{(1-s)}{4}^{-1}\hat{\lambda}^{-1}\Omega^{-2}|\fbar|^2|{\nabla} U|^2\right),
	\end{align*}
	which implies
	\begin{align*}
		\sqrt{-\g(\D\wbar,\D\wbar)}\mathbf{F}_{\wbar}^{(N)}[U]\geq &s\frac{\poids}{2}\hat{\lambda}\Omega^{-2}|\nabla_3 U|^2+\frac{\poids}{2}\hat{\lambda}^{-1}|\kappa_-|\Deltahat|\nabla_4U|^2\nn\\
		&+\frac{\poids}{2}\hat{\lambda}^{-1}\left(|\kappa_-|\Deltahat+\frac{1-(1-s)^{-1}}{4}|\fbar|^2\Omega^{-2}\right)|\nabla U|^2\nn\\
		&+\poids O(\Omega^2(|\nabla_3U|^2+|\nabla_4U|^2+|\nabla U|^2)).
	\end{align*}
	We choose the constant $s>0$ such that
	$$0<s\leq 1-\left({\frac{|\kappa_-|\Deltahat}{|\fbar|^2\Omega^{-2}}+1}\right)^{-1},$$
	which is a well-defined choice because $|\kappa_-|\Deltahat\sim\Omega^{2}\sim |\fbar|^2\Omega^{-2} $ which implies that the function ${|\kappa_-|\Deltahat}/{|\fbar|^2\Omega^{-2}}$ is bounded above and below by positive constants. This choice of $s$ implies
	$$|\kappa_-|\Deltahat+\frac{1-(1-s)^{-1}}{4}|\fbar|^2\Omega^{-2}\geq\frac{3}{4}|\kappa_-|\Deltahat,$$
	hence the bound
	\begin{align*}
		\sqrt{-\g(\D\wbar,\D\wbar)}\mathbf{F}_{\wbar}^{(N)}[U]\geq &s\frac{\poids}{2}\hat{\lambda}\Omega^{-2}|\nabla_3 U|^2+\frac{\poids}{2}\hat{\lambda}^{-1}|\kappa_-|\Deltahat|\nabla_4U|^2\nn\\
		&+\frac{3\poids}{8}\hat{\lambda}^{-1}|\kappa_-|\Deltahat|\nabla U|^2+\poids O(\Omega^2(|\nabla_3U|^2+|\nabla_4U|^2+|\nabla U|^2)).
	\end{align*}
	For $C_R(a,M)\gg 1$ large such that $\sup_\deux\Omega^2$ is small, and using $\sqrt{-\g(\D\wbar,\D\wbar)}\sim 1$, we thus get 
	$$\mathbf{F}_{\wbar}^{(N)}[U]\gtrsim\poids\left(|\nabla_3 U|^2+|\nabla_4U|^2+|\nabla U|^2\right).$$
	Since the reverse bound also clearly holds, we obtain $\mathbf{F}_{\wbar}^{(N)}[U]\sim\poids\left(|\nabla_3 U|^2+|\nabla_4U|^2+|\nabla U|^2\right)$, which implies the following estimate for the boundary terms on $\{\wbar=\wbar_a\}$, $a=1,2$:
	\begin{align*}
		\int_{\deux\cap\{\wbar=\wbar_a\}}\mathbf{F}^{(N)}[U]_\mu n_{\wbar}^\mu\sim\int_{\deux\cap\{\wbar=\wbar_a\}}\Deltahat\varpi^N\left(|\nabla_3U|^2+|\nabla_4U|^2+|{\nabla} U|^2\right).
	\end{align*}
	
	\noindent\textbf{Boundary term on $\Sigma_0$.} We now estimate the boundary term on $\Sigma_0=\{u+\ubar=C_R\}$. We have by \eqref{eq:DuDubardoublenull} the identities $\D\ubar=-\frac{1}{2}\Omega^{-2}\ering_3$, $\D u=-\frac{1}{2}\ering_4$ in $\deux$, which imply 
	$$\D(u+\ubar)=-\frac{1}{2}\Omega^{-2}\ering_3-\frac{1}{2}\ering_4,$$
	and $\g(\D(u+\ubar),\D(u+\ubar))\sim_{C_R} -1$ on $\Sigma_0$. Using $\Omega^2\sim_{C_R} 1$ on $\Sigma_0$, this proves the following estimate for the boundary term on $\Sigma_0$,
	\begin{align*}
		\int_{\Sigma_0\cap\{\wbar_1\leq\wbar\leq\wbar_2\}}\mathbf{F}_0^{(N)}[A]&=\int_{\Sigma_0\cap\{\wbar_1\leq\wbar\leq\wbar_2\}}\mathbf{F}^{(N)}[A]_\mu n^\mu_{\Sigma_0}\\
		&\lesssim_{C_R}\int_{\Sigma_0\cap\{\wbar_1\leq\wbar\leq\wbar_2\}}\varpi^N\left(|\nabla_3U|^2+|\nabla_4U|^2+|{\nabla} U|^2)\right).
	\end{align*}
	
	\noindent\textbf{Boundary term on $\Gamma$.} We will only use positivity of the boundary term on $\Gamma$ and then drop it\footnote{We will not use any coercivity property that it might satisfy because, contrary as in the linearized setting in \cite{spin-2,spin+2}, we do not use any energy estimates in region $\trois$ to analyze the Teukolsky equation.}. Recall the definition of $\Gamma$ in \eqref{eq:defiGammahyp}. We define the function $h:=u+\ubar-\ubar^\gamma$ and we compute its gradient. We have $\partial_{\theta^A}h=0$, thus 
	$$\D h=-\frac{1}{2}\ering_3(h)\ering_4-\frac{1}{2}\ering_4(h)\ering_3=-\frac{1}{2}\ering_4-\frac{1}{2}\Omega^{-2}(1-\gamma\ubar^{\gamma-1})\ering_3.$$
	This gives
	$$\g(\D h,\D h)=-\Omega^{-2}(1-\gamma\ubar^{\gamma-1})<0$$
	for $0<\gamma<1$ and $|u_f|$, hence $\ubar$, large enough. To prove positivity of the boundary term on $\Gamma$, it is enough to prove pointwise positivity of
	\begin{align*}
		\sqrt{-\g(\D h,\D h)}\mathbf{F}_{\Gamma}^{(N)}[U]=&\frac{\poids}{2}\left(|\nabla_3U|^2\g(e_4,\D h)+|\nabla_4U|^2\g(e_3,\D h)+|\nabla U|^2\g(e_3+e_4,\D h)\right)\nn\\
		&-\poids(\nu^{(3)}[U]+\nu^{(4)}[U])^a\g(e_a,\D h).
	\end{align*}
	We compute:
	\begin{align*}
		\g(e_4,\D h)&=e_4(h)=\hat{\lambda}\left(\Omega^{-2}(1-\gamma\ubar^{\gamma-1})+\frac{1}{4}|f|^2\right)=\hat{\lambda}\Omega^{-2}(1-\gamma\ubar^{\gamma-1})+O(\Omega^2),\\
		\g(e_3,\D h)&=e_3(h)=\hat{\lambda}^{-1}\left(\left(1+\frac{1}{2}f\cdot\fbar+\frac{1}{16}|f|^2|\fbar|^2\right)+\frac{1}{4}|\fbar|^2\Omega^{-2}\left(1-\gamma\ubar^{\gamma-1}\right)\right)=\hat{\lambda}^{-1}+O(1),\\
		\g(e_3+e_4,\D h)&=\hat{\lambda}^{-1}+O(1).
	\end{align*}
	This yields, for $\gamma>0$ small enough:
	\begin{align}
		\frac{\Deltahat\varpi^N}{2}\Big[|\nabla_3A|^2\g(e_4&,\D h)+|\nabla_4A|^2\g(e_3,\D h)+|{\nabla} A|^2\g(e_3+e_4,\D h)\Big]\nn\\
		&\gtrsim\Deltahat\varpi^N\Big[|\nabla_3 A|^2+\Omega^{-2}|\nabla_4 A|^2+\Omega^{-2}|{\nabla} A|^2\Big].\label{eq:absgamma}
	\end{align}
	We now prove that we can absorb $-\poids(\nu^{(3)}[U]+\nu^{(4)}[U])^a\g(e_a,\D h)$ in \eqref{eq:absgamma}. We have:
	\begin{align*}
		-\poids(\nu^{(3)}[U]+\nu^{(4)}[U])^a\g(e_a,\D h)=-\Deltahat\varpi^N\Real((\overline{\nabla_3 U+\nabla_4 U})\cdot{\nabla}^aU)\g(e_a,\D h),
	\end{align*}
	where
	\begin{align*}
		\g(e_a,\D h)=e_a(h)=\frac{1}{2}\fbar_a\Omega^{-2}(1+\gamma\ubar^{\gamma-1})+\frac{1}{2}f_a+\frac{1}{8}|f|^2\fbar_a=O(1).
	\end{align*}
	We can now write the bounds:
	\begin{align*}
		&|\Deltahat\varpi^N\Real((\overline{\nabla_4 A})\cdot{\nabla}^aA)\g(e_a,\D h)|\lesssim\Deltahat\varpi^N(|\nabla_4A|^2+|{\nabla}A|^2),\\
		&|\Deltahat\varpi^N\Real((\overline{\nabla_3 A})\cdot{\nabla}^aA)\g(e_a,\D h)|\lesssim\Deltahat\varpi^N(\Omega|\nabla_3A|^2+\Omega^{-1}|{\nabla}A|^2),
	\end{align*}
	which prove that the term $-\poids(\nu^{(3)}[U]+\nu^{(4)}[U])^a\g(e_a,\D h)$ can be absorbed in \eqref{eq:absgamma} for $C_R(a,M)$ large enough, hence $\mathbf{F}_{\Gamma}^{(N)}[U]\geq 0$, which concludes the proof of the proposition.
\end{proof}

\begin{prop}\label{prop:enerteukcomm}
	Let ${U}\in\fraks_k(\mathbb{C})$, $k\in\{0,1,2\}$ be such that \eqref{eq:teukmod} holds in $\deux$, where $\wh{\mcl}$ is as in \eqref{eq:operateurLhat}. We assume that $0<\gamma<1$, and for some constant $C>0$:
	\begin{itemize}
		\item The following estimates hold in $\deux$,
		\begin{equation}\label{eq:hypLhat}
			\begin{aligned}
			&\Real(h)\sim 1,\quad |h|\lesssim 1,\quad |\underline{h}|\lesssim 1+\Omega^{-2}\ubar^{-2-\delta/5},\\
			&|L_1[U]|\lesssim |\nabla U|,\quad |L[U]|\lesssim \left(1+{\Omega^{-2}}{\ubar^{-2-\delta/5}}\right)|U|.				
			\end{aligned}
		\end{equation}
		\item We have the initial energy bound on $\Sigma_0$,
		\begin{align}\label{eq:initialenergyhyp}
			\int_{\Sigma_0\cap\{\wbar_1\leq\wbar\leq\wbar_2\}}\left(|\nabla_3U|^2+|\nabla_4U|^2+|{\nabla} U|^2\right)\lesssim C\left(\wbar_1^{-6-\delta}+\int_{\wbar_1}^{\wbar_2}\wbar^{-6-\delta}\dee\wbar\right).
		\end{align}
		\item We have the bound 
		\begin{align}\label{eq:hyppointwise}
			\Ldeux{\nabla_4^{\leq 1}{U}}\lesssim{\ubar^{-2-\delta/5}},\quad\text{in}\:\:\deux.
		\end{align}
		\item For any $1\leq\wbar_1\leq\wbar_2$,
		\begin{align}\label{eq:hyperrhat}
			\int_{\deux[\wbar_1,\wbar_2]}\Omega^2|\err[\wh{\mcl}({U})]|^2\lesssim C\left(\frac{1}{\wbar_1^{6+\delta}}+\int_{\wbar_1}^{\wbar_2}\frac{1}{\wbar^{6+\delta}}\dee\wbar\right).
		\end{align}
		\item For any $1\leq \wbar_1\leq\wbar_2$,
		\begin{align}\label{eq:hypenerprecedente}
			\int_{\deux[\wbar_1,\wbar_2]}\Omega^2|{U}|^2\lesssim C\left(\frac{1}{\wbar_1^{6+\delta}}+\int_{\wbar_1}^{\wbar_2}\frac{1}{\wbar^{6+\delta}}\dee\wbar\right).
		\end{align}
	\end{itemize}
	Then, recalling the definition \eqref{eq:defenerdeg} of $\Ener[U]$, we have the following energy decay in $\deux$,
	$$\Ener[U](\wbar)\lesssim C'\wbar^{-6-\delta},$$
	where $C'=C'(a,M,u_f,C,C_R)$ depends on $a,M,u_f,C,C_R$.
\end{prop}

\begin{rem}
	Assumption \eqref{eq:hypenerprecedente} will be satisfied in practice for ${U}$ being some derivatives of $A$. Indeed, we will be able to bound the bulk integral of ${U}$ by energies of lower-order derivatives of $A$ that we previously controled\footnote{We will use a different result to control $A$, where the zero-order term will be dealt with by Proposition \ref{prop:hardy}.}.
\end{rem}
\begin{proof}[Proof of Proposition \ref{prop:enerteukcomm}]	 In this proof we denote by $\lesssim$ the $(a,M,u_f,C,C_R)$-dependent bounds. We multiply \eqref{eq:teukmod} with $\Deltahat\varpi^N(\overline{\nabla_3{U}+\nabla_4{U}}),$
	where $N$ is large enough as chosen later in the proof, and we integrate on $\deux[\wbar_1,\wbar_2]$ for $\wbar_2\geq\wbar_1\geq\wbar|_{\Gamma\cap\{u=u_f\}}$ with respect to spacetime volume, and take the real part. By Proposition \ref{prop:calculcommunenergy}, we get 
	\begin{align}\label{eq:step1Psi}
		&\int_{\deux[\wbar_1,\wbar_2]}\mathbf{B}^{(N)}[U]+\int_{\deux\cap\{\wbar=\wbar_2\}}\mathbf{F}_{\wbar}^{(N)}[U]\\
		&\leq\int_{\deux\cap\{\wbar=\wbar_1\}}\mathbf{F}_{\wbar}^{(N)}[U]+\int_{\Sigma_0\cap\{\wbar_1\leq\wbar\leq\wbar_2\}}\mathbf{F}_{0}^{(N)}[U]+\left|\int_{\deux[\wbar_1,\wbar_2]}\Deltahat\varpi^N\Real\left((\overline{\nabla_3{U}+\nabla_4{U}})\cdot\err[\wh{\mcl}({U})]\right)\right|,\nn
	\end{align}
	where we used in particular \eqref{eq:estimbdGamma}, namely the fact that the boundary term on $\Gamma$ is non-negative. Next, using Proposition \ref{prop:bulkpos} in appendix, we get that for $N(a,M), C_R(a,M)\gg 1$,
	\begin{align*}
		\Omega^2\left|{{\nabla_3}{U}}\right|^2&+\left|{\nabla_4{U}}\right|^2+|{{\nabla}U}|^2\lesssim \mathbf{B}^{(N)}[{U}]+\Deltahat^{-1}\ubar^{-4-2\delta/5}|\nabla_4^{\leq 1}U|^2+\Deltahat |U|^2,
	\end{align*}
	in $\deux$, which implies, together with \eqref{eq:hyppointwise},
	\begin{align*}
		\int_{\deux[\wbar_1,\wbar_2]}\Big(\Omega^2\left|{{\nabla_3}{U}}\right|^2+\left|{\nabla_4{U}}\right|^2+|{{\nabla}U}|^2\Big)\lesssim &\int_{\deux[\wbar_1,\wbar_2]}\left(\mathbf{B}^{(N)}[{U}]+\Omega^2|{U}|^2+\Omega^{-2}\ubar^{-8-4\delta/5}\right).
	\end{align*}
	Moreover, using $\dee\wbar=(1-|\kappa_-|\Deltahat)\dee\ubar-|\kappa_-|\Deltahat\dee u$ and \eqref{eq:volumeendoublenull}, we get for any $A>1$,
	\begin{align*}
		\int_{\deux[\wbar_1,\wbar_2]}\Omega^{-2}\ubar^{-A}\lesssim \int_{\deux[\wbar_1,\wbar_2]} \ubar^{-A}\mathrm{vol}_\gamma\dee u\dee\ubar\lesssim \int_{\wbar_1}^{\wbar_2}\int_{C_R-\wbar}^{u_\Gamma(\wbar)}\wbar^{-A}\dee u\dee\wbar,
	\end{align*}
	where we also used $\wbar|_{\Sigma_0}=\ubar|_{\Sigma_0}$, and where $u_\Gamma(\wbar)$ is such that  $u_\Gamma(\wbar)+\ubar(u_\Gamma(\wbar),\wbar)=\ubar(u_\Gamma(\wbar),\wbar)^\gamma$. Note that we have by definition $\wbar=\ubar+O(1)$, which implies $u_\Gamma(\wbar)=-\wbar+O(1)+(\wbar+O(1))^\gamma$,
	hence the estimate
	\begin{align}
		u_\Gamma(\wbar)-(C_R-\wbar)\lesssim (\wbar+1)^\gamma+1\lesssim \wbar^\gamma
	\end{align}
	for $\wbar\gtrsim 1$. This yields, for any $A>1$,
	\begin{align}\label{eq:bornimportane}
		\int_{\deux[\wbar_1,\wbar_2]}\Omega^{-2}\ubar^{-A}\lesssim \int_{\wbar_1}^{\wbar_2}\wbar^{-A+\gamma}\dee\wbar.
	\end{align}
	Since $\Omega^2\left|{{\nabla_3}{U}}\right|^2+\left|{\nabla_4{U}}\right|^2+|{{\nabla}U}|^2\gtrsim\Omega^2(\left|{{\nabla_3}{U}}\right|^2+\left|{\nabla_4{U}}\right|^2+|{{\nabla}U}|^2)$ we infer, recalling \eqref{eq:defenerdeg} and Proposition \ref{prop:toutsurwbar},
	\begin{align*}
		\int_{\deux[\wbar_1,\wbar_2]}\Big(\Omega^2\left|{{\nabla_3}{U}}\right|^2+\left|{\nabla_4{U}}\right|^2+|{{\nabla}U}|^2\Big)\gtrsim\int_{\wbar_1}^{\wbar_2}\Ener[U](\wbar)\dee\wbar.
	\end{align*}
	Combining this with \eqref{eq:hypenerprecedente}, \eqref{eq:bornimportane} with $A=8+4\delta/5$ and the estimates \eqref{eq:estimbdwbar} for the boundary terms on $\wbar=\wbar_1,\wbar_2$ and \eqref{eq:initialenergyhyp} on $\Sigma_0$, we deduce from \eqref{eq:step1Psi},
	\begin{align}\label{eq:stepestepU}
		\Ener[U](\wbar_2)+\int_{\wbar_1}^{\wbar_2}\Ener[U](\wbar)\dee\wbar\lesssim & \Ener[U](\wbar_1)+\frac{1}{\wbar_1^{6+\delta}}+\int_{\wbar_1}^{\wbar_2}\frac{1}{\wbar^{6+\delta}}\dee\wbar\\
		&+\left|\int_{\deux[\wbar_1,\wbar_2]}\Deltahat\varpi^N\Real\left((\overline{\nabla_3{U}+\nabla_4{U}})\cdot\err[\wh{\mcl}({U})]\right)\right|.\nn
	\end{align}
	Now we bound the last term on the RHS of the bound above, which we denote by $L$. Using $\Deltahat\lesssim\Omega^2$, $\varpi\lesssim 1$, $N=N(a,M)\lesssim 1$, we get for any $s>0$:
	\begin{align*}
	L&\lesssim\int_{\deux[\wbar_1,\wbar_2]}s\Omega^2\left(|\nabla_3{U}|^2+|\nabla_4{U}|^2\right)+\int_{\deux[\wbar_1,\wbar_2]}s^{-1}\Omega^2|\err[\wh{\mcl}({U})]|^2\\
	&\lesssim s\int_{\wbar_1}^{\wbar_2}\Ener[U](\wbar)\dee\wbar+s^{-1}\left(\frac{1}{\wbar_1^{6+\delta}}+\int_{\wbar_1}^{\wbar_2}\frac{1}{\wbar^{6+\delta}}\dee\wbar\right),\nn
	\end{align*}
	where we used assumption \eqref{eq:hyperrhat} in the last step. Choosing $s>0$ small enough ensures that the first term on the RHS of the bound above can be absorbed in the LHS of \eqref{eq:stepestepU}, which yields the following integrated energy decay estimate,
	$$\Ener[U](\wbar_2)+\int_{\wbar_1}^{\wbar_2}\Ener[U](\wbar)\dee\wbar\lesssim \Ener[U](\wbar_1)+\frac{1}{\wbar_1^{6+\delta}}+\int_{\wbar_1}^{\wbar_2}\frac{1}{\wbar^{6+\delta}}\dee\wbar.$$
	We conclude the proof by using Lemma \ref{lem:decay} with $p=6+\delta$, $b=\wbar|_{\Gamma\cap\{u=u_f\}}$ (which depends on $u_f$) and $f(\wbar)=\Ener[U](\wbar)$. \end{proof}

\subsubsection{Energy method for the Teukolsky equation for $A$}\label{section:energyA}
Recall from \eqref{eq:teukA} the Teukolsky equation $\mcl(A)=\err[\mcl(A)]$ for $A$, where $\err[\mcl(A)]$ is defined in \eqref{eq:teukerror}, and where the wave operator $\mcl$ is defined in \eqref{eq:teukop}. By \eqref{eq:igno}, denoting
\begin{align}\label{eq:defhhbar}
	h:=\frac{1}{2}tr X+2\overline{tr X}+2\omega+\frac{i}{2}\atrchi,\quad \underline{h}:=-
	\frac{1}{2}tr \underline{X}-4\omegabar+\frac{i}{2}\atrchibar,\\
	L_0[U]:=- H\hot\parentheses{(\overline{2 Z+\underline{H}})\cdot U}+VU+2{}^{(h)}\!KU,\label{eq:zerorderA}\\
	F:=-\parentheses{2Z+\Hbar+\overline{2Z+\Hbar}+4H},\label{eq:oneformteukA}
\end{align}
the Teukolsky operator rewrites
\begin{align*}
	\mcl(U)=&\nabla_4\nabla_3 U-\triangle_2 U+h\nabla_3 U+\underline{h} \nabla_4U+F\cdot\nabla U+L_0[U].
\end{align*}

\begin{prop}\label{prop:enerteukAerrA}
	Let ${U}\in\fraks_2(\mathbb{C})$ which satisfies a inhomogeneous Teukolsky equation
	\begin{align}\label{eq:teukolskylike}
		|{\mcl(U)}|\lesssim C\Omega^{-2}\ubar^{-\beta},
	\end{align}
	in $\deux$, where $\beta>1$, $C>0$. We assume the following :
	\begin{itemize}
		\item We have the initial energy and $L^2$ bounds on $\Sigma_0$,
		\begin{align}\label{eq:initialenergyhypA}
			\int_{\Sigma_0\cap\{\wbar_1\leq\wbar\leq\wbar_2\}}\left(|(\nabla_3,\nabla_4,\nabla)U|^2+|U|^2\right)\lesssim C\left(\wbar_1^{-2\beta}+\int_{\wbar_1}^{\wbar_2}\wbar^{-2\beta}\dee\wbar\right).
		\end{align}
		\item For some $\beta',\beta''>1$, we have the bounds 
		\begin{align}\label{eq:hyppointwiseA}
			\Ldeux{U}\lesssim C\ubar^{-\beta'},\quad\Ldeux{\nabla_4{U}}\lesssim C{\ubar^{-\beta''}},\quad\text{in}\quad\deux.
		\end{align}
	\end{itemize}
	Then, recalling the definition \eqref{eq:defenerdeg} of $\Ener[U]$, we have the following energy decay in $\deux$,
	$$\Ener[U](\wbar)\lesssim C'(a,M,u_f,C,C_R)\wbar^{-\min(2\beta, 2\beta'+4+2\delta/5,2\beta''+8+4\delta/5)+\gamma}.$$
\end{prop}

\begin{rem}
	We will use Proposition \ref{prop:enerteukAerrA} with $U=A$, $\beta=4+2\delta/5$, $\beta'=\beta''=2+\delta/5$, and also with $U=\err[A]:=A-\Psi$ (see \eqref{eq:defansatzdeux}), $\beta=6+\delta/2$, $\beta'=4+\delta/3$, $\beta''=2+\delta/5$.
\end{rem}

\begin{proof}[Proof of Proposition \ref{prop:enerteukAerrA}]
	The proof is similar as the one of Proposition \ref{prop:enerteukcomm}\footnote{Here, we also denote by $\lesssim$ the $(a,M,u_f,C,C_R)$-dependent bounds.}, except that we have to deal with the zero-order terms\footnote{We do not have an assumption of the type \eqref{eq:hypenerprecedente} here, but we have an initial $L^2(\Sigma_0)$ assumption in \eqref{eq:initialenergyhypA}.}. By Propositions \ref{prop:calculcommunenergy} and \ref{prop:bulkpos} where we use the case $k=2$, $\widehat{\mcl}=\mcl$, and by the estimates
	\eqref{eq:estimbdwbar}, \eqref{eq:estimbdsigmazero}, \eqref{eq:estimbdGamma}, we get for $N(a,M),C_R(a,M)\gg1$,
	\begin{align*}
		\Ener[U](\wbar_2)+\int_{\wbar_1}^{\wbar_2}\Ener[U](\wbar)\dee\wbar\lesssim & \Ener[U](\wbar_1)+\int_{\deux[\wbar_1,\wbar_2]}\Bigg[\frac{\Omega^{-2}}{\ubar^{8+4\delta/5}}|\nabla_4U|^2+\frac{\Omega^{-2}}{\ubar^{4+2\delta/5}}|U|^2\Bigg]\\
		&+\int_{\Sigma_0\cap\{\wbar_1\leq\wbar\leq\wbar_2\}}\left(|\nabla_3U|^2+|\nabla_4U|^2+|{\nabla} U|^2+|U|^2\right)\\
		&+\left|\int_{\deux[\wbar_1,\wbar_2]}\Deltahat\varpi^N\Real\left((\overline{\nabla_3{U}+\nabla_4{U}})\cdot{\mcl}({U})\right)\right|.
	\end{align*}
	Here we used that \eqref{eq:hypLhat} holds for $h,\underline{h},F,L_0$ defined in \eqref{eq:defhhbar}, \eqref{eq:zerorderA}, \eqref{eq:oneformteukA}, since
	\begin{align*}
		&|trX|\lesssim\Omega^2+\ubar^{-2-\delta/5},\quad|tr\Xbar|+|\omegabar|\lesssim 1+\Omega^{-2}\ubar^{-2-\delta/5},\quad\omega\gtrsim 1,\\
		&|H|+|\Hbar|+|Z|\lesssim 1,\quad |V|+|{}^{(h)}\!K|\lesssim 1+\Omega^{-2}\ubar^{-2-\delta/5},
	\end{align*}
	which hold by Proposition \ref{prop:coeffin}, \eqref{eq:omegapos}, and Corollary \ref{cor:coeffinsanscheck}. Now, writing 
	\begin{align*}
		\Bigg|\int_{\deux[\wbar_1,\wbar_2]}&\Deltahat\varpi^N\Real((\overline{\nabla_3{U}+\nabla_4{U}})\cdot\err[\wh{\mcl}({U})])\Bigg|\lesssim\\
		&s\int_{\wbar_1}^{\wbar_2}\Ener[U](\wbar)\dee\wbar+\int_{\deux[\wbar_1,\wbar_2]}s^{-1}\Omega^2|\mcl(U)|^2
	\end{align*}
	for sufficiently small $s>0$, and using \eqref{eq:teukolskylike}, \eqref{eq:initialenergyhypA} and \eqref{eq:hyppointwiseA}, we get
	\begin{align*}
		\Ener[U](\wbar_2)+\int_{\wbar_1}^{\wbar_2}\Ener[U](\wbar)\dee\wbar\lesssim & \Ener[U](\wbar_1)+\frac{1}{\wbar_1^{2\beta}}+\int_{\wbar_1}^{\wbar_2}\frac{1}{\wbar^{2\beta}}\dee\wbar\nn\\
		&+\int_{\deux[\wbar_1,\wbar_2]}{\Omega^{-2}}\left({\ubar^{-2\beta'-4-2\delta/5}}+\ubar^{-2\beta''-8-4\delta/5}+\ubar^{-2\beta}\right).
	\end{align*}
	Moreover, by \eqref{eq:bornimportane} we have
	\begin{align*}
		\int_{\deux[\wbar_1,\wbar_2]}{\Omega^{-2}}\left({\ubar^{-2\beta'-4-2\delta/5}}+\ubar^{-2\beta''-8-4\delta/5}+\ubar^{-2\beta}\right)\lesssim \int_{\wbar_1}^{\wbar_2}\wbar^{-Q+\gamma}\dee\wbar,
	\end{align*}
		where $Q:=\min(2\beta, 2\beta'+4+2\delta/5,2\beta''+8+4\delta/5)$, hence the estimate
	\begin{align*}
		\Ener[U](\wbar_2)+\int_{\wbar_1}^{\wbar_2}\Ener[U](\wbar)\dee\wbar\lesssim & \Ener[U](\wbar_1)+\frac{1}{\wbar_1^{Q-\gamma}}+\int_{\wbar_1}^{\wbar_2}\frac{1}{\wbar^{Q-\gamma}}\dee\wbar.
	\end{align*}
 We conclude the proof by Lemma \ref{lem:decay} with $p=Q-\gamma$, $b=\wbar|_{\Gamma\cap\{u=u_f\}}$, $f(\wbar)=\Ener[U](\wbar)$.\end{proof}

\subsubsection{$L^2(S(u,\ubar))$ decay for generalized Teukolsky fields}\label{section:decayL2generalizedteukolsky}
The following result will be applied to deduce $L^2(S(u,\ubar))$ decay for some derivatives of $A$.
\begin{prop}\label{prop:decayL2genteukfield}
	Let ${U}\in\fraks_k(\mathbb{C})$, $k\in\{0,1,2\}$, be such that the assumptions of Proposition \ref{prop:enerteukcomm} are satisfied. If, in addition, we have the initial estimate
	$$\|U\|_{L^2(S(C_R-\ubar,\ubar))}\lesssim C\ubar^{-3-\delta/2}$$
	on $\Sigma_0$, then we have in $\deux$ the bound
	$$\Ldeux{U}\lesssim C'(a,M,u_f,C,C_R)\ubar^{-3-\delta/2+\gamma/2}.$$
\end{prop}
\begin{proof}
	This a direct consequence of Proposition \ref{prop:enerteukcomm} and \eqref{prop:inegwbarcst}: we have in $\deux$
	\begin{align*}
		\Ldeux{U}&\lesssim \|U\|_{L^2(S(C_R-\wbar(u,\ubar),\wbar(u,\ubar)))}+\ubar^{\gamma/2}\Ener[U](\wbar(u,\ubar))^{1/2}\\
		&\lesssim C'(\wbar(u,\ubar)^{-3-\delta/2}+\ubar^{\gamma/2}\wbar(u,\ubar)^{-3-\delta})\lesssim C'\ubar^{-3-\delta/2+\gamma/2},
	\end{align*}
	where we used $\ubar\sim\wbar(u,\ubar)$ in $\deux$.
\end{proof}
The following result will eventually be used to deduce sharp $L^2(S(u,\ubar))$ decay for $A$ in $\deux$.
\begin{prop}\label{prop:decayL2AerrA}
	Let ${U}\in\fraks_2(\mathbb{C})$ be such that the assumptions of Proposition \ref{prop:enerteukAerrA} are satisfied. If, in addition, we have the initial estimate 
	$$\|U\|_{L^2(S(C_R-\ubar,\ubar))}\lesssim C\ubar^{-\min(\beta, \beta'+2+\delta/5,\beta''+4+2\delta/5)}$$
	on $\Sigma_0$, then we have in $\deux$ the bound
	$$\Ldeux{U}\lesssim C'(a,M,u_f,C,C_R)\ubar^{-\min(\beta, \beta'+2+\delta/5,\beta''+4+2\delta/5)+\gamma}.$$
\end{prop}
\begin{proof}
	The proof is exactly the same as the one of Proposition \ref{prop:decayL2genteukfield}.
\end{proof}
\subsection{Preliminary decay estimates for $A$ in $\deux$}\label{section:prelimdecayA}
The goal of this section is to prove non-sharp $\ubar^{-3-\delta}$ decay for $A$ and some $\df:=\{\hat{\lambda}\nabla_3,\nabla_4,\nabla\}$ derivatives of $A$. More precisely, in Section \ref{section:L2smallderA} we will prove decay estimates  of the type\footnote{Note that this is an improvement from the bounds for $A$ in Proposition \ref{prop:coeffin}.}: 
\begin{align}\label{eq:schematicannounced}
	\left\|(\hat{\lambda}\nabla_3,\nabla_4,\barre{\df})^{\leq 2}\barre{\df}^{\leq 2}A\right\|_{L^2(S(u,\ubar))}\lesssim\ubar^{-3-\delta},\quad\text{in}\:\: \deux,
\end{align}
where $\barre{\df}$ represents appropriate horizontal Hodge operators. This relies on the commutation computations done in Proposition \ref{prop:nouvellescommut}. Next, in Section \ref{section:pointwisehorizA}, from \eqref{eq:schematicannounced} and the Sobolev embedding on the spheres $S(u,\ubar)$ we will get pointwise decay for some $\barre{\df}^{\leq 2}$ derivatives of $A$\footnote{We do note obtain the sharp asymptotics for $A$ yet, this is delayed to Section \ref{section:sharpforAdeux}.}.

\subsubsection{Teukolsky-like equations satisfied by derivatives of $A$}\label{section:waveequationsderivees}

Before proving decay estimates, in this section we summarize the commutator computations proven in Proposition \ref{prop:nouvellescommut} in the case of region $\deux$.
\begin{defi}
For any horizontal tensor $U$, we define the energy density
$$\mathbf{e}[U]:=|\nabla_3U|^2+|\nabla_4U|^2+|\nabla U|^2.$$
\end{defi}
\begin{prop}\label{prop:teukcommutepleindefois}
We define the ordered set of horizontal operators
$$\mathcal{O}=(S_1,S_2,S_3,S_4,S_5,S_6),$$
where $S_1=\divc$, $S_2=\divc\divc$, $S_3=\mcd\hot\divc$, $S_4=\divc\mcd\hot\divc$, $S_5=\mcd\divc\divc$, $S_6=\divc\mcd\divc\divc$. Then for $i=1,\ldots,6$ we have the Teukolsky-like equation
$$\mcl^{(i)}(S_i A)=\err[\mcl^{(i)}(S_i A)],$$
where the Teukolsky-like operators $\mcl^{(i)}$ satisfy, recalling \eqref{eq:dfhorizontal}, $$\mcl^{(i)}=\nabla_4\nabla_3-\triangle+\omega_\mck\nabla_3+O(1)\df^{\leq 1},$$ and where the error terms satisfy in $\deux$, for $i=1,\ldots,6$, and denoting $S_0A:=A$ by convention,
\begin{align}\label{eq:errmco}
	\Omega^2|\err[\mcl^{(i)}(S_i A)]|^2\lesssim \sum_{j=0}^{i-1}\Omega^2\mathbf{e}[S_j A]+\Omega^2|A|^2+\Omega^{-2}\ubar^{-8-4\delta/5}.
\end{align}
\end{prop}
\begin{proof}
	This is a direct consequence of Proposition \ref{prop:nouvellescommut}, where we notice that in that result the error term satisfy $|\err|\lesssim\Omega^{-2}\ubar^{-4-2\delta/5}$ by Proposition \ref{prop:coeffin}, hence $\Omega^2|\err|^2\lesssim \Omega^{-2}\ubar^{-8-4\delta/5}$.\end{proof}
Now, we write the equations satisfied by $\nabla_4$ derivatives of $A$.
\begin{prop}\label{prop:teukcommutepleindefois4}
	We define the ordered set of horizontal operators
	$$\mathcal{O}_4=(S_1^{(4)},S_2^{(4)},S_3^{(4)},S_4^{(4)},S_5^{(4)},S_6^{(4)},S_7^{(4)}),$$
	where $S_1^{(4)}=\nabla_4$, $S_2^{(4)}=\nabla_4\divc$, $S_3^{(4)}=\nabla_4\divc\divc$, $S_4^{(4)}=\nabla_4\mcd\divc\divc$, $S_5^{(4)}=\nabla_4\nabla_4$, $S_6^{(4)}=\nabla_4\nabla_4\divc$, $S_7^{(4)}=\nabla_4\nabla_4\divc\divc$. Then for $i=1,\ldots,7$ we have the Teukolsky-like equation
	$$\mcl^{(i)}_4(S_i^{(4)} A)=\err[\mcl^{(i)}_4(S_i A)],$$
	with  $\mcl^{(i)}_4=\nabla_4\nabla_3-\triangle+\omega_\mck\nabla_3+O(1)\df^{\leq 1}$, and where in $\deux$, for $i=1,\ldots,7$, denoting $S_0^{(4)}A:=A$,
\begin{align}\label{eq:errmco4}
	\Omega^2|\err[\mcl^{(i)}_4(S_i^{(4)} A)]|^2\lesssim \sum_{j=0}^{i-1}\Omega^2\mathbf{e}[S_j^{(4)} A]+\sum_{j=1}^{6}\Omega^2\mathbf{e}[S_j A]+\Omega^2|A|^2+\Omega^{-2}\ubar^{-8-4\delta/5}.
\end{align}
\end{prop}
\begin{proof}
This is a direct consequence of Proposition \ref{prop:nouvellescommut}, similarly as Proposition \ref{prop:teukcommutepleindefois}.
\end{proof}
Recalling $\hat{\lambda}\sim\Omega^2$ from Remark \ref{rem:remlabel}, we write the equations satisfied by $\hat{\lambda}\nabla_3$ derivatives.
\begin{prop}\label{prop:teukcommutepleindefois3}
	We define the ordered set of horizontal operators
	$$\mathcal{O}_3=(S_1^{(3)},S_2^{(3)},S_3^{(3)},S_4^{(3)},S_5^{(3)},S_6^{(3)},S_7^{(3)}),$$
	where $S_1^{(3)}=\hat{\lambda}\nabla_3$, $S_2^{(3)}=\hat{\lambda}\nabla_3\divc$, $S_3^{(3)}=\hat{\lambda}\nabla_3\divc\divc$, $S_4^{(3)}=\hat{\lambda}\nabla_3\mcd\divc\divc$, $S_5^{(3)}=\hat{\lambda}\nabla_3(\hat{\lambda}\nabla_3)$, $S_6^{(3)}=\hat{\lambda}\nabla_3(\hat{\lambda}\nabla_3)\divc$, $S_7^{(3)}=\hat{\lambda}\nabla_3(\hat{\lambda}\nabla_3)\divc\divc$. Then for $i=1,\ldots,7$ we have
	$$\mcl^{(i)}_3(S_i^{(3)} A)=\err[\mcl^{(i)}_3(S_i A)],$$
with $\mcl^{(i)}_3=\nabla_4\nabla_3-\triangle+\omega_\mck\nabla_3+O(1)\df^{\leq 1}$, and where, in $\deux$, for $i=1,\ldots,7$, denoting $S_0^{(3)}A:=A$,
	\begin{align}\label{eq:errmco3}
		\Omega^2|\err[\mcl^{(i)}_3(S_i^{(3)} A)]|^2\lesssim \sum_{j=0}^{i-1}\Omega^2\mathbf{e}[S_j^{(3)} A]+\sum_{j=1}^{6}\Omega^2\mathbf{e}[S_j A]+\Omega^2|A|^2+\Omega^{-2}\ubar^{-8-4\delta/5}.
	\end{align}
\end{prop}
\begin{proof}
This is a direct consequence of Proposition \ref{prop:nouvellescommut}, similarly as Proposition \ref{prop:teukcommutepleindefois}.
\end{proof}
\begin{rem}From Propositions \ref{prop:teukcommutepleindefois}, \ref{prop:teukcommutepleindefois4}, \ref{prop:teukcommutepleindefois3}, we see that we will be able to control $\mco,\mco_4,\mco_3$ derivatives of $A$ by the energy estimates for generalized Teukolsky fields (Proposition \ref{prop:enerteukcomm}) in the following order: we first control $\mco$ derivatives by induction and then we control $\mco_3$ and $\mco_4$ derivatives by induction as well, relying on the energy control of $\mco$ derivatives.
\end{rem}
\subsubsection{ Energy and $L^2(S(u,\ubar))$ decay for derivatives of $A$}\label{section:L2smallderA}
We will now state and prove the schematic $L^2(S(u,\ubar))$ decay in $\deux$ stated in \eqref{eq:schematicannounced}. Note that by Proposition \ref{prop:enerpointwisefromISA}, for $S$ in the following set of operators, 
\begin{align}\label{eq:setofopS}
	S\in\mathcal{S}=\Big\{&\nabla_4^{\leq 2},\:\nabla_4^{\leq 2}\divc,\:\nabla_4^{\leq 2}\divc\divc,\:\nabla_4^{\leq 1}\mcd\divc\divc,\:\mcd\hot\divc,\:\divc\mcd\hot\divc,\:\divc\mcd\divc\divc,\\
	&(\hat{\lambda}\nabla_3)^{\leq 2},\:(\hat{\lambda}\nabla_3)^{\leq 2}\divc,\:(\hat{\lambda}\nabla_3)^{\leq 2}\divc\divc,\:(\hat{\lambda}\nabla_3)^{\leq 1}\mcd\divc\divc\Big\}=\mco\cup\mco_3\cup\mco_4,\nn
\end{align} 
we have the (non-sharp) initial energy decay on $\Sigma_0$,
\begin{align}\label{eq:initialenergyA}
	\int_{\Sigma_0\cap\{\wbar_1\leq\wbar\leq\wbar_2\}}\left(|\nabla_3 SA|^2+|\nabla_4SA|^2+|{\nabla} SA|^2\right)\lesssim\wbar_1^{-2(3+\delta/2)}+\int_{\wbar_1}^{\wbar_2}\wbar^{-2(3+\delta/2)}\dee\wbar,
\end{align}
and the following initial $L^2(S(u,\ubar))$ decay on $\Sigma_0$,
\begin{align}\label{eq:nonsharpL2decayinitialA}
	\|SA\|_{L^2(S(C_R-\ubar,\ubar))}\lesssim\ubar^{-3-\delta/2}.
\end{align}
Note that by \eqref{eq:initA4plusdelta}, for $SA=A$, up to a constant we can replace the exponent $3+\delta/2$ in the bounds above by $4+\delta$. By \eqref{eq:boundalphaprimin}, we also have the following pointwise decay in $\deux$, for $S\in\mathcal{S}$,
\begin{align}\label{eq:pointwiseA}
	\Ldeux{\nabla_4^{\leq 1}SA}\lesssim{\ubar^{-2-\delta/5}},\quad\text{in}\quad\deux.
\end{align}
\begin{defi}\label{defi:anyy}
	We denote by $\lesssim_C$ any $(a,M,\delta_+,\delta,u_f,C_R,(Q_m)_{|m|\leq 2})$-dependent bound.
\end{defi}
We now prove the following non-sharp energy and $L^2(S(u,\ubar))$ decay result for $A$ in $\deux$.
\begin{prop}
	We have, in $\deux$, 
	\begin{align}\label{eq:prelimdecayAdeux}
		\Ener[A](\wbar)\lesssim_C \wbar^{-8-4\delta/5+\gamma},\quad \Ldeux{A}\lesssim_C\ubar^{-4-2\delta/5+\gamma}.
	\end{align}
\end{prop}
\begin{proof}
	This is a simple application of Propositions \ref{prop:enerteukAerrA} and \ref{prop:decayL2AerrA}. Indeed, by \eqref{eq:teukA} we have
	$$|\mcl(A)|\lesssim |(\divc)^{\leq 1} B|^2+|(\divc)^{\leq 1}\wh{X}|^2+|\wh{X}||\wh{\Xbar}||A|\lesssim\ubar^{-4-2\delta/5}+\Omega^{-2}\ubar^{-6-3\delta/5}\lesssim\Omega^{-2}\ubar^{-4-2\delta/5},$$
	in $\deux$, where we used the bounds for $B$ and $\wh{X}$ in Proposition \ref{prop:coeffin}. Together with \eqref{eq:initA4plusdelta}, this proves that the assumptions of Propositions \ref{prop:enerteukAerrA} and \ref{prop:decayL2AerrA} are satisfied with 
	$$\beta=4+2\delta/5,\quad\beta'=\beta''=2+\delta/5,\quad C=C(a,M,u_f),$$
	where we used $C(a,M,\delta_\pm,w_f,(Q_m)_{|m|\leq 2})=C(a,M,\delta_+,u_f,(Q_m)_{|m|\leq 2})$ in this context\footnote{$\delta_-=\delta_-(a,M)$ and $w_f=w_f(a,M,u_f)$ are chosen independently of the estimates for $A$.}. Here, 
	$$\min(\beta,\beta'+2+\delta/5,\beta''+4+2\delta/5)=4+2\delta/5,$$
	which concludes the proof, by applying Propositions \ref{prop:enerteukAerrA} and \ref{prop:decayL2AerrA}.
\end{proof}
Recalling the set of operators $\mco$ defined in Proposition \ref{prop:teukcommutepleindefois}, we now use the decay results for $A$ and the structure of the successive equations satisfied by $\mco$ derivatives of $A$ to deduce the similar decay results in $\deux$ for $\mco A$, replacing the exponent $4+2\delta/5$ by $3+\delta/2$. 
\begin{prop}\label{prop:prelimdecaydivcAdeux}
	Provided $0<\gamma<\delta/5$, we have in $\deux$ for any $S_i\in\mco$,
	\begin{align}\label{eq:prelimdecaydivcAdeux}
		\Ener[S_i A](\wbar)\lesssim_C\wbar^{-6-\delta},\quad \Ldeux{S_i A}\lesssim_C\ubar^{-3-\delta/2+\gamma/2}.
	\end{align}
\end{prop}
\begin{proof}
Recall the operators $S_i$ defined in Proposition \ref{prop:teukcommutepleindefois}. We prove by induction on $i=0,\ldots,7$ that \eqref{eq:prelimdecaydivcAdeux} holds in $\deux$. It holds for $i=0$ by \eqref{eq:prelimdecayAdeux}. Now, let $1\leq i\leq 7$ and let us assume that it holds for all $j\leq i-1$. By Proposition \ref{prop:teukcommutepleindefois} we have for any $C_R-u_f\leq\wbar_1\leq\wbar_2,$
\begin{align*}
	\int_{\deux[\wbar_1,\wbar_2]}\Omega^2|\err[\mcl^{(i)}(S_i A)]|^2&\lesssim\int_{\deux[\wbar_1,\wbar_2]}\left(\sum_{j=0}^{i-1}\Omega^2\mathbf{e}[S_j A]+\Omega^{-2} \ubar^{-8-2\delta}+\Omega^2|A|^2\right)\\
	&\lesssim_C\int_{\wbar_1}^{\wbar_2}\left(\sum_{j=0}^{i-1}\Ener[S_jA](\wbar)+\wbar^{-8-4\delta/5+3\gamma}\right)\dee\wbar\lesssim_C\int_{\wbar_1}^{\wbar_2}\wbar^{-6-\delta}\dee\wbar,
\end{align*}
where we used \eqref{eq:prelimdecayAdeux}, \eqref{eq:bornimportane}, and \eqref{eq:prelimdecaydivcAdeux} at rank $j\leq i-1$, and $0<\gamma<\delta/5$. This proves assumption \eqref{eq:hyperrhat} of Proposition \ref{prop:enerteukcomm}, for $\wh{\mcl}=\mcl^{(i)}$, $U=S_iA$. Using \eqref{eq:prelimdecaydivcAdeux} at rank $j\leq i-1$ again, and $S_iA=_s\nabla(S_{\leq i-1}A)$, we get
$$\int_{\deux[\wbar_1,\wbar_2]}\Omega^2|S_i A|^2\lesssim\sum_{j=0}^{i-1}\int_{\wbar_1}^{\wbar_2}\Ener[S_jA](\wbar)\dee\wbar\lesssim_C\int_{\wbar_1}^{\wbar_2}\wbar^{-6-\delta}\dee\wbar,$$
which proves that assumption \eqref{eq:hypenerprecedente} is also satisfied. Moreover, assumptions \eqref{eq:hyppointwise} and \eqref{eq:initialenergyhyp} are satisfied by \eqref{eq:pointwiseA}, \eqref{eq:initialenergyA}. Finally, by Proposition \ref{prop:teukcommutepleindefois} the operators $\mcl^{(i)}=\nabla_4\nabla_3-\triangle+\omega_\mck\nabla_3+O(1)\df^{\leq 1}$ satisfy assumption \eqref{eq:hypLhat}, where we use in particular $\omega_\mck\gtrsim 1$ in $\deux$. Thus, by Proposition \ref{prop:enerteukcomm}, we deduce the energy decay in \eqref{eq:prelimdecaydivcAdeux} at rank $i$, and we also deduce the $L^2(S(u,\ubar))$ decay in $\deux$ by using \eqref{eq:nonsharpL2decayinitialA} and by applying Proposition \ref{prop:decayL2genteukfield}. This concludes the induction and hence the proof.
\end{proof}
We now prove the similar decay estimates for more general $\mathcal{S}$ derivative of $A$ in $\deux$.
\begin{prop}\label{prop:premlimdecaySAdeux}
	We have in $\deux$, for $S\in\mathcal{S}$ as defined in \eqref{eq:setofopS},
	\begin{align}\label{eq:prelimdecaySAdeux}
		\Ener[SA](\wbar)\lesssim_C\wbar^{-6-\delta},\quad \Ldeux{SA}\lesssim_C\ubar^{-3-\delta+\gamma/2}.
	\end{align}
\end{prop}
\begin{proof}
Relying on Propositions \ref{prop:teukcommutepleindefois4} and 	\ref{prop:teukcommutepleindefois3} we prove by induction on $1\leq i\leq 7$ that $S_i^{(4)}A$ and $S_i^{(3)}A$ satisfy \eqref{eq:prelimdecaySAdeux}. Using the energy bounds of Proposition \ref{prop:prelimdecaydivcAdeux} to bound the integral of the terms $\sum_{j=1}^{6}\Omega^2\mathbf{e}[S_j A]$ which appear on the RHS of \eqref{eq:errmco4} and \eqref{eq:errmco3}, we proceed exactly like in the proof of Proposition \ref{prop:prelimdecaydivcAdeux}, so we omit the details.
\end{proof}
\subsubsection{Pointwise decay for some horizontal derivatives of $A$}\label{section:pointwisehorizA}
In this section, we prove polynomial decay for $(\divc )^{\leq 2}A$ in $\deux$. This will follow from Proposition \ref{prop:premlimdecaySAdeux}, combined with the Sobolev embedding of Proposition \ref{prop:sovolinfty}, which we recall here : given a horizontal tensor $U\in\fraks_k(\C)$ in $\deux$, $k\in\{0,1,2\}$, we have the estimate
\begin{align*}
	\|{U}\|_{L^\infty(S(u,\ubar))}^2\lesssim\intS\Big(&|\triangle_k{U}|^2+|\mcd_k{U}|^2+|{U}|^2+|\hat{\lambda}\nabla_3\mcd_k{U}|^2+|\nabla_4\mcd_k{U}|^2\\
	&+|\hat{\lambda}\nabla_3{U}|^2+|\nabla_4{U}|^2+|(\hat{\lambda}\nabla_3)^2{U}|^2+|\Omega^2\nabla_4\nabla_3{U}|^2+|\nabla_4^2{U}|^2\Big),
\end{align*}
where the operator $\mcd_k$ is given by $\mcd_2=\divc,\:\mcd_1=\divc,\:\mcd_0=\mcd$.
\begin{prop}\label{prop:pointwisenablaleq2A}
	For $0<\gamma<\delta/3$, recalling Definition \ref{defi:anyy}, we have the estimate
	$$|A|+|\divc A|+|\divc(\divc A)|\lesssim_C\frac{1}{\ubar^{3+\delta/3}},\quad\text{ in }\deux.$$
\end{prop}
\begin{proof}
	We begin by proving the bound for $A$. By Proposition \ref{prop:sovolinfty} we have
	\begin{align*}
		\|{A}\|_{L^\infty(S(u,\ubar))}^2\lesssim\intS\Big(&|\triangle_k{A}|^2+|\mcd_2{A}|^2+|{A}|^2+|\hat{\lambda}\nabla_3\mcd_2{A}|^2+|\nabla_4\mcd_2{A}|^2\\
		&+|\hat{\lambda}\nabla_3{A}|^2+|\nabla_4{A}|^2+|(\hat{\lambda}\nabla_3)^2{A}|^2+|\Omega^2\nabla_4\nabla_3{A}|^2+|\nabla_4^2{A}|^2\Big).
	\end{align*}
	By Lemma \ref{lem:ignocoucou} combined with	$\Omega^2\sim\hat{\lambda},\: |\atrchibar|\lesssim 1,\: |\atrchi|\lesssim\Omega^2,\:|{}^{(h)}K|\lesssim 1$, which come from Proposition \ref{prop:gausscurvbornee} and \eqref{eq:borneantitracesin}, we deduce
	\begin{align*}
		\|{A}\|_{L^\infty(S(u,\ubar))}^2\lesssim\intS\Big(&|\mcd\hot(\divc {A})|^2+|\divc{A}|^2+|{A}|^2+|\hat{\lambda}\nabla_3\divc{A}|^2+|\nabla_4\divc{A}|^2\\
		&+|\hat{\lambda}\nabla_3{A}|^2+|\nabla_4{A}|^2+|(\hat{\lambda}\nabla_3)^2{A}|^2+|\Omega^2\nabla_4\nabla_3{A}|^2+|\nabla_4^2{A}|^2\Big).
	\end{align*}
	Moreover, by the Teukolsky equation \ref{eq:teukA} and Proposition \ref{prop:coeffin} we get
	\begin{align*}
		|\Omega^2\nabla_4\nabla_3{A}|\lesssim & |\mcd\hot(\divc {A})|+\Omega^2 |\nabla_3 A|+|\nabla_4 A|+ |\nabla A|+ |A|+ \Omega^2|\err[\mcl(A)]|\\
		\lesssim & |\mcd\hot(\divc {A})|+ |\hat{\lambda}\nabla_3 A|+|\nabla_4 A|+ |\nabla A|+ |A|+\ubar^{-4-2\delta/5}.
	\end{align*}
	Thus, using also Lemma \ref{lem:619} to deal with the term $|\nabla A|$ above, we deduce 
	\begin{align*}
		\|{A}\|_{L^\infty(S(u,\ubar))}^2\lesssim \ubar^{-8-4\delta/5}+\intS\Big(&|\mcd\hot(\divc {A})|^2+|\divc{A}|^2+|{A}|^2+|\hat{\lambda}\nabla_3\divc{A}|^2+|\nabla_4\divc{A}|^2\\
		&+|\hat{\lambda}\nabla_3{A}|^2+|\nabla_4{A}|^2+|(\hat{\lambda}\nabla_3)^2{A}|^2+|\nabla_4^2{A}|^2\Big),
	\end{align*}
	hence the bound  $\|{A}\|_{L^\infty(S(u,\ubar))}\lesssim_C \ubar^{-3-\delta/2+\gamma/2}\lesssim_C\ubar^{-3-\delta/3}$ by Proposition \ref{prop:premlimdecaySAdeux}, for $\gamma<\delta/3$. Similarly, by Proposition \ref{prop:sovolinfty}, Lemmas \ref{lem:unlemsympa}, \ref{lem:laplaciens1(C)}, \ref{lem:unlemsympascalair} and Proposition \ref{prop:teukcommutepleindefois} we get
		\begin{align*}
		\|\divc{A}\|_{L^\infty(S(u,\ubar))}^2\lesssim&\:\ubar^{-8-4\delta/5}+\intS\Big(|\mcd(\divc (\divc{A}))|^2+|\divc(\divc{A})|^2+|\hat{\lambda}\nabla_3\divc(\divc{A})|^2\\
		&+|\nabla_4\divc(\divc{A})|^2+|\hat{\lambda}\nabla_3\divc{A}|^2+|\nabla_4\divc{A}|^2+|\hat{\lambda}\nabla_3(\hat{\lambda}\nabla_3\divc{A})|^2\\
		&+|\nabla_4\nabla_4\divc{A}|^2+|\hat{\lambda}\nabla_3A|^2+|\nabla_4A|^2+|\divc A|^2+|A|^2\Big)\lesssim_C\ubar^{-6-2\delta/3},
	\end{align*}
and 
	\begin{align*}
	&\|\divc(\divc{A})\|_{L^\infty(S(u,\ubar))}^2\lesssim\\
	&\intS\Big(|\divc(\mcd(\divc (\divc{A})))|^2+|\mcd\divc(\divc{A})))|^2+|\divc(\divc{A})|^2+|\hat{\lambda}\nabla_3\mcd\divc(\divc{A})|^2\\
	&+|\nabla_4\mcd\divc(\divc{A})|^2+|\hat{\lambda}\nabla_3\divc(\divc{A})|^2+|\nabla_4\divc(\divc{A})|^2+|\hat{\lambda}\nabla_3(\hat{\lambda}\nabla_3\divc(\divc{A}))|^2\\
	&+|\nabla_4\nabla_4\divc(\divc{A})|^2+|\hat{\lambda}\nabla_3\divc A|^2+|\nabla_4\divc A|^2+|\divc(\divc A)|^2+|\hat{\lambda}\nabla_3A|^2+|\nabla_4 A|^2\\
	&+|\divc A|^2+|A|^2\Big)+\ubar^{-8-4\delta/5}\lesssim_C\ubar^{-6-\delta+\gamma}\lesssim_C\ubar^{-6-2\delta/3},
\end{align*} 
where we used Proposition \ref{eq:prelimdecaySAdeux} and $\gamma<\delta/3$ in the last steps above, concluding the proof.
\end{proof}

\subsection{Sharp asymptotics for $A$ in $\deux$}\label{section:sharpforAdeux}

In this section, we use the estimates proven in Section \ref{section:energyteukgeneralizedsectionpassubsection} to deduce the sharp asymptotics for $A$ in region $\deux$. First, in Section \ref{section:betterdecayteukerr}, from Proposition \ref{prop:pointwisenablaleq2A} and some null structure and Bianchi equations, we deduce improved decay for $(\divc)^{\leq 1}\wh{X}$ and $(\divc)^{\leq 1}B$, and hence for the Teukolsky error term \eqref{eq:teukerror}, in $\deux$. Then, in Section \ref{section:teukappliedtoansatz} we prove an upper bound for the Teukolsky operator applied to the ansatz $\Psi$. Combining these two results, the energy method of Section \ref{section:energyteukgeneralizedsectionpassubsection}, and the initial precise estimates for $A$ on $\Sigma_0$, we obtain in Section \ref{section:preciseasymptotics} the sharp asymptotics of $A$ in region $\deux$.

\subsubsection{Better decay for the Teukolsky error term}\label{section:betterdecayteukerr}
We recall that the primed quantities denote geometric quantities defined with respect to the \emph{outgoing} non-integrable frame $(e_\mu')$ defined in Section \ref{section:defprincipalframes}, which is such that $\omega'=0$, and that the unprimed notation denotes geometric quantities defined with respect to the \emph{ingoing} non-integrable frame $(e_\mu)$ which satisfies $\omegabar=0$.
\begin{lem}\label{lem:goui}
	Let ${U}$ be a horizontal tensor such that $|{U}|\lesssim{\ubar^{-3-\delta/3}}\text{ on }\Sigma_0$ and $|\nabla_{4'}{U}|\lesssim{\Omega^{-4}}{\ubar^{-3-\delta/3}}\text{ in }\deux$. Then, in $\deux$,
	$$|{U}|\lesssim\frac{\Omega^{-2}}{\ubar^{3+\delta/3}}.$$
\end{lem}
\begin{proof}
By Proposition \ref{prop:transporte4'} we have
	\begin{align*}
		|U|(p)&\lesssim |U|(u_{Y,\Sigma_0}(p),\ubar_{Y,\Sigma_0}(p),\theta^A_{Y,\Sigma_0}(p))+\int_{\ubar_{Y,\Sigma_0}(p)}^\ubar|\Omega^2\nabla_{4'}U|(u_Y^p(\ubar'),\ubar',(\theta^A)_Y^p(\ubar'))\dee\ubar'\\
		&\lesssim\frac{1}{|u|^{3+\delta/3}}+\int_{\ubar_{Y,\Sigma_0}(p)}^\ubar\frac{e^{|\kappa_-|(u+\ubar')}}{(\ubar')^{3+\delta/3}}\dee\ubar'\lesssim\frac{\Omega^{-2}}{\ubar^{3+\delta/3}},
	\end{align*}
	where we used \eqref{eq:equivtransporte4'} in the second step to write in $\deux$, $\ubar_{Y,\Sigma_0}(p)=C_R-u_{Y,\Sigma_0}(p)\sim |u_{Y,\Sigma_0}(p)|\sim |u|\sim\ubar$, $e^{|\kappa_-|(u_Y^p(\ubar')+\ubar')}\sim e^{|\kappa_-|(u+\ubar')},$ and the bound
	$$\int_{\ubar_{Y,\Sigma_0}(p)}^\ubar\frac{e^{|\kappa_-|(u+\ubar')}}{(\ubar')^{3+\delta/3}}\dee\ubar'\lesssim\frac{e^{|\kappa_-|(u+\ubar)}}{\ubar^{3+\delta/3}}$$
	in the last step, which comes from an integration by parts. This concludes the proof.
\end{proof}
\begin{prop}\label{prop:betterdecayteukerrterm}
	Recalling Definition \ref{defi:anyy}, we have the following estimate in $\deux$,
	$$|(\divc)^{\leq 1}\wh{X}|+|(\divc)^{\leq 1}B|\lesssim_C\frac{1}{\ubar^{3+\delta/3}}.$$
\end{prop}
\begin{proof}
	By the null structure and Bianchi identities in the outgoing non-integrable frame (see Section \ref{section:ricciandcurvcomplex}) for $\nabla_{4'}B',\nabla_{4'}\wh{X}'$ combined with $\omega'=0$, we have in $\deux$,
	\begin{align}
		&\nabla_{4'}\wh{X}'+\Real(tr X')\wh{X}'=-A',\label{eq:nullxhat}\\
		&\nabla_{4'}B'+2\overline{trX'}B'=\frac{1}{2}\divc A'+\frac{1}{2}A'\cdot (\overline{2Z'+\Hbar'}).\label{eq:bianchiB}
	\end{align}
	Using \eqref{eq:commnab4divc} in the outgoing non-integrable frame we obtain
	\begin{align*}
		[\nabla_{4'},\divc]\wh{X}'&=-\frac{1}{2}\overline{trX'}\divc\wh{X}'+(\overline{\Hbar'+Z'})\cdot\nabla_{4'}\wh{X}-\wh{\chi}'\cdot\overline{\mcd}\wh{X}'+D_4^{(2)}[\wh{X}']\\
		&=-\frac{1}{2}\overline{trX'}\divc\wh{X}'-\Real(trX')(\overline{\Hbar'+Z'})\cdot\wh{X}'+D_4^{(2)}[\wh{X}']-(\overline{\Hbar'+Z'})\cdot A'+O\left(\frac{\Omega^{-4}}{\ubar^{4+2\delta/5}}\right),\\
		[\nabla_{4'},\divc]B'&=-\frac{1}{2}\overline{trX'}\divc B'-2\overline{trX'}(\overline{\Hbar'+Z'})\cdot B'+D_4^{(1)}[B']+O\left(\frac{\Omega^{-4}}{\ubar^{4+2\delta/5}}\right)\\
		&\quad +(\overline{\Hbar'+Z'})\cdot \left(\frac{1}{2}\divc A'+\frac{1}{2}A'\cdot (\overline{2Z'+\Hbar'})\right).
	\end{align*}
where we also used Proposition \ref{prop:coeffout}. Thus commuting \eqref{eq:nullxhat} and \eqref{eq:bianchiB} with $\divc$ gives
	\begin{align}
		&\left|\nabla_{4'}(\divc\wh{X}')+\left(\frac{1}{2}\overline{tr X'}+\Real(trX')\right)\divc\wh{X}'\right|\lesssim |(\divc)^{\leq 1}A'|+|\wh{X}'|+\frac{\Omega^{-4}}{\ubar^{4+2\delta/5}},\\
		&\left|\nabla_{4'}(\divc B')+\frac{5}{2}\overline{tr X'}(\divc B')\right|\lesssim|(\divc)^{\leq 2}A'|+|B'|+\frac{\Omega^{-4}}{\ubar^{4+2\delta/5}}.
	\end{align}
Or, equivalently,
	\begin{align*}
		|\nabla_{4'}(h_1\wh{X}')|&\lesssim |A'|,\quad|\nabla_{4'}(h_2 B')|\lesssim |(\divc)^{\leq 1}A'|,\\
		|\nabla_{4'}(h_3\divc\wh{X}')|&\lesssim|(\divc)^{\leq 1}A'|+|\wh{X}'|+\frac{\Omega^{-4}}{\ubar^{4+2\delta/5}},\quad|\nabla_{4'}(h_4\divc B')|\lesssim|(\divc)^{\leq 2}A'|+|B'|+\frac{\Omega^{-4}}{\ubar^{4+2\delta/5}}.
	\end{align*}
where the scalar functions $h_1$, $h_2$, $h_3$, $h_4$ satisfy $h_k|_{\Sigma_0}=1$, and the transport equations
	\begin{align*}
		&\Omega^2\nabla_{4'}h_1-\Omega^2\Real(trX')h_1=0,\quad\Omega^2\nabla_{4'}h_2-2\Omega^2\overline{trX'}h_2=0,\\
		&\Omega^2\nabla_{4'}h_3-\Omega^2\left(\frac{1}{2}\overline{tr X'}+\Real(trX')\right)h_3=0,\quad\Omega^2\nabla_{4'}h_4-\frac{5}{2}\Omega^2\overline{trX'}h_4=0,
	\end{align*}
 so that $|h_k|\sim 1$ for $k=1,2,3,4$ in $\deux$ by the bound $|\Omega^2trX'|\lesssim\Omega^2+\ubar^{-2-\delta/5}$ (see Proposition \ref{prop:coeffout}). Moreover, by $|\nabla^{\leq 2}\hat{\Omega}^{-4}|\lesssim \Omega^{-4}$, $A'=\hat{\Omega}^{-4}A$, and Proposition \ref{prop:pointwisenablaleq2A}, we have in $\deux$, $|(\divc)^{\leq 2} A'|\lesssim \hat{\Omega}^{-4}|(\divc)^{\leq 2} A| \lesssim_C\Omega^{-4}\ubar^{-3-\delta/3}$. The proof of Proposition \ref{prop:betterdecayteukerrterm} then follows from Lemma \ref{lem:goui}, which can be directly applied for $B,\wh{X}$, and then for $\divc B,\divc\wh{X}$ using the previously proven bounds for $B,\wh{X}$. 
\end{proof}

\subsubsection{Upper bound for Teukolsky operator applied to the ansatz}\label{section:teukappliedtoansatz}
To ensure that assumption \eqref{eq:teukolskylike} of Proposition \ref{prop:enerteukAerrA} is satisfied for $U=\err[A]$ with $\beta>6$, we need to bound $\mcl(\Psi)$. First, we define for $|m|\leq 2$, 
$$\hat{\Psi}_m=\frac{ A_m(r_\mck)}{\qbar_\mck^2}(\mcd\hot(\mcd(Y_{m,2}(\cos\theta)e^{im\phi_+}))_\mck^S)^{H}\in\fraks_2(\C),$$
such that, recalling \eqref{eq:defansatzdeux}, $\Psi={\ubar_\mck'^{-6}}\sum_{|m|\leq 2}Q_m \hat{\Psi}_m$.
\begin{lem}\label{lem:lemoui}
	We have the followings bounds in $\deux$,
	\begin{align*}
		|e_4(\ubar_\mck')|\lesssim 1,\quad |e_3(\ubar_\mck')|\lesssim\Omega^{-2},\quad |\nabla^{\leq 2}\ubar_\mck'|\lesssim 1,\quad |e_4(e_3(\ubar_\mck'))|\lesssim 1.
	\end{align*}
\end{lem}
\begin{proof}
	Using the following identities\footnote{These come from simply computing all the scalar products between $e_\mu$ and $e'_\nu$ defined by a frame transformation as in \eqref{eq:frametransfo}.} in Kerr,
	\begin{align*}
		(\ering_3)_\mck&=\left(\frac{1}{4}\hat{\lambda}^{-1}|\fbar|^2e_4+\hat{\lambda} e_3-\fbar^a e_a\right)_\mck,\\
		(\ering_4)_\mck&=\Big(\hat{\lambda}^{-1}\left(1+\frac{1}{2}f\cdot\fbar+\frac{1}{6}|f|^2|\fbar|^2\right)e_4+\hat{\lambda}|f|^2 e_3-\Big(f^a+\frac{1}{4}|f|^2\fbar_a\Big)e ^a\Big)_\mck,\\
		(\ering_a)_\mck&=\Big(-\frac{1}{2}\hat{\lambda}^{-1}\Big(\fbar_a+\frac{1}{4}|\fbar|^2 f_a\Big)e_4-\frac{1}{2}\hat{\lambda} f_a e_3+\Big(\delta_{a}^b+\frac{1}{2}\fbar_a f^b\Big)e_b\Big)_\mck,
	\end{align*}
	as well as $\partial_\ubar=\Omega^2_\mck (\ering_4)_\mck-b^A_\mck\partial_{\theta^A}$ and the identities $(e_3)_\mck(\ubar_\mck')=0,\: (e_4)_\mck(\ubar_\mck')=2,\: |\nabla_\mck\ubar_\mck'|\lesssim 1, $ in Kerr, we compute the following derivatives
	\begin{align*}
		\partial_u (\ubar_\mck')&=\frac{1}{2}\hat{\lambda}_\mck^{-1}|\fbar|_\mck^2-\fbar_\mck\cdot_\mck (\nabla\ubar_{\dagger})_\mck=O(\Omega^2),\\
		\nabring(\ubar_\mck')&=-\hat{\lambda}_\mck^{-1}\Big(\fbar+\frac{1}{4}|\fbar|^2 f\Big)_\mck+(\nabla\ubar_{\dagger})^S_\mck+\frac{1}{2}(f\cdot\nabla\ubar_{\dagger})_\mck \fbar_\mck=O(1),\\
		\partial_\ubar(\ubar_\mck')&=\Omega_\mck^2\left( 2\hat{\lambda}_\mck^{-1}\left(1+\frac{1}{2}f\cdot\fbar+\frac{1}{6}|f|^2|\fbar|^2\right)_\mck-\Big(f+\frac{1}{4}|f|^2\fbar\Big)_\mck\cdot_\mck (\nabla\ubar_{\dagger})_\mck\right)-b_\mck\cdot\nabring\ubar_\mck'=O(1).
	\end{align*}
	Combining this with the expression of $e_3,e_4,e_a$ with respect to $\ering_3,\ering_4,\ering_a$, $f,\fbar,\hat{\lambda}$, and the bounds $|\df^{\leq 1}f|\lesssim 1$, $|\df^{\leq 1}\fbar|\lesssim \Omega^2$, $|\df^{\leq 1}\hat{\lambda}|\lesssim\Omega^2$, we easily conclude the proof.
\end{proof}
Recall $V_0':=V-4(\eta\cdot(2\zeta+\etabar)-i\eta\wedge(2\zeta+\etabar)),$ defined in the second item of Remark \ref{rem:boundteukerr} whic is such that the Teukolsky operator \eqref{eq:teukop} rewrites
\begin{align}\label{eq:expressioncooldeteuk}
	\mcl(U)=\nabla_4\nabla_3U-\frac{1}{4}\mcd\hot(\divc U)+h_0\nabla_3U+\hhbar_0\nabla_4U+(F+q)\cdot\nabla U+V_0'U,
\end{align}
where $h_0=\frac{1}{2}trX+2\overline{trX}+2\omega,\: \hhbar_0=-\frac{1}{2}tr\Xbar,\: F=4H,\: q=4\zeta+2\etabar$.
\begin{prop}\label{prop:teukansatzz}
	Let $\Psi$ be as in \eqref{eq:defansatzdeux}. We have in $\deux$,
	$$|\mcl(\Psi)|\lesssim_Q\Omega^{-2}\ubar^{-7}. $$
\end{prop}

\begin{proof}
We denote here by $\lesssim$ the $(Q_m)$-dependent bounds. By \eqref{eq:teukop} we compute
	\begin{align*}
		\mcl({\Psi})=&\left(\frac{-6}{\ubar_\mck'^7}e_4(e_3(\ubar_\mck'))+\frac{42}{\ubar_\mck'^{8}}e_4(\ubar_\mck')e_3(\ubar_\mck')\right)\sum_{|m|\leq 2}Q_m\hatpsim-\frac{6e_3(\ubar_\mck')}{\ubar_\mck'^7}\sum_{|m|\leq 2}Q_m\nabla_4\hatpsim\\
		&-\frac{6e_4(\ubar_\mck')}{\ubar_\mck'^7}\sum_{|m|\leq 2}Q_m\nabla_3\hatpsim+3\frac{\mcd\cdot(\overline{\mcd}(\ubar_\mck'))}{\ubar_\mck'^7}\sum_{|m|\leq 2}Q_m\hatpsim+6\frac{\overline{\mcd}(\ubar_\mck')}{\ubar_\mck'^7}\cdot\sum_{|m|\leq 2}Q_m\nabla\hatpsim\\
		&+\frac{6}{\ubar_\mck'^7}\left[(2Z+\Hbar+\overline{2Z+\Hbar})\cdot\nabla\ubar_\mck'-\left(\frac{1}{2}tr\Xbar-4\omegabar\right){e_4(\ubar_\mck')}\right]\sum_{|m|\leq 2}Q_m\hatpsim\\
		&-\frac{6}{\ubar_\mck'^7}\left(\frac{1}{2}trX+2\overline{trX}+2\omega\right){e_3(\ubar_\mck')}\sum_{|m|\leq 2}Q_m\hatpsim\\
		&-\frac{3}{2}\frac{{\mcd}(\ubar_\mck')}{\ubar_\mck'^7}\hot\sum_{|m|\leq 2}Q_m\overline{\mcd}\cdot\hatpsim+\frac{1}{\ubar_\mck'^6}\sum_{|m|\leq 2}Q_m\mcl(\hatpsim).
	\end{align*}
	Moreover, by Propositions \ref{prop:changebasisder} and \ref{prop:linderdoublenull}, and $\hatpsim^S=(\hatpsim)_\mck^S$, we get the following bounds,
	\begin{align*}
		|\nabla_4 \hatpsim|=|(\nabla_4 \hatpsim)^S|&\lesssim \Omega^2\left(|\mathring{\nabla}_4\hatpsim^S|+|\nabring\hatpsim^S|+|\mathring{\nabla}_3\hatpsim^S|\right)+|\hatpsim|\\
		&\lesssim 1+ \Omega^2\left(|\widecheck{\nabring}_4\hatpsim^S|+|\widecheck{\nabring}\hatpsim^S|+|\widecheck{\nabring}_3\hatpsim^S|\right)\lesssim 1,\\
		|\nabla_3 \hatpsim|=|(\nabla_3 \hatpsim)^S|&\lesssim \Omega^{-2}\left(|\mathring{\nabla}_3\hatpsim^S|+|\nabring\hatpsim^S|+\Omega^4|\mathring{\nabla}_4\hatpsim^S|\right)+\Omega^{-2}|\hatpsim|\\
		&\lesssim \Omega^{-2}+ \Omega^{-2}\left(|\widecheck{\nabring}_3\hatpsim^S|+|\widecheck{\nabring}\hatpsim^S|+\Omega^4|\widecheck{\nabring}_4\hatpsim^S|\right)\lesssim \Omega^{-2}
	\end{align*}
	which yields, by Lemma \ref{lem:lemoui},
	\begin{align}\label{eq:eqinteravantconc}
		\mcl({\Psi})=\frac{1}{\ubar_\mck'^6}\sum_{|m|\leq 2}Q_m\mcl(\hatpsim)+O_Q(\Omega^{-2}\ubar^{-7}).
	\end{align}
	Now, by Corollary \ref{cor:teukansatzdansdeux}, which uses the fact that $\mcl(\Psi_m)$ vanishes in exact Kerr (see Proposition \ref{prop:teukansatzkerr}), we have $\mcl(\hatpsim)^S_\mck=0$, hence the following identity
	\begin{align}\label{eq:pasforecemment}
		|\mcl(\hatpsim)|&=|\mcl(\hatpsim)^S|=|\mcl(\hatpsim)^S-\mcl(\hatpsim)^S_\mck|.
	\end{align}
It remains only to check that $\mcl(\hatpsim)^S-\mcl(\hatpsim)^S_\mck$ can be expressed with the linearized double null quantities, by dealing with every term in the expression of the Teukolsky operator \eqref{eq:expressioncooldeteuk}. This is straightforward relying on Proposition \ref{prop:changebasisder} and \ref{prop:linderdoublenull}; for instance we have
	\begin{align*}
		(\nabla_4\nabla_3\hatpsim)^S&=_s\nabring_{e_4}(\nabla_3\hatpsim)^S+\hat{\mct}^{(4)}\cdot(\nabla_3\hatpsim)^S\\
		&=_s\nabring_{e_4}\nabring_{e_3}\hatpsim^S+\hat{\mct}^{(3)}\nabring_{e_4}\hatpsim^S+\hat{\mct}^{(4)}\nabring_{e_3}\hatpsim^S+\left(\nabla_{e_4}^S\hat{\mct}^{(3)}+\hat{\mct}^{(3)}\hat{\mct}^{(4)}\right)\cdot\hatpsim^S,
	\end{align*}
so that, using $\hatpsim^S=(\hatpsim^S)_\mck$,
	\begin{align*}
	|\nabring_{e_4}\nabring_{e_3}\hatpsim^S&
	-(\nabring_{e_4}\nabring_{e_3}\hatpsim^S)_\mck|\\ &\lesssim|\widecheck{\hat{\lambda}}|+\Omega^2(|\widecheck{\nabring}_4(\nabring_{e_3}\hatpsim^S)_\mck|+|\widecheck{\nabring}(\nabring_{e_3}\hatpsim^S)_\mck|+|\widecheck{\nabring}_3(\nabring_{e_3}\hatpsim^S)_\mck)|+|\fc|)\\
	&\quad+|(\Omega^2\mathring{\nabla}_4,\Omega^2\nabring,\Omega^2\mathring{\nabla}_3)^{\leq 1}(\widecheck{\hat{\lambda}}^{-1},{\hat{\lambda}}^{-1}\widecheck{\nabring}_3\hatpsim^S,{\hat{\lambda}}^{-1}\fbar\widecheck{\nabring}\hatpsim^S,{\hat{\lambda}}^{-1}|\fbar|^2\widecheck{\nabring}_4\hatpsim^S,{\hat{\lambda}}^{-1} \widecheck{\fbar})|\\
	&\lesssim \Omega^{-2}|(\Omega^{-2}\df^{\leq 1}\widecheck{\hat\lambda},\df^{\leq 2}\widecheck{g},\Omega^2\df^{\leq 1}\widecheck{\chi},\df^{\leq 1}\widecheck{b},\df^{\leq 1}\widecheck{\chibar})|\lesssim \Omega^{-2}\ubar^{-2-\delta/5},
\end{align*}
by \eqref{eq:voilauneborne}, \eqref{eq:voilabornefbarcheckkk}. All the other terms in $\mcl(\hatpsim)^S-\mcl(\hatpsim)^S_\mck$ satisfy similar bounds, hence 
\begin{align*}
	|\mcl(\hatpsim)|&\lesssim |\mcl(\hatpsim)^S-\mcl(\hatpsim)^S_\mck|\lesssim \Omega^{-2}\ubar^{-2-\delta/5},
\end{align*}
by \eqref{eq:pasforecemment}, which concludes the proof by \eqref{eq:eqinteravantconc}.
\end{proof}
\begin{rem}
We will use at some point the following bound in $\deux$,
	\begin{align}\label{eq:manquaitcellela}
		|\nabla_4^{\leq 1}\Psi|\lesssim_Q\ubar^{-6}.
	\end{align}
	which follows easily from Lemma \ref{lem:lemoui} and Propositions \ref{prop:changebasisder}, \ref{prop:linderdoublenull}, which yield $|\nabla_4\hatpsim|\lesssim_Q 1$.
\end{rem}

\subsubsection{Precise asymptotics of $A$ in $\deux$}
\label{section:preciseasymptotics}
We define, in $\deux$,
$$\err[A]:=A-\Psi,$$
with $\Psi$ defined in \eqref{eq:defansatzdeux}. The following result, which is the main result of Section \ref{section:teukolskydeux}, is then a simple application of Proposition \ref{prop:decayL2AerrA} combined with the previously proven estimates.
\begin{prop}\label{prop:preciseasymptA}
	Provided $0<\gamma<\delta/6$, we have the following $L^2(S(u,\ubar))$ decay in $\deux$,
	$$\Ldeux{\err[A]}\lesssim_C\ubar^{-6-\delta/3}.$$
\end{prop}
\begin{proof}
	Using \eqref{eq:teukerror} combined with $\Xi=0$ here by \eqref{eq:xiprimxibarprimomegaprimzeroin} and the bounds in Proposition \ref{prop:coeffin}, we get the inhomogeneous Teukolsky equation in $\deux$,
	\begin{align*}
		|\mcl(\err[A])|&\lesssim|\err[\mcl(A)]|+|\mcl(\Psi)|\lesssim|(\divc)^{\leq 1}\wh{X}|^2+|{(\divc)^{\leq 1}B}|^2+|\wh{X}||\wh{\Xbar}||A|+|\mcl(\Psi)|\\
		&\lesssim_C\Omega^{-2}\ubar^{-6-2\delta/3},
	\end{align*}
where we also used Propositions \ref{prop:teukansatzz} and \ref{prop:betterdecayteukerrterm}, in the last step above.This shows that assumption \eqref{eq:teukolskylike} of Proposition \ref{prop:enerteukAerrA} is satisfied with $U=\err[A]$ and $\beta=6+\delta/2$. Next, by \eqref{eq:energyhyputilisabledansII} and \eqref{eq:pointwiseutilisabledansII}, assumption \eqref{eq:initialenergyhypA} of Proposition \ref{prop:enerteukAerrA} is also satisfied with $\beta=6+\delta/2$.  Now, by \eqref{eq:prelimdecayAdeux}, \eqref{eq:boundalphaprimin} and \eqref{eq:manquaitcellela} we get in $\deux$,
	\begin{align*}
		\Ldeux{\err[A]}&\lesssim\Ldeux{A}+\Ldeux{\Psi}\lesssim_C\ubar^{-4-2\delta/5+\gamma},\\
		\Ldeux{\nabla_4\err[A]}&\lesssim_C\Ldeux{\nabla_4A}+\Ldeux{\nabla_4\Psi}\lesssim\ubar^{-2-\delta/5},
	\end{align*}
	so that the assumption \eqref{eq:hyppointwiseA} of Proposition \ref{prop:enerteukAerrA} is satisfied with $\beta'=4+2\delta/5-\gamma>1$ and $\beta''=2+\delta/5$, in which case
	$$\min(\beta,\beta'+2+\delta/5,\beta''+4+2\delta/5)=\min(6+\delta/2,6+3\delta/5-\gamma, 6+3\delta/5)=6+\delta/2,$$
	for $\gamma<\delta/5$. Thus, by \eqref{eq:pointwiseutilisabledansII}, we can apply Proposition \ref{prop:decayL2AerrA} to get the bound
	$$\Ldeux{\err[A]}\lesssim_C\ubar^{-6-\delta/2+\gamma}\lesssim_C\ubar^{-6-\delta/3},$$
	 in $\deux$ for $\gamma<\delta/6$, which concludes the proof.
\end{proof}

\section{Inextendibility across the Cauchy horizon}\label{section:regionIII}

\subsection{Preliminary estimates in region $\trois$}
Recall that we denote with the ring $\:\ring{}\:$ notation the geometric quantities defined in the the double null frame $(\ering_\mu)$ in $\deux\cup\trois$, and with the unprimed notation the geometric quantities defined in to the ingoing non-integrable frame $(e_\mu)$. Since $\trois=\{u+\ubar\geq\ubar^\gamma\}$ and $\Omega^2\sim\Omega^2_\mck\sim e^{-|\kappa_-|(u+\ubar)}$ where $(u,\ubar)$ are the double null retarded and advanced time, we have 
\begin{align}\label{eq:expodecayIII}
	\Omega^2\lesssim e^{-|\kappa_-|\ubar^\gamma},\quad\text{in}\:\:\trois.
\end{align}
\subsubsection{Estimates on $\Gamma$ for $\mathring{\alpha}$, $\mathring{tr\chi}$, $\mathring{\chihat}$ from the frame transformation formulas}

By \eqref{eq:outtoin} where $(e_\mu')$ is the outgoing non-integrable frame, we have $\chihat'=\hat{\Omega}^{-2}\chihat,\: \alpha'=\hat{\Omega}^{-4}\alpha$. Thus, by \eqref{eq:wantedd}, and Propositions \ref{prop:betterdecayteukerrterm}, \ref{prop:preciseasymptA} we deduce the following estimates in the non-integrable outgoing frame, on $\Gamma$ (recall Definition \ref{defi:anyy}) :
$$|{tr\chi}'|\lesssim_C\Omega^{-2}\ubar^{-3-\delta/3},\quad|\chihat'|\lesssim_C\Omega^{-2}\ubar^{-3-\delta/3},\quad\Ldeux{\hat{\Omega}^4\alpha'-\Real(\Psi)}\lesssim_C \ubar^{-6-\delta/3},\quad\text{on}\:\:\Gamma,$$
where $\Psi$ is defined in \eqref{eq:defansatzdeux}. We also have the less sharp (but pointwise) bound on $\Gamma$ for $\alpha'$, by Proposition \ref{prop:pointwisenablaleq2A}:
$$|\alpha'|\lesssim_C\Omega^{-4}\ubar^{-3-\delta/3},\quad\text{on}\:\:\Gamma.$$
Now, recall from \eqref{eq:bornecoeffout} that $|\df^{\leq 1}\widecheck{f}|\lesssim{\varepsilon}{\ubar^{-2-\delta/5}},\quad|\df^{\leq 1}\widecheck{\fbar}|\lesssim{\varepsilon\Omega^2}{\ubar^{-2-\delta/5}}$ on $\Gamma$. Thus, by the transformation formulas \eqref{eq:changealpha}, \eqref{eq:changechihat}, \eqref{eq:changetrchi} from the double null frame $(\ering_\mu)$ to the outgoing non-integrable frame $(e'_\mu)$ with coefficients $(f,\fbar,\lambda)$, and by the estimates on the connection and curvature coefficients in the double null frame in $\deux\cup\trois$ in \eqref{eq:onlyused}, we obtain on $\Gamma$ :
\begin{align}\label{eq:fromomo}
	&|\mathring{\alpha}-\lambda^{-2}(\alpha')^S|\lesssim \Omega^{-2},\quad|\mathring{\chihat}-\lambda^{-1}(\chihat')^S|\lesssim 1,\quad |\mathring{tr\chi}-\lambda^{-1}tr\chi'|\lesssim 1.
\end{align}
Thus, using $\lambda\sim 1$ and $\Omega^2|_\Gamma\sim\exp(-|\kappa_-|\ubar^\gamma)$ which implies $1\lesssim\Omega^{-2}\ubar^{-3-\delta/3}$ and $\Omega^2\lesssim\ubar^{-6-\delta/3}$ on $\Gamma$, we obtain the following estimates in the double null frame, 
\begin{equation}\label{eq:estimatesongamma}
	\begin{gathered}
		|\mathring{tr\chi}|\lesssim_C\Omega^{-2}\ubar^{-3-\delta/3},\quad|\mathring{\chihat}|\lesssim_C\Omega^{-2}\ubar^{-3-\delta/3},\quad|\mathring{\alpha}|\lesssim_C\Omega^{-4}\ubar^{-3-\delta/3},\quad\text{on}\:\:\Gamma,\\
		\|{\hat{\lambda}^2\mathring{\alpha}-\Real(\Psi)^S}\|_{L^2(S(u_\Gamma(\ubar),\ubar))}\lesssim_C\ubar^{-6-\delta/3},
	\end{gathered}
\end{equation}
{on $\Gamma$}\footnote{Note that this bound holds for the \emph{non-linearized} double null quantities as all Kerr values are bounded, and hence bounded by $\Omega^{-2}\ubar^{-3-\delta/3}$ and $\Omega^{-4}\ubar^{-6}$ in $\trois$. This observation allows to bypass the linearization of double null quantities in $\trois$.}, where we recall $\hat{\lambda}=\lambda\hat{\Omega}^2$ and where $u_\Gamma(\ubar)=\ubar^\gamma-\ubar$.
\subsubsection{Consequences of the Bianchi equation for $\nabring_3\mathring{\alpha}$ in $\trois$}\label{section:blowupalpha}
\begin{prop}
	Let $h_{\mathring{tr\chibar}}$ be the scalar function defined by
	\begin{align}\label{eq:htrchibarringdef}
		h_{\mathring{tr\chibar}}(u,\ubar,\theta^A):=\exp\left(\int_{C_R-\ubar}^u\frac{1}{2}\mathring{tr\chibar}(u',\ubar,\theta^A)\dee u'\right)\sim 1.
	\end{align}
	We have the following bound in $\trois$,    \begin{align}\label{eq:thebound}
		|\nabring_3(h_{\mathring{tr\chibar}}\Omega^4\mathring{\alpha})|\lesssim_C\frac{1}{|u|^{2+\delta/3}\ubar^{6+2\delta/3}}.
	\end{align}
\end{prop}
\begin{proof}In this proof we simplify the notation $\lesssim_C$ and write instead $\lesssim$. Recall from Section \ref{section:bianchii} the Bianchi identity for $\nabring_3\mathring{\alpha}$, which takes the following form in the double null frame 
	$$\nabring_3\mathring{\alpha}-4\mathring{\omegabar}\mathring{\alpha}+\frac{1}{2}\mathring{tr\chibar}\mathring{\alpha}=\mathring{\nabla}\hot\mathring{\beta}+(\mathring{\zeta}+4\mathring{\eta})\hot\mathring{\beta}-3(\mathring{\rho}\mathring{\chihat}+\hodge{\mathring{\rho}}\hodge\mathring{{\chihat}}).$$
	\textbf{Step 1.}\textit{ The key estimate.} From the identity $\mathring{\omegabar}=-\nabring_3\log\Omega$ (see \eqref{eq:omegabarendoublenul}) we get 
	$$\nabring_3\mathring{\alpha}-4\mathring{\omegabar}\mathring{\alpha}=\Omega^{-4}\nabring_3(\Omega^4\mathring{\alpha})$$
which implies
	$$\nabring_3(\Omega^4\mathring{\alpha})+\frac{1}{2}\mathring{tr\chibar}\Omega^4\mathring{\alpha}=\Omega^4\left(\mathring{\nabla}\hot\mathring{\beta}+(\mathring{\zeta}+4\mathring{\eta})\hot\mathring{\beta}-3(\mathring{\rho}\mathring{\chihat}+\hodge{\mathring{\rho}}\hodge{\mathring{\chihat}})\right).$$
	Moreover, by the estimates \eqref{eq:onlyused} in the double null frame we get in $\trois$,
	\begin{align}\label{eq:bornebetaIII}
		|\Omega^4(\mathring{\nabla}\hot\mathring{\beta}+(\mathring{\zeta}+4\mathring{\eta})\hot\mathring{\beta})|\lesssim\Omega^2\lesssim\frac{1}{\ubar^{6+2\delta/3}|u|^{2+\delta/3}}
	\end{align}
	where we used the exponential decay \eqref{eq:expodecayIII} of $\Omega^2$ in $\ubar$ in $\trois$, as well the inequality $\ubar\geq |u|$ in $\deux\cup\trois$ (see Lemma \ref{lem:usimubarII}). Next, using the bounds $|\mathring{K}|+|\hodge{\mathring{K}}|\lesssim 1$ from \eqref{eq:onlyused}, combined with the definitions \eqref{eq:KKstarDNdeff} of $\mathring{K},\hodge{\mathring{K}}$, we infer in $\trois$ 
	\begin{align*}
		&\mathring{\rho}=-\mathring{K}+\frac{1}{2}\mathring{\chihat}\cdot\mathring{\wh{\chibar}}-\frac{1}{4}\mathring{tr\chi} \mathring{tr\chibar}=O(1)+\frac{1}{2}\mathring{\chihat}\cdot\mathring{\wh{\chibar}}-\frac{1}{4}\mathring{tr\chi}\mathring{tr\chibar},\\
		&\hodge{\mathring{\rho}}=\hodge{\mathring{K}}-\frac{1}{2}\mathring{\chihat}\wedge\mathring{\wh{\chibar}}=O(1)-\frac{1}{2}\mathring{\chihat}\wedge\mathring{\wh{\chibar}},
	\end{align*}
which implies
$$\Omega^4|(\mathring{\rho}\mathring{\chihat}+\hodge{\mathring{\rho}}\hodge{\mathring{\chihat}})|\lesssim \Omega^4|\mathring{\chihat}|+\Omega^4|\mathring{\chihat}|(|\mathring{\wh{\chibar}}||\mathring{\chihat}|+|\mathring{tr\chi}||\mathring{tr\chibar}|),$$
where by \eqref{eq:onlyused}, $\Omega^4|\mathring{\chihat}|\lesssim\Omega^2\lesssim\ubar^{-6-\delta}|u|^{-2-\delta}$ in $\trois$. Moreover, $|\mathring{\wh{\chibar}}_\mck|\lesssim \Omega^2$ and $|\mathring{tr\chibar}_\mck|\lesssim\Omega^2$ (see \eqref{eq:bornesdanskerr}), hence by \eqref{eq:expodecayIII} and \eqref{eq:onlyused} we get in $\trois$, $|\mathring{\wh{\chibar}}|,|\mathring{tr\chibar}|\lesssim{|u|^{-2-\delta/3}}$, which yields
	\begin{align*}
		\Big|\nabring_3(\Omega^4\mathring{\alpha})+\frac{1}{2}\mathring{tr\chibar}\Omega^4\mathring{\alpha}\Big|\lesssim \frac{1}{\ubar^{6+2\delta/3}|u|^{2+\delta/3}}+\Omega^4\frac{|\mathring{\chihat}|^2+|\mathring{\chihat}||\mathring{tr\chi}|}{|u|^{2+\delta/3}}.
	\end{align*}
in $\trois$. Now, notice that the definition \eqref{eq:htrchibarringdef} of $h_{\mathring{tr\chibar}}$ implies
	\begin{align}\label{eq:defhtrchibar}
		\frac{1}{2}\mathring{tr\chibar}(u,\ubar,\theta^A)=\frac{\nabring_3h_{\mathring{tr\chibar}}}{h_{\mathring{tr\chibar}}}.
	\end{align}
	Moreover, by \eqref{eq:onlyused} we have $|\mathring{tr\chibar}|\lesssim\Omega^2+|u|^{-2-\delta/3}$, which implies $h_{\mathring{tr\chibar}}\sim 1$. Thus, in $\trois$,
	\begin{align}\label{eq:reutiliser}
		\Big|\nabring_3(h_{\mathring{tr\chibar}}\Omega^4\mathring{\alpha})\Big|\lesssim \frac{1}{\ubar^{6+2\delta/3}|u|^{2+\delta/3}}+\Omega^4\frac{|\mathring{\chihat}|^2+|\mathring{\chihat}||\mathring{tr\chi}|}{|u|^{2+\delta/3}}.
	\end{align}
	\textbf{Step 2.} \textit{Proving $|\Omega^4\mathring{\alpha}|\lesssim \ubar^{-3-\delta/2}$ in $\trois$.} We have, using $|\mathring{tr\chi}_\mck|,|\mathring{\chihat}_\mck|\lesssim 1$ and \eqref{eq:onlyused}, in $\trois$,
	\begin{align}\label{eq:eqagainnee}
		|\mathring{tr\chi}|,|\mathring{\chihat}|\lesssim \frac{\Omega^{-2}}{\ubar^{2+\delta/3}}.
	\end{align}
	Thus by \eqref{eq:reutiliser} we get $|\nabring_3(h_{\mathring{tr\chibar}}\Omega^4\mathring{\alpha})|\lesssim {\ubar^{-4-2\delta/3}|u|^{-2-\delta/3}}$, and hence integrating from $\Gamma$ along the flow of $\ering_3$ this implies the following bound in $\trois$,
	\begin{align}\label{eq:usdeddd}
		|\Omega^4\mathring{\alpha}|(u,\ubar,\theta^A)\lesssim |\Omega^4\mathring{\alpha}|(u_\Gamma(\ubar),\ubar,\theta^A)+\int_{u_\Gamma(\ubar)}^u\frac{\dee u'}{\ubar^{4+2\delta/3}|u'|^{2+\delta}}\lesssim \frac{1}{\ubar^{3+\delta/3}},
	\end{align}
	where we used $h_{\mathring{tr\chibar}}\sim 1$ and the pointwise bound in \eqref{eq:estimatesongamma} on $\Gamma$ for $\mathring{\alpha}$.
	\\\\
	\textbf{Step 3.} \textit{Improved decay for $\mathring{tr\chi}$, $\mathring{\chihat}$ in $\trois$.} We first use the null structure equation for $\nabring_4 \mathring{tr\chi}$ (see Section \ref{section:nullstructureeq}) in the double null gauge which writes
	$$\Omega^2\nabring_4\mathring{tr\chi}=-\frac{1}{2}\Omega^2(\mathring{tr\chi})^2-\Omega^2|\mathring{\chihat}|^2=O\left(\frac{\Omega^{-2}}{\ubar^{4+2\delta/3}}\right)=O\left(\frac{\Omega^{-2}}{\ubar^{3+\delta/3}}\right).$$
	in $\trois$, where we also used \eqref{eq:eqagainnee} to bound the RHS. Thus integrating from $\Gamma$ along the flow of $\Omega^2\ering_4$ we get\footnote{Here, $\wt{\theta}^A_\Gamma$ is the value on $\Gamma$ of $(\wt{\theta}^A)'=(\wt{\theta}^A)(u,\ubar',\theta^A)$ such that $(\wt{\theta}^A)(u,\ubar,\theta^A)=\theta^A,\:(\partial_\ubar+b^B\partial_{\theta^B})\wt{\theta}^A=0$.}, using \eqref{eq:estimatesongamma},
	\begin{align}
		|\mathring{tr\chi}|(u,\ubar,{\theta}^A)\lesssim|\mathring{tr\chi}|(u,\ubar_\Gamma(u),\wt{\theta}^A_\Gamma)+\int_{\ubar_\Gamma(u)}^\ubar \frac{\Omega^{-2}}{(\ubar')^{3+\delta/3}}\dee\ubar'\lesssim \frac{\Omega^{-2}(u,\ubar_\Gamma(u))}{\ubar_\Gamma(u)^{3+\delta/3}}+\frac{\Omega^{-2}}{\ubar^{3+\delta/3}},\label{eq:todeducerob}
	\end{align}
	where $u+\ubar_\Gamma(u)=\ubar_\Gamma(u)^\gamma$ and where we used $\Omega^{-2}\sim e^{|\kappa_-|(u+\ubar)}$ and an integration by parts to bound the integral term above. We now proceed similarly as in \cite[(6.13)]{spin+2} to prove, in $\trois$, 
	\begin{align}\label{eq:prouven}
		\frac{\Omega^{-2}(u,\ubar_\Gamma(u))}{\ubar_\Gamma(u)^{3+\delta/3}}\lesssim\frac{\Omega^{-2}}{\ubar^{3+\delta/3}}.
	\end{align}
	Let $\gamma'\in(\gamma,1)$. We introduce 
	$$\mathbf{A}=\trois\cap\{u+\ubar\leq \ubar^{\gamma'}\},\quad\mathbf{B}=\trois\cap\{u+\ubar\geq \ubar^{\gamma'}\}.$$
	First, in $\mathbf{A}$ we have $|u|\sim \ubar$ (see Lemma \ref{lem:usimubarII}) thus as $\Gamma\subseteq\mathbf{A}$, $\ubar_\Gamma(u)\sim |u|\sim \ubar$ and we get
	$$\frac{\Omega^{-2}(u,\ubar_\Gamma(u))}{\ubar_\Gamma(u)^{3+\delta/3}}\lesssim \frac{\Omega^{-2}}{\ubar_\Gamma(u)^{3+\delta/3}}\sim \frac{\Omega^{-2}}{\ubar^{3+\delta/3}},$$
	where we used $\Omega^{-2}\sim e^{|\kappa_-|(u+\ubar)}$ and $\ubar_\Gamma(u)\leq\ubar$ in the first step above. Next, in $\mathbf{B}$ we have
	$$\Omega^2(u,\ubar)\lesssim \exp(-|\kappa_-|\ubar^{\gamma'}).$$
	We also have on $\Gamma$, 
	$$\Omega^{-2}(u,\ubar_\Gamma(u))\sim\exp(|\kappa_-|\ubar_\Gamma(u)^\gamma)\leq\exp(|\kappa_-|\ubar^\gamma),$$
where we used $\ubar_\Gamma(u)\leq\ubar$. Thus since $\gamma'>\gamma$ we deduce 
	\begin{align*}
		\frac{\Omega^{-2}(u,\ubar_\Gamma(u))\ubar^{3+\delta/3}}{\Omega^{-2}\ubar_\Gamma(u)^{3+\delta/3}}\lesssim \ubar^{3+\delta/3}\exp(|\kappa_-|(\ubar^\gamma-\ubar^{\gamma'}))\lesssim 1,
	\end{align*}
which concludes the proof of \eqref{eq:prouven}. By \eqref{eq:todeducerob}, we have thus proven, in $\trois$,
	\begin{align}\label{eq:trchitrois}
		|\mathring{tr\chi}|\lesssim\frac{\Omega^{-2}}{\ubar^{3+\delta/3}}.
	\end{align}
	We now prove the same bound for $|\mathring{\chihat}|$ in $\trois$. We use the null structure identity, in $\trois$,
	$$\Omega^2\nabring_4\mathring{\chihat}=-\Omega^2\mathring{tr\chi}\mathring{\chihat}-\Omega^2\mathring{\alpha}=O\left(\frac{\Omega^{-2}}{\ubar^{3+\delta/3}}\right),$$
	where we used \eqref{eq:eqagainnee} and \eqref{eq:usdeddd} to bound the RHS. We then proceed exactly as above to get
	\begin{align}\label{eq:chihattrois}
		|\mathring{\chihat}|\lesssim\frac{\Omega^{-2}}{\ubar^{3+\delta/3}},\quad\text{in}\:\:\trois.
	\end{align}
	\textbf{Step 4.} \textit{Conclusion.} We go back to \eqref{eq:reutiliser} which, together with \eqref{eq:trchitrois} and \eqref{eq:chihattrois} yields
	$$\Big|\nabring_3(h_{\mathring{tr\chibar}}\Omega^4\mathring{\alpha})\Big|\lesssim \frac{1}{\ubar^{6+2\delta/3}|u|^{2+\delta/3}}+\Omega^4\frac{|\mathring{\chihat}|^2+|\mathring{\chihat}||\mathring{tr\chi}|}{|u|^{2+\delta/3}}\lesssim \frac{1}{\ubar^{6+2\delta/3}|u|^{2+\delta/3}},$$
	 in $\trois$, which concludes the proof.
\end{proof}
\begin{cor}\label{cor:corboundsharp}
	We have in $\trois$ the bound
	$$\Ldeux{\mathring{\alpha}}\lesssim_C \frac{e^{2|\kappa_-|(u+\ubar)}}{\ubar^6}.$$
\end{cor}
\begin{proof}
	This is a simple consequence of \eqref{eq:thebound}, \eqref{eq:htrchibarringdef},  and Proposition \ref{prop:13.2DL} combined with the initial data estimate \eqref{eq:estimatesongamma} which implies $\|\Omega^4\mathring{\alpha}\|_{L^2(S(u_\Gamma(\ubar),\ubar))}\lesssim\ubar^{-6}$ on $\Gamma$. 
\end{proof}

\subsection{$C^{0,1}_{loc}$-inextendibility across $\ch$ (proof of Theorems \ref{thm:alphaGamma} and \ref{thm:inextensibilite})}\label{section:knextextex}

In this section, we show how to obtain from \eqref{eq:thebound} (and the estimates on $\Gamma$) the blow-up of an integrated curvature quantity which, by a result of Sbierski \cite{sbierskiinextdernier}, implies Lipschitz inextendibility of the spacetime across the Cauchy horizon $\ch$. We introduce below the perturbed Eddington-Finkelstein type coordinates that we will use in the rest of the proof.

\begin{defi}\label{defi:cccva?}
	In Kerr spacetime, let $(u'_\mck,\ubar'_\mck)$ be the Eddington-Finkelstein retarded and advanced time, and $(\phi_-)_\mck$ the azimuthal coordinate regular on $\ch$, namely
	$$u'_\mck=r^*_\mck-t_\mck,\quad\ubar'_\mck=r^*_\mck+t_\mck,\quad(\phi_-)_\mck=\phi_\mck-r_{mod}(r_\mck),$$
	with $(r_\mck,t_\mck,\theta_\mck,\phi_\mck)$ being the usual Boyer-Lindquist coordinates of Kerr spacetime. In region $\deux\cup\trois$, we define the coordinates 
	$$(u',\ubar',\theta,\phi_-):=(u'_\mck(u,\ubar,\theta^A),\ubar'_\mck(u,\ubar,\theta^A),\theta_\mck(u,\ubar,\theta^A),(\phi_-)_\mck(u,\ubar,\theta^A)),$$
as the Kerr values of the coordinates $(u'_\mck,\ubar'_\mck,\theta_\mck,(\phi_-)_\mck)$ with respect to the double null coordinates $(u,\ubar,\theta^A)$ in $\deux\cup\trois$.
\end{defi}

\subsubsection{Leading-order term for $\Omega^4\mathring{\alpha}$ in $\trois$}\label{eq:deuzhhh}
By \eqref{eq:fromomo} we have
\begin{align}\label{eq:lineline}
	\|\Omega^4\mathring{\alpha}-\Omega^4\hat{\lambda}^{-2}\Real(\Psi)^S\|_{L^2(S(u_\Gamma(\ubar),\ubar))}\lesssim_C{\ubar^{-6-\delta/3}}.
\end{align}
Note that by \eqref{eq:deflambdahattt} we have the identity
$$\Omega^4\hat{\lambda}^{-2}=\lambdabar^2\left(1-\Omega^2 f\cdot\gbar+\frac{\Omega^4}{16}|f|^2|\gbar|^2\right)^2\sim 1,$$
so that, by Theorems \ref{thm:controllambdabarcheck}, \ref{thm:controlfc}, \ref{thm:controlgbarc} we have on $\Gamma$, 
\begin{align}\label{eq:linelineline}
	\Omega^4\hat{\lambda}^{-2}\Real(\Psi)^S=\psi+O(\ubar^{-6-\delta/3}),
\end{align}
where
\begin{align}\label{eq:defpsiansatz}
	\psi&:=\frac{W_\mck}{\ubar'^6}\sum_{|m|\leq 2}\Real\left(\frac{Q_mA_m(r_\mck)}{\qbar_\mck^2}\Big(\mcd\hot(\mcd(Y_{m,2}(\cos\theta)e^{im\phi_+}))\Big)^S_\mck\right),\\
	W_\mck&:=\lambdabar_\mck^2\left(1-\Omega^2 f\cdot\gbar+\frac{\Omega^4}{16}|f|^2|\gbar|^2\right)^2_\mck\sim 1.\label{eq:defWlambdabarahah}
\end{align}
We deduce from \eqref{eq:lineline} and \eqref{eq:linelineline} the estimate
\begin{align}\label{eq:initalpharinggamma}
	\|\Omega^4\mathring{\alpha}-\psi\|_{L^2(S(u_\Gamma(\ubar),\ubar))}\lesssim_C\frac{1}{\ubar^{6+\delta/3}},\quad\text{on }\Gamma.
\end{align}
Now, defining in $\trois$ the following $S(u,\ubar)$-tangent 2-tensor
\begin{align}\label{eq:defalpharing}
	\err[\Omega^4\mathring{\alpha}]:=\Omega^4\mathring{\alpha}-\psi,
\end{align}
our goal is to prove good decay estimates for $\err[\Omega^4\mathring{\alpha}]$, such that $\psi$ is the leading order term for $\Omega^4\mathring{\alpha}$ in $\trois$.
\begin{prop}\label{prop:nab3ansatzbetter}
	Let $\psi$ be defined by \eqref{eq:defpsiansatz}. We have, in $\trois$, 
	$$|\nabring_3\psi|\lesssim_Q\frac{1}{\ubar^6|u|^{2+\delta/3}}.$$
\end{prop}
\begin{rem}
	In exact Kerr, we actually have the identity $(|\nabring_3\psi|)_\mck\lesssim\Omega_\mck^2$ which comes from $e_3(\phi_+)=0$. This is closely related to the oscillating term $e^{im\phi_+}$ which appears in the expression \eqref{eq:defpsiansatz} of the ansatz $\psi$ for $\Omega^4\mathring{\alpha}$. If this term was replaced by $e^{im\phi_-}$ instead, the bound above would not hold, highlighting that the-leading order term for $\mathring{\alpha}$ necessarily oscillates (see \eqref{eq:thebound}).
\end{rem}
\begin{proof}[Proof of Proposition \ref{prop:nab3ansatzbetter}]
In this proof, we simplify the notation $\lesssim_Q$ and write instead $\lesssim$. First, as $\psi=\psi_\mck$ by \eqref{eq:defpsiansatz}, we get using \eqref{eq:nab3diff},
	\begin{align}\label{eq:appldllin}
		\nabring_3\psi_{AB}=((\nabring_3\psi)_\mck)_{AB}+(\gamma^{CD}\ring{\chibar}_{CA}-\gamma_\mck^{CD}(\ring{\chibar}_\mck)_{DA})\psi_{CB}+(\gamma^{CD}\ring{\chibar}_{CB}-\gamma_\mck^{CD}(\ring{\chibar}_\mck)_{DB})\psi_{AC}.
	\end{align}
	Moreover, we have\footnote{To see this, we write for $S=\nabring_3\psi$, $$|S|^2=\gamma_\mck^{AB}\gamma_\mck^{CD}S_{AC}S_{BD}+2\widecheck{\gamma}^{AB}\gamma_\mck^{CD}S_{AC}S_{BD}+\widecheck{\gamma}^{AB}\widecheck{\gamma}^{CD}S_{AC}S_{BD}\lesssim\gamma_\mck^{AB}\gamma_\mck^{CD}S_{AC}S_{BD}+|\widecheck{\gamma}||S|^2+|\widecheck{\gamma}|^2|S|^2,$$
		and the two last terms on the RHS can be absorbed in the LHS for $\varepsilon>0$ small enough.\label{footnote:normes}} $|\nabring_3\psi|^2\lesssim\gamma_\mck^{AB}\gamma_\mck^{CD}\nabring_3\psi_{AC}\nabring_3\psi_{BD}$, which implies, together with \eqref{eq:appldllin} and \eqref{eq:onlyused}, in $\trois$,
	\begin{align}
		|\nabring_3\psi|^2\lesssim (|\nabring_3\psi|)_\mck^2+|(\nabring_3\psi)_\mck||(\widecheck{\gamma},\widecheck{\ring{\chibar}})\cdot\psi|+|(\widecheck{\gamma},\widecheck{\ring{\chibar}})\cdot\psi|^2\lesssim (|\nabring_3\psi|)_\mck^2+\varepsilon^2|u|^{-4-2\delta/3}\ubar^{-12}.\label{eq:dernierepourconc234}
	\end{align}
	We now bound the term $(|\nabring_3\psi|)_\mck$. \textbf{In the rest of the proof of Proposition \ref{prop:nab3ansatzbetter}, the computation is done in Kerr, and we drop the subscripts $\mck$ for conciseness.} Recalling the analog of \eqref{eq:grosPhi} in Kerr, we introduce the following $S(u,\ubar)$-tangent orthonormal frame:
	$$\ering_1:=\Phi^{-1}\left(\frac{1}{|q|}\partial_\theta\right),\quad \ering_2:=\Phi^{-1}\left(\frac{1}{|q|\sin\theta}\partial_\phi+\frac{a\sin\theta}{|q|}\partial_t\right).$$
	Note that in Kerr, $\partial_\phi=\partial_{\phi^*}$, $2\partial_t=\Omega^2e_4-\ering_3-b^A\partial_{\theta^A}$, thus by \eqref{eq:grosPhi} and using
	$$\frac{1}{|q|\sin\theta}\partial_\phi+\frac{a\sin\theta}{|q|}\partial_t=\Phi(\ering_2)=\ering_2+\frac12\fbar(\ering_2)f^A\partial_{\theta^A}+\frac12\fbar(\ering_2)\ering_4+\left(\frac12 f(\ering_2)+\frac18|f|^2\fbar(\ering_2)\right)\ering_3,$$
	we get that $\ering_2$ is solution of the equation 
	\begin{align}\label{eq:eqe2S}
		\ering_2=\frac{1}{|q|\sin\theta}\partial_{\phi^*}-\frac{a\sin\theta}{2|q|}b^A\partial_{\theta^A}-\frac{1}{2}\fbar(\ering_2)f^A\partial_{\theta^A}.
	\end{align}
	In what follows, if $T$ is a $S(u,\ubar)$-tangent 2-tensor in Kerr, we define $T_{ab}:=T(\ering_a,\ering_b)$ for $a,b=1,2$,  such that $|\nabring_3\psi|^2=(\nabring_3\psi_{11})^2+(\nabring_3\psi_{22})^2+2(\nabring_3\psi_{12})^2$, where, for $a,b=1,2$,
	\begin{align}\label{eq:palalal}
		\nabring_3\psi_{ab}=\ering_3(\psi_{ab})-\gamma(\nabring_3\ering_a,\ering_c)\psi_{cb}-\gamma(\nabring_3\ering_b,\ering_c)\psi_{ac}.
	\end{align}
	We begin by bounding the last term on the RHS of \eqref{eq:palalal}. Note that using $\gamma_{11}=\gamma_{22}=1$, $\gamma_{12}=0$, and $\nabring_3\gamma=0$ we get 
	$$\gamma(\nabring_3\ering_1,\ering_c)\psi_{c1}=-\gamma(\nabring_3\ering_2,\ering_1)\psi_{21},\quad\gamma(\nabring_3\ering_2,\ering_c)\psi_{c2}=\gamma(\nabring_3\ering_2,\ering_1)\psi_{12}, $$
	and $\gamma(\nabring_3\ering_1,\ering_c)\psi_{c2}=-\gamma(\nabring_3\ering_2,\ering_1)\psi_{22}$
	so we only need to control the term $\gamma(\nabring_3\ering_2,\ering_1)$ in order to bound $\gamma(\nabring_3\ering_a,\ering_c)\psi_{cb}$ for $a,b=1,2$. By \eqref{eq:eqe2S}, and using 
	$b^A\partial_{\theta^A}=\frac{2Mar}{|q|^2 R^2}\partial_{\phi^*}$
	in Kerr (see \eqref{eq:bDNdanskerr}) we get
	\begin{align}
		\nabring_3\ering_2=&\ering_3\left(\frac{1}{|q|\sin\theta}\right)\partial_{\phi^*}+\frac{1}{|q|\sin\theta}\nabring_3\partial_{\phi^*}-\ering_3\left(\frac{a^2\sin\theta Mr}{R^2|q|^3}\right)\partial_{\phi^*}-\frac{a^2\sin\theta Mr}{R^2|q|^3}\nabring_3\partial_{\phi^*}\nn\\
		&-\frac{1}{2}\ering_3(\fbar(\ering_2))f^A\partial_{\theta^A}-\fbar(\ering_2)\nabring_3(f^A\partial_{\theta^A}).\label{eq:combinertoutcahah}
	\end{align}
	Moreover, from the bounds $|\chibar|+|\nabring_3^{\leq 1}\fbar|\lesssim\Omega^2$, $|\partial_{\phi^*}|\lesssim\sin\theta$, $|\nabring_3^{\leq 1}f|\lesssim 1$ in Kerr, we infer
	\begin{align*}
		|\gamma(\nabring_3\partial_{\phi^*},\ering_1)|&=|\g(\D_{u}\partial_{\phi^*},\ering_1)|=|\g(\D_{\phi^* }\ering_3,\ering_1)|\lesssim|\mathring{\chibar}_{\phi^* 1}|\lesssim\Omega^2\sin\theta,\\
		|\ering_3(\fbar(\ering_2))|&=|\nabring_3\fbar_2+\fbar_a\gamma(\nabring_3 \ering_2,\ering_a)|=|\nabring_3\fbar_2+\fbar_1\gamma(\nabring_3 \ering_2,\ering_1)|\lesssim\Omega^2+\Omega^2|\gamma(\nabring_3 \ering_2,\ering_1)|,\\
		|\nabring_3(f^A\partial_{\theta^A})|&=|\nabring_3f|\lesssim 1.
	\end{align*}
	Moreover, using \eqref{eq:derpardanslautresens} which gives $|{\partial\theta}/{\partial{r^*}}|\lesssim\Omega^2\sin\theta^*\lesssim \Omega^2\sin\theta$, $|{\partial r}/{\partial{r^*}}|\lesssim\Omega^2$, we deduce 
	$$\left|\ering_3\left(\frac{1}{|q|\sin\theta}\right)\right|\lesssim\frac{\Omega^2}{\sin\theta},\quad \left|\ering_3\left(\frac{a^2\sin\theta Mr}{R^2|q|^3}\right)\right|\lesssim\Omega^2.$$
	All these estimates combined with \eqref{eq:combinertoutcahah} yields 
	$$|\gamma(\nabring_3 \ering_2,\ering_1)|\lesssim \Omega^2\left(\frac{\gamma_{\phi^*1}}{\sin\theta}+1+|\gamma(\nabring_3 \ering_2,\ering_1)|\right)\lesssim\Omega^2\left(1+|\gamma(\nabring_3 \ering_2,\ering_1)|\right)\lesssim\Omega^2$$
	in $\deux\cup\trois$ after choosing $C_R(a,M)\gg 1$ such that the term $\Omega^2|\gamma(\nabring_3 \ering_2,\ering_1)|$ on the RHS above is absorbed in the LHS. From the bound above combined with \eqref{eq:palalal} we deduce, in Kerr,
	\begin{align}
		|\nabring_3\psi|^2\lesssim(\ering_3(\psi_{11}))^2+(\ering_3(\psi_{22}))^2+(\ering_3(\psi_{12}))^2+\Omega^4.\label{eq:concconcconc}
	\end{align}
	Now we bound the terms $|\ering_3(\psi_{ab})|^2$. Recalling the definition \eqref{eq:defpsiansatz} of $\psi$, we have for $a,b=1,2$,
	\begin{align*}
		\psi_{ab}&=\frac{W_\mck}{\ubar'^6}\sum_{|m|\leq 2}\Real\left(\frac{Q_mA_m(r_\mck)}{\qbar_\mck^2}\Big(\mcd\hot(\mcd(Y_{m,2}(\cos\theta)e^{im\phi_+}))\Big)^S_\mck(\ering_a,\ering_b)\right)\\
		&=\frac{W_\mck}{\ubar'^6}\sum_{|m|\leq 2}\Real\left(\frac{Q_mA_m(r_\mck)}{\qbar_\mck^2}\mcd\hot(\mcd(Y_{m,2}(\cos\theta)e^{im\phi_+}))_{ab}\right),
	\end{align*}
	where $\mcd\hot(\mcd(Y_{m,2}(\cos\theta)e^{im\phi_+}))_{ab}=\mcd\hot(\mcd(Y_{m,2}(\cos\theta)e^{im\phi_+}))(e_a,e_b)$ by definition of the supscript ${}^S$, where $(e_1,e_2)$ is defined in \eqref{eq:horizontalframekerr}. Recall the notation $(e_\mu')$ for the outgoing principal frame in Kerr. Note that we have from \eqref{eq:frametransfo} with coefficients $(f,\fbar,\lambda)$ the identities 
	\begin{align}\label{eq:proddskerr}
		|\g(\ering_3,e_3')|\lesssim|\fbar|^2\lesssim\Omega^4,\quad |\g(\ering_3,e_a)|\lesssim |\fbar_a|\lesssim \Omega^2,\quad |\g(\ering_3,e_4')|\lesssim 1.
	\end{align}
	Thus, we have, using $e_3'(\ubar')=0$, $|e_4'(\ubar')|\lesssim\Omega^{-2}$, $|e_a(\ubar')|\lesssim 1$, $|e_3'(r)|\lesssim\Omega^2$, $|e_4'(r)|\lesssim1$, $e_a(r)=0$, $|e_3'(\theta)|=|e_4'(\theta)|=0$, $|e_a(\theta)|\lesssim 1$, the bounds
	\begin{align*}
		&|\ering_3(\ubar')|\lesssim |\g(\ering_3,e_3')||e_4'(\ubar')|+|\g(\ering_3,e_a)||e_a(\ubar')|\lesssim \Omega^2,\\
		&|\ering_3(r)|\lesssim |\g(\ering_3,e_3')||e_4'(r)|+|\g(\ering_3,e_4')||e_3'(r)|\lesssim \Omega^2,\\
		&|\ering_3(\theta)|\lesssim |\g(\ering_3,e_a)||e_a(\theta)|\lesssim \Omega^2.
	\end{align*}
	This implies, using also $\mcd\hot(\mcd(Y_{m,2}(\cos\theta)e^{im\phi_+})\in\fraks_2(\C)$,
	\begin{align}
		&(\ering_3(\psi_{11}))^2+(\ering_3(\psi_{22}))^2+(\ering_3(\psi_{12}))^2\nn\\
		&\lesssim\sum_{a,b=1,2}\left|\frac{W_\mck}{\ubar'^6}\sum_{|m|\leq 2}\Real\left(\frac{Q_mA_m(r_\mck)}{|q|^2\qbar_\mck^2}\ering_3\left[|q|^2\mcd\hot(\mcd(Y_{m,2}(\cos\theta)e^{im\phi_+}))_{ab}\right]\right)\right|^2+O(\Omega^4)\nn\\
		&\lesssim \frac{1}{\ubar'^6}\sum_{|m|\leq 2}\left|\ering_3\left[|q|^2\mcd\hot(\mcd(Y_{m,2}(\cos\theta)e^{im\phi_+}))_{11}\right]\right|^2+O(\Omega^4).\label{eq:jelutilisepourdeduce}
	\end{align}
Now, recalling Lemma \ref{lem:DhotDYm2} and $e_3'(\theta)=e_3'(\phi_+)=0$, we get for $|m|\leq 2$
	\begin{align*}
		\left|\ering_3\left[|q|^2\mcd\hot(\mcd(Y_{m,2}(\cos\theta)e^{im\phi_+}))_{11}\right]\right|\lesssim &|\g(\ering_3,e_3')|\left|e_4'\left(Y_{m,2}^{+2}(\cos\theta)e^{im\phi_+}\right)\right|\\
		&+\sum_{a=1,2}|\g(\ering_3,e_a)|\left|e_a\left(Y_{m,2}^{+2}(\cos\theta)e^{im\phi_+}\right)\right|.
	\end{align*}
	Combining this with \eqref{eq:proddskerr} and the following estimates in Kerr\footnote{Note that in Kerr, $e_4'(\theta)=0$ while $e_4'(\phi_+)=-2a\Delta^{-1}$.}
	$$\left|e_a\left(Y_{m,2}^{+2}(\cos\theta)e^{im\phi_+}\right)\right|\lesssim 1,\quad \left|e_4'\left(Y_{m,2}^{+2}(\cos\theta)e^{im\phi_+}\right)\right|\lesssim\Delta^{-1}\lesssim\Omega^{-2}, $$
	we deduce $(\ering_3(\psi_{11}))^2+(\ering_3(\psi_{22}))^2+(\ering_3(\psi_{12}))^2\lesssim\Omega^4$ from \eqref{eq:jelutilisepourdeduce}, hence $(|\nabring_3\psi|)_\mck\lesssim\Omega^2_\mck$
	by \eqref{eq:concconcconc}, which concludes the proof by \eqref{eq:dernierepourconc234} since $\Omega^4\lesssim |u|^{-4-2\delta/3}\ubar^{-12}$ in $\trois$ by \eqref{eq:expodecayIII}.\end{proof}

\begin{defi}
	For some $\ubar_0,\ubar'_0\geq 1$, $u_0<u_1$, $u_0'<u_1'$ and $\ubar_1\geq\ubar_0$, $\ubar'_1\geq\ubar'_0$ we define the following spacetime regions (where we omit the dependence in $u_0,u_0',u_1,u_1',\ubar_0,\ubar_0'$),
	\begin{equation}\label{eq:defregiondoublenull}
		\begin{aligned}
			\mcr[\ubar_1]&:=\trois\cap\{u_0<u<u_1\}\cap\{\ubar_0\leq\ubar\leq\ubar_1\},\\ \mcr'[\ubar_1']&:=\trois\cap\{u_0'<u'<u_1'\}\cap\{\ubar'_0\leq\ubar'\leq\ubar_1'\}.
		\end{aligned}
	\end{equation}
\end{defi}

The following result proves that $\psi$ is indeed the leading-order term for $\Omega^4\mathring{\alpha}$ in $\trois$ in a suitable sense.
\begin{prop}\label{prop:exploR}
	Let $X,Y$ be bounded $S(u,\ubar)$-tangent vector fields in $\trois$. Then we have
	\begin{align*}
		\int_{\mcr'[\ubar_1']}\left|\mathring{\alpha}(X,Y)-\Omega^{-4}\psi(X,Y)\right|\lesssim_Q e^{|\kappa_-|\ubar_1'}\left(\frac{C}{(\ubar'_1)^{6+\delta/3}}+\frac{e^{|\kappa_-|u'_1}-e^{|\kappa_-|u'_0}}{(\ubar_1')^{6}|u_f|}\right),
	\end{align*}
where $C$ depends on $u_f,{a,M,(Q_m)_{|m|\leq 2},\delta,\delta_+}$.
\end{prop}
\begin{proof}
By \eqref{eq:thebound}, recalling $\err[\Omega^4\mathring{\alpha}]=\Omega^4\mathring{\alpha}-\psi$, \eqref{eq:htrchibarringdef}, and Definition \ref{defi:anyy}, we have in $\trois$
	$$|\nabring_3(h_{tr\chibar}\err[\Omega^4\mathring{\alpha}])|\lesssim |\nabring_3\psi |+|\mathring{tr\chibar}\psi|+\frac{C}{|u|^{2+\delta/3}\ubar^{6+2\delta/3}},$$
	where $C$ depends on $u_f,{a,M,Q_m,C_R,\delta,\delta_+}$. The definition \eqref{eq:defpsiansatz} of $\psi$ and the bound $|\mathring{tr\chibar}|\lesssim|u|^{-2-\delta/3}$ in $\trois$ (where we use $|\mathring{tr\chibar}_\mck|\lesssim\Omega^2\lesssim|u|^{-2-\delta/3}$ in $\trois$ and \eqref{eq:onlyused}) yields
	$$|\mathring{tr\chibar}\psi|\lesssim_Q{|u|^{-2-\delta/3}\ubar^{-6}},$$
	in $\trois$. Thus, by Proposition \ref{prop:nab3ansatzbetter}, we deduce
	\begin{align}\label{eq:fautintegrerca}
		|\nabring_3(h_{\mathring{tr\chibar}}\err[\Omega^4\mathring{\alpha}])|\lesssim_Q\frac{1}{|u|^{2+\delta/3}\ubar^{6}}+\frac{C}{|u|^{2+\delta/3}\ubar^{6+2\delta/3}}.
	\end{align}
Integrating \eqref{eq:fautintegrerca} from $\Gamma$ yields in $\trois$ the estimate
	\begin{align}
		|\err[\Omega^4\mathring{\alpha}]|(u,\ubar,\theta^A)\lesssim_Q |\err[\Omega^4\mathring{\alpha}]|(u_\Gamma(\ubar),\ubar,\theta^A)+\frac{1}{|u_f|\ubar^{6}}+\frac{C}{\ubar^{6+2\delta/3}}.\label{eq:enoughforldeux00}
	\end{align}
Moreover, integrating on $S(u,\ubar)$ we have
\begin{align}
	\intS|\err[\Omega^4\mathring{\alpha}]|(u_\Gamma(\ubar),\ubar,\theta^A)|&\lesssim\sum_{i=1,2}\int_{\mcv_i\cap S(u,\ubar)}|\err[\Omega^4\mathring{\alpha}]|(u_\Gamma(\ubar),\ubar,\theta^1_{(i)},\theta^2_{(i)})\dee\theta^1_{(i)}\dee\theta^2_{(i)}\nn\\
	&\lesssim\int_{S(u_\Gamma(\ubar),\ubar)}|\err[\Omega^4\mathring{\alpha}]|\lesssim_C\ubar^{-6-\delta/3},\label{eq:enoughforldeux01}
\end{align}
where in the first and second steps above we used $\sqrt{\det\gamma_{A_{(i)}B_{(i)}}}\sim 1$ in $\mcv_i$ for $i=1,2$ by the bound for $\gcheck$ in \eqref{eq:onlyused}, and in the third step we used the Cauchy-Schwarz inequality on $S(u_\Gamma(\ubar),\ubar)$ and \eqref{eq:initalpharinggamma}. 
	\begin{rem}
		Using the same argument as above we also obtain 
	\begin{align}\label{eq:enoughforldeux}
	\Ldeux{\err[\Omega^4\mathring{\alpha}]}\lesssim_Q\frac{1}{|u_f|\ubar^{6}}+\frac{C}{\ubar^{6+2\delta/3}}.
\end{align}		
		Note that using $\Omega^2\sim e^{-|\kappa_-|(u+\ubar)}$, this concludes the proof of Theorem \ref{thm:alphaGamma}. More precisely, we chose the parameters $C_R(a,M)$, $|u_f(a,M,C_R)|$ large enough such that Propositions \ref{prop:recupfbarcheck}, \ref{prop:coeffout}, \ref{prop:hardy}, \ref{prop:bulkpos}, \ref{prop:sovolinfty} and Lemma \ref{lem:619} can be applied\footnote{From this point, $C_R=C_R(a,M)$ is fixed so the constant $C$ depends on $u_f,{a,M,(Q_m)_{|m|\leq 2},\delta,\delta_+}$.}. Concerning the choice of $\gamma$, at some points in the analysis we chose $\gamma$ small enough depending on $\delta$, the most constraining choice being $0<\gamma<\delta/12$ in Propositions \ref{prop:bootstrapfcdfrak} and \ref{prop:fbarbootstrapfcdfrak}.
	\end{rem}
Now, recalling \eqref{eq:defregiondoublenull} and Lemma \ref{lem:ubarssontpareils}, we see that we have
$$\mcr'[\ubar'_1]\subset \mcr[\ubar_1] $$
for some $u_0,u_1,\ubar_{0},\ubar_1$ satisfying $|u_0- u_0'|,\: |u_1- u_1'|,\:|\ubar_0-\ubar'_0|\lesssim 1,\: |\ubar_1-\ubar'_1|\lesssim 1$. Thus by \eqref{eq:enoughforldeux00} we deduce the bounds
\begin{align*}
	\int_{\mcr'[\ubar_1']}\left|\mathring{\alpha}(X,Y)-\Omega^{-4}\psi(X,Y)\right|&\lesssim_Q \int_{\mcr[\ubar_1]}\Omega^{-4}|\err[\Omega^4\mathring{\alpha}]|(u_\Gamma(\ubar),\ubar,\theta^A)+\int_{\mcr'[\ubar_1']}\left(\frac{1}{|u_f|\ubar^{6}}+\frac{C}{\ubar^{6+2\delta/3}}\right)\\
	&\lesssim\int_{u_0}^{u_1}\int_{\ubar_0}^{\ubar_1}e^{|\kappa_-|(u+\ubar)}\int_{S(u,\ubar)}|\err[\Omega^4\mathring{\alpha}]|(u_\Gamma(\ubar),\ubar,\theta^A)\mathrm{vol}_\gamma\dee\ubar\dee u\\
	&\quad\quad+\int_{\mcr'[\ubar_1']}\Omega^{-4}\left(\frac{1}{|u_f|(\ubar')^{6}}+\frac{C}{(\ubar')^{6+2\delta/3}}\right).
\end{align*}
Moreover, by \eqref{eq:enoughforldeux01} the second line on the RHS above is bounded by
\begin{align*}
	C\int_{u_0}^{u_1}\int_{\ubar_0}^{\ubar_1}e^{|\kappa_-|(u+\ubar)}\ubar^{-6-\delta/3}&\lesssim \frac{Ce^{|\kappa_-|\ubar_1'}}{(\ubar'_1)^{6+\delta/3}},
\end{align*}
where we used $\int_{u_0}^{u_1}e^{|\kappa_-|u}\dee u=(e^{|\kappa_-|u_1}-e^{|\kappa_-|u_0})/{|\kappa_-|}\lesssim 1$ since $u_1<0$, an integration by parts in $\ubar$, and $|\ubar_1-\ubar'_1|\lesssim 1$. Also, we have 
\begin{align}\label{eq:arguinglikein}
	\int_{\mcr'[\ubar_1']}\Omega^{-4}\left(\frac{1}{|u_f|(\ubar')^{6}}+\frac{C}{(\ubar')^{6+2\delta/3}}\right)\lesssim \int_{u_0'}^{u_1'}\int_{\ubar_0'}^{\ubar_1'}e^{|\kappa_-|(u'+\ubar')}\left(\frac{1}{|u_f|(\ubar')^{6}}\dee u\dee\ubar+\frac{C}{(\ubar')^{6+2\delta/3}}\right)\dee u'\dee\ubar'.
\end{align}
where we used \eqref{eq:volumeendoublenull} and the exact Kerr change of variables from the double null coordinates $(u,\ubar,\theta^A)$ to the coordinates $(u',\ubar',\theta,\phi_-)$, which has bounded Jacobian\footnote{Recall from Definition \ref{defi:cccva?} that $u',\ubar',\theta,\phi_+$ are defined as their Kerr values with respect to $u,\ubar,\theta^A$ so that the Jacobian coincides with its Kerr value.}. Finally, another integration by parts in $\ubar'$ proves that the RHS above is bounded by 
	$$e^{|\kappa_-|\ubar_1'}\left(\frac{C}{(\ubar'_1)^{6+\delta/3}}+\frac{e^{|\kappa_-|u'_1}-e^{|\kappa_-|u'_0}}{(\ubar_1')^{6}|u_f|}\right),$$
which concludes the proof.
\end{proof}
\subsubsection{Integral estimates and $C^{0,1}_{loc}$-inextendibility across $\ch$}\label{section:troizhhh}

\begin{prop}\label{prop:explosionintegrale}
	Let $\psi$ be defined by \eqref{eq:defpsiansatz}. 	We assume that
	\begin{align}\label{eq:conditionQm2}
		(Q_{-2},Q_{-1},Q_{+1},Q_{+2})\neq (0,0,0,0).
	\end{align}
We define the following sequence for $k\geq 1$,
$$\ubar'_{(k)}:=\frac{4Mr_- }{|a|}k\pi\underset{k\to+\infty}{\longrightarrow}+\infty.$$ 
There exists vector fields $X_1,X_2,Y_1,Y_2\in TS(u,\ubar)$ in $\trois$, which extend continuously to $\ch$ in the $C^0$ coordinates $(u,\ubar_\ch,\theta^A_\ch)$, such that provided $|u_f(a,M,(Q_m)_{|m|\leq 2})|\gg 1$ is large enough, for any $u'\leq u_f$ such that the corresponding level set of $u'$ is included in $\trois$, and for any $u_0'<u'<u_1'$ and $\ubar'_0$ large enough such that $\mcr'[\ubar'_{(k)}]\subset\trois$ for $k$ large enough, we have
	\begin{align*}
		\left|\int_{\mcr'[\ubar'_{(k)}]}\Omega^{-4}\left(\psi(X_1,Y_1)-\psi(X_2,Y_2)+i(\psi(X_1,Y_2)+\psi(X_2,Y_1))\right)\right|\gtrsim  \frac{e^{|\kappa_-|\ubar'_{(k)}}}{(\ubar'_{(k)})^6}(e^{|\kappa_-|u_1'}-e^{|\kappa_-|u_0'}),
	\end{align*}
for $k\gg 1$, where the implicit constant above depends on $a,M,(Q_m)_{|m|\leq 2}$.
\end{prop}
\begin{proof}\noindent\textbf{Step 1.}\textit{ Definition of $X_1,X_2,Y_1,Y_2$ and continuity up to $\ch$.}
	We define 
\begin{align}\label{eq:defXetY}
	{X}:=\left(1-\frac{Ma^2r_-\sin^2\theta_\mck(u,\ubar,\theta^A_\ch)}{(r_-^2+a^2)^2}\right)\partial_{\phi^*},
\end{align}
where $\partial_{\phi_*}$ is the $\phi_*$-coordinate vector field in double null coordinates $(u,\ubar,\theta_*,\phi_*)$ in $\trois$, and where $\sin^2\theta_\mck(u,\ubar,\theta^A_\ch)$ is the Kerr value of $\sin^2\theta$ with respect to the $C^0$ coordinates $(u,\ubar,\theta^A_\ch)$, see Section \ref{section:C0extension}. For a given complex bounded scalar function 
$$h:=x+iy,\quad x,y:\trois\longrightarrow\mathbb{R},$$
in $\trois$ that we will choose later in the proof, we define
$$X_1:=X,\quad X_2=\hodge{X},\quad Y_1:=xX,\quad Y_2:=yX, $$
where $\hodge{X}^A:=\in^A_C X^C$, where we recall that $\in$ denotes the volume form of $(S(u,\ubar),\gamma)$. The function $h(u,\ubar,\theta^A_\ch)$ will eventually be chosen to be continuous in the $C^0$ coordinates up to $\ch$ and $\gamma,\in$ extend continuously also, so for now we only have to prove that $X$ extends continuously to $\ch$. We know from exact Kerr (where $\sin^2\theta$ is smooth up to $\ch$) that the function $\sin^2\theta_\mck(u,\ubar,\theta^A_\ch)$ extends smoothly to the boundary $\{\ubar=+\infty\}$. By \eqref{eq:defXetY} this proves that $X$ extends continuously to $\ch$ in the $(u,\ubar_{\ch},\theta_*,\phi_*)$ coordinate system\footnote{And hence also in both coordinate systems $(u,\ubar_{\ch},\theta^1_{(i)},\theta^2_{(i)})$ in $\mcv_i$, $i=1,2$, as defined in Section \ref{section:C0extension}, since $\partial_{\phi_*}=-\theta^2_{(i)}\partial_{\theta^1_{(i)}}+\theta^1_{(i)}\partial_{\theta^2_{(i)}}$.}.  By \cite[Corollary 16.13]{stabC0}, we deduce that $X$ extends continuously to $\ch$ in the $(u,\ubar_{\ch},\theta^A_{\ch})$ coordinate system.

	\noindent\textbf{Step 2.}\textit{ Computation of $\psi(X_1,Y_1)-\psi(X_2,Y_2)+i(\psi(X_1,Y_2)+\psi(X_2,Y_1))$ in $\trois$.} We have
\begin{align}\label{eq:toutrevenir}
	\psi(X_1,Y_1)-\psi(X_2,Y_2)+i(\psi(X_1,Y_2)+\psi(X_2,Y_1))=h(\psi(X,X)+i\psi(\hodge{X},X))=h(\psi+i\hodge{\psi})(X,X),
\end{align}
where $\hodge{\psi}_{AB}=\in^{C}_A\psi_{CB}$. Also, defining $\mathring{\Psi}:=\psi+i\hodge{\psi}$ then by the bound for $\gcheck$ in \eqref{eq:onlyused}, from \eqref{eq:defpsiansatz} we get
\begin{align}\label{eq:grospsiring}
	\mathring{\Psi}=\frac{W_\mck}{\ubar'^6}\sum_{|m|\leq 2}\frac{Q_mA_m(r_\mck)}{\qbar_\mck^2}\Big(\mcd\hot(\mcd(Y_{m,2}(\cos\theta)e^{im\phi_+}))\Big)^S_\mck+O_Q(\ubar'^{-6}|u|^{-2-\delta/3}).
\end{align}
We now define the auxiliary vector field
		$$		\wt{X}:=[\Phi^{-1}(\partial_\phi+a\sin^2\theta\partial_t)]_\mck\in TS(u,\ubar).$$
Note that we have, in Kerr, $\partial_\phi+a\sin^2\theta\partial_t=\partial_{\phi^*}-{a\sin^2\theta}b^A\partial_{\theta^A}/2+{a\sin^2\theta}(\Omega^2\ering_4-\ering_3)/2$. Thus $\wt{X}=\wt{X}_\mck^A\partial_{\theta^A}$ where, in Kerr, by \eqref{eq:grosPhi} the identity $\Phi_\mck(\wt{X}_\mck^A\partial_{\theta^A})=\partial_\phi+a\sin^2\theta\partial_t$ implies
$$\wt{X}^A_\mck=\partial_{\phi^*}^A-\frac{a\sin^2\theta}{2}b^A_\mck-\frac{1}{2}(\fbar_\mck)_B\wt{X}_\mck^Bf^A_\mck.$$
Moreover, as $\Phi_\mck$ is an isometry we have $|\wt{X}|\lesssim |{\wt{X}}|_{\gamma_\mck}=|\partial_\phi+a\sin^2\theta\partial_t|_\mck\lesssim 1$, see also Footnote \ref{footnote:normes} for the first bound. Thus using $|\fbar_\mck|_\mck\lesssim\Omega_\mck^2$, $|f_\mck|_\mck\lesssim 1$ and \eqref{eq:bDNdanskerr} we deduce
\begin{align}\label{eq:lienXXtilde}
	\wt{X}=\partial_{\phi^*}-\frac{a\sin^2\theta}{2}b^A_\mck\partial_{\theta^A}+O(\Omega_\mck^2)=\left(1-\frac{Ma^2r_-\sin^2\theta}{(r_-^2+a^2)^2}\right)\partial_{\phi^*}+O(\Omega_\mck^2),
\end{align}	
where we recall that $\theta$ is defined as its Kerr value with respect to $u,\ubar,\theta^A$. Now, notice that in Kerr, $\sin^2\theta$ depends only on $u,\ubar,\theta_*$, thus we have 
\begin{align}\label{eq:commencertt}
	\sin\theta_\mck(u,\ubar,\theta^A_\ch)=\sin\theta_\mck(u,\ubar,(\theta_*)_\mck(u,\ubar,\theta^A_\ch)),
\end{align}
where $(\theta_*)_\mck(u,\ubar,\theta^A_\ch)$ is the Kerr value of $\theta^*$ with respect to $(u,\ubar,\theta^A_\ch)$. Moreover by definition \eqref{eq:defstereocoord} we have the identities
\begin{align*}
	\theta_*&=2\mathrm{cot}^{-1}\left(\sqrt{(\theta^1_{\ch,(1),\mck})^2+(\theta^2_{\ch,(1),\mck})^2}\right)\quad\text{in }\:\mcv'_1,\nn\\
	\theta_*&=2\mathrm{tan}^{-1}\left(\sqrt{(\theta^1_{\ch,(2),\mck})^2+(\theta^2_{\ch,(2),\mck})^2}\right)\quad\text{in }\:\mcv'_2.
\end{align*}
Thus, we deduce by \eqref{eq:oulalaclutchbetter}\footnote{Using also the fact that $\tan^{-1},\cot^{-1}$ are Lipschitz, and the triangle inequality in $\mathbb{R}^2$ for $\|\cdot\|_2$.} the bound
\begin{align}\label{eq:toiaussssi}
	|\theta_*-(\theta_*)_\mck(u,\ubar,\theta^A_\ch)|\lesssim\varepsilon|u_f|^{-2-\delta/3}\lesssim|u_f|^{-2-\delta/3} ,
\end{align}
(which holds because it hold on each $\mcv'_i$). This implies by \eqref{eq:commencertt} the bound
\begin{align}\label{eq:attdesuitetoi}
	|\sin^2\theta-\sin^2\theta_\mck(u,\ubar,\theta^A_\ch)|\lesssim|u_f|^{-2-\delta/3},
\end{align}
	and hence, by \eqref{eq:lienXXtilde} and \eqref{eq:defXetY} we infer $|X-\wt{X}|\lesssim|u_f|^{-2-\delta/3}+\Omega^2$. Thus, using \eqref{eq:grospsiring} and \eqref{eq:expodecayIII}, we deduce in $\trois$
\begin{align*}
	\mathring{\Psi}(X,X)=\mathring{\Psi}(\wt{X},\wt{X})+O_Q\left({\ubar'^{-6}|u_f|^{-2-\delta/3}}\right).
\end{align*}
	 Also, we have by definition of the subscript $\mck$ and of the supscript $S$,
	\begin{align*}
		\Big(&\mcd\hot(\mcd(Y_{m,2}(\cos\theta)e^{im\phi_+}))\Big)^S_\mck(\wt{X},\wt{X})\\
		&=\left[\Big(\mcd\hot(\mcd(Y_{m,2}(\cos\theta)e^{im\phi_+}))\Big)^S(\Phi^{-1}(\partial_\phi+a\sin^2\theta\partial_t),\Phi^{-1}(\partial_\phi+a\sin^2\theta\partial_t)\right]_\mck\\
		&=\left[\Big(\mcd\hot(\mcd(Y_{m,2}(\cos\theta)e^{im\phi_+}))\Big)(\partial_\phi+a\sin^2\theta\partial_t,\partial_\phi+a\sin^2\theta\partial_t)\right]_\mck\\
		&=\sin^2\theta_\mck|q_\mck|^2\left[\Big(\mcd\hot(\mcd(Y_{m,2}(\cos\theta)e^{im\phi_+}))\Big)_{22}\right]_\mck,
	\end{align*}
	with $e_2$ as in \eqref{eq:horizontalframekerr} in Kerr. Thus by Lemma \ref{lem:DhotDYm2} in Kerr, combined with \eqref{eq:grospsiring}, we get
	\begin{align}\label{eq:jppdlabelss}
		\mathring{\Psi}(X,X)=\frac{1}{\ubar'^6}\sum_{|m|\leq 2}\frac{-4\sqrt{6}Q_m{W_\mck}A_m(r_\mck)\sin^2\theta_\mck}{\qbar_\mck^2}Y_{m,2}^{+2}(\cos\theta_\mck)e^{im(\phi_+)_\mck}+O_Q\left({\ubar'^{-6}|u_f|^{-2-\delta/3}}\right),
	\end{align}
 in $\trois$. Using $2(r^*)_\mck=u'+\ubar'$ and \eqref{eq:rstarexpre}, \eqref{eq:rmodexpre}, we also have the identity 
	\begin{align*}
		(\phi_+)_\mck&=\phi_-+2(r_{mod})_\mck=\phi_-+\frac{a}{2Mr_-}(u'+\ubar')+h_{reg}(r_\mck),
	\end{align*}
	where $h_{reg}(r_\mck)$ is a function of $r_\mck$ which admits a finite limit as $r_\mck\to r_-$. This implies 
	\begin{align}
		\mathring{\Psi}(X,X)=&\frac{1}{\ubar'^6}\sum_{|m|\leq 2}Q_mz_m(r_\mck,\theta)\sin^2\theta e^{\frac{aim}{2Mr_-}(u'+\ubar')}Y_{m,2}^{+2}(\cos\theta)e^{im\phi_-}+O_Q\left({\ubar'^{-6}|u_f|^{-2-\delta/3}}\right),\nn\\		z_m(r_\mck,\theta):=&\frac{-4\sqrt{6}A_m(r_\mck)W_\mck}{\qbar_\mck^2}e^{imh_{reg}(r_\mck)}.\label{eq:smallzmm}
	\end{align}
	\noindent\textbf{Step 3.} \textit{Preliminary integral computations.} We will eventually choose $u'_1,u'_0$ such that $\mcr'[\ubar'_1]\subset\trois$ for any $\ubar'_1\geq\ubar'_0$. Thus denoting
	$$I[\ubar'_1]:=\int_{\mcr'[\ubar'_{(k)}]}\Omega^{-4}\left(\psi(X_1,Y_1)-\psi(X_2,Y_2)+i(\psi(X_1,Y_2)+\psi(X_2,Y_1))\right),$$
	 by \eqref{eq:toutrevenir} we compute, using $|\widecheck{\Omega^2}|\lesssim\Omega^2|\gcheck|\lesssim \varepsilon\Omega^2|u|^{-2-\delta/3}$ in $\trois$,
	\begin{align}
		I[\ubar'_1]&=2\int_{\mcr'[\ubar'_1]}\Omega^{-2}h\mathring{\Psi}(X,X)\mathrm{vol}_\gamma\dee u\dee\ubar\nn\\
		&=2\int_{\mcr'[\ubar'_1]}\left(\Omega_\mck^{-2}h\mathring{\Psi}(X,X)+O_Q\left(\frac{\Omega^{-2}}{\ubar^6|u|^{2+\delta/3}}\right)\right)\mathrm{vol}_\gamma\dee u\dee\ubar.\label{eq:salutcestmoi0}
	\end{align}
Moreover, we have $\mathrm{vol}_\gamma=\sqrt{|\det\gamma_{AB}|}\dee\theta^1\dee\theta^2$, where, in each $\mcv'_i$, $i=1,2$,
	\begin{align*}
		|\sqrt{|\det\gamma_{AB}|}-\sqrt{|\det(\gamma_\mck)_{AB}|}|\lesssim{\varepsilon}|u|^{-2-\delta/3}\sqrt{|\det\gamma_{AB}|},
	\end{align*}
since $|\gammacheck|\lesssim\varepsilon|u|^{-2-\delta/3}$ by \eqref{eq:onlyused}. Integrating and summing over $i=1,2$ we infer from \eqref{eq:salutcestmoi0},
	\begin{align}
	I[\ubar'_1]=2\int_{\mcr'[\ubar'_1]}\left(\Omega_\mck^{-2}h\mathring{\Psi}(X,X)+O_Q\left(\frac{\Omega^{-2}}{\ubar^6|u_f|^{2+\delta/3}}\right)\right)(\mathrm{vol}_\gamma)_\mck\dee u\dee\ubar.\label{eq:salutcestmoi}
\end{align}
	Moreover, arguing exactly like in \eqref{eq:arguinglikein} we get
	\begin{align}
		\int_{\mcr'[\ubar'_1]}\frac{\Omega^{-2}}{\ubar^6|u_f|^{2+\delta/3}}(\mathrm{vol}_\gamma)_\mck\dee u\dee\ubar\lesssim \frac{e^{|\kappa_-|\ubar_1'}(e^{|\kappa_-|u'_1}-e^{|\kappa_-|u'_0})}{(\ubar_1')^{6}|u_f|}.\label{eq:simililirara}
	\end{align}
 Thus, by \eqref{eq:salutcestmoi} we get
	\begin{align*}
		I[\ubar'_1]&=2\int_{\mcr'[\ubar'_1]}\left[\Omega_\mck^{-4}h\mathring{\Psi}(X,X)\right]\Omega_\mck^{2}\mathrm{vol}_{\gamma_\mck}\dee u\dee\ubar+O_Q\left(\frac{e^{|\kappa_-|\ubar_1'}(e^{|\kappa_-|u'_1}-e^{|\kappa_-|u'_0})}{(\ubar_1')^{6}|u_f|}\right).
	\end{align*}
Now, by the first line of \eqref{eq:smallzmm} we infer
	\begin{align}\label{eq:needrhsonly}
		I[\ubar'_1]=&\int_{\mcr'[\ubar'_1]}\frac{\Omega_\mck^{-4}}{\ubar'^6}\sum_{|m|\leq 2}hQ_mz_m(r_\mck,\theta_\mck)\sin^2\theta_\mck Y_{m,2}^{+2}(\cos\theta_\mck)e^{im(\phi_-)_\mck}e^{\frac{aim}{2Mr_-}(u'+\ubar')}(\vol)_\mck\nn\\
		&+O_Q\left(\frac{e^{|\kappa_-|\ubar_1'}(e^{|\kappa_-|u'_1}-e^{|\kappa_-|u'_0})}{(\ubar_1')^{6}|u_f|}\right).
	\end{align}
	Now, applying the exact Kerr coordinate transform from the double null coordinates to the coordinates $(u',\ubar',\theta,\phi_-)$, and omitting some subscripts ${}_\mck$ in what follows, we find that
	\begin{align*}
		I[\ubar_1']=&\int_{\mcr'[\ubar'_1]}\frac{-\Delta^{-1}}{\ubar'^6}\sum_{|m|\leq 2}\frac{8hQ_mz_m(r,\theta)R_\mck^4|q_\mck|^2}{r^2+a^2}\sin^2\theta Y_{m,2}^{+2}(\cos\theta)e^{im\phi_-}e^{\frac{aim}{2Mr_-}(u'+\ubar')}\sin\theta\dee\theta\dee\phi_-\dee u'\dee \ubar'\\
		&+O_Q\left(\frac{e^{|\kappa_-|\ubar_1'}(e^{|\kappa_-|u'_1}-e^{|\kappa_-|u'_0})}{(\ubar_1')^{6}|u_f|}\right),
	\end{align*}
where we used : 
\begin{itemize}
	\item The identity $\Omega_\mck^2=-\Delta_\mck/(4R_\mck^2)$ where $R_\mck^2=r^2+a^2+2Ma^2r\sin^2\theta/|q|^2$ in Kerr, see \eqref{eq:bDNdanskerr}.
	\item The expression \eqref{eq:kerrvolumeformexact} for the Kerr spacetime volume form in $(u',\ubar',\theta,\phi_-)$ coordinates.
\end{itemize}
	Also recall the expression \eqref{eq:deltaenexp} in Kerr which implies $-\Delta=e^{-|\kappa_-|(u'+\ubar')}e^{h_-(r)}$, where $h_-(r)$ as a finite limit as $r\to r_-$. From this, and using $aim/(2Mr_-)=2|\kappa_-|aim/(r_+-r_-)$ we infer
	\begin{align}\label{eq:allercoreeheh}
		I[\ubar_1']=&\sum_{|m|\leq 2}\int_{\mcr'[\ubar'_1]}\frac{e^{|\kappa_-|\left(1+\frac{2aim}{r_+-r_-}\right)(u'+\ubar')}}{\ubar'^6}hQ_mZ_m(r,\theta) Y_{m,2}^{+2}(\cos\theta)e^{im\phi_-}\sin\theta\dee\theta\dee\phi_-\dee u'\dee \ubar'\nn\\
		&+O_Q\left(\frac{e^{|\kappa_-|\ubar_1'}(e^{|\kappa_-|u'_1}-e^{|\kappa_-|u'_0})}{(\ubar_1')^{6}|u_f|}\right),
	\end{align}
	where, recalling \eqref{eq:smallzmm},
	$$Z_m(r,\theta):=\frac{8z_m(r,\theta)R_\mck^4|q_\mck|^2e^{-h_-(r)}}{r^2+a^2}\sin^2\theta.$$
	Now, notice that we have for any $m=0,\pm1,\pm 2$,
	\begin{align}
		Z_m(r,\theta)&=F_m(\theta)+O(|\Delta|),\label{eq:extredeltadomco}\\
		F_m(\theta):&=-\frac{32\sqrt{6}A_m(r_-)W_\mck(r_-,\theta) R_\mck^4(r_-,\theta)(r_-^2+a^2\cos^2\theta)e^{imh_{reg}(r_-)-h_-(r_-)}}{(r_-^2+a^2)\qbar_\mck^2(r_-,\theta)}\sin^2\theta.\nn
	\end{align}
Now, note that using \eqref{eq:defWlambdabarahah} which implies $W_\mck(r_-,\theta)=\lambdabar^2_\mck(r_-,\theta)$, combined with the definition \eqref{eq:enfaitsijutilise} of $\lambdabar_\mck$ and the identity \eqref{eq:alatoutefinnn}, we get $W_\mck(r_-,\theta)={(r_-^2+a^2)^2}/(4{R_\mck^4(r_-,\theta)})$, and hence
$$F_m(\theta)=-\frac{8\sqrt{6}A_m(r_-)(r_-^2+a^2)(r_-^2+a^2\cos^2\theta)e^{imh_{reg}(r_-)-h_-(r_-)}}{\qbar_\mck^2(r_-,\theta)}\sin^2\theta,$$
which shows by \eqref{eq:amdermoins} that for $m\neq 0$ we have $F_m(\theta)\neq 0$ for $\theta\neq 0,\pi$, and $F_0(\theta)=0$ for any $\theta$.

\textbf{Step 4.}\textit{ Choice of $h$ and end of the proof.} By assumption \eqref{eq:conditionQm2}, there is $m_0=\pm 1,\pm 2$ such that $Q_{m_0}\neq 0$. Then, let $h_0$ be the following smooth function $h_0:\mathbb{S}^2\rightarrow\mathbb{C}$ on $\mathbb{S}^2$:
\begin{align}\label{defi:bonbahdefh0}
	h_0(\theta,\phi):=\overline{F_{m_0}(\theta) Y_{{m_0},2}^{+2}(\cos\theta)e^{i{m_0}\phi}}.
\end{align}
Note that thanks to the factor $\sin^2\theta$ in $F_m(\theta)$ in \eqref{eq:extredeltadomco}, $h_0$ is indeed a smooth function on $\mathbb{S}^2$. Then the function $h(u,\ubar,\theta^A_\ch)$ is defined by
$$h(u,\ubar,\theta^A_\ch):=h_0(\wt{\theta},\wt{\phi}_-),$$
where
$$\wt{\theta}:=\theta_\mck(u,\ubar,\theta^A_\ch),\quad \wt{\phi}_-:=(\phi_-)_\mck(u,\ubar,\theta^A_\ch),$$
are respectively the Kerr values of $\theta,\phi_-$ with respect to the $C^0$ coordinates $(u,\ubar,\theta^A_\ch)$. Note that by smoothness of $h_0$, this choice ensures that $h(u,\ubar,\theta^A_\ch)$ extends continuously to $\ch$. Now recalling from Definition \ref{defi:cccva?} the values 
$$\theta=\theta_\mck(u,\ubar,\theta^A),\quad\phi_-=(\phi_-)_\mck(u,\ubar,\theta^A),$$
we have the following bound in $\trois$,
\begin{align}\label{eq:toutcaprcalol}
	|h_0(\wt{\theta},\wt{\phi}_-)-h_0({\theta},{\phi}_-)|\lesssim|u_f|^{-2-\delta/3},
\end{align}
which follows from \eqref{eq:oulalaclutchbetter}. More precisely, by the definition \eqref{defi:bonbahdefh0} of $h_0$ we have
	\begin{align*}
		|h_0(\wt{\theta},\wt{\phi}_-)-h_0({\theta},{\phi}_-)|&\lesssim |\sin^2\theta-\sin^2\wt{\theta}|+\sin\theta |e^{i{m_0}\phi_-}-e^{i{m_0}\wt{\phi}_-}|\\
		&\lesssim |u_f|^{-2-\delta/3}+\sin\theta |e^{i{m_0}\phi_-}-e^{i{m_0}\wt{\phi}_-}|,
	\end{align*}
where we used \eqref{eq:attdesuitetoi} in the last step. Now, recalling the definition of $\phi^*_{\ch,\mck}$ in \eqref{eq:DNthetaCH} we get that there is a function $F_\mck(u,\ubar,\theta_*)$ such that
$$\phi_-=\phi^*_{\ch,\mck}+F_\mck(u,\ubar,\theta_*),$$
where we know from the Kerr case that $F_\mck(u,\ubar,\theta_*)$ extends smoothly to $\{\ubar=+\infty\}$. We deduce
$$|e^{i{m_0}\phi_-}-e^{i{m_0}\wt{\phi}_-}|\lesssim |\theta_*-(\theta_*)_\mck(u,\ubar,\theta^A_\ch)|+|e^{i{m_0}\phi^*_{\ch,\mck}}-e^{i{m_0}\phi^*_{\ch}}|\lesssim |u_f|^{-2-\delta/3}+|e^{i\phi^*_{\ch,\mck}}-e^{i\phi^*_{\ch}}|$$
by \eqref{eq:toiaussssi}. Finally, expressing the $\mathbb{S}^2$ coordinates $(\sin\theta^*_{\ch,\mck}\cos\phi^*_{\ch,\mck},\sin\theta^*_{\ch,\mck}\sin\phi^*_{\ch,\mck})$ and the analog ones $(\sin\theta^*_{\ch}\cos\phi^*_{\ch},\sin\theta^*_{\ch}\sin\phi^*_{\ch})$ with respect to the stereographic projection coordinates $\theta^A_{\ch,(i),\mck}$ and $\theta^A_{\ch,(i)}$ on each $\mcv'_i$, and using \eqref{eq:toiaussssi} one more time and $\theta^*_{\ch,\mck}=\theta_*$, we finally get
\begin{align*}
	\sin\theta |e^{i\phi^*_{\ch,\mck}}-e^{i\phi^*_{\ch}}|&\lesssim\sin\theta_*|e^{i\phi^*_{\ch,\mck}}-e^{i\phi^*_{\ch}}|\\
	&\lesssim |u_f|^{-2-\delta/3}+|\sin\theta^*_{\ch,\mck}e^{i\phi^*_{\ch,\mck}}-\sin\theta^*_{\ch}e^{i\phi^*_{\ch}}|\\
	&\lesssim |u_f|^{-2-\delta/3}+\sup_{i=1,2,A=1,2}|\theta^A_{\ch,(i)}-\theta^A_{\ch,(i),\mck}|\lesssim|u_f|^{-2-\delta/3},
\end{align*}
by \eqref{eq:oulalaclutchbetter}, hence the proof of \eqref{eq:toutcaprcalol}. Combining this with \eqref{eq:allercoreeheh} and \eqref{eq:extredeltadomco} thus yields
	\begin{align}
		I[\ubar_1']&=\sum_{|m|\leq 2}\int_{u_0'}^{u_1'}\int_{0}^{\pi}\int_{0}^{2\pi}{e^{|\kappa_-|\left(1+\frac{2aim}{r_+-r_-}\right)u'}}h_0(\theta,\phi_-)Q_mF_m(\theta) Y_{m,2}^{+2}(\cos\theta)e^{im\phi_-}\label{eq:Iexre}\\
		&\quad\quad\quad\quad\times\left[\int_{\ubar_0'}^{\ubar_1'}\frac{e^{|\kappa_-|\left(1+\frac{2aim}{r_+-r_-}\right)\ubar'}}{\ubar'^6}\dee \ubar'\right]\sin\theta\dee\theta\dee\phi_-\dee u'+O_Q\left(\frac{e^{|\kappa_-|\ubar_1'}(e^{|\kappa_-|u'_1}-e^{|\kappa_-|u'_0})}{(\ubar_1')^{6}|u_f|}\right),\nn
	\end{align}
where we used \eqref{eq:expodecayIII} to bound the term with an extra $\Delta$ (which comes from the last term on the RHS of \eqref{eq:extredeltadomco}), and estimates similar to \eqref{eq:simililirara} to bound the integral of the error term \eqref{eq:toutcaprcalol}, by the last term on the RHS above. Moreover, we have by integration by parts
	\begin{align*}
		&\int_{\ubar_0'}^{\ubar_1'}\frac{e^{|\kappa_-|\left(1+\frac{2aim}{r_+-r_-}\right)\ubar'}}{\ubar'^6}\dee \ubar'\\
		&=\frac{1}{|\kappa_-|\left(1+\frac{2aim}{r_+-r_-}\right)}\left(\left[\frac{e^{|\kappa_-|\left(1+\frac{2aim}{r_+-r_-}\right)\ubar'}}{\ubar'^6}\right]_{\ubar_0'}^{\ubar_1'}-\int_{\ubar_0'}^{\ubar_1'}e^{|\kappa_-|\left(1+\frac{2aim}{r_+-r_-}\right)\ubar'}\frac{\partial}{\partial\ubar'}\left[\frac{1}{\ubar'^6}\right]\dee \ubar'\right)\\
		&=\frac{e^{|\kappa_-|\left(1+\frac{2aim}{r_+-r_-}\right)\ubar_1'}}{|\kappa_-|\left(1+\frac{2aim}{r_+-r_-}\right)(\ubar_1')^6}-\frac{e^{|\kappa_-|\left(1+\frac{2aim}{r_+-r_-}\right)\ubar_0'}}{|\kappa_-|\left(1+\frac{2aim}{r_+-r_-}\right)(\ubar_0')^6}+\frac{6}{|\kappa_-|\left(1+\frac{2aim}{r_+-r_-}\right)}\int_{\ubar_0'}^{\ubar_1'}\frac{e^{|\kappa_-|\left(1+\frac{2aim}{r_+-r_-}\right)\ubar'}}{\ubar'^7}\dee \ubar'.
	\end{align*}
	Keeping $\ubar'_0$ fixed and performing another integration by parts to bound the absolute value of the last term on the RHS above, this yields
	$$\int_{\ubar_0'}^{\ubar_1'}\frac{e^{|\kappa_-|\left(1+\frac{2aim}{r_+-r_-}\right)\ubar'}}{\ubar'^6}\dee \ubar'=\frac{e^{|\kappa_-|\left(1+\frac{2aim}{r_+-r_-}\right)\ubar_1'}}{|\kappa_-|\left(1+\frac{2aim}{r_+-r_-}\right)(\ubar_1')^6}+O\left(\frac{e^{|\kappa_-|\ubar_1'}}{(\ubar_1')^7}\right).$$
	We deduce by \eqref{eq:Iexre},
	\begin{align*}
		I[\ubar_1']=&\sum_{|m|\leq 2}\frac{e^{|\kappa_-|\left(1+\frac{2aim}{r_+-r_-}\right)\ubar_1'}}{|\kappa_-|\left(1+\frac{2aim}{r_+-r_-}\right)(\ubar_1')^6}Q_mJ_m(u_0', u_1')I_m(h_0)\\
		&+O_Q\left(\frac{e^{|\kappa_-|\ubar'_1}}{(\ubar_1')^7}\right)+O_Q\left(\frac{e^{|\kappa_-|\ubar_1'}(e^{|\kappa_-|u'_1}-e^{|\kappa_-|u'_0})}{(\ubar_1')^{6}|u_f|}\right),
	\end{align*}
	where for $|m|\leq 2$,
	\begin{align}\label{eq:notincne}
		J_m(u_0', u_1')&:=\int_{u_0'}^{u_1'}e^{|\kappa_-|\left(1+\frac{2aim}{r_+-r_-}\right)u'}\dee u'=\frac{e^{|\kappa_-|\left(1+\frac{2aim}{r_+-r_-}\right)u'_1}-e^{|\kappa_-|\left(1+\frac{2aim}{r_+-r_-}\right)u'_0}}{|\kappa_-|\left(1+\frac{2aim}{r_+-r_-}\right)},\\
		I_{m}(h_0)&:=\int_{0}^{\pi}\int_{0}^{2\pi} h_0(\theta,\phi_-)F_{m}(\theta) Y_{m_,2}^{+2}(\cos\theta)e^{im\phi_-}\sin\theta\dee\theta\dee\phi_-.
	\end{align}
By the choice \eqref{defi:bonbahdefh0} for $h_0$ we deduce
\begin{align}\label{eq:nonzerh}
	I_m(h_0)=0\:\:\text{for}\:\:m\neq m_0,\quad 	I_{m_0}(h_0)\neq 0,
\end{align} 
since $F_{m_0}(\theta) Y_{{m_0},2}^{+2}(\cos\theta)e^{i{m_0}\phi}$ is non-zero in $L^2(\mathbb{S}^2)$. We thus deduce for any $\ubar'_1$, 
	\begin{align*}
		I[\ubar'_1]=&Q_{m_0}I_{m_0}(h_0)\frac{e^{|\kappa_-|\left(1+\frac{2aim_0}{r_+-r_-}\right)\ubar_1'}}{|\kappa_-|\left(1+\frac{2aim_0}{r_+-r_-}\right)(\ubar_1')^6}\times\frac{e^{|\kappa_-|\left(1+\frac{2aim_0}{r_+-r_-}\right)u'_1}-e^{|\kappa_-|\left(1+\frac{2aim_0}{r_+-r_-}\right)u'_0}}{|\kappa_-|\left(1+\frac{2aim_0}{r_+-r_-}\right)}\\
		&+O_Q\left(\frac{e^{|\kappa_-|\ubar'_1}}{(\ubar_1')^7}\right)+O_Q\left(\frac{e^{|\kappa_-|\ubar_1'}(e^{|\kappa_-|u'_1}-e^{|\kappa_-|u'_0})}{(\ubar_1')^{6}|u_f|}\right).
	\end{align*}
Moreover, defining for $k\geq 1$, $\ubar'_{(k)}:=\frac{4Mr_- }{|a|}k\pi$ we get for $m=0,\pm 1,\pm 2$ and $k\geq 1$,
$$\frac{2aim}{r_+-r_-}|\kappa_-|\ubar'_{(k)}=\pm2im k\pi\in2\pi\mathbb{Z},$$
which yields
	\begin{align*}
	I[\ubar'_{(k)}]=&Q_{m_0}I_{m_0}(h_0)\frac{e^{|\kappa_-|\ubar_{(k)}'}}{|\kappa_-|\left(1+\frac{2aim_0}{r_+-r_-}\right)(\ubar_{(k)}')^6}\times\frac{e^{|\kappa_-|\left(1+\frac{2aim_0}{r_+-r_-}\right)u'_1}-e^{|\kappa_-|\left(1+\frac{2aim_0}{r_+-r_-}\right)u'_0}}{|\kappa_-|\left(1+\frac{2aim_0}{r_+-r_-}\right)}\\
	&+O_Q\left(\frac{e^{|\kappa_-|\ubar_{(k)}'}}{(\ubar_{(k)}')^7}\right)+O_Q\left(\frac{e^{|\kappa_-|\ubar_{(k)}'}(e^{|\kappa_-|u'_1}-e^{|\kappa_-|u'_0})}{(\ubar_{(k)}')^{6}|u_f|}\right).
\end{align*}
Now, from the triangle inequality we get
$$\left|e^{|\kappa_-|\left(1+\frac{2aim_0}{r_+-r_-}\right)u'_1}-e^{|\kappa_-|\left(1+\frac{2aim_0}{r_+-r_-}\right)u'_0}\right|\geq \left|e^{|\kappa_-|\left(1+\frac{2aim_0}{r_+-r_-}\right)u'_1}\right|-\left|e^{|\kappa_-|\left(1+\frac{2aim_0}{r_+-r_-}\right)u'_0}\right|=e^{|\kappa_-|u_1'}-e^{|\kappa_-|u_0'},$$
which proves that for some $(Q_m)_{|m|\leq 2}$-dependent constant $C_Q$, we have
$$|I[\ubar'_{(k)}]|\gtrsim |Q_{m_0}|I_{m_0}(h_0)\frac{e^{|\kappa_-|\ubar'_{(k)}}}{(\ubar'_{(k)})^6}(e^{|\kappa_-|u_1'}-e^{|\kappa_-|u_0'})-C_Q\frac{e^{|\kappa_-|\ubar'_{(k)}}}{(\ubar'_{(k)})^6}\left((\ubar'_{(k)})^{-1}+(e^{|\kappa_-|u_1'}-e^{|\kappa_-|u_0'})|u_f|^{-1}\right).$$
Since $|Q_{m_0}|I_{m_0}(h_0)>0$ by \eqref{eq:nonzerh} the definition of $m_0$, this concludes the proof of Proposition \ref{prop:explosionintegrale}, choosing $|u_f(a,M,(Q_m)_{|m|\leq 2})|\gg 1$ large enough.
\end{proof}

\begin{cor}\label{cor:curvintblowup}
	Let $u'\ll -1$ be negative enough as in the assumptions of Proposition \ref{prop:explosionintegrale}, and $u_0'<u'<u_1'$. Recall the quantities $X_1,X_2,Y_1,Y_2\in TS(u,\ubar)$, $\ubar'_{(k)}$ given by Proposition \ref{prop:explosionintegrale}. We define, for $j=1,2,3,4$,
	\begin{equation*}
		\begin{gathered}
			\overline{X}_1^{(j)}=\overline{X}_3^{(j)}=\ering_4,\\
			\overline{X}_2^{(1)}=\overline{X}_2^{(3)}:=X_1,\quad \overline{X}_4^{(1)}=\overline{X}_4^{(4)}:=Y_1,\quad \overline{X}_2^{(2)}=\overline{X}_2^{(4)}:=X_2,\quad \overline{X}_4^{(2)}=\overline{X}_4^{(3)}:=Y_2,
		\end{gathered}
	\end{equation*}
and the constants $z_1:=1,z_2:=-1,z_3:=z_4:=i$. Then, provided that \eqref{eq:conditionQm2} holds and that $|u_f|(a,M,(Q_m)_{|m|\leq 2})\gg 1$ is large enough, we have for $k$ large enough,
	\begin{align*}
		\left|\int_{\mcr'[\ubar'_{(k)}]}\sum_{j=1}^4 z_j\mathbf{R}(\overline{X}_1^{(j)},\overline{X}_2^{(j)},\overline{X}_3^{(j)},\overline{X}_4^{(j)})\right|\gtrsim\frac{e^{|\kappa_-|\ubar'_{(k)}}}{(\ubar'_{(k)})^6} \underset{k\to+\infty}{\longrightarrow}+\infty,
	\end{align*}
where the implicit constant above depends on $a,M,(Q_m)_{|m|\leq 2},u_0',u_1'$.
\end{cor}
\begin{proof}
By Proposition \ref{prop:exploR}, recalling the definition $\mathring{\alpha}_{AB}=\mathbf{R}(\ering_4,\partial_{\theta^A},\ering_4,\partial_{\theta^B})$, for any $j$ we get
	\begin{align*}
		\int_{\mcr'[\ubar'_{(k)}]}\bigg|\mathbf{R}(\overline{X}_1^{(j)},\overline{X}_2^{(j)},\overline{X}_3^{(j)},\overline{X}_4^{(j)})-\Omega^{-4}\psi(\overline{X}_2^{(j)},&\overline{X}_4^{(j)})\Bigg|= \int_{\mcr'[\ubar'_{(k)}]}\left|\mathring{\alpha}(\overline{X}_4^{(j)},\overline{X}_4^{(j)})-\Omega^{-4}\psi(\overline{X}_4^{(j)},\overline{X}_4^{(j)})\right|\\
		&\lesssim_Q e^{|\kappa_-|\ubar'_{(k)}}\left(\frac{C}{(\ubar'_{(k)})^{6+\delta/3}}+\frac{e^{|\kappa_-|u'_1}-e^{|\kappa_-|u'_0}}{(\ubar_{(k)}')^{6}|u_f|}\right).
	\end{align*}
This implies 
	\begin{align*}
		\Bigg|\int_{\mcr'[\ubar'_{(k)}]}\sum_{j=1}^4 z_j\mathbf{R}(\overline{X}_1^{(j)},&\overline{X}_2^{(j)},\overline{X}_3^{(j)},\overline{X}_4^{(j)})\Bigg|\gtrsim_Q-e^{|\kappa_-|\ubar'_{(k)}}\left(\frac{C}{(\ubar'_{(k)})^{6+\delta/3}}+\frac{e^{|\kappa_-|u'_1}-e^{|\kappa_-|u'_0}}{(\ubar_{(k)}')^{6}|u_f|}\right)\\
		&+\left|\int_{\mcr'[\ubar'_{(k)}]}\Omega^{-4}\left(\psi(X_1,Y_1)-\psi(X_2,Y_2)+i(\psi(X_1,Y_2)+\psi(X_2,Y_1))\right)\right|,
	\end{align*}
	thus by Proposition \ref{prop:explosionintegrale} we get
	\begin{align*}
		&\Bigg|\int_{\mcr'[\ubar'_{(k)}]}\sum_{j=1}^4 z_j\mathbf{R}(\overline{X}_1^{(j)},\overline{X}_2^{(j)},\overline{X}_3^{(j)},\overline{X}_4^{(j)})\Bigg|\\
		&\gtrsim_Q\frac{e^{|\kappa_-|\ubar_{(k)}'}}{(\ubar_{(k)}')^6}\left((e^{|\kappa_-|u'_1}-e^{|\kappa_-|u'_0})(1-C'(a,M,(Q_{m})_{|m|\leq 2})|u_f|^{-1})-C(\ubar'_{(k)})^{-\delta/3}\right),
	\end{align*}
for some $(a,M,(Q_{m})_{|m|\leq 2})$-dependent constant $C'(a,M,(Q_{m})_{|m|\leq 2})$ which concludes the proof, taking $|u_f|(a,M,(Q_m)_{|m|\leq 2})\gg 1$ and $k\gg 1$.
\end{proof}

We now prove another assumption of \cite[Theo. 3.14]{sbierskiinextdernier}, which is that the curvature blow-up of Proposition \ref{cor:curvintblowup} still holds for small perturbations of the vector fields $\overline{X}_i^{(j)}$. Recall the following Riemannian metric introduced in \cite{sbierskiinextdernier}\footnote{Note that \cite{sbierskiinextdernier} denotes as $\ubar$ the coordinate denoted here $\ubar_{\ch}$, defined in Section \ref{section:C0extension} by $\frac{\dee\ubar_{\ch}}{\dee\ubar}=e^{-|\kappa_-|\ubar}$.}: 
$$h:=\dee u^2+e^{-2|\kappa_-|\ubar}\dee\ubar^2+\gamma.$$
\begin{prop}\label{prop:curvintpertblowup}
We consider the setting of Corollary \ref{cor:curvintblowup}. Then, for any $\varepsilon_0>0$ small enough\footnote{Depending on all parameters $a,M,\delta_+,\delta,u_f,(Q_m)_{|m|\leq 2},u_1',u_0'$.}, and for any vector fields $\wh{X}_1^{(j)},\wh{X}_2^{(j)},\wh{X}_3^{(j)},\wh{X}_4^{(j)}$, $j=1,2,3,4$ continuous up to $\ch$ such that
	\begin{align}\label{eq:lasthyp}
		\sum_{i,j=1}^4\sup_{\mcr'[+\infty]}|\overline{X}_i^{(j)}-\wh{X}_i^{(j)}|_h\leq\varepsilon_0,
	\end{align}
where $\mcr'[+\infty]:=\trois\cap\{\ubar_0'\leq\ubar'\}\cap\{u_0'<u'<u_1'\}$, then for some $j=1,2,3,4$ we have
	\begin{align*}
		\left|\int_{\mcr'[\ubar_{(k)}']}\mathbf{R}(\wh{X}_1^{(j)},\wh{X}_2^{(j)},\wh{X}_3^{(j)},\wh{X}_4^{(j)})\right|\underset{k\to+\infty}{\longrightarrow}+\infty,
	\end{align*}
possibly along a subsequence of $\ubar_{(k)}'$.
\end{prop}
\begin{proof}
In this proof, since we are interested in a blow-up result which is not quantitative, we denote by $\lesssim$ any $(a,M,(Q_m)_{|m|\leq 2},u'_1,u'_0,\delta,\delta_+, u_f)$-dependent bound. Let $X=X^\mu\partial_\mu$ be a vector field. We have $X=X^u\partial_u+X^\ubar\partial_\ubar+X^A\partial_{\theta^A}=X^u\ering_3+X^\ubar\Omega^2\ering_4+(X^A-X^\ubar b^A)\partial_{\theta^A}$, thus
	$$|X|_h^2=(X^u)^2+e^{-2|\kappa_-|\ubar}(X^\ubar)^2+\gamma_{AB}(X^A-X^\ubar b^A)(X^B-X^\ubar b^B).$$
	We deduce that $|X|_h\leq\varepsilon_0$ in $\mcr'[+\infty]$ if and only if 
	$$X=O(\varepsilon_0)\ering_3+O(\varepsilon_0)\ering_4+X^S_{\varepsilon_0},$$
	where $X^S_{\varepsilon_0}\in TS(u,\ubar)$ is such that $|X^S_{\varepsilon_0}|_\gamma\lesssim\varepsilon_0$, and where we used the fact that $\Omega^2\sim e^{-|\kappa_-|\ubar}$ in $\mcr'[+\infty]$, as $u\sim u'\sim -1$ there. Thus, being given vector fields $\wh{X}_i^{(j)}$, $i,j=1,2,3,4$ such that \eqref{eq:lasthyp} holds, then we can write in $\mcr'[+\infty]$ for $i,j=1,2,3,4$:
	$$\wh{X}_i^{(j)}=\overline{X}_i^{(j)}+O(\varepsilon_0)\ering_3+O(\varepsilon_0)\ering_4+X^S_{\varepsilon_0,i,j},$$
	where $X^S_{\varepsilon_0,i,j}\in TS(u,\ubar)$ is such that $|X^S_{\varepsilon_0,i,j}|_\gamma\lesssim\varepsilon_0$. We now expand the term $\mathbf{R}(\wh{X}_1^{(j)},\wh{X}_2^{(j)},\wh{X}_3^{(j)},\wh{X}_4^{(j)})$ to get\footnote{Here, we use the expressions $\mathbf{R}_{a3b4}=-\rho\delta_{ab}+\hodge{\rho}\in_{ab}$, $\mathbf{R}_{ab34}=2\hodge{\rho}\in_{ab}$, $\mathbf{R}_{abcd}=-\in_{ab}\in_{cd}\rho$, $\mathbf{R}_{abc3}=\in_{ab}\hodge{\betabar}_c$, $\mathbf{R}_{abc4}=-\in_{ab}\hodge{\beta}_c$ (which follow from the identities in Section \ref{section:ricciandcurvdef}) to express the terms that cannot be directly written with $\alpha,\alphabar,\beta,\betabar,\rho,\hodge{\rho}$. These expressions show that all the curvature terms in the expansion can be expressed with $\alpha,\alphabar,\beta,\betabar,\rho,\hodge{\rho}$, multiplied with an appropriate power of $\varepsilon_0$.}, denoting $\delta\mathbf{R}^{(j)}:=\mathbf{R}(\wh{X}_1^{(j)},\wh{X}_2^{(j)},\wh{X}_3^{(j)},\wh{X}_4^{(j)})-\mathbf{R}(\overline{X}_1^{(j)},\overline{X}_2^{(j)},\overline{X}_3^{(j)},\overline{X}_4^{(j)})$,
	\begin{align*}
		\delta\mathbf{R}^{(j)}=_sO(\varepsilon_0)\cdot\mathring{\alpha}+O(\varepsilon_0^2)\cdot\mathring{\alphabar}+O(\varepsilon_0)\cdot\mathring{\beta}+O(\varepsilon_0^2)\cdot\mathring{\betabar}+O(\varepsilon_0)(\mathring{\rho},\hodge{\mathring{\rho}}).
	\end{align*}
	We also have by \eqref{eq:onlyused} and \eqref{eq:atsomepointtkt} the bounds $|\mathring{\alphabar}|+|\mathring{\betabar}|\lesssim 1$, $|\mathring{\beta}|\lesssim\Omega^{-2}$, $|\mathring{\rho}|+|\hodge{\mathring{\rho}}|\lesssim\Omega^{-2}$ in $\trois$, so that recalling the first line of \eqref{eq:defregiondoublenull} and Lemma \ref{lem:ubarssontpareils},
	\begin{align*}
		\Bigg|\int_{\mcr'[\ubar_{(k)}']}\delta\mathbf{R}^{(j)}\Bigg|\lesssim\int_{\mcr[\ubar_{(k)}]}\varepsilon_0\left(\Omega^{-2}+|\mathring{\alpha}|\right),
	\end{align*}
for some $u_0,u_1,\ubar_0,\ubar_{(k)}$ which satisfy $|\ubar'_0-\ubar_0|\lesssim 1,|\ubar'_{(k)}-\ubar_{(k)}|\lesssim 1,|u'_0-u|\lesssim 1,|u'_1-u_1|\lesssim 1$. Thus, by Corollary \ref{cor:corboundsharp} and the Cauchy-Schwarz inequality on the spheres $S(u,\ubar)$, we get
	\begin{align*}
		\Bigg|\int_{\mcr'[\ubar_{(k)}']}\delta\mathbf{R}^{(j)}\Bigg|&\lesssim\varepsilon_0\int_{u_0}^{u_1}\int_{\ubar_0}^{\ubar_{(k)}}\left(1+e^{-|\kappa_-|\ubar}\Ldeux{\mathring{\alpha}}\right)\dee u\dee\ubar\lesssim\varepsilon_0\left(\ubar_{(k)}'+\frac{e^{|\kappa_-|\ubar_{(k)}'}}{(\ubar_{(k)}')^6}\right),
	\end{align*}
where we used an integration by parts in $\ubar$ and $|\ubar_{(k)}'-\ubar_{(k)}|\lesssim 1$ in the last step. By Corollary \ref{cor:curvintblowup} we thus get for some constant $C$,
	\begin{align*}
		\left|\int_{\mcr'[\ubar'_{(k)}]}\sum_{j=1}^4 z_j\mathbf{R}(\widehat{X}_1^{(j)},\widehat{X}_2^{(j)},\widehat{X}_3^{(j)},\widehat{X}_4^{(j)})\right|\gtrsim\frac{e^{|\kappa_-|\ubar_{(k)}'}}{(\ubar_{(k)}')^6}(1-C\varepsilon_0)-\varepsilon_0\ubar'_{(k)} \underset{k\to+\infty}{\longrightarrow}+\infty,
	\end{align*}
where we used $\varepsilon_0>0$ small enough in the last step, which directly concludes the proof.
\end{proof}
The following Lipschitz inextendibility result, combined with Corollary \ref{cor:curvintblowup} and Proposition \ref{prop:curvintpertblowup}, concludes the proof of Theorem \ref{thm:inextensibilite}.
\begin{thm}\label{thm:lipinext}
	We assume that \eqref{eq:conditionQm2} holds. For $|u_f|\gg1$ large enough depending on $a,M,(Q_m)_{|m|\leq 2}$, the spacetime $\mcm$ is $C^{0,1}_{loc}$-inextendible across the Cauchy horizon $\ch$ in the sens of \cite[Theorem 3.14]{sbierskiinextdernier}: there is no $C^{0,1}_{loc}$-extension $\hat{\iota}:\mcm\rightarrow\hat{\mcm}$ with the property that for $t_0>0$ there is an affinely parametrised, future directed and future inextendible timelike geodesic $\tau:(-t_0,0)\rightarrow\mcm$ for $t_0>0$ with $\lim_{s\to 0}\ubar_\ch(\tau(s))=0$, $\lim_{s\to 0}u(\tau(s))<u_f$, and such that $\lim_{s\to 0}(\hat{\iota}\circ\tau)(s)\in\hat{\mcm}$ exists. 
\end{thm}
\begin{proof}
As proven in Section \ref{section:C0extension}, the assumptions of \cite[Prop. 3.6]{sbierskiinextdernier} hold by \eqref{eq:hypsbierski1}, and assumption \cite[(3.15)]{sbierskiinextdernier} holds by \eqref{eq:hypsbierski2}. Note however that Proposition \ref{prop:curvintpertblowup} does not exactly imply assumption \cite[(3.16)]{sbierskiinextdernier}. Indeed, the blow-up in Proposition \ref{prop:curvintpertblowup} holds on a spacetime region approaching $\ch$ which is not of the form $\mcr[\ubar_{(k)}]$, but which is of the form $\mcr'[\ubar_{(k)}']$ as defined in \eqref{eq:defregiondoublenull} where we recall that $u',\ubar'$ correspond to the Kerr values with respect to the double null coordinates $(u,\ubar,\theta^A)$ of the usual Eddington-Finkelstein $u=r^*-t$, 
	$\ubar=r^*+t$ in the Kerr black hole interior. We now show that this difference is irrelevant and that the conclusion (namely the Lipschitz inextendibility) is the same. 
	
The proof of \cite[Theo. 3.14]{sbierskiinextdernier} does not use the exact form of the spacetime region of integration which approaches $\ch$, except in the last step, where the assumption that there is a $C^{0,1}_{loc}$ extension implies boundedness of the sequence of spacetime curvature integrals by using an integration by parts, and we prove that this last step also holds for our spacetime regions of the form $\mcr'[\ubar_{(k)}']$ with almost same proof. More precisely, assuming the existence of the extension, we perform below the same integration by parts as in \cite{sbierskiinextdernier} and we deal with bulk terms and boundary terms as follows:
	\begin{itemize}
		\item The bulk terms are bounded exactly as in \cite{sbierskiinextdernier}, simply by the existence of the $C^{0,1}_{loc}$ extension.
		\item The boundary terms are bounded by computation and by using the estimates proven in the previous sections\footnote{Note that this step is different from \cite{sbierskiinextdernier} where the boundary terms are automatically bounded by existence of the extension. Here, our boundary terms are on $\ubar'=cst$ and $u'=cst$ hypersurfaces, and hence we prefer to directly bound the terms by computing them, instead of invoking some regularity arguments.}.
	\end{itemize}
	From now on, we use the same notations as in the proof of \cite[Theo. 3.14]{sbierskiinextdernier}, and we assume the existence of a $C^{0,1}_{loc}$ extension $\hat{\iota}:\mcm\rightarrow\hat{\mcm}$ as stated in Theorem \ref{thm:lipinext} around a boundary point $p$. We define the vectors $\overline{X}^{(j)}_i$ as in Corollary \ref{cor:curvintblowup}, which extend up to $\ch$ in the $C^0$ coordinates $(u,\ubar_\ch,\theta^A_\ch)$ by Proposition \ref{prop:explosionintegrale}, and by \eqref{eq:e4estC0} together with the fact that $\Omega^{-2}_\ch,b^A_\ch$ extend continuously to $\ch$ in the $(u,\ubar_\ch,\theta^A_\ch)$ coordinate system (see Theorem 16.14 in \cite{stabC0}).
We now denote 
	$$\wh{X}_i^{(j)},\quad i,j=1,2,3,4$$
	the $C^0$ vector fields constructed from the vectors $\overline{X}_i^{(j)}$ in Step 3 of the proof of \cite[Theo. 3.14]{sbierskiinextdernier}, which satisfy
	\begin{align*}
	\sum_{i=1}^4\sup_{\mcr'[+\infty]}|\overline{X}_i^{(j)}-\wh{X}_i^{(j)}|\leq\varepsilon_0,
\end{align*}
	where we choose $\varepsilon_0>0$ as in Proposition \ref{prop:curvintpertblowup}. More precisely, the construction is such that 
	$$\wh{X}_i^{(j)}=\left(\mathrm{id}|_{\tilde{W}}^{-1}\right)_*\hat{Y}_i^{(j)},$$
	where the vector fields $\hat{Y}_i^{(j)}$ are smooth on $\hat{W}$, which is an open neighborhood of the boundary point $p$ in the $C^{0,1}_{loc}$ extension $\hat{\mcm}$, and where $\tilde{W}$ is an open neighborhood of $p$ in the $C^0$ extension $\tilde{\mcm}$ in coordinates $(u,\ubar_\ch,\theta^A_\ch)$ ($\hat{W}$ and $\tilde{W}$ are precisely defined in Step 2 of the proof of \cite[Theo. 3.14]{sbierskiinextdernier}). In what follows, we denote by $C$ any large enough $k$-independent constant, and we fix some $\ubar'_0,u'_1,u'_0$ with $u'_1-u'_0>0$ sufficiently small and $\ubar'_0$ sufficiently large such that $\mcr'[\ubar'_1]$ as defined in \eqref{eq:defregiondoublenull} is included in $\hat{W}$ for any $\ubar'_1\geq \ubar'_0$. Now, by Proposition \ref{prop:curvintpertblowup}, there is $j\in\{1,2,3,4\}$ and a sequence $\ubar_{(k)}'\to+\infty$ such that, denoting $\widehat{X}_i:=\widehat{X}^{(j)}_i$ for $i=1,2,3,4$, 
	\begin{align}\label{eq:contrad}
		I_k:=\left|\int_{\mcr'[\ubar_{(k)}']}\mathbf{R}(\wh{X}_1,\wh{X}_2,\wh{X}_3,\wh{X}_4)\right|\underset{k\to+\infty}{\longrightarrow}+\infty,
	\end{align}
with $\mcr'[\ubar_{(k)}']$ defined as in \eqref{eq:defregiondoublenull}. However, denoting $\widehat{Y}_i:=\widehat{Y}^{(j)}_i$ and $\hat{\mathrm{vol}}$ the volume form such that $(\mathrm{id}|_{\tilde{W}})^*\hat{\mathrm{vol}}=\vol$, we have the following bounds (the term $+[1\leftrightarrow 2]$ corresponds to adding the term directly on the left of the $+$ sign where the indices $1$ and $2$ are switched):
	\begin{align*}
		I_k&=\left|\int_{\mathrm{id}(\mcr'[\ubar_{(k)}'])}\hat{\mathbf{R}}(\hat{Y}_1,\hat{Y}_2,\hat{Y}_3,\hat{Y}_4)\hat{\mathrm{vol}}\right|\\
		&=\left|\int_{\mathrm{id}(\mcr'[\ubar_{(k)}'])}\hat{\mathbf{g}}\left(\hat{\nabla}_{\hat{Y}_1}(\hat{\nabla}_{\hat{Y}_2}\hat{Y}_3)-\hat{\nabla}_{\hat{Y}_2}(\hat{\nabla}_{\hat{Y}_1}\hat{Y}_3)-\hat{\nabla}_{[\hat{Y}_1,\hat{Y}_2]}\hat{Y}_3,\hat{Y}_4\right)\hat{\mathrm{vol}}\right|\\
		&\leq C+\left|\int_{\mathrm{id}(\mcr'[\ubar_{(k)}'])}\left[\hat{Y}_1\left(\hat{\mathbf{g}}\left(\hat{\nabla}_{\hat{Y}_2}\hat{Y}_3,\hat{Y}_4\right)\right)-\hat{\mathbf{g}}\left(\hat{\nabla}_{\hat{Y}_2}\hat{Y}_3,\hat{\nabla}_{\hat{Y}_1}\hat{Y}_4\right)\right]\hat{\mathrm{vol}}\right|+[1\leftrightarrow 2]\\
		&\leq C+\left|\int_{\mathrm{id}(\mcr'[\ubar_{(k)}'])}\hat{Y}_1\left(\hat{\mathbf{g}}\left(\hat{\nabla}_{\hat{Y}_2}\hat{Y}_3,\hat{Y}_4\right)\right)\hat{\mathrm{vol}}\right|+[1\leftrightarrow 2].
	\end{align*}
Here, just like in \cite{sbierskiinextdernier} we used the smoothness of the vector fields $\hat{Y}_i$ and the definition of the curvature tensor in the second step. In the third step, we used the smoothness of $[\hat{Y}_1,\hat{Y}_2]$, $\hat{Y}_3$, $\hat{Y}_4$, the bounds \cite[(3.32),(3.33)]{sbierskiinextdernier}, and the compactness of $\overline{\hat{W}}$ to bound the last term, as well as the product rule to transform the first two terms. In the fourth step, we bounded the first order terms as before using smoothness of the $\hat{Y}_i$. Thus, using Cartan's formula, we deduce
	\begin{align*}
		I_k&\leq C+\left|\int_{\mathrm{id}(\mcr'[\ubar_{(k)}'])}\mathrm{d}\left(\hat{\mathbf{g}}\left(\hat{\nabla}_{\hat{Y}_2}\hat{Y}_3,\hat{Y}_4\right)\cdot \hat{Y}_1\lrcorner\hat{\mathrm{vol}}\right)-\hat{\mathbf{g}}\left(\hat{\nabla}_{\hat{Y}_2}\hat{Y}_3,\hat{Y}_4\right)\cdot\diver\hat{Y}_1\cdot\hat{\mathrm{vol}}\right|+[1\leftrightarrow 2]\\
		&\leq C+\left|\int_{\mathrm{id}(\mcr'[\ubar_{(k)}'])}\mathrm{d}\left(\hat{\mathbf{g}}\left(\hat{\nabla}_{\hat{Y}_2}\hat{Y}_3,\hat{Y}_4\right)\cdot \hat{Y}_1\lrcorner\hat{\mathrm{vol}}\right)\right|+[1\leftrightarrow 2],
	\end{align*}
	where we bounded the first order terms as before. Now, using Stokes theorem for manifolds with corners yields
	\begin{align*}
		I_k&\leq C+\left|\int_{\partial\mathrm{id}(\mcr'[\ubar_{(k)}'])}\hat{\mathbf{g}}\left(\hat{\nabla}_{\hat{Y}_2}\hat{Y}_3,\hat{Y}_4\right)\cdot \hat{Y}_1\lrcorner\hat{\mathrm{vol}}\right|+[1\leftrightarrow 2].
	\end{align*}
	Next, recalling the definition of $\mcr'[+\infty]=\trois\cap\{\ubar_0'\leq\ubar'\}\cap\{u_0'<u'<u_1'\}$, we have 
	\begin{align*}
		&\partial\mathrm{id}(\mcr'[\ubar_{(k)}'])\subset \mathrm{id}\left(\mcr'[+\infty]\cap\{\ubar'=\ubar'_{(k)}\}\right) \cup\mathcal{B
},\quad\text{where}\\
&\mathcal{B}=\mathrm{id}(\{u'=u_0'\}\cap\{\ubar'\geq\ubar_0'\})\cup\mathrm{id}(\{u'=u_1'\}\cap\{\ubar'\geq\ubar_0'\})\cup\mathrm{id}(\{\ubar'=\ubar_0'\}\cap\{u_0'\leq u'\leq u_1'\}),
	\end{align*}
	which splits the boundary $\partial\mathrm{id}(\mcr'[\ubar_{(k)}'])$ into a $k$-dependent part and a $k$-independent part. Now, $\hat{\mathbf{g}}\left(\hat{\nabla}_{\hat{Y}_{1,2}}\hat{Y}_3,\hat{Y}_4\right)$ is bounded on the compact $\mathrm{id}(\mcr'[+\infty])$ by smoothness of the $\hat{Y}_i$, so we deduce 
	\begin{align}\label{eq:finito0}
		I_k\leq C+\sum_{i=1,2}\left(\int_{\mathrm{id}\left(\mcr'[+\infty]\cap\{\ubar'=\ubar'_{(k)}\}\right)}|\hat{Y}_i\lrcorner\hat{\mathrm{vol}}|+\int_{\mathcal{B}}|\hat{Y}_i\lrcorner\hat{\mathrm{vol}}|\right).
	\end{align}
We now bound all the terms on the RHS above:

\noindent\textbf{Bound for the boundary term on $\{\ubar'=\ubar'_0\}$.} $\mathrm{id}(\{\ubar'=\ubar_0'\}\cap\{u_0'\leq u'\leq u_1'\})$ is clearly a compact $C^1$ hypersurface and $|\hat{Y}_i\lrcorner\hat{\mathrm{vol}}|$ is a continuous 3-density thus
\begin{align}\label{eq:finitopipau}
	\int_{\mathrm{id}(\{\ubar'=\ubar_0'\}\cap\{u_0'\leq u'\leq u_1'\})}|\hat{Y}_i\lrcorner\hat{\mathrm{vol}}|\leq C.
\end{align}
\noindent\textbf{Bound for the boundary terms on $\{u'=cst\}$.} Pulling back via id we get
\begin{align*}
	\int_{\mathrm{id}(\{u'=u_1'\}\cap\{\ubar'\geq\ubar_0'\})}|\hat{Y}_i\lrcorner\hat{\mathrm{vol}}|&=\int_{\{u'=u_1'\}\cap\{\ubar'\geq\ubar_0'\}}|\wh{X}_i\lrcorner\mathrm{vol}_{{\g}}|\\
	&=\int_{\{u'=u_1'\}\cap\{\ubar'\geq\ubar_0'\}}\sqrt{-\det(\mathbf{g}'_{\mu\nu})}|\wh{X}_i(u')|\dee \ubar'\dee\theta\dee\phi_-
\end{align*}
	where $\mathbf{g}'_{\mu\nu}$ is the matrix of $\g$ written in the coordinates $(y^\mu)=(u',\ubar',\theta,\phi_-)$. To compute $\sqrt{-\det(\mathbf{g}'_{\mu\nu})}$ we use the change of variables
\begin{align}\label{eq:equivalentexactkerr}
	\sqrt{-\det(\mathbf{g}'_{\mu\nu})}=\sqrt{-\det(\mathbf{g}_{\mu\nu})}\left|\det\left(\frac{\partial x^\nu}{\partial y^\mu}\right)_{\mu,\nu=1,2,3,4}\right|,
\end{align}
where the last factor on the RHS above is the Jacobian of the change of variables from the coordinates $(x^\nu)=(u,\ubar,\theta^A)$ to the coordinates $(y^\mu)$, and where, see \eqref{eq:volumeendoublenull},
$$\det(\mathbf{g}_{\mu\nu})=-4\Omega^4\det\gamma$$
is the determinant of the matrix of $\g$ in the coordinates $(x^\mu)$. As we defined the coordinates $(y^\mu)$ by their Kerr values with respect to the coordinates $(x^\mu)$, the Jacobian above is equal to its Kerr value with respect to $(u,\ubar,\theta^A)$, namely
\begin{align*}
	\left|\det\left(\frac{\partial x^\nu}{\partial y^\mu}\right)_{\mu,\nu=1,2,3,4}\right|=\left|\det\left(\frac{\partial x^\nu}{\partial y^\mu}\right)_{\mu,\nu=1,2,3,4}\right|_\mck.    
\end{align*}
Also, to compute the term on the RHS above, we can use the analog formula of \eqref{eq:equivalentexactkerr} in {exact} Kerr, which gives, using \eqref{eq:detgammaKsim} and \eqref{eq:kerrvolumeformexact}, 
$$\left|\det\left(\frac{\partial x^\nu}{\partial y^\mu}\right)_{\mu,\nu=1,2,3,4}\right|_\mck=\sqrt{\frac{\det(\mathbf{g}'_{\mu\nu})_\mck}{\det(\mathbf{g}_{\mu\nu})_\mck}}\sim\frac{|\Delta_\mck|\sin\theta}{\Omega^2_\mck\sin\theta}\sim 1.$$
Using also $|\det\gamma|\lesssim 1$ we deduce
\begin{align}\label{eq:finito2u}
	\int_{\mathrm{id}(\{u'=u_1'\}\cap\{\ubar'\geq\ubar_0'\})}|\hat{Y}_i\lrcorner\hat{\mathrm{vol}}|\leq C\int_{\{u'=u_1'\}\cap\{\ubar'\geq\ubar_0'\}} \Omega^2 |\wh{X}_i(u')| \dee \ubar'\dee\theta\dee\phi_-.
\end{align}
Now we bound the term $|\wh{X}_i(u')|$. From the estimate $|\wh{X}_i-\overline{X}_i^{(j)}|_h\lesssim \varepsilon_0,\: |\overline{X}_i^{(j)}|_h\lesssim 1$ we get 
\begin{align}\label{eq:getgetoui}
	\wh{X}_i=O(1)\ering_3+O(1)\ering_4+\wh{X}_i^S,
\end{align}
where $\wh{X}_i^S\in TS(u,\ubar)$ is such that $|\wh{X}_i^S|_\gamma\lesssim 1$. Note that we have $\ering_3(u')=(\partial_u u')_\mck,\nabring u'=(\nabring u')_\mck$,
where we get in exact Kerr, recalling the outgoing principal frame $(e'_\mu)_\mck$ and using $(e_4'({u'}))_\mck=0$ and $|e_3'({u'})|_\mck\lesssim1$, $|\nabla u'|_\mck\lesssim 1$, denoting $(\ering_a)_\mck$ a frame of $TS(u,\ubar)$ and $(e_a)_\mck$ the horizontal frame such that \eqref{eq:frametransfo} holds with coefficients $(\lambda_\mck,f_\mck,\fbar_\mck)$,
\begin{align*}
	|\partial_u{u'}|_\mck&\lesssim \left(|\g(\ering_3,e_4')||e_3'(u')|+\sum_{a=1,2}|\g(\ering_3,e_a')||e_a'({u'})|\right)_\mck\lesssim |f|^2_\mck+\sum_{a=1,2}|\fbar_a|_\mck\lesssim 1,\\
	|\nabring{u'}|_\mck&\lesssim \sum_{a=1,2}\left(|\g(\ering_a,e_4')||e_3'(u')|+\sum_{b=1,2}|\g(\ering_a,e_b')||e_b'({u'})|\right)_\mck\lesssim 1,
\end{align*}
hence $|\ering_3(u')|+|\nabring u'|\lesssim 1$. We also have $\partial_\ubar u'=(\partial_\ubar u')_\mck= \Omega^2_\mck(\ering_4(u'))_\mck-b_\mck^A\partial_{\theta^A}u'$ which implies, noting the cancellation of the terms $b_\mck^A$,
\begin{align*}
	\ering_4(u')&=\Omega^{-2}\left(\partial_\ubar u'+b^A\partial_{\theta^A}u'\right)=\Omega^{-2}\left(\Omega^2_\mck(\ering_4(u'))_\mck+\widecheck{b}^A\partial_{\theta^A}u'\right)=O(1)(\ering_4(u'))_\mck+O\left(\frac{\Omega^{-2}}{\ubar^{2+\delta/3}}\right),
\end{align*}
where we used the bound for $\bcheck$ in \eqref{eq:onlyused} and $|\nabring u'|\lesssim 1$. Moreover we have in Kerr
\begin{align*}
	|\ering_4(u')|_\mck\lesssim \left(|\g(\ering_4,e_4')||e_3'(u')|+\sum_{a=1,2}|\g(\ering_4,e_a)||e_a(u')|\right)_\mck\lesssim 1.
\end{align*}
From the bounds above combined with \eqref{eq:getgetoui} we get
$$|\wh{X}_i(u')|\lesssim 1+\frac{\Omega^{-2}}{\ubar^{2+\delta/3}},$$
which implies by \eqref{eq:finito2u},
\begin{align}\label{eq:finitopipopo}
	\int_{\mathrm{id}(\{u'=u_1'\}\cap\{\ubar'\geq\ubar_0'\})}|\hat{Y}_i\lrcorner\hat{\mathrm{vol}}|\leq C\int_{\ubar'_0}^{\infty}\left(e^{-|\kappa_-|\ubar'}+\frac{1}{\ubar'^{2+\delta/3}}\right)\dee\ubar'\leq C,
\end{align}
where we used $\Omega^2\sim\Delta_\mck\sim e^{-|\kappa_-|(u'+\ubar')}\sim e^{-|\kappa_-|\ubar'}$ in $\mcr'[+\infty]$. For the same reasons, the boundary term on $\{u'=u_1'\}\cap\{\ubar'\geq\ubar_0'\}$ is also bounded.
\\
\noindent\textbf{Bound for the boundary term on $\{\ubar'=\ubar'_{(k)}\}$.} Similarly as in \eqref{eq:finito2u} this term satisfies
\begin{align}
	\int_{\mathrm{id}\left(\mcr'[+\infty]\cap\{\ubar'=\ubar'_{(k)}\}\right)}|\hat{Y}_i\lrcorner\hat{\mathrm{vol}}|&=\int_{\mcr'[+\infty]\cap\{\ubar'=\ubar'_{(k)}\}}|\wh{X}_i\lrcorner\mathrm{vol}_{{\g}}|\nn\\
	&\leq C\int_{\mcr'[+\infty]\cap\{\ubar'=\ubar'_{(k)}\}} \Omega^2 |\wh{X}_i(\ubar')| \dee u'\dee\theta'\dee\phi'_-.\label{eq:finito2}
\end{align}
Now, we bound the term $|\wh{X}_i(\ubar')|$. We first estimate the derivatives $\partial_u\ubar'=(\partial_u\ubar')_\mck$, $\partial_\ubar\ubar'=(\partial_\ubar\ubar')_\mck$, $\nabring\ubar'=(\nabring\ubar')_\mck$, which all equal their Kerr values. We have, in exact Kerr, using $(e_3'(\ubar'))_\mck=0$, $|e_4'(\ubar')|_\mck\lesssim\Omega^{-2}_\mck$ and $|\nabla \ubar'|_\mck\lesssim 1$,
	\begin{align*}
		|\partial_u\ubar'|_\mck&\lesssim \left(|\g(\ering_3,e_3')||e_4'(\ubar')|+\sum_{a=1,2}|\g(\ering_3,e_a')||e_a'(\ubar')|\right)_\mck\lesssim |\fbar|_\mck^2\Omega^{-2}_\mck+\sum_{a=1,2}|\fbar_a|_\mck\lesssim\Omega_\mck^2,\\
		|\nabring\ubar'|_\mck&\lesssim \sum_{a=1,2}\left(|\g(\ering_a,e_3')||e_4'(\ubar')|+\sum_{b=1,2}|\g(\ering_a,e_b')||e_b'(\ubar')|\right)_\mck\lesssim |\fbar|_\mck\Omega^{-2}_\mck+1\lesssim 1,\\
		|\partial_\ubar\ubar'|_\mck&\lesssim \Omega^2_\mck|\ering_4(\ubar')|_\mck+|b|_\mck|\nabring\ubar'|_\mck\lesssim \Omega^2_\mck\left(|\g(\ering_4,e_3')||e_4'(\ubar')|+\sum_{a=1,2}|\g(\ering_4,e_a')||e_a'(\ubar')|\right)_\mck+1\\
		&\lesssim (1+|f|_\mck|\fbar|_\mck+|f|_\mck^2|\fbar|_\mck^2)+1\lesssim 1.
	\end{align*}
	This implies in $\trois$,
	\begin{align*}
		|\ering_3(\ubar')|\lesssim\Omega^2_\mck\lesssim 1,\quad |\nabring\ubar'|\lesssim 1,\quad|\ering_4(\ubar')|\lesssim \Omega^{-2}(|\partial_\ubar\ubar'|+|b||\nabring\ubar'|)\lesssim\Omega^{-2}.
	\end{align*}
	Now, recalling \eqref{eq:getgetoui} we deduce $|\wh{X}_i(\ubar')|\lesssim \Omega^{-2}$. Thus, by \eqref{eq:finito2} we infer
\begin{align}\label{eq:finitopipopopopopo}
	\int_{\mathrm{id}\left(\mcr'[+\infty]\cap\{\ubar'=\ubar'_{(k)}\}\right)}|\hat{Y}_i\lrcorner\hat{\mathrm{vol}}|\leq C\int_{\mcr'[+\infty]\cap\{\ubar'=\ubar'_{(k)}\}} \dee u'\dee\theta'\dee\phi'_-\leq C(u_1'-u_0')\leq C.
\end{align}
Combining \eqref{eq:finito0}, \eqref{eq:finitopipau}, \eqref{eq:finitopipopo} and \eqref{eq:finitopipopopopopo} we finally get $I_k\leq C$, which is in contradiction with \eqref{eq:contrad}, and concludes the proof of Theorem \ref{thm:lipinext}.
\end{proof}

\appendix
\section{General computations}
\subsection{Derivation of the Teukolsky equation}\label{appendix:teukderivation}

\begin{proof}[Proof of Proposition \ref{prop:teuk}]
We apply $\nabla_4$ to the Bianchi identity for $\nabla_4A$ in Proposition \ref{prop:bianchicomplex}:
	\begin{align}
		\nabla_4\nabla_3 A &=\frac{1}{2}\parentheses{\nabla_4\left(\mathcal{D} \widehat{\otimes} B\right)+(Z+4 H) \widehat{\otimes} \nabla_4B}+\nabla_4\parentheses{-\frac{1}{2}tr \underline{X}+4\omegabar} A\nn\\
		&\quad+\parentheses{-\frac{1}{2}tr \underline{X}+4\omegabar} \nabla_4A+\frac{1}{2}(\nabla_4(Z+4 H))\hot B-3 (\nabla_4\overline{P}) \widehat{X}-3\overline{P}\nabla_4\wh{X}.\label{eq:teuk00}
	\end{align}
First, using the null structure equation for $\nabla_4tr\Xbar$ in Proposition \ref{prop:nullstructurecomplex} we have
\begin{align}
	\nabla_4&\parentheses{-\frac{1}{2}tr \underline{X}+4\omegabar} A+\parentheses{-\frac{1}{2}tr \underline{X}+4\omegabar} \nabla_4A\nn\\
	&=\parentheses{\frac{1}{4}tr X tr\Xbar-\omega tr\Xbar-\frac{1}{2}\parentheses{\mcd\cdot\overline{\Hbar}+\Hbar\cdot\overline{\Hbar}}-\overline{P}+4\nabla_4\omegabar}A+\parentheses{-\frac{1}{2}tr \underline{X}+4\omegabar} \nabla_4A\nn\\
	&\quad-\frac{1}{2}\parentheses{\Xi\cdot\overline{\Xibar}-\frac{1}{2}\wh{X}\cdot\overline{\wh{\Xbar}}}A.\label{eq:teuk2}
\end{align}
Next, in view of Proposition \ref{prop:commnab34mcdhot} we get 
	\begin{align*}
		\nabla_4&\left(\mathcal{D} \widehat{\otimes} B\right) +(Z+4H)\hot \nabla_4 B=[\nabla_4,\mcd\hot]B+\mcd\hot\nabla_4 B+(Z+4H)\hot \nabla_4 B\nn\\
		&=\mcd\hot\nabla_4 B+(2Z+4H+\Hbar)\hot \nabla_4 B-\frac{1}{2} tr X(\mathcal{D} \widehat{\otimes} B+\underline{H} \widehat{\otimes} B)+\Xi \widehat{\otimes} \nabla_3 B \nn\\
		&\quad-B \widehat{\otimes} B-\frac{1}{2}tr \Xbar\Xi \hot B-\frac{1}{2}\widehat{X} \cdot \overline{\mathcal{D}} B+\frac{1}{2}\widehat{X}(\overline{\Hbar} \cdot B)+\frac{1}{2}\underline{\widehat{X}}(\overline{\Xi} \cdot B)+\frac{1}{2}\Xi\hot(\overline{\wh{\Xbar}}\cdot B).
	\end{align*}
	Reinjecting the following Bianchi identity for $\nabla_4B$ in Proposition \ref{prop:bianchicomplex}, this yields
	\begin{align}
		\nabla_4&\left(\mathcal{D} \widehat{\otimes} B\right) +(Z+4H)\hot \nabla_4 B\nn\\
		&=\frac{1}{2}\mcd\hot\parentheses{\overline{\mcd}\cdot A+(\overline{2Z+\Hbar})\cdot A}+\parentheses{-\frac{1}{2}tr X-2\overline{tr X}-2\omega}\mcd\hot B\nn\\
		&\quad+\parentheses{Z+2H+\frac{1}{2}\Hbar}\hot\parentheses{\overline{\mcd}\cdot A+ (\overline{2 Z+\underline{H}})\cdot A}\nn\\
		&\quad+2\parentheses{-\mcd(\overline{tr X}+\omega)-(\overline{tr X}+\omega)(2Z+4H+\Hbar)-\frac{1}{4}tr X\Hbar}\hot B\nn\\
		&\quad+3\overline{P}\parentheses{\mcd\hot\Xi+(2Z+4H+\Hbar)\hot\Xi}+3\mcd(\overline{P})\hot\Xi+\Xi \widehat{\otimes} \nabla_3 B \nn\\
		&\quad-B \widehat{\otimes} B-\frac{1}{2}tr \Xbar\Xi \hot B-\frac{1}{2}\widehat{X} \cdot \overline{\mathcal{D}} B+\frac{1}{2}\widehat{X}(\overline{\Hbar} \cdot B)+\frac{1}{2}\underline{\widehat{X}}(\overline{\Xi} \cdot B)+\frac{1}{2}\Xi\hot(\overline{\wh{\Xbar}}\cdot B).\label{eq:teuk1}
	\end{align}

Also, by the null structure equations for $\nabla_4H$ and $\nabla_4Z$  in Proposition \ref{prop:nullstructurecomplex},
	\begin{align}
		&\nabla_4(Z+4H)\hot B\nn\\
		&=\parentheses{-2\overline{tr X}(H-\Hbar)-\frac{1}{2}tr X(Z-\Hbar)+2\omega(Z+\Hbar)+2\mcd(\omega)}\hot B\label{eq:teuk3}\\
		&\:\: +\parentheses{4\nabla_3\Xi-2\wh{X}\cdot(\overline{H}-\overline{\Hbar})-16\omegabar\Xi-5B+\frac{1}{2}\wh{X}\cdot(-\overline{Z}+\overline{\Hbar})-\frac{1}{2}tr\Xbar\Xi-2\omegabar\Xi-\frac{1}{2}\overline{\Xi}\wh{\Xbar}}\hot B,\nn
	\end{align}
and by the null structure equation for $\nabla_4\wh{X}$ and Bianchi equation for $\nabla_4P$,
	\begin{align}
		-3 &(\nabla_4\overline{P}) \widehat{X}-3\overline{P}\nabla_4\wh{X}\nn\\
		&=3\parentheses{\frac{3}{2} \overline{tr X} \overline{P}}\wh{X}-3\overline{P}\parentheses{-\Real(tr X)\wh{X}-2\omega\wh{X}+\frac{1}{2}\mcd\hot\Xi+\frac{1}{2}\Xi\hot(\Hbar+H+2Z)-A}\nn\\
		&\quad-3\parentheses{\frac{1}{2} \overline{\mathcal{D}} \cdot {B}+\frac{1}{2}(2 \overline{\underline{H}+Z}) \cdot {B}-{\Xi} \cdot \overline{\underline{B}}}\wh{X}+\frac{3}{4}\parentheses{\overline{\wh{\Xbar}}\cdot A}\wh{X}\nn\\
		&=\parentheses{\frac{3}{2}tr X+6\overline{tr X}+6\omega}\overline{P}\wh{X}+3\overline{P}A-\frac{3}{2}\overline{P}\parentheses{\mcd\hot\Xi+\Xi\hot(\Hbar+H+2Z)}\nn\\
		&\quad-3\parentheses{\frac{1}{2} \overline{\mathcal{D}} \cdot {B}+\frac{1}{2}(2 \overline{\underline{H}+Z}) \cdot {B}-{\Xi} \cdot \overline{\underline{B}}}\wh{X}+\frac{3}{4}\parentheses{\overline{\wh{\Xbar}}\cdot A}\wh{X}.\label{eq:teuk4}
	\end{align}
	Thus the identity \eqref{eq:teuk00} rewrites, summing $1/2$\eqref{eq:teuk1}, \eqref{eq:teuk2}, $1/2$\eqref{eq:teuk3}, \eqref{eq:teuk4},
	\begin{align}
		\nabla_4\nabla_3A&=\parentheses{-\frac{1}{2}tr X-2\overline{tr X}-2\omega}\frac{1}{2}\mcd\hot B+\parentheses{\frac{3}{2}tr X+6\overline{tr X}+6\omega}\overline{P}\wh{X}\label{eq:teuk5}\\
		&+\Bigg(-\mcd(\overline{tr X}+\omega)-(\overline{tr X}+\omega)(2Z+4H+\Hbar)-\frac{1}{4}tr X\Hbar\nn\\
		&-\overline{tr X}(H-\Hbar)-\frac{1}{4}tr X(Z-\Hbar)+\omega(Z+\Hbar)+\mcd(\omega)\Bigg)\hot B\nn\\
		&+\parentheses{\frac{1}{4}tr X tr\Xbar-\omega tr\Xbar-\frac{1}{2}\parentheses{\mcd\cdot\overline{\Hbar}+\Hbar\cdot\overline{\Hbar}}+2\overline{P}+4\nabla_4\omegabar}A+\parentheses{-\frac{1}{2}tr \underline{X}+4\omegabar} \nabla_4A\nn\\
		&+\frac{1}{4}\mcd\hot\parentheses{\overline{\mcd}\cdot A+(\overline{2Z+\Hbar})\cdot A}+\parentheses{\frac{1}{2}Z+H+\frac{1}{4}\Hbar}\hot\parentheses{\overline{\mcd}\cdot A+ (\overline{2 Z+\underline{H}})\cdot A}\nn\\
		&+\frac{3}{2}\overline{P}\parentheses{(2Z+4H+\Hbar)\hot\Xi}+\frac{3}{2}\mcd(\overline{P})\hot\Xi+\frac{1}{2}\Xi \widehat{\otimes} \nabla_3 B-\frac{3}{2}\overline{P}\parentheses{\Xi\hot(\Hbar+H+2Z)}+\err_0,\nn
	\end{align}
where we defined
\begin{align}
	&\err_0\nn:=\\
	&\frac{1}{2}\parentheses{-B \widehat{\otimes} B-\frac{1}{2}tr \Xbar\Xi \hot B-\frac{1}{2}\widehat{X} \cdot \overline{\mathcal{D}} B+\frac{1}{2}\widehat{X}(\overline{\Hbar} \cdot B)+\frac{1}{2}\underline{\widehat{X}}(\overline{\Xi} \cdot B)+\frac{1}{2}\Xi\hot(\overline{\wh{\Xbar}}\cdot B)}\nn\\
	&-\frac{1}{2}\parentheses{\Xi\cdot\overline{\Xibar}-\frac{1}{2}\wh{X}\cdot\overline{\wh{\Xbar}}}A-3\parentheses{\frac{1}{2} \overline{\mathcal{D}} \cdot {B}+\frac{1}{2}(2 \overline{\underline{H}+Z}) \cdot {B}-{\Xi} \cdot \overline{\underline{B}}}\wh{X}+\frac{3}{4}\parentheses{\overline{\wh{\Xbar}}\cdot A}\wh{X}\nn\\
	&+\frac{1}{2}\parentheses{4\nabla_3\Xi-2\wh{X}\cdot(\overline{H}-\overline{\Hbar})-16\omegabar\Xi-5B+\frac{1}{2}\wh{X}\cdot(-\overline{Z}+\overline{\Hbar})-\frac{1}{2}tr\Xbar\Xi-2\omegabar\Xi-\frac{1}{2}\overline{\Xi}\wh{\Xbar}}\hot B.\nn
\end{align}

	We now simplify \eqref{eq:teuk5}. The 6th line of \eqref{eq:teuk5} is 
	\begin{align}
		&\frac{3}{2}\overline{P}\parentheses{(2Z+4H+\Hbar)\hot\Xi}+\frac{3}{2}\mcd(\overline{P})\hot\Xi+\frac{1}{2}\Xi \widehat{\otimes} \nabla_3 B-\frac{3}{2}\overline{P}\parentheses{\Xi\hot(\Hbar+H+2Z)}\label{eq:teuk7}\\
		&=\frac{1}{2}\Xi \widehat{\otimes} \nabla_3 B+\frac{3}{2}\parentheses{\mcd(\overline{P})+3\overline{P}H}\hot\Xi=2\Xi \widehat{\otimes} \nabla_3 B+\frac{3}{2}\parentheses{tr \underline{X} B-2 \underline{\omega} B-\underline{\overline{B}} \cdot \widehat{X}-\frac{1}{2} A \cdot \overline{\underline{\Xi}}}\hot\Xi\nn
	\end{align}
where in the last step we used the Bianchi identity for $\nabla_3B$. We now simplify the second and third ligns of \eqref{eq:teuk5}. We have 
	\begin{align}
		&-\mcd(\overline{tr X}+\omega)-(\overline{tr X}+\omega)(2Z+4H+\Hbar)-\frac{1}{4}tr X\Hbar-\overline{tr X}(H-\Hbar)-\frac{1}{4}tr X(Z-\Hbar)\nn\\
		&+\omega(Z+\Hbar)+\mcd(\omega) =-\mcd(\overline{tr X})-5\overline{tr X} H-2Z\overline{tr X}-\frac{1}{4}Ztr X-\omega(Z+4H).\label{eq:teuk8}
	\end{align}
	Combining \eqref{eq:teuk5}, \eqref{eq:teuk7} and \eqref{eq:teuk8} yields
	\begin{align}
		\nabla_4\nabla_3A&=\parentheses{-\frac{1}{2}tr X-2\overline{tr X}-2\omega}\frac{1}{2}\mcd\hot B+\parentheses{\frac{3}{2}tr X+6\overline{tr X}+6\omega}\overline{P}\wh{X}\nn\\
		&+\left(-\mcd(\overline{tr X})-5\overline{tr X} H-2Z\overline{tr X}-\frac{1}{4}Ztr X-\omega(Z+4H)\right)\hot B\nn\\
		&+\parentheses{\frac{1}{4}tr X tr\Xbar-\omega tr\Xbar-\frac{1}{2}\parentheses{\mcd\cdot\overline{\Hbar}+\Hbar\cdot\overline{\Hbar}}+2\overline{P}+4\nabla_4\omegabar}A+\parentheses{-\frac{1}{2}tr \underline{X}+4\omegabar} \nabla_4A\nn\\
		&+\frac{1}{4}\mcd\hot\parentheses{\overline{\mcd}\cdot A+(\overline{2Z+\Hbar})\cdot A}+\parentheses{\frac{1}{2}Z+H+\frac{1}{4}\Hbar}\hot\parentheses{\overline{\mcd}\cdot A+ (\overline{2 Z+\underline{H}})\cdot A}\nn\\
		&+\err_1,\label{eq:teuk9}
	\end{align}
	where we defined
	\begin{align}
		\err_1:=&2\parentheses{\Xi \widehat{\otimes} \nabla_3 B+\nabla_3\Xi\hot B}+\frac{3}{2}\parentheses{tr \underline{X} B-2 \underline{\omega} B-\underline{\overline{B}} \cdot \widehat{X}-\frac{1}{2} A \cdot \overline{\underline{\Xi}}}\hot\Xi\nn\\
		&+\frac{1}{2}\parentheses{-B \widehat{\otimes} B-\frac{1}{2}tr \Xbar\Xi \hot B-\frac{1}{2}\widehat{X} \cdot \overline{\mathcal{D}} B+\frac{1}{2}\widehat{X}(\overline{\Hbar} \cdot B)+\frac{1}{2}\underline{\widehat{X}}(\overline{\Xi} \cdot B)+\frac{1}{2}\Xi\hot(\overline{\wh{\Xbar}}\cdot B)}\nn\\
		&-\frac{1}{2}\parentheses{\Xi\cdot\overline{\Xibar}-\frac{1}{2}\wh{X}\cdot\overline{\wh{\Xbar}}}A-3\parentheses{\frac{1}{2} \overline{\mathcal{D}} \cdot {B}+\frac{1}{2}(2 \overline{\underline{H}+Z}) \cdot {B}-{\Xi} \cdot \overline{\underline{B}}}\wh{X}+\frac{3}{4}\parentheses{\overline{\wh{\Xbar}}\cdot A}\wh{X}\nn\\
		&+\frac{1}{2}\parentheses{-2\wh{X}\cdot(\overline{H}-\overline{\Hbar})-16\omegabar\Xi-5B+\frac{1}{2}\wh{X}\cdot(-\overline{Z}+\overline{\Hbar})-\frac{1}{2}tr\Xbar\Xi-2\omegabar\Xi-\frac{1}{2}\overline{\Xi}\wh{\Xbar}}\hot B.\nn
	\end{align}
	We continue by simplifying the first line of \eqref{eq:teuk9}. By Proposition \ref{prop:bianchicomplex} again,
	\begin{align}
		\parentheses{-\frac{1}{2}tr X-2\overline{tr X}-2\omega}&\frac{1}{2}\mcd\hot B+\parentheses{\frac{3}{2}tr X+6\overline{tr X}+6\omega}\overline{P}\wh{X}\nn\\
		&=\parentheses{-\frac{1}{2}tr X-2\overline{tr X}-2\omega}\parentheses{\frac{1}{2}\mcd\hot B-3\overline{P}\wh{X}}\nn\\
		&=\parentheses{-\frac{1}{2}tr X-2\overline{tr X}-2\omega}\parentheses{\nabla_3 A+\frac{1}{2} tr \underline{X} A-4 \underline{\omega} A-\frac{1}{2}(Z+4 H) \widehat{\otimes} B}.\nn
	\end{align}
	This yields 
	\begin{align}
		\nabla_4\nabla_3A&=\parentheses{-\frac{1}{2}tr X-2\overline{tr X}-2\omega}\nabla_3 A+\parentheses{-\frac{1}{2}tr \underline{X}+4\omegabar} \nabla_4A\label{eq:teuk10}\\
		&+\Bigg(-\mcd(\overline{tr X})+(tr X-\overline{tr X}) H-Z\overline{tr X}\Bigg)\hot B\nn\\
		&+\Bigg(-\overline{tr X}tr\Xbar+2\omegabar trX-2\omega tr\Xbar+8\omegabar(\omega+\overline{tr X})-\frac{1}{2}\parentheses{\mcd\cdot\overline{\Hbar}+\Hbar\cdot\overline{\Hbar}}+2\overline{P}+4\nabla_4\omegabar\Bigg)A\nn\\
		&+\frac{1}{4}\mcd\hot\parentheses{\overline{\mcd}\cdot A+(\overline{2Z+\Hbar})\cdot A}+\parentheses{\frac{1}{2}Z+H+\frac{1}{4}\Hbar}\hot\parentheses{\overline{\mcd}\cdot A+ (\overline{2 Z+\underline{H}})\cdot A}+\err_1.\nn
	\end{align}
	We now simplify the second line of \eqref{eq:teuk10}. By the null structure equation for $\divc\wh{X}$, we get 
	\begin{align}
		\nabla_4\nabla_3A&=\parentheses{-\frac{1}{2}tr X-2\overline{tr X}-2\omega}\nabla_3 A+\parentheses{-\frac{1}{2}tr \underline{X}+4\omegabar} \nabla_4A\label{eq:teuk11}\\
		&+\Bigg(-\overline{tr X}tr\Xbar+2\omegabar trX-2\omega tr\Xbar+8\omegabar(\omega+\overline{tr X})-\frac{1}{2}\parentheses{\mcd\cdot\overline{\Hbar}+\Hbar\cdot\overline{\Hbar}}+2\overline{P}+4\nabla_4\omegabar\Bigg)A\nn\\
		&+\frac{1}{4}\mcd\hot\parentheses{\overline{\mcd}\cdot A+(\overline{2Z+\Hbar})\cdot A}+\parentheses{\frac{1}{2}Z+H+\frac{1}{4}\Hbar}\hot\parentheses{\overline{\mcd}\cdot A+ (\overline{2 Z+\underline{H}})\cdot A}+\err[\mcl(A)],\nn
	\end{align}
	where, using $\widehat{X} \cdot \overline{\mathcal{D}} B=(\divc B)\wh{X}$ by Lemma \ref{lem:leibnizchilll}, $\err[\mcl(A)]$ coincides with \eqref{eq:teukerror}. We now further simplify the third line of \eqref{eq:teuk11}:
	\begin{align}
		\frac{1}{4}&\mcd\hot\parentheses{\overline{\mcd}\cdot A+(\overline{2Z+\Hbar})\cdot A}+\parentheses{\frac{1}{2}Z+H+\frac{1}{4}\Hbar}\hot\parentheses{\overline{\mcd}\cdot A+ (\overline{2 Z+\underline{H}})\cdot A}\nn\\
		&=\frac{1}{4}\mcd\hot\parentheses{\overline{\mcd}\cdot A}+\frac{1}{4}\mcd\hot\parentheses{(\overline{2Z+\Hbar})\cdot A}+\parentheses{\frac{1}{2}Z+H+\frac{1}{4}\Hbar}\hot\parentheses{\overline{\mcd}\cdot A}\nn\\
		&\quad+ \parentheses{\frac{1}{2}Z+H+\frac{1}{4}\Hbar}\hot\parentheses{(\overline{2 Z+\underline{H}})\cdot A}\nn\\
		&=\frac{1}{4}\mcd\hot\parentheses{\overline{\mcd}\cdot A}+\frac{1}{2}(\mcd\cdot(\overline{2Z+\Hbar}))A+\frac{1}{2}((\overline{2Z+\Hbar})\cdot\mcd)A+\parentheses{\frac{1}{2}Z+H+\frac{1}{4}\Hbar}\hot\parentheses{\overline{\mcd}\cdot A}\nn\\
		&\quad+\frac{1}{2}\parentheses{\parentheses{2Z+\Hbar}\cdot\parentheses{\overline{2Z+\Hbar}}}A+ H\hot\parentheses{(\overline{2 Z+\underline{H}})\cdot A}.\label{eq:anginject}
	\end{align}
	where we used Lemmas \ref{lem:monamitoitcool} and \ref{lem:nouveauleibniz} which yield
	\begin{align*}
		&\mcd\hot\parentheses{(\overline{2Z+\Hbar})\cdot A}=2(\mcd\cdot(\overline{2Z+\Hbar}))A+2((\overline{2Z+\Hbar})\cdot\mcd)A,\\
		&\parentheses{2Z+\Hbar}\hot\parentheses{(\overline{2 Z+\underline{H}})\cdot A}=2\parentheses{\parentheses{2Z+\Hbar}\cdot\parentheses{\overline{2Z+\Hbar}}}A.
	\end{align*}
	Reinjecting \eqref{eq:anginject} in \eqref{eq:teuk11} yields
	\begin{align}
		\nabla_4\nabla_3A&=\frac{1}{4}\mcd\hot\parentheses{\overline{\mcd}\cdot A}+\parentheses{-\frac{1}{2}tr X-2\overline{tr X}-2\omega}\nabla_3 A+\parentheses{-\frac{1}{2}tr \underline{X}+4\omegabar} \nabla_4A\label{eq:teuk12}\\
		&+\Big(-\overline{tr X}tr\Xbar+2\omegabar trX-2\omega tr\Xbar+8\omegabar(\omega+\overline{tr X})\nn\\
		&\quad\quad+\mcd\cdot\overline{Z}+2\parentheses{Z\cdot\overline{Z}+\Real(Z\cdot\overline{\Hbar})}+2\overline{P}+4\nabla_4\omegabar\Big)A\nn\\
		&+\frac{1}{2}((\overline{2Z+\Hbar})\cdot\mcd)A+\parentheses{\frac{1}{2}Z+H+\frac{1}{4}\Hbar}\hot\parentheses{\overline{\mcd}\cdot A}+ H\hot\parentheses{(\overline{2 Z+\underline{H}})\cdot A}+\err[\mcl(A)].\nn
	\end{align}
	We finally simplify the third line of \eqref{eq:teuk12} by Lemmas \ref{lem:leibnizchilll} and \ref{lem:nouveauleibniz} twice which yield
	\begin{align*}
		\frac{1}{2}((\overline{2Z+\Hbar})\cdot\mcd)A+\parentheses{\frac{1}{2}Z+H+\frac{1}{4}\Hbar}\hot\parentheses{\overline{\mcd}\cdot A}=\parentheses{2Z+\Hbar+\overline{2Z+\Hbar}+4H}\cdot\nabla A,
	\end{align*}
thereby concluding the proof of Proposition \ref{prop:teuk}.
\end{proof}

\subsection{Computations related to frame transformations}
We assume in this subsection that we are in the setting described in Section \ref{section:derivativechangeofframe}.
\begin{prop}\label{prop:diffchristo}
	Let $U$ be a $\mch$-horizontal $k$-tensor. Then for $b_1,\ldots,b_k=1,2$ we have
	$$\nabla'_{e_a}(\Phi_*U)_{b_1\cdots b_k}=\nabla_{a}U_{b_1\cdots b_k}+\sum_{i=1}^k\mathcal{T}_{ab_ic}U_{b_1\cdots c\cdots b_k},$$
	where the $\mch$-horizontal tensor $\mathcal{T}_{abc}$ is defined by
	\begin{align*}
		\mathcal{T}_{abc}=\frac{1}{2}\fbar_{[c}\nabla_af_{b]}+\frac{1}{2}\fbar_{[c}\chi_{ab]}+\left(\frac{1}{2}f_{[c}+\frac{1}{8}|f|^2\fbar_{[c}\right)\chibar_{ab]}+\frac{1}{2}f_d\fbar_{[b}f_{c]}\chibar_{ad}+\frac{1}{2}f_{[b}\fbar_{c]}\eta_a.
	\end{align*}
\end{prop}
\begin{proof}
	We treat the case of a horizontal 1-form as the argument can be trivially extended for horizontal $k$-tensors with $k\geq 1$. First, we have by definition of the horizontal covariant derivatives $\nabla,\nabla'$,
	\begin{align*}
		\nabla'_{e_a}(\Phi_*U)_b=e_a((\Phi_*U)_b)-(\Phi_*U)_c\g(\D_{e_a}e_b',e_c')&=e_a(U_b)-U_{c}\g(\D_{e_a}e_b',e_c')\\
		&=\nabla_{a}U_b+U_c\left(\g(\D_{e_a}e_b,e_c)-\g(\D_{e_a}e_b',e_c')\right).
	\end{align*}
	Moreover, by \eqref{eq:frametransfo} we have
	\begin{align*}
		\g(\D_{e_a}e_b',e_c')=\g&(\D_{e_a}e_b',e_c)+\frac{1}{2}\fbar_c f_d\g(\D_{e_a}e_b',e_d)+\frac{1}{2}\fbar_c\g(\D_{e_a}e_b',e_4)\\
		&+\left(\frac{1}{2}f_c+\frac{1}{8}|f|^2\fbar_c\right)\g(\D_{e_a}e_b',e_3).
	\end{align*}
	We first compute
	\begin{align*}
		\g(\D_{e_a}e_b',e_c)&=\g(\D_{e_a}e_b,e_c)+\frac{1}{2}\fbar_b f_d\g(\D_{e_a}e_d,e_c)+\frac{1}{2}f_ce_a(\fbar_b )+\frac{1}{2}\fbar_be_a(f_c)\\
		&\quad+\frac{1}{2}\fbar_b\chi_{ac}+\left(\frac{1}{2}f_b+\frac{1}{8}|f|^2\fbar_b\right)\chibar_{ac},
	\end{align*}
which also implies
	\begin{align*}
		\frac{1}{2}\fbar_c f_d\g(\D_{e_a}e_b',e_d)&=\frac{1}{2}\fbar_c f_d\g(\D_{e_a}e_b,e_d)+\frac{1}{4}\fbar_c |f|^2e_a(\fbar_b )+\frac{1}{8}\fbar_c\fbar_be_a(|f|^2)\\
		&\quad+\frac{1}{4}\fbar_c f_d\fbar_b\chi_{ad}+\frac{1}{2}\fbar_c f_d\left(\frac{1}{2}f_b+\frac{1}{8}|f|^2\fbar_b\right)\chibar_{ad},
	\end{align*}
	where we used $f_df_g\g(\D_{e_a}e_g,e_d)=0$ by antisymetry of $\g(\D_{e_a}e_g,e_d)$ in $d,g$, and $f_d e_a(f_d)=e_a(|f|^2)/2$. We now compute successively
	\begin{align*}
		\frac{1}{2}\fbar_c\g(\D_{e_a}e_b',e_4)&=\frac{1}{2}\fbar_c\g(\D_{e_a}e_b,e_4)+\frac{1}{4}\fbar_c\fbar_bf_d\g(\D_{e_a}e_d,e_4)+\frac{1}{4}\fbar_c\fbar_b\g(\D_{e_a}e_4,e_4)\\
		&\quad+\frac{1}{2}\fbar_c\left(\frac{1}{2}f_b+\frac{1}{8}|f|^2\fbar_b\right)\g(\D_{e_a}e_3,e_4)-\fbar_c\left(\frac{1}{2}e_a(f_b)+\frac{1}{8}e_a(|f|^2\fbar_b)\right)\\
		&=-\frac{1}{2}\fbar_c\chi_{ab}-\frac{1}{4}\fbar_c\fbar_b f_d\chi_{ad}-\fbar_c\left(\frac{1}{2}f_b+\frac{1}{8}|f|^2\fbar_b\right)\eta_a\\
		&\quad-\fbar_c\left(\frac{1}{2}e_a(f_b)+\frac{1}{8}e_a(|f|^2\fbar_b)\right),
	\end{align*}
	and \begin{align*}
		\left(\frac{1}{2}f_c+\frac{1}{8}|f|^2\fbar_c\right)\g(\D_{e_a}e_b',e_3)&=-\left(\frac{1}{2}f_c+\frac{1}{8}|f|^2\fbar_c\right)\chibar_{ab}-\fbar_b f_d\left(\frac{1}{2}f_c+\frac{1}{8}|f|^2\fbar_c\right)\chibar_{ad}\\
		&\quad+\left(\frac{1}{2}f_c+\frac{1}{8}|f|^2\fbar_c\right)\fbar_b\eta_a-\left(\frac{1}{2}f_c+\frac{1}{8}|f|^2\fbar_c\right)e_a(\fbar_b).
	\end{align*}
	Putting it all together yields
	\begin{align*}
		\g(\D_{e_a}e_b',e_c')=&\g(\D_{e_a}e_b,e_c)+\frac{1}{2}\fbar_b\nabla_af_c-\frac{1}{2}\fbar_c\nabla_af_b+\frac{1}{2}\fbar_b\chi_{ac}+\left(\frac{1}{2}f_b+\frac{1}{8}|f|^2\fbar_b\right)\chibar_{ac}\\
		&+\frac{1}{4}\fbar_c f_d\fbar_b\chi_{ad}+\frac{1}{2}\fbar_c f_d\left(\frac{1}{2}f_b+\frac{1}{8}|f|^2\fbar_b\right)\chibar_{ad}-\frac{1}{2}\fbar_c\chi_{ab}-\frac{1}{4}\fbar_c\fbar_b f_d\chi_{ad}\\
		&-\fbar_c\left(\frac{1}{2}f_b+\frac{1}{8}|f|^2\fbar_b\right)\eta_a-\left(\frac{1}{2}f_c+\frac{1}{8}|f|^2\fbar_c\right)\chibar_{ab}-\frac{1}{2}\fbar_b f_d\left(\frac{1}{2}f_c+\frac{1}{8}|f|^2\fbar_c\right)\chibar_{ad}\\
		&+\left(\frac{1}{2}f_b+\frac{1}{8}|f|^2\fbar_b\right)\eta_a,
	\end{align*}
	and simplifying gives $\g(\D_{e_a}e_b',e_c')=\g(\D_{e_a}e_b,e_c)-\mathcal{T}_{abc}$, hence the proof.
\end{proof}

\begin{prop}\label{prop:diffchristo4}
	Let $U$ be a $\mch$-horizontal $k$-tensor. Then for $b_1,\ldots,b_k=1,2$ we have
	$$\nabla'_{e_4}(\Phi_*U)_{b_1\cdots b_k}=\nabla_{4}U_{b_1\cdots b_k}+\sum_{i=1}^k\mct^{(4)}_{b_ic}U_{b_1\cdots c\cdots b_k},$$
	where the $\mch$-horizontal tensor $\mct^{(4)}_{bc}$ is defined by
	\begin{align*}
		\mct^{(4)}_{bc}=&\frac{1}{2}\fbar_{[b}\nabla_4f_{c]}+\left(f_{[b}+\frac{1}{4}|f|^2\fbar_{[b}\right)\left(\etabar_{c]}+\frac{1}{2}(f\cdot\etabar)\fbar_{c]}\right)+\fbar_{[b}\xi_{c]}+\omega\fbar_{[b}\left(f_{c]}+\frac{1}{4}|f|^2\fbar_{c]}\right).
	\end{align*}
\end{prop}
\begin{proof}
	We have in the case $k=1$,
	\begin{align*}
		\nabla'_{e_4}(\Phi_*U)_{b}=e_4((\Phi_*U)_{b})-(\Phi_*U)_{c}\g(\D_{e_4}e_b',e_c')&=e_4(U_b)-U_{c}\g(\D_{e_4}e_b',e_c')\\
		&=\nabla_{4}U_b+U_c\left(\g(\D_{e_4}e_b,e_c)-\g(\D_{e_4}e_b',e_c')\right).
	\end{align*}
	Moreover,
	\begin{align*}
		\g(\D_{e_4}e_b',e_c')=\g&(\D_{e_4}e_b',e_c)+\frac{1}{2}\fbar_c f_d\g(\D_{e_4}e_b',e_d)+\frac{1}{2}\fbar_c\g(\D_{e_4}e_b',e_4)\\
		&+\left(\frac{1}{2}f_c+\frac{1}{8}|f|^2\fbar_c\right)\g(\D_{e_4}e_b',e_3),
	\end{align*}
where
	\begin{align*}
		\g(\D_{e_4}e_b',e_c)&=\g(\D_{e_4}e_b,e_c)+\frac{1}{2}\fbar_b f_d\g(\D_{e_4}e_d,e_c)+\frac{1}{2}f_ce_4(\fbar_b )+\frac{1}{2}\fbar_be_4(f_c)\\
		&\quad+\fbar_b\xi_c+\left(f_b+\frac{1}{4}|f|^2\fbar_b\right)\etabar_{c},\\
		\frac{1}{2}\fbar_c f_d\g(\D_{e_4}e_b',e_d)&=\frac{1}{2}\fbar_c f_d\g(\D_{e_4}e_b,e_d)+\frac{1}{4}\fbar_c |f|^2e_4(\fbar_b )+\frac{1}{8}\fbar_c\fbar_be_4(|f|^2)\\
		&\quad+\frac12\fbar_c\fbar_b (f\cdot\xi)+\frac{1}{2}\fbar_c f_d\left(f_b+\frac{1}{4}|f|^2\fbar_b\right)\etabar_{d},
	\end{align*}
and
	\begin{align*}
		\frac{1}{2}\fbar_c\g(\D_{e_4}e_b',e_4)&=-\fbar_c\left(\frac{1}{2}e_4(f_b)+\frac{1}{8}\fbar_be_4(|f|^2)+\frac{1}{8}|f|^2e_4(\fbar_b)\right)-\fbar_c\xi_b-\frac12\fbar_c\fbar_b (f\cdot\xi)\\
		&\quad-2\omega\fbar_c\left(\frac12 f_b+\frac18|f|^2\fbar_b\right),\\
		\left(\frac{1}{2}f_c+\frac{1}{8}|f|^2\fbar_c\right)\g(\D_{e_4}e_b',e_3)&=-\left(f_c+\frac{1}{4}|f|^2\fbar_c\right)\etabar_{b}-\fbar_b f_d\left(\frac{1}{2}f_c+\frac{1}{8}|f|^2\fbar_c\right)\etabar_{d}\\
		&\quad-\left(\frac{1}{2}f_c+\frac{1}{8}|f|^2\fbar_c\right)e_4(\fbar_b)+\left(f_c+\frac{1}{4}|f|^2\fbar_c\right)\fbar_b\omega,
	\end{align*}
	which concludes the proof by putting it all together.
\end{proof}

\begin{prop}\label{prop:diffchristo3}
	Let $U$ be a $\mch$-horizontal $k$-tensor. Then for $b_1,\ldots,b_k=1,2$ we have
	$$\nabla'_{e_3}(\Phi_*U)_{b_1\cdots b_k}=\nabla_{3}U_{b_1\cdots b_k}+\sum_{i=1}^k\mct^{(3)}_{b_ic}U_{b_1\cdots c\cdots b_k},$$
	where the $\mch$-horizontal tensor $\mct^{(3)}_{bc}$ is defined by
	\begin{align*}
		\mct^{(3)}_{bc}=&\frac{1}{2}\fbar_{[b}\nabla_3f_{c]}+\fbar_{[b}\eta_{c]}+\omegabar\left(f_{[b}+\frac14|f|^2\fbar_{[b}\right)\fbar_{c]}+\left(f_{[b}+\frac{1}{4}|f|^2\fbar_{[b}\right)\left(\xibar_{c]}+\frac{1}{2}(f\cdot\xibar)\fbar_{c]}\right).
	\end{align*}
\end{prop}
\begin{proof}
The proof is very similar to the one of Proposition \ref{prop:diffchristo4}, so we omit it.
\end{proof}

\section{Derivative estimates in the double null gauge}

\subsection{Bounds for background Kerr quantities}
\textbf{All the implicit constants in the bounds stated in this subsection depend on the number $N$ of derivatives considered, and on the black hole parameters $a,M$.} Let
\begin{align}\label{eq:dfrakkerrrr}
	\df_\mck=\{\Omega^2_\mck(\nabring_4)_\mck,(\nabring_3)_\mck,\nabring_\mck\}.
\end{align}
\noindent\textbf{Bounds for derivatives of Kerr double null metric, Ricci and curvature components.}
Recall from \eqref{eq:bDNdanskerr} the metric components in coordinates $(\theta_*,\phi_*)$. We recall the identities \eqref{eq:trucsnulendoublenul}, \eqref{eq:omegabarendoublenul}, \eqref{eq:parubA}, \eqref{eq:pargammadn} in Kerr which write as follows,
\begin{align}
	\mathring{\omega}_\mck&=0,\quad\mathring{\xi}_\mck=\mathring{\xibar}_\mck=0,\quad\mathring{\atrchi}_\mck=\mathring{\atrchibar}_\mck=0,\label{eq:kerrtrucsnulendoublenulkerr}\\
	\mathring{\omegabar}_\mck&=-\partial_u(\log\Omega_\mck),\quad\mathring{\nabla}_\mck\log\Omega_\mck=\frac12(\mathring{\eta}_\mck+\mathring{\etabar}_\mck),\quad \mathring{\eta}_\mck=\mathring{\zeta}_\mck,\label{eq:kerromegabarendoublenul}
\end{align}
as well as
\begin{align}
	\partial_u b^A_\mck&=2\Omega^2(\mathring{\eta}_\mck^A-\mathring{\etabar}_\mck^A)=4\Omega^2_\mck(\mathring{\zeta}^A_\mck-\nabring^A_\mck\log\Omega_\mck),\label{eq:kerrparubA}\\
	\partial_u(\gamma_\mck)_{AB}&=2(\mathring{\chibar}_\mck)_{AB},\quad(\partial_\ubar+b_\mck^C\partial_{\theta^C})(\gamma_\mck)_{AB}+(\gamma_\mck)_{AC}\partial_{\theta^B}b^C_\mck+(\gamma_\mck)_{BC}\partial_{\theta^A}b_\mck^C=2\Omega^2_\mck(\mathring{\chi}_\mck)_{AB}.\label{eq:kerrpargammadn}
\end{align}

\begin{prop}\label{prop:boundtoutkerr}
	Let $C_R\geq 1$. Dropping the ring notation, and denoting $\psi_\mck=\{{\eta}_\mck, {\etabar}_\mck\}$, we have in $\{u+\ubar\geq C_R\}\cap\{u\leq-1\}$, 
\begin{equation}\label{eq:bouboubvjkeee}
	\begin{aligned}
				&|\df_\mck^{\leq N}\log\Omega_\mck|{\lesssim} 1,\quad |\df_\mck^{\leq N}K_\mck|{\lesssim} 1,\quad |\df_\mck^{\leq N-1}\hodge{K}_\mck|{\lesssim} 1,\quad|\df_\mck^{\leq N}b_\mck|{\lesssim} 1,\\
		& |\df_\mck^{\leq N}\chi_\mck|\lesssim 1,\quad |\df_\mck^{\leq N}\chibar_\mck|\lesssim\Omega_\mck^2,\quad |\df_\mck^{\leq N}\psi_\mck|\lesssim 1\quad|\df^{\leq N-1}\omegabar_\mck|\lesssim 1,\\
		& |\df_\mck^{\leq N-1}\beta_\mck|{\lesssim} 1,\quad |\df_\mck^{\leq N-1}\betabar_\mck|{\lesssim} \Omega_\mck^2,\quad|\df_\mck^{\leq N-1}\rho_\mck|{\lesssim} 1,\quad |\df_\mck^{\leq N}\hodge{\rho}_\mck|{\lesssim} 1, \\
		&|\df^{\leq N-1}\alphabar_\mck|\lesssim\Omega^2_\mck,\quad |\df^{\leq N-1}\alpha_\mck|\lesssim\Omega^{-2}_\mck.
	\end{aligned}
\end{equation}
\end{prop}
\begin{proof}
This can be proven by induction on $N$. For $N=0$ - and for pure $\nabring_\mck^k$ derivatives - this holds by Appendix A of \cite{stabC0}. Next, we assume that \eqref{eq:bouboubvjkeee} holds at rank $N-1$. First, note that to obtain the required bounds at rank $N$ for $\log\Omega_\mck$, $b_\mck$, $\chi_\mck$, $\chibar_\mck$, $\psi_\mck$, it is enough to prove
	\begin{align}
	&|\nabring_3^k\nabring^{k'}\log\Omega_\mck|\lesssim1,\quad |\nabring_3^k\nabring^{k'}b_\mck|{\lesssim} 1,\quad |\nabring_3^k\nabring^{k'}\chi_\mck|\lesssim 1,\quad |\nabring_3^k\nabring^{k'}\chibar_\mck|\lesssim\Omega_\mck^2,\quad |\nabring_3^k\nabring^{k'}\psi_\mck|\lesssim 1, \label{eq:interemejsais}
\end{align}
for any $k'+k=N$. Indeed, for any background Kerr quantity $U_\mck$ we have $\lieT U_\mck=0$ thus by \eqref{eq:expreT} and Lemma \ref{lem:lienablaT} we get that the $\Omega^2\nabring_4$ derivatives can be expressed with the operators $\nabring_3$ , $b\cdot\nabring$, $\lieT$, $(\Omega^2\chi,\chibar,\nabring b)\cdot$, where we control the derivatives which act on $\Omega^2\chi,\chibar,b$ by the induction assumption. It is thus enough to control mixed $\nabring_3,\nabring$ derivatives, and by Proposition \ref{prop:commnablaboudnenulll} the commutators $[\nabring_3,\nabring]$ only produce lower-order terms (involving coefficients of the schematic form $\df_\mck^{\leq N-2}(\chibar,\psi,\betabar)$) which are bounded by the induction assumption, thus it is enough to control $\nabring_3^k\nabring^{k'}$ derivatives. Now, for $\log\Omega_\mck$, we simply use $\Omega_\mck^2=-\Delta(r)/(4R^2)$ combined with the estimates for $\frac{\partial}{\partial r_*'}^k(r,\theta)$ in \cite[Prop. A.5,A.6]{stabC0} which imply that
$$\partial_u^k\log\Omega_\mck=\left(\frac{\partial}{\partial r_*'}\right)^k\log\Omega_\mck(r'_*,\theta_*)$$
is a bounded function such that its angular derivatives are bounded. This yields $|\nabring^{k'}\nabring_3^k\log\Omega_\mck|\lesssim 1$, and hence the stated bound for $\log\Omega_\mck$ by commuting $\nabring,\nabring_3$ and using the induction assumption to control the commutator terms. The bound for $K_\mck$ holds for the sames reasons using the definition of the Gauss curvature which implies $K_\mck=_s\partial_{\theta^C}^{\leq 2}(\gamma_\mck)_{AB}$ and the expression of $\gamma_\mck$ in \eqref{eq:bDNdanskerr}. Now we deal with $b_\mck$, $\chi_\mck$, $\chibar_\mck$, $\psi_\mck$. First, note that for any $S(u,\ubar)$-tangent tensor, using the expression of $(\gamma_\mck)_{AB}$ to bound the $\partial_u$ and angular derivatives of the spherical Christoffel symbols, we get by iterating \eqref{eq:nab3doublenul}, \eqref{eq:nabdoublenul} the following schematic bound in local $\theta^A$ coordinates,
\begin{align}
	\nabring_3^{k}\nabring^{k'}U_{A_1\cdots A_r}=_{rs} \partial_u^k\partial_{\theta^C}^{k'}(U_{A_1\cdots A_r})+\left(\df^{\leq k'-1}\chibar^{\leq M(k,k')}\df^{\leq k+k'-1}U^{\leq L(k,k')}\right)_{A_1\cdots A_r},
\end{align}
where $M(k,k'),L(k,k')$ are numbers of factors which depend on $k,k'$. Moreover, by the induction assumption we have for any $U\in\{b_\mck, \chi_\mck, \chibar_\mck,\psi_\mck\}$ the bound $$|\df^{\leq k'-1}\chibar|^{\leq M(k,k')}|\df^{\leq k+k'-1}U|^{\leq L(k,k')}|\lesssim 1,$$
where the bound above holds with $\Omega^2_\mck$ on the RHS for $U=\chibar$. Also we have the following bounds in coordinates
\begin{align}
 |\partial_u^k\partial_{\theta^C}^{k'}b_\mck^A|{\lesssim} 1,\quad |\partial_u^k\partial_{\theta^C}^{k'}(\chi_\mck)_{AB}|\lesssim 1,\quad |\partial_u^k\partial_{\theta^C}^{k'}(\chibar_\mck)_{AB}|\lesssim\Omega_\mck^2,\quad |\partial_u^k\partial_{\theta^C}^{k'}\psi_\mck^A|{\lesssim} 1,\label{eq:vraiementktle}
\end{align}
which follow from the identities \eqref{eq:kerrpargammadn}, \eqref{eq:kerrparubA}, $\eta_\mck=\zeta_\mck$, combined with the explicit expression of $b_\mck^A$ and $(\gamma_\mck)_{AB}$ in \eqref{eq:bDNdanskerr} and the bounds for the derivatives or $(r,\theta)$ with respect to $r'_*,\theta_*$ in \cite[Prop. A.5,A.6]{stabC0}\footnote{Here for $\chibar_\mck$ we use $(\mathring{\chibar}_\mck)_{AB}=_s\partial_u(\gamma_\mck)_{AB}=\partial_{r'_*}(\gamma_\mck)_{AB}$ and we notice that by the results invoked, the derivatives of this function are bounded by $\Delta$. This fact was proven for angular derivatives in \cite[Prop. A.12]{stabC0}, and the extension to $\partial_{r'_*}$ derivatives is a consequence of \cite[Prop. A.5,A.6]{stabC0}.}. We deduce that \eqref{eq:interemejsais} holds. We then conclude that the stated bounds for $\beta_\mck$, $\betabar_\mck$, $\omegabar_\mck$, $\rho_\mck$, $\hodge{K}_\mck$, $\hodge{\rho}_\mck$ also hold by the constraint equations \eqref{eq:constraintDNouioui}, \eqref{eq:KKstarDNdeff} and \eqref{eq:kerromegabarendoublenul}. Finally, we use the null structure equations
\begin{align*}
	{\alpha}_\mck=_s\nabring_4\chi_\mck+\chi_\mck^2&=_s \Omega^{-2}_\mck\df_\mck\chi_\mck+\chi_\mck^2,\\
	{\alphabar}_\mck=_s\nabring_3\chibar_\mck+\chibar_\mck^2+\omegabar_\mck\chibar_\mck&=_s\df_\mck\chibar_\mck+\chibar_\mck^2+\omegabar_\mck\chibar_\mck.
\end{align*}
By the previous bounds on $\chi_\mck,\chibar_\mck$, and on $\log\Omega_\mck$ which implies $|\df_\mck^{\leq N}\Omega_\mck^{-2}|=|\df_\mck^{\leq N}e^{-2\log\Omega_\mck}|\lesssim\Omega_\mck^{-2}$, we get $|\df^{\leq N-1}\alphabar_\mck|\lesssim\Omega^2_\mck$, $|\df^{\leq N-1}\alpha_\mck|\lesssim\Omega^{-2}_\mck$, concluding the induction, hence the proof.
\end{proof}
\begin{rem}
 The induction is needed because of the following facts: from identities \eqref{eq:kerrpargammadn}, \eqref{eq:kerrparubA}, $\eta_\mck=\zeta_\mck$, combined with the explicit expression of $b_\mck^A$ and $(\gamma_\mck)_{AB}$ in \eqref{eq:bDNdanskerr} and the bounds for the derivatives or $(r,\theta)$ with respect to $r'_*,\theta_*$ in \cite[Prop. A.5,A.6]{stabC0}, we get \eqref{eq:vraiementktle}. To recover bounds for covariant derivatives $\df^{\leq N}_\mck$, we need to control the Christoffel symbols which appear when iterating \eqref{eq:nab3doublenul}, \eqref{eq:nabdoublenul}, and the commutator $[\nabring_3,\nabring]$ (see Proposition \ref{prop:commnablaboudnenulll}), which produce lower-order derivatives of $\gamma_\mck$, $\chibar$, $\psi$, $\betabar$. This is why we need the induction.
\end{rem}

\noindent\textbf{Bounds for derivatives of $f_\mck,\widehat{\gbar}_\mck$ (see Proposition \ref{prop:transfodanskerr}).}

\begin{prop}\label{prop:annexeptddd}
	Let $\lambda_\mck$, $f_\mck$, $\widehat{\gbar}_\mck$ be the quantities defined in the proof of Proposition \ref{prop:transfodanskerr}. We have the following bounds in $\{u+\ubar\geq C_R\}\cap\{u\leq-1\}$, 
	$$	|\df^{\leq N}_\mck\lambda_\mck|\lesssim 1,\quad |\df_\mck^{\leq N}\wt{\lambda}_\mck|\lesssim 1,\quad|\df_\mck^{\leq N}f_\mck|\lesssim 1,\quad |\df_\mck^{\leq N}\widehat{\gbar}_\mck|\lesssim\Omega^2_\mck,$$
	where the implicit constants above depend only on $a,M,N$ provided $C_R\geq 1$.
\end{prop}
\begin{proof}
Note that one could invoke the smoothness of $(e_4')_\mck$ and $(\ering_4)_\mck$ at $\ch$ to prove the bounds for the (regular) derivatives of $\lambda_\mck$. Alternatively, recalling the expression \eqref{eq:lambdamckkerr} together with \eqref{eq:bDNdanskerr} we get
	\begin{align*}
	\lambda_\mck=\frac{\sqrt{(r^2+a^2)^2-\Delta a^2\sin^2\theta_*}+r^2+a^2}{4(r^2+a^2)+8Ma^2r\sin^2\theta/|q|^2},
\end{align*}
which we see as a function of $r,\theta$. Then, the bound $|\df_\mck^{\leq N}\lambda_\mck|\lesssim 1$ follows from \eqref{eq:derpardanslautresens} and \cite[Prop. A.6]{stabC0} which bounds successive $\partial_{r'_*},\partial_{\theta_*}$ derivatives of $r,\theta$, and similarly for $\wt{\lambda}_\mck$ using instead \eqref{eq:alatoutefinnn}. Now we prove the bound for $\widehat{\gbar}_\mck$. First note that defining 
\begin{align}\label{eq:jeveuxrentrerrr}
	\gbar_\mck=\Omega^{-2}_\mck\widehat{\gbar}_\mck,
\end{align}
 then recalling \eqref{eq:principalingoinkerr} we have the identity
 \begin{align}\label{eq:elienkerrhj}
 	(e_3)_\mck=\lambdabar_\mck\left(	(\hat{e}_3)_\mck+\gbar_\mck^A\partial_{\theta^A}+\frac14|\gbar|_\mck^2(\hat{e}_4)_\mck\right),
 \end{align} 
where, recalling \eqref{eq:alatoutefinnn},
\begin{align*}
	(\hat{e}_3)_\mck=\Omega^{-2}_\mck (\ering_3)_\mck,\quad(\hat{e}_4)_\mck=\Omega^{2}_\mck (\ering_4)_\mck ,\quad\lambdabar_\mck=\frac{|q_\mck|^2\wt{\lambda}_\mck}{R^2}\sim 1.
\end{align*}
We see \eqref{eq:elienkerrhj} as a frame transformation \eqref{eq:frametransfo} with coefficients $(\lambdabar_\mck,f=0,\gbar_\mck)$. Thus, exchanging the role of $e_3$ and $e_4$ and using the frame transformation formula \eqref{eq:changexi} for $\xi$, and using that both $(e_3)_\mck$ and $(\ering_3)_\mck$ are geodesic we get that $\gbar_\mck$ satisfies the transport equation
\begin{align}
	&(\nabring_{	(\hat{e}_3)_\mck+\gbar_\mck^A\partial_{\theta^A}+\frac14|\gbar|_\mck^2(\hat{e}_4)_\mck})_\mck\gbar_\mck+\frac{1}{4}\wh{tr\chibar}_\mck\gbar_\mck+\frac{1}{2}\gbar_\mck\cdot_\mck\wh{\wh{\chibar}}_\mck+\frac{1}{4}|\gbar_\mck|^2_\mck(\wh{\etabar}_\mck-\wh{\eta}_\mck)\nn\\
	&+\frac{1}{2}(\gbar_\mck\cdot_\mck\wh{\zeta}_\mck)\gbar_\mck-\frac{1}{4}\wh{\omega}_\mck|\gbar_\mck|^2_\mck\gbar_\mck-\frac{1}{8}|\gbar_\mck|^2\gbar_\mck\cdot_\mck\wh{\chihat}_\mck-\frac{1}{8}|\gbar_\mck|^2\wh{tr\chi}_\mck\gbar_\mck=0,\label{eq:2kerredofbar}
\end{align}
where in the equation above we denote with a hat the Ricci coefficients with respect to the rescaled double null pair $(\hat{e}_3,\hat{e}_4)_\mck$, i.e.
\begin{align*}
	&\wh{tr\chibar}_\mck=\Omega_\mck^{-2}\mathring{tr{\chibar}}_\mck,\quad\wh{\wh{\chibar}}_\mck=\Omega_\mck^{-2}\mathring{{\wh{\chibar}}}_\mck,\quad \wh{\etabar}_\mck,\wh{\eta}_\mck=\mathring{\etabar}_\mck,\mathring{\eta}_\mck,\quad \hat{\zeta}_\mck=-\hat{\etabar}_\mck,\\
	&\wh{tr\chi}_\mck=\Omega_\mck^{2}\mathring{tr{\chi}}_\mck,\quad\wh{\wh{\chi}}_\mck=\Omega_\mck^{2}\mathring{{\wh{\chi}}}_\mck,\quad\wh{\omega}_\mck=-\frac12\Omega^2\ering_4(\log\Omega_\mck).
\end{align*}
Note that \eqref{eq:2kerredofbar} can be rewritten as 
\begin{align}\label{eq:eqpourgbarmcke}
	&(\nabring_{\Ybar})_\mck\gbar_\mck=F,\quad F=_s\Gamma\gbar_\mck^{\leq 2}\gbar_\mck,\\
	\Ybar_\mck:=\Omega_\mck^2\Big(	(\hat{e}_3)_\mck+\gbar_\mck^A\partial_{\theta^A}+&\frac14|\gbar|_\mck^2(\hat{e}_4)_\mck\Big),\quad \Gamma=_s(\mathring{\chibar}_\mck,\Omega^2_\mck(\mathring{\eta}_\mck,\mathring{\etabar}_\mck),\Omega^2_\mck\hat{\omega}_\mck,\Omega^4_\mck\mathring{\chi}_\mck).\nn
\end{align}
Moreover, by \eqref{eq:bornesdanskerr} we get 
\begin{align}\label{eq:Gammasurlasoupe}
	|\df_\mck^{\leq N}\Gamma|\lesssim\Omega^2_\mck.
\end{align}
Also, notice that from \eqref{eq:caonsaitdejafacile} we already have $|\gbar_\mck|\lesssim 1$, which allows us to use the transport estimates along the flow of $\Ybar_\mck$ of Section \ref{section:transpoestperturbingoing}. From these observations, using the commutation formulas in Proposition \ref{prop:commnablaboudnenulll} in exact Kerr, in the style of Section \ref{section:controlgbarlambdabar} we obtain by induction
\begin{align}\label{eq:egegegcdetrop}
	|\nabring^{k'}_\mck(\nabring_{\Ybar})_\mck^k\gbar_\mck|\lesssim 1.
\end{align}
We skip the details as this is very similar to the proof of Theorem \ref{thm:controlgbarc}, but we provide a few remarks on how to obtain the bound above:
\begin{itemize}
	\item The bound \eqref{eq:egegegcdetrop} for $k'=0$ follows directly from \eqref{eq:eqpourgbarmcke}, \eqref{eq:Gammasurlasoupe}, the fact that $\Ybar_\mck$ can be expressed with $\df_\mck$, and an induction in $k$.
	\item The bound \eqref{eq:egegegcdetrop} for general $k'$ and $k\geq 1$ goes similarly by induction on $k'$, differentiating by $\nabring$ the equation  $(\nabring_{\Ybar})_\mck^{k+1}\gbar_\mck=(\nabring_{\Ybar})_\mck^kF$ without commuting $\nabring_\mck$ and $\Ybar_\mck$.
	\item To prove \eqref{eq:egegegcdetrop} for $k'\geq 1$ and $k=0$, we commute \eqref{eq:eqpourgbarmcke} with $\nabring_\mck^k$ and we proceed by induction on $k'$. Using Proposition \ref{prop:commnablaboudnenulll} we get for any $U$
	\begin{align*}
		[(\nabring_{\Ybar})_\mck,\nabring_\mck]U=_s&(\mathring{\chibar}_\mck,|\gbar|^2\Omega^4_\mck\mathring{\chi}_\mck)\nabring_\mck U+((\mathring{\chibar}_\mck,|\gbar|^2\Omega^4_\mck\mathring{\chi}_\mck)(\mathring{\eta}_\mck,\mathring{\etabar}_\mck)+\mathring{\betabar}_\mck+\Omega^4_\mck|\gbar|^2\mathring{\beta}_\mck)U\\
		&+\nabring_\mck(\Omega^2_\mck\gbar_\mck)\nabring_\mck U+\Omega^2_\mck\gbar_\mck \mathring{K}_\mck U+\nabring_\mck(|\gbar_\mck|^2\Omega^2_\mck)\Omega^2_\mck(\nabring_4)_\mck U.
	\end{align*}
Using Lemma \ref{lem:lienablaT} and \eqref{eq:expreT} we can rewrite the last term on the RHS above with $\Omega^2_\mck(\nabring_4)_\mck U=_s(\nabring_{\Ybar})_\mck U+(\Omega^2_\mck\gbar_\mck,b_\mck)\nabring_\mck U+(\Omega^2_\mck\mathring{\chi}_\mck,\mathring{\chibar}_\mck,\nabring_\mck b_\mck)U$, where we used $\lieT U=0$ which holds in this context. Then by the induction assumption, we see that for $k'\geq 2$\footnote{For $k'=1$, by the commutation identity above, there is $\Omega^2_\mck(\nabring_\mck\gbar_\mck)^2$ on the RHS of the transport equation for $\nabring_\mck\gbar_\mck$, which can not be treated by Grönwall (there is no such problematic squared top-order term for $k'\geq 2$).}, $(\nabring_{\Ybar})_\mck\nabring_\mck^{k'}\gbar_\mck$ is bounded by $\Omega_\mck^2|\nabring^{k'}\gbar|$ (which can be treated by a Grönwall inequality) plus some already controlled lower-order terms bounded by $\Omega^2_\mck$. Thus, provided $|\nabring_\mck\gbar_\mck|\lesssim 1$ holds, we conclude the proof by induction on $k'$ and the transport estimates of Section \ref{section:transpoestperturbingoing}. Moreover, $|\nabring_\mck\gbar_\mck|\lesssim 1$ holds by the computation in \eqref{eq:gbarhattheta}, \eqref{eq:gbarhatphi} in coordinates.
\end{itemize}
This concludes the proof of \eqref{eq:egegegcdetrop}. Using $\lieT\gbar_\mck=0$, we deduce $|\lieT^{k''}\nabring^{k'}_\mck(\nabring_{\Ybar})_\mck^k\gbar_\mck|\lesssim 1$. Finally, as we can express $\df_\mck$ derivatives with $\lieT,\nabring,\nabring_{\Ybar}$, this concludes the proof of the bound $|\df_\mck^{\leq N}\gbar_\mck|\lesssim 1$,
which yields $|\df_\mck^{\leq N}\widehat{\gbar}_\mck|\lesssim \Omega^2_\mck$ by \eqref{eq:jeveuxrentrerrr} and by the bounds for $\log\Omega_\mck$ in \eqref{eq:bornesdanskerr}. Finally, the bound for $\df_\mck^{\leq N}f_\mck$ can be argued to hold by invoking the smoothness of the vector field $(e_4')_\mck$ and of the foliation $S(u,\ubar)$ at the Kerr Cauchy horizon $\ch$. Alternatively, one can prove it similarly as for $\gbar_\mck$ using that $e_4'$ and and $\ering_4$ are both geodesic, and using the transformation formula for $\xi$ with coefficients $\lambda_\mck,f_\mck,\fbar=0$, which yields a transport equation for $f_\mck$ allowing us to bound $\df_\mck^{\leq N}f_\mck$ by induction.
\end{proof}
\subsection{Reduced schematic equations for $\nabring^k$ derivatives of linearized quantities}\label{section:proofsschemIInbabla}
We derive the schematic equations satisfied by $\nabring^k$ derivatives of the double null linearized quantities, \textbf{and drop the ring notation in this section}. We begin with the following result, which computes the commutator between $\nabring_3,\nabring_4,\nabring$ and the check operator. 

\begin{prop}\label{prop:linderdoublenull}
	We have the following formulas, for any $S(u,\ubar)$-tangent $r$-tensor $\phi$,
	\begin{align}
		(\nabring_3\widecheck{\phi})_{A_1\cdots A_r}=&({\nabring_3\phi}-(\nabring_3\phi)_\mck)_{A_1\cdots A_r}+\sum_{i=1}^r\left(\left(\gamma_{\mathcal{K}}^{-1}\right)^{B C}\left(\underline{\chi}_{\mathcal{K}}\right)_{C A_i}-\left(\gamma^{-1}\right)^{B C} \underline{\chi}_{C A_i}\right)\left(\phi_{\mathcal{K}}\right)_{A_1 \ldots  B \ldots A_r},\label{eq:nab3diff}\\
		(\nabring_4\widecheck{\phi})_{A_1\cdots A_k}=&({\nabring_4\phi}-(\nabring_4\phi)_\mck)_{A_1\cdots A_r}-\frac{\Omega_\mck^2-\Omega^2}{\Omega^2}((\nabring_4\phi)_\mck)_{A_1\cdots A_k}\nn\\
		&-\sum_{i=1}^r\frac{\Omega_\mck^2-\Omega^2}{\Omega^2}(\gamma_\mck^{-1})^{BC}(\chi_\mck)_{CA_i}(\phi_\mck)_{A_1\ldots B\ldots A_r}-\frac{b^B-(b_\mck)^B}{\Omega^2}\nabring_B(\phi_\mck)_{A_1\cdots A_r}\nn\\
		&-\sum_{i=1}^r\left((\gamma^{-1})^{BC}\chi_{CA_i}-(\gamma_\mck^{-1})^{BC}(\chi_\mck)_{CA_i}\right)(\phi_\mck)_{A_1\cdots B\cdots A_r}\nn\\
		&-\sum_{i=1}^r\frac{1}{\Omega^2}\nabring_{A_i}(b^B-(b_\mck)^B)(\phi_\mck)_{A_1\cdots B\cdots A_r},\label{eq:nab4diff}\\
		\nabring_B\widecheck{\phi}_{A_1\cdots A_r}=&({\nabring_B\phi}-(\nabring_B\phi)_\mck)_{BA_1\cdots A_r}-\sum_{i=1}^r\left(\mathring{\Gamma}_{A_i B}^C-(\mathring{\Gamma}_{A_i B}^C)_\mck\right)(\phi_\mck)_{A_1\ldots C\ldots A_r},\label{eq:nabladiff}
	\end{align}
where $\mathring{\Gamma}_{A B}^C-(\mathring{\Gamma}_{A B}^C)_\mck$ is the following $S(u,\ubar)$-tangent tensor,
\begin{align*}
	\mathring{\Gamma}_{A B}^C-(\mathring{\Gamma}_{A B}^C)_\mck=\frac12(\gamma^{-1}_\mck)^{CD}\left(\nabring_A\gammacheck_{BD}+\nabring_B\gammacheck_{AD}-\nabring_D\gammacheck_{AB}\right).
\end{align*}
	Note that in the three identities above, the terms with sums $\sum_{i=1}^r$ on the RHS are not present if $\phi$ is a scalar. We define the operators $\widecheck{\nabring}_{3,4,B}$ such that \eqref{eq:nab3diff}, \eqref{eq:nab4diff}, \eqref{eq:nabladiff} rewrite
	$$\nabring_3\widecheck{\phi}={\nabring_3\phi}-(\nabring_3\phi)_\mck+\widecheck{\nabring}_3[\phi_\mck],\quad\nabring_4\widecheck{\phi}={\nabring_4\phi}-(\nabring_4\phi)_\mck+\widecheck{\nabring}_4[\phi_\mck],\quad \nabring_B\widecheck{\phi}={\nabring_B\phi}-(\nabring_B\phi)_\mck+\widecheck{\nabring}_B[\phi_\mck].$$
\end{prop}
\begin{proof}
	This follows from \cite[Prop. 7.4, Prop. 7.6, (2.9), Prop. 7.3]{stabC0}.
\end{proof}
We recall below Proposition 7.2 in \cite{stabC0}.
\begin{prop}\label{prop:7.2DL}
	Suppose $\nabring_4\phi=F_0$. Then for $i\geq 0$,
	$$\nabring_4\nabring^i\phi=_s\sum_{i_1+i_2+i_3=i}\nabring^{i_1}\psi^{i_2}\nabring^{i_3}F_0+\sum_{i_1+i_2+i_3+i_4=i}\nabring^{i_1}\psi^{i_2}\nabring^{i_3}\chi\nabring^{i_4}\phi.$$
	Similarly, suppose $\nabring_3\phi=G_0$. Then for $i\geq 0$,
	$$\nabring_3\nabring^i\phi=_s\sum_{i_1+i_2+i_3=i}\nabring^{i_1}\psi^{i_2}\nabring^{i_3}G_0+\sum_{i_1+i_2+i_3+i_4=i}\nabring^{i_1}\psi^{i_2}\nabring^{i_3}\chibar\nabring^{i_4}\phi.$$
\end{prop}

\begin{prop}\label{prop:schemnab33}
	Let $U$ be a $S(u,\ubar)$-tangent tensor. Then we have for $i\geq 0$,
	\begin{align*}
		\nabring_3\nabring^i\widecheck{U}=_s&\sum_{i_1+i_2+i_3=i}\nabring^{i_1}\psi^{i_2}\Big[\nabring^{i_3}(\nabring_3U-(\nabring_3U)_\mck)+\sum_{i_4+i_5+i_6=i_3}\Big(\nabring^{i_4}\gcheck\nabring^{i_5}\chibar_\mck\nabring^{i_6}U_\mck\\
		&+\sum_{i_4+i_5=i_3}\nabring^{i_4}\widecheck{\chibar}\nabring^{i_5}U_\mck\Big)\Big]+\sum_{i_1+i_2+i_3+i_4=i}\nabring^{i_1}\psi^{i_2}\nabring^{i_3}\chibar\nabring^{i_4}\widecheck{U}.
	\end{align*}
Moreover, if $U$ is a scalar,
	\begin{align*}
	\nabring_3\nabring^i\widecheck{U}=_s&\sum_{i_1+i_2+i_3=i}\nabring^{i_1}\psi^{i_2}\nabring^{i_3}(\nabring_3U-(\nabring_3U)_\mck)+\sum_{i_1+i_2+i_3+i_4=i}\nabring^{i_1}\psi^{i_2}\nabring^{i_3}\chibar\nabring^{i_4}\widecheck{U}.
\end{align*}
\end{prop}
\begin{proof}
	This follows from \eqref{eq:nab3diff} combined with Proposition \ref{prop:7.2DL} and the Leibniz rule.
\end{proof}
\begin{prop}\label{prop:schemnab44}
		Let $U$ be a $S(u,\ubar)$-tangent tensor. Then we have for $i\geq 0$,
	\begin{align*}
		\nabring_4\nabring^i\widecheck{U}=_s&\sum_{i_1+i_2+i_3=i}\nabring^{i_1}\psi^{i_2}\Bigg[\nabring^{i_3}(\nabring_4U-(\nabring_4U)_\mck)+\sum_{i_4+i_5+i_6+i_7=i_3}\Big(\nabring^{i_4}\gcheck\nabring^{i_5}(\nabring_4 U)_\mck\\
		&+(\nabring^{i_4}\gcheck\nabring^{i_5}\chi_\mck(1+\nabring^{i_6}\gamma_\mck^{-1})+\nabring^{i_4}\widecheck{\chi})\nabring^{i_7} U_\mck+\nabring^{i_4}\Omega^{-2}\nabring^{i_5}\bcheck\nabring^{i_6+1} U_\mck\\
		&+\nabring^{i_4}\Omega^{-2}\nabring^{i_5+1}\bcheck\nabring^{i_6} U_\mck\Big)\Bigg]+\sum_{i_1+i_2+i_3+i_4=i}\nabring^{i_1}\psi^{i_2}\nabring^{i_3}\chi\nabring^{i_4}\widecheck{U}.
	\end{align*}
	Morover, if $U$ is a scalar, 
	\begin{align*}
		\nabring_4\nabring^i\widecheck{U}=_s&\sum_{i_1+i_2+i_3=i}\nabring^{i_1}\psi^{i_2}\Big[\nabring^{i_3}(\nabring_4U-(\nabring_4U)_\mck)+\sum_{i_4+i_5+i_6=i_3}\Big(\nabring^{i_4}\gcheck\nabring^{i_5}(\nabring_4 U)_\mck\\
		&+\nabring^{i_4}\Omega^{-2}\nabring^{i_5}\bcheck\nabring^{i_6+1} U_\mck\Big)\Big]+\sum_{i_1+i_2+i_3+i_4=i}\nabring^{i_1}\psi^{i_2}\nabring^{i_3}\chi\nabring^{i_4}\widecheck{U}.
	\end{align*}
\end{prop}
\begin{proof}
	This follows from \eqref{eq:nab4diff} combined with Proposition \ref{prop:7.2DL} and the Leibniz rule.
\end{proof}

We are now ready to prove Propositions \ref{prop:DNmetricnabla} to \ref{prop:DNbianchinabla}. We will actually prove more precise schematic equations satisfied by the $\nabring^i$ derivatives of the linearized quantities.  This will allow us later to commute these schematic equations with respect to $\lieT$, see Section \ref{section:notethtatt}. We begin by defining the following schematic inhomogeneous terms for $i\geq 0$:
\begin{align*}
	F_\gamma^{(i)}=_s&\sum_{i_1+i_2+i_3=i}\nabring^{i_1}\psi^{i_2}\left(\sum_{i_4+i_5+i_6=i_3}\nabring^{i_4}\gammacheck \nabring^{i_5}\chibar_\mck\nabring^{i_6}\gamma_\mck+\sum_{i_4+i_5=i_3}\nabring^{i_4}\widecheck{\chibar}\nabring^{i_5}\gamma_\mck\right)\\
	&+\sum_{i_1+i_2+i_3+i_4=i}\nabring^{i_1}\psi^{i_2}\nabring^{i_3}\chibar\nabring^{i_4}\gammacheck,
\end{align*}
\begin{align*}
	F_\Omega^{(i)}=_s&\sum_{i_1+i_2+i_3=i}\nabring^{i_1}\psi^{i_2}\left(\nabring^{i_3}\widecheck{\omegabar}+\sum_{i_4+i_5+i_6=i_3}\nabring^{i_4}\gammacheck \nabring^{i_5}\chibar_\mck\nabring^{i_6}\log\Omega_\mck+\sum_{i_4+i_5=i_3}\nabring^{i_4}\widecheck{\chibar}\nabring^{i_5}\log\Omega_\mck\right)\\
	&+\sum_{i_1+i_2+i_3+i_4=i}\nabring^{i_1}\psi^{i_2}\nabring^{i_3}\chibar\nabring^{i_4}\logomegacheck,
\end{align*}
\begin{align}\label{eq:Fb}
	F_b^{(i)}=_s&\sum_{i_1+i_2+i_3=i}\nabring^{i_1}\psi^{i_2}\nabring^{i_3}\nabring_3\bcheck+\sum_{i_1+i_2+i_3+i_4=i}\nabring^{i_1}\psi^{i_2}\nabring^{i_3}\chibar\nabring^{i_4}\bcheck,
\end{align}
where $\nabring_3\bcheck=_s\Omega^2(\zetacheck-\nabring\logomegacheck)+\widecheck{\Omega^2}(\zeta_\mck-\nabring_\mck\log\Omega_\mck)+\Omega^2\widecheck{\gamma^{-1}}\cdot\nabring\log\Omega_\mck+\chibar\cdot\bcheck$ (see \cite[Eq. (7.23)]{stabC0}). Next we define
\begin{align*}
	F_{\etabar}^{(i)}=_s&\sum_{i_1+i_2+i_3=i}\nabring^{i_1}\psi^{i_2}\Big[\nabring^{i_3}(\nabring_3\etabar-(\nabring_3\etabar)_\mck)+\sum_{i_4+i_4+i_6=i_3}\nabring^{i_4}\gammacheck\nabring^{i_5}\chibar_\mck\nabring^{i_6}\etabar_\mck+\sum_{i_4+i_5=i_3}\nabring^{i_4}\widecheck{\chibar}\nabring^{i_5}\etabar_\mck\Big]\\
	&+\sum_{i_1+i_2+i_3+i_4=i}\nabring^{i_1}\psi^{i_2}\nabring^{i_3}\chibar\nabring^{i_4}\etabarcheck,
\end{align*}
where $\nabring_3\etabar=_s\sum_{i_1+i_2=1}\psi^{i_1}\nabring^{i_2}\psi_{\Hbar}$. Similarly we define
\begin{align*}
	F_{\mubar}^{(i)}=_s&\sum_{i_1+i_2+i_3=i}\nabring^{i_1}\psi^{i_2}\Big[\nabring^{i_3}(\nabring_3\mubar-(\nabring_3\mubar)_\mck)+\sum_{i_4+i_4+i_6=i_3}\nabring^{i_4}\gammacheck\nabring^{i_5}\chibar_\mck\nabring^{i_6}\mubar_\mck+\sum_{i_4+i_5=i_3}\nabring^{i_4}\widecheck{\chibar}\nabring^{i_5}\mubar_\mck\Big]\\
	&+\sum_{i_1+i_2+i_3+i_4=i}\nabring^{i_1}\psi^{i_2}\nabring^{i_3}\chibar\nabring^{i_4}\mubarcheck,
\end{align*}
where $\nabring_3\mubar=_s\psi_{\Hbar}K+\sum_{i_1+i_2+i_3=1}\psi^{i_1}\nabring^{i_2}\psi\nabring^{i_3}\psi_{\Hbar}$ (see \cite[Eq. (3.20)]{stabC0}). Next we define
\begin{align*}
	F_{\psi_H}^{(i)}=_s&\sum_{i_1+i_2+i_3=i}\nabring^{i_1}\psi^{i_2}\Big[\nabring^{i_3}(\nabring_3(\Omega_\mck^2\psi_H)-(\nabring_3(\Omega_\mck^2\psi_H))_\mck)\\
	&+\sum_{i_4+i_4+i_6=i_3}\nabring^{i_4}\gammacheck\nabring^{i_5}\chibar_\mck\nabring^{i_6}(\Omega_\mck^2(\psi_H)_\mck)+\sum_{i_4+i_5=i_3}\nabring^{i_4}\widecheck{\chibar}\nabring^{i_5}(\Omega_\mck^2(\psi_H)_\mck)\Big]\\
	&+\sum_{i_1+i_2+i_3+i_4=i}\nabring^{i_1}\psi^{i_2}\nabring^{i_3}\chibar\nabring^{i_4}(\Omega_\mck^2\psicheck_H),
\end{align*}
where $\nabring_3(\Omega_\mck^2\psi_H)=_s\Omega_\mck^2(K+\nabring\eta+\psi_{\Hbar}\psi_H+\psi\psi+\widecheck{\omegabar}\psi_H)$ (see \cite[Eq. (3.10)]{stabC0}). We also define
\begin{align*}
	F_\eta^{(i)}=_s&\sum_{i_1+i_2+i_3=i}\nabring^{i_1}\psi^{i_2}\Big[\nabring^{i_3}(\nabring_4\eta-(\nabring_4\eta)_\mck)+\sum_{i_4+i_5+i_6+i_7=i_3}\Big(\nabring^{i_4}\gcheck\nabring^{i_5}(\nabring_4 \eta)_\mck\\
	&+(\nabring^{i_4}\gcheck\nabring^{i_5}\chi_\mck(1+\nabring^{i_6}\gamma_\mck^{-1})+\nabring^{i_4}\widecheck{\chi})\nabring^{i_7} \eta_\mck+\nabring^{i_4}\Omega^{-2}\nabring^{i_5}\bcheck\nabring^{i_6+1} \eta_\mck+\nabring^{i_4}\Omega^{-2}\nabring^{i_5+1}\bcheck\nabring^{i_6} \eta_\mck\Big)\Big]\\
	&+\sum_{i_1+i_2+i_3+i_4=i}\nabring^{i_1}\psi^{i_2}\nabring^{i_3}\chi\nabring^{i_4}\widecheck{\eta},
\end{align*}
where $\nabring_4\eta=_s\sum_{i_1+i_2=1}\psi^{i_1}\nabring^{i_2}\psi_H$ (see \cite[Eq. (3.18)]{stabC0}). Similarly we define
\begin{align*}
	F_\mu^{(i)}=_s&\sum_{i_1+i_2+i_3=i}\nabring^{i_1}\psi^{i_2}\Big[\nabring^{i_3}(\nabring_4\mu-(\nabring_4\mu)_\mck)+\sum_{i_4+i_5+i_6+i_7=i_3}\Big(\nabring^{i_4}\gcheck\nabring^{i_5}(\nabring_4 \mu)_\mck\\
	&+(\nabring^{i_4}\gcheck\nabring^{i_5}\chi_\mck(1+\nabring^{i_6}\gamma_\mck^{-1})+\nabring^{i_4}\widecheck{\chi})\nabring^{i_7} \mu_\mck+\nabring^{i_4}\Omega^{-2}\nabring^{i_5}\bcheck\nabring^{i_6+1} \mu_\mck+\nabring^{i_4}\Omega^{-2}\nabring^{i_5+1}\bcheck\nabring^{i_6} \mu_\mck\Big)\Big]\\
	&+\sum_{i_1+i_2+i_3+i_4=i}\nabring^{i_1}\psi^{i_2}\nabring^{i_3}\chi\nabring^{i_4}\widecheck{\mu},
\end{align*}
where $\nabring_4\mu=_s\psi_HK+\sum_{i_1+i_2+i_3=1}\psi^{i_1}\nabring^{i_2}\psi\nabring^{i_3}\psi_H$ (see \cite[Eq. (3.21)]{stabC0}). Next we define 
\begin{align*}
	F_{\psi_{\Hbar}}^{(i)}=_s&\sum_{i_1+i_2+i_3=i}\nabring^{i_1}\psi^{i_2}\Big[\nabring^{i_3}(\nabring_4\psi_{\Hbar}-(\nabring_4\psi_{\Hbar})_\mck)+\sum_{i_4+i_5+i_6+i_7=i_3}\Big(\nabring^{i_4}\gcheck\nabring^{i_5}(\nabring_4 \psi_{\Hbar})_\mck\\
	&+(\nabring^{i_4}\gcheck\nabring^{i_5}\chi_\mck(1+\nabring^{i_6}\gamma_\mck^{-1})+\nabring^{i_4}\widecheck{\chi})\nabring^{i_7} (\psi_{\Hbar})_\mck+\nabring^{i_4}\Omega^{-2}\nabring^{i_5}\bcheck\nabring^{i_6+1} (\psi_{\Hbar})_\mck\\
	&+\nabring^{i_4}\Omega^{-2}\nabring^{i_5+1}\bcheck\nabring^{i_6} (\psi_{\Hbar})_\mck\Big)\Big]+\sum_{i_1+i_2+i_3+i_4=i}\nabring^{i_1}\psi^{i_2}\nabring^{i_3}\chi\nabring^{i_4}\widecheck{\psi}_{\Hbar},
\end{align*}
where $\nabring_4\psi_{\Hbar}=_sK+\nabring\etabar+\psi\psi+\psi_H\psi_{\Hbar}$ (see \cite[Eq. (3.11)]{stabC0}). Similarly, we define 
\begin{align*}
	F_{\omegabar}^{(i)}=_s&\sum_{i_1+i_2+i_3=i}\nabring^{i_1}\psi^{i_2}\Big[\nabring^{i_3}(\nabring_4\omegabar-(\nabring_4\omegabar)_\mck)+\sum_{i_4+i_5+i_6=i_3}\Big(\nabring^{i_4}\gcheck\nabring^{i_5}(\nabring_4 \omegabar)_\mck\\
	&+\nabring^{i_4}\Omega^{-2}\nabring^{i_5}\bcheck\nabring^{i_6+1} \omegabar_\mck\Big)\Big]+\sum_{i_1+i_2+i_3+i_4=i}\nabring^{i_1}\psi^{i_2}\nabring^{i_3}\chi\nabring^{i_4}\widecheck{\omegabar},
\end{align*}
where $\nabring_4\omegabar=_sK+\psi\psi+\psi_H\psi_{\Hbar},$ (see \cite[Eq. (3.14)]{stabC0}). And also
\begin{align*}
	F_{{\barred{\omegabar}}}^{(i)}=_s&\sum_{i_1+i_2+i_3=i}\nabring^{i_1}\psi^{i_2}\Big[\nabring^{i_3}(\nabring_4{\barred{\omegabar}}-(\nabring_4{\barred{\omegabar}})_\mck)+\sum_{i_4+i_5+i_6=i_3}\Big(\nabring^{i_4}\gcheck\nabring^{i_5}(\nabring_4 {\barred{\omegabar}})_\mck\\
	&+\nabring^{i_4}\Omega^{-2}\nabring^{i_5}\bcheck\nabring^{i_6+1} {\barred{\omegabar}}_\mck\Big)\Big]+\sum_{i_1+i_2+i_3+i_4=i}\nabring^{i_1}\psi^{i_2}\nabring^{i_3}\chi\nabring^{i_4}\widecheck{{\barred{\omegabar}}},
\end{align*}
where (see \cite[Eq. (3.22)]{stabC0})
\begin{align*}
	\nabring_4{\barred{\omegabar}}=_s\sum_{i_1+i_2+i_3=2}\psi^{i_1}\nabring^{i_2}\psi\nabring^{i_3}\psi+\sum_{i_1+i_2+i_3=2}\psi^{i_1}\nabring^{i_2}\psi_H\nabring^{i_3}\psi_{\Hbar}+\sum_{i_1+i_2+i_3=1}\psi^{i_1}\nabring^{i_2}\psi\nabring^{i_3}K.
\end{align*}
Next, we define
\begin{align*}
	F_{tr\chibar}^{(i)}=_s&\sum_{i_1+i_2+i_3=i}\nabring^{i_1}\psi^{i_2}\nabring^{i_3}(G-G_\mck)+\sum_{i_1+i_2+i_3+i_4=i}\nabring^{i_1}\psi^{i_2}\nabring^{i_3}\chibar\nabring^{i_4}\widecheck{tr\chibar}\\
	&+\sum_{i_1+i_2\leq i, i_1\geq 1}\nabring^{i_1}\omegabar_\mck\nabring^{i_2}\widecheck{tr\chibar}+\sum_{i_1+i_2\leq i}\nabring^{i_1}\widecheck{\omegabar}\nabring^{i_2}tr\chibar,
\end{align*}
where $G=-\frac12 (tr\chibar)^2-|\wh{\chibar}|^2$. Next, we define
\begin{align*}
	F_{tr\chi}^{(i)}=_s&\sum_{i_1+i_2+i_3=i}\nabring^{i_1}\psi^{i_2}\Big[\nabring^{i_3}(F-F_\mck)+\sum_{i_4+i_5+i_6+i_7=i_3}\Big(\nabring^{i_4}\gcheck\nabring^{i_5}(\nabring_4 tr\chi)_\mck\\
	&+\nabring^{i_4}\Omega^{-2}\nabring^{i_5}\bcheck\nabring^{i_6+1} tr\chi_\mck\Big)\Big]+\sum_{i_1+i_2+i_3+i_4=i}\nabring^{i_1}\psi^{i_2}\nabring^{i_3}\chi\nabring^{i_4}\widecheck{tr\chi},
\end{align*}
where $F=-\frac12 (tr\chi)^2-|\wh{\chi}|^2$. Now, we define
\begin{align*}
	F_\beta^{(i)}=_s&\sum_{i_1+i_2+i_3=i}\nabring^{i_1}\psi^{i_2}\Big[\nabring^{i_3}(G-G_\mck)+\sum_{i_4+i_5+i_6=i_3}\Big(\nabring^{i_4}\gcheck\nabring^{i_5}\chibar_\mck\nabring^{i_6}(\Omega_\mck^2\beta_\mck)\\
	&+\sum_{i_4+i_5=i_3}\nabring^{i_4}\widecheck{\chibar}\nabring^{i_5}(\Omega_\mck^2\beta_\mck)\Big)\Big]+\sum_{i_1+i_2+i_3+i_4=i}\nabring^{i_1}\psi^{i_2}\nabring^{i_3}\chibar\nabring^{i_4}(\Omega\Omega_\mck\betacheck)\\
	&+\sum_{i_1+i_2+i_3+i_4\leq i,i_4\leq i-1}(1+\nabring^{i_1}\psi^{i_2})\nabring^{i_3}(\Omega\Omega_\mck)\nabring^{i_4+1}(\Kcheck,\hodge{\Kcheck})\\
	&+\sum_{i_1+i_2+i_3+i_4+i_5\leq i,i_4+i_5\leq i-1}(1+\nabring^{i_1}\psi^{i_2})\nabring^{i_3}(\Omega\Omega_\mck)\nabring^{i_4}(\nabring\hodge{K})_\mck\nabring^{i_5}\gcheck\\
	&+\Omega\Omega_\mck\sum_{i_1+i_2\leq i-1}\nabring^{i_1}K\nabring^{i_2}(\Kcheck,\hodge{\Kcheck})
\end{align*}
where $G=_s\Omega_\mck\Omega\left(\sum_{i_1+i_2+i_3=1}\psi^{i_1}\nabring^{i_2}(\psi_{\Hbar},\widecheck{\omegabar})\nabring^{i_3}\psi_H+\psi K+\sum_{i_1+i_2=1}\psi^{i_1}\psi\nabring^{i_2}\psi\right)$, as well as
\begin{align*}
	F_K^{(i)}=_s&\sum_{i_1+i_2+i_3=i}\nabring^{i_1}\psi^{i_2}\Big[\nabring^{i_3}(F-F_\mck)+\sum_{i_4+i_5+i_6=i_3}\Big(\nabring^{i_4}\gcheck\nabring^{i_5}F_\mck\\
	&+(\nabring^{i_4}\gcheck\nabring^{i_5}\chi_\mck+\nabring^{i_4}\widecheck{\chi})\nabring^{i_6} K_\mck+\nabring^{i_4}\Omega^{-2}\nabring^{i_5}\bcheck\nabring^{i_6+1} K_\mck+\nabring^{i_4}\Omega^{-2}\nabring^{i_5+1}\bcheck\nabring^{i_6} K_\mck\Big)\Big]\\
	&+\sum_{i_1+i_2+i_3+i_4=i}\nabring^{i_1}\psi^{i_2}\nabring^{i_3}\chi\nabring^{i_4}\widecheck{K}+\sum_{i_1+i_2+i_3+i_4\leq i}(1+\nabring^{i_1}\psi^{i_2})\nabring^{i_3}(\nabring\beta)_\mck\nabring^{i_2}\gcheck\\
	&+\sum_{i_1+i_2\leq i-1}\nabring^{i_1}K\nabring^{i_2}\betacheck+\sum_{i_1+i_2+i_3\leq i,i_3\leq i-1}\nabring^{i_1}\psi^{i_2}\nabring^{i_3+1}\betacheck
\end{align*}
where $F=_s\psi_H K+\sum_{i_1+i_2+i_3=1}\psi^{i_1}\nabring^{i_2}\psi\nabring^{i_3}\psi_H$, and
\begin{align*}
	F_{\hodge{K}}^{(i)}=_s&\sum_{i_1+i_2+i_3=i}\nabring^{i_1}\psi^{i_2}\Big[\nabring^{i_3}(F'-F'_\mck)+\sum_{i_4+i_5+i_6=i_3}\Big(\nabring^{i_4}\gcheck\nabring^{i_5}F'_\mck\\
	&+(\nabring^{i_4}\gcheck\nabring^{i_5}\chi_\mck+\nabring^{i_4}\widecheck{\chi})\nabring^{i_6} \hodge K_\mck+\nabring^{i_4}\Omega^{-2}\nabring^{i_5}\bcheck\nabring^{i_6+1} \hodge K_\mck+\nabring^{i_4}\Omega^{-2}\nabring^{i_5+1}\bcheck\nabring^{i_6} \hodge K_\mck\Big)\Big]\\
	&+\sum_{i_1+i_2+i_3+i_4=i}\nabring^{i_1}\psi^{i_2}\nabring^{i_3}\chi\nabring^{i_4}\hodge\widecheck{K}+\sum_{i_1+i_2+i_3+i_4\leq i}(1+\nabring^{i_1}\psi^{i_2})\nabring^{i_3}(\nabring\beta)_\mck\nabring^{i_2}\gcheck\\
	&+\sum_{i_1+i_2\leq i-1}\nabring^{i_1}K\nabring^{i_2}\betacheck+\sum_{i_1+i_2+i_3\leq i,i_3\leq i-1}\nabring^{i_1}\psi^{i_2}\nabring^{i_3+1}\betacheck,
\end{align*}
where $F'=_s\sum_{i_1+i_2+i_3=1}\psi^{i_1}\nabring^{i_2}\psi\nabring^{i_3}\psi_H$. Finally, we define
\begin{align}
	F_{\betabar}^{(i)}=_s&\sum_{i_1+i_2+i_3=i}\nabring^{i_1}\psi^{i_2}\Bigg[\nabring^{i_3}(H-H_\mck)+\sum_{i_4+i_5+i_6+i_7=i_3}\Big(\nabring^{i_4}\gcheck\nabring^{i_5}H_\mck\nn\\
	&+(\nabring^{i_4}\gcheck\nabring^{i_5}\chi_\mck(1+\nabring^{i_6}\gamma_\mck^{-1})+\nabring^{i_4}\widecheck{\chi})\nabring^{i_7} \betabar_\mck+\nabring^{i_4}\Omega^{-2}\nabring^{i_5}\bcheck\nabring^{i_6+1} \betabar_\mck\nn\\
	&+\nabring^{i_4}\Omega^{-2}\nabring^{i_5+1}\bcheck\nabring^{i_6} \betabar_\mck\Big)\Bigg]+\sum_{i_1+i_2+i_3+i_4=i}\nabring^{i_1}\psi^{i_2}\nabring^{i_3}\chi\nabring^{i_4}\widecheck{\betabar}\nn\\
	&+\sum_{i_1+i_2+i_3\leq i,i_3\leq i-1}\nabring^{i_1}\psi^{i_2}\nabring^{i_3+1}(\Kcheck,\hodge{\Kcheck})+\sum_{i_1+i_2\leq i-1}\nabring^{i_1}K\nabring^{i_2}(\Kcheck,\hodge{\Kcheck})\nn\\
	&+\sum_{i_1+i_2+i_3+i_4\leq i,i_3+i_4\leq i-1}(1+\nabring^{i_1}\psi^{i_2})\nabring^{i_3}(\nabring\hodge{K})_\mck\nabring^{i_4}\gcheck\nn
\end{align}
where $H=_s\sum_{i_1+i_2+i_3=1}\psi^{i_1}\nabring^{i_2}\psi_H\nabring^{i_3}\psi_{\Hbar}+\psi K+\sum_{i_1+i_2=1}\psi^{i_1}\psi\nabring^{i_2}\psi$. Next we define
\begin{align*}
	G_K^{(i)}=_s&\sum_{i_1+i_2+i_3=i}\nabring^{i_1}\psi^{i_2}\Big[\nabring^{i_3}(G-G_\mck)+\sum_{i_4+i_5+i_6=i_3}\Big(\nabring^{i_4}\gcheck\nabring^{i_5}\chibar_\mck\nabring^{i_6}K_\mck\\
	&+\sum_{i_4+i_5=i_3}\nabring^{i_4}\widecheck{\chibar}\nabring^{i_5}K_\mck\Big)\Big]+\sum_{i_1+i_2+i_3+i_4=i}\nabring^{i_1}\psi^{i_2}\nabring^{i_3}\chibar\nabring^{i_4}\widecheck{K}\\
	&+\sum_{i_1+i_2+i_3+i_4\leq i}(1+\nabring^{i_1}\psi^{i_2})\nabring^{i_3}(\nabring\betabar)_\mck\nabring^{i_2}\gcheck\\
	&+\sum_{i_1+i_2\leq i-1}\nabring^{i_1}K\nabring^{i_2}\betabarcheck+\sum_{i_1+i_2+i_3\leq i,i_3\leq i-1}\nabring^{i_1}\psi^{i_2}\nabring^{i_3+1}\betabarcheck,
\end{align*}
where $G=_s\psi_{\Hbar} K+\sum_{i_1+i_2+i_3=i}\psi^{i_1}\nabring^{i_2}\psi\nabring^{i_3}\psi_{\Hbar}$. We also define 
\begin{align*}
	G_{\hodge{K}}^{(i)}=_s&\sum_{i_1+i_2+i_3=i}\nabring^{i_1}\psi^{i_2}\Big[\nabring^{i_3}(G'-G'_\mck)+\sum_{i_4+i_5+i_6=i_3}\Big(\nabring^{i_4}\gcheck\nabring^{i_5}\chibar_\mck\nabring^{i_6}\hodge{K}_\mck\\
	&+\sum_{i_4+i_5=i_3}\nabring^{i_4}\widecheck{\chibar}\nabring^{i_5}\hodge{K}_\mck\Big)\Big]+\sum_{i_1+i_2+i_3+i_4=i}\nabring^{i_1}\psi^{i_2}\nabring^{i_3}\chibar\nabring^{i_4}\widecheck{\hodge{K}}\\
	&+\sum_{i_1+i_2+i_3+i_4\leq i}(1+\nabring^{i_1}\psi^{i_2})\nabring^{i_3}(\nabring\betabar)_\mck\nabring^{i_2}\gcheck\\
	&+\sum_{i_1+i_2\leq i-1}\nabring^{i_1}K\nabring^{i_2}\betabarcheck+\sum_{i_1+i_2+i_3\leq i,i_3\leq i-1}\nabring^{i_1}\psi^{i_2}\nabring^{i_3+1}\betabarcheck,
\end{align*}
where $G'=_s\sum_{i_1+i_2+i_3=i}\psi^{i_1}\nabring^{i_2}\psi\nabring^{i_3}\psi_{\Hbar}$.
\begin{prop}\label{prop:equationsschdoublenull}
	Let $i\geq 0$. We have the following schematic equations for the linearized metric coefficients,
\begin{align*}
	&\nabring_3\nabring^i\widecheck{\gamma}=_sF_\gamma,\quad\nabring_3\nabring^i\widecheck{\log\Omega}=_sF_\Omega,\quad\nabring_3\nabring^i\widecheck{b}=_sF_b.
\end{align*}
Next, we have the following schematic equations for the linearized Ricci coefficients,
\begin{alignat*}{6}
	\nabring_3\nabring^i\widecheck{\etabar}&=_sF_{\etabar}^{(i)},\quad&\nabring_3\nabring^i\widecheck{\mubar}&=_sF_{\mubar}^{(i)},\quad&\nabring_4\nabring^i\widecheck{\eta}&=_sF_\eta^{(i)},\quad&\nabring_4\nabring^i\widecheck{\mu}&=_sF_\mu^{(i)},\\
	\nabring_3\nabring^i(\Omega^2_\mck\widecheck{\psi}_H)&=_sF_{\psi_H}^{(i)},\quad&\nabring_4\nabring^i\widecheck{\psi}_{\Hbar}&=_sF_{\psi_{\Hbar}}^{(i)},\quad&\nabring_4\nabring^i\widecheck{\omegabar}&=_s F_{\omegabar}^{(i)},\quad&
	\nabring_4\nabring^{i}\widecheck{\barred{\omegabar}}&=_sF_{\barred{\omegabar}}^{(i)}\:.
\end{alignat*}
Moreover, we have the following additional identities for $\widecheck{tr\chi},\widecheck{tr\chibar}$,
\begin{align*}
	\nabring_3(\Omega_\mck^{-2}\nabring^{N_2}\widecheck{tr\chibar})= \Omega_\mck^{-2} F_{tr\chibar}^{(N_2)},\quad\nabring_4(\nabring^{(N_2)}\widecheck{tr\chi})= F_{tr\chi}^{N_2}.
\end{align*}
Finally, we have the following schematic commuted Bianchi identities,
\begin{align*}
	\nabring_3\nabring^i_{A_1\cdots A_i}(\Omega\Omega_\mck\widecheck{\beta})_C+\nabring_C(\Omega\Omega_\mck\nabring^i_{A_1\cdots A_i}\widecheck{K})-\in_{CD}\nabring^D(\Omega\Omega_\mck\nabring^i_{A_1\cdots A_i}\hodge{\widecheck{K}})&=_sF_\beta^{(i)},\\
	\nabring_4\nabring^i_{A_1\cdots A_i}\hodge{\widecheck{K}}+\nabring^C\nabring^i_{A_1\cdots A_i}(\in_{CD}\widecheck{\beta}^D)&=_sF_{\hodge{K}}^{(i)},\\
	\nabring_4\nabring^i_{A_1\cdots A_i}{\widecheck{K}}+\nabring^C\nabring^i_{A_1\cdots A_i}\widecheck{\beta}_C&=_sF_{{K}}^{(i)},
\end{align*}
and 
\begin{align*}
	\nabring_4\nabring^i_{A_1\cdots A_i}\widecheck{\betabar}_C-\nabring_C\nabring^i_{A_1\cdots A_i}\widecheck{K}-\in_{CD}\nabring^D\nabring^i_{A_1\cdots A_i}\hodge{\widecheck{K}}&=_sF_{\betabar}^{(i)},\\
	\nabring_3\nabring^i_{A_1\cdots A_i}\hodge{\widecheck{K}}+\nabring^C\nabring^i_{A_1\cdots A_i}(\in_{CD}\widecheck{\betabar}^D)&=_sG_{\hodge{K}}^{(i)},\\
	\nabring_3\nabring^i_{A_1\cdots A_i}{\widecheck{K}}-\nabring^C\nabring^i_{A_1\cdots A_i}\widecheck{\betabar}&=_sG_{{K}}^{(i)}.
\end{align*}
\end{prop}
	
\begin{proof}
The commuted equations for the linearized metric and Ricci coefficients are direct consequences of Propositions \ref{prop:schemnab33} and \ref{prop:schemnab44}, except the one for $tr\chibar$ which is obtained similarly as in \cite[Prop. 7.19]{stabC0}, keeping the term $\nabring^{N_2}(\omegabar tr\chibar-\omegabar_\mck tr\chibar_\mck)$ on the LHS and noticing that in the commutator $[\nabring_3,\nabring_A]$ in Proposition \ref{prop:commnablaboudnenulll} there is no term with $\nabring_3$ derivative since $\mathring{\eta}=\mathring{\zeta}$, so that there is no additional term involving $\omegabar$ of $\omegabar_\mck$ in the RHS of the commuted equation for $tr\chibar$. The commuted Bianchi equations are obtained similarly as in \cite{stabC0}, using Propositions \ref{prop:schemnab33} and \ref{prop:schemnab44} and leaving the highest order $\nabring$ derivative of $\Kcheck,\hodge{\Kcheck},\betacheck,\betabarcheck$ on the LHS.
\end{proof}
\begin{proof}[Proof of Propositions \ref{prop:DNmetricnabla} to \ref{prop:DNbianchinabla}]
Combining Proposition \ref{prop:equationsschdoublenull} combined with the bootstrap assumption \eqref{eq:BAII}, and inspecting each terms in the schematic RHS terms $F$ as defined above, we observe that  for $i\leq N_2$ and $j\leq N_2-1$, we have the following reduced schematic identities 
\begin{align*}
	&F_\gamma^{(i)},F_\Omega^{(i)},F_{\etabar}^{(j)}=_{rs}\mcf_3',\: F_b^{(i)}=_{rs}\mcf_b,\: F_{\underline{\mu}}^{(j)}=_{rs}\mcf_3,\: F_\eta^{(j)}=_{rs}\mcf'_2,\: F_{\mu}^{(j)}=_{rs}\mcf_2,\: F_{\psi_H}^{(j)}=_{rs}\mcf'_1,\: \\
	&F_{\psi_{\Hbar}}^{(j)},F_{\omegabar}^{(j)}=_{rs}\mcf_4',\: F_{\barred{\omegabar}}^{(N_2-2)}=_{rs}\mcf_4,\: F_\beta^{(j)}=_{rs}\mcf_1,\: F_{\hodge{K}}^{(j)},F_K^{(j)}=_{rs}\mcf_2,\: F_{\betabar}^{(j)}=_{rs}\mcf_4,\: G_{\hodge{K}}^{(j)},G_K^{(j)}=_{rs}\mcf_3,\\
	& \Omega_\mck^{-2}F_{tr\chibar}^{(N_2)}=_{rs}\Omega_\mck^{-2}\sum_{{i_1+i_2+i_3\leq N_2}}(1+\nabring^{i_1}\gcheck+\nabring^{i_1-1}\psicheck)(\nabring^{i_2}\psicheck_{\Hbar}+\Omega_\mck^2)(\nabring^{i_3}(\psicheck_{\Hbar},\widecheck{\omegabar})+\Omega_\mck^2\nabring^{i_3}\gcheck),\\
	&F_{tr\chi}^{(N_2)}=_{rs}\sum_{{i_1+i_2+i_3\leq N_2}}(1+\nabring^{i_1}\gcheck+\nabring^{i_1-1}\psicheck)(\nabring^{i_2}\psicheck_H+1)(\nabring^{i_3}(\psicheck_H,\gcheck)+\Omega_\mck^{-2}\nabring^{i_3}\bcheck),
\end{align*}
which concludes the proof.
\end{proof}

\subsection{Reduced schematic equations and decay estimates for $\lieT^{k'}\nabring^k$ derivatives}\label{section:proofsschemIIlieT}

\subsubsection{Schematic equations satisfied by $\lieT^{k'}\nabring^k$ derivatives}\label{section:schematicequationslieT}
We begin with commutation identities between covariant derivatives and $\lieT$.
\begin{prop}\label{prop:commlieT}
	Let $U$ be a $S(u,\ubar)$-tangent $k$-tensor. We have the commutation formulas
	\begin{align}
		[\lieT,\nabring_3]U_{A_1\cdots A_k}&=-\sum_{i=1}^k\left((\lieT\widecheck{\gamma})^{BC}\mathring{\chibar}_{A_iC}+\gamma^{BC}(\lieT\widecheck{\mathring{\chibar}})_{A_iC}\right)U_{A_1\cdots B\cdots A_k},\label{eq:commlieTnab3}\\
		[\lieT,\nabring]U_{BA_1\cdots A_k}&=-\frac12\gamma^{CD}\sum_{i=1}^k\left(\nabring_{A_i}(\lieT\widecheck{\gamma})_{BD}+\nabring_{B}(\lieT\widecheck{\gamma})_{A_iD}-\nabring_{D}(\lieT\widecheck{\gamma})_{A_iB}\right)U_{A_1\cdots C\cdots A_k},\label{eq:commlieTnab}\\
		[\lieT,\Omega^2\nabring_4]U_{A_1\cdots A_k}&=\lieT\widecheck{b}\cdot\nabring U_{A_1\cdots A_k}\label{eq:commlieTnab4zero}\\
		&\quad-\sum_{i=1}^k\left((\lieT\widecheck{\gamma})^{BC}\Omega^2\mathring{\chi}_{A_iC}+\gamma^{BC}\lieT(\Omega^2\mathring{\chi})_{A_iC}-\nabring_{A_i}\lieT\widecheck{b}^B\right)U_{A_1\cdots B\cdots A_k}.\nn
	\end{align}
\end{prop}

\begin{rem}
	In particular, \eqref{eq:commlieTnab4zero} implies the identity
	\begin{align}
		[\lieT,\nabring_4]U_{A_1\cdots A_k}&=\Omega^{-2}\lieT\widecheck{b}\cdot\nabring U_{A_1\cdots A_k}-2\T(\widecheck{\log\Omega})\nabring_4U\label{eq:commlieTnab4}\\
		&\quad-\Omega^{-2}\sum_{i=1}^k\left((\lieT\widecheck{\gamma})^{BC}\Omega^2\mathring{\chi}_{A_iC}+\gamma^{BC}\lieT(\Omega^2\mathring{\chi})_{A_iC}-\nabring_{A_i}\lieT\widecheck{b}^B\right)U_{A_1\cdots B\cdots A_k}.\nn
	\end{align}
\end{rem}

\begin{proof}[Proof of Proposition \ref{prop:commlieT}]
	Note that for any $S(u,\ubar)$-tangent tensor $V$, we have 
	\begin{align}\label{eq:lieTA}
		\T(V_A)=(\lieT \widecheck{V})_A,
	\end{align}
	where we used $\mcl_\T(\partial_{\theta^A})=0$ and $\T(V_\mck^A)=0$ for any exact Kerr metric, Ricci, and curvature coefficient in the Kerr double null gauge since $\T$ is Killing in Kerr. Then, \eqref{eq:commlieTnab3} follows easily from \eqref{eq:nab3doublenul} and \eqref{eq:lieTA}. Next, from \eqref{eq:nabdoublenul} and \eqref{eq:lieTA} we get
	\begin{align*}
		[\lieT,\nabring]U_{BA_1\cdots A_k}=-\sum_{i=1}^k\T(\mathring{\Gamma}_{A_iB}^C)U_{A_1\cdots C\cdots A_k},
	\end{align*}
	where $\mathring{\Gamma}_{A_iB}^C$ is defined in \eqref{eq:christodoublenulspheres}. We now compute $\T(\mathring{\Gamma}_{A_iB}^C)$. By the identity above applied to $U=\gamma$, together with $\nabring\gamma=0$, we get $\nabring_A\lieT\widecheck{\gamma}_{BC}=\nabring_A\lieT\gamma_{BC}=\T(\mathring{\Gamma}_{AB}^D)\gamma_{CD}+\T(\mathring{\Gamma}_{AC}^D)\gamma_{BD}$, where we used the symetry of $\mathring{\Gamma}_{AB}^C$ in $A,B$ by \eqref{eq:christodoublenulspheres}. This implies the computation
	\begin{align*}
		&\gamma^{CD}\left(\nabring_A\lieT\widecheck{\gamma}_{BD}+\nabring_B\lieT\widecheck{\gamma}_{AD}-\nabring_D\lieT\widecheck{\gamma}_{AB}\right)=2\T(\mathring{\Gamma}_{AB}^C),
	\end{align*}
	which concludes the proof of \eqref{eq:commlieTnab}. Now, combining \eqref{eq:nab4doublenul} and \eqref{eq:lieTA} we get 
	\begin{align*}
		[\lieT,\Omega^2\nabring_4]U_{A_1\cdots A_k}&=\T(b^C)\partial_{\theta^C}U_{A_1\cdots A_k}\nn\\
		&\quad-\sum_{i=1}^k\left((\lieT\widecheck{\gamma})^{BC}\Omega^2\mathring{\chi}_{A_iC}+\gamma^{BC}\lieT(\Omega^2\mathring{\chi})_{A_iC}-\T(\partial_{\theta^{A_i}}\widecheck{b}^B)\right)U_{A_1\cdots B\cdots A_k}\\
		&=\lieT \widecheck{b}^C\left(\nabring_C U_{A_1\cdots A_k}+\sum_{i=1}^k\mathring{\Gamma}_{A_iC}^BU_{A_1\cdots B\cdots A_k}\right)\\
		&\quad-\sum_{i=1}^k\left((\lieT\widecheck{\gamma})^{BC}\Omega^2\mathring{\chi}_{A_iC}+\gamma^{BC}\lieT(\Omega^2\mathring{\chi})_{A_iC}-\T(\partial_{\theta^{A_i}}\widecheck{b}^B)\right)U_{A_1\cdots B\cdots A_k},
	\end{align*}
and $\T(\partial_{\theta^{A_i}}\widecheck{b}^B)+\lieT \widecheck{b}^C\mathring{\Gamma}_{A_iC}^B=\partial_{\theta^{A_i}}(\lieT \widecheck{b}^B)+\lieT \widecheck{b}^C\mathring{\Gamma}_{A_iC}^B=\nabring_{A_i}\lieT \widecheck{b}^B$ concludes the proof.
\end{proof}
\noindent Note that the results of Proposition \ref{prop:commlieT} can be rewritten schematically as follows,
\begin{align}
	[\lieT,\nabring_3]U&=_s\left(\lieT\widecheck{\gamma^{-1}}\chibar+\lieT\widecheck{\chibar}\right)\cdot U,\label{eq:commlieTschnab3}\\
	[\lieT,\nabring]U&=_s \nabring\lieT\widecheck{\gamma}\cdot U,\label{eq:commlieTschnab}\\
	[\lieT,\Omega^2\nabring_4]U&=_s\lieT\widecheck{b}\cdot\nabring U+\left(\lieT\widecheck{\gamma^{-1}}\Omega^2\chi+\lieT\widecheck{\Omega^2\chi}+\nabring\lieT\widecheck{b}\right)\cdot U.\label{eq:commlieTschnab4}
\end{align}
Moreover, the tensor $\lieT\gammacheck_{AB}$ can actually be expressed with the linearized quantities without $\lieT$ derivatives. Indeed we have
	\begin{align}\label{eq:expresssionlieTgamma}
		\lieT\widecheck{\gamma}_{AB}=\lieT\gamma_{AB}=\frac12\left(\widecheck{\Omega^2}(\chi_\mck)_{AB}+\Omega^2\widecheck{\chi}_{AB}-\widecheck{\chibar}_{AB}-\frac12\widecheck{\nabring}_{(A}(b_\mck)_{B)}-\frac12\nabring_{(A}\bcheck_{B)}\right),
	\end{align}
which comes from the expression for $\lie_L\gamma$, $\lie_{\underline{L}}\gamma$ in \cite[(2.15)]{stabC0}, combined with $\lie_b\gamma_{AB}=\nabring_A b_B+\nabring_B b_A$, the fact that $\lieT\gamma_\mck=0$, and the identity \eqref{eq:expreT}. We also have the following result for higher order commutations between $\lieT$ and $\nabring^k$.
\begin{lem}\label{lem:lemB21}
	Let $U$ be a $S(u,\ubar)$-tangent tensor. We have for $k\leq N_2$,
	$$[\lieT,\nabring^k]U=_s\sum_{i_1+i_2=k-1}\nabring^{i_1+1}\lieT\gammacheck\cdot\nabring^{i_2}U.$$
	In particular, in reduced schematic notation we have
	$$\lieT\nabring^kU=_{rs}\nabring^k\lieT U+\nabring^{\leq k} (\nabring^{\leq 1}\gcheck,\Omega^2\widecheck{\chi},\widecheck{\chibar},\nabring^{\leq 1}\bcheck)\cdot\nabring^{\leq k-1}U.$$
\end{lem}
\begin{proof}
	The proof of the first identity goes easily by induction using \eqref{eq:commlieTnab}, and the second identity comes from \eqref{eq:expresssionlieTgamma}.
\end{proof}
\begin{cor}\label{cor:gaufinaltcool}
	Let $U=U_\mck+\widecheck{U}$ be a $S(u,\ubar)$-tangent tensor, where $U_\mck$ is some background Kerr quantity satisfying $\lieT U_\mck=0$. Then we have 
	$$\lieT\nabring^k U=_s\lieT\nabring^k\widecheck{U}+\sum_{i_1+i_2=k-1}\nabring^{i_1+1}\lieT\gammacheck\cdot\nabring^{i_2}U_\mck.$$
\end{cor}
We also define the tensor 
$$\mathcal{W}:=\lieT\gammacheck,$$
and we recall the following schematic notation, for $i,j\geq 0$,
$$\lieT^{i}\gamma^{j}=_s\sum_{i_1+\cdots+i_j=i}\lieT^{i_1}\gamma\cdots\lieT^{i_j}\gamma,$$
where it is understood in the notations (including in the one for $\mathcal{W}$) above that $\gamma$ can represent either the metric $\gamma$ or the inverse metric $\gamma^{-1}$. Then we can write the following commutation formulas with higher-order $\lieT$ derivatives:
\begin{prop}\label{prop:commmmmmm}
	Let $k\geq 0$. Then, for a $S(u,\ubar)$-tangent tensor field $U$,
	\begin{align}
		[\lieT^{k},\nabring_3]U=_s&\sum_{i_1+i_2+i_3+i_4\leq k, i_4\leq k-1}\lieT^{i_1}\gamma^{i_2}\lieT^{i_3}\chibar\lieT^{i_4}U,\nn\\
		[\lieT^{k},\nabring]U=_s&\sum_{i_1+i_2+i_3+i_4\leq k-1}\lieT^{i_1}\gamma^{i_2}\lieT^{i_3}\nabring\mathcal{W}\lieT^{i_4}U,\nn\\
		[\lieT^{k},\Omega^2\nabring_4]U=_s&\sum_{i_1+i_2+i_3+i_4\leq k-1}\lieT^{i_1}\gamma^{i_2}\left(\lieT^{i_3+1}\bcheck\underline{\lieT^{i_4}\nabring U}+\underline{\lieT^{i_3+1}\nabring\bcheck}\lieT^{i_4}U\right)\label{eq:hellounderlined}\\
		&+\sum_{i_1+i_2+i_3+i_4\leq k-1}\lieT^{i_1}\gamma^{i_2}\lieT^{i_3+1}(\Omega^2\chi)\lieT^{i_4}U\nn\\
		&+\sum_{i_1+i_2+i_3+i_4+i_5\leq k-1}\lieT^{i_1}\gamma^{i_2}\lieT^{i_3}(\Omega^2\chi)\lieT^{i_4}\mathcal{W}\lieT^{i_5}U.\nn
 	\end{align}
\end{prop}
\begin{proof}
	The formulas above for $k=1$ reduce to \eqref{eq:commlieTschnab3}, \eqref{eq:commlieTschnab}, \eqref{eq:commlieTschnab4} thus hold. The formulas for general $k$ are proven relying on \eqref{eq:commlieTschnab3}, \eqref{eq:commlieTschnab}, \eqref{eq:commlieTschnab4} and the formula
	$$[\lieT^{k},\nabring_\mu]U=\sum_{i=0}^{k-1}\lieT^{k-1-i}[\lieT,\nabring_\mu]\lieT^i$$
	for $\mu=A,3,4$. Note that the terms of the form $\lieT^{i_1}\gamma^{i_2}$ are generated at each step because of interaction between the schematic notation $=_s$ and the $\lieT$ derivative, see Section \ref{section:schematicnotations}.
\end{proof}
We now write down for future reference the equations satisfied by $\lieT^{j}\nabring^i $ derivatives\footnote{We leave the commutators on the RHS but these can be expressed more precisely using Proposition \ref{prop:commmmmmm}.}.
\begin{prop}\label{prop:eqschemaliet}
	We have the following schematic equations,
	\begin{align*}
		&\nabring_3\lieT^{j}\nabring^i\widecheck{\gamma}=_s\lieT^{j} F_\gamma+[\nabring_3,\lieT^{j}]\nabring^i\widecheck{\gamma},\quad\nabring_3\lieT^{j}\nabring^i\widecheck{\log\Omega}=_s\lieT^{j} F_\Omega+[\nabring_3,\lieT^{j}]\nabring^i\widecheck{\log\Omega},\\
		&\nabring_3\lieT^{j}\nabring^i\widecheck{b}=_s\lieT^{j} F_b+[\nabring_3,\lieT^{j}]\nabring^i\widecheck{b},\quad\nabring_3\lieT^{j}\nabring^i\widecheck{\etabar}=_s\lieT^{j} F_{\etabar}+[\nabring_3,\lieT^{j}]\nabring^i\widecheck{\etabar},\\
		&\nabring_3\lieT^{j}\nabring^i(\Omega^2_\mck\widecheck{\psi}_H)=_s\lieT^{j} F_{\psi_H}+[\nabring_3,\lieT^{j}]\nabring^i(\Omega^2_\mck\widecheck{\psi}_H),\quad\Omega^2\nabring_4\lieT^{j}\nabring^i\widecheck{\eta}=_s\lieT^{j}(\Omega^2 F_\eta)+[\Omega^2\nabring_4,\lieT^{j}]\nabring^i\widecheck{\eta},\\
		&\Omega^2\nabring_4\lieT^{j}\nabring^i\widecheck{\psi}_{\Hbar}=_s\lieT^{j}(\Omega^2 F_{\psi_{\Hbar}})+[\Omega^2\nabring_4,\lieT^{j}]\nabring^i\widecheck{\psi}_{\Hbar},\quad\Omega^2\nabring_4\lieT^{j}\nabring^i\widecheck{\omegabar}=_s\lieT^{j}(\Omega^2 F_{\omegabar})+[\Omega^2\nabring_4,\lieT^{j}]\nabring^i\widecheck{\omegabar},\\
		&\Omega^2\nabring_4\lieT^{j}\nabring^i\widecheck{\barred{\omegabar}}=_s\lieT^{j}(\Omega^2 F_{\barred{\omegabar}})+[\Omega^2\nabring_4,\lieT^{j}]\nabring^i\widecheck{\barred{\omegabar}}.
	\end{align*}
We also have the additional reduced schematic equations for $\widecheck{tr\chi}$ and $\widecheck{tr\chibar}$:
\begin{align*}
	\nabring_3(\Omega_\mck^{-2}\lieT^{ j}\nabring^{i}\widecheck{tr\chibar})&=_s\Omega_\mck^{-2}\lieT^{j} F_{tr\chibar}+[\nabring_3,\lieT^{j}]\nabring^i\widecheck{tr\chibar},\\
	\Omega^2\nabring_4(\lieT^{ j}\nabring^{i}\widecheck{tr\chi})&=_s\lieT^{j}(\Omega^2 F_{tr\chi})+[\Omega^2\nabring_4,\lieT^{j}]\nabring^i\widecheck{tr\chibar}.
\end{align*}
We now write the schematic Bianchi equations:
\begin{align*}
	&\nabring_3\lieT^{j}\nabring^{i}_{A_1\cdots A_{i}}(\Omega\Omega_\mck\widecheck{\beta})_C+\nabring_C(\Omega\Omega_\mck\lieT^{j}\nabring^{i}_{A_1\cdots A_{i}}\widecheck{K})-\in_C^D\nabring_D(\Omega\Omega_\mck\lieT^{j}\nabring^{i}_{A_1\cdots A_{i}}\hodge{\widecheck{K}})=_s\\
	&\lieT^j F_{\beta}+[\nabring_3,\lieT^j]\nabring^i(\Omega\Omega_\mck\betacheck)+[\nabring\Omega\Omega_\mck,\lieT^j]\nabring^{i}\widecheck{K}+[\in^D_C\nabring_D\Omega\Omega_\mck,\lieT^j]\nabring^{i}\widecheck{\hodge{K}},\\
	&\Omega^2\nabring_4\lieT^{j}\nabring^{i}_{A_1\cdots A_{i}}\hodge{\widecheck{K}}+\Omega^2\nabring^C\lieT^{j}\nabring^{i}_{A_1\cdots A_{i}}(\in_C^D\widecheck{\beta}_D)=_s\lieT^j(\Omega^2 F_{K})+[\Omega^2\nabring_4,\lieT^j]\nabring^{i}\widecheck{K}+[\Omega^2\nabring,\lieT^j]\nabring^i\betacheck,\\
	&\Omega^2\nabring_4\lieT^{j}\nabring^{i}_{A_1\cdots A_{i}}{\widecheck{K}}+\Omega^2\nabring^C\lieT^{j}\nabring^{i}_{A_1\cdots A_{i}}\widecheck{\beta}_C=_s\lieT^j(\Omega^2 F_{\hodge{K}})+[\Omega^2\nabring_4,\lieT^j]\nabring^{i}\widecheck{\hodge{K}}+[\Omega^2\nabring,\lieT^j]\nabring^i\betacheck,
\end{align*}
as well as 
\begin{align*}
	&\Omega^2\nabring_4\lieT^{j}\nabring^{i}_{A_1\cdots A_{i}}\widecheck{\betabar}_C+\Omega^2\nabring_C\lieT^{j}\nabring^{i}_{A_1\cdots A_{i}}\widecheck{K}-\Omega^2\in_C^D\nabring_D\lieT^{j}\nabring^{i}_{A_1\cdots A_{i}}\hodge{\widecheck{K}}=_s\\
	&\lieT^j(\Omega^2 F_{\betabar})+[\Omega^2\nabring_4,\lieT^j]\nabring^i\betabarcheck+[\Omega^2\nabring,\lieT^j]\nabring^{i}\widecheck{K}+[\Omega^2\in^D_C\nabring_D,\lieT^j]\nabring^{i}\widecheck{\hodge{K}},\\
	&\nabring_3\lieT^{j}\nabring^{i}_{A_1\cdots A_{i}}\hodge{\widecheck{K}}+\nabring^C\lieT^{j}\nabring^{i}_{A_1\cdots A_{i}}(\in_C^D\widecheck{\betabar}_D)=_s\lieT^j G_{\hodge{K}}+[\nabring_3,\lieT^j]\nabring^{i}\widecheck{\hodge{K}}+[\nabring,\lieT^j]\nabring^i\betabarcheck,\\
	&\nabring_3\lieT^{j}\nabring^{i}_{A_1\cdots A_{i}}{\widecheck{K}}+\nabring^C\lieT^{j}\nabring^{i}_{A_1\cdots A_{i}}\widecheck{\betabar}_C=_s\lieT^{j}G_{{K}}+[\nabring_3,\lieT^j]\nabring^{i}\widecheck{{K}}+[\nabring,\lieT^j]\nabring^i\betabarcheck.
\end{align*}
\end{prop}
\begin{proof}
	This is a trivial consequence of Proposition \ref{prop:equationsschdoublenull}.
\end{proof}

We also define, for $k,k'\geq 0$,
\begin{align*}
	\mcn_{int}^{(k',k)}:=\sum_{j\leq k',i\leq k}\Big(&\||u|^{\frac{5}{2}+\frac{\delta}{3}}\varpi^N\Omega_\mck\lieT^{j}\nabring^i(\psicheck_{\Hbar},\widecheck{\omegabar})\|_{L^2_u L^2_\ubar L^2_S}^2+\|\ubar^{\frac{5}{2}+\frac{\delta}{3}}\varpi^N\Omega_\mck^3\lieT^{j}\nabring^i\psicheck_{H}\|_{L^2_u L^2_\ubar L^2_S}^2\\
	&+\|\ubar^{\frac{5}{2}+\frac{\delta}{3}}\varpi^N\Omega_\mck(\lieT^{j}\nabring^i(\psicheck,\gcheck,\bcheck),\lieT^{j}\nabring^{\min(i,k-1)}\Kcheck)\|_{L^2_u L^2_\ubar L^2_S}^2\Big),
\end{align*}
as well as
\begin{align*}
	&\mcn_{hyp}^{(k',k)}:=\\
	&\sum_{j\leq k',i\leq k-1}\Big(\||u|^{\frac{5}{2}+\frac{\delta}{3}}\varpi^N\lieT^{j}\nabring^i(\psicheck_{\Hbar},\widecheck{\omegabar})\|^2_{L^2_u L^\infty_\ubar L^2_S}+\|\ubar^{\frac{5}{2}+\frac{\delta}{3}}\varpi^N\Omega_\mck\lieT^{j}\nabring^i\etacheck\|^2_{L^2_u L^\infty_\ubar L^2_S}\\
	&+\||u|^{\frac{5}{2}+\frac{\delta}{3}}\varpi^N\Omega_\mck(\lieT^{j}\nabring^i\etabarcheck,\lieT^{j}\nabring^{\min(i,k-2)}\Kcheck)\|^2_{L^\infty_\ubar L^2_u  L^2_S}+\|\ubar^{\frac{5}{2}+\frac{\delta}{3}}\varpi^N\Omega_\mck^2\lieT^{j}\nabring^i\psicheck_H\|^2_{L^2_\ubar L^\infty_u L^2_S}\\
	&+ \|\ubar^{\frac{5}{2}+\frac{\delta}{3}}\varpi^N\Omega_\mck(\lieT^{j}\nabring^i\etacheck,\lieT^{j}\nabring^i\etabarcheck,\lieT^{j}\nabring^{\min(i,k-2)}\Kcheck)\|^2_{L^2_\ubar L^\infty_u  L^2_S}\Big)\\
	&+\||u|^{\frac{5}{2}+\frac{\delta}{3}}\varpi^N\lieT^{j}\nabring^{k}(\psicheck_{\Hbar},\widecheck{\omegabar})\|^2_{L^\infty_\ubar L^2_u  L^2_S}+\|\ubar^{\frac{5}{2}+\frac{\delta}{3}}\varpi^N\Omega_\mck^2\lieT^{j}\nabring^{k}\psicheck_H\|^2_{ L^\infty_u L^2_\ubar L^2_S}\\
	&+\||u|^{\frac{5}{2}+\frac{\delta}{3}}\varpi^N\Omega_\mck(\lieT^{j}\nabring^{k}\etabarcheck,\lieT^{j}\nabring^{k-1}\Kcheck)\|^2_{L^\infty_u L^2_\ubar  L^2_S}+\|\ubar^{\frac{5}{2}+\frac{\delta}{3}}\varpi^N\Omega_\mck(\lieT^{j}\nabring^{k}\etacheck,\lieT^{j}\nabring^{k-1}\Kcheck)\|^2_{L^\infty_\ubar L^2_u L^2_S}\\
	&+\sum_{i\leq k}\Big(\|\ubar^{\frac{5}{2}+\frac{\delta}{3}}\varpi^N\Omega_\mck\lieT^{j}\nabring^i\gcheck\|_{L^2_\ubar L^\infty_u L^2_S}^2+\||u|^{\frac{5}{2}+\frac{\delta}{3}}\varpi^N\Omega_\mck\lieT^{j}\nabring^i\gcheck\|_{L^\infty_\ubar L^2_u  L^2_S}^2+\|\ubar^{\frac{5}{2}+\frac{\delta}{3}}\varpi^N\lieT^{j}\nabring^i\bcheck\|_{L^2_\ubar L^\infty_u  L^2_S}^2\Big),
\end{align*}
and
\begin{align*}
	\mcn_{sph}^{(k',k)}:=&\|\lieT^{\leq k'}\nabring^{\leq k}\gcheck\|^2_{L^\infty_u L^\infty_\ubar L^2_S}+\|\lieT^{\leq k'}\nabring^{\leq k-1}(\psicheck_{\Hbar},\widecheck{\omegabar},\psicheck,\bcheck)\|^2_{L^\infty_u L^\infty_\ubar L^2_S}\\
	&+\|\Omega_\mck^2\lieT^{\leq k'}\nabring^{\leq k-1}\psicheck_H\|^2_{L^\infty_u L^\infty_\ubar L^2_S}+\|\lieT^{\leq k'}\nabring^{\leq k-2}\Kcheck\|^2_{L^\infty_u L^\infty_\ubar L^2_S}.
\end{align*}

\subsubsection{Proof of Proposition \ref{prop:inductionliekprim}}\label{section:proofinducliekprim}

In all this section, we assume $H(k'-1,k_0)$ and we introduce some $k\leq k_0$ (which will be chosen later as $k=k_0-5$ to show that $H(k',k)$ holds). To this end we compute the reduced schematic form of the RHS of the schematic equations for $\lieT^{k'}\nabring^k$ derivatives in Proposition \ref{prop:eqschemaliet}. We define the following inhomogeneous terms which generalize the ones in Section \ref{section:schematicequationsangular},

\begin{align*}
	\mathcal{F}_1^{(k',k)}=&\Omega_\mck^2\sum_{\underset{j_1+j_2\leq k'}{i_1+i_2\leq k}}\left(1+\lieT^{j_1}\nabring^{i_1}(\gcheck,\psicheck)\right)\Big(\lieT^{j_2}\nabring^{i_2}(\gcheck,\psicheck,\psicheck_{\Hbar},\widecheck{\omegabar})+\Omega_\mck^2\lieT^{j_2}\nabring^{i_2}\psicheck_H+\lieT^{j_2}\nabring^{i_2-1}\Kcheck\Big)\\
	&+\Omega_\mck^2\sum_{\underset{j_1+j_2+j_3\leq k'}{i_1+i_2+i_3\leq k}}(1+\lieT^{j_1}\nabring^{i_1}\gcheck+\lieT^{j_1}\nabring^{\min(i_1,k-1)}\psicheck)\lieT^{j_2}\nabring^{i_2}(\psicheck_{\Hbar},\widecheck{\omegabar})\lieT^{j_3}\nabring^{i_3}\psicheck_H,\\
	\mcf_2^{(k',k)}=&\sum_{\underset{j_1+j_2\leq k'}{i_1+i_2\leq k}}(1+\LN{1}(\gcheck,\psicheck))(\Omega_\mck^{-2}\LN{2}\bcheck+\LN{2}(\gcheck,\psicheck,\psicheck_H)+\lieT^{j_2}\nabring^{i_2-1}\Kcheck)\\
	&+\sum_{\underset{j_1+j_2\leq k'}{i_1+i_2\leq k-1}}\lieT^{j_1}\nabring^{i_1}\Kcheck\lieT^{j_2}\nabring^{i_2}\psicheck_{H},\\
	\mathcal{F}_3^{(k',k)}=&\sum_{\underset{j_1+j_2\leq k'}{i_1+i_2\leq k}}\left(1+\lieT^{j_1}\nabring^{i_1}(\gcheck,\psicheck)\right)\left(\Omega_\mck^2\lieT^{j_2}\nabring^{i_2}(\gcheck,\psicheck)+\lieT^{j_2}\nabring^{i_2}(\psicheck_{\Hbar},\widecheck{\omegabar})+\Omega_\mck^2\lieT^{j_2}\nabring^{i_2-1}\Kcheck\right)\\
	&+\sum_{\underset{j_1+j_2\leq k'}{i_1+i_2\leq k-1}}\lieT^{j_1}\nabring^{i_1}\Kcheck\lieT^{j_2}\nabring^{i_2}\psicheck_{\Hbar},\\
	\mcf_4^{(k',k)}&=\\
	&\sum_{\underset{j_1+j_2\leq k'}{i_1+i_2\leq k}}(1+\LN{1}\gcheck+\lieT^{j_1}\nabring^{\min(i_2,k-1)}\psicheck)(\LN{2}(\gcheck,\bcheck,\psicheck,\psicheck_{\Hbar},\widecheck{\omegabar})+\Omega_\mck^2\LN{2}\psicheck_H+\lieT^{j_2}\nabring^{i_2-1}\Kcheck)\\
	&+\sum_{\underset{j_1+j_2+j_3\leq k'}{i_1+i_2+i_3\leq k}}(1+\LN{1}\gcheck+\lieT^{j_1}\nabring^{\min(i_1,k-1)}\psicheck)\LN{2}(\psicheck_{\Hbar},\widecheck{\omegabar})(\LN{3}\psicheck_H+\Omega_\mck^{-2}\LN{3}\bcheck),
\end{align*}
as well as the slightly modified inhomogeneous terms,
\begin{align*}
	\mcf_1'^{(k',k)}=&\Omega_\mck^2\lieT^{\leq k'}\nabring^{k}(\gcheck,\etacheck)+\Omega^2_\mck\lieT^{\leq k'}\nabring^{k-1}\Kcheck\\
	&+\Omega_\mck^2\sum_{\underset{j_1+j_2\leq k'}{i_1+i_2\leq k-1}}\big(1+\LN{1}\gcheck\big)\big(\LN{2}(\gcheck,\psicheck,\psicheck_{\Hbar},\widecheck{\omegabar})+\Omega_\mck^2\LN{2}\psicheck_H\big)\\
	&+\Omega_\mck^2\sum_{\underset{j_1+j_2+j_3\leq k'}{i_1+i_2+i_3\leq k-1}}(1+\LN{1}(\gcheck,\psicheck))\LN{2}(\psicheck_{\Hbar},\widecheck{\omegabar})\LN{3}\psicheck_H,\\
	\mcf_2'^{(k',k)}=&\sum_{\underset{j_1+j_2\leq k'}{i_1+i_2\leq k}}(1+\LN{1}\gcheck+\lieT^{j_1}\nabring^{\min(i_1,k-1)}\psicheck)\\
	&\quad\quad\times(\Omega_\mck^{-2}\LN{2}\bcheck+\lieT^{j_2}\nabring^{\min(i_2,k-1)}\psicheck+\LN{2}(\gcheck,\psicheck_H)),
\end{align*}
\begin{align*}
	\mathcal{F}_3'^{(k',k)}=&\sum_{\underset{j_1+j_2\leq k'}{i_1+i_2\leq k}}\left(1+\lieT^{j_1}\nabring^{i_1}\gcheck+\lieT^{j_1}\nabring^{\min(i_1,k-1)}\psicheck\right)\\
	&\quad\quad\times\left(\Omega_\mck^2\lieT^{j_2}\nabring^{i_2}\gcheck+\Omega_\mck^2\lieT^{j_2}\nabring^{\min(i_2,k-1)}\psicheck+\lieT^{j_2}\nabring^{i_2}(\psicheck_{\Hbar},\widecheck{\omegabar})\right),\\
	\mcf_4'^{(k',k)}=&\lieT^{\leq k'}\nabring^{k}(\gcheck,\etabarcheck)+\lieT^{\leq k'}\nabring^{k-1}\Kcheck\\
	&+\sum_{\underset{j_1+j_2\leq k'}{i_1+i_2\leq k-1}}(1+\LN{1}(\gcheck,\psicheck))(\LN{2}(\gcheck,\bcheck,\psicheck,\psicheck_{\Hbar},\widecheck{\omegabar})+\Omega_\mck^2\LN{2}\psicheck_H)\\
	&+\sum_{\underset{j_1+j_2+j_3\leq k'}{i_1+i_2+i_3\leq k-1}}(1+\LN{1}\gcheck+\lieT^{j_1}\nabring^{\min(i_1,k-1)}\psicheck)\LN{2}(\psicheck_{\Hbar},\widecheck{\omegabar})(\LN{3}\psicheck_H+\Omega^{-2}_\mck\LN{3}\bcheck).
\end{align*}
We also introduce a specific inhomogeneous term for the equation for $\bcheck$,
\begin{align*}
	\mcf_b^{(k',k)}:=&\sum_{\underset{j_1+j_2\leq k'}{i_1+i_2\leq k}}(1+\LN{1}(\gcheck,\psicheck))(\Omega^2_\mck\LN{2}(\gcheck,\psicheck,\bcheck))\\
	&+\sum_{\underset{j_1+j_2+j_3\leq k'}{i_1+i_2+i_3\leq k}}(1+\LN{1}\gcheck+\lieT^{j_1}\nabring^{\min(i_1,k-1)}\psicheck)\LN{2}\psicheck_{\Hbar}\LN{3}\bcheck.
\end{align*}

We can now write the reduced schematic equations satisfied by $\lieT$ derivatives of the linearized quantities. For every equation, we use the schematic equations for $\lieT^{k'}\nabring^k$ derivatives written in Proposition \ref{prop:eqschemaliet} and we carefully inspect each RHS by using the commutation identities in Proposition \ref{prop:commmmmmm} combined with the following observations :
\begin{itemize}
	\item  The terms with $\lieT^{k'}\nabring^{\leq k}$ derivatives are left untouched as they are the one to be controlled (for a good choice of $k\leq k_0$).
	\item The terms with $\lieT^{k'-1}\nabring^{\leq k}$ derivatives (which are factors in front of the linear terms mentioned in the item above) can be controlled in $L^\infty$ by the assumption $H(k'-1,k_0)$ in Proposition \ref{prop:inductionliekprim}, because $k\leq k_0$.
	\item The terms with one extra $\nabring$ derivative, that are only present for quantities which satisfy a $\nabring_4$ derivative, and which come from the first underlined term in the commutator \eqref{eq:hellounderlined}, are error terms which can be treated by non-linear estimates. Note that in the commuted equations for $\etacheck,\mucheck,\hodge{\Kcheck},\Kcheck$, using $H(k'-1,k_0)$ to get ${\lieT^{i_4}\nabring U}=O(1)$ in that context in \eqref{eq:hellounderlined} yields that these terms take the form $\mcf_2'^{(k',k)}$. But for $(\psicheck_{\Hbar},\widecheck{\omegabar}),\widecheck{tr\chi},\widecheck{\barred{\omegabar}},\widecheck{\betabar}$ these error terms do not directly take the form $\mcf_4'^{(k',k)},\mcf_{tr\chi}^{(k',k)}$ so we keep them on the RHS and bound them  by non-linear estimates.\footnote{Note however that the second term underlined in \eqref{eq:hellounderlined}, (and the other terms in $\Omega_\mck^{-2}[\Omega^2\nabring_4,\lieT^{k'}]$) is always acceptable because $i_1,i_2,i_4\leq k'-1$ so $\lieT^{i_1}\gamma^{i_4}\lieT^{i_4}U$ is bounded by $H(k'-1,k_0)$, so that this term is linear in $\lieT^{\leq k'}\nabring^{\leq 1}\bcheck$ which can be written as $\mcf_i'^{(k',k)}$ with appropriate $i$.}
\end{itemize}
Using these observations, and defining the specific commutator term
$$\mathrm{Com}[U]:=\Omega_\mck^{-2}\sum_{i_1+i_2+i_3+i_4\leq k'-1}\lieT^{i_1}\gamma^{i_2}\lieT^{i_3+1}\bcheck{\lieT^{i_4}\nabring U}$$
 we get the following reduced schematic equations: For $i\leq k,j\leq k'$,
	\begin{align*}
		\nabring_3\lieT^j\nabring^i\gammacheck=_{rs}\mathcal{F}_3'^{(k',k)},\quad\nabring_3\lieT^j\nabring^i\logomegacheck=_{rs}\mathcal{F}_3'^{(k',k)},\quad\nabring_3\lieT^j\nabring^i\bcheck=_{rs}\mcf_b^{(k',k)}.
	\end{align*}
	For $i\leq k-2,j\leq k'$, $\Kcheck$ satisfies
	$$\lieT^j\nabring^i\Kcheck=_{rs}\lieT^j\nabring^{\leq i+2}\gcheck.$$
	For $i\leq k-1,j\leq k'$, 
	\begin{alignat*}{3}
		\nabring_3\lieT^j\nabring^i\etabarcheck&=_{rs}\mcf_3'^{(k',k)},\quad&\nabring_3\lieT^j\nabring^i\widecheck{\underline{\mu}}&=_{rs}\mcf_3^{(k',k)},\\
		\nabring_4\lieT^j\nabring^i\etacheck&=_{rs}\mcf_2'^{(k',k)},\quad&\nabring_4\lieT^j\nabring^i\widecheck{\mu}&=_{rs}\mcf_2^{(k',k)},\\
		\nabring_3\lieT^j\nabring^i(\Omega_\mck^2\psicheck_H)&=_{rs}\mcf_1'^{(k',k)},\quad&\nabring_4\lieT^j\nabring^i(\psicheck_{\Hbar},\widecheck{\omegabar})&=_{rs}\mcf_4'^{(k',k)}+\mathrm{Com}[\nabring^i(\psicheck_{\Hbar},\widecheck{\omegabar})].
	\end{alignat*}
	Moreover, $\lieT^{\leq k'}\nabring^{k-2}\widecheck{\barred{\omegabar}}$ satisfies the following reduced schematic equation,
	$$	\nabring_4\lieT^{\leq k'}\nabring^{k-2}\widecheck{\barred{\omegabar}}=_{rs}\mcf_4^{(k',k)}+\mathrm{Com}[\nabring^{k-2}\widecheck{\barred{\omegabar}}].$$
	We also have the additional reduced schematic equations for $\widecheck{tr\chi}$ and $\widecheck{tr\chibar}$:
	\begin{align*}
		\nabring_3(\Omega_\mck^{-2}\lieT^{\leq k'}\nabring^{k}\widecheck{tr\chibar})=_{rs}\Omega_\mck^{-2}\sum_{\underset{j_1+j_2+j_3\leq k'}{i_1+i_2+i_3\leq k}}&(1+\LN{1}\gcheck+\lieT^{j_1}\nabring^{i_1-1}\psicheck)\\
		&\times(\LN{2}\psicheck_{\Hbar}+\Omega_\mck^2)(\LN{3}(\psicheck_{\Hbar},\widecheck{\omegabar})+\Omega_\mck^2\LN{3}\gcheck),
	\end{align*}
	as well as
	\begin{align*}
		\nabring_4\lieT^{\leq k'}\nabring^{k}\widecheck{tr\chi}=_{rs}\sum_{\underset{j_1+j_2+j_3\leq k'}{i_1+i_2+i_3\leq k}}&(1+\LN{1}\gcheck+\lieT^{j_1}\nabring^{i_1-1}\psicheck)\\
		&\times(\LN{2}\psicheck_H+1)(\LN{3}(\psicheck_H,\gcheck)+\Omega_\mck^{-2}\LN{3}\bcheck)+\mathrm{Com}[\nabring^{k}\widecheck{tr\chi}],
	\end{align*}
We now write the reduced schematic Bianchi equations:
	\begin{align*}
		\nabring_3\lieT^{\leq k'}\nabring^{k-1}_{A_1\cdots A_{k-1}}(\Omega\Omega_\mck\widecheck{\beta})_C+\nabring_C(\Omega\Omega_\mck\lieT^{\leq k'}\nabring^{k-1}_{A_1\cdots A_{k-1}}\widecheck{K})&\\
		-\in_C^D\nabring_D(\Omega\Omega_\mck\lieT^{\leq k'}\nabring^{k-1}_{A_1\cdots A_{k-1}}\hodge{\widecheck{K}})&=_{rs}\mcf_1^{(k',k)},\\
		\nabring_4\lieT^{\leq k'}\nabring^{k-1}_{A_1\cdots A_{k-1}}\hodge{\widecheck{K}}+\nabring^C\lieT^{\leq k'}\nabring^{k-1}_{A_1\cdots A_{k-1}}(\in_C^D\widecheck{\beta}_D)&=_{rs}\mcf_2^{(k',k)},\\
		\nabring_4\lieT^{\leq k'}\nabring^{k-1}_{A_1\cdots A_{k-1}}{\widecheck{K}}+\nabring^C\lieT^{\leq k'}\nabring^{k-1}_{A_1\cdots A_{k-1}}\widecheck{\beta}_C&=_{rs}\mcf_2^{(k',k)},
	\end{align*}
	as well as 
	\begin{align*}
		\nabring_4\lieT^{\leq k'}\nabring^{k-1}_{A_1\cdots A_{k-1}}\widecheck{\betabar}_C+\nabring_C\lieT^{\leq k'}\nabring^{k-1}_{A_1\cdots A_{k-1}}\widecheck{K}-\in_C^D\nabring_D\lieT^{\leq k'}\nabring^{k-1}_{A_1\cdots A_{k-1}}\hodge{\widecheck{K}}&=_{rs}\mcf_4^{(k',k)}+\mathrm{Com}[\nabring^{k-1}\widecheck{\betabar}],\\
		\nabring_3\lieT^{\leq k'}\nabring^{k-1}_{A_1\cdots A_{k-1}}\hodge{\widecheck{K}}+\nabring^C\lieT^{\leq k'}\nabring^{k-1}_{A_1\cdots A_{k-1}}(\in_C^D\widecheck{\betabar}_D)&=_{rs}\mcf_3^{(k',k)},\\
		\nabring_3\lieT^{\leq k'}\nabring^{k-1}_{A_1\cdots A_{k-1}}{\widecheck{K}}+\nabring^C\lieT^{\leq k'}\nabring^{k-1}_{A_1\cdots A_{k-1}}\widecheck{\betabar}_C&=_{rs}\mcf_3^{(k',k)}.
	\end{align*}
Notice that this is a system of equations which is written exactly like the one for the $\nabring^k$ derivatives in Section \ref{section:DLmoredecaynabring}, except that there are some additional commutator terms in the RHS. However, although these Com terms mix the quantities with $\lieT^{\leq k'}\nabring^{\leq k}$ derivatives (which we want to control) and the ones with $\lieT^{\leq k'-1}\nabring^{\leq k+1}$ derivatives (which are controlled by $H(k'-1,k_0)$), these terms for $(\psicheck_{\Hbar},\widecheck{\omegabar}),\widecheck{\barred{\omegabar}},\widecheck{\betabar}$ give rise to error terms $\mathcal{T}'_{1}$ analog to $\mathcal{T}_1$, that can be bounded by the non-linear estimates for $\mathcal{T}_{1}$ in \cite[Section 12.3]{stabC0}, as in Proposition \ref{prop:controlerrorIInabla}. More precisely, using $H(k'-1,k_0)$, via Hölder inequalities as in the proof of \cite[Prop. 12.4]{stabC0} we get for $k=k_0-1$,
$$\||u|^{5+\frac{2\delta}{3}}\varpi^{2N}\Omega_\mck^2\mathcal{T}'_1\|_{L^1_u L^1_\ubar L^1_S}\lesssim(\mcn_{hyp}^{(k'-1, k_0)})^{1/2}\mcn_{hyp}^{(k', k_0-1)}\lesssim\varepsilon\mcn_{hyp}^{(k', k_0-1)}.$$
Concerning $\widecheck{tr\chi}$, the corresponding Com term can be treated with estimates as in \cite[Prop. 11.3]{stabC0} to control $\lieT^{k'}\nabring^k\widecheck{tr\chi}$ using $H(k'-1,k_0)$. These observations show that applying the same Dafermos-Luk estimates \cite{stabC0} as we did in Section \ref{section:DLmoredecaynabring} for controlling pure $\nabring^k$ derivatives, for $\varepsilon$ small enough we get the same conclusion as in Theorem \ref{thm:controlangulairedoublenull}:
\begin{align}\label{eq:mololo}
	\mcn_{int}^{(k', k_0-1)}+\mcn_{hyp}^{(k', k_0-1)}+\mcn_{sph}^{(k', k_0-1)}\lesssim\varepsilon^2,
\end{align}
(which also implies $\mcn_{int}^{(k', k_0-5)}+\mcn_{hyp}^{(k', k_0-5)}+\mcn_{sph}^{(k', k_0-5)}\lesssim\varepsilon^2$). Now, proceeding exactly as in the proof of Proposition \ref{prop:L2decaydoublenullangular} (commuting some $\lieT$ and angular derivatives which only produces lower-order terms by \eqref{eq:commlieTschnab}), re-integrating the equations satisfied by $\gcheck$, $\psicheck_H$, $\psicheck_{\Hbar}$, $\psicheck$, $\bcheck$, $\Kcheck$ and using \eqref{eq:mololo} we get

\begin{align*}
	\|\lieT^{\leq k'}\nabring^{\leq k_0-3}\gcheck\|_{L^2(S(u,\ubar))}+\|\lieT^{\leq k'}\nabring^{\leq k_0-4}(\psicheck_{\Hbar},\widecheck{\omegabar},\psicheck)\|_{L^2(S(u,\ubar))}+\|\lieT^{\leq k'}\nabring^{\leq k_0-5}\Kcheck\|_{L^2(S(u,\ubar))}&\lesssim\frac{\varepsilon}{|u|^{2+\frac{\delta}{3}}},\\
	\|\Omega_\mck^2\lieT^{\leq k'}\nabring^{\leq k_0-4}\psicheck_H\|_{L^2(S(u,\ubar))}+\|\lieT^{\leq k'}\nabring^{\leq k_0-4}\bcheck\|_{L^2(S(u,\ubar))}&\lesssim\frac{\varepsilon}{\ubar^{2+\frac{\delta}{3}}}.
\end{align*}

Finally, using the Sobolev embedding of Proposition \ref{prop:sobolevdoublenullII} (commuting once again some $\lieT$ and $\nabring$ derivatives) which loses two more derivatives, we get that $H(k',k_0-5)$ holds, which conclude the proof of Proposition \ref{prop:inductionliekprim}.

\begin{rem}
	 Note that here, to control the initial data for $\lieT^{k'}\nabring^k$ derivatives of the linearized quantities on $\Sigma_0$, we used the fact that $\T=\frac12(\partial_\ubar-\partial_u)$ is initially tangent to $\Sigma_0=\{u+\ubar=C_R\}$. Indeed, using also \eqref{eq:pourrelatergeompasgeom} to express the linearized quantities with respect to $\hat{g},\hat{k}$ on $\Sigma_0$, this implies that the $\lieT^{k'}\nabring^k$ derivatives of the linearized
	quantities are initially bounded by the tangential $\hat{\nabring}_\mck$ derivatives of $\hat{g},\hat{k}$, thus decays like $\varepsilon\ubar^{-3-\delta/2}$ by Propositions \ref{prop:bornegchecksig0}, \ref{prop:bornekchecksig0}.
\end{rem}

\subsection{Proof of Theorem \ref{thm:controldoublenull}}\label{section:proofthmdoublenull}
We define the following set of renormalized bounded linearized quantities,
		$$\widecheck{S}=\big\{\gammacheck,\logomegacheck, \bcheck,\etabarcheck,\etacheck,\Omega_\mck^2\psicheck_H,\psicheck_{\Hbar},\widecheck{\omegabar},\Omega\Omega_\mck\betacheck,\Kcheck,\hodge{\Kcheck},\betabarcheck\big\},$$
and well as the non-linearized version
		$${S}=\big\{\gamma,\log\Omega, b,\etabar,\eta,\Omega_\mck^2\psi_H,\psi_{\Hbar},{\omegabar},\Omega\Omega_\mck\beta,K,\hodge{K},\betabar\big\}.$$
Let $r,k,k'\geq0$. We define the following quantity which will be the schematic form of successive applications of $(\nabring_3,\Omega^2\nabring_4)$ derivatives to quantities $\phi\in\nabring^i\widecheck{S}$,
\begin{align*}
	RHS[r,k',k]=_s \sum_{m=1}^{r}\sum_{n=1}^m\sum_{\widecheck{s}_1,\cdots,\widecheck{s}_{n}\in\widecheck{S}}\sum_{s_1,\cdots, s_{m-n}\in S}\left(\prod_{j=1}^{n}\lieT^{\leq k'}\nabring^{\leq k}\widecheck{s}_j\right)\left(\prod_{i=1}^{m-n}\nabring^{\leq k}s_i\right),
\end{align*} 
namely $RHS[r,k',k]$ is the sum of products of a number $m\leq r$ of factors of the type $\lieT^{\leq k'}\nabring^{\leq k}\widecheck{S}$ or $\nabring^{\leq k}{S}$, where at least one of the factors is of the type $\lieT^{\leq k'}\nabring^{\leq k}\widecheck{S}$. We note that by Corollary \ref{cor:n0primprimprim}, for $k+k'\leq N_{max}+1$ we have
\begin{align}\label{eq:lefinaltoutsimplement}
	|RHS[r,k',k]|\lesssim_{r,k,k'}\frac{\varepsilon}{\ubar^{2+\frac{\delta}{3}}},\quad\text{in}\:\deux,
\end{align}
where we used Lemma \ref{lem:usimubarII} and the fact that the $\lieT^{k'}\nabring^{k}$ derivatives of exact Kerr $S_\mck$ components in the double null gauge are bounded, see \eqref{eq:bornesdanskerr}. Moreover, by the schematic equations satisfied by $\lieT^{\leq k'}\nabring^{\leq k}$ derivatives of the linearized quantities in Proposition \ref{prop:eqschemaliet}, we get that each $\phi\in\lieT^{\leq k'}\nabring^{\leq k}\widecheck{S}$ satisfies either a $\nabring_3$ or a $\Omega^2\nabring_4$ schematic equation with RHS which can be written as\footnote{Here, the number $r=k+k'+4$ of factors, which is not optimal, comes from both the terms $\lieT^j F_{\cdot}$ (where most of the factors come from the terms $\nabring^{i_1}\psi^{i_2}$) and the commutator terms $[\nabring_{3,\Omega^2 4},\lieT^j]$.} $RHS[k+k'+4,k',k+1]$ (note the loss of one angular derivative here). Moreover, for any such $\phi$ we get by Lemma \ref{lem:lienablaT} and \eqref{eq:expreT} the schematic identity
$$\lieT \phi-\frac12\Omega^2\nabring_4\phi+\frac12\nabring_3\phi=_s b\nabring \phi+(\Omega^2\chi,\chibar,\nabring b)\cdot\phi.$$
Combining these two observations, we deduce for $\phi\in\lieT^{\leq k'}\nabring^{\leq k}\widecheck{S}$,
\begin{align}
	(\nabring_3,\Omega^2\nabring_4)\phi&=_sRHS[k+k'+4,k',k+1]+\lieT\phi\nn\\
	&=_sRHS[k+k'+4,k',k+1]+RHS[1,k'+1,k].\label{eq:mlmlmloo1}
\end{align}
Moreover, commuting the non-linearized system satisfied by the quantities $S$ (see Section \ref{section:equationsDNnonlinearize}) with $\nabring^k$, we obtain similarly
\begin{align}
	(\nabring_3,\Omega^2\nabring_4)\nabring^kS&=_s\sum_{m=1}^{k+4}\sum_{s_1,\cdots, s_{m}\in S}\prod_{i=1}^{m-n}\nabring^{\leq k+1}s_i+\lieT\nabring^{\leq k}S\nn\\
	&=_s\sum_{m=1}^{k+4}\sum_{s_1,\cdots, s_{m}\in S}\prod_{i=1}^{m-n}\nabring^{\leq k+1}s_i+RHS[2,1,k],\label{eq:mlmlmloo2}
\end{align}
where we used Corollary \ref{cor:gaufinaltcool} to treat the last term on the RHS in the second equality. Combining \eqref{eq:mlmlmloo1} and \eqref{eq:mlmlmloo2} and using the Leibniz rule we deduce
$$(\nabring_3,\Omega^2\nabring_4)RHS[r,k',k]=_sRHS[r+k+k'+4,k',k+1]+RHS[r,k'+1,k],$$
which implies, using $RHS[r_1,k'_1,k_1]=_sRHS[r_2,k'_2,k_2]$ whenever $r_1\leq r_2,k'_1\leq k'_2, k_1\leq k_2$, 
\begin{align}\label{eq:ffimportanteui}
	(\nabring_3,\Omega^2\nabring_4)RHS[r,k',k]=_sRHS[r+k+k'+4,k'+1,k+1].
\end{align}
Now, let us fix $\phi\in\widecheck{S}$ and $i\geq 0$. Then by \eqref{eq:mlmlmloo1} for $k'=0$ we have
$$(\nabring_3,\Omega^2\nabring_4)\nabring^i\phi=_s RHS[i+4,1,i+1].$$
Thus, applying successively the identity \eqref{eq:ffimportanteui}, we obtain by induction for $l\geq 1$,
\begin{align}
	(\nabring_3,\Omega^2\nabring_4)^l\nabring^i\phi=_s RHS\left[i+4l+\sum_{j=1}^{l-1}j+\sum_{m=i+1}^{i+l-1}m,\quad l,\quad i+l\right].
\end{align}
Thus, by \eqref{eq:lefinaltoutsimplement} we infer that for $i+l\leq N_{max}+1$, in $\deux$,
$$\sum_{\phi\in\widecheck{S}}|(\nabring_3,\Omega^2\nabring_4)^l\nabring^i\phi|\lesssim\frac{\varepsilon}{\ubar^{2+\frac{\delta}{3}}}.$$
Up to commutations (which are lower-order terms treated by Proposition \ref{prop:commnablaboudnenulll}), we deduce
\begin{align}\label{eq:onserapprocheallerr}
	\sum_{\phi\in\widecheck{S}}|\df^{\leq N_{max}+1}\phi|\lesssim\frac{\varepsilon}{\ubar^{2+\frac{\delta}{3}}},\quad\text{in}\:\:\deux.
\end{align}
It is now only left to control $\alpha,\alphabar$. Recall from Section \ref{section:nullstructureeq} the null structure equations for $\nabring_3\wh{\chibar}$ and $\nabring_4\wh{\chi}$ which rewrite $\alpha=-\nabring_4\wh{\chi}-tr\chi\wh{\chi}$, $\alphabar=-\nabring_3\wh{\chibar}-tr\chibar\:\wh{\chibar}-2\omegabar\wh{\chibar} $ in the double null gauge. By \eqref{eq:nab4diff}, \eqref{eq:nab3diff}, this implies
\begin{align*}
	\widecheck{\alpha}=_s&\nabring_4\psicheck_H+\psicheck_H^2+\psicheck_H(\psi_H)_\mck+\gcheck((\nabring_4\psi_H)_\mck,\gamma_\mck\chi_\mck^2)+(\gcheck\chi_\mck,\psicheck_H)\chi_\mck+\Omega^{-2}\nabring^{\leq 1}\bcheck\nabring^{\leq 1}\chi_\mck,\\
	\widecheck{\alphabar}=_s&\nabring_3\psicheck_{\Hbar}+\psicheck_{\Hbar}^2+(\psicheck_{\Hbar},\widecheck{\omegabar})((\psi_{\Hbar})_\mck,\omegabar_\mck)+\gcheck\chibar_\mck^2.
\end{align*}
Thus, using \eqref{eq:onserapprocheallerr}, we infer the bounds
$$|\df^{\leq N_{max}}\widecheck{\alpha}|\lesssim\frac{\varepsilon\Omega^{-4}}{\ubar^{2+\frac{\delta}{3}}},\quad|\df^{\leq N_{max}}\widecheck{\alphabar}|\lesssim\frac{\varepsilon}{\ubar^{2+\frac{\delta}{3}}},$$
which concludes the proof of Theorem \ref{thm:controldoublenull}.

\section{Auxiliary results for analysing the Teukolsky equation}

\subsection{Proof of Lemma \ref{lem:teukpsimfrakF}}\label{appendix:teukpsimfrakF}
We begin by expressing $\mcd\hot(\mcd(Y_{m,2}(\cos\theta)e^{im\phi_+}))$ with $\frakJ,\frakJ_\pm$ up to $\err$ terms as in \eqref{eq:asinflemme}.
\begin{lem}\label{lem:absolutt}
We have the following expressions in $\un$:
	\begin{align*}
		C_{\pm2}\mcd\hot(\mcd(Y_{\pm2,2}(\cos\theta)e^{\pm2i\phi_+}))&=(\frakJ_+\pm i\frakJ_-)\hot(\frakJ_+\pm i\frakJ_-)+\err,\\
		C_{\pm1}\mcd\hot(\mcd(Y_{\pm1,2}(\cos\theta)e^{\pm i\phi_+}))&=\frakJ\hot(\frakJ_+\pm i\frakJ_-)+\err,\\
		C_{0}\mcd\hot(\mcd(Y_{0,2}(\cos\theta))&=\frakJ\hot\frakJ+\err,
	\end{align*}
	where
	$$C_{\pm 2}^{-1}=\frac{1}{4}\sqrt{\frac{15}{2\pi}},\quad C_{\pm 1}^{-1}=\mp\frac{1}{4}\sqrt{\frac{15}{2\pi}},\quad C_0^{-1}=\frac{1}{4}\sqrt{\frac{5}{2\pi}}.$$
\end{lem}
\begin{proof}
	This basically follows from the linearizations of the derivatives of $J^{(\pm)},J^{(0)}:=\cos\theta$, and the linearizations of the derivatives of $\frakJ,\frakJ_\pm$, see Definition \ref{def:linearizedquantities}, combined with the expressions of $Y_{m,2}e^{im\phi_+}$ with respect to $J^{(\pm)}$, $\cos\theta$. First of all, the standard expressions of the spherical harmonics $Y_{m,2}$ yields
	\begin{align}
		C_{\pm2}Y_{\pm 2,2}(\cos\theta)e^{\pm2i\phi_+}=\sin^2\theta e^{\pm2i\phi_+}&=(\Jplus)^2-(\Jmoins)^2\pm2i \Jplus\Jmoins,\nn\\
		C_{\pm1}Y_{\pm1,2}(\cos\theta)e^{\pm i\phi_+}=\sin\theta\cos\theta e^{\pm i\phi_+}&=\Jzero(\Jplus\pm i\Jmoins)\label{eq:bcpplustardsisdonc},\\
		C_{0}Y_{0,2}(\cos\theta)=3\cos^2\theta-1&=3(\Jzero)^2-1.\nn
	\end{align}
	Moreover, using \eqref{eq:controlI} we compute, recalling the $\err$ terms defined in \eqref{eq:asinflemme},
	\begin{align*}
		\mcd\hot\mcd((\Jplus)^2-(\Jmoins)^2\pm2i \Jplus\Jmoins)&=\mcd\hot\left(2\Jplus\frakJ_+-2\Jmoins\frakJ_-\pm 2i(\Jplus\frakJ_-+\Jmoins\frakJ_+)+\err\right)\\
		&=2\left(\frakJ_+\hot\frakJ_+-\frakJ_-\hot\frakJ_-\pm 2i \frakJ_+\hot\frakJ_-\right)+\err\\
		&=2(\frakJ_+\pm i\frakJ_-)\hot(\frakJ_+\pm i\frakJ_-)+\err,
	\end{align*}
	as well as 
	\begin{align*}
		\mcd\hot\mcd(\Jzero(\Jplus\pm i\Jmoins))&=\mcd\hot\left(i\frakJ(\Jplus\pm i\Jmoins)+J^{(0)}(\frakJ_+\pm i\frakJ_-)+\err\right)\\
		&=2i\frakJ\hot(\frakJ_+\pm i\frakJ_-)+\err,
	\end{align*}
	and 
	$$\mcd\hot\mcd(3(\Jzero)^2-1)=6i\mcd\hot(\Jzero\frakJ+\err)=-6\frakJ\hot\frakJ+\err,$$
	which concludes the proof.
\end{proof}
\begin{lem}\label{lem:calculschiants}
	Let $\frakJ_1,\frakJ_2\in\fraks_1(\C)$ such that $\mcd\hot\frakJ_1=\err$ and $\mcd\hot\frakJ_2=\err$. Then we have
	\begin{align*}
		-\mcd\hot\divc(\frakJ_1\hot\frakJ_2)=&i(\atrchi\nabla_3\frakJ_1+\atrchibar\nabla_4\frakJ_1)\hot\frakJ_2+i(\atrchi\nabla_3\frakJ_2+\atrchibar\nabla_4\frakJ_2)\hot\frakJ_1\\
		&+4{}^{(h)}K\frakJ_1\hot\frakJ_2+\err,
	\end{align*}
	and for any $F=f+i\hodge{f}\in\fraks_1(\C)$ we have
	\begin{align*}
		F\cdot\nabla(\frakJ_1\hot\frakJ_2)&=\frac12 \divc\frakJ_1 F\hot\frakJ_2+\frac12 \divc\frakJ_2 F\hot\frakJ_1,\\
		(F+\overline{F})\cdot\nabla(\frakJ_1\hot\frakJ_2)&=\frac12 \divc\frakJ_1 F\hot\frakJ_2+\frac12 \divc\frakJ_2 F\hot\frakJ_1+F\cdot\err.
	\end{align*}
\end{lem}
\begin{proof}
	By Lemma \ref{prop:cestpourteukansatz} we have
	\begin{align}\label{eq:divcdehothmm}
		\divc(\frakJ_1\hot\frakJ_2)=2((\divc\frakJ_1)\frakJ_2+(\divc\frakJ_2)\frakJ_1),
	\end{align}
	which implies
	\begin{align*}
		\mcd\hot\divc(\frakJ_1\hot\frakJ_2)=&2(\mcd(\divc\frakJ_1)\hot\frakJ_2+\mcd(\divc\frakJ_2)\hot\frakJ_1)+\err\\
		=&i(\atrchi\nabla_3\frakJ_1+\atrchibar\nabla_4\frakJ_1)\hot\frakJ_2+i(\atrchi\nabla_3\frakJ_2+\atrchibar\nabla_4\frakJ_2)\hot\frakJ_1\\
		&+4{}^{(h)}K\frakJ_1\hot\frakJ_2+\err,
	\end{align*}
	where we used Lemma \ref{lem:unlemsympa} in the last step. Next, by Lemma \ref{lem:leibnizchilll} and \eqref{eq:divcdehothmm}, we get
	\begin{align*}
		F\cdot\nabla(\frakJ_1\hot\frakJ_2)=\frac14 F\hot(\divc(\frakJ_1\hot\frakJ_2))=\frac12 \divc\frakJ_1 F\hot\frakJ_2+\frac12 \divc\frakJ_2 F\hot\frakJ_1.
	\end{align*}
	Finally, denoting $j_1=\Real(\frakJ_1),j_2=\Real(\frakJ_2)$, we have
	$$(F+\overline{F})\cdot\nabla(\frakJ_1\hot\frakJ_2)=2 f\cdot\nabla(\frakJ_1\hot\frakJ_2)=4(f\cdot\nabla(j_1\hot j_1)+i\hodge{(f\cdot\nabla(j_1\hot j_1))}),$$
	where, using $2\nabla_a j_b=\nabla\hot j_{ab}+\delta_{ab}\diver j+\in_{ab}\curl j$ we compute
	\begin{align*}
		f\cdot\nabla(j_1\hot j_1)&=(f\cdot\nabla j_1)\hot j_2+j_1\hot (f\cdot\nabla j_2)\\
		&=\frac12\diver j_1 f\hot j_2-\frac12\curl j_1 f\hot\hodge{j_2}+\frac12\diver j_2 f\hot j_1-\frac12\curl j_2 f\hot\hodge{j_1}+F\cdot\err.
	\end{align*}
	This yields $(F+\overline{F})\cdot\nabla(\frakJ_1\hot\frakJ_2)=\frac12 \divc\frakJ_1 F\hot\frakJ_2+\frac12 \divc\frakJ_2 F\hot\frakJ_1+F\cdot\err$,	concluding the proof.
\end{proof}

\begin{lem}\label{lem:notationbien}
	We denote
	\begin{align*}
		Y_{\pm 2}:=\frac{1}{\qbar^2}(\frakJ_+\pm i\frakJ_-)\hot(\frakJ_+\pm i\frakJ_-),\quad
		Y_{\pm 1}:=\frac{1}{\qbar^2}\frakJ\hot(\frakJ_+\pm i\frakJ_-),\quad
		Y_0:=\frac{1}{\qbar^2}\frakJ\hot\frakJ.
	\end{align*}
	Then we have the following computations: for any $m=0,\pm 1,\pm 2$,
	\begin{align*}
		\nabla_4 Y_{m}&=\left(-\frac{2\Delta\qbar}{|q|^4}-\frac{2\Delta}{\qbar|q|^2}+\frac{2iam}{|q|^2}\right)Y_{m}+\err,\quad \nabla_3(Y_{m})=\frac{4}{\qbar}Y_{m}\\
		\nabla_4\nabla_3 Y_{m}&=\left(-\frac{8\Delta}{|q|^4}-\frac{12\Delta}{\qbar^2|q|^2}+\frac{8iam}{\qbar|q|^2}\right)Y_{m}+\err,\\
		-\mcd\hot\divc Y_m&=\frac{2ia\cos\theta}{|q|^2}\left(-\frac{2\Delta\qbar}{|q|^4}+\frac{2\Delta}{\qbar|q|^2}+\frac{2iam}{|q|^2}\right)Y_{m}+4{}^{(h)}K Y_m+\err.
	\end{align*}	
	Moreover, for $F\in\fraks_1(\C)$ we have
	\begin{align*}
		F\cdot\nabla Y_{\pm 2}&=\left(-\frac{4}{\qbar r^2}\mp\frac{4a^2\cos\theta}{|q|^4\qbar^2}+\frac{2ia}{|q|^2}(\cos\theta\pm 1)\right)(\Jplus\pm i \Jmoins)F\hot(\frakJ_+\pm i\frakJ_-)+\err,\\
		F\cdot\nabla Y_{\pm 1}&=\left(-\frac{2}{\qbar r^2}\mp\frac{2a^2\cos\theta}{|q|^4\qbar^2}+\frac{2ia}{|q|^2}(\cos\theta\pm 1)\right)(\Jplus\pm i \Jmoins)F\hot\frakJ\\
		&\quad +\left(\frac{a\sin^2\theta}{|q|^2\qbar^3}+\frac{2i(r^2+a^2)\cos\theta}{\qbar^2|q|^4}\right)F\hot(\frakJ_+\pm i\frakJ_-)+\err,\\
		F\cdot\nabla Y_0&=\left(\frac{2a\sin^2\theta}{|q|^2\qbar^3}+\frac{4i(r^2+a^2)\cos\theta}{\qbar^2|q|^4}\right)F\hot\frakJ+\err.
	\end{align*}
	Also, the three identities above also hold replacing $F$ by $F+\overline{F}$.
\end{lem}
\begin{proof}
	The identities for $\nabla_3$ derivatives are immediate consequences of $\nabla_3(\frakJ,\frakJ_\pm)={\qbar}^{-1}(\frakJ,\frakJ_\pm)$, while the identities for $\nabla_4$ derivatives are consequences of the linearizations of $\nabla_4(\frakJ,\frakJ_\pm)$ and \eqref{eq:controlI}. The identity for $\nabla_4\nabla_3$ derivatives then follow from combining the results for $\nabla_3$ and $\nabla_4$ derivatives and the computation $\nabla_4(1/\qbar)=-\frac{\Delta}{|q|^2\qbar^2}+\err$. The remaining identities involving horizontal derivatives are also simple consequences of Lemma \ref{lem:calculschiants}.
\end{proof}
The proof of Lemma \ref{lem:teukpsimfrakF} then follows from Lemma \ref{lem:notationbien} and \eqref{eq:teukop}.
\subsection{Commuting the Teukolsky operator with $(\nabla_3,\nabla_4,\barred{\df})$ derivatives}\label{section:commutingteukwith}
In this section, we prove some general commutation identities which will be used in both regions $\un$ and $\deux$.  Here, in $\un$ we consider the horizontal distribution associated to the PT frame, and in $\deux$ we consider the one associated to the ingoing non-integrable frame (see Section \ref{section:defprincipalframes}).  We define the following ordered sets of operators: in $\un\cup\deux$,
$$\mathcal{O}:=(S_1,S_2,S_3,S_4,S_5,S_6),$$
where $S_1=\divc$, $S_2=\divc\divc$, $S_3=\mcd\hot\divc$, $S_4=\divc\mcd\hot\divc$, $S_5=\mcd\divc\divc$, $S_6=\divc\mcd\divc\divc$, as well as, in $\un\cup\deux$,
$$\mathcal{O}_4:=(S_1^{(4)},S_2^{(4)},S_3^{(4)},S_4^{(4)},S_5^{(4)},S_6^{(4)},S_7^{(4)}),$$
where $S_1^{(4)}=\nabla_4$, $S_2^{(4)}=\nabla_4\divc$, $S_3^{(4)}=\nabla_4\divc\divc$, $S_4^{(4)}=\nabla_4\mcd\divc\divc$, $S_5^{(4)}=\nabla_4\nabla_4$, $S_6^{(4)}=\nabla_4\nabla_4\divc$, $S_7^{(4)}=\nabla_4\nabla_4\divc\divc$, and, in $\deux$,
$$\mathcal{O}_3:=(S_1^{(3)},S_2^{(3)},S_3^{(3)},S_4^{(3)},S_5^{(3)},S_6^{(3)},S_7^{(3)}),$$
where $S_1^{(3)}=\hat{\lambda}\nabla_3$, $S_2^{(3)}=\hat{\lambda}\nabla_3\divc$, $S_3^{(3)}=\hat{\lambda}\nabla_3\divc\divc$, $S_4^{(3)}=\hat{\lambda}\nabla_3\mcd\divc\divc$, $S_5^{(3)}=\hat{\lambda}\nabla_3(\hat{\lambda}\nabla_3)$, $S_6^{(3)}=\hat{\lambda}\nabla_3(\hat{\lambda}\nabla_3)\divc$, $S_7^{(3)}=\hat{\lambda}\nabla_3(\hat{\lambda}\nabla_3)\divc\divc$. 
We also denote $p(k),p^4(k),p^3(k)\in\{0,1,2\}$ the indices such that 
$$S_kU\in\fraks_{p(k)}(\C),\quad S_k^{(4)}U\in\fraks_{p^4(k)}(\C),\quad S_k^{(3)}U\in\fraks_{p^3(k)}(\C).$$
\begin{prop}\label{prop:nouvellescommut}
	Let $U\in\fraks_2(\C)$. We assume that $U$ satisfies
	$$\mcl(U)=\err[\mcl(U)],$$
	where $\mcl$ is the Teukolsky operator \eqref{eq:teukop}. Then we have 
	\begin{align*}
		\nabla_4\nabla_3S_kU-\triangle_{p(k)}S_kU&=F_k[U],\quad (1\leq k\leq 6),\\
		\nabla_4\nabla_3S_k^{(4)}U-\triangle_{p^4(k)}S_k^{(4)}U&=F^{(4)}_k[U],\quad (1\leq k\leq 7),\\
		\nabla_4\nabla_3S_k^{(3)}U-\triangle_{p^3(k)}S_k^{(3)}U&=F^{(3)}_k[U],\quad (1\leq k\leq 7),
	\end{align*}
where the right-hand sides above satisfy the following:
\begin{itemize}
	\item In $\un$, provided that \eqref{eq:controlI} holds and recalling \eqref{eq:setofderivatives}, we have 
	\begin{align*}
		F_k[U]&=_{rs}\sum_{k'=0}^k\df^{\leq 1}S_{k'}U+\df^{\leq k}(\Gammacheck,\Rcheck)\cdot\df^{\leq k+1}U+\df^{\leq k}\err[\mcl(U)],\\
		F_k^{(4)}[U]&=_{rs}\sum_{k'=0}^k\df^{\leq 1}(S_{k'}^{(4)}U,S_{k'}U)+\df^{\leq k}(\Gammacheck,\Rcheck)\cdot\df^{\leq k+1}U+\df^{\leq k}\err[\mcl(U)].
	\end{align*}
		\item In $\deux$, provided that the bounds in Proposition \ref{prop:coeffin} hold and recalling \eqref{eq:dfhorizontal}, we have 
	\begin{align*}
		F_k[U]+\omega_\mck\nabla_3S_kU&=_{rs}\df^{\leq 1} S_k[U]+\sum_{k'=0}^{k-1}(\nabla_3,\nabla_4,\nabla)^{\leq 1}S_{k'}U+\err,\\
		F_k^{(4)}[U]+\omega_\mck\nabla_3S_k^{(4)}U&=_{rs}\df^{\leq 1}(S_k^{(4)}U,S_kU)+\sum_{k'=0}^{k-1}(\nabla_3,\nabla_4,\nabla)^{\leq 1}(S_{k'}^{(4)},S_{k'}U)+\err,\\
		F_k^{(3)}[U]+\omega_\mck\nabla_3S_k^{(3)}U&=_{rs}\df^{\leq 1}(S_k^{(3)}U,S_kU)+\sum_{k'=0}^{k-1}(\nabla_3,\nabla_4,\nabla)^{\leq 1}(S_{k'}^{(3)},S_{k'}U)+\err,
	\end{align*}
	where, recalling the notation $\Gamma_p,R_p$ defined in \eqref{eq:gammasignature}, \eqref{eq:rsignature}, the error terms above are
	$$\err=_{rs}\df^{\leq k}\err[\mcl(U)]+\left(\hat{\lambda}^{-1}\df^{\leq k}(\Gammacheck_{+1},\Rcheck_{+1}), \df^{\leq k}(\Gammacheck_{-1}\Rcheck_{-1},\Gammacheck_0,\Rcheck_0)\right)\cdot\df^{\leq k+1}U.$$
\end{itemize}
\end{prop}
\begin{rem}We remark the following:
	\begin{itemize}
		\item In the commutations above, we use structural identities in the commutators $[\nabla_{3,4},\divc]$ and $[\nabla_{3,4},\mcd\hot]$ which allow for the expressions for $F_k[U],F_k^{(4)}[U],F_k^{(3)}[U]$ to hold. In the commutations, there are terms which are not acceptable (namely which are not some $\mco,\mco_4,\mco_3$ derivative of $A$), which arise from the terms with $(\chihat,\wh{\chibar})$ in \eqref{eq:commnab4divc}, \eqref{eq:commnab3divc}, \eqref{eq:commnab4mcdhot}, \eqref{eq:commnab3mcdhot}, but which nonetheless are quadratic error terms.
		\item The terms $\omega_\mck\nabla_3$ are necessary in region $\deux$ to use the energy estimate for Teukolsky-like equations which is based on an effective blueshift effect taking advantage of $\omega_\mck\gtrsim 1$ in $\deux$, see the first estimate in assumption \eqref{eq:hypLhat}. This is not needed in $\un$ where these terms are discarded to the RHS.
		\item All terms in the expressions of $F_k[U],F_k^{(4)}[U],F_k^{(3)}[U]$ above are lower-order terms which can be treated by induction, except the terms $\df S_kU,\df S_k^{(4)}U$ corresponding to $k'=k$ in $\un$, and the analog terms $\df(S_kU,S_k^{(4)}U,S_k^{(3)}U)$ in $\deux$.  The terms $\df S_kU,\df S_k^{(4)}U,\df S_k^{(3)}U$ in the equations for, respectively, $S_kU,S_k^{(4)}U,S_k^{(3)}U$ will be put on the LHS as part of the equation and are treated in the energy estimates. The terms $\df S_kU$ appearing in the equations for $S_k^{(4)}U,S_k^{(3)}U$ are bounded using the previously proven estimates for $\mco$ derivatives. 
	\end{itemize}

\end{rem}
\begin{proof}
 This follows from commuting sucessively the equation $\mcl(U)=\err[\mcl(U)]$. More precisely, by \eqref{eq:teukop} and Lemma \ref{lem:ignocoucou} we have
\begin{align}\label{eq:teukdebasequoi}
	\nabla_4\nabla_3U-\triangle_2U+\omega_\mck\nabla_3U=_s(\widecheck{\omega},\Gamma_{+1})\nabla_3U+\Gamma_{-1}\nabla_4U+\Gamma_0\nabla U+(V_0',{}^{(h)}K)U+\err[\mcl(U)].
\end{align}
Then, relying on \eqref{eq:controlI} in $\un$ and Proposition \ref{prop:coeffin} in $\deux$, it is straightforward to check that $S_k,S_k^{(4)},S_k^{(3)}$ derivatives applied to the RHS above yield terms which take the reduced schematic form stated for $F_k[U],F_k^{(4)}[U],F_k^{(3)}[U]$\footnote{To check that these terms take reduced schematic forms exactly as stated in Proposition \ref{prop:nouvellescommut}, one must use the commutation formulas in Section \ref{section:commutationformulas}. We note in particular that to check that the derivatives of the term $\Gamma_0\cdot\nabla U$ takes an acceptable form, one can use the formulas in Lemma \ref{lem:divcfcdotnablaU}.}. The crucial point in the proof of Proposition \ref{prop:nouvellescommut} is instead the computation of the top-order terms appearing in the commutator between the LHS $\nabla_4\nabla_3-\triangle$ above and the operators $S_k,S_k^{(4)},S_k^{(3)}$. This relies on the following computations: for any $V\in\fraks_2(\C)$,
\begin{align*}
	\divc (\nabla_4\nabla_3V-\triangle_2V+\omega_\mck\nabla_3 V)&=\divc\left(\nabla_4\nabla_3-\frac{1}{4}\mcd\hot\divc V+\omega_\mck\nabla_3V\right)+good\\
	&=\nabla_4\nabla_3\divc V-\frac14\divc\mcd\hot\divc V+\omega_\mck\nabla_3\divc V\\
	&\quad +[\divc,\nabla_4]\nabla_3V+\nabla_4([\divc,\nabla_3]V)+good\\
	&=\nabla_4\nabla_3\divc V-\triangle_1\divc V+\omega_\mck\nabla_3\divc V+good,
\end{align*}
where we denote by $good$ any term which, at each step, take reduced schematic form compatible with the RHS for $F_k[U],F_k^{(4)}[U],F_k^{(3)}[U]$ stated in Proposition \ref{prop:nouvellescommut} in both regions $\un$ and $\deux$. Here, in the first step we used Lemma \ref{lem:ignocoucou}, we commuted $\nabla_4\nabla_3$ and $\omega_\mck\nabla_3$ with $\divc$ in the second step, and we used Lemma \ref{lem:laplaciens1(C)} in the third step combined with the following computation based on Proposition \ref{prop:commnabladivc}:
	\begin{align*}
	[\divc,&\nabla_4]\nabla_3 V+\nabla_4([\divc,\nabla_3]V)=\nn\\
	&\frac{1}{2}\overline{trX}\nabla_3\divc V+\frac{1}{2}\overline{tr\Xbar}\nabla_4\divc V-(\overline{H+\Hbar})\cdot\nabla_4\nabla_3V+\wh{\chi}\cdot\overline{\mcd}\nabla_3 V-D_4^{(2)}[\nabla_3 V]\nn\\
	& -\frac{1}{2}\overline{trX}\left(-\frac{1}{2}\overline{tr\Xbar}\divc V+(\overline{H-Z})\cdot\nabla_3 V-\wh{\chibar}\cdot\overline{\mcd}V+D_3^{(2)}[V]\right)\nn\\
	&+\frac{1}{2}\nabla_4tr\Xbar \divc V-\nabla_4(\overline{H-Z})\cdot\nabla_3V+\nabla_4\wh{\chibar}\cdot\overline{\mcd}V+\wh{\chibar}\cdot\nabla_4\overline{\mcd}V-\nabla_4 D_3^{(2)}[V]=good,
\end{align*}
where in particular the term $(\overline{H+\Hbar})\cdot\nabla_4\nabla_3V$ is good by re-expressing it using the Teukolsky-like equation $\nabla_4\nabla_3V=\triangle_2V-\omega_\mck\nabla_3V+F$\footnote{Also note that the terms with $\hat{\lambda}^{-1}$ in the error terms for $F_k[U],F_k^{(4)}[U],F_k^{(3)}[U]$ in $\deux$ arise from manipulations of the type $\overline{trX}\nabla_3\divc V=\hat{\lambda}^{-1}\overline{\widecheck{trX}}(\hat{\lambda}\nabla_3\divc V)+O(1)\hat{\lambda}\nabla_3\divc V$.}. Relying on Proposition \ref{prop:commnabladivc} again and Lemmas \ref{lem:unlemsympa}, \ref{lem:unlemsympascalair}, we also get for any $G\in\fraks_1(\C)$,
\begin{align*}
	\divc (\nabla_4\nabla_3G-\triangle_1G+\omega_\mck\nabla_3 G)=\nabla_4\nabla_3\divc G-\triangle_0\divc G+\omega_\mck\nabla_3\divc G+good,
\end{align*}
and relying on Proposition \ref{prop:commnab34mcdhot}, and Lemmas \ref{lem:laplaciens1(C)}, \ref{lem:ignocoucou} we also get 
\begin{align*}
	\mcd\hot(\nabla_4\nabla_3G-\triangle_1G+\omega_\mck\nabla_3 G)=\nabla_4\nabla_3\mcd\hot G-\triangle_2\mcd\hot G+\omega_\mck\nabla_3\mcd\hot G+good.
\end{align*}
Also, if $h\in\fraks_0(\C)$ is a function we have by Proposition \ref{prop:commnab34mcd} and Lemmas \ref{lem:unlemsympa}, \ref{lem:unlemsympascalair}, 
\begin{align*}
	\mcd (\nabla_4\nabla_3h-\triangle_0 h+\omega_\mck\nabla_3 h)=\nabla_4\nabla_3\mcd h -\triangle_1\mcd h+\omega_\mck\nabla_3\mcd h+good.
\end{align*}
This concludes the proof of the reduced schematic form for $F_k[U]$ in both regions $\un$ and $\deux$\footnote{Note that in region $\un$ the term $\omega_\mck\nabla_3$ can be discarded on the RHS since it takes an acceptable reduced schematic form.}. Now we deal with the reduced schematic expressions for $F_k^{(4)}[U],F_k^{(3)}[U]$. Similarly as for $F_k[U]$, the lower-order terms which correspond to taking $\mco_4$ derivatives of the RHS of \eqref{eq:teukdebasequoi} are easily shown to take acceptable reduced schematic form. To deal with the commutations between the LHS $\nabla_4\nabla_3-\triangle+\omega_\mck\nabla_3$ and $\nabla_4$ we use the commutation identities in Section \ref{section:commutationformulas} and Lemma \ref{lem:ignocoucou} to get for any $U\in\fraks_2(\C)$:
\begin{align*}
	[\nabla_4&\nabla_3-\triangle_2+\omega_\mck\nabla_3,\nabla_4]U\\
	=&\nabla_4([\nabla_3,\nabla_4]U)-\frac14[\mcd\hot\divc,\nabla_4]U+good\\
	=&\nabla_4(2\omega\nabla_3U-2\omegabar\nabla_4U+2(\etabar-\eta)\cdot\nabla U+C_k[U])-\frac14\mcd\hot[\divc,\nabla_4]U-\frac14[\mcd\hot,\nabla_4]\divc U\\
	=&2\omega\nabla_4\nabla_3U-2\omegabar\nabla_4\nabla_4U+2(\etabar-\eta)\cdot\nabla\nabla_4U+2e_4(\omega)\nabla_3U-2e_4(\omegabar)\nabla_4U\\
	&(tr\chi,\atrchibar)(\etabar-\eta)\cdot\nabla U-(\etabar-\eta)\cdot(\etabar+\zeta)\nabla_4U+(\etabar-\eta)\cdot(\wh{\chi}\cdot\nabla U)\\
	&+(\etabar-\eta)\cdot B_k[U]+2\nabla_4(\etabar-\eta)\cdot\nabla U+\nabla_4 C_k[U]+O(\df^{\leq 2}(\Xi\cdot U))\\
	&+\frac{1}{4}\Big(tr\chi \mcd\hot\divc U-(\Hbar+Z)\hot\nabla_4\divc U+\wh{\chi}\cdot\overline{\mcd}\divc U-E_4[\divc U]\\
	&+\frac{1}{2}\mcd(\overline{trX})\hot(\divc U)-2(\mcd\cdot(\overline{\Hbar+Z}))\nabla_4U-8(\etabar+\zeta)\cdot\nabla\nabla_4U\\
	&+(\overline{\Hbar+Z})\hot(\divc\nabla_4U)-\mcd\hot(\wh{\chi}\cdot\overline{\mcd}U)+\mcd\hot(D_4^{(2)}[U])+O(\df^{\leq 2}(\Xi\cdot U))\Big)+good=good,
\end{align*}
where we used that:
\begin{itemize}
	\item the term $2\omega\nabla_4\nabla_3U$ is good by expressing with the Teukolsky-like equation $\nabla_4\nabla_3U=\triangle_2 U-\omega_\mck\nabla_3U+F$,
	\item the terms with $\nabla_4\nabla_4$ and $\nabla\nabla_4$ derivatives yield reduced schematic terms of the form $(\nabla_4,\nabla)S_k^{(4)}[U]$ and hence are good,
	\item the terms with $\mcd\hot\divc$ and $\nabla_4\divc$ derivatives yield reduced schematic terms of the form $(\nabla_4,\nabla)S_k[U]$ and hence are good,
	\item the other terms are lower-order terms or quadratic error terms which are easily seen to be good.
\end{itemize}
Similarly we also get for $G\in\fraks_1(\C)$ and $h\in\fraks_0(\C)$,
$$[\nabla_4\nabla_3-\triangle_1+\omega_\mck\nabla_3,\nabla_4]G=good,\quad [\nabla_4\nabla_3-\triangle_0+\omega_\mck\nabla_3,\nabla_4]h=good,$$
thereby concluding the proof for $F_k^{(4)}[U]$. We now deal with $F_k^{(3)}[U]$ in $\deux$. Here, the choice of normalization $\hat{\lambda}\nabla_3$ is crucial in order to cancel bad terms. Recall from Remark \ref{rem:remlabel} that $\hat{\lambda}$ is the $\lambda$ coefficient in the change of frame from the double null frame to the non-integrable ingoing frame. Thus, by the change of frame formula \eqref{eq:changeomega} for $\omega$, we have the identity
$$\hat{\lambda}^{-1}\omega=-\frac{1}{2}\hat{\lambda}^{-1}e_4(\log(\hat{\lambda}))+K_4[f],$$
where
$$K_4[f]=\frac{1}{2}f\cdot(\mathring{\zeta}-\mathring{\etabar})-\frac{1}{4}|f|^2\mathring{\omegabar}-\frac18\mathring{tr\chibar}|f|^2-\frac14f\cdot(f\cdot\wh{\chibar})=O(1).$$
Multiplying the identity above by $-2(\hat{\lambda})^2$ we thus get 
\begin{align}\label{eq:e4Deltahat}
	e_4(\hat{\lambda})=-2\omega\hat{\lambda}+2\hat{\lambda}^2K_4[f].
\end{align}
	This implies that for $U\in\fraks_k(\C)$, $k=0,1,2$, using Lemma \ref{lem:comm34} and \eqref{eq:e4Deltahat}, 
\begin{align*}
	[\hat{\lambda}\nabla_3,\nabla_4\nabla_3]U=&[\hat{\lambda}\nabla_3,\nabla_4]\nabla_3U+\nabla_4([\hat{\lambda}\nabla_3,\nabla_3]U)\\
	=&\hat{\lambda}[\nabla_3,\nabla_4]\nabla_3U-e_4(\hat{\lambda})\nabla_3\nabla_3U-\nabla_4(e_3(\hat{\lambda})U)\\
	&=\hat{\lambda}\Big(-2\omega\nabla_3\nabla_3U+2\omegabar\nabla_4\nabla_3U+2(\eta-\etabar)\cdot\nabla\nabla_3U+C_k[\nabla_3U]\Big)\\
	&-e_4(\hat{\lambda})\nabla_3\nabla_3U-\nabla_4(e_3(\hat{\lambda})U)\\
	=& -2\hat{\lambda}^2K_4[f]\nabla_3\nabla_3U+2\omegabar\hat{\lambda}\nabla_4\nabla_3U+2\hat{\lambda}(\eta-\etabar)\cdot\nabla\nabla_3U+\hat{\lambda} C_k[\nabla_3U]\\
	&-e_4(e_3(\hat{\lambda}))U-e_3(\hat{\lambda})\nabla_4U\\
	=&-2K_4[f]\Big(\hat{\lambda}\nabla_3(\hat{\lambda}\nabla_3U)-\hat{\lambda} e_3(\hat{\lambda})\nabla_3U\Big)+2\omegabar\hat{\lambda}\nabla_4\nabla_3U\\
	&+2(\eta-\etabar)\cdot\nabla(\hat{\lambda}\nabla_3U)-2(\eta-\etabar)\cdot\nabla(\hat{\lambda})\nabla_3U+\hat{\lambda} C_k[\nabla_3U]\\
	&-e_4(e_3(\hat{\lambda}))U-e_3(\hat{\lambda})\nabla_4U=good.
\end{align*}
	Note in particular the crucial partial cancellation of the terms $-2\omega\hat{\lambda}\nabla_3\nabla_3U$ and $-e_4(\hat{\lambda})\nabla_3\nabla_3U$ (which lack a $\Omega^2$ factor compared to $O(\Omega^4)\nabla_3\nabla_3U$). Next, by \eqref{eq:commnab3mcdhot}, \eqref{eq:commnab3divc}, and Lemma \ref{lem:nouveauleibniz},
	\begin{align*}
		[\hat{\lambda}\nabla_3,\mcd\hot\divc]U&=[\hat{\lambda}\nabla_3,\mcd\hot]\divc U+\mcd\hot([\hat{\lambda}\nabla_3,\divc]U)\\
		&=\hat{\lambda}[\nabla_3,\mcd\hot]\divc U-\mcd(\hat{\lambda})\hot\nabla_3\divc U+\mcd\hot(\hat{\lambda}[\nabla_3,\divc]U-\overline{\mcd}\hat{\lambda}\cdot\nabla_3U)\\
		&=\hat{\lambda}\Big(-\frac{1}{2}tr\Xbar\mcd\hot\divc U+(H-Z)\hot\nabla_3\divc U-\wh{\chibar}\cdot\overline{\mcd}\divc U+E_3[\divc U]\Big)\\
		&\quad+\mcd\hot\Big(\hat{\lambda}\Big(-\frac{1}{2}\overline{tr\Xbar}\divc U+(\overline{H-Z})\cdot\nabla_3 U-\wh{\chibar}\cdot\overline{\mcd}U+D_3^{(2)}[U]\Big)\Big)\\
		&\quad-2\mcd\cdot(\overline{\mcd}\hat{\lambda})\nabla_3U-8\nabla\hat{\lambda}\cdot\nabla\nabla_3U+\mcd\hat{\lambda}\hot(\divc\nabla_3U)-\mcd(\hat{\lambda})\hot\nabla_3\divc U\\
				=&\hat{\lambda}\Big[-tr\chibar \mcd\hot\divc U+(H-Z)\hot\nabla_3\divc U-\wh{\chibar}\cdot\overline{\mcd}\divc U+E_3[\divc U]-\frac{1}{2}\mcd(\overline{tr\Xbar})\hot(\divc U)\\
		&+2(\mcd\cdot(\overline{H-Z}))\nabla_3U-\mcd\hot(\wh{\chibar}\cdot\overline{\mcd}U)+\mcd\hot(D_3^{(2)}[U])\Big]+8(\eta-\zeta)\cdot\nabla({\hat{\lambda}\nabla_3}U)\\
		&-({H-Z})\hot(\divc({\hat{\lambda}\nabla_3}U))-8(\eta-\zeta)\cdot\nabla(\hat{\lambda})\nabla_3U+({H-Z})\hot(\overline{\mcd}(\hat{\lambda})\cdot\nabla_3U)\\
		&-2\mcd\cdot(\overline{\mcd}\hat{\lambda})\nabla_3U-8\nabla\log\hat{\lambda}\cdot\nabla({\hat{\lambda}\nabla_3}U)+8\nabla\log\hat{\lambda}\cdot\nabla(\hat{\lambda})\nabla_3U\\
		&+\mcd(\hat{\lambda})\hot\Big(-\frac{1}{2}\overline{tr\Xbar}\divc U+(\overline{H-Z})\cdot\nabla_3 U-\wh{\chibar}\cdot\overline{\mcd}U+D_3^{(2)}[U]\Big)=good.
	\end{align*}
	Similarly we also have $[\hat{\lambda}\nabla_3,\divc\mcd\hot]G=good$, $[\hat{\lambda}\nabla_3,\divc\mcd]h=good$ for $G\in\fraks_1(\C)$ and $h\in\fraks_0(\C)$, which implies by Lemmas \ref{lem:ignocoucou}--\ref{lem:ignocoucoulast}:
	$$[\nabla_4\nabla_3-\triangle_k+\omega_\mck\nabla_3,\hat{\lambda}\nabla_3]U=good,$$
for any $U\in\fraks_k(\C)$, $k=0,1,2$, thereby concluding the proof of Proposition \ref{prop:nouvellescommut}.
\end{proof}

\subsection{Coercivity results for energy estimate bulk terms in $\deux$}\label{appendix:bulkpos}
\begin{prop}\label{prop:bulkpos}
	Let ${U}\in\fraks_k(\mathbb{C})$, $k\in\{0,1,2\}$. We assume that the estimates \eqref{eq:hypLhat} hold. Let $\mathbf{B}^{(N)}[U]$ be defined as in Proposition \ref{prop:calculcommunenergy}. Then we have, provided that $N=N(a,M)$ and $C_R=C_R(a,M)$ are large enough, in $\deux$,
	\begin{align*}
		\Omega^2\left|{{\nabla_3}{U}}\right|^2+\left|{\nabla_4{U}}\right|^2+|{{\nabla}U}|^2\lesssim \mathbf{B}^{(N)}[{U}]+\Deltahat|U|^2+\Deltahat^{-1}\ubar^{-4-2\delta/5}|\nabla_4^{\leq 1}U|^2.
	\end{align*}
	Moreover, if $k=2$ and $\widehat{\mcl}=\mcl$ is the Teukolsky operator \eqref{eq:teukop}, then we have for $\wbar_2\geq\wbar_1\gtrsim 1$ large enough,
	\begin{align*}
		&\int_{\deux\cap\{\wbar_1\leq\wbar\leq\wbar_2\}}\left(\Omega^2\left|{{\nabla_3}{U}}\right|^2+\left|{\nabla_4{U}}\right|^2+|{{\nabla}U}|^2\right)\lesssim\\ &\int_{\deux\cap\{\wbar_1\leq\wbar\leq\wbar_2\}}\left(\mathbf{B}^{(N)}[{U}]+\Deltahat^{-1}\ubar^{-4-2\delta/5}|U|^2+\Deltahat^{-1}\ubar^{-8-4\delta/5}|\nabla_4U|^2\right)+\int_{\Sigma_0\cap\{\wbar_1\leq\wbar\leq\wbar_2\}}|U|^2.
	\end{align*}
\end{prop}
\begin{rem}
	Note that in the bounds above we consider a fixed large enough value of $N$ which depends only on the black hole parameters $(a,M)$, so that in the end the implicit constants do not depend on $N$.
\end{rem}
\begin{proof}[Proof of Proposition \ref{prop:bulkpos}]
	We have $\mathbf{B}^{(N)}[{U}]=\mathbf{B}^{(N)}_{pr}[U]+(\mathbf{B}^{(N)}[{U}]-\mathbf{B}^{(N)}_{pr}[U])$ where, from Proposition \ref{prop:calculcommunenergy} the principal bulk term is
	\begin{align*}
		\mathbf{B}_{pr}^{(N)}[U]=&\Big(\frac{1}{2}{\Deltahat \varpi^N}(2\omega-tr\chi)-\frac{1}{2}e_4(\Deltahat \varpi^N)+\poids\Real(h)\Big)\left|{{\nabla_3}{{U}}}\right|^2\\
		&+\left(\Deltahat\varpi^N\left(-\frac{1}{2}tr\chibar+\Real(\underline{h})\right)-\frac{1}{2}e_3(\Deltahat \varpi^N)\right)\left|{{\nabla_4}{{U}}}\right|^2\\
		&-\frac{1}{2}\left(e_4(\poids)+e_3(\poids)+\poids (tr\chi -2\omega)+\poids tr\chibar\right)|\nabla{U}|^2.
	\end{align*}
	By \eqref{eq:weightder} we have 
	\begin{align*}
		&e_4(\poids)=e_4(\Deltahat)\varpi^N+N\Deltahat\varpi^{N-1}e_4(\varpi)\sim -\Deltahat\varpi^N(1+N\Deltahat),\\
		&e_3(\poids)=e_3(\Deltahat)\varpi^N+N\Deltahat\varpi^{N-1}e_3(\varpi)\sim -\varpi^N(1+N\Deltahat).
	\end{align*}
	Combining these estimates with
	$$\Deltahat\sim\Omega^2,\quad\omega\gtrsim 1,\quad |tr\chi|\lesssim\Omega^2+\ubar^{-2-\delta/5},\quad |tr\chibar|\lesssim 1+\Omega^{-2}\ubar^{-2-\delta/5}$$
	which hold in $\deux$ by \eqref{eq:boundtrxbarchibarhatin}, \eqref{eq:boundtrxchihatin} and \eqref{eq:hypLhat}, we obtain for $N(a,M)\gg 1$,
	\begin{align}\label{eq:bulkpr}
		\mathbf{B}^{(N)}_{pr}[U]\gtrsim&\:\Omega^2\varpi^N\left(1+N\Omega^2\right)\left|{{\nabla_3}{U}}\right|^2+\varpi^N\Big(1+N\Omega^2\Big)\left|{\nabla_4{U}}\right|^2+\varpi^N\Big(1+N\Omega^2\Big)|{{\nabla}U}|^2.
	\end{align}
	Now we bound $\mathbf{B}^{(N)}[{U}]-\mathbf{B}^{(N)}_{pr}[U]$, and we prove that this term can be absorbed in \eqref{eq:bulkpr} for $N,C_R$ sufficiently large. Note that by \eqref{eq:weightder} we have
	\begin{align}
		\nabla_a(\Deltahat\varpi^N)=\Deltahat N\varpi^{N-1}\nabla_a\varpi+\varpi^N\nabla_a\Deltahat=O(N\varpi^N\Omega^4)+O(\varpi^N\Omega^2)
		.\label{eq:relouluimaiscava}
	\end{align} 
	We deduce
	\begin{align*}
		\left|\nabla^a(\Deltahat\varpi^N)\cdot\Real(\overline{{\nabla_3}{U}}\cdot{\nabla}_a{U})\right|&\lesssim \Omega^2\varpi^N\left(1+N\Omega^2\right)\left(s|\nabla_3U|^2+s^{-1}|{\nabla}U|^2\right),
	\end{align*}
	for any $s>0$, thus we can absorb this term in \eqref{eq:bulkpr} by taking $s(a,M)>0$ small and then $C_R(a,M)$ large. We now use the following bounds: by Proposition \ref{prop:commnabla}, 
	\begin{align*}
		|\poids\Real(\overline{[\nabla_3,{\nabla}_a]U}\cdot\nabla^aU)|&\lesssim  \poids\left(|(tr\Xbar,\wh{\Xbar},\Bbar){\nabla}^{\leq 1}U|+|\nabla_3U|\right)|\nabla U|\\
		&\lesssim \Omega^2\varpi^N \left(|\nabla U|^2+|U|^2\right)+\varpi^N\ubar^{-2-\delta/5}|\nabla U|^2+\varpi^N\ubar^{-4-2\delta/5}|U|^2\\
		&\quad\quad + s\Omega^2\varpi^N |\nabla_3 U|^2+s^{-1}\Omega^2\varpi^N |\nabla U|^2,\\
		|\poids\Real(\underline{h}\overline{\nabla_3 U}\cdot\nabla_4 U)|&\lesssim\poids (1+\Omega^{-2}\ubar^{-2-\delta/5})|\nabla_3U||\nabla_4U|\\
		&\lesssim s\Omega^2\varpi^N |\nabla_3 U|^2+s^{-1}\Omega^2\varpi^N |\nabla_4 U|^2+s^{-1}\Omega^{-2}\varpi^N\ubar^{-4-2\delta/5}|\nabla_4U|^2.
	\end{align*}
	Now, in the case $\widehat{\mcl}=\mcl$, we have $\hhbar=tr\Xbar/2$, see \eqref{eq:teukop}. Thus by Proposition \ref{prop:conseqraych} we have the more precise bound
	\begin{align}
		|\poids\Real(tr\underline{X}\overline{\nabla_3 U}\cdot\nabla_4 U)|&\lesssim\poids (1+\Omega^{-2}\ubar^{-4-2\delta/5})|\nabla_3U||\nabla_4U|\nn\\
		&\lesssim s\Omega^2\varpi^N |\nabla_3 U|^2+s^{-1}\Omega^2\varpi^N |\nabla_4 U|^2+s^{-1}\Omega^{-2}\ubar^{-8-4\delta/5}|\nabla_4U|^2.\label{eq:boundmclhategalmcl}
	\end{align}
	We continue for now in the general case with the bounds
	\begin{align*}
		|\poids \Real(\overline{\nabla_3 U}\cdot L_1[U])|&\lesssim s\Omega^2\varpi^N |\nabla_3 U|^2+s^{-1}\Omega^2\varpi^N |\nabla U|^2,\\
		|\poids\Real(\overline{\nabla_3 U}\cdot L[U])|&\lesssim \poids |\nabla_3 U| (1+\Omega^{-2}\ubar^{-2-\delta/5})|U|\\
		&\lesssim s\Omega^2\varpi^N |\nabla_3 U|^2+\varpi^Ns^{-1}\Omega^2|U|^2+2^Ns^{-1}\Omega^{-2}\ubar^{-4-2\delta/5}|U|^2,\\
		|\poids(\eta+\etabar)\cdot\nu^{(3)}[U]|&\lesssim\poids |\nabla U||\nabla_3 U|\\
		&\lesssim s\Omega^2\varpi^N |\nabla_3 U|^2+s^{-1}\Omega^2\varpi^N |\nabla U|^2.
	\end{align*}
	Next, we have
	\begin{align*}
		\left|\nabla^a(\Deltahat\varpi^N)\cdot\Real(\overline{{\nabla_4}{U}}\cdot{\nabla}_a{U})\right|&\lesssim \Omega^2\varpi^N\left(1+N\Omega^2\right)\left(|\nabla_4U|^2+|{\nabla}U|^2\right),\\
		|\poids\Real(\overline{[\nabla_4,{\nabla}_a]U}\cdot\nabla^aU)|&\lesssim  \poids\left(|(trX,\wh{X},B){\nabla}^{\leq 1}U|+|\nabla_4U|\right)|\nabla U|\\
		&\lesssim \Omega^2\varpi^N \left(|\nabla U|^2+|U|^2+|\nabla_4 U|^2\right),\\
		|\poids\Real((h+2\omega)\overline{\nabla_4 U}\cdot\nabla_3 U)|&\lesssim\Omega^2\varpi^N(s|\nabla_3 U|^2+s^{-1}|\nabla_4 U|^2),
	\end{align*}
	\begin{align*}
		|\poids \Real(\overline{\nabla_4 U}\cdot (L_1[U]+2(\eta-\etabar)\cdot\nabla U))|&\lesssim \Omega^2\varpi^N |\nabla_4 U|^2+\Omega^2\varpi^N |\nabla U|^2,\\
		|\poids\Real(\overline{\nabla_4 U}\cdot (L[U]-C_k[U])|&\lesssim \poids |\nabla_4 U| (1+\Omega^{-2}\ubar^{-2-\delta/5})|U|\\
		&\lesssim \Omega^2\varpi^N |\nabla_4 U|^2+\varpi^N\Omega^2|U|^2+2^N\Omega^{-2}\ubar^{-4-2\delta/5}|U|^2,\\
		|\poids(\eta+\etabar)\cdot\nu^{(4)}[U]|&\lesssim\poids |\nabla U||\nabla_4 U|\\
		&\lesssim \Omega^2\varpi^N |\nabla_4 U|^2+\Omega^2\varpi^N |\nabla U|^2.
	\end{align*}
	Together with the expression of $\mathbf{B}^{(N)}[{U}]-\mathbf{B}^{(N)}_{pr}[U]$ in Proposition \ref{prop:calculcommunenergy}, the bounds above prove that for $C_R(a,M),N(a,M)\gg 1$, and choosing $s(a,M)>0$ small enough such that all the terms in the RHS of the bounds above are absorbed in \eqref{eq:bulkpr} except the terms $\varpi^N\Omega^2 |U|^2$, $\varpi^N\ubar^{-4-2\delta/5}|U|^2$, $\Omega^{-2}\ubar^{-4-2\delta/5}|\nabla_4 U|^2$ (replaced with $\Omega^{-2}\ubar^{-8-4\delta/5}|\nabla_4 U|^2$ in the case $k=2$, $\widehat{\mcl}=\mcl$), $\Omega^{-2}\ubar^{-4-2\delta/5}|U|^2$, we have
	\begin{align*}
		\mathbf{B}^{(N)}[U]\gtrsim &\Omega^2\varpi^N\left(1+N\Omega^2\right)\left|{{\nabla_3}{U}}\right|^2+\varpi^N\Big(1+N\Omega^2\Big)\left|{\nabla_4{U}}\right|^2+\varpi^N\Big(1+N\Omega^2\Big)|{{\nabla}U}|^2\nn\\
		&-\varpi^N\Omega^2 |U|^2-\varpi^N\Omega^{-2}\ubar^{-4-2\delta/5}|\nabla_4 U|^2-\varpi^N\Omega^{-2}\ubar^{-4-2\delta/5}|U|^2,
	\end{align*}
	which concludes the first part of the proposition, fixing $N=N(a,M)\gg 1$ sufficiently large. Now that $N$ is fixed, if $k=2$ and $\widehat{\mcl}=\mcl$, then integrating the estimate above with $\Omega^{-2}\ubar^{-4-2\delta/5}|\nabla_4 U|^2$ replaced with $\Omega^{-2}\ubar^{-8-4\delta/5}|\nabla_4 U|^2$ (see \eqref{eq:boundmclhategalmcl}) on $\deux[\wbar_1,\wbar_2]=\deux\cap\{\wbar_1\leq\wbar\leq\wbar_2\}$, we get
	\begin{align}
		&\int_{\deux[\wbar_1,\wbar_2]}\left(\Omega^2\left|{{\nabla_3}{U}}\right|^2+\left|{\nabla_4{U}}\right|^2+|{{\nabla}U}|^2\right)\label{eq:onabsorbelhs}\\
		&\lesssim \int_{\deux[\wbar_1,\wbar_2]}\left(\mathbf{B}^{(N)}[{U}]+\Deltahat^{-1}\ubar^{-4-2\delta/5}|U|^2+\Deltahat^{-1}\ubar^{-8-4\delta/5}|\nabla_4U|^2\right)+\int_{\deux[\wbar_1,\wbar_2]}\Omega^2|U|^2.\nn
	\end{align}
	Moreover, by the Hardy-type inequality in Proposition \ref{prop:hardy}, we deduce the bound
	\begin{align*}
		\int_{\deux[\wbar_1,\wbar_2]}\Omega^2|U|^2&\lesssim \int_{\wbar_1}^{\wbar_2}\left(\int_{\deux\cap\{\wbar=\wbar'\}}\Omega^2|U|^2\right)\dee\wbar'\\
		&\lesssim \int_{\deux[\wbar_1,\wbar_2]}\Omega^4\left(|\nabla_3 U|^2+|\nabla_4 U|^2+|\nabla U|^2\right)+\int_{\Sigma_0\cap\{\wbar_1\leq\wbar\leq\wbar_2\}}|U|^2.
	\end{align*}
	We conclude the proof of the case $k=2,\hat{\mcl}=\mcl$ by taking $C_R(a,M)\gg 1$ large enough so that the first term on the RHS above is absorbed in the LHS of \eqref{eq:onabsorbelhs}.
\end{proof}

\subsection{Elliptic and Sobolev estimates reformulated with non-integrable frames}

Recall that we denote $(e_\mu)$ the ingoing non-integrable frame and $(\ering_\mu)$ the double null frame in $\deux$. In what follows, we denote $(\ering_a)_{a=1,2}$ any local orthonormal basis of $TS(u,\ubar)$, end $(e_a)_{a=1,2}$ the corresponding horizontal frame given by \eqref{eq:frametransfo} with coefficients $f,\fbar,\hat{\lambda}$.
\begin{lem}\label{prop:expreinv}
	We have for $a=1,2$:
	\begin{align}\label{eq:inversetransfo}
		\ering_a=\left(\delta_{ab}+\frac{1}{2}f_a\fbar^b\right)e_b-\frac{1}{2}\hat{\lambda} f_ae_3-\frac{1}{2}\left(\fbar_a+\frac{1}{4}|\fbar|^2f_a\right)\hat{\lambda}^{-1}e_4.
	\end{align}

\end{lem}
\begin{proof}
	From \eqref{eq:frametransfo} with coefficients $f,\fbar,\hat{\lambda}$ we easily compute $\g(\ering_a,e_b)$, $\g(\ering_a,e_3)$, $\g(\ering_a,e_4)$, which concludes the proof.
\end{proof}

Let $U\in\fraks_k(\C)$ be a horizontal tensor in $\deux$, $k=0,1,2$. Notice that we have schematically by Proposition \ref{prop:diffchristo} and Lemma \ref{prop:expreinv}, omitting the indices $b_1\cdots b_k=1,2$,
\begin{align}\label{eq:nabnabS}
	\nabring_{a}{U}^S=\nabla_a{U}+\frac{1}{2}f_a\fbar\cdot\nabla{U}-\frac{1}{2}\hat{\lambda} f_a\nabla_3{U}-\frac{1}{2}\left(\fbar_a+\frac{1}{4}|\fbar|^2f_a\right)\hat{\lambda}^{-1}\nabla_4{U}+({\mct}\cdot{U})_a.
\end{align}
Considering the cases $k=0,1,2$ yields the following:
\\\\
\noindent\underline{If ${U}=\psi+i\hodge{\psi}\in\mathfrak{s}_2(\C)$ :} $\mathring{\divc}\:{U}^S=2(\mathring{\diver}\:\psi^S+i\hodge{(\mathring{\diver}\:\psi^S)})$, where 
\begin{align*}
	\mathring{\diver}\:\psi^S_b&=\nabring_a\psi^S_{ab}\\
	&=\diver\psi_{b}+\frac{1}{2}f_a\fbar\cdot\nabla\psi_{a b}-\frac{1}{2}\hat{\lambda} f_a\nabla_3\psi_{a b}-\frac{1}{2}\left(\fbar_a+\frac{1}{4}|\fbar|^2f_a\right)\hat{\lambda}^{-1}\nabla_4\psi_{a b}+({\mct}\cdot\psi)_b.
\end{align*}

\noindent\underline{If ${U}=\psi+i\hodge{\psi}\in\mathfrak{s}_1(\C)$ :} $\mathring{\divc}\:{U}^S=\mathring{\diver}\:\psi^S+i\hodge{(\mathring{\curl}\:\psi^S)}$, where 
\begin{align*}
	\mathring{\diver}\:\psi^S&=\nabring_a\psi^S_{a}=\diver\psi+\frac{1}{2}f_a\fbar\cdot\nabla\psi_{a}-\frac{1}{2}\hat{\lambda} f_a\nabla_3\psi_{a}-\frac{1}{2}\left(\fbar_a+\frac{1}{4}|\fbar|^2f_a\right)\hat{\lambda}^{-1}\nabla_4\psi_{a}+({\mct}\cdot\psi),
\end{align*}
similarly as in the case $k=2$, and
\begin{align*}
	\mathring{\curl}\:\psi^S&=\in_{ab}\nabring_a\psi^S_{b}=\curl\psi+\frac{1}{2}\hodge{f}_b\fbar\cdot\nabla\psi_{b}-\frac{1}{2}\hat{\lambda} \hodge{f}_b\nabla_3\psi_{b}-\frac{1}{2}\left(\hodge{\fbar}_b+\frac{1}{4}|\fbar|^2\hodge{f}_b\right)\hat{\lambda}^{-1}\nabla_4\psi_{b}+({\mct}\cdot\psi).
\end{align*}

\noindent\underline{If ${U}=\psi+i\psi_*$ is a function :} $\mathring{\mcd}U=\nabring\psi-\hodge{\:\nabring}\psi_*+i\hodge{(\nabring\psi-\hodge{\:\nabring}\psi_*)}$, where
\begin{align*}
	&\nabring_a\psi-\in_{ab}\nabring_b\psi_*\\
	&=\nabla_a\psi+\frac{1}{2}f_a\fbar\cdot\nabla\psi-\frac{1}{2}\hat{\lambda} f_a\nabla_3\psi-\frac{1}{2}\left(\fbar_a+\frac{1}{4}|\fbar|^2f_a\right)\hat{\lambda}^{-1}\nabla_4\psi+({\mct}\cdot\psi)_a\\
	&\quad\quad-\in_{ab}\nabla_b\psi_*-\frac{1}{2}\in_{ab}f_b\fbar\cdot\nabla\psi_*+\frac{1}{2}\hat{\lambda} \in_{ab}f_b\nabla_3\psi_*+\frac{1}{2}\in_{ab}\left(\fbar_b+\frac{1}{4}|\fbar|^2f_b\right)\hat{\lambda}^{-1}\nabla_4\psi_*.
\end{align*}
In any case, using the estimates $|f|\lesssim 1$, $|\fbar|\lesssim\Omega^2$, $\hat{\lambda}\sim\Omega^2$ (see \eqref{eq:bornecoeffin}) we infer for $k=0,1,2$,
\begin{align}\label{eq:divcsdiv}
	|\mathring{\mcd}_k{U}^S|\lesssim |\mcd_k{U}|+\Omega^2|\nabla_3{U}|+|\nabla_4{U}|+\Omega^2|\nabla{U}|+|{U}|,
\end{align}
where in both non-integrable and double null frames,
\begin{equation}\label{eq:defiopellip}
	\begin{gathered}
		\mcd_0=\mcd,\quad \mcd_1=\divc,\quad \mcd_2=\divc,\\
		\mathring{\mcd}_0=\mathring\mcd,\quad \mathring{\mcd}_1=\mathring{\overline{\mcd}}\cdot,\quad \mathring{\mcd}_2=\mathring{\overline{\mcd}}\cdot.
	\end{gathered}
\end{equation}
\begin{rem}\label{rem:ellipticsuuubarr}
	Let $\phi$ be a $S(u,\ubar)$-tangent tensor in $\deux$ which is either a scalar, a 1-form, or a symmetric traceless 2-tensor. Then, as a consequence of the boundedness of the Gauss curvature of the double null spheres $S(u,\ubar)$ (see \eqref{eq:onlyused}), the identities in Proposition \ref{prop:ellipticestimates} can be rewritten in the following complexified way: for any $S(u,\ubar)\subset\deux$,
	$$\intS|\nabring \phi|^2\lesssim\intS |\mathring{\mcd}_k^{\leq 1} \phi|^2.$$
\end{rem}
We can now prove a preliminary result which should be interpreted as the horizontal projection of the above elliptic estimates on $S(u,\ubar)$.
\begin{lem}\label{lem:619}
	Let ${U}\in\fraks_k(\C)$ in $\deux$, $k=0,1,2$. We have for any $S(u,\ubar)\subset\deux$,
	$$\intS|\nabla{U}|^2\lesssim\intS\left(|{\mcd}_k{U}|^2+\Omega^4|\nabla_3{U}|^2+|\nabla_4{U}|^2+|{U}|^2\right),$$
	where ${\mcd}_k$ is defined in \eqref{eq:defiopellip}, provided $C_R(a,M)\gg 1$ is large enough.
\end{lem}
\begin{proof}
	By \eqref{eq:nabnabS} and the bounds $|f|\lesssim 1$, $|\fbar|\lesssim\Omega^2$, $\hat{\lambda}\sim\Omega^2$, $|{\mct}|\lesssim 1$ (see \eqref{eq:voilauneborne}, \eqref{eq:voilabornefbarcheckkk}),
	\begin{align*}
		\intS|\nabla{U}|^2&\lesssim \intS\left(|\nabring{U}^S|^2+\Omega^4|\nabla{U}|^2+\Omega^4|\nabla_3{U}|^2+|\nabla_4{U}|^2+|{U}|^2\right).
	\end{align*}
	Using the elliptic estimate in Remark \ref{rem:ellipticsuuubarr} with $\phi=U^S$ we get 
	\begin{align*}
		\intS|\nabla{U}|^2&\lesssim \intS\left(|\mathring{\mcd}_k{U}^S|^2+\Omega^4|\nabla{U}|^2+\Omega^4|\nabla_3{U}|^2+|\nabla_4{U}|^2+|{U}|^2\right)\\
		&\lesssim \intS\left(|\mcd_k{U}|^2+\Omega^4|\nabla{U}|^2+\Omega^4|\nabla_3{U}|^2+|\nabla_4{U}|^2+|{U}|^2\right),
	\end{align*}
	where we used \eqref{eq:divcsdiv} for the last step. Choosing $C_R(a,M)\gg 1$ such that the term $\intS\Omega^4|\nabla{U}|^2$ on the RHS above is absorbed in the LHS, we conclude the proof.
\end{proof}

The following result is the second order equivalent of formula \eqref{eq:nabnabS}.
\begin{prop}\label{prop:doublediffchristo}
	Let $U$ be a horizontal $k$-tensor. We have, in $\deux$,
	\begin{align*}
		\nabring_{a}&\nabring_{b}U^S_{c_1\cdots c_k}=\\
		&\nabla_{\ering_a}\nabla_{b}U_{c_1\cdots c_k}+\frac{1}{2}(\fbar_c\nabring_af_b+f_b\nabring_a\fbar_c)\nabla_{c}U_{c_1\cdots c_k}+\frac{1}{2}f_b\fbar_c\nabla_{\ering_a}\nabla_{c}U_{c_1\cdots c_k}\\
		&-\frac{1}{2}f_b\fbar_c\left(\sum_{i=1}^k{\mct}_{ac_id}\nabla_{c}U_{c_1\cdots d\cdots c_k}+{\mct}_{acd}\nabla_{d}U_{c_1\cdots c_k}\right)+\sum_{i=1}^k\nabring_{a}{\mct}_{bc_ic}U_{c_1\cdots  c\cdots c_k}\\
		&+\sum_{i=1}^k{\mct}_{bc_ic}\left(\nabla_{\ering_a}U_{c_1\cdots  c\cdots c_k}-\sum_{j=1}^k{\mct}_{ac_jd}U_{c_1\cdots c\cdots d\cdots c_k}\right)-\sum_{i=1}^k{\mct}_{ac_ic}\nabla_{b}U_{c_1\cdots  c\cdots c_k}\\
		&-{\mct}_{abc}\nabla_{c}U_{c_1\cdots c_k}+\frac{1}{2}(\nabring_a\hat{\lambda}f_b+\hat{\lambda}\nabring_af_b)\nabla_{3}U_{c_1\cdots c_k}+\frac{1}{2}\hat{\lambda}f_b\nabla_{\ering_a}\nabla_{3}U_{c_1\cdots c_k}\\
		&-\frac{1}{2}\hat{\lambda} f_b\sum_{i=1}^k{\mct}_{ac_ic}\nabla_{3}U_{c_1\cdots c\cdots c_k}+\frac{1}{2}\Bigg[\nabring_a\hat{\lambda}^{-1}\left(\fbar_b+\frac{1}{4}|\fbar|^2f_b\right)\\
		&+\hat{\lambda}^{-1}\left(\nabring_a\fbar_b+\frac{1}{4}\nabring_{a}(|\fbar|^2f)_b\right)\Bigg]\nabla_{4}U_{c_1\cdots c_k}+\frac{1}{2}\hat{\lambda}^{-1}\left(\fbar_b+\frac{1}{4}|\fbar|^2f_b\right)\nabla_{\ering_a}\nabla_{4}U_{c_1\cdots c_k}\\
		&-\frac{1}{2}\hat{\lambda}^{-1}\left(\fbar_b+\frac{1}{4}|\fbar|^2f_b\right)\sum_{i=1}^k{\mct}_{ac_ic}\nabla_{4}U_{c_1\cdots c\cdots c_k}.
	\end{align*}
\end{prop}
\begin{proof}
	This is a simple computation which relies on a double application of Proposition \ref{prop:diffchristo}: first for $U$, and then for the horizontal $(k+1)$-tensor $V=\nabla U$, we omit the details. 
\end{proof}

We now turn to re-expressing the $L^\infty-L^2$ spherical Sobolev embedding of Proposition \ref{prop:sobolevdoublenullII} with respect to the non-integrable frame.

\begin{prop}\label{prop:sovolinfty}
	Let ${U}\in\fraks_k(\C)$, $k\in\{0,1,2\}$ . Recalling \eqref{eq:defiopellip}, we have for $S(u,\ubar)\subset\deux$,
	\begin{align*}
		\|{U}\|_{L^\infty(S(u,\ubar))}^2\lesssim\intS\Big(&|\triangle_k{U}|^2+|\mcd_k{U}|^2+|{U}|^2+|\hat{\lambda}\nabla_3\mcd_k{U}|^2+|\nabla_4\mcd_k{U}|^2\\
		&+|\hat{\lambda}\nabla_3{U}|^2+|\nabla_4{U}|^2+|(\hat{\lambda}\nabla_3)^2{U}|^2+|\Omega^2\nabla_4\nabla_3{U}|^2+|\nabla_4^2{U}|^2\Big),
	\end{align*}
provided $C_R(a,M)\gg 1$ is large enough.
\end{prop}
\begin{proof}
	Combining Proposition \ref{prop:comparisisom} and Corollary \ref{cor:sobolaplace} we get the bound 
	\begin{align}\label{eq:nontoipremiere}
		\|{U}\|_{L^\infty(S(u,\ubar))}=\|{U}^S\|_{L^\infty(S(u,\ubar))}\lesssim\Ldeux{
			\mathring{\triangle}^{\leq 1}{U}^S}. 
	\end{align}
	Moreover, by Lemma \ref{prop:expreinv} which yields
	\begin{align*}
	\nabla_{\ering_a}\nabla_{a}{U}=\triangle_k{U}+\frac{1}{2}f_a\fbar_b\nabla_{b}\nabla_{a}{U}-\frac{1}{2}\hat{\lambda} f_a\nabla_3\nabla_{a}{U}-\frac{1}{2}\hat{\lambda}^{-1}\left(\fbar_a+\frac{1}{4}|\fbar|^2f_a\right)\nabla_4\nabla_{a}{U},
\end{align*}
	 and by Proposition \ref{prop:doublediffchristo} with $a=b$ and the bounds on $f,\fbar,\hat{\lambda}$ in \eqref{eq:bornecoeffin}, we get
	\begin{align}
		|\mathring{\triangle}{U}^S|^2=|\nabring^{a}\nabring_{a}{U}^S|^2\lesssim&|\triangle_k{U}|^2+\Omega^4|\nabla^2 U|^2+|\nabla^{\leq 1}{U}|^2+|\hat{\lambda}\nabla_3{U}|^2+|\nabla_4{U}|^2+|\nabla(\hat{\lambda}\nabla_3{U})|^2\nn\\
		&+|\nabla\nabla_4{U}|^2+|(\hat{\lambda}\nabla_3)^2{U})|^2+|\nabla_4^2{U}|^2+\hat{\lambda}|\nabla_3\nabla_4{U}|^2+|\nabla_4(\hat{\lambda}\nabla_3{U})|^2\nn\\
		&+|\hat{\lambda}\nabla_3\nabla{U}|+|\nabla_4\nabla{U}|.\label{eq:premiereee}
	\end{align}
	Moreover, using Proposition \ref{prop:doublediffchristo} again we infer, using again $|\fbar|\lesssim\Omega^2$, $|f|\lesssim 1$,
	\begin{align*}
		|\nabla^2{U}|^2\lesssim\sum_{a,b=1,2}|\nabla_a\nabla_b U|^2\lesssim&|\nabring^2{U}^S|^2+\Omega^4|\nabla^2U|^2+|\nabla^{\leq 1}{U}|^2+|\hat{\lambda}\nabla_3{U}|^2+|\nabla_4{U}|^2\nn\\
		&+|\nabla(\hat{\lambda}\nabla_3{U})|^2+|\nabla\nabla_4{U}|^2+|(\hat{\lambda}\nabla_3)^2{U})|^2+|\nabla_4^2{U}|^2\nn\\
		&+\hat{\lambda}|\nabla_3\nabla_4{U}|^2+|\nabla_4(\hat{\lambda}\nabla_3{U})|^2+|\hat{\lambda}\nabla_3\nabla{U}|^2+|\nabla_4\nabla{U}|^2.
	\end{align*}
	Now, using the elliptic estimate \eqref{eq:thebound222} for $\phi=U^S$, combined with \eqref{eq:premiereee}, we deduce
	\begin{align*}
		\intS|\nabla^2{U}|^2\lesssim&\intS\Bigg(|\triangle_k{U}|^2+\Omega^4|\nabla^2U|^2+|\nabla^{\leq 1}{U}|^2+|\hat{\lambda}\nabla_3{U}|^2+|\nabla_4{U}|^2+|\nabla(\hat{\lambda}\nabla_3{U})|^2\nn\\
		&+|\nabla\nabla_4{U}|^2+|(\hat{\lambda}\nabla_3)^2{U})|^2+|\nabla_4^2{U}|^2+\hat{\lambda}|\nabla_3\nabla_4{U}|^2+|\nabla_4(\hat{\lambda}\nabla_3{U})|^2\nn\\
		&+|\hat{\lambda}\nabla_3\nabla{U}|^2+|\nabla_4\nabla{U}|^2\Bigg).
	\end{align*}
	Thus, choosing $C_R(a,M)\gg 1$ large enough such that the term $\Omega^4|\nabla^2U|^2$ on the RHS above is absorbed in the LHS and going back to \eqref{eq:premiereee} we deduce
	\begin{align*}
		\intS|\mathring{\triangle}{U}^S|^2\lesssim&\intS\Bigg(|\triangle_k{U}|^2+|\nabla^{\leq 1}{U}|^2+|\hat{\lambda}\nabla_3{U}|^2+|\nabla_4{U}|^2+|\nabla(\hat{\lambda}\nabla_3{U})|^2\nn\\
		&+|\nabla\nabla_4{U}|^2+|(\hat{\lambda}\nabla_3)^2{U})|^2+|\nabla_4^2{U}|^2+\hat{\lambda}|\nabla_3\nabla_4{U}|^2+|\nabla_4(\hat{\lambda}\nabla_3{U})|^2\nn\\
		&+|\hat{\lambda}\nabla_3\nabla{U}|^2+|\nabla_4\nabla{U}|^2\Bigg).
	\end{align*}
	We now commute $\nabla_3,\nabla_4$ with $\nabla$ to deal with the last terms on the RHS above by Lemma \ref{lem:commutnablageneral}, and use Lemma \ref{lem:619} and commute $\mcd_k$ with $\hat{\lambda}\nabla_3$, $\nabla_4$, and $\hat{\lambda}\nabla_3$ and $\nabla_4$, which yields\footnote{We note that by $|\df\hat{\lambda}|\lesssim\Omega^2\sim\hat{\lambda}$ and Lemmas \ref{lem:comm34} and Propositions \ref{prop:commnabladivc}, \ref{prop:commnab34mcd}, these commutators only produces lower-order terms with appropriate $\hat{\lambda}$ factors.}
	\begin{align*}
		\intS|\mathring{\triangle}{U}^S|^2\lesssim\intS\Big(&|\triangle_k{U}|^2+|\mcd_k{U}|^2+|{U}|^2+|\hat{\lambda}\nabla_3\mcd_k{U}|^2+|\nabla_4\mcd_k{U}|^2\\
		&+|\hat{\lambda}\nabla_3{U}|^2+|\nabla_4{U}|^2+|(\hat{\lambda}\nabla_3)^2{U}|^2+|\Omega^2\nabla_4\nabla_3{U}|^2+|\nabla_4^2{U}|^2\Big),
	\end{align*}
	which concludes the proof by \eqref{eq:nontoipremiere}.
\end{proof}

\section{Computation of Teukolsky operator applied to the ansatz}\label{section:teukansatzannex}
\subsection{In Kerr spacetime}

In this subsection, \textbf{where we omit the subscrit ${}_\mck$ for conciseness}, we prove the fact that the Teukolsky operator applied to the horizontal tensor $\Psi_m$ defined below vanishes in exact Kerr, which relies on computations which can be found in \cite{formalismkerr} and that establish a direct link between the classical Teukolsky equation written in the Newman-Penrose formalism, and the Teukolsky equation written in the tensorial form \eqref{eq:teukA} in Kerr.

For $|m|\leq 2$ and denoting $Y_{m,2}(\cos\theta)e^{im\phi_+}$ the standard spherical harmonics (with renormalized coordinate $\phi_+$ regular on $\mch_+$), we define the horizontal tensor $\Psi_m$ in Kerr by
\begin{align}\label{eq:defpsrrrrim}
	\Psi_m:=\frac{A_m(r)}{\qbar^2}\mcd\hot(\mcd(Y_{m,2}(\cos\theta)e^{im\phi_+})),
\end{align}
where we recall the definition \eqref{eq:dedfAm(r)} of $A_m(r)$. We also denote by $L_{[s]}$ the usual spin $s=\pm 2$ Teukolsky operator of the linearized setting \underline{obtained in the principal frame regular on $\mch_+$} (see \cite[Section 2.4]{spin+2}), which is 
\begin{align*}
	L_{[s]}\psi=&-\left[\frac{\left(r^2+a^2\right)^2}{\Delta}-a^2 \sin ^2 \theta\right] \partial_t^2\psi -\frac{4 M a r}{\Delta}\partial_t\partial_\phi\psi  -\left[\frac{a^2}{\Delta}-\frac{1}{\sin ^2 \theta}\right] \partial_\phi^2\psi \nonumber \\
	& +\Delta^{-s} \partial_r\left(\Delta^{s+1} \partial_r \psi\right)+\frac{1}{\sin \theta} \partial_\theta\left(\sin \theta \partial_\theta \psi\right)+2 s\left[\frac{a(r-M)}{\Delta}+\frac{i \cos \theta}{\sin ^2 \theta}\right] \partial_\phi\psi  \nonumber\\
	& +2 s\left[\frac{M\left(r^2-a^2\right)}{\Delta}-r-i a \cos \theta\right]\partial_t\psi -\left[\frac{s^2 \cos ^2 \theta}{\sin ^2 \theta}+s\right]\psi-4s(r-M)\partial_r\psi.
\end{align*}
We also denote
\begin{align*}
	L'_{[s]}\psi=&-\left[\frac{\left(r^2+a^2\right)^2}{\Delta}-a^2 \sin ^2 \theta\right] \partial_t^2\psi -\frac{4 M a r}{\Delta}\partial_t\partial_\phi\psi  -\left[\frac{a^2}{\Delta}-\frac{1}{\sin ^2 \theta}\right] \partial_\phi^2\psi \nonumber \\
	& +\Delta^{-s} \partial_r\left(\Delta^{s+1} \partial_r \psi\right)+\frac{1}{\sin \theta} \partial_\theta\left(\sin \theta \partial_\theta \psi\right)+2 s\left[\frac{a(r-M)}{\Delta}+\frac{i \cos \theta}{\sin ^2 \theta}\right] \partial_\phi\psi  \nonumber\\
	& +2 s\left[\frac{M\left(r^2-a^2\right)}{\Delta}-r-i a \cos \theta\right]\partial_t\psi -\left[\frac{s^2 \cos ^2 \theta}{\sin ^2 \theta}-s\right]\psi,
\end{align*}
which is the standard Teukolsky operator obtained in the frame regular at $\ch$. Then we have the following identity (see \cite[Section 2.4]{spin+2})
\begin{align}\label{eq:jedoisyaller}
	L'_{[+2]}=\Delta^{-2}L_{[+2]}(\Delta^{2}\cdot).
\end{align}
We also have the following expression for $L'_{[+2]}$, which coincides with (114) in \cite{formalismkerr} and ensures that we are in position to apply the computations done in that paper.
\begin{lem}\label{lem:exprecooool}
	We have for any $\psi$,
	\begin{align*}
		L'_{[+2]}\psi=&|q|^2\Box_{\g_{a,M}}\psi+4(r-M)\partial_r\psi+4\left(\frac{M(r^2-a^2)}{\Delta}-r-ia\cos\theta\right)\partial_t\psi\\
		&+4\left(\frac{a(r-M)}{\Delta}+i\frac{\cos\theta}{\sin^2\theta}\right)\partial_\phi\psi-(4\cot^2\theta-2)\psi.
	\end{align*}
\end{lem}
\begin{proof}
	By \cite[(1.22a)]{Ma:energymorawetz} we have 
	$$L_{[+2]}\psi=|q|^2\Box_{\g_{a,M}}\psi+{4i}\left(\frac{\cos\theta}{\sin^2\theta}\Z-a\cos\theta\T\right)\psi-(4\cot^2\theta+2)\psi+4((r-M)e_3-2r\T)\psi,$$
which concludes the proof by the identity $L'_{[+2]}=L_{[+2]}+4+8(r-M)\partial_r$.
\end{proof}
Recall the notation $Y_{m,2}^{+2}(\cos\theta)e^{im\phi_+}$ for the $\ell=2$ spin $+2$ spherical harmonics.
\begin{prop}\label{prop:dejaprovelol}
	We have for any $|m|\leq 2$, 
	$$L_{[+2]}\left(A_m(r)Y_{m,2}^{+2}(\cos\theta)e^{im\phi_+}\right)=0.$$
\end{prop}
\begin{proof}
	By \cite[Prop. 2.18]{spin+2} (where $\mathbf{T}_{+2}$ denotes $L_{[+2]}$, and where $e_3$ is rescaled by a factor $1/2$ from the $e_3$ defined in \eqref{eq:principalingoinkerr}) we have
	\begin{align*}
		L_{[+2]}&\left(A_m(r)Y_{m,2}^{+2}(\cos\theta)e^{im\phi_+}\right)=\\
		&\left(\Delta A_m''(r)+2(iam-(r-M))A_m'(r)-4A_m(r)\right)Y_{m,2}^{+2}(\cos\theta)e^{im\phi_+}=0
	\end{align*}
	since $A_m''(r)+2(iam-(r-M))A_m'(r)-4A_m(r)=0$, see Appendix B.2 in \cite{spin+2}.
\end{proof}
\begin{lem}\label{lem:DhotDYm2}
	Recalling \eqref{eq:horizontalframekerr} we have, for any $|m|\leq 2$, in Kerr,
	$$Y_{m,2}^{+2}(\cos\theta)e^{im\phi_+}=\frac{|q|^2}{4\sqrt{6}}(\mcd\hot(\mcd(Y_{m,2}(\cos\theta)e^{im\phi_+})))_{11},$$
	where $Y_{m,2}(\cos\theta)e^{im\phi}$ is the standard $(\ell=2,m)$ spherical harmonic on $\mathbb{S}^2$.
\end{lem}
\begin{proof}
	The classical relation between $Y_{m,\ell}^{s}(\cos\theta)e^{im\phi}$ and $Y_{m,\ell}=Y_{m,\ell}(\cos\theta)e^{im\phi}$ is
	$$Y_{m,\ell}^{s}(\cos\theta)e^{im\phi}=\sqrt{\frac{(\ell-s)!}{(\ell+s)!}}\drond^2Y_{m,\ell},$$
	where the spin-dependent spherical eth operator $\drond$ is defined by $\drond=\partial_\theta+\frac{i}{\sin\theta}\partial_\phi-s\cot\theta$. This yields in particular
	\begin{align}\label{eq:conbiannep1}
		2\sqrt{6}Y_{m,2}^{+2}(\cos\theta)e^{im\phi}=\left(\partial_\theta^2-\frac{1}{\sin^2\theta}\partial_\phi^2-\cot\theta\partial_\theta\right)Y_{m,2}+\frac{2i}{\sin\theta}\left(\partial_\theta-\cot\theta\right)\partial_\phi Y_{m,2}.
	\end{align}
Now, by the identity
\begin{align}\label{eq:conbiannep2}
	\mcd\hot\mcd Y_{m,2}=2(\nabla\hot\nabla Y_{m,2}+i\hodge{(\nabla\hot\nabla Y_{m,2})}),
\end{align}
we compute $|q|^2(\nabla\hot\nabla Y_{m,2})_{11}$ and $|q|^2(\nabla\hot\nabla Y_{m,2})_{12}$, where $(e_1,e_2)$ is the horizontal frame given in \eqref{eq:horizontalframekerr}. Denoting $\Lambda=(r^2+a^2)\cot\theta/|q|^3$ and recalling from \cite[(3.3.11)]{KSwaveeq} the identities
$$\nabla_1 e_1=\nabla_1 e_2=0,\quad\nabla_2 e_1=\Lambda e_2,\quad\nabla_2 e_2=-\Lambda e_1,$$
we get for any scalar $h(\theta,\phi_+)$ on $\mathbb{S}^2$,
\begin{align*}
	|q|^2(\nabla\hot\nabla h)_{11}&=|q|^2(\nabla_1\nabla_1 h-\nabla_2\nabla_2 h)=|q|^2\left(e_1(e_1(h))-e_2(e_2(h))+\g(\nabla_2 e_2,e_1)e_1(h)\right)\\
	&=|q|^2\left(\frac{1}{|q|}\partial_\theta\left(\frac{1}{|q|}\partial_\theta h\right)-\frac{1}{|q|\sin\theta}\partial_\phi\left(\frac{1}{|q|\sin\theta}\partial_\phi h\right)-\frac{r^2+a^2}{|q|^4}\cot\theta\partial_\theta h\right)\\
	&=\left(\partial_\theta^2-\frac{1}{\sin^2\theta}\partial_\phi^2-\cot\theta\partial_\theta\right) h-F\partial_\theta h,
\end{align*}
where
\begin{align*}
	F&=\frac{1}{|q|}\partial_\theta|q|+\left(\frac{r^2+a^2}{|q|^2}-1\right)\cot\theta=\frac{1}{2|q|^2}\partial_\theta|q|^2+\frac{r^2+a^2-r^2-a^2\cos^2\theta}{|q|^2}\cot\theta=0.
\end{align*}
Similarly, we compute 
\begin{align*}
	|q|^2(\nabla\hot\nabla Y_{m,2})_{12}&=|q|^2(\nabla_1\nabla_2 h+\nabla_2\nabla_1 h)=|q|^2\left(e_1(e_2(h))+e_2(e_1(h))-\g(\nabla_2 e_1,e_2)e_2(h)\right)\\
	&=|q|^2\left(\frac{1}{|q|}\partial_\theta\left(\frac{1}{|q|\sin\theta}\partial_\phi h\right)+\frac{1}{|q|\sin\theta}\partial_\phi\left(\frac{1}{|q|}\partial_\theta h\right)-\frac{r^2+a^2}{|q|^4}\cot\theta\frac{1}{\sin\theta}\partial_\phi h\right)\\
	&=\frac{2}{\sin\theta}\left(\partial_\theta-\cot\theta\right)\partial_\phi h-\frac{1}{\sin\theta}F\partial_\theta h=\frac{2}{\sin\theta}\left(\partial_\theta-\cot\theta\right)\partial_\phi h,
\end{align*}
which concludes the proof, combining \eqref{eq:conbiannep1} and \eqref{eq:conbiannep2}.
\end{proof}

\begin{prop}\label{prop:teukansatzkerr}
	Recalling the definition \eqref{eq:defpsrrrrim} of $\Psi_m$, we have in Kerr
	$$\mcl(\Psi_m)=0.$$
\end{prop}

\begin{proof}
	We will prove the equivalent identity $\mcl(\Psi_m)_{11}=0$, with the horizontal frame $(e_1,e_2)$ defined in \eqref{eq:horizontalframekerr}. We use the following computation which is the main result of \cite[Section A.3]{formalismkerr} : for any $U\in\fraks_2(\C)$, denoting ${\mcl}'$ the Teukolsky operator defined in \eqref{eq:teukop} in exact Kerr with respect to the \emph{outgoing} principal frame\footnote{Namely, replacing $e_3$ and $e_4$ by $e_3'$ and $e_4'$ in \eqref{eq:teukop}, and replacing appropriately all Ricci and curvature coefficients $(\Gamma,R)$ by the analog ones $(\Gamma',R')$. Note that the outgoing principal null pair $(e'_3,e'_4)$ used in \cite{formalismkerr} presents a factor -1 compared to \eqref{eq:principaloutgoinkerr}, which does not change the corresponding Teukolsky operator.} \eqref{eq:principaloutgoinkerr},
	\begin{align*}
		-\mcl'(U)_{11}=&\Box_{\g_{a,M}}(U_{11})+(4\omegabar'-i\atrchibar')e_4'(U_{11})+(2tr\chi'+3i\atrchi')e_3'(U_{11})+4i\etabar' e_1(U_{11})\\
		&+(8\etabar'-4i(\Lambda-\etabar'_1))e_2(U_{11})+VU_{11},
	\end{align*}
with $\Lambda=(r^2+a^2)\cot\theta/|q|^3$ and $V$ precisely computed in the end of \cite[Section A.3]{formalismkerr}. Moreover by Lemma \ref{lem:exprecooool} and the results of \cite[Sections A.1,A.2]{formalismkerr}, we also get for any spin $+2$ scalar $\psi$:
	\begin{align*}
	L'_{[+2]}\psi=-\qbar^2\Bigg(&\Box_{\g_{a,M}}\left(-\frac{q}{\qbar}\psi\right)+(4\omegabar'-i\atrchibar')e_4'\left(-\frac{q}{\qbar}\psi\right)+(2tr\chi'+3i\atrchi')e_3'\left(-\frac{q}{\qbar}\psi\right)\\
	&+4i\etabar' e_1\left(-\frac{q}{\qbar}\psi\right)+(8\etabar'-4i(\Lambda-\etabar'_1))e_2\left(-\frac{q}{\qbar}\psi\right)+V\left(-\frac{q}{\qbar}\psi\right)\Bigg).
\end{align*}
Also, as the principal outgoing and ingoing frames differ by a factor $-\Delta/|q|^2$, we have
$$\mcl=\frac{\Delta^2}{|q|^4}\mcl'\left(\frac{|q|^4}{\Delta^2}\cdot\right).$$
Indeed, this can be seen from the expression of the operator $-\mcl'$ with conformal derivatives at the top of \cite[Section A.3]{formalismkerr}. Thus, we can compute \begin{align*}
	\mcl(\Psi_m)_{11}&=\frac{\Delta^2}{|q|^4}\mcl'\left(\frac{|q|^4}{\Delta^2}\Psi_m\right)_{11}=\qbar^{-2}\frac{\Delta^2}{|q|^4}L'_{[+2]}\left(-\frac{\qbar}{q}\frac{|q|^4}{\Delta^2}(\Psi_m)_{11}\right)=-\frac{1}{\qbar^2|q|^4}L_{[+2]}\left(\frac{\qbar|q|^4}{q}(\Psi_m)_{11}\right),
\end{align*}
where we used \eqref{eq:jedoisyaller} in the last step. Reinjecting the definition \eqref{eq:defpsrrrrim} of $\Psi_m$ we finally get
\begin{align*}
	\mcl(\Psi_m)_{11}&=-\frac{1}{\qbar^2|q|^4}L_{[+2]}\left(A_m(r)|q|^2(\mcd\hot(\mcd(Y_{m,2})))_{11}\right)\\
	&=-\frac{4\sqrt{6}}{\qbar^2|q|^4}L_{[+2]}\left(A_m(r)Y_{m,2}^{+2}(\cos\theta)e^{im\phi_+}\right)=0,
\end{align*}
where we used Proposition \ref{prop:dejaprovelol} in the last step, which concludes the proof.
\end{proof}

\subsection{In a perturbation}\label{section:teukansatzperturbed}
We project the exact Kerr identity of Proposition \ref{prop:teukansatzkerr} to the perturbed Kerr black hole interior spacetime studied in this paper in regions $\un$ and $\deux$ as follows.

\noindent\underline{\textbf{Case of region $\un$.}} For a scalar $f$ in Kerr\footnote{This procedure could also be done for tensors in $\un$ but we do not need this here.}, we define the scalar $f_\mck$ in $\un$ by $f_\mck:=f\circ\mcf^{-1}$, where
$$\mathcal{F}:(r,\ubar,\theta,\phi_+)\in\un_\mck\longmapsto (r,\ubar,\theta,\phi_+)\in\un$$
is the identification of the coordinates $(r,\ubar,\theta,\phi_+)$ of regions $\un_\mck$ in Kerr and $\un$.

\begin{cor}\label{cor:teukansatzdansun}
	As a consequence of Proposition \ref{prop:teukansatzkerr}, we have in $\un$ for $|m|\leq 2$,
	$$|\mcl(\Psi_m)|^2_\mck(r,\ubar,\theta,\phi_+)=0.$$
\end{cor}
\noindent\underline{\textbf{Case of region $\deux$.}} Recall the isometry in Kerr
\begin{align*}
	&\Phi_\mck:X\in TS_\mck(u,\ubar)\longmapsto \Phi_\mck(X)\in \mch_\mck,\\
	&\Phi_\mck(X)=\dfrac{1}{2}\fbar_\mck(X)f_\mck^A\partial_{\theta^A}+\dfrac{1}{2}\fbar_\mck(X)(\ering_4)_\mck+\left(\dfrac{1}{2}f_\mck(X)+\dfrac{1}{8}|f_\mck|_\mck^2\fbar_\mck(X)\right)(\ering_3)_\mck
\end{align*}
between the Kerr double null spheres $TS_\mck(u,\ubar)$ and $\mch_\mck$, where the $S_\mck(u,\ubar)$-tangent 1-forms $f_\mck,\fbar_\mck$ are defined in Proposition \ref{prop:transfodanskerr}. Then, for any horizontal tensor $U_\mck$ in Kerr, we define the $S(u,\ubar)$-tangent tensor $U_\mck^S$ in $\deux$ by 
$$U_\mck^S:=\mathcal{G}_*((\Phi_\mck)^*U_\mck),$$
where
$$\mathcal{G}:(u,\ubar,\theta^A)\in\deux_\mck\longmapsto (u,\ubar,\theta^A)\in\deux$$
is the identification of the coordinates $(u,\ubar,\theta^A)$ of regions $\deux_\mck$ and $\deux$, and where $(\Phi_\mck)^*U_\mck$ is the pullback $U\circ\Phi_\mck$. In the paper, we drop the push-forward $\mathcal{G}_*$ for conciseness in the notations.
\begin{cor}\label{cor:teukansatzdansdeux}
	As a consequence of Proposition \ref{prop:teukansatzkerr}, we have in $\deux$ for $|m|\leq 2$,
	$$\mcl(\Psi_m)^S_\mck=0,$$
	or equivalently recalling \eqref{eq:teukop},
	\begin{align*}
		&\left(\nabla_4\nabla_3 \Psi_m\right)_\mck^S-\frac{1}{4}\left(\mcd\hot\parentheses{\overline{\mcd}\cdot \Psi_m}\right)_\mck^S+\parentheses{\frac{1}{2}tr X_\mck+2\overline{tr X}_\mck}\left(\nabla_3 \Psi_m\right)_\mck^S+\parentheses{\frac{1}{2}tr \underline{X}_\mck-4\omegabar_\mck} \left(\nabla_4\Psi_m\right)_\mck^S\\
		&-\left(\parentheses{2Z+\Hbar+\overline{2Z+\Hbar}+4H}\cdot\nabla \Psi_m\right)_\mck^S- \left(H\hot\parentheses{(\overline{2 Z+\underline{H}})\cdot \Psi_m}\right)_\mck^S+V_\mck\left(\Psi_m\right)_\mck^S=0.
	\end{align*}
\end{cor}

	\printbibliography
\end{document}